%% file: DMreview.tex
\newcommand{\beq}{\begin{equation}}
\newcommand{\eeq}{\end{equation}}
\newcommand{\bea}{\begin{eqnarray}}
\newcommand{\eea}{\end{eqnarray}}

\newcommand{\gsim}{\lower.7ex\hbox{$\;\stackrel{\textstyle>}{\sim}\;$}}
\newcommand{\lsim}{\lower.7ex\hbox{$\;\stackrel{\textstyle<}{\sim}\;$}}

\documentclass[12pt]{article}
\pdfoutput=1
\usepackage{geometry,amsmath,amsfonts}
\usepackage{slashed}
\usepackage{graphicx}
\usepackage{epstopdf}
\usepackage{mathrsfs}
\usepackage{amssymb}
\usepackage{verbatim}
\usepackage{color}
\usepackage{multirow}
\usepackage{subfig}
\usepackage{cite}

\def\stacksymbols #1#2#3#4{\def\theguybelow{#2}
    \def\vp{\lower#3pt}
    \def\sp{\baselineskip0pt\lineskip#4pt}
    \mathrel{\mathpalette\intermediary#1}}

\def\intermediary#1#2{\vp\vbox{\sp
     \everycr={}\tabskip0pt
     \halign{$\mathsurround0pt#1\hfil##\hfil$\crcr#2\crcr
              \theguybelow\crcr}}}

\def\be{\begin{equation}}
\def\ee{\end{equation}}
\def\bea{\begin{eqnarray}}
\def\eea{\end{eqnarray}}

\def\lsim{\raise0.3ex\hbox{$\;<$\kern-0.75em\raise-1.1ex\hbox{$\sim\;$}}}
\def\gsim{\raise0.3ex\hbox{$\;>$\kern-0.75em\raise-1.1ex\hbox{$\sim\;$}}}

\def\tb{\tan\beta}
\def\s{\smallskip}

\def\inbar{\,\vrule height1.5ex width.4pt depth0pt}

\def\IC{\relax\hbox{$\inbar\kern-.3em{\rm C}$}}
\def\IQ{\relax\hbox{$\inbar\kern-.3em{\rm Q}$}}
\def\IR{\relax{\rm I\kern-.18em R}}
 \font\cmss=cmss10 \font\cmsss=cmss10 at 7pt
\def\IZ{\relax\ifmmode\mathchoice
 {\hbox{\cmss Z\kern-.4em Z}}{\hbox{\cmss Z\kern-.4em Z}}
 {\lower.9pt\hbox{\cmsss Z\kern-.4em Z}}
 {\lower1.2pt\hbox{\cmsss Z\kern-.4em Z}}\else{\cmss Z\kern-.4em Z}\fi}

\def\comment#1{}
\def\to{\rightarrow}

\def\u1x{U(1)_X}
\newcommand{\nc}{\newcommand}
\nc{\LL}{L}
\nc{\vv}{\tilde{v}}
\nc{\ccdot}{\!\cdot\!}
\nc{\gsm}{G_{SM}}
\nc{\vfive}{\mathbf{5}\oplus\mathbf{\overline{5}}}
\nc{\vten}{\mathbf{10}\oplus\mathbf{\overline{10}}}
\nc{\zhol}{Z^{\rm hol}}
\nc{\xfb}{\,{\rm fb}}

\begin{document}
\vspace*{7mm}

\begin{flushright}
LAPTH--010/19 
\end{flushright}

\vspace*{2mm}

\begin{center}

{\large\bf  Dark Matter through the Higgs portal}

\vspace*{9mm}

{\sc Giorgio Arcadi$^{1,2}$},  {\sc Abdelhak~Djouadi$^{3,4}$}   
and  {\sc Martti Raidal$^4$} 

\vspace*{9mm}

$^1$ Max-Planck-Institut  fur  Kernphysik, Saupfercheckweg  1,  69117  Heidelberg,  Germany. \\
\vspace{0.15cm}

$^2$ Dipartimento di Matematica e Fisica, Universit\`a di Roma
3, Via della Vasca Navale 84, 00146, Roma, Italy.\\
\vspace{0.15cm}

$^3$ Universit\'e Savoie--Mont Blanc, USMB, CNRS, LAPTh, F-74000 Annecy, France.\\
\vspace{0.15cm}

$^4$ NICPB, R{\"a}vala pst. 10, 10143 Tallinn, Estonia.\\
\vspace{0.15cm}

\end{center}

\vspace*{4mm}

\begin{abstract}

We review scenarios in which the particles that account for the Dark Matter (DM)
in the Universe interact only through their couplings with the Higgs sector of
the theory, the so-called Higgs-portal models.  In a first step, we use a 
general and model-independent approach in which the DM particles are singlets
with spin $0,\frac12$ or $1$, and assume a minimal Higgs sector with the
presence of only the Standard Model (SM) Higgs particle observed at the LHC.  
In a second step, we discuss non-minimal scenarios in which the spin-$\frac12$
DM particle is accompanied by additional lepton partners and consider several 
possibilities like sequential, singlet-doublet and vector-like leptons.  In a
third step, we examine the case in which it is the Higgs sector of the theory
which is enlarged either by a singlet scalar or pseudoscalar field,  an
additional two Higgs doublet field or by both; in this case, the matter content
is also extended in several ways.    Finally, we investigate the case of
supersymmetric extensions of the SM   with neutralino DM, focusing on the
possibility that the latter couples mainly to the neutral Higgs particles of the
model which then serve as the main portals for DM phenomenology.   In all these
scenarios, we summarize and update the present constraints and future prospects
from the collider physics perspective, namely from the determination of the SM
Higgs properties at the LHC and the search for its invisible decays into DM,
and  the search for heavier Higgs bosons and the  DM companion particles at
high-energy colliders.  We then compare these results with the constraints and
prospects obtained  from the cosmological relic abundance as well as from direct
and indirect DM searches in astroparticle physics experiments. The
complementarity of collider and astroparticle DM searches is investigated in all
the considered models.  

\end{abstract}

\newpage

\setlength{\parskip}{0.8mm}
\tableofcontents
\setlength{\parskip}{0.14cm}

\newpage


\section{Introduction}  


Particle physics is presently facing at least two majors issues. A first one is
the exploration of the fundamental mechanism that generates the elementary
particle masses and leads to the existence of a new type of particles, the Higgs
bosons \cite{Higgs:1964pj,Englert:1964et,Guralnik:1964eu}. The discovery in 2012
of such a particle at the CERN Large Hadron Collider (LHC)
\cite{Aad:2012tfa,Chatrchyan:2012xdj} with a mass of \cite{Aad:2015zhl}\\[-3mm]
\beq
M_H=125~{\rm GeV} \, , \\[-.1mm]
\label{eq:Hmass}
\eeq
is acknowledged to be of very high relevance but an equally important
undertaking would be the precise determination of its basic properties
\cite{Khachatryan:2016vau,ATLAS-web,CMS-web}. In particular, we need to answer
to the question whether this new state is the one predicted by the Standard
Model (SM)
\cite{Glashow:1961tr,Weinberg:1967tq,Salam:1968rm,Gross:1973id,Politzer:1973fx},
the theory that describes in a minimal way the electromagnetic, weak and strong
interactions, or it is part of the extended structure of a more fundamental
theory; for reviews of the SM Higgs sector, see for instance 
Refs.~\cite{Gunion:1989we,Spira:1997dg,Djouadi:2005gi,Dittmaier:2011ti,Dittmaier:2012vm,Heinemeyer:2013tqa,deFlorian:2016spz,Spira:2016ztx,Dawson:2018dcd}.
This is a particularly important question as the SM has many shortcomings, a
crucial one being due to the Higgs sector itself which is considered to be
highly unnatural from a theoretical perspective, as it does not warrant a
protection against the extremely high scales that contribute to the Higgs boson
mass and make it in principle close to the Planck scale rather than to the weak
scale.   Whether or not there is New Physics beyond the SM is vital for particle
physics.

A second major issue, which provides at the same time a decisive hint for the
existence of New Physics beyond the SM, is related to the longstanding problem
\cite{Zwicky:1933gu} of the existence and the nature of the Dark Matter (DM) in
the Universe. Indeed, cosmological considerations and astrophysical observations
point toward the existence of a matter component, distinct from ordinary
baryonic matter, whose cosmological relic abundance according to the recent
extremely precise measurements from the PLANCK satellite~\cite{Ade:2015xua} is
given by  
\beq      
\Omega_{\rm DM} h^2 = 0.1188 \pm 0.0010 \, , \label{eq:omegah}   
  \eeq      
with $h$ being the reduced Hubble constant, and corresponds to approximately
$25\%$ of the energy budget of the Universe. It is commonly believed that this
DM component is accounted for by a new particle, stable at least on cosmological
scales, with very suppressed interactions with the SM states and cold, i.e.\
non--relativistic at the time of matter--radiation equality in the Universe. 
Particle physics proposes a compelling solution to this puzzle in terms of a
colorless, electrically neutral, weakly interacting, absolutely stable particle
with a mass in the vicinity of the electroweak scale. While the observed matter
content in the SM does not involve such a state, the neutrinos being too light
to offer a viable solution, many of its extensions predict the occurrence of new
weakly interacting massive particles (WIMPs)  that could naturally account for
this phenomenon; see for instance 
Refs.~\cite{Jungman:1995df,Drees:1998ra,Bergstrom:2000pn,Munoz:2003gx,Bertone:2004pz,Feng:2010gw,Drees:2012ji,Roszkowski:2017nbc,Arcadi:2017kky,Kahlhoefer:2017dnp,Tanabashi:2018oca}
for some general reviews on the possible candidates. 

In fact, in many extensions of the SM, the naturalness and DM problems can be
solved at once, sometimes in a rather elegant manner. This is, for instance, the
case of supersymmetric theories 
\cite{Wess:1974tw,Golfand:1971iw,Drees:2004jm,Baer:2006rs,Martin:1997ns} which
postulate the existence of a new partner to every SM particle and the lightest
superparticle was considered for a long time as the ideal candidate
\cite{Ellis:1983wd,Ellis:1983ew,Goldberg:1983nd,Krauss:1983ik,Griest:1988ma,Drees:1992am}
for Dark Matter\footnote{The two other theoretical constructions that address
the problem of the hierarchy of scales in the SM Higgs sector, namely extra
space--time dimensions and composite models have also their DM candidates,
respectively, the lightest Kaluza--Klein \cite{Servant:2002aq,Cheng:2002ej} and
the lightest T--odd \cite{Cheng:2004yc} states.}. It is extremely tempting and,
in fact, rather natural to consider that these two important issues are
intimately related and the Higgs bosons serve as mediators or portals to the DM.
As a matter of fact, in order to make the DM states absolutely stable, one has
to invoke a discrete symmetry under which they (and their eventual companions in
an extended DM sector) are odd while all SM particles are even, forbidding the
DM to decay into ordinary fermions and gauge bosons. If the DM particle is not
charged under the electroweak group, the Higgs sector of the theory allows to
accommodate in a minimal way the interaction among pairs of DM and of SM
particles~\cite{Silveira:1985rk,McDonald:1993ex,Burgess:2000yq,Kim:2006af,Kanemura:2010sh,Djouadi:2012zc,Djouadi:2011aa,LopezHonorez:2012kv,Andreas:2010dz,Lebedev:2011iq,Mambrini:2011ik,Davoudiasl:2004be,Schabinger:2005ei,Patt:2006fw,OConnell:2006rsp,Barger:2007im,He:2008qm,He:2009yd,Barger:2010mc,Clark:2009dc,Lerner:2009xg,Goudelis:2009zz,Yaguna:2008hd,Cai:2011kb,Biswas:2011td,Farina:2011bh,Hambye:2008bq,Hambye:2009fg,Hisano:2010yh,Englert:2011yb,Englert:2011aa,Andreas:2008xy,Foot:1991bp,Melfo:2011ie,Raidal:2011xk,He:2011de,Mambrini:2011ri,Chu:2011be,Ghosh:2011qc,Greljo:2013wja,Cline:2013gha}.
These Higgs--portal models can then describe in an economic manner a most
peculiar feature of the DM particles,  namely their generation mechanism which
is based on the freeze--out paradigm and relates the DM cosmological relic
density to a single particle physics input, their thermally averaged
annihilation cross section.   Indeed, in  these scenarios, the relic density would be  induced when pairs of DM states annihilate into SM fermions and gauge
bosons, through the $s$--channel exchange of the Higgs bosons. These Higgs
bosons  will also be the mediators of the mechanisms that allow for the
experimental detection of the DM states.

The simplest of the Higgs--portal scenarios is when the Higgs sector of the
theory is kept minimal and identical to the one postulated in the SM, namely the
single doublet Higgs field structure that leads to the unique $H$ boson which
has been observed so far. Mindful of William of Occam, one could then extend the
model by simply adding only one new particle to the spectrum, the DM state, as
an isosinglet under the electroweak gauge group.  Nevertheless, the DM particle
can have the three possible spin assignments, that is, can be a spin--zero or
scalar particle, a spin--1 vector boson or a Dirac or Majorana spin--$\frac12$
fermion (a spin--2 DM state has been also proposed \cite{Babichev:2016bxi}). 
Although only effective and eventually non--renormalisable, one can adopt this
approach as it is rather model--independent and does not make any assumption on
the very nature of the DM
\cite{Kim:2006af,Kanemura:2010sh,Djouadi:2012zc,Djouadi:2011aa,LopezHonorez:2012kv,Goodman:2010ku,Fox:2011pm,Buckley:2014fba,Abdallah:2015ter,Alanne:2017oqj}.
In addition, such a scheme can be investigated in all facets as it has a very
restricted number of extra parameters in addition to the SM ones, namely the
mass of the DM particle and its coupling to the Higgs boson\footnote{These two
parameters can be further related by the requirement that the cosmological relic
density takes a value that is very close to the experimentally measured one,
eq.~(\ref{eq:omegah}).  However, as will be seen later, one could consider more
general scenarios in which the DM particle is not absolutely stable and/or does
not account for the entire DM in the Universe.}. This effective, simple and
economical SM Higgs--portal scenario can be considered to be, in some sense, a
prototype WIMP model.

A most interesting realization of the SM--like Higgs--portal discussed above is
when the DM particle is an  electroweak singlet fermion.  However, a coupling
between this DM candidate and the SM Higgs doublet field is necessarily not
renormalizable and this theory can only be effective and valid at the low energy
scale. In order to cure this drawback and make the theory complete in the
ultraviolet regime while keeping the Higgs sector as minimal as in the SM, the
DM state should be accompanied by some fermionic partners that are non--singlets
under the SU(2) electroweak group. The spin--$\frac12$ DM particle could then be
part of an isodoublet or, if it is still an isosinglet, could mix with it.
Hence, the possibility of further extending the fermionic sector of the theory
should be considered. 

Besides the option of a fourth generation of fermions with a massive
right--handed neutrino \cite{Belotsky:2002ym,Kribs:2007nz,Denner:2011vt}, which
is now completely excluded by the LHC Higgs data in the context of a SM--like
Higgs sector \cite{Djouadi:2012ae,Kuflik:2012ai},  two other possibilities have
been advocated. A first one is the introduction of a Majorana neutral fermion
that is part of a singlet--doublet lepton extension of the SM, the so--called
singlet--doublet model
\cite{Cohen:2011ec,Cheung:2013dua,Calibbi:2015nha,Yaguna:2015mva}.  A second
option for such an extended  fermionic sector would be a Dirac heavy neutrino
that belongs to an entire vector--like fermion family added to the SM fermionic
spectrum
\cite{Fujikawa:1994we,Hambye:2008bq,Hambye:2009fg,Hisano:2010yh,Lebedev:2011iq,Angelescu:2015uiz,Angelescu:2016mhl}.
A renormalizable Higgs--DM interaction is then generated through mixing, even if
the DM particle is the isosinglet neutral state in the two constructions. The
fermionic Higgs--portal discussed before can be then interpreted as an effective
limit of such a framework in which the extra fermionic fields, except from the
one of the DM, are assumed to be very heavy and integrated out  (though the
scheme is rather constrained by electroweak precision data).

In the case of scalar and vector DM states, the model--independent approach
mentioned above can, instead, be  made renormalisable. In the vector case, the
DM can be identified as  the stable gauge boson of a dark U(1) gauge symmetry
group that is spontaneously broken by the vacuum expectation value of an
additional complex scalar field
\cite{Hambye:2008bq,Lebedev:2011iq,Baek:2012se,Farzan:2012hh,Arcadi:2016qoz}. 
In the scalar case, one can either add simply a gauge singlet field
\cite{Silveira:1985rk,McDonald:1993ex,Burgess:2000yq} or invoke the possibility
of an additional scalar doublet field that does not develop a vacuum expectation
value and, hence, does not participate to electroweak symmetry breaking
\cite{Deshpande:1977rw,LopezHonorez:2006gr,Barbieri:2006bg,Ma:2006km,Arhrib:2013ela}.
The four degrees of freedom of the inert doublet field would then correspond to
four scalar particles and the lightest of them, when electrically neutral, could
be the DM candidate. Hence, in both the vector and scalar cases,  the DM
particle comes with additional beyond the SM states that can also be considered
to be heavy in an effective framework. Nevertheless, there are theoretical
constraints on these scenarios, as well as experimental ones  that are mainly
due to the  high precision electroweak data, which make that the extra states
associated with the DM particle should have a comparable mass and thus, can be
searched for and observed at present or future collider experiments. 

Another possibility for having a  Higgs--portal model which remains
theoretically consistent up to very high energy scales, is when it is the  Higgs
sector itself that is enlarged. For instance, an additional Higgs singlet field
that acquires a vacuum expectation value and mixes with the SM--like Higgs field
allows for a renormalisable coupling with an isosinglet fermion state
\cite{Schabinger:2005ei,Patt:2006fw,OConnell:2006rsp,Barger:2007im,Profumo:2007wc,Baek:2011aa,Bertolini:2012gu,Robens:2015gla,Godunov:2015nea,Falkowski:2015iwa}.
Such a scheme remains minimal compared to the SM effective scenario since  the
DM mass can be generated dynamically by the extra singlet field, hence  relating
it to the DM coupling to the Higgs bosons. More generally, many extensions which
were considered in the past to address some of the shortcomings of the SM
involve a Higgs sector that is extended by a singlet scalar field. Another
possibility of the additional singlet scalar would be  that it does not mix with
the SM Higgs doublet, as it often appears in (partially) composite Higgs models
\cite{Eichten:1979ah,Kaplan:1983sm}  thus opening the possibility that the new
singlet could also correspond to a  pseudoscalar Higgs state 
\cite{Mambrini:2015wyu,DiChiara:2015vdm,Backovic:2015fnp,Falkowski:2015swt,Franceschini:2015kwy,Barducci:2015gtd,DEramo:2016aee,Djouadi:2016eyy}.
The new scalar or pseudoscalar particles, together with the SM Higgs boson, will
then serve as a double portal to the DM. The latter can be again the neutral
component of a vector--like fermion family, for instance. Extensions in which
both scalar and pseudoscalar Higgs states are simultaneously present have also
been considered and lead to a rather interesting phenomenology in the DM
context, in particular when the pseudoscalar state is very light compared to the
scalar one or when the two states are almost degenerate in mass. 

Among the theories with an extended scalar sector, two--Higgs doublet models
have a special status and are, by far, the most studied ones  in the last
decades; for a review, see Ref.~\cite{Branco:2011iw}. Compared to the SM  with 
its  unique  Higgs  particle,  the Higgs sector of the model involves five
physical states after electroweak symmetry breaking:  two CP--even neutral
ones that mix and share the properties of the SM Higgs boson, a CP--odd or
pseudoscalar neutral and two charged Higgs states with properties that are
completely different from those of the SM Higgs boson.  The presence of the
additional particle lead to a very rich  phenomenology and interesting new
signatures, in particular, as  a result of the many possibilities for the
structure of the couplings of the Higgs bosons to standard fermions
\cite{Glashow:1976nt}. Two Higgs--doublet models appear naturally in very well
motivated extensions of the SM, such as the minimal supersymmetric model, and
provide a very good benchmark for investigating physics beyond the SM. 

These models should be extended to incorporate the DM particles and this can be
done in a way analogous to what has been mentioned previously, by introducing a
full sequential family of vector--like fermions 
\cite{Djouadi:2016eyy,Angelescu:2015uiz,Bizot:2015qqo} or a singlet--doublet of
heavy leptons \cite{Berlin:2015wwa,Arcadi:2018pfo} for instance. As also
noted above, there is the possibility that only one of the Higgs doublets is
responsible of electroweak symmetry breaking, while the other doublet does not
acquire a vacuum expectation value nor couple to SM fermions as a result of a
discrete symmetry, the  so--called inert doublet model in which the DM candidate
is the lightest neutral state of the inert field~\cite{Deshpande:1977rw}.
Another scenario which recently gained a wide interest in the context of DM, as
it represents a useful limit of some theoretically well motivated  models and
leads to a very interesting phenomenology, is the one in which the two--doublet
Higgs sector is further extended to incorporate a light pseudoscalar singlet
field that can serve as an additional Higgs--portal to the DM 
\cite{Ipek:2014gua,Goncalves:2016iyg,Bauer:2017ota,Tunney:2017yfp,Abe:2018bpo}.

Finally, to close this tentative list of possible extended Higgs and DM models,
there are supersymmetric extensions of the SM
\cite{Wess:1974tw,Golfand:1971iw,Drees:2004jm,Baer:2006rs,Martin:1997ns} which
solve what was for a long time considered as the most notorious problem of the
SM, the hierarchy problem mentioned in the beginning of our discussion: the
cancellation of the quadratic divergences that appear when calculating the
radiative corrections to the Higgs boson mass is highly unnatural in the SM and
needs an extreme fine--tuning. Supersymmetric theories postulate the existence
of a new partner to every SM particle with couplings that are related in such a
way that these quadratic divergences are naturally cancelled. 

In the Minimal Supersymmetric Standard Model (MSSM)
\cite{Martin:1997ns,Haber:1984rc,Djouadi:1998di,Chung:2003fi}, in which the
Higgs sector is extended to contain two doublet fields
\cite{Gunion:1989we,Djouadi:2005gj,Carena:2002es,Heinemeyer:2004gx}, there is an
ideal candidate for the weakly interacting massive particle which is expected to
form the cold DM: the lightest supersymmetric particle, which is in general a
neutralino, a mixture of the  superpartners of the neutral gauge and Higgs bosons
\cite{Ellis:1983wd,Ellis:1983ew,Goldberg:1983nd,Krauss:1983ik,Griest:1988ma,Drees:1992am}.
This particle is absolutely stable when a symmetry called R--parity
\cite{Farrar:1978xj} is conserved and, in a wide and natural range of the MSSM
parameter space, its annihilation rate into SM particles fulfills the
requirement that the resulting cosmological relic density is within the measured
range~
\cite{Jungman:1995df,Drees:1998ra,Bergstrom:2000pn,Munoz:2003gx,Bertone:2004pz,Feng:2010gw,Drees:2012ji,Roszkowski:2017nbc}.
In order to circumvent some shortcomings of the MSSM, the so--called
$\mu$--problem \cite{Nilles:1982mp,Frere:1983ag,Kim:1983dt}, a further extension
that is becoming popular by now, is the so--called next--to--minimal MSSM
(NMSSM) \cite{Ellwanger:2009dp,Maniatis:2009re,Djouadi:2008uw,Baum:2017enm} in
which a complex isosinglet field is added thus extending the two--Higgs doublet
Higgs sector of the theory by  an extra CP--even and one CP--odd Higgs particles
that could be very light and have a quite interesting phenomenology. 

In most cases, in particular when the superpartners of the fermionic spectrum
are very heavy as  indicated by current LHC data, the neutral states of the
extended Higgs sector of these models can serve as the privileged portals to the
DM neutralino in a large area of the parameter space. In fact, the 
singlet--doublet lepton model and the models with two--Higgs doublets  and  a
pseudoscalar field introduced previously can be seen as representing simple
limiting cases of the MSSM and the NMSSM, respectively. 

Hence, there is  broad variety of models, with various degrees of complexity, 
in which the relevant interactions of the DM particles that are present in the
Universe are mediated by the Higgs sector of the theory. The aim of this review
is to analyze these models  and to study their phenomenology in both collider 
and astroparticle physics experiments. 

Actually, a fundamental and interesting aspect of all these Higgs--portal DM
models, is that they can be probed not only in direct
detection~\cite{Goodman:1984dc,Wasserman:1986hh,Drukier:1986tm} in astrophysical
experiments, i.e.\ in elastic scattering of the DM with nuclei, or in indirect
detection, when one looks in the sky for some clean products of their
annihilation processes such as gamma rays
\cite{Silk:1984zy,Turner:1986vr,Rudaz:1987ry,Ellis:1988qp,Primack:1988zm,Bergstrom:1988fp,Bouquet:1989sr,Ellis:2001hv},
but also at  colliders. There, and in contrast to astroparticle experiments, one
can search at the same time for the DM particles by looking for instance at
invisible Higgs decays 
\cite{Shrock:1982kd,Belotsky:2002ym,Joshipura:1992hp,Choudhury:1993hv,Frederiksen:1994me,Belanger:2001am,Godbole:2003it,Eboli:2000ze,Davoudiasl:2004aj,Batell:2011pz,Belanger:2013kya,Low:2011kp,Espinosa:2012vu}
and other missing transverse energy signatures~\cite{Goodman:2010ku,Fox:2011pm},
as well as for the possible companions of these particles, the new fermions or
new bosons that belong to the same representation or mix with it, and the
mediators of the DM interactions, the Higgs bosons including those that
eventually appear in extended scenarios.  These distinct types of searches are
hence highly  complementary and in many different ways.

 During the last decade, the experimental community, with the lead of the
intense effort at the LHC, complemented by an impressive array of other
experiments, from low energy experiments in the neutrino and B--meson sectors
for instance to cosmology and astroparticle physics experiments searching for DM
such as XENON \cite{Aprile:2015uzo,Aprile:2017iyp,Aprile:2018dbl}, has
challenged the SM from all imaginable corners. While brilliant and historical
successes have been achieved, like the discovery of the Higgs boson, no sign of
a departure from the SM predictions has emerged so far. This is particularly the
case at the high--energy frontier, where the first tests of the Higgs boson
properties at the LHC  have shown that the particle is approximately SM--like
\cite{Khachatryan:2016vau,ATLAS-web,CMS-web}.   Furthermore, direct searches for
new particles have been performed in many topologies, covering a large number of
new physics possibilities, and turned out to be unsuccessful for the time being
\cite{ATLAS-web,CMS-web}. On the other hand, the absence of signals in
astrophysical experiments searching for DM particles is  putting the paradigm of
a weakly interacting massive DM particle under increasing pressure. For
instance,  the XENON1T experiment
\cite{Aprile:2015uzo,Aprile:2017iyp,Aprile:2018dbl} has set strong bounds on the
mass and couplings of the DM, excluding  large areas of the natural parameter
space of the beyond the  SM schemes that predict them.     To achieve a better
sensitivity to these extended Higgs--portal scenarios, a significantly larger
data sample is required and, eventually, new experiments that are capable of 
exploring higher DM mass scales or smaller couplings are needed. 

Particle physics is undergoing a crucial moment where a strategy for the future
is being decided and choices for the next generation of experiments are to be
made \cite{EU-Strategy}.  Besides the high--luminosity option of the LHC 
\cite{ATLAS:2013hta,CMS:2013xfa,Cepeda:2019klc} in which an extremely large data
sample than presently should be collected at the slightly higher center of mass
energy of 14 TeV and which should be the natural next step, another subsequent 
possibility will be to move to higher energies, and doubling the LHC energy is
under serious consideration  \cite{Baur:2002ka,Cepeda:2019klc}. In a longer run,
proton colliders with energies up to 100 TeV are currently envisaged both at
CERN  \cite{Contino:2016spe} and in China \cite{Tang:2015qga}. A preliminary
step at these colliders would be to run in the much cleaner $e^+e^-$ mode at an
energy of about 250 GeV and with high luminosity, allowing them to be true
Higgs  boson factories
\cite{Gomez-Ceballos:2013zzn,Mangano:2018mur,CEPC-SPPCStudyGroup:2015csa,CEPCStudyGroup:2018ghi}. 
Such a plan is also under discussion in Japan with a linear $e^+e^-$  collider
that can be possibly extended up to 1 TeV  \cite{Djouadi:2007ik,Baer:2013cma}
and at CERN where a multi--TeV $e^+e^-$ machine is contemplated 
\cite{Battaglia:2004mw,Linssen:2012hp}. 

On the astrophysical front also, several experiments are planed in a near and
medium future with a significant increase in sensitivity in the search for the
DM particles. Some examples of experiments in direct detection are the XENONnT 
\cite{Aprile:2015uzo} and the LUX--ZEPLIN (LZ)  \cite{Akerib:2015cja} detectors
and, later, the DARWIN experiment  \cite{Aalbers:2016jon}  which would be  the
``ultimate" DM detector as it could reach a sensitivity close to the irreducible
background represented by the $Z$--boson mediated coherent scattering of SM
neutrinos on nucleons, the so--called neutrino floor. Very powerful and
sensitive indirect detection experiments are also planed in a near future, such
as the Cherenkov Telescope Array (CTA) \cite{Acharya:2013sxa} and the High
Altitude Water Cherenkov (HAWC) \cite{Abeysekara:2014ffg},  the next generation
ground--based observatories for gamma--ray astronomy at very high energies. 

At this stage, we believe that it is an appropriate time to summarize and update
the large amount of analyses that have been performed in the last decade at both
collider and astrophysics experiments and infer the constraints that they
impose on these Higgs--portal to the DM particles scenarios. It also seems
opportune to investigate the potential of the upgrades planed at present
machines, now that we have a relatively clear idea of the near and medium
future, and at the facilities that are planed for the more remote future, in
pursuing the search for the DM particle and the possible new spectra which is
associated to it. This is the aim of this review: an extensive and comprehensive
account of the present constraints and the future prospects on the various
Higgs--portal scenarios for Dark Matter and the possible complementarity between
the different experiments and approaches.   

The organization of this review follows the classification of the numerous 
Higgs--portal models for DM introduced above. The next section is  devoted to
the minimal Higgs--portal model with a SM--like Higgs sector that mediates the
interactions of an isosinglet spin--0, $\frac12$ and spin--1 DM state in an
effective approach. Section 3 is dedicated to scenarios in which the Higgs
sector is kept minimal  but the DM one is extended to incorporate additional
states; we  specialize to renormalizable models in which the DM is a
spin--$\frac12$ fermion that is part of a fourth generation family, a
singlet-doublet lepton or a full family of vector--like fermions.  The
subsequent sections are instead devoted to scenarios in which it is the  Higgs
sector of the theory which is extended to incorporate additional singlet or
doublet fields. In section 4, we analyze extensions with additional scalar
singlets that mix or not with the SM Higgs boson and couple to the DM, either in
the general effective approach  or when it is an isosinglet heavy neutrino.
Section 5 is for two--Higgs doublet models that couple to a lepton in a
singlet--doublet or a vector--like representation; we also consider the cases in
which one of the scalar doublet is inert and when an additional light
pseudoscalar Higgs state is present. In section 6, we consider two
supersymmetric models, the MSSM and NMSSM, in which the partners of the fermions
and the gluons are assumed to be very heavy and the DM phenomenology is
essentially mediated by the Higgs bosons.  Each of these sections is structured
as follows. In a first part, we introduce the various models and summarize the
eventual theoretical constraints to which they are subject. It is followed by an
extensive discussion of the most relevant collider aspects of the Higgs and DM
sectors and the constraints or prospects in their searches. We conclude the
sections by an updated analysis  of the DM phenomenology, the relic density and
the constraints/prospects in direct and indirect detection experiments, 
including the eventual complementarity with colliders. Our conclusions will be
briefly stated in section 7. The review includes three appendices: one for the
analytical material describing Higgs and DM production at colliders in
the various models, one for analytical approximations for DM annihilation
cross sections and another for expressions of the renormalization group
evolution of some Higgs couplings.

\setcounter{section}{1}
\renewcommand{\thesection}{\arabic{section}}
\include{sec-SM}

\setcounter{section}{2}
\renewcommand{\thesection}{\arabic{section}}
\include{sec-NF}

\setcounter{section}{3}
\renewcommand{\thesection}{\arabic{section}}
\include{sec-Sing}

\setcounter{section}{4}
\renewcommand{\thesection}{\arabic{section}}
\include{sec-2HDM}

\setcounter{section}{5}
\renewcommand{\thesection}{\arabic{section}}
\include{sec-MSSM}

\setcounter{section}{6}
\renewcommand{\thesection}{\arabic{section}}

\section{Conclusions}

The absence of explanation for one of the most important contemporary 
scientific puzzles, the origin and the nature of the observed Dark Matter
component in the Universe,  strongly suggests to extend the Standard Model of
particle physics by at least one weakly interacting and massive particle that
would account for it. The interaction between this DM particle and the SM
fermions and gauge bosons, which is at the base of the mechanism that generates
the DM and allows to detect it experimentally, can be accommodated through the
Higgs sector of the theory. The latter hence serves as a privileged ``portal''
between the visible sector of the SM and the DM sector. In general, not only the
dark sector should be extended in order to comprise companions of the DM
particle that would permit renormalisable interactions among other features, but
also, the Higgs sector of the theory can be enlarged, hence allowing for
additional Higgs--portals to the DM states to be present. 

In the present work, we have reviewed a multitude of elaborated theoretical
realizations, with various degrees of complexity, of such Higgs--portal
scenarios. We have summarized the important theoretical elements that allow to
describe them, discussed the most relevant collider aspects of the Higgs and DM
sectors including present constraints  on the spectra and future prospects for
observation and, finally, analysed and updated the two most important
characteristics of the phenomenology of the DM state, namely its cosmological
relic abundance and its rates in direct and indirect detection in astroparticle
physics experiments. We have paid a particular attention to the complementarity
between, on the one hand,  the collider searches for the DM states and their
companions as well as to the extra Higgs bosons and, on the other hand,  the
dedicated direct and indirect DM searches. 

The minimal way of realizing a Higgs--portal scenario would be simply to extend the SM with a single particle, the DM candidate, which couples to the
unique Higgs boson of the theory through an effective and possibly
non--renormalizable interaction. Although the DM can have three different spin
assignments, namely be a spin--0 scalar, a spin--$\frac12$ Dirac or Majorana
fermion and a spin--1 vector, the resulting model is rather simple as it has
only two free parameters, the DM mass and its coupling to the $H$ boson, and is
thus easily testable. We have thoroughly analysed such a scenario, starting with
the possibility of searching for the DM particles at high energy colliders and,
in particular, at the LHC. Being electrically neutral and stable, they are
essentially undetectable and would appear only as missing transverse energy when
produced in association with visible SM particles which should be then tagged.
In the context of this SM Higgs--portal scenario, there are two main ways to
observe such elusive states.  First, if they are lighter than half of the
Higgs mass, $m_{\rm DM} \lsim \frac12  M_H \approx 62$ GeV, they will appear as
decay products  of the observed Higgs boson.  For slightly  heavier DM
particles, $m_{\rm DM} \gsim \frac12  M_H$, the produced Higgs boson should be
virtual or off--shell and would split into a pair of DM states, which results
into much  smaller production cross sections.  Still for light DM particles,
$m_{\rm DM} \lsim \frac12  M_H$, a second possibility would be to search
indirectly for the invisible Higgs decays into DM particles by measuring
precisely the total decay width of the Higgs boson and, alternatively, its
various visible decay branching fractions. If any additional decay mode like the
invisible one is present at a substantial level, it will affect the two types of
observables. This is one of the primary reasons to perform the high--precision
Higgs measurements that are planned, for instance, in the high--luminosity
option of the LHC or at future collider facilities. 

Of course, DM particles can be experimentally probed also through dedicated
search strategies, namely direct and indirect detection. In this review, we have
summarised the constraints on Higgs--portal models from astroparticle physics 
experiments  and compared them with what is  obtained at high--energy colliders
like the LHC. In this context, direct detection typically sets the most
stringent limits. We have updated the presently existing ones,  in particular by
the XENON1T experiment which provides the strongest bounds, and discussed the
projected sensitivities of future experiments like XENONnT, LZ and DARWIN
detectors. Current exclusion limits already rule out large regions of the
theoretically viable parameter space of the SM Higgs--portal model and the
absence of a signal at the next generation detectors will rule out thermal DM
states for masses up to about 1 TeV. 

One should note that the collider constraints from the invisible width of the
Higgs boson, although relatively weak compared to the above astrophysical ones,
are nevertheless complementary to them, in particular at low DM masses when the
sensitivity of direct detection experiments degrades. Furthermore, the collider 
searches do not rely on a specific hypothesis for the DM abundance and are thus
more general, applying also for particles that are stable on detector but not
necessarily on cosmological scales. It is thus extremely important to further
exploit the potential of searches of additional exotic decay channels of the
Higgs boson at the LHC including the high--luminosity option and at future
higher--energy hadron and lepton colliders, covering in particular the region
$M_H \lesssim 2 m_{\rm DM}$. 

These colliders are also useful in searching for the possible companions of the
DM particle. Indeed, while the scalar and vector effective DM Higgs--portals are
renormalizable, the singlet fermionic effective one is not, being realized
through a dimension--5 operator. The first type of extension of the SM
Higgs--portal scenario considered in this review hence consisted into enlarging 
the DM sector to permit renormalizable interactions of a fermionic DM with the
SM Higgs sector. Two simple examples of extensions have been studied, in
addition to the possibility of a fourth generation of chiral fermions which was
shown to be excluded by present LHC and astrophysical data:  the so--called
singlet--doublet lepton model with a Majorana DM and the addition of a   full
``family'' of vector--like fermions with its Dirac singlet neutrino being the DM
candidate. Hence, in both scenarios the DM is accompanied by fermionic partners
that are non--singlets under the electroweak group.

Adopting an analogous strategy as the minimal Higgs--portal scenario, we have
summarized the present constraints and the expectations for these two scenarios
from both the collider and astroparticle physics perspectives, as well as from 
theoretical considerations such as perturbativity, stability of the electroweak
potential and conformity with the precision data. Concerning the  phenomenology
of the DM state, the singlet--doublet lepton model is in similar tension with
direct detection as the effective Higgs--portal, with the exception of the
so--called blind-spots in which the Higgs--DM coupling vanishes. The case of a
vector--like Dirac DM is even more constrained since the vectorial coupling with
the $Z$ boson further enhances the DM spin--independent interactions. The only
viable solution is represented by coannihilations of a mostly singlet--like DM
state that is nearly mass degenerate with the extra leptons that are present in
the spectrum.  Indirect detection constraints are not competitive with the ones
from direct detection and have been often omitted. On the other hand, collider
phenomenology is enriched by the possibility of searching for the fermionic, non
isosinglet partners of the DM state. Current limits on their masses and
couplings have been presented and the prospects for future detection at the
HL--LHC, as well as at future proton or $e^+e^-$  colliders have been examined
in detail.

A third class of models considered in this review consisted into extensions of
the Higgs sector of the theory with the incorporation of additional scalar
fields that could also serve as portals to the DM particles. We have first
studied a minimal extension with a singlet scalar Higgs field that develops a
vacuum expectation value and mixes with the SM Higgs state. The DM sector
consisted again on a particle with the three possible spin assignments, namely
spin--0, $\frac12$ and 1, and coupling effectively to the Higgs bosons  but this
time in a renormalizable way even in the spin--$\frac12$ DM case. Analogously to
the effective Higgs--portal model, a strong correlation between the DM
annihilation into SM states and its scattering on nucleon cross sections is
present. This implies very strong constrains from DM direct detection which can
be evaded only in proximity of $s$--channel resonances or in the so--called
``secluded'' regime, corresponding to the annihilation of the DM into pairs of
the additional mediator. A quite orthogonal scenario that we have then examined,
is when the additional scalar state does not participate to electroweak symmetry
breaking and does not mix with  the SM--like Higgs boson and, thus, can be also
of a pseudoscalar nature. These scalar and pseudoscalar resonances have been
assumed either to have direct couplings only to the heavy top quark, or have
exclusively one--loop couplings with the SM gauge bosons, induced by
vector--like fermions for instance. In such a case, a  nice complementarity
between the requirements of a correct relic density and LHC searches for
resonances decaying into gauge bosons, in particular diphotons  or heavy top
quarks, can be established. Concerning DM phenomenology, in the case of a scalar
mediator, despite the weakness of the limits from direct detection as a result
of the small resonance couplings to the $c$ and $b$ quarks, the favoured regions
correspond again to the cases in which the mass of the DM lies close to the
$s$--channel resonance or it is greater than the mass of the mediator. The
interactions of the DM with a heavy pseudoscalar mediator are, in turn,  left
unconstrained by direct detection and are only moderately sensitive to indirect
detection. A precise assessment of the collider constraints is thus crucial in order to properly probe this scenario.

We have also briefly discussed the option in which both a scalar and a
pseudoscalar resonances are present which is very interesting in two limiting
cases: when the two states are almost degenerate in mass and  appear as a single
resonance when produced at colliders and when  the pseudoscalar is much lighter
than the scalar and even the DM particle. The DM phenomenology of this last
scenario has been studied in detail and features new efficient DM annihilation
channels without altering the direct detection signals.  The correct relic
density is achieved in a region of the parameter space which can be probed by
searches of collimated photons from the decays of the light pseudoscalar.

Further increasing the degree of complexity of the models, we have then
considered the case in which the Higgs sector is extended to incorporate 
two--Higgs doublet fields and, possibly, further augmented by a pseudoscalar
SU(2) singlet. While keeping again most of the focus on scenarios with fermionic
DM, extending the singlet--doublet and the vector--like fermion models to the
2HDM case,  we have nevertheless also considered a popular model in which  the
second scalar doublet is inert and enclose a scalar DM state and its partners.
The former types of models are particularly interesting for two reasons.  First,
they can be seen a special and simple limits of more complete theories, namely
the MSSM and NMSSM. On the other hand, they offer a richer Higgs spectrum with a
broad variety of collider signatures that are not fully explored by the
experimental collaborations.  Concerning DM phenomenology, different scenarios
have been considered for the various models. In the singlet--doublet case for
instance, we have adopted a   set--up that is similar to the MSSM, with heavy
Higgs bosons that are degenerate in mass and having Type--II couplings with to
SM fermions, and shown that strong constraints from LHC searches and flavor 
physics apply on it. The case of a Type--I 2HDM coupled with a singlet--doublet
DM is in turn more interesting in this regard as the CP--odd $A$ state can be
kept light enough to impact the DM and open new viable regions of parameter
space for it. For a vector--like DM, the phenomenology is much more contrived
because of the strong spin--independent interactions generated by the DM
vectorial coupling with the $Z$ boson. Viable DM regions can nevertheless open
up, e.g.\ when the  DM is heavy enough to annihilate into channels involving
charged Higgs bosons. On the contrary, bounds from DM direct detection are
significantly relaxed (though not absent) when the 2HDM Higgs sector is further
extended with a  pseudoscalar singlet. Collider probes hence play a crucial role
to test these models.  

As a final step, we have studied the Higgs and the DM sectors of the most
popular ultraviolet complete extensions of the SM, namely supersymmetric
extensions such as the MSSM. We have first characterized the Higgs sector of
the model, reviewing the  so--called hMSSM in which the information that the
mass of the lightest $h$ state  is $M_h\!=\!125$ GeV, allows to simply describe
it in terms of two input parameters and, hence, simplifies the discussion to a
large extent. In  this simple framework, we have summarized the results of
present collider searches for the extra neutral and charged Higgs bosons.
Assuming that the scalar partners of the SM fermions and the partner of the
gluon are very heavy, as indicated by LHC data,  we have focused on the chargino
and neutralino sectors of the theory, which incorporate the DM as the lightest
of the  neutral particles. Under the assumption that it interacts mostly with
the MSSM Higgs sector, the correct relic density can be achieved for DM masses
below the  scale of 1 TeV that keeps the model natural, either around the
``poles'' at the neutral Higgs boson masses or, by invoking a suitable
bino--higgsino admixture of the DM neutralino. Similarly to the singlet--doublet
lepton model, the current and eventual future absence of signals in DM direct
detection experiments will exclude increasingly large regions of the natural and
viable DM parameter space. 

The same type of study has been repeated in the case of the NMSSM in which the
Higgs sector is further extended by a complex singlet scalar field which leads
to an extra scalar and pseudoscalar states that can be relatively light.  The
model can also be described in terms of a limited set of input parameters. The
DM sector of the NMSSM is enriched as well with the presence of an additional SM
singlet component, the singlino, which increases the number of neutralinos to
five. A suitable admixture of singlino and higgsino components for the DM,
together with the presence of a  light pseudoscalar particle, allow to have the
required cosmological relic density for DM masses of few hundreds of GeV and, at
the same time, evade constraints from direct detection and from the LHC. 

In summary, we have reviewed in a rather comprehensive way the Higgs--portal
scenarios for DM, which are very interesting and natural realizations of the
WIMP paradigm. We have considered a series of increasingly refined models and
summarized and updated the present constraints to which they are subject at
high--energy colliders and in astroparticle physics experiments. While some of
these models are  severely constrained, other scenarios are still viable and
call for a further  probing and exploration at present and future
facilities.\bigskip


\noindent {\bf Acknowledgements:} This work is supported by the Estonian 
Mobilitas Pluss Grant. Part of this review is based on recent and less recent
work with many colleagues that we would like to thank for fruitful and enjoyable
collaborations. AD would like to thank colleagues at the University of Granada
for their hospitality and for discussions.  

\newpage

\setcounter{section}{0}
\renewcommand{\thesection}{A}
\setcounter{equation}{0}
\renewcommand{\theequation}{A.\arabic{equation}}

\include{sec-Appendix-Higgs}

\newpage

\setcounter{section}{0}
\renewcommand{\thesection}{B}
\setcounter{equation}{0}
\renewcommand{\theequation}{B.\arabic{equation}}

\include{sec-Appendix-DM}

\newpage

\setcounter{section}{0}
\renewcommand{\thesection}{C}
\setcounter{equation}{0}
\renewcommand{\theequation}{C.\arabic{equation}}

\include{sec-Appendix-RGE}

\newpage

\bibliographystyle{unsrt}
\bibliography{bibfile}

\end{document}

%% file: sec-SM.tex
\section{The Standard Model with DM particles}

In this section we summarize the theoretical elements that allow to describe the
scenario where the DM particles, which can be made stable by invoking a
$\mathbb{Z}_2$ symmetry, interact only with the Higgs sector. As stated
previously, these  Higgs--portal  scenarios for DM can be of several kinds,
depending on whether the models contain  additional Higgs multiplets and/or new
matter particles or not, but the simplest one  would be clearly when the SM is
extended to contain only one new particle, the DM state\footnote{We will see
later that in most cases, to have a renormalisable interaction with the SM Higgs
sector, the DM particles cannot be  alone and should appear together with some
accompanying particles which electroweak charges. However, the latter can be
considered as rather heavy and can be integrated out so that in practice, only
the DM particle would affect the phenomenology that is of interest to us.},  and
its minimal  Higgs sector is kept unchanged and hence, contains a unique Higgs
boson with a mass of 125 GeV as observed at the LHC. The DM particles will then
interact only with this $H$ state and their annihilation into SM particles,  for
instance, can occur only through $H$ boson exchange in the $s$--channel. This is
the  scenario that we will consider in this section  using an effective and
model--independent approach.  

\subsection{The minimal model in an effective approach}

\subsubsection{The SM Higgs sector}

To set the notation which will be used throughout this review,  we start by   briefly describing the Higgs sector in the SM. In this context, a doublet  of complex scalar fields with hypercharge $Y_\Phi=+1$
\begin{eqnarray}
\Phi= \left( \begin{array}{c} {\Phi^+} \\ {\Phi^0} \end{array} \right) \, , 
\end{eqnarray}
is introduced, to which one associates the ${\rm SU(2)_L \times U(1)_Y}$ invariant scalar potential  
\begin{eqnarray} 
V(\Phi)=\mu^2 \Phi^\dagger \Phi+  \lambda ( \Phi^\dagger \Phi)^2\, ,     \label{Vscalar}
\end{eqnarray} 
where the quartic coupling $\lambda$ is positive and the mass squared term is negative, $\lambda>0$ and $\mu^2<0$. In this case, the neutral component of the field develops a non--zero vacuum expectation value (vev)
\begin{equation} 
\Phi \to \frac{1}{\sqrt 2} \bigg( \begin{array}{c} {0} \\ {v + H} \end{array} \bigg)~~{\rm with}~v\!=\!\sqrt{- \mu^2/\lambda} =1/(\sqrt{2} G_F)^{1/2}=246~{\rm GeV} \, ,
\end{equation} 
with $G_F$ the Fermi constant. Three degrees of freedom, the  would be Goldstone bosons,  make the longitudinal components of the $W$ and $Z$ gauge bosons which get their masses
\begin{eqnarray}
M_W = \frac12  vg, ~~M_Z= \frac12 v  \sqrt{g^2_2+g'^2} \, ,
\end{eqnarray}
where $g$ and $g'$ are the ${\rm SU(2)_L}$ and ${\rm U(1)_Y}$ coupling constants that are related to the electroweak mixing angle $\theta_W$ by ($e$ is the proton charge) 
\begin{eqnarray}
\sin^2\theta_W = \frac{g^2}{g^ 2 + g'^2} = \frac{e^2}{g^2} =1 - \frac{M_W^2}{M_Z^2} \, .
 \end{eqnarray}

The remaining degree of freedom will correspond to the Higgs boson $H$
with a mass $M_H\!=\!125$ GeV observed at the LHC. The fermion
masses are generated by introducing the Yukawa Lagrangian for the field $\Phi$
\begin{eqnarray}
{\cal L}_{\rm Yukawa}= -\lambda_{e} \, \bar{L} \, \Phi \, e_{R}  
- \lambda_{d} \, \bar{Q} \, \Phi \, d_{R}
- \lambda_u \, \bar{Q} \, \tilde{\Phi} \, u_R    \ + \ {\rm  h.c.} \, ,
\end{eqnarray}
where the left-- and right--handed fermion fields correspond to the SU(2)
multiplets 
\begin{eqnarray}
L= \left( \begin{array}{c} \nu_e \\ e^- \end{array} \right)_L \, \ \  
e^-_R  \, , \ \ Q= \left( \begin{array}{c} u \\ d \end{array} \right)_L,\ \ 
u_R \, ,\ d_R \, ,
\end{eqnarray}
using the notation for the first generation. The Higgs effective  Lagrangian 
will be then 
\begin{eqnarray} 
{\cal L}_H  &\! =\! &   g_{HWW}  H  W_{\mu}^+ W^{- \mu} +  g_{HZZ}  
H Z_{\mu} Z^{\mu}- \sum_f   g_{Hff}  H  \bar f_L  f_R  \! + \! {\rm h.c.} \, .
\end{eqnarray}
The Higgs interactions with particles increase with their masses
and the couplings to gauge bosons and fermions  can be written as
\begin{eqnarray}
\label{higgs_coupling}
g_{Hff} = m_f/v \, ,  \ g_{HWW}= 2 M^2_W/v \, , \  g_{HZZ}=M^2_Z/v\, .
\end{eqnarray}

To be complete, there are also Higgs  self--interactions, residual of those of
the original field $\Phi$ appearing in the potential of eq.~(\ref{Vscalar}), and the magnitude of the Higgs  triple and quartic self--interactions are proportional to $M_H^2$ and are given by
\begin{eqnarray}
{\cal L}_{HHH} \propto g_{HHH}= 3 M_H^2/v~, ~~~{\cal L}_{HHHH} \propto g_{HHHH}= 3 M_H^2/v^2.
\end{eqnarray}

In this review, we will also often need the couplings of the gauge bosons to fermions and we 
introduce them here.  For a given fermion $f$, in terms of its electric charge
$e_f$ in units of the proton charge $e$, its left--handed and right--handed weak
isospin  $I_f^{3L,3R}$ (in the case of SM fermions one has $I_f^{3L} = \pm
\frac{1}{2}$ while $I_f^{3R} = 0$ but we keep the latter no-zero to generalize
to other fermion isospin assignments), and the weak mixing parameter $\sin^2 \theta_W$ that we will often denote $s_W^2\!=\!1\!-\!c_W^2 \!\equiv \!\sin^2 \theta_W$, one can write the vector and axial--vector fermion couplings  to the $Z$ boson as
\beq
v_f = \frac{ \hat{v}_f} { 4 s_W c_W} = \frac{ 2I_f^{3L}+ 2I_f^{3R}-4 e_f s_W^2}{ 4 s_W c_W} \ \ , \ \ a_f = \frac{ \hat{a}_f} { 4 s_W c_W} = \frac{ 2I_f^{3L}+
2I_f^{3R}}{ 4 s_W c_W} \, ,
\label{Zffcouplings}
\eeq
where we also defined the reduced $Z f\bar f$ couplings $\hat{v}_f, \hat{a}_f$
which, for instance in the case of the electron read  $\hat{v}_e= -1+4
s_W^2$ and $\hat{a}_e= -1$. As for the $W$ boson, its vector and axial--vector
couplings to fermions of a same doublet are simply given by (ignoring  CKM mixing in the case of quarks and generalizing to arbitrary isospin) 
\beq
v_f = a_f = \frac{2I_f^{3L}+ 2I_f^{3R}}{2 \sqrt{2}s_W} = \frac{\hat a_f}{4 s_W} = \frac{\hat{v}_f}{4 s_W} \, .
\label{Wffcouplings}
\eeq
In the numerical analyses that we will conduct in this review, we will always
use  the following SM input parameters \cite{Tanabashi:2018oca}: 
\beq
\alpha (M_Z^ 2)= 128.95 \ , \ \  G_F=1.66 \times 10^{-5}\, {\rm GeV}^{-2} ,  \ \ \alpha_s (M_Z^ 2)= 0.118 \, ,
\eeq
for the electromagnetic, Fermi and strong coupling constants and, 
\beq
M_Z\! =\! 91.19\; {\rm GeV}, \  M_W\!=\!80.38\; {\rm GeV}, \ m_t\!=\! 173\,{\rm GeV}, \ \bar m_b\!= \! 4.18\; {\rm GeV}, \ m_\tau\!=\!1.78\, {\rm GeV} \, ,
\eeq 
for the masses of the weak gauge bosons, the heavy top and bottom quarks and $\tau$ lepton. 

\subsubsection{The DM sector in an effective approach}

To the particle content of the SM, whose interactions with the Higgs field have
been described above, we add now weakly interacting massive particles (WIMPs)
that account for the Dark Matter. To describe the phenomenology of the minimal
Higgs--portal scenario for these DM particles,  it is convenient to  work in an
effective and model--independent framework in which these are a real scalar $S$,
a vector $V$ or a fermion $f$ which can be either of the Dirac or Majorana
types. One can then write the interactions of the DM particle with the Higgs
boson in a general, quite model--independent and  simple manner.   In this case,
the phenomenology of the model would be described,   besides the three possible
spin assignments, only by two  parameters in addition to those of the SM: the 
mass of the DM state and its effective coupling to the $H$ boson.  

The relevant  terms of the Lagrangians describing the spin--0, the
spin--$\frac12$ and the spin--1 DM particles  interacting with the SM Higgs
field $\Phi$ can be simply written as: 
\begin{eqnarray} 
\label{Lag:DM}  
\!&&\Delta {\cal L}_S = -\frac12 M_S^2 S^2 -\frac14 \lambda_S S^4 - \frac14 \lambda_{HSS}  \Phi^\dagger \Phi S^2 \;, \nonumber \\ 
\!&&\Delta {\cal L}_V = \frac12 M_V^2 V_\mu V^\mu\! +\! \frac14
\lambda_{V}  (V_\mu V^\mu)^2\! +\! \frac14 \lambda_{HVV}  \Phi^\dagger \Phi
V_\mu V^\mu , \nonumber \\  
\!  &&\Delta {\cal L}_\chi = - \frac12 M_\chi \bar \chi \chi - 
\frac14 {\lambda_{H\chi\chi}\over \Lambda} \Phi^\dagger \Phi \bar \chi \chi \;. 
\end{eqnarray} 
The self--interaction terms $S^4$ in the scalar and the $(V_\mu V^\mu)^2$ term 
in the vector cases are not essential for our discussion and can be ignored. For
fermionic DM, one can consider the cases of both Dirac and Majorana  DM and,
in many cases, the phenomenology is quite similar. Unless otherwise specified,
we assume  in this section the DM to be of Dirac type. 

The  models described by the Lagrangians above, and in fact all the models that
we will consider in this review,    involve a discrete $\mathbb{Z}_2$ symmetry
or parity which ensures the stability of the DM particle\footnote{The origin of
the $\mathbb{Z}_2$ parity is, however, model--dependent. The symmetry can be
connected with other discrete symmetries such as parity and CP symmetries; in
the case of some U(1) vector DM portal for instance, the $\mathbb{Z}_2$ symmetry
could be  related  to charge  conjugation. In fact, discrete 
symmetries should have a gauge origin in order not to be broken by gravitational
effects at high scale and, hence, all these $\mathbb{Z}_2$ that stabilize the DM
particles should be the remnant of the spontaneous breaking of a gauge
symmetry.}. Under this symmetry, the new fields consisting  of the DM particle
and its eventual companions are odd and transform as $\phi_{\rm new} \to
-\phi_{\rm new}$, while the SM fields   are even and transform like $\phi_{\rm
SM} \to \phi_{\rm SM}$. This makes that the new particles can only appear in
even number in interaction vertices, with the important consequence that all new
particles will decay into some lighter partners and gauge or Higgs bosons. The
lightest of these $\mathbb{Z}_2$--odd new particles, since it cannot decay into
SM particles, would be stable and would constitute a potentially good candidate
for the DM.

On should also note that the scalar DM effective model above is theoretically
consistent as long as the $\mathbb{Z}_2$ symmetry is unbroken: it is
renormalizable and can be extrapolated up to extremely high energies provided
that no Landau pole is reached. In turn,  in the  fermionic case, the form that
we adopt for the Higgs--DM coupling is not  renormalisable; the effective
coupling $\lambda_{Hff}$ is damped by a New Physics scale $\Lambda$ and, for
definiteness, we will implicitly assume for it a value of 1 TeV (as we will see
below, it is possible to get rid of the dependence on the scale $\Lambda$
through a coupling redefinition). Although non--renormalisable, the effective
approach is rather minimal and convenient and we will use it in our discussion.
Renormalisable models for fermions, which could in some limit reduce to the
present effective model, will be discussed in the next section.  Finally,
concerning the effective model with a vector DM particle $V_\mu$, a
renormalisable Lagrangian can also be generated by considering the possibility
that it is a gauge boson associated with an Abelian U(1) dark gauge symmetry as
will be also discussed later.

To determine the basic parameters that characterize the Higgs--portal models, after electroweak symmetry breaking, i.e. when the original field $\Phi$ is shifted to $(v + H)/ \sqrt{2}$,   the physical masses of the DM particles $X$ in terms of the Lagrangian mass parameters $M_X$ and their couplings to the Higgs boson $\lambda_{HXX}$,  are then given by
\begin{eqnarray}
m_S^2 &=& M_S^2 + \frac{1}{4}\lambda_{HSS} v^2 \; , \nonumber \\ 
m_V^2 &=& M_V^2 + \frac{1}{4}\lambda_{HVV} v^2 \; , \nonumber \\
m_\chi &=& M_\chi + \frac{1}{4}{\lambda_{H\chi \chi}\over \Lambda} v^2 \;.
\end{eqnarray}
Hence, there are only two free parameters in each spin configuration for the  DM
particles $X$, the mass $m_X$ and the coupling $\lambda_{HXX}$, a set which is
quite minimal. As the DM relic density is determined by its annihilation cross
section which is a function of the DM mass and coupling, a strict
imposition of the PLANCK constraint \cite{Ade:2015xua} would be translated into
a relation between these two parameters. One could however consider, for a more
general discussion from the collider physics perspective, the case that the $X$
particle accounts only for a fraction of the total DM relic density or,
alternatively, it is not cosmologically stable. In this case, constraints from
the DM  relic density can be alleviated or even ignored. 

As already mentioned previously, in the fermionic DM case, the Higgs--portal
operator is of dimension--5 and, hence, it is non--renormalisable and  depends on the New Physics scale $\Lambda$. However, one can get rid of the additional free parameter $\Lambda$ through the coupling redefinition, $\lambda_{H\chi \chi} \rightarrow \lambda_{H\chi \chi} \times v/\Lambda$ which, if the scale is
assumed to be $\Lambda=1$ TeV for instance, would simply lead approximately to a
shift $\lambda_{H\chi \chi} \rightarrow \frac14 \lambda_{H\chi \chi}$. 

We note that if the Higgs boson is integrated out, one obtains effective interactions of the DM particles to fermions and gauge bosons and, in the case 
of light $u,d,s$ quarks of current mass $m_q$ and gluons with $G_{\mu \nu}$ field strength, one obtains 
\begin{eqnarray}
{\cal L}_S^{\rm eff} &= \frac{\lambda_{HSS} } {4M_H^2} |S|^2 \left[ \sum_q m_q 
\bar qq -\frac{\alpha_s}{4\pi} G_{\mu \nu} G^{\mu \nu} \right] \, , \nonumber \\
{\cal L}_\chi^{\rm eff} &= \frac{\lambda_{H\chi\chi} } {4M_H^2} \bar \chi \chi  \left[ \sum_q m_q  \bar q q -\frac{\alpha_s}{4\pi} G_{\mu \nu} G^{\mu \nu} \right] \, , \nonumber \\
{\cal L}_V^{\rm eff} &= - \frac{\lambda_{HVV} } {4M_H^2} V_\mu V^\mu \left[ \sum_q m_q \bar qq -\frac{\alpha_s}{4\pi} G_{\mu \nu} G^{\mu \nu} \right]
\, .
\end{eqnarray}
These expressions will be useful when we will discuss the interaction of the DM with nucleons and their direct detection in astrophysical experiments. 

We have now the elements that allow us to study the phenomenology of  the Higgs
portal to DM scenarios and, in the following subsections, we  discuss the
constraints on the DM states from LHC data in the SM--like Higgs boson searches
and measurements and then, the constraints from Astroparticle physics
experiments. The analytical elements that allow to describe collider Higgs
phenomenology which will be used throughout this review are presented in
Appendix A, while the astrophysical ones,  related to DM annihilation through
Higgs exchange are given in Appendix B.   

\subsection{Collider constraints on DM}

\subsubsection{Higgs production at the LHC}

At hadron colliders such as the  LHC, the  special mass value $M_H=125$ GeV
allows  to observe the SM Higgs particle in many redundant production channels
and to detect it in a  variety of decay modes
\cite{Gunion:1989we,Spira:1997dg,Djouadi:2005gi,Dittmaier:2011ti,Dittmaier:2012vm,Heinemeyer:2013tqa,deFlorian:2016spz,Baglio:2015wcg,Spira:2016ztx,Dawson:2018dcd}.
It is this mass value that  enabled the very detailed studies of the Higgs
properties, which have been performed by the ATLAS and CMS collaborations
already in the first LHC run with $\sqrt s=7$ and 8 TeV center of mass energies
\cite{Khachatryan:2016vau}.   The analytical elements that allow to describe the
Higgs boson decays and production mechanisms at hadron colliders have been
relegated to Appendices B1 and B2, respectively, and we simply summarize the
main features here.  

\begin{figure}[h]
\begin{center}
\vspace*{-2.4cm}
\mbox{\hspace*{-.6cm} \resizebox{0.9\textwidth}{!}{\includegraphics{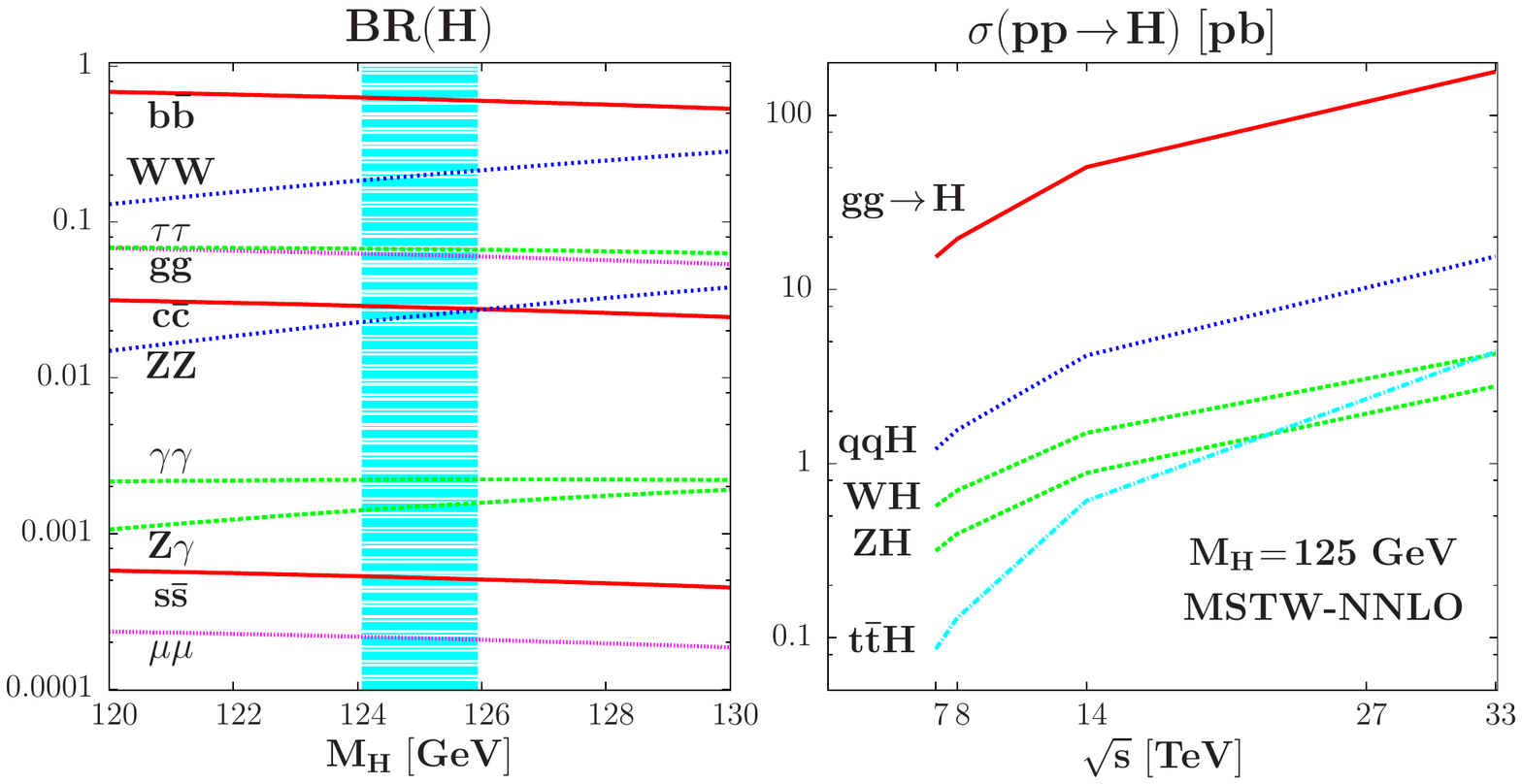}} }
\vspace*{-12cm}
\end{center}
\caption{The SM Higgs branching ratios in the mass range $M_H=120$--130 GeV
obtained using the program {\tt HDECAY} \cite{Djouadi:1997yw,Djouadi:2006bz,Djouadi:2018xqq} (left) and its cross sections at proton colliders as a function of  the c.m. energy in the various production modes including higher order effects, obtained using the programs of 
Refs.~\cite{Michael-web,Spira:1997dg} (right).}
\label{fig:allH}
\vspace*{-.2cm}
\end{figure}

Considering first the decay modes, for $M_H = 125$ GeV, the Higgs mainly decays 
into $b \bar b$ pairs but the channels with $WW^*$ and $ZZ^*$ final states,
before  allowing the gauge bosons to decay leptonically $W \! \to \! \ell \nu$
and $Z\! \to  \! \ell \ell$ ($\ell\! =\! e,\mu$), are also significant. The $H\!
\to \! \tau^+\tau^-$  channel (as well  as the $gg$ and $c\bar c$ decays that
are not detectable at the LHC) is also of significance,  while the clean loop
induced $H\to \gamma \gamma$ mode can be easily detected albeit  its small
rates. The very rare $H\to Z\gamma$  and even $H\to \mu^+\mu^-$ channels  should
be accessible at the LHC but only with a much larger data sample
\cite{ATLAS:2013hta,CMS:2013xfa,Cepeda:2019klc}. These features are 
illustrated  in the left--hand side of Fig.~\ref{fig:allH} where the  decay
branching fractions  of a SM--like Higgs are displayed for the mass range
$M_H=120$--130 GeV.  For this purpose, we have used the program {\tt HDECAY}
\cite{Djouadi:1997yw,Djouadi:2006bz,Djouadi:2018xqq} which calculates the
partial widths and the branching ratios of all Higgs decays (in the SM but also
in some of its extensions like the 2HDM and the MSSM  as will be seen later in
this review) including all relevant higher order effects. 

On the other hand, many Higgs production processes have significant cross
sections as is shown in the right--hand side of Fig.~\ref{fig:allH} where, for a
125 GeV SM Higgs boson,  they are displayed  at a proton collider at various
past, present and foreseen center of mass energies. They have been obtained
using the programs of Refs.~\cite{Michael-web,Spira:1997dg} which include all
relevant higher order QCD corrections; the MSTW parton densities
\cite{Martin:2009iq} have been  used. While the by far dominant gluon  fusion
mechanism $gg\to H$  (ggF) has extremely large rates, the subleading channels,
i.e. vector boson fusion (VBF)  $qq\! \to\! Hqq$ and Higgs--strahlung (HV)
$q\bar q \!\to \!HV$ with $V\!=\!W,Z$ mechanisms, have cross sections which
allow for the Higgs study already at $\sqrt  s \!=\!7$ and $8$ TeV with the
$\approx \!  5$ and 20 fb$^{-1}$ of integrated luminosity  collected there  by
each experiment. Associated Higgs production with top quark pairs (ttH), $p p\!\to \!t\bar  tH$,  requires higher luminosity to be precisely probed. 

This production/decay pattern already allows the ATLAS and CMS experiments to
observe the Higgs boson in several channels and to measure some of its couplings
in a reasonably accurate way \cite{Khachatryan:2016vau}.  The main
topologies in which the Higgs boson  has been searched for at the LHC are the
following:\vspace*{-1mm}

-- $H \! \to \! ZZ$ with a real and a virtual $Z$ decaying into leptons  $ZZ^* \! \to \! 4\ell^\pm$ with $\ell=e+\mu$;\vspace*{-1mm}

-- $H\! \to \! WW$ with a real and virtual $W$ decaying into leptons
$WW^* \! \to \! 2\ell 2 \nu$ with $\ell=e+\mu$;\vspace*{-1mm}  

-- $H \! \to \! \gamma\gamma$ which proceeds through loops involving the $W$ boson and the top quark;\vspace*{-1mm} 

\noindent in these three processes, the Higgs is  mainly produced in ggF with
subleading contributions from $Hjj$ in the VBF process and, to a lesser extent, 
VH as well as ttH production;\vspace*{-1mm} 

-- $H \! \to \! \tau \tau$  with $H$  produced in  association with one (in ggF) and two (in VBF) jets;\vspace*{-1mm}

-- $H\to b \bar b$  with the Higgs boson mainly produced in the HV process with
$V \to \ell=e+\mu$.\vspace*{-1mm}  

As already mentioned, the additional decay channels $H\! \to \! \mu \mu$ and
$H\! \to \! Z\gamma$ as well as the production channel $pp\to t\bar t H$ play a little role for the time being.

A convenient way to study the couplings of the $H$ boson at the LHC is to look
at its deviations from the SM expectation. One then considers for a given
search  channel, the signal strength modifier $\mu$ which can  be  identified as
the Higgs production cross section times  the decay branching fraction 
normalized to the  values expected in the SM  \cite{Khachatryan:2016vau}.
For the $H\! \to \! XX$ decay channel for instance, one would have  
\beq
\mu_{XX}={\sigma( pp \to H \to XX)} /{\sigma( pp \to H\to XX)|_{\rm SM}}\, , 
\eeq
which, in the narrow width approximation, can be simply rewritten as 
\beq
\mu_{XX}=  \frac{ \sigma( pp \to H)\times {\rm BR} (H \to
XX)} {\sigma( pp \to H)|_{\rm SM} \times {\rm BR} (H \to XX)|_{\rm SM} }.
\eeq

\begin{figure}[!h]
\vspace*{-1mm}
\begin{center}
\begin{tabular}{ll}
\begin{minipage}{4cm}
\hspace*{-4.5cm}
\resizebox{1.6\textwidth}{!}{\includegraphics{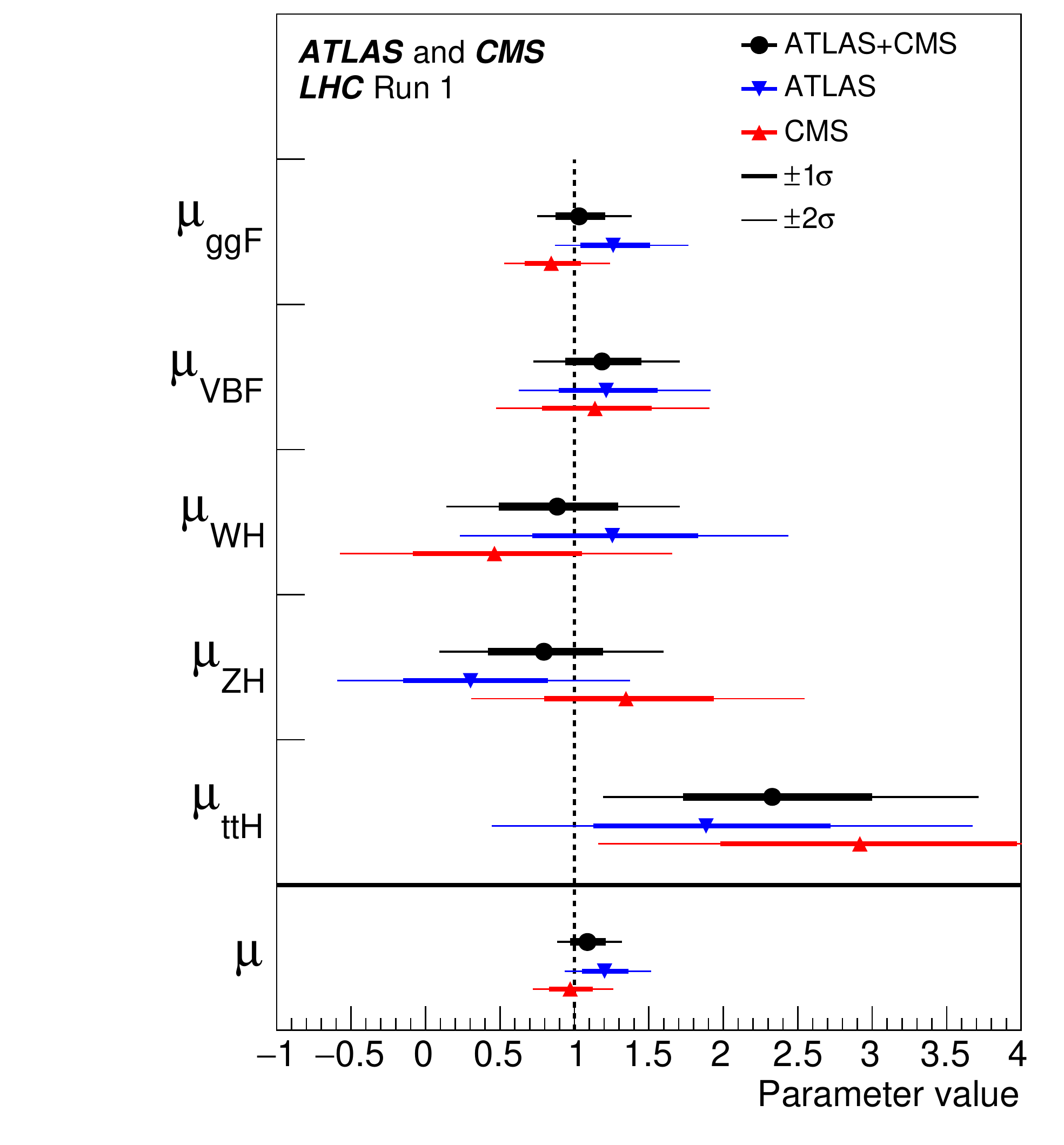}}\hspace*{-.3cm}
\end{minipage}
& \hspace*{-1.9cm} 
\begin{minipage}{4cm}
\resizebox{1.95\textwidth}{!}{\includegraphics{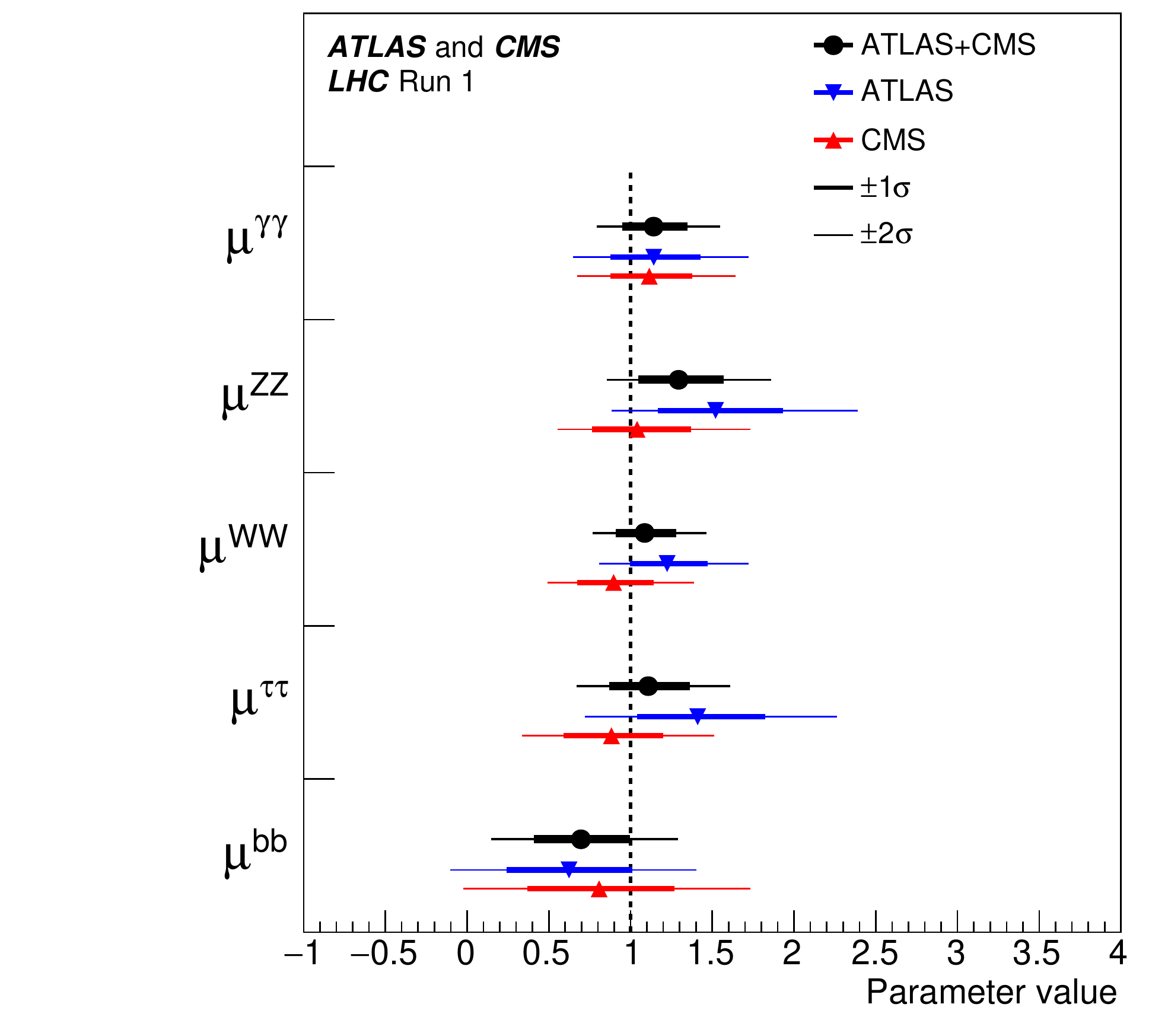}}\hspace*{.3cm}
\end{minipage}
\end{tabular}
\end{center}
\vspace*{-6mm}
\caption{
Best fit results for the production (left) and  decay (right) signal strengths in the ATLAS and CMS data and their combination.  In both cases, the error bars indicate the $1\sigma$ (thick lines) and $2\sigma$ (thin lines) intervals.  From
Ref.~\cite{Khachatryan:2016vau}. }
\label{Fig:constraints}
\vspace*{-5mm}
\end{figure}

ATLAS and CMS have provided the signal strengths for the various final states at
the RunI of the LHC, i.e. at $\sqrt s=7$ TeV with $\approx 5$ fb$^{-1}$ data and
at $\sqrt s=8$ TeV with $\approx 20$ fb$^{-1}$ data, and they combined their
results in Ref.~\cite{Khachatryan:2016vau}. The individual constraints of ATLAS
and CMS and their combinations are shown in Fig.~\ref{Fig:constraints} for the
signal strengths in the  production (left) and decay (right) channels. The 
$1\sigma$  and  $2\sigma$ error bars are indicated.  As can be seen, no
deviation from the SM expectation is observed. This is particularly the case in
the $H\to ZZ,WW$ and $H\to \gamma\gamma$ channels in which the measurements have
been performed at the level of 20\% accuracy with combined ATLAS+CMS results of
\cite{Khachatryan:2016vau}
\beq 
\mu_{\gamma\gamma}= 1.14 ^{+0.19}_{-0.18} \ , \ \ 
\mu_{ZZ}= 1.29 ^{+0.26}_{-0.23} \ , \ \ 
\mu_{WW}= 1.09 ^{+0.18}_{-0.16} \ , \ \ 
\eeq
where the uncertainty is for the combination of the statistical, systematical
and theoretical errors. More accurate results in some channels  have been
obtained by ATLAS and CMS at $\sqrt s=13$ TeV (in particular in the fermionic
ones, where e.g. observations at $5\sigma$ have been finally made in the $b\bar
b$ mode \cite{Aaboud:2018zhk,Sirunyan:2018kst}), but a global combination of
their results was not performed. Here, we thus restrict to RunI results which
for our purpose, are sufficient.

The result which summarizes these studies and that we will use later on, is the total $\mu$ value at RunI obtained from a combined ATLAS and CMS fit of all Higgs production and decay channels, when the various uncertainties are combined \cite{Khachatryan:2016vau}
\beq 
\mu_{\rm tot}= 1.09 ^{+0.11}_{-0.10} \ \ \Rightarrow \ \ \mu_{\rm tot} \geq 0.89~{\rm at}~95\%{\rm CL} \, , 
\label{eq:mutotLHC}
\eeq  
which shows that the observed Higgs boson deviates from the SM--like behaviour by less than one standard deviation. In our context of invisible Higgs decays, it is the 95\% confidence level (CL) limit above which is relevant and that we will use.  

The previous ATLAS and CMS constraints can be turned into limits on the
couplings modifiers of the Higgs boson, defined as the production cross sections
or the decay rates in a specific channel normalised to the SM values.
Ultimately, they would correspond to the deviations of the $H$ coupling to a
given particle $X$ from the SM expectation\footnote{We consider here the
possibility of non--standard Higgs couplings, in anticipation of the next
sections in which  we will discuss New Physics scenarios where this would
occur.},  
\beq
\kappa_X^2= \frac{\sigma (X) }{\sigma (X) |_{\rm SM}} \ , \ 
\kappa_X^2 = \frac{\Gamma( H\to XX) }{\Gamma (H \to XX) |_{\rm SM}} \  \ 
\Rightarrow \kappa_X= \frac{g_{HXX}} {g_{HXX}|_{\rm SM}} \, .
\eeq
For most production and decay channels, only one Higgs coupling is modified at a time
\begin{eqnarray}
\Gamma(H\to ZZ) \to  \kappa_Z^2 \ , \ \Gamma(H\to WW) \to  \kappa_W^2 \ , \ \ 
\Gamma(H\to bb) \to  \kappa_b^2 \ , \ \Gamma(H\to \tau\tau) \to  \kappa_\tau^2
\ ,  \nonumber \\
\sigma (WH) \to \kappa_W^2 \ , \ \ \sigma (ZH) \to \kappa_Z^2 \ , \ \ 
\sigma (ttH) \to \kappa_t^2 \ , \ \ \sigma ({\rm VBF}) \to 0.74 \kappa_W^2 + 0.26 \kappa_Z^2 \, ,~~~ 
\label{kappa-all}
\end{eqnarray}  
but there are exceptions: for Higgs production in VBF where both the $WW$ and $ZZ$ fusion processes are present, and especially for the loop induced $gg\to H$ production mechanism and the $H\to \gamma \gamma$ decay channel which proceed through the exchange of mainly top and bottom quarks in the first case and  top quarks and $W$ bosons in the second one:
\begin{eqnarray}
\Gamma(H\to \gamma\gamma) & \to &  \kappa_\gamma^2 = 1.56 \kappa_W^2 +0.07 
\kappa_t^2 -0.66 \kappa_t \kappa_W \, , \nonumber \\
\sigma(gg \to H) & \to &  \kappa_g^2 = 1.06 \kappa_t^2 +0.01 
\kappa_b^2 -0.07 \kappa_t \kappa_b \, .
\label{kappa-loop}
\end{eqnarray}
The total decay width of the Higgs boson, where all channels contribute,  is then modified by the amount (which explicitly gives the branching ratios for the various channels) 
\begin{eqnarray}
\Gamma_H  \to  \kappa_H^2 &= &0.57 \kappa_b^2 +0.22 \kappa_W^2 + 
0.06 \kappa_\tau^2 + 0.03 \kappa_Z^2 + 0.03 \kappa_c^2 + 0.0023 \kappa_\gamma^2
\nonumber \\
&& \hspace*{-4mm} + 0.0016 \kappa_{(Z\gamma)}^2 + 0.00022 \kappa_\mu^2 + 0.0001 \kappa_s^2 \, .
\label{kappaH}
\end{eqnarray}

\begin{figure}[!h]
\vspace*{-1mm}
\begin{center}
\begin{tabular}{ll}
\begin{minipage}{4cm}
\hspace*{-4cm}
\resizebox{1.75\textwidth}{!}{\includegraphics{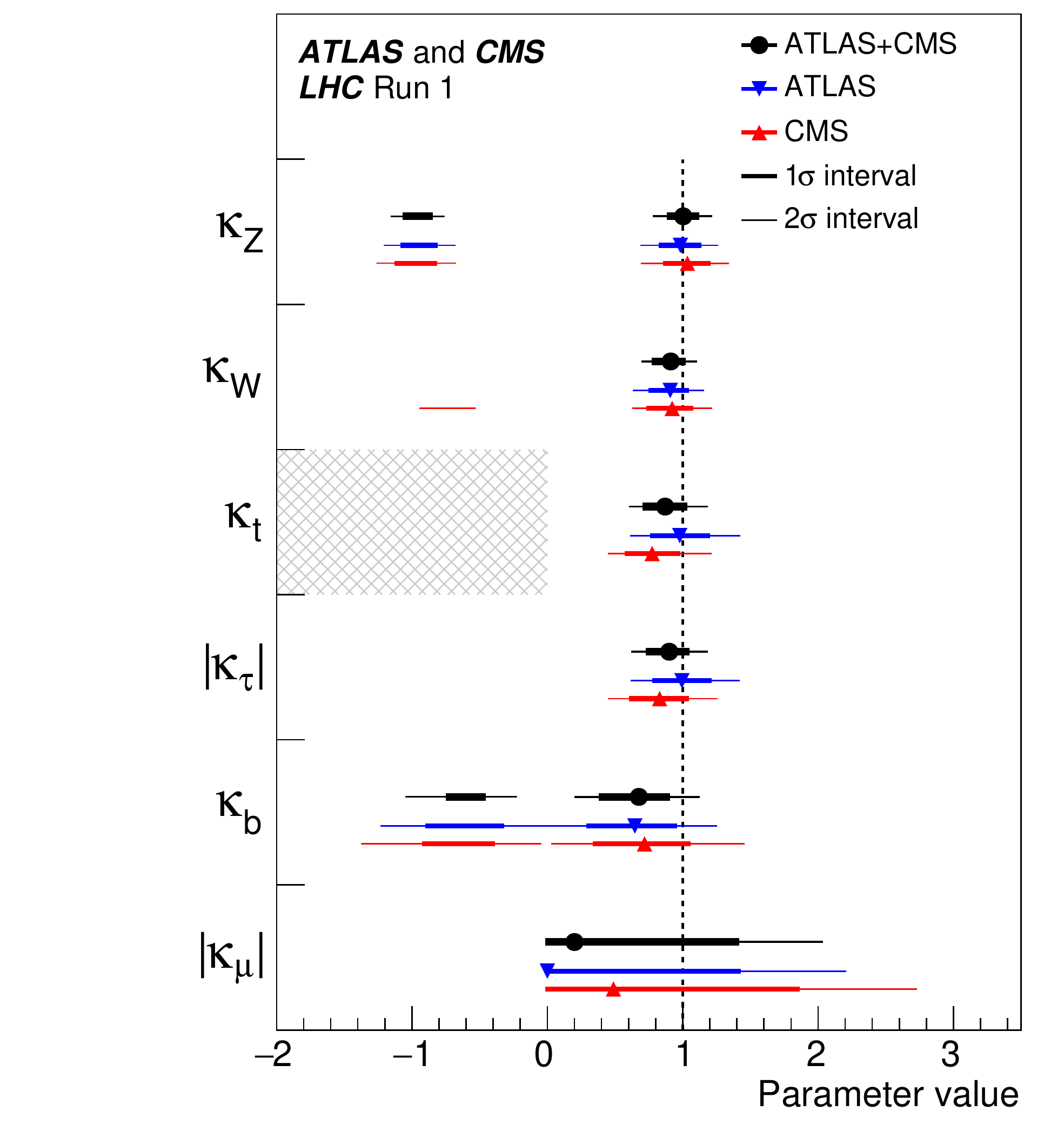}}\hspace*{-.3cm}
\end{minipage}
& \hspace*{-1cm} 
\begin{minipage}{4cm}
\resizebox{1.73\textwidth}{!}{\includegraphics{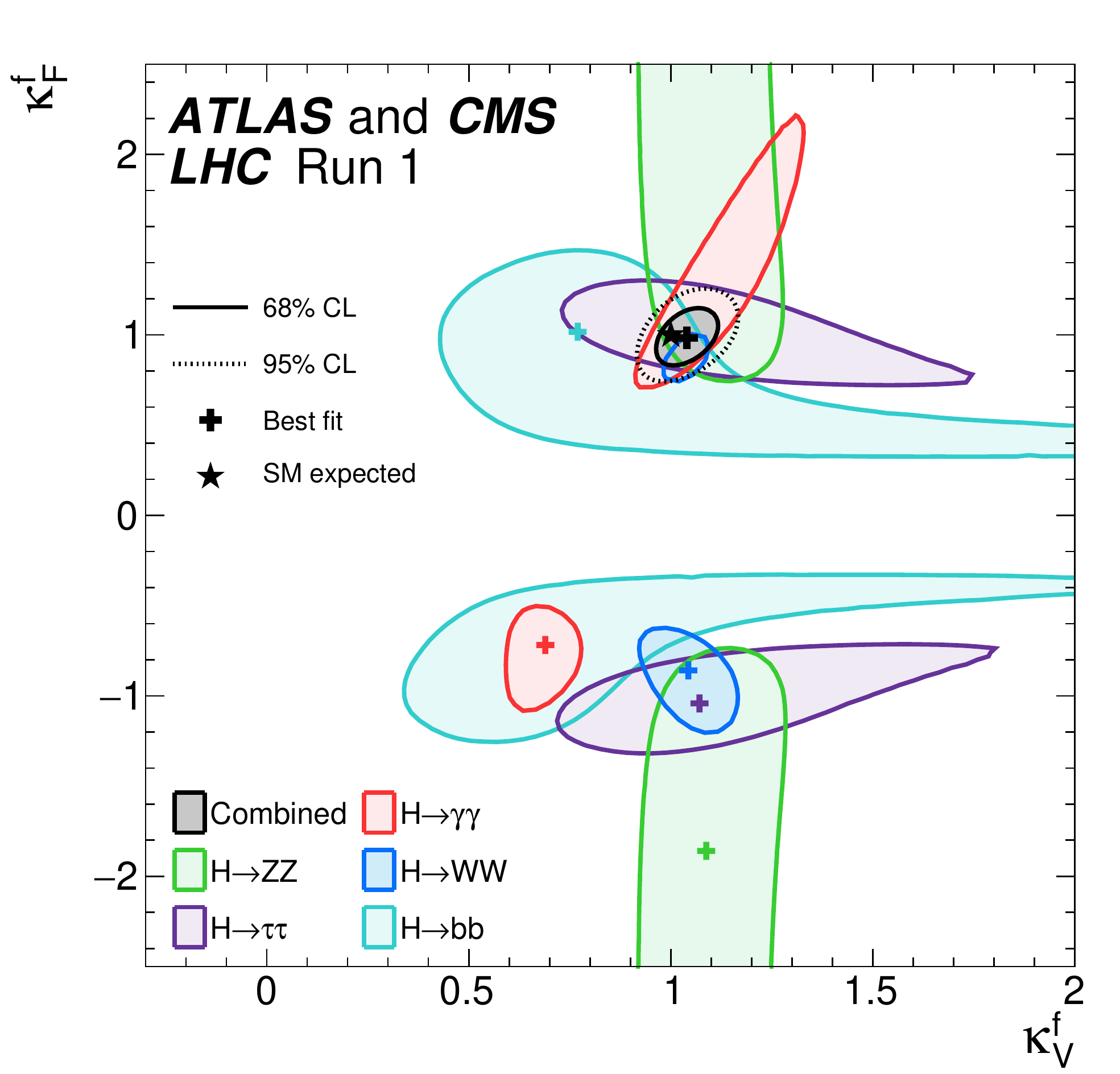}}\hspace*{-.3cm}
\end{minipage}
\end{tabular}
\end{center}
\vspace*{-6mm}
\caption{
Left: best fit values of the coupling modifiers $\kappa_X$ in the ATLAS and CMS RunI data and their combination, assuming no new particles in the loops; the error bars indicate the $1\sigma$ (thick lines) and $2\sigma$ (thin lines) intervals. Right: log--likelihood contours at 68\% and 95\%CL in the ($\kappa_f, \kappa_V)$ plane for the combination of ATLAS and CMS and for the individual decay channels and their combination (in black). From Ref.~\cite{Khachatryan:2016vau}. }
\label{Fig:constraints-kappa}
\vspace*{-1mm}
\end{figure}

These couplings modifiers, as determined by ATLAS only, CMS only and by the
combined results of the two collaborations  are shown in the left--hand side of
Fig.~\ref{Fig:constraints-kappa} for the RunI LHC. All channels in the
production and in the decays have been included, using the expressions
eqs.~(\ref{kappa-all}-\ref{kappaH}) for the various contributions and, hence,
assuming the absence of additional non--standard particles in the loops. All
couplings were left free with some minimal assumptions. The $1\sigma$ and
$2\sigma$ intervals for the  error bars are indicated.  As can be seen, some
couplings like $\kappa_W, \kappa_Z$ are measured with an accuracy of about 10\%
which is in line with the previous discussion (since the cross sections and
decay signal strengths are proportional to $\kappa_X^2$, their error is hence
twice the one that affects the reduced couplings). 

In the right--hand side of Fig.~\ref{Fig:constraints-kappa}, negative  68\% and
95\% confidence level (CL) log--likelihood contours are displayed in the
$(\kappa_f, \kappa_V)$ plane for the combined RunI ATLAS and CMS measurements in
various channels  and their combination (in black) with no assumption on the
sign of the couplings. Two other quadrants, symmetric with respect to
the point (0,0), are not shown.  For the upper quadrant, the SM expectation
falls in the middle of the combined measurement which sets strong constraints 
on $|\kappa_f|$ and   $|\kappa_V|$. 

From this discussion, one concludes that both signal strengths and coupling
modifiers as determined by the ATLAS and CMS collaborations, are rather close to
unity, implying that the Higgs boson is SM--like at the level of roughly $10\%$.
This leaves only little room for non standard Higgs decays, such as the
invisible modes  to which we now turn.

\subsubsection{Collider constraints on invisible Higgs decays}

If the DM particles $X$ are light enough, $m_{X} \leq \frac12 M_H$, the 
invisible Higgs decays would occur at the two--body level and, for the  three
spin cases discussed in the previous subsection,  the Higgs partial decay  widths can be simply written as
\begin{eqnarray}
 &&\Gamma_{{\rm inv}} ( H\rightarrow SS) = \frac{\lambda_{HSS}^2 v^2 \beta_S}{64 \pi  M_H}  \;, \nonumber\\
&& \Gamma_{\rm inv} (H \rightarrow V V) = \frac{\lambda^2_{HVV} v^2 M_H^3
\beta_V }{256 \pi   M_{V}^4}
\left( 1-4 \frac{M_V^2}{M_H^2}+12\frac{M_V^4}{M_H^4}
\right), \nonumber\\
&&  \Gamma_{\rm inv} (H \rightarrow f f  ) = {\lambda_{H ff}^2 v^2 M_H \beta_f^{3}\over 32 \pi \Lambda^2 }  \;, 
\label{GammaInv}
\end{eqnarray}
where $\beta_X=\sqrt{1-4m_X^2/M_H^2}$ is the velocity of the DM state. To
precisely evaluate the invisible Higgs decay branching ratios, BR($H\to {\rm
inv}) =  \Gamma_{\rm inv} (H \rightarrow XX)/\Gamma_H $,  we have adapted the program HDECAY \cite{Djouadi:1997yw,Djouadi:2006bz,Djouadi:2018xqq} to incorporate them.

\underline{Invisible Higgs decays from the total  decay width.}

Invisible decays could be constrained if the total decay width of the 125 GeV
Higgs boson, given in  eq.~(\ref{kappaH}) up to a normalization which would lead
to the value  $\Gamma_H^{\rm SM}\!=\! 4.07\,$MeV in the SM, could be
determined.  A direct measurement of  $\Gamma_H$  would have been possible for a
heavy Higgs boson by  exploiting the process $H \!\to \!ZZ \!\to \! 4\ell^\pm$
for instance: beyond the value $M_H \! \gsim \! 180$ GeV, $\Gamma_H \! \gsim \!
1$ GeV, and would have been large enough to be measured. In  fact, for even
higher masses, the total width is so large, $\Gamma_H\! \gsim\! 100$ GeV for
$M_H\! \gsim\! 500$ GeV as a result of the longitudinal components of the
massive gauge bosons which make the partial widths $\Gamma(H \to VV)\! \propto\!
M_H^3$, that it contributes significantly to the cross section.  In  contrast,
the total decay width of the SM--Higgs is far too small to be resolved
experimentally.

However, rather recently, it was noticed \cite{Kauer:2012hd,Caola:2013yja} that
in the production channel  $pp \! \to VV  \! \to \! 4f$ with $V\!=\!W,Z$, a
large fraction of the Higgs--mediated cross  section lies in the high--mass tail
where the invariant mass of the $VV$ system is  larger than $2M_V$. For instance, at $\sqrt s=8$ TeV, approximately 15\% of the total cross section in the $pp \to H \to ZZ^*$ process has an invariant mass of $M_{ZZ} \gsim  140$ GeV, so that off--shell Higgs events can be measured and the Higgs total width can be probed. The main idea is that  the Breit--Wigner Higgs propagator in the above process being $1 / [( M^2_{ZZ} - M^2_H)^2 + M^2_H \Gamma_H^2]$, one can extract information on $\Gamma_H$ by measuring the cross section at the Higgs resonance and above. Indeed, assuming a Higgs dominantly produced in $gg$ fusion, the  cross section reads 
\beq
\frac{d \sigma_{gg \to H \to ZZ}}{ d M_{ZZ}^2} \propto \frac{g^2_{ggH} g^2_{HZZ}} {(M^2_{ZZ } - M^2_H )^2 + M^2_ H \Gamma^2_H} \, , 
\eeq
where $g_{ggH}$ and $g_{HZZ}$ are the Higgs couplings  to gluons and $Z$ bosons, respectively. Integrating either in a small region around $M_H$, or above the mass threshold $M_{ZZ} > 2M_Z$ with  $M_{ZZ} -M_H \gg \Gamma_H$, the on--shell and off--shell cross sections are, respectively, $\sigma_{gg \to H \to ZZ}^{\rm on\!-\!shell} \propto  \frac{g^2_{ggH} g^2_{HZZ}} {M_H\Gamma_H}$ and $ \sigma_{gg \to H \to ZZ}^{\rm off\!-\!shell} \propto  \frac{g^2_{ggH} g^2_{HZZ}} {2M_Z^2}$. This means that a measurement of the two observables provides direct information on $\Gamma_H$ if the coupling ratios remain the same\footnote{For instance, one should assume that gluon fusion production is dominated by the top--quark loop and there are no new particles contributing to the process. Note that the on--shell cross section is unchanged under a common scaling of the squared product of the couplings and of the total width $\Gamma_H$, while the off--shell production cross section increases linearly with this scaling factor.}.  

The dominant contribution in the $pp\to ZZ$ production process is due to  the
tree--level quark--initiated  process  $q q \to ZZ$ but there is also a gluon
induced mechanism  $gg \to ZZ$ from a one--loop box diagram which has a large
rate at high energies. There is a significant and destructive interference
between the $gg\to H\to ZZ$ signal and the continuum $gg\to ZZ$ background in
the off--shell region that is mainly due to  two  threshold  effects, one near
$2M_Z$ from  the $H\to ZZ$ decay and the other near $2m_t$ from $gg \to H$
production. 

\begin{figure}[!ht]
\vspace*{-1mm}
\begin{center}
\begin{tabular}{ll}
\begin{minipage}{4cm}
\hspace*{-4cm}
\resizebox{1.86\textwidth}{!}{\includegraphics{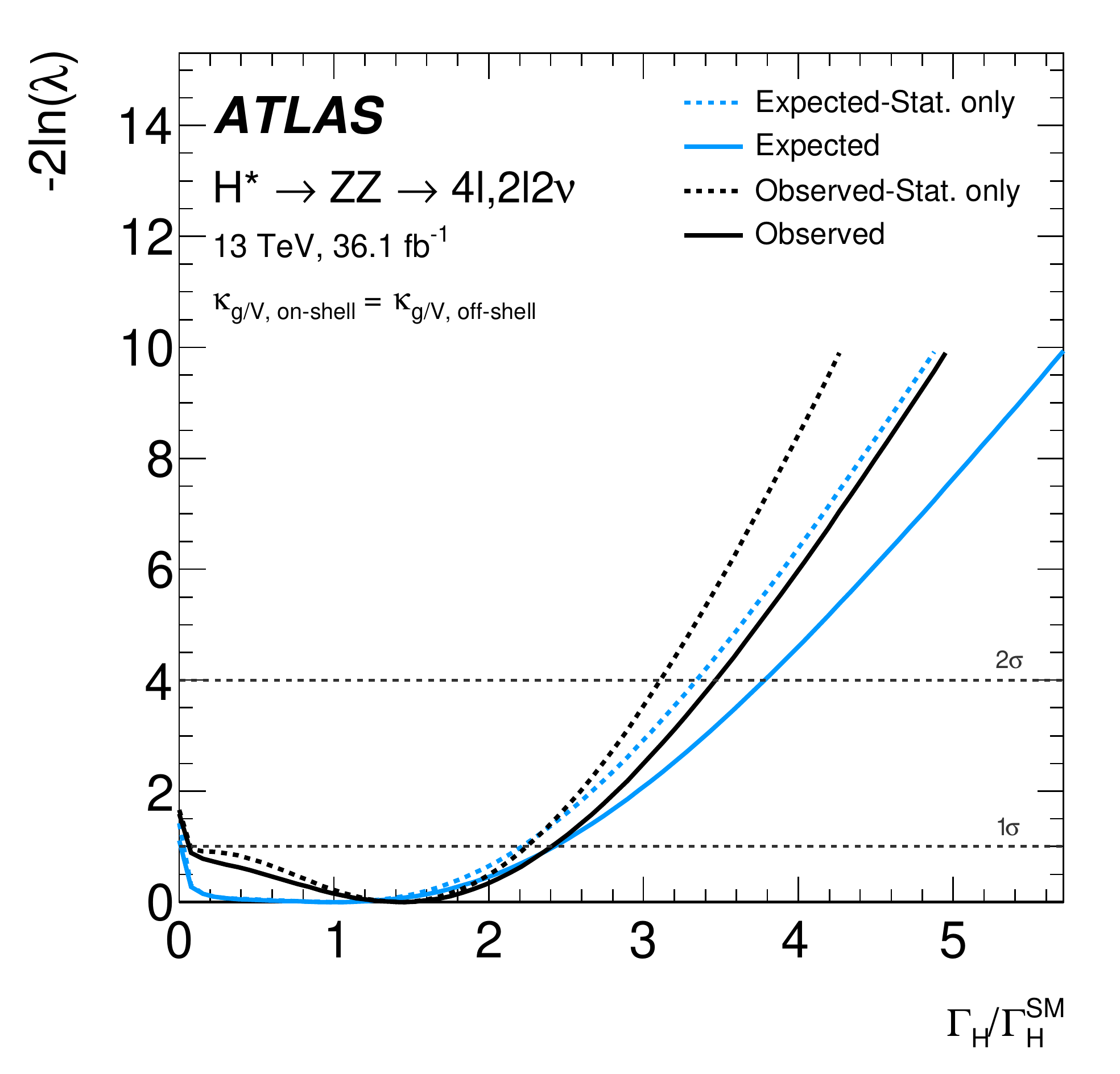}}\hspace*{-.3cm}
\end{minipage}
& \hspace*{-.1cm} 
\begin{minipage}{4cm}
\resizebox{1.84\textwidth}{!}{\includegraphics{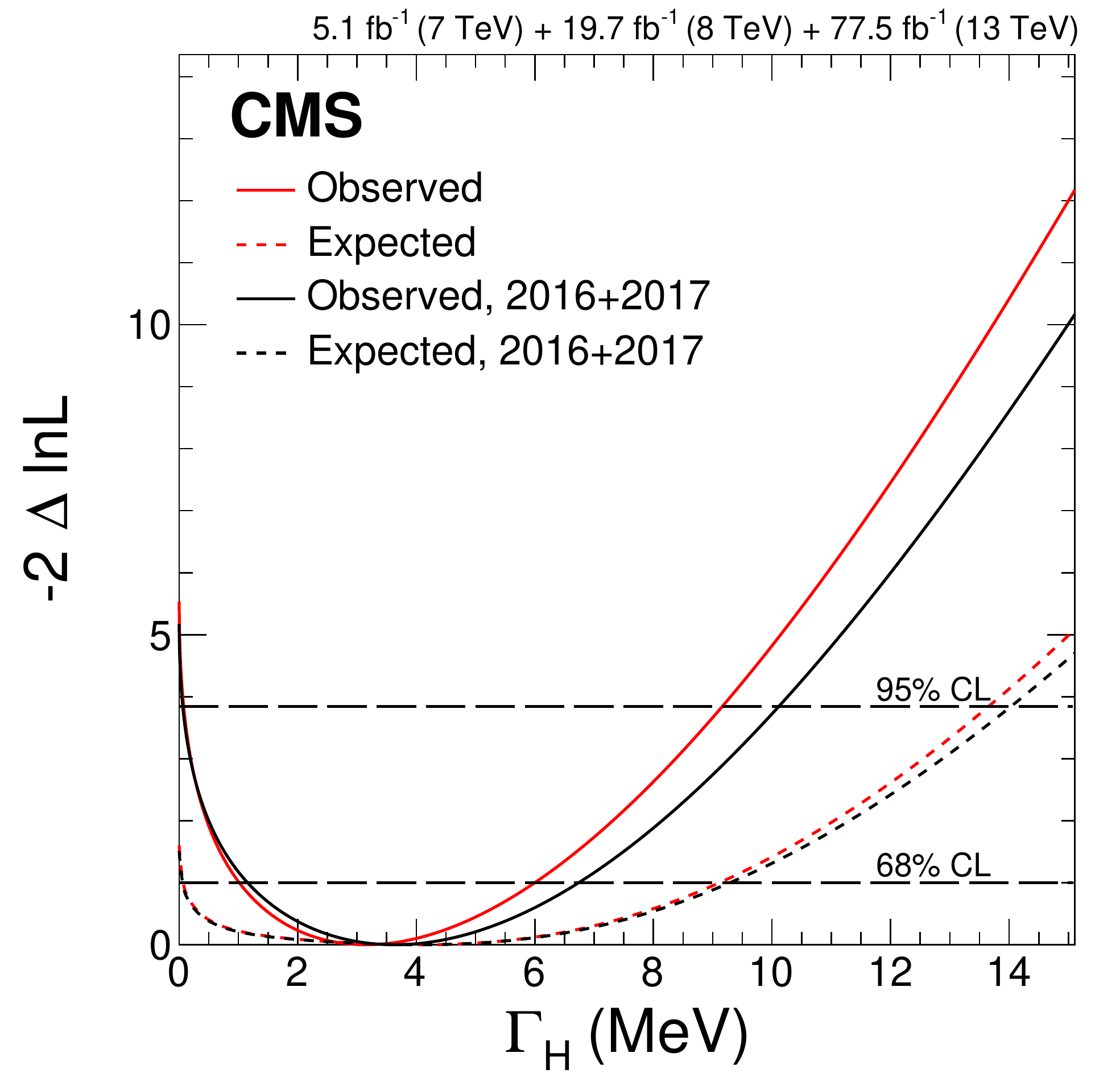}}\hspace*{-.3cm}
\end{minipage}
\end{tabular}
\end{center}
\vspace*{-7mm}
\caption{
Left: log--likelihood contours for expected and observed  values of the Higgs total width compared to the SM expectation  $\Gamma_H / \Gamma_H^{\rm SM}$ from ATLAS at $\sqrt s=13$ TeV \cite{Aaboud:2018puo}. 
Right: log--likelihood contours as a function of $\Gamma_H$ from off--shell measurements in the $pp\to H \to ZZ^* \to 4\ell$ process with all production channels included performed by CMS when full Run I data are combined with 77.5 fb$^{-1}$ data collected at $\sqrt s =13$ TeV  \cite{Sirunyan:2019twz}.}
\label{Fig:gammaH}
\vspace*{-1mm}
\end{figure}

It is this feature that allows to  constrain the total width $\Gamma_H$ at the
LHC.  For instance,  off--shell measurements made by ATLAS  in the channel $pp\!
\to \! H \! \to \! ZZ^* \! \to \! 4\ell^\pm, 2 \ell 2\nu$  at $\sqrt s\!=\!13$
TeV with a luminosity of 36 fb$^{-1}$, combined with signal strength
measurements in on--shell processes,  lead to an observed upper limit on the
Higgs total width  of $\Gamma_H < 14.4\,$MeV at the 95\%CL
\cite{Aaboud:2018puo}, which means a limit $\kappa_H \lsim 3.6$ for the signal
strength or coupling modifier. This is exemplified in Fig.~\ref{Fig:gammaH}
(left)  where the negative log--likelihood $-2\log \lambda$, is shown for the
ratio $\Gamma_H/\Gamma_H^{\rm SM} \equiv \kappa_H$ when all measurements are
combined  \cite{Aaboud:2018puo}. 

Slightly better results have been obtained by CMS  using only the 7 and 8 TeV
data when  combining the $H \to ZZ^*$ and $H\to WW^*$ off--shell decay channels 
with the Higgs produced in both the ggF and VBF processes. Assuming an SU(2)
custodial symmetry which provides the constraint  $\mu_{ZZ}\!=\!\mu_{WW}$, one
obtains a total decay width $\Gamma_H <13$ MeV at the 95\%CL (but an expected
limit of only 26 MeV)  \cite{Khachatryan:2016ctc}.

However, in a very recent analysis \cite{Sirunyan:2019twz}, CMS has considered
the $pp\to H \to ZZ^* \to 4\ell$ process with all production channels, namely
ggF, VBF, HV and ttH production, using 77.5 fb$^{-1}$ data collected at $\sqrt s
=13$ TeV  and combined it with the 25 fb$^{-1}$ data  collected at RunI. The
value obtained for the total Higgs width is amazingly precise, $\Gamma_H = 3.2
^{+2.8}_{-2.2}$ MeV with an expected measurement based on simulation of
$\Gamma_H = 4.1^{+5.0}_{-4.0}$ MeV, leading to a 95\%CL upper limit on the Higgs
width of $\Gamma_H < 2.15\, \Gamma_H^{\rm SM}$ again at 95\%CL. This is shown in
the right--hand side of Fig.~\ref{Fig:gammaH} where the  log--likelihood as a
function of the Higgs total width is displayed for  observation (solid lines)
and expectation (dashed lines) using RunII data only (black lines) and a
combination of RunI and RunII data (red lines). 

Such a precise result was really unexpected and there is now a hope that one
could reach a precision of less than 100\% and maybe even a few 10\% on the total Higgs width. Nevertheless, we should again emphasize the fact that the
measurement is model dependent and  strongly rely  on the
assumption that the off--shell  Higgs couplings are exactly the same as the
on--shell ones, which has been shown  not to be the case  in many scenarios
where New Physics is involved.  These bounds should be thus taken with care in beyond the SM scenarios as they can be relaxed in many cases.

\underline{Direct searches for invisible Higgs decays.}

A more direct and less model--dependent approach would be to perform direct
searches for topologies involving missing transverse energy that would signal
the production of a Higgs boson which then decays into invisible particles
\cite{Choudhury:1993hv,Eboli:2000ze,Godbole:2003it}.  Such searches have been
conducted  by ATLAS and CMS  in particular in the processes
\cite{Aaboud:2017bja,Khachatryan:2016whc}
\begin{eqnarray}
pp & \to & q \bar q  \to HV \to V E_T \hspace*{-3mm}\slash  \to f \bar f~  ,   \nonumber \\
pp & \to& qq \to qqV^*V^* \to qq E_T \hspace*{-3mm}\slash ~. 
\end{eqnarray}

\begin{figure}[!h]
\vspace*{-2mm}
\begin{tabular}{ll}
\begin{minipage}{8cm}
\hspace*{-.5cm}
\resizebox{1.06\textwidth}{!}{\includegraphics{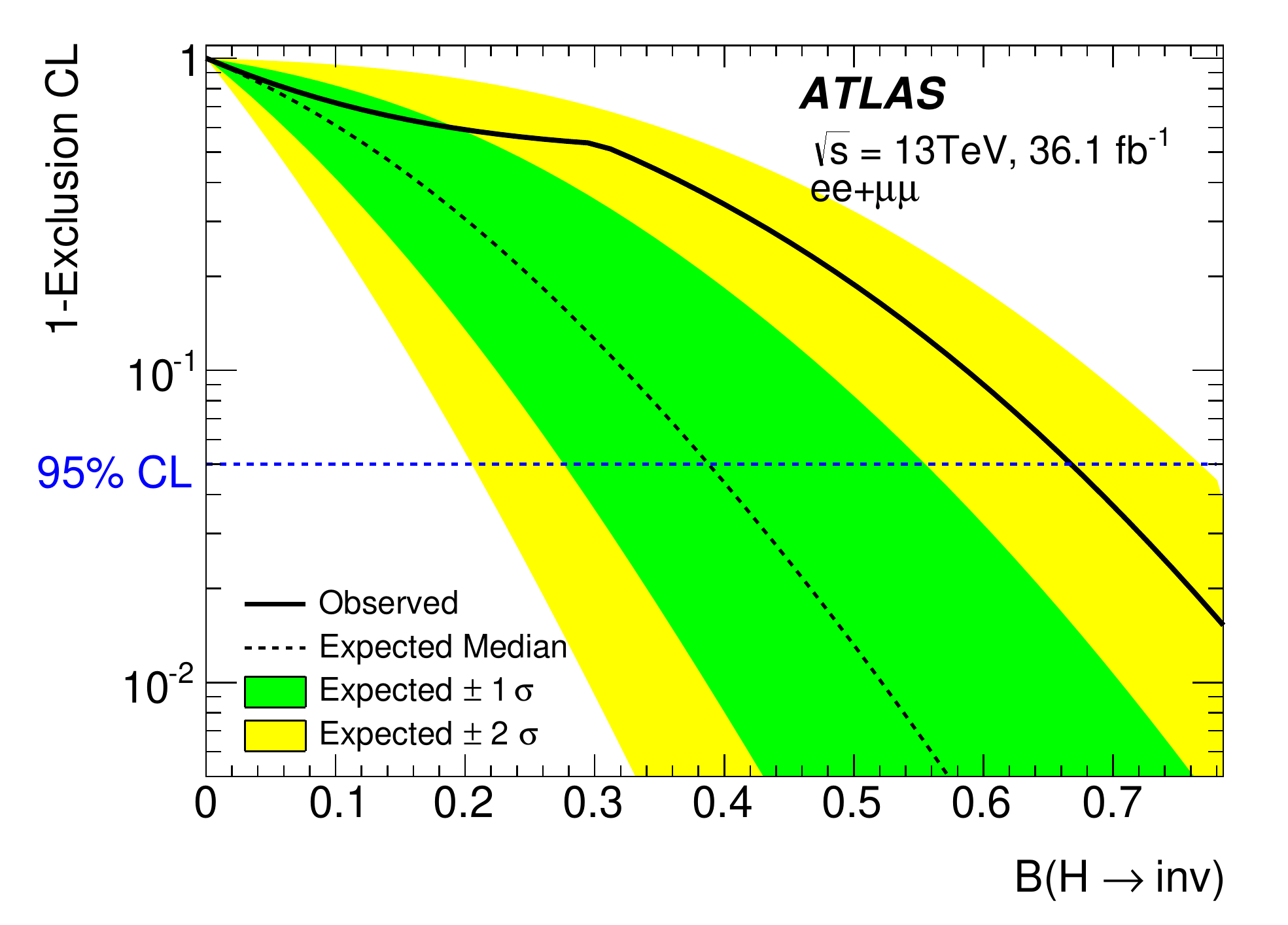}}
\end{minipage}
& \hspace*{-.6cm} 
\begin{minipage}{8cm}
\vspace*{-.5cm}
\resizebox{1.1\textwidth}{!}{\includegraphics{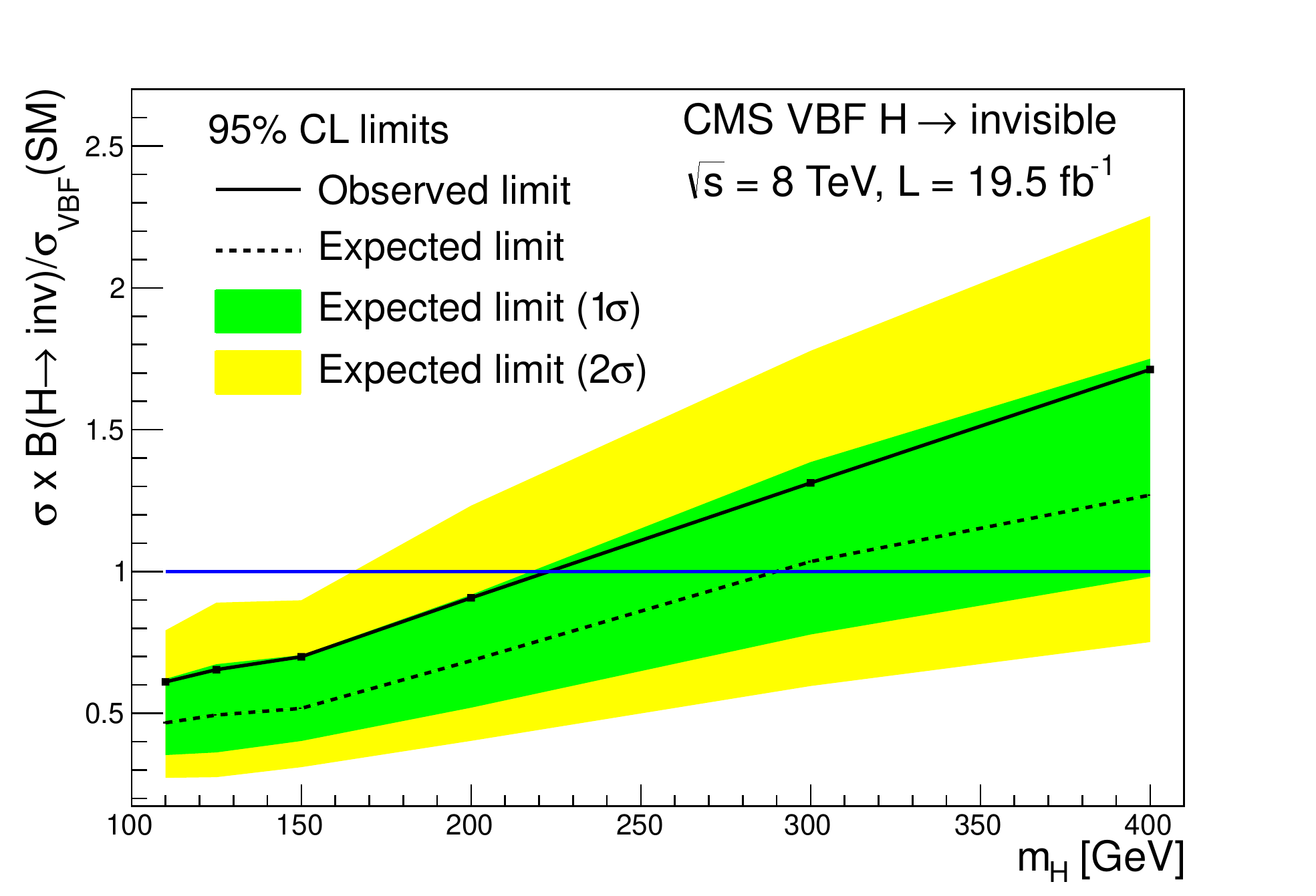}}
\end{minipage}
\end{tabular}]
\vspace*{-.2cm}
\caption{ Left: limits on the invisible Higgs branching ratio  from ATLAS
searches in the channel $pp\! \to\! HZ$ with $H\! \to$inv and the combined $Z \to ee+ \mu\mu$ modes~\cite{Aaboud:2017bja}.  Right: observed (expected) CMS limits on $\sigma({\rm VBF})\times {\rm BR}(H\! \to {\rm inv})$ normalized to the SM value as a function of the mass of a heavy SM--like Higgs boson using the 8 TeV data~\cite{Chatrchyan:2014tja}.}  \label{inv-exp}
\vspace*{-2mm}
\end{figure}

As an example, we show in Fig.~\ref{inv-exp} (left) the ATLAS results for the
Higgs invisible branching ratio in  searches performed at $\sqrt s=13$ TeV with
a luminosity of 36 fb$^{-1}$ in the Higgs--strahlung process $pp \to HZ$ with
the clean decay channel $Z\to \ell^+ \ell^-$ with $\ell=e,\mu$
\cite{Aaboud:2017bja}. Shown are the observed (solid black) and expected (dashed
black)  curves of ``1--exclusion CL'' as a function of BR($H \to$inv) with
$M_H=125$ GeV using the combined $Z \to ee+ \mu\mu$ search channels.  The $\pm 1
\sigma  (\pm  2 \sigma)$ error band on the expectation is shown in green
(yellow) and the crossing point between the dashed blue line and the scan curve
gives the observed (expected) upper limit on BR($H \to$inv) at 95\%CL. The
turning points in the observed lines correspond to the best--fit  values.  As
can be seen,  values BR($H \to $inv)$\lsim 30\%$ are excluded at 95\%CL only
using this channel. 

Similar limits have been obtained in the VBF mode where heavier Higgs bosons
have been searched for by ATLAS \cite{Aaboud:2018sfi}  as well as by CMS
\cite{Chatrchyan:2014tja,Sirunyan:2018owy}. This is illustrated in
Fig.~\ref{inv-exp} (right) where the 95\%CL upper limits on the VBF production
cross section times the invisible Higgs branching fraction, normalized to the SM
VBF rate, is shown as a function of the mass $M_H$,  assuming a SM  Higgs--like
state; the full $\sqrt s=8$ TeV data has been used. The observed (expected)
limit for $M_H=125$ GeV is BR($H\! \to\! {\rm inv})\!=\! 0.63\, (0.48)$ at 95\%CL in VBF only; when combined with the ZH channel, again at $\sqrt s=8$ TeV,  the limits become  BR($H\! \to\! {\rm inv})\! =\! 0.55\, (0.41)$. 

A promising search for invisible decays is the monojet channel
\cite{Djouadi:2011aa,Bai:2011wz,Englert:2011us}. In the ggF mode, an additional
jet can be emitted at NLO leading to $gg\!\to\! Hj$ final states and, because
the QCD corrections  are large, $\sigma(H\!+\!1j$) is not considerably smaller
than $\sigma(H\!+\!0j$) \cite{Djouadi:1991tka,Dawson:1990zj,Spira:1995rr}.  The
NNLO corrections
\cite{Harlander:2002wh,Anastasiou:2002yz,Ravindran:2003um,Anastasiou:2015ema},
besides significantly increasing the $H\!+\!0j$ and  $H\!+\!1j$ rates, lead to
$H\!+\!2j$ events that also occur in VBF and VH with $V\to jj$.  Hence, if the
Higgs is coupled to invisible particles, it may recoil against  hard QCD
radiation, leading to monojets or dijets. Already in Ref.~\cite{Djouadi:2011aa},
it has been shown that the monojet signature carries a good  potential to
constrain the Higgs invisible decay branching ratio.  In a model independent
fashion,  constraints can be placed on the process
\begin{eqnarray}
R_{\rm inv}^{\rm ggF}  =  \sigma (g g \to  H)/ \sigma (g g \to H)_{\rm SM} \times {\rm BR}  (H \to {\rm inv})  \, , 
\end{eqnarray}
even if the Higgs couplings to fermions and gauge bosons are not SM--like, 
$\kappa_f, \kappa_V \neq 1$. 

\begin{figure}[!h]
\begin{center}
\mbox{
\resizebox{.48\textwidth}{!}{\includegraphics{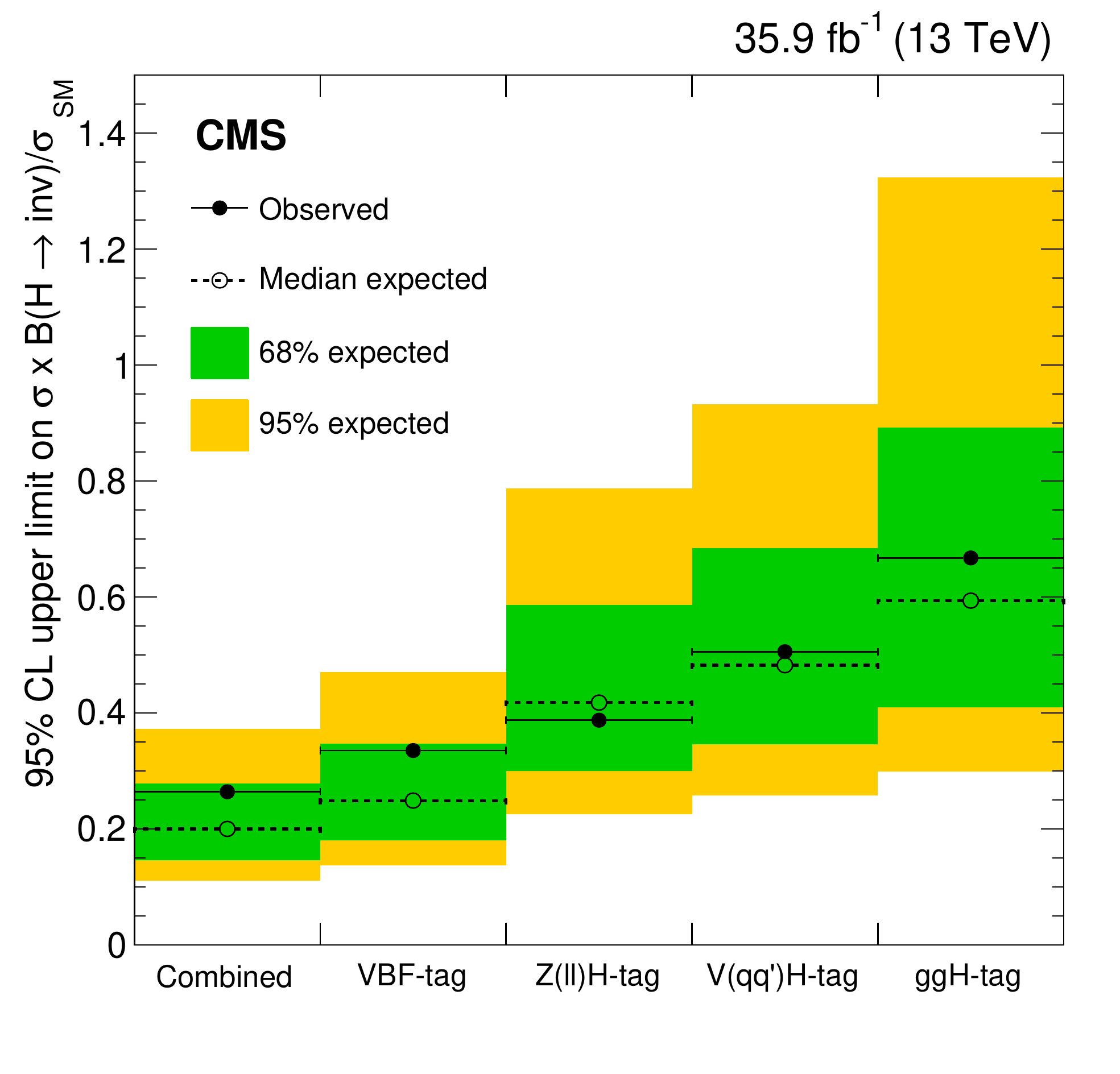}}
\resizebox{.48\textwidth}{!}{\includegraphics{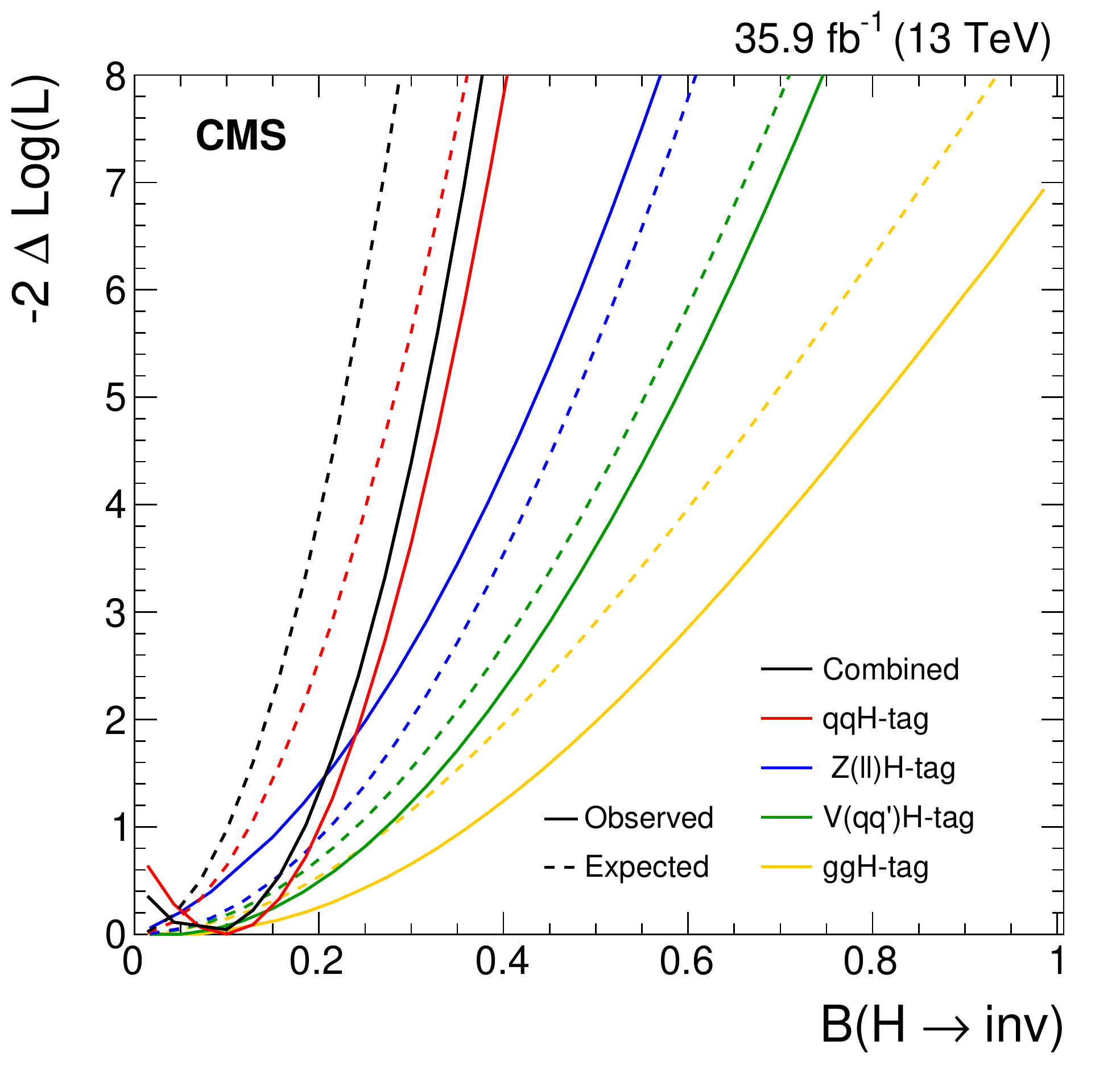}}
}
\end{center}
\vspace*{-.6cm}
\caption{
Left: observed and expected 95\%CL upper limits on $\sigma/\sigma_{\rm
SM}\times {\rm BR}(H \to {\rm inv})$ for the individual channels VBF, 
$HZ \to H\ell\ell$, $HV\to H q\bar q$ and ggF production modes and their combination for the 125 GeV SM Higgs boson \cite{Khachatryan:2016whc}. Right: the observed and expected  profile likelihood ratios as a function of BR($H \to$ inv)  in the same searches; the results are obtained at $\sqrt s=13$ TeV with about 36 fb$^{-1}$ data \cite{Sirunyan:2018owy}.}
\label{inv-ggF} 
\vspace*{-1mm}
\end{figure}

This is illustrated in Fig.~\ref{inv-ggF}  where CMS results, obtained at $\sqrt
s=13$ TeV  with a luminosity of 35.9 fb$^{-1}$ are shown when all channels
above,  namely VBF, Higgs--strahlung with $HZ \to  H\ell^+ \ell^-$ and $HV \to
Hq\bar q$ and $gg\to H$+jets are considered and then combined for the SM--like 
125 GeV Higgs boson.  The observed and expected 95\%CL upper limit on  the cross
section times branching ratio normalized to the SM value (left) and  the profile
likelihood ratios as a function of BR($H \to$inv) are displayed.  As can be
seen, the VBF channel is the most constraining, followed by Higgs--strahlung
and, for this luminosity, the ggF mode where one only obtains the
observed weaker limit of BR($H\! \to\! {\rm inv})\!\leq\! 0.66$.  When all
channels are combined, the  observed (expected) 95\%CL upper limit of $0.26\,
(0.20)$ is set on the Higgs invisible branching ratio at this energy assuming a
SM production rate.

One should note for completeness that despite of the predicted low rates,
invisible Higgs decays have also been discussed in the ttH process, see e.g.
Refs.~\cite{Zhou:2014dba,Haisch:2015ioa,Arina:2016cqj}.  In
Ref.~\cite{Zhou:2014dba} a search performed by CMS with the full 8 TeV data on
stop squark pair production in the MSSM, with the stops decaying into top quarks
and the lightest stable neutralinos leading to a topology similar  to the one we
are discussing here, namely  $pp \to t \bar t H \to t \bar t + E_T^{\rm mis}$,
has been recast in order to feature the Higgs--portal scenario. As expected, the
resulting constraints from this process are much weaker than those derived in
the other channels: the observed   upper limits on $\sigma(t\bar{t}H )/ \sigma(
t\bar{t}H)|_{\rm SM}  \times {\rm BR} (H \to \textrm{inv})$ obtained from  the
analysis above was 1.9 at the 95\%CL for a 125 GeV SM--like Higgs 
\cite{Zhou:2014dba} (because of a fluctuation in the data, the $<1.9$ observed
95\%CL limit is tighter than the expected one, $<3$). 


\underline{Indirect constraints on invisible Higgs decays from the signal strengths.}

\begin{figure}[!h]
\vspace*{-4.8cm}
\begin{tabular}{ll}
\begin{minipage}{11cm}
\hspace*{-2.5cm}
\resizebox{1.1\textwidth}{!}{\includegraphics{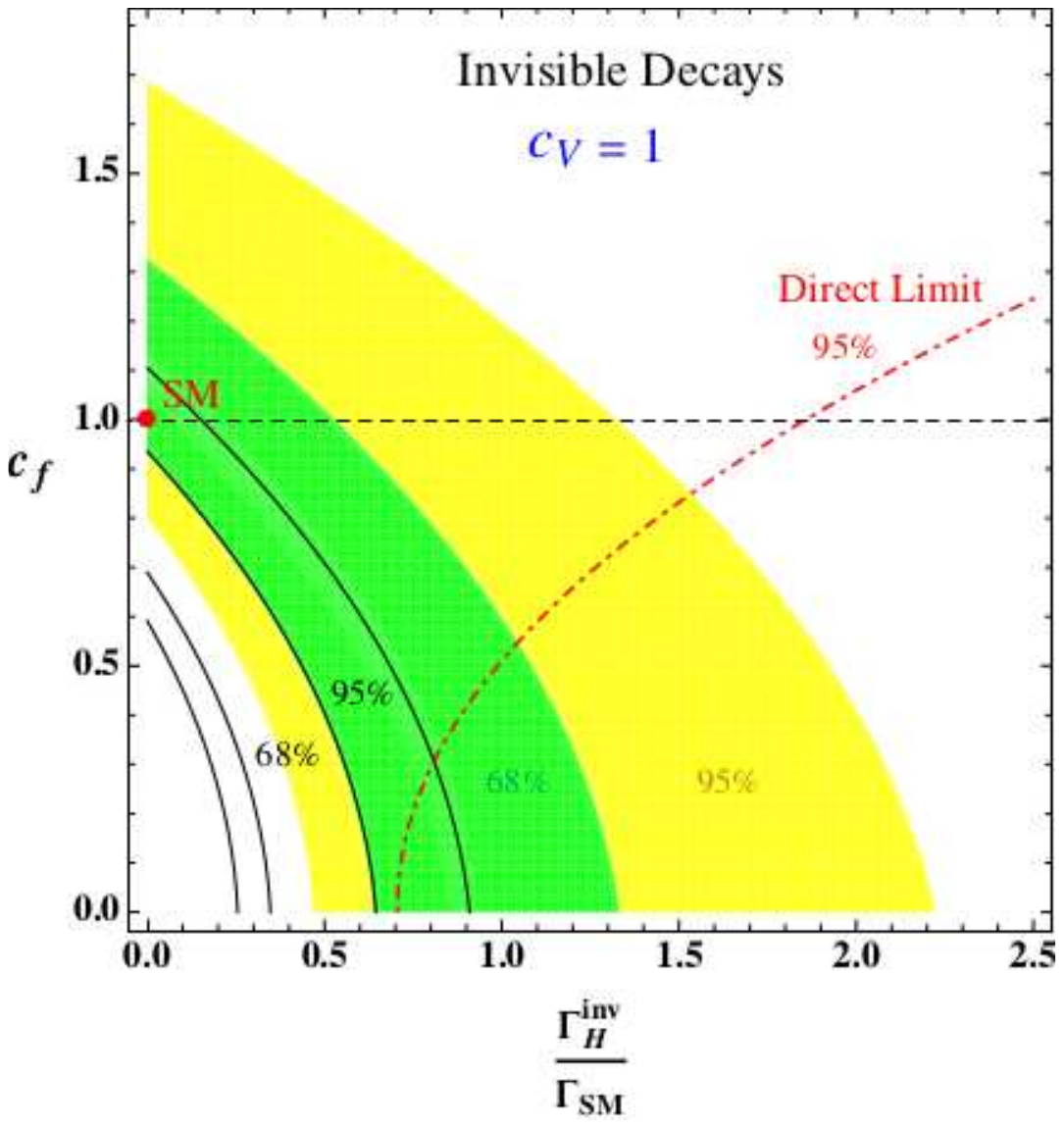} }
\vspace*{-1.5cm}
\end{minipage}
& \hspace*{-4cm} 
\begin{minipage}{11cm}
\resizebox{.68\textwidth}{!}{\includegraphics{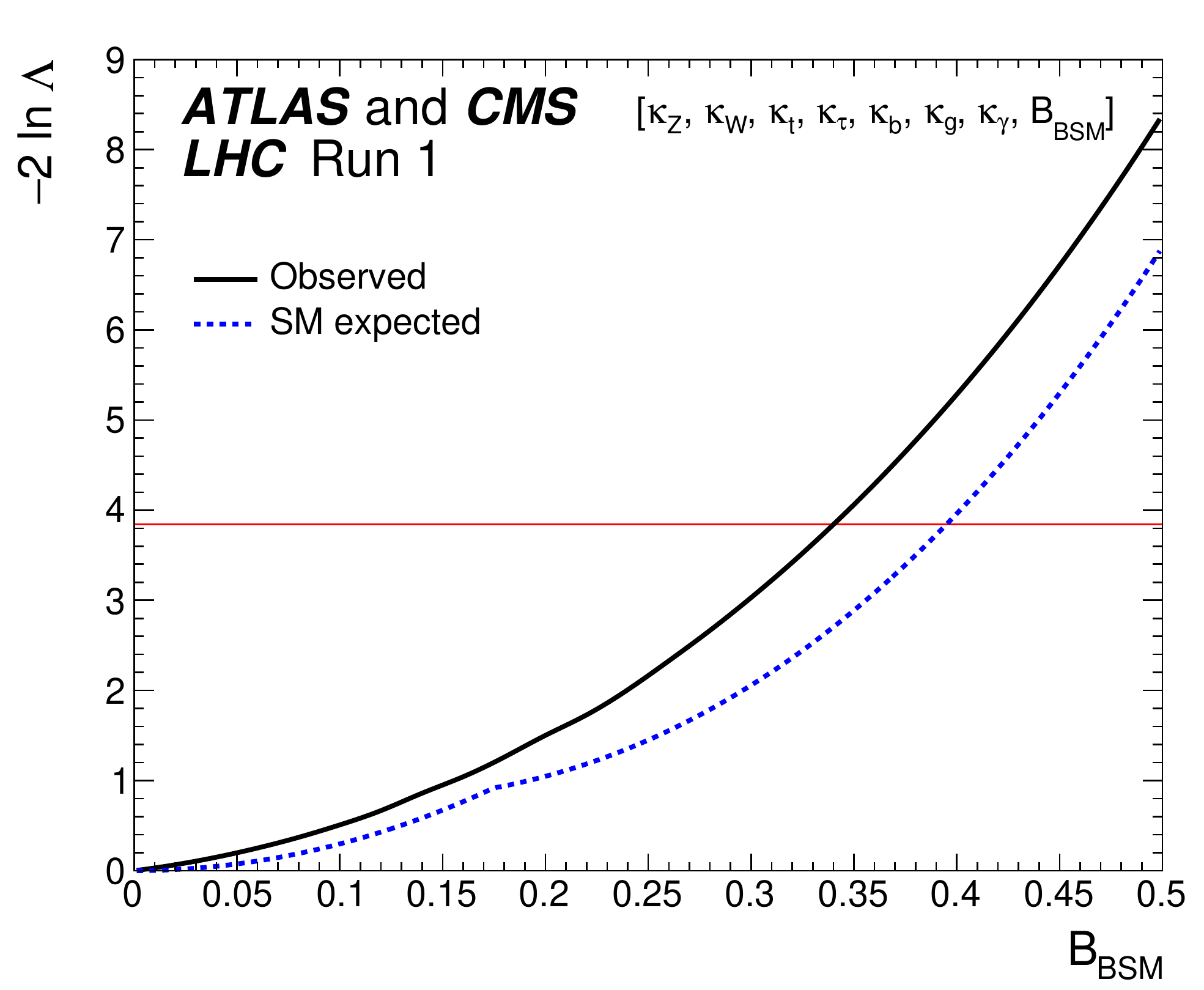}}
\end{minipage}
\end{tabular}]
\vspace*{-42mm}
\caption{
Left: $1\sigma$ and $2\sigma$ domains from $\mu_{ZZ}$ for $c_V\!=\!1$ in the 
plane $[c_f,  \Gamma_{\rm inv}/\Gamma_{H}^{\rm SM}]$ \cite{Djouadi:2013qya}; the dependence on the theory uncertainties are shown by the black curves and the upper limit on  $\Gamma_{\rm inv}$  from  direct searches at LHC for $c_V\!=c_f\!=\!1$~(from an old ATLAS study \cite{ATLAS:2013pma}) is also shown. Right: negative log--likelihood scan of the additional Higgs branching ratio $B_{\rm BSM}$ performed by ATLAS and CMS using the full RunI data when all Higgs couplings listed on top  are varied; the red horizontal line corresponds to the 95\%CL limit; from Ref.~\cite{Khachatryan:2016vau}. } \label{fig:Inv-exp}
\vspace*{-.2cm}
\end{figure}

On the other hand,  the invisible Higgs decay width can be constrained
indirectly by a fit  of the Higgs couplings and, in particular, with the signal
strength $\mu_{ZZ}$ which is one of the  most accurate ones and has the least
theoretical ambiguities. $\Gamma_{\rm inv}$  enters in the signal strength
through the total width $\Gamma_{H}$,  $\mu_{ZZ}\! \propto\! \Gamma (H\! \to \!
ZZ)/\Gamma_{H}$ with $\Gamma_{H} \! = \! \Gamma_{\rm inv} \! + \! \Gamma_H^{\rm
SM} $ and $\Gamma_H^{\rm SM}$ calculated with free coefficients $c_f$ and $c_V$
of the Higgs couplings to fermions and massive gauge bosons.  The resulting
$1\sigma$ or $2\sigma$ ranges are shown in Fig.~\ref{fig:Inv-exp} (left) using
early RunI data. Here, $c_f$ is freely varied while $c_V=1$,  and the
theoretical uncertainties  on the various production processes were supposed to
be about 30\% \cite{Djouadi:2012rh,Djouadi:2013qya,Djouadi:2013uqa}. This gives
$\Gamma_{\rm inv} /  \Gamma_H^{\rm SM} \lsim 50\%$ at the $95\%~{\rm CL}$  for 
$c_f=c_V=1$.

With the full set of RunI  data and a smaller theory uncertainties as estimated
by ATLAS and CMS, one would obtain  much better limits on the invisible Higgs
width from coupling measurements. Indeed, a combined ATLAS+CMS analysis using
the full set of data collected at RunI has been performed assuming that New
Physics will enter both directly in the decays of the Higgs boson, and
indirectly by modifying the Higgs couplings to fermions and gauge bosons and the
loop induced processes $gg\to H$ and $H\to \gamma\gamma$.  The right--hand side
of Fig.~\ref{fig:Inv-exp} shows a negative log--likelihood scan of the branching
ratio $B_{\rm BSM}$ when additional (and positive) contributions to the Higgs
total width are allowed, contributions which can be thus identified with the
invisible Higgs branching ratio. The analysis has been performed allowing all
Higgs couplings to freely vary with some very mild assumptions,  $\kappa_Z,
\kappa_W, \kappa_t, \kappa_b, \kappa_g, \kappa_\gamma \neq 1$. As can be seen,
an upper limit BR$(H \to {\rm inv}) <0.34$ on the invisible Higgs width can be
set at the 95\%CL in this general case \cite{Aad:2015pla}.  

\begin{figure}[!h]
\begin{center}
\vspace*{-4.5cm}
\resizebox{.55\textwidth}{!}{\includegraphics{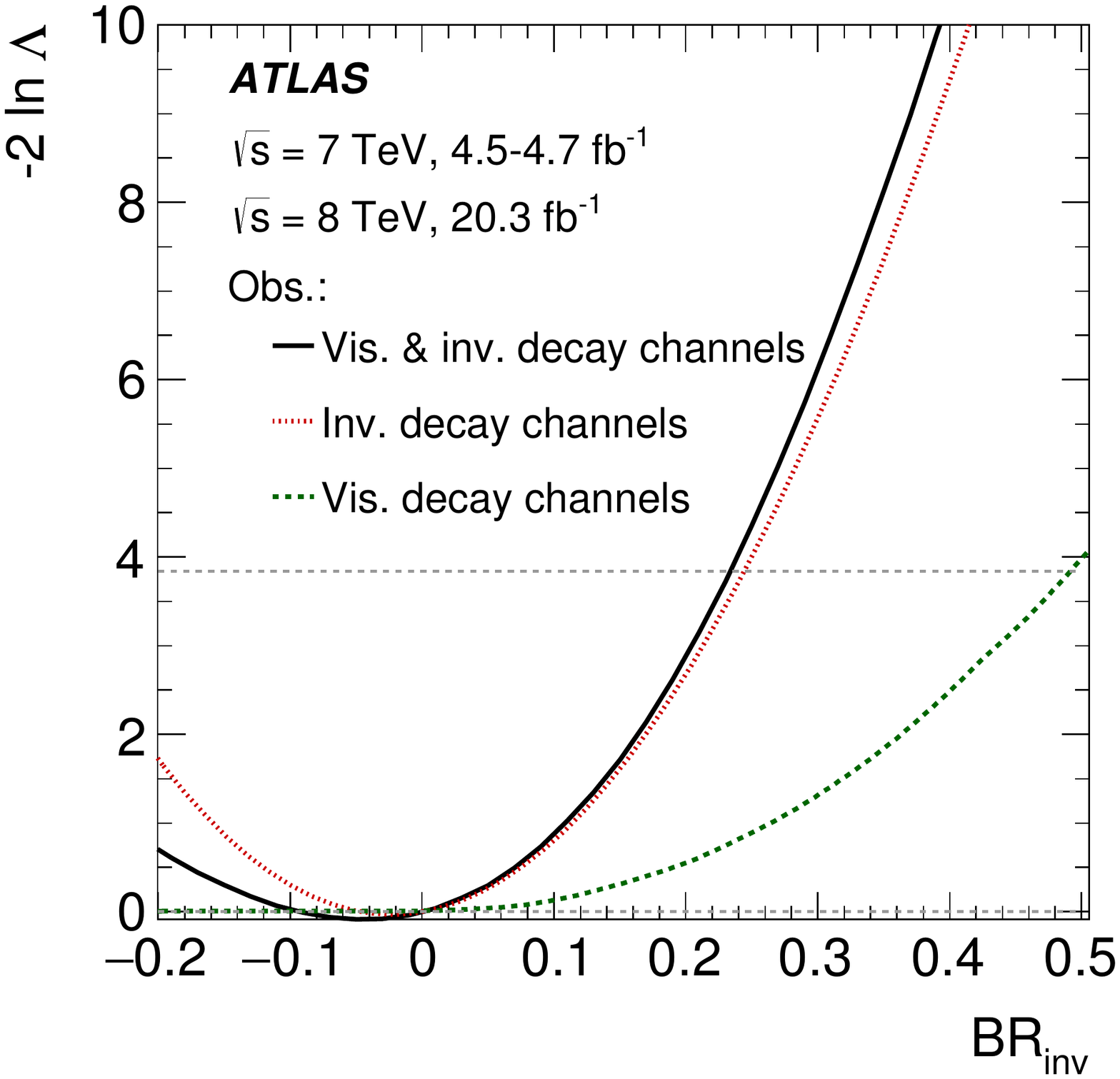}}
\end{center}
\vspace*{-.7cm}
\caption{Likelihood scans of the Higgs invisible branching ratio using direct and indirect searches made by ATLAS at RunI and their combination; from Ref.~\cite{Aad:2015pla}.  \label{fig:Inv-exp1}}
\vspace*{-.3cm}
\end{figure}

 One can finally combine direct and indirect measurements as, for instance,  has
been done by ATLAS~\cite{Aad:2015pla} using only RunI data. This is illustrated
in Fig.~\ref{fig:Inv-exp1} where observed likelihood scans of BR($H\to {\rm
inv})$ are shown  using direct searches for missing energy, rate measurements in
visible Higgs decays as well as the overall combination of invisible and visible
channels. The line at $-2\ln\Lambda=0$ corresponds to the most likely value of
BR($H\to$inv) within the physical region in which it is positive,  while the
line at $-2\ln\Lambda=3.84$ corresponds to the one--sided upper limit at  the
95\%CL. As can be seen, the combination of visible and invisible channels gives
the constrain BR($H\to$inv)$<0.23$ at the 95\%CL.

\underline{Summary.}

In conclusion, both the direct and indirect searches for invisible Higgs decays
performed presently at the LHC allow to exclude branching ratios of the order 
of 20\% to 30\% depending on whether the Higgs couplings to fermions and gauge
bosons are also modified or not. In our minimal Higgs--portal scenario, the
Higgs couplings are assumed to be SM--like so that ultimately, one obtains a
limit on the invisible Higgs branching ratio of 
\beq 
{\rm BR}(H \to {\rm inv}) < 20\% \, , 
\eeq 
that will be assumed from now on.  

\subsubsection{Prospects for future measurements}

Most of the results presented in the previous subsection were obtained by  the
ATLAS and CMS collaborations for the data collected at the first run of the LHC
with c.m. energies of $\sqrt s\!=\! 7$ and 8 TeV, with a total luminosity of
$\approx 25$ fb$^{-1}$. A few analyses were performed at some early stage of
RunII with an energy of 13 TeV and a luminosity below 36 fb$^{-1}$. In this
case, only individual channels have been considered and no combination of the
ATLAS and CMS results has been made. At the end of the present LHC RunII with a
c.m. energy of $\sqrt s=13$ TeV, the ATLAS and CMS have collected about 150
fb$^{-1}$ each so that the previous analyses on both  direct and indirect
determination of the Higgs invisible branching fraction will clearly improve.
Another upgrade is currently underway, which would allow two years from now to
collect 300 fb$^{-1}$ data. A major upgrade of the LHC is planed in a near
future and there is  a wide consensus that it should be the priority for
particle physics in the next decade: the LHC high--luminosity option (HL--LHC)
in which one would collect up to 3 ab$^{-1}$ of data at a slightly larger c.m.
energy, $\sqrt s= 14$ TeV. 

\begin{figure}[!h]
\vspace{-2mm}
\centerline{
\includegraphics[width=5.3cm,clip]{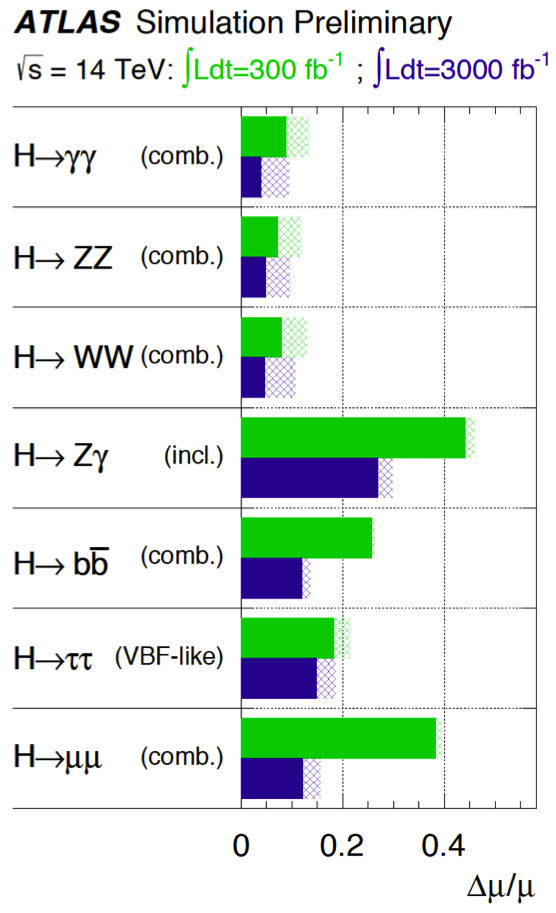}~~~~~~~~
\includegraphics[width=5.6cm,clip]{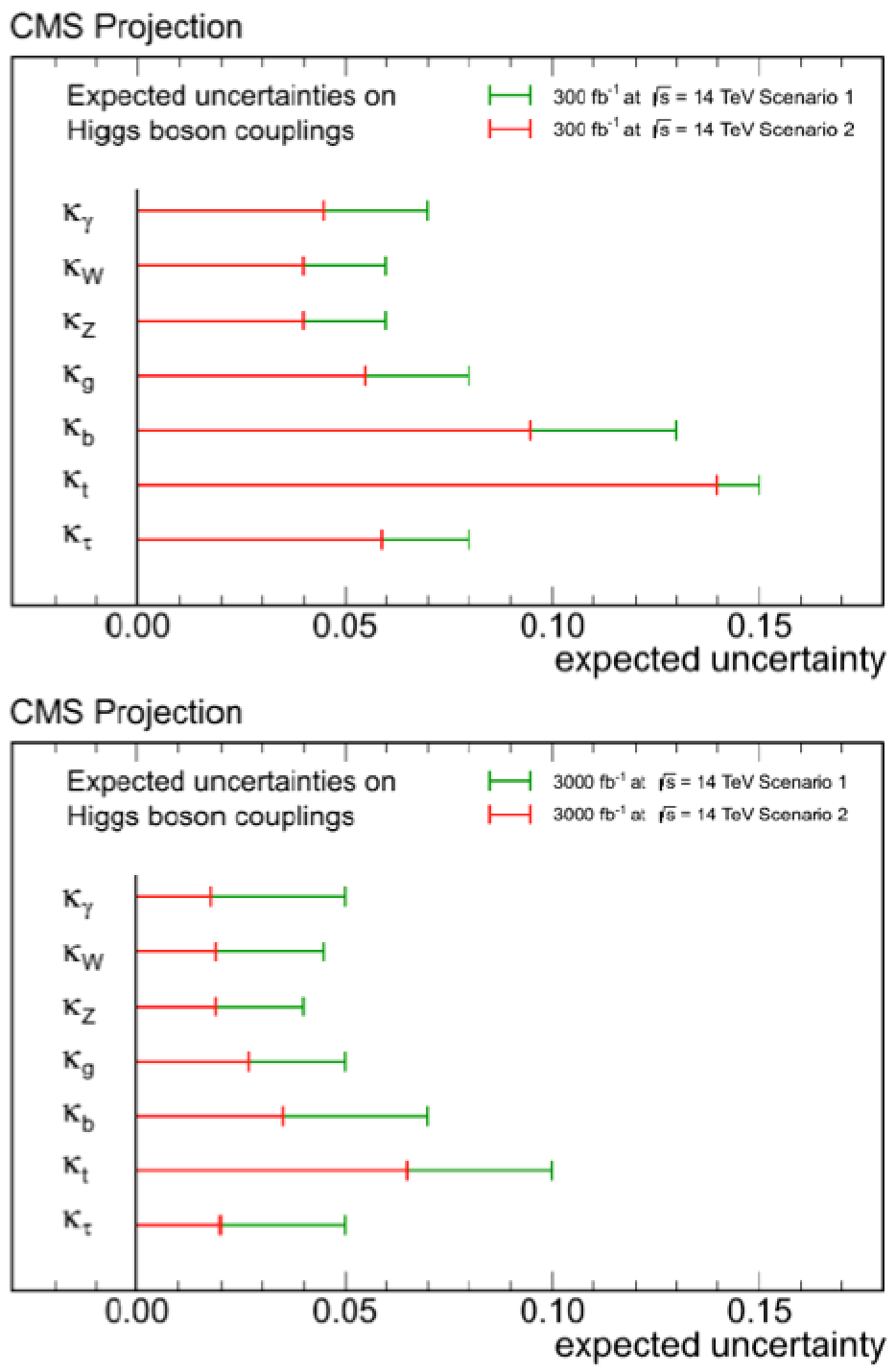}
}
\caption{The projected uncertainties on the signal strengths in the main decay
channels for ATLAS \cite{ATLAS:2013hta} (left) and on the reduced Higgs couplings to fermions and gauge bosons for CMS \cite{CMS:2013xfa} (right) with 300 fb$^{-1}$ and 3 ab$^{-1}$ of data at LHC with $\sqrt s=14$ TeV. }
\label{Fig:BLHC}
\vspace*{-3mm}
\end{figure}

For a SM--like  Higgs boson, the ATLAS and CMS collaborations have studied  the
projected performances of the two
upgrades~\cite{ATLAS:2013hta,CMS:2013xfa,Cepeda:2019klc}, i.e. with  300 and
3000 fb$^{-1}$ both at $\sqrt s=14$ TeV, by scaling the signal and background
events from the measurements at RunI, and the results are shown in
Fig.~\ref{Fig:BLHC}.

In the left--hand side, the projected accuracy on the Higgs signal strengths as
measured by ATLAS in different decay channels is shown  and one sees that, at
least for the bosonic $ZZ,WW,\gamma\gamma$ channels, the experimental accuracies
will reach the 10\% level with 300 fb$^{-1}$ and about 5\% with 3 ab$^{-1}$
data. The  Higgs reduced couplings $\kappa_X$ as measured by CMS are shown in
the right--hand  plot in two scenarios: a first one in which all systematic
uncertainties are left unchanged from the  RunI case, while  in scenario 2, the
theoretical and systematical uncertainties are scaled by a factor 1/2 and by 
$\sqrt{\cal L}$, respectively.  In this optimistic scenario 2, accuracies of a
few percent could be reached  at HL--LHC in the bosonic case for instance, an
uncertainty that is two times smaller than in scenario 1.   

As for the invisible Higgs decays, CMS has also performed a likelihood scan and
expected 95\%CL limits of BR($H \to {\rm inv}) \!< \! 18\, (11)\%$  for scenario
$1(2)$ are obtained with 300 fb$^{-1}$ data. At the HL--LHC  with 3 ab$^{-1}$
data, the more precise limits BR($H \to {\rm inv}) < 14\, (7)\%$ are achieved 
\cite{CMS:2013xfa}. A direct search for invisible Higgs decays has also been
done by CMS in associated HZ production and the 95\%CL upper limits BR($H \to
{\rm inv}) < 28 \, (17)\%$ for scenario 1 and  $17 \, (6.4)\%$ for scenario 2
were set with 300 (3000) fb$^{-1}$ data.  

\begin{figure}[!h]
\vspace*{-2mm}
\centerline{
\includegraphics[width=0.45\textwidth]{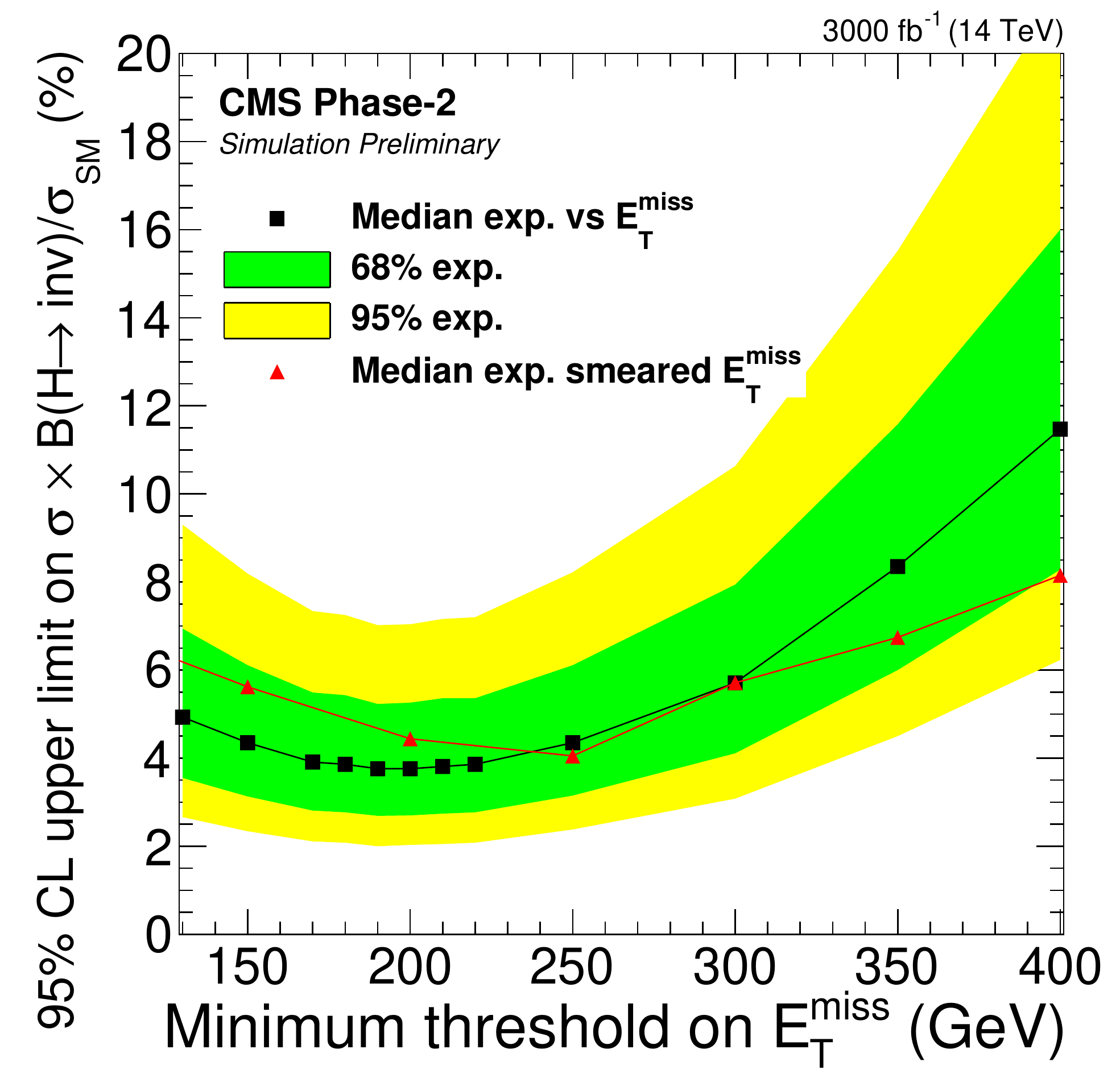}
\includegraphics[width=0.49\textwidth]{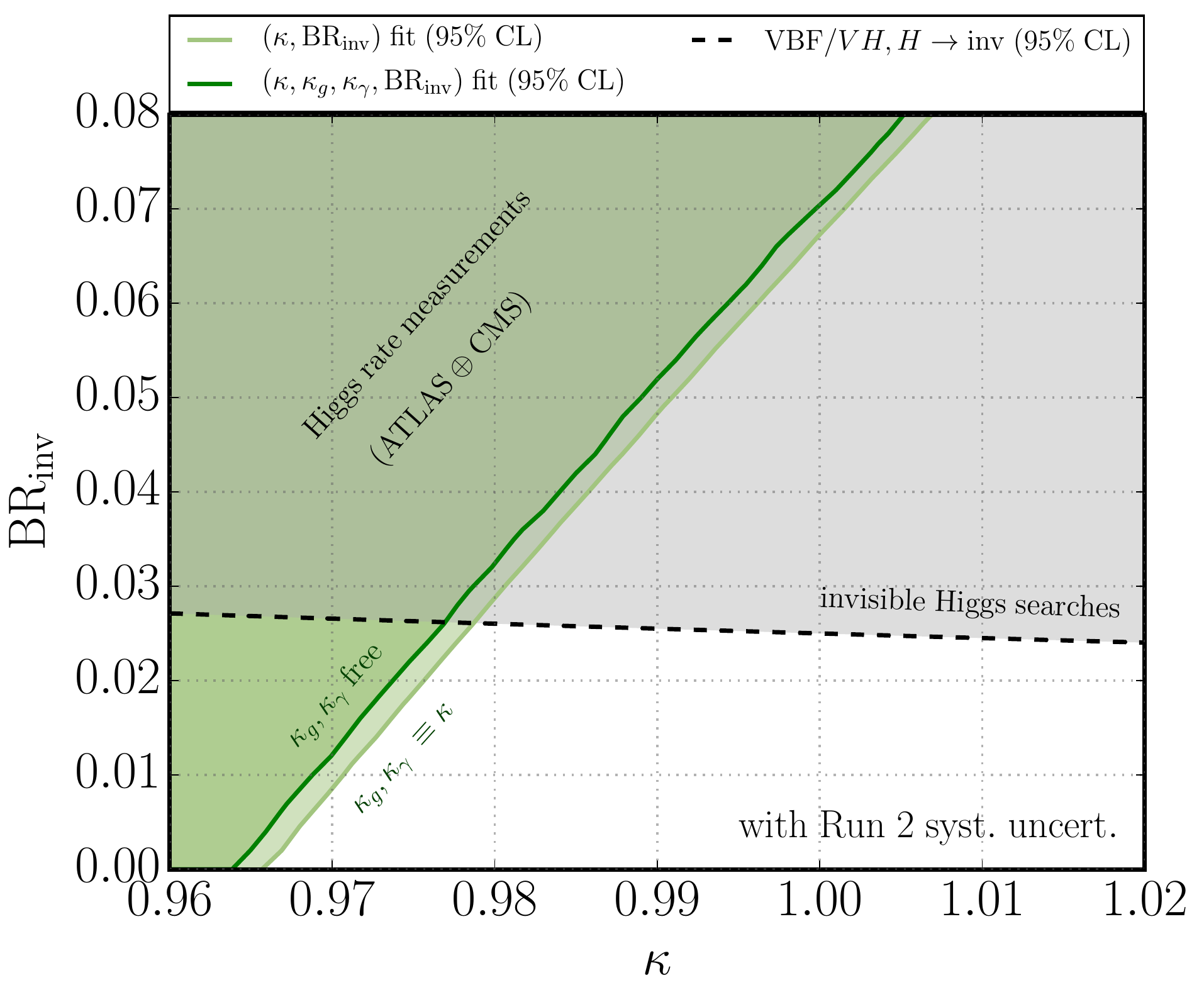}
}
\caption{Left: 95\% CL limits on $\sigma$(VBF)$\times$BR($H\!\to$inv) at the HL--LHC as a function of the minimum threshold on $E_T^{\rm mis}$ for $M_{jj}>2.5$ TeV and a luminosity of 3 ab$^{-1}$ \cite{CMS-PAS-FTR-18-016}. 
Right: projected 95\%CL limits in the $[\kappa$, BR($H\!\to$inv)] plane
obtained from the measurement of the Higgs signal strengths (green regions) and
from direct invisible Higgs searches (the black dashed line) at the HL--LHC,
assuming  RunII systematic uncertainties \cite{Cepeda:2019klc}. }
\label{Fig:HL-LHC-inv}
\vspace*{-3mm}
\end{figure}

Very recently, the report of the physics working group on Higgs physics at the
HL--LHC has appeared \cite{Cepeda:2019klc} and it constrains an updated analysis 
of the prospects for measuring the invisible Higgs branching ratio either
directly in $E_T^{\rm mis}$ searches or indirectly through the Higgs signal
strengths. The outcome of this study is summarized in
Fig.~\ref{Fig:HL-LHC-inv}. In the left--hand side of the figure, shown is the
projection for the 95\%CL upper limit on the Higgs  cross section in the VBF
channel (which provides the best sensitivity) times the invisible branching
fraction as obtained in an analysis of a search for missing transverse energy
with 3 ab$^{-1}$ of data \cite{CMS-PAS-FTR-18-016}. For  an invariant mass
$M_{jj}>2.5$ TeV of the two VBF jets and $E_T^{\rm mis} \approx 200$ GeV, a
sensitivity of 4\% can be reached on BR($H\!\to$inv). Assuming a similar
performance by ATLAS, a combined 95\%CL limit of BR($H\!\to$inv)$\lsim 3\%$ can
be set for a SM--like $H$ boson. In the right--hand side of the figure, this
limit is compared to what can be obtained indirectly from the Higgs signal
strengths when the ATLAS and CMS  measurements are combined, conservatively
assuming that the systematical uncertainties will remain the same as in RunII. 
The limits in the plane [BR($H\! \to$inv),$\kappa$] are shown in the case where 
the $\kappa$ factor is universal (light green) and when there are additional
loop contributions to the $Hgg$ and $H\gamma\gamma$ vertices (dark green).
Depending whether the global $\kappa$ will be smaller or larger than unity, the
indirect constraint could be tighter or looser. 

\begin{figure}[!h]
\centerline{
\includegraphics[scale=0.84]{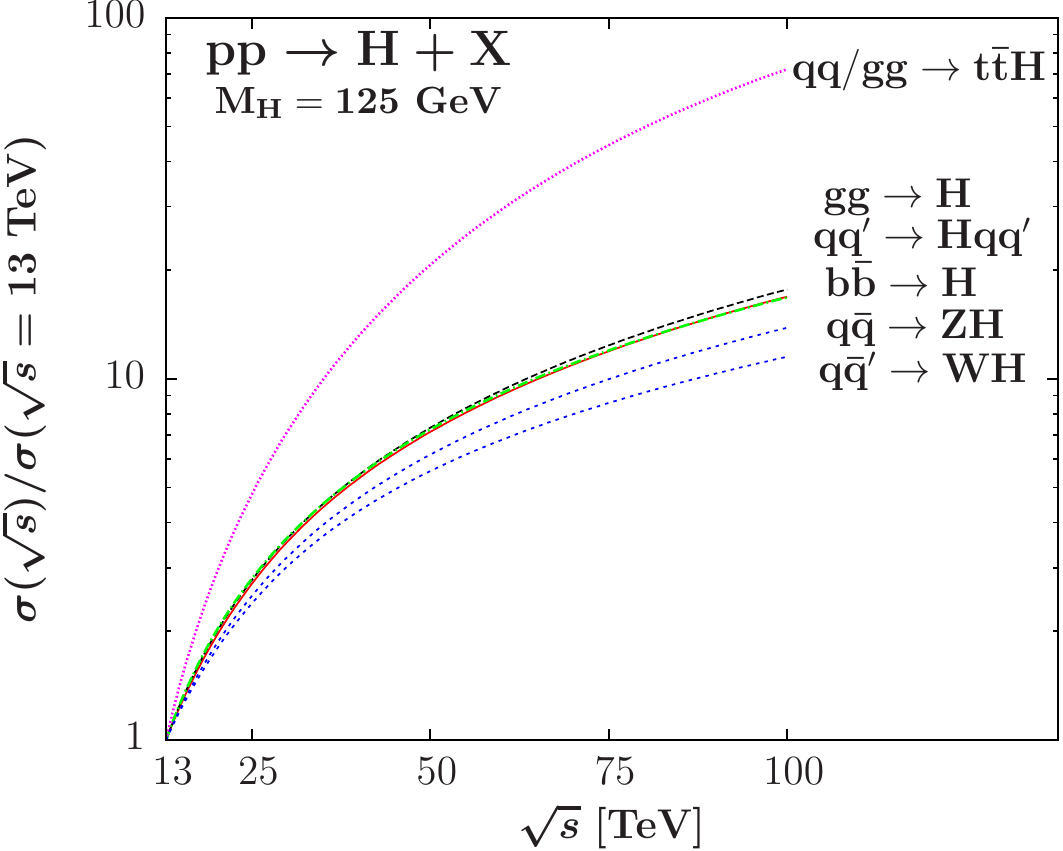}
}
\caption{The SM Higgs production cross sections in pp collisions in the main channels as a function of $\sqrt s$ normalized to their values at $\sqrt s=13$ TeV  \cite{Baglio:2015wcg}. }
\vspace*{-2mm}
\label{Fig:BLHC2}
\end{figure}

A more radical option would be a significant increase of the  c.m. energy. In
this context, an upgrade of the LHC to an energy  about 2 times higher has been
discussed and, for instance, detailed studies of the physics of a $\sqrt s=33$
TeV collider have been performed~\cite{Baur:2002ka} (see also the recent review 
\cite{Cepeda:2019klc}). More recently, a Future Circular Collider (FCC--hh), a
hadron collider with a c.m. energy of 100 TeV, has been proposed as a potential
follow-up of the LHC at CERN  \cite{Contino:2016spe}; such a very high energy
machine is also under study in China \cite{Tang:2015qga}. In the context of the
SM--like Higgs boson, such energies would allow an  increase in the production
cross sections y more than an order of magnitude compared to the 13 TeV LHC.
This is exemplified in Fig.~\ref{Fig:BLHC2} where the variation of the
production cross sections for the main Higgs production channels in pp
collisions with the c.m. energy, relative to their values at $\sqrt s=13$ TeV. 
As can be seen, at $\sqrt s=100$ TeV, the rates increase by a factor of about 20
in the ggF and VBF processes and even a factor  of 70 for the ttH process,
compared to $\sqrt s=13$ TeV. If high luminosities are available at the same
time,  huge samples of Higgs particles  could be collected, allowing  to make
detailed studies of the Higgs properties and accuracies for the determination 
of the invisible Higgs branching ratio at least as good as those obtained at the
HL--LHC would be achievable. 

Turning to future $e^+e^-$
colliders~\cite{Djouadi:2007ik,Baer:2013cma,Gomez-Ceballos:2013zzn,CEPC-SPPCStudyGroup:2015csa,Battaglia:2004mw,Linssen:2012hp,Moortgat-Picka:2015yla,AguilarSaavedra:2001rg,Accomando:1997wt,Djouadi:1994mr,Li:2012taa}
more precise measurements of the SM--like Higgs boson properties can be
achieved  already  with an energy of $\sqrt s=240$ GeV and a luminosity at the
ab$^{-1}$ level,  thanks to the clean environment and the low backgrounds. Many
proposals for such machines have been put forward:  the International Linear
Collider (ILC) in Japan which can start with an energy of 250 GeV and be
hopefully extended to $\sqrt s=1$ TeV \cite{Djouadi:2007ik,Baer:2013cma},  the
electron--positron stage of the Future Circular Collider (FCC--ee) at CERN,
previously known as TLEP  \cite{Gomez-Ceballos:2013zzn,Mangano:2018mur} and the
Circular Electron Positron Collider (CEPC) in China
\cite{CEPC-SPPCStudyGroup:2015csa,CEPCStudyGroup:2018ghi}. The two last
colliders would mainly operate at an energy of $\sqrt s=240$ GeV or slightly
above. There is also a plan at CERN for a very high energy $e^+ e^-$ linear
collider, the CLIC machine with $\sqrt s$ up to 3 TeV \cite{Battaglia:2004mw}.

\begin{figure}[!h]
\vspace*{-3cm}
\centerline{\hspace*{-2cm}
\includegraphics[scale=0.89]{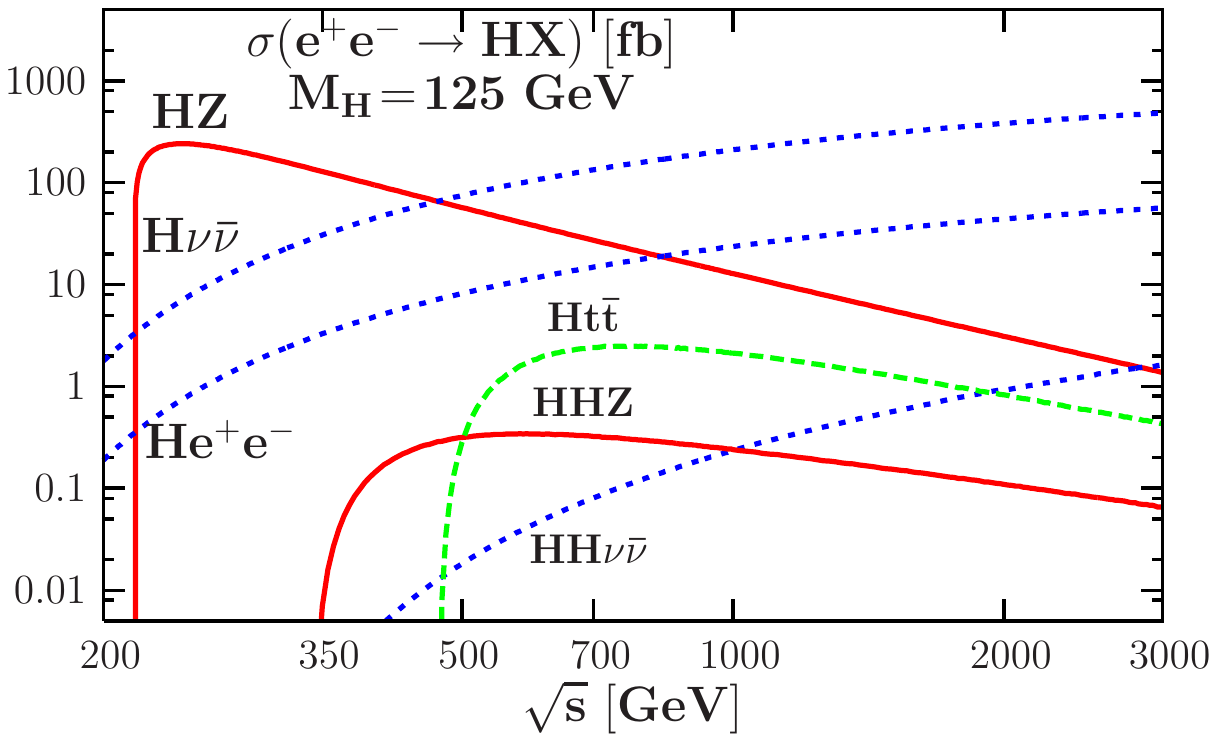}
}
\vspace*{-16.9cm}
\caption{The production cross sections of the SM--like Higgs boson in $e^+e^-$ collisions as a function of the center of mass energy in all the dominant channels for $M_H=125$ GeV.}
\vspace*{-1mm}
\label{Fig:BLHC3}
\end{figure}

At these machines, the Higgs production processes have been discussed in
Appendix A3 and, at not too high energies,  the main role is played by the
Higgs--strahlung  process $e^+e^- \to HZ$ for which the cross section is maximal
at $\sqrt s \simeq 240$ GeV for a state with $M_H\!=\! 125$ GeV. Other
production processes such as $WW$ and $ZZ$ fusion leading to $e^+e^- \to H \nu
\bar \nu$ and $e^+e^-  \to H e^+e^-$ final states, have too low cross sections
at this moderate energy  but become extremely important at the higher energies
of CLIC and ILC beyond $\sqrt s \approx 500$ GeV; see Fig.~\ref{Fig:BLHC3} where
the Higgs cross  sections as shown as a function of $\sqrt s$. 

In Higgs--strahlung,  one tags the $Z$ boson through e.g. its clean leptonic
modes and, by simply studying the recoiling $Z$ boson, one can measure the
production cross section independently of the Higgs decays. Indeed,  because the
energy of the initial $e^+e^-$ state is precisely known, the Higgs can be
reconstructed from its recoil against the $Z$ boson  and, hence, regardless to
its decays. One can then get a direct access  to the coupling $HZZ$ in a
model--independent way, as $\sigma(e^+e^- \! \to \! HZ) \propto g^2_{HZZ}$, 
with a precision at the percent level. The various Higgs branching ratios,
including the ones for the decays $H\to gg$ and $H\to c\bar c$  which are not
accessible at the LHC, can be accurately determined and  for instance a
precision of less than half a percent is expected for BR($H\!\to \!b\bar b)$.
The various couplings are then unambiguously extracted as $g_{HZZ}$ can serve as
an absolute normalization.  

\begin{figure}[!h]
\vspace*{-2.9cm}
\begin{tabular}{ll}
\hspace*{-4.8cm}
\begin{minipage}{8cm}
\vspace*{-10.5cm}
\resizebox{2.3\textwidth}{!}{\includegraphics{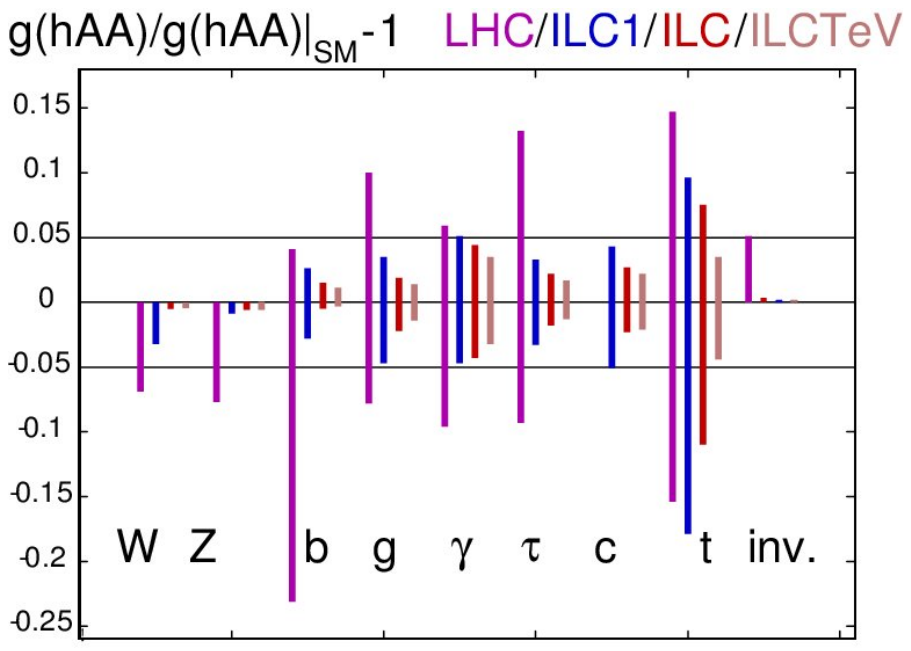}}~~ 
\vspace*{-10.5cm}
\end{minipage}
& \hspace*{3.6cm}
\begin{minipage}{8cm}
\resizebox{1.15\textwidth}{!}{\includegraphics{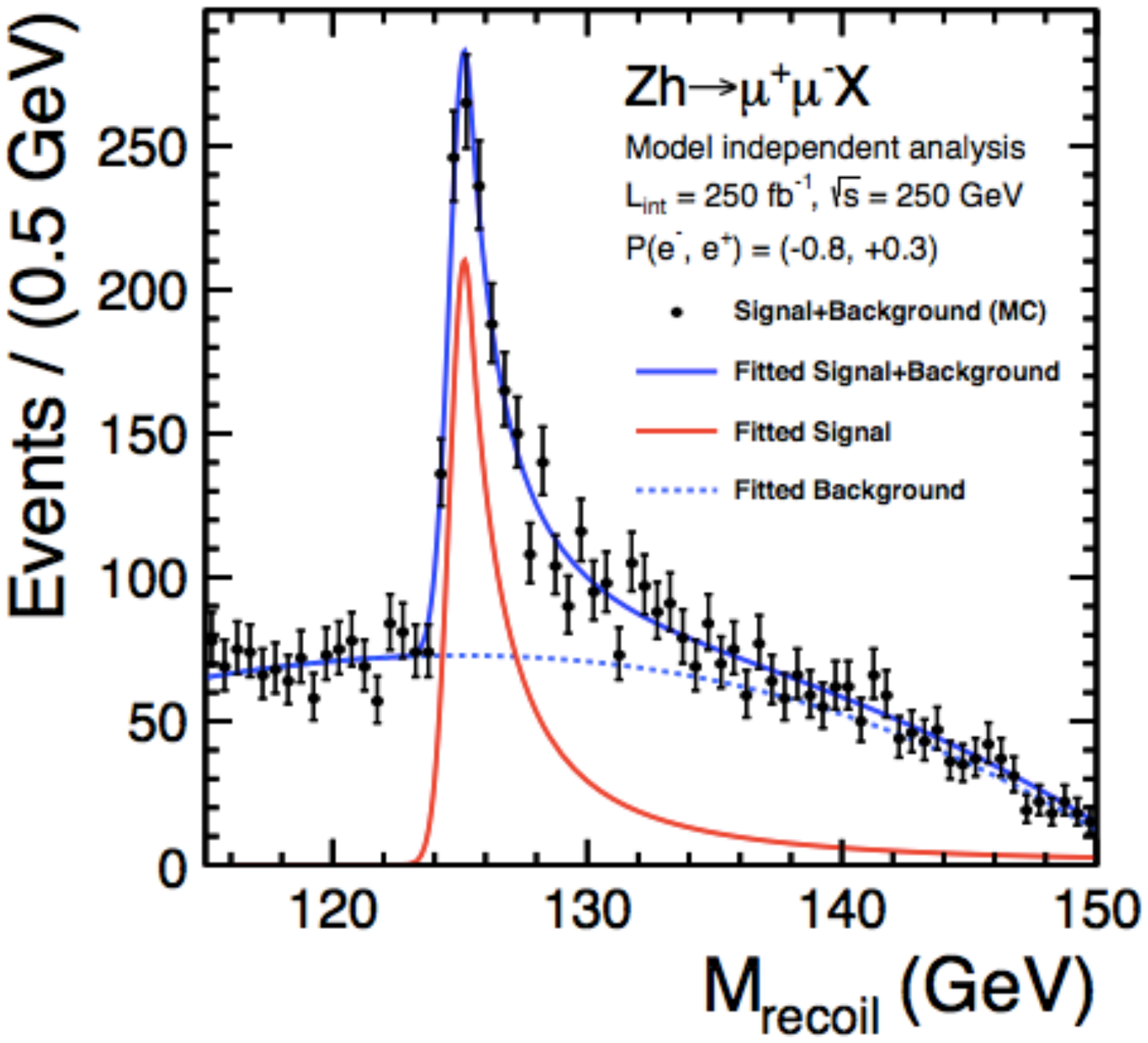} }
\end{minipage}
\end{tabular}
\vspace*{-3.5cm}
\caption{ Left: sensitivities at various energy and luminosity
stages  ILC1:$\sqrt s\!=\!250$ GeV with 250 fb$^{-1}$,  ILC:$\sqrt s\!=\!500$ GeV with 500 fb$^{-1}$ and  ILCTeV:$\sqrt s\!=\!1$ TeV with 1 ab$^{-1}$ data on the reduced Higgs couplings to fermions and gauge bosons, when cumulatively added to the sensitivity of the LHC with $\sqrt s\!=\!14$ TeV and 300 fb$^{-1}$ data \cite{Moortgat-Picka:2015yla}. Right: recoil  mass  distribution  for  the  process  $e^ + e^- \to HZ \to H \mu^+ \mu^-$ at the ILC with $\sqrt s=250$ GeV, 250 fb$^{-1}$ data and initial beam polarizations of  $P_{e^-}=-0.8$ and $P_{e^+}=+0.3$ \cite{Li:2012taa}.}
\label{Fig:kappaILC}
\vspace*{-.2cm}
\end{figure}

Examples of the capability of $e^+e$ colliders in the determination of the
SM Higgs boson couplings, or more precisely the $\kappa_X$ parameters, is shown
in Fig.~\ref{Fig:kappaILC} (left) at various c.m. energies  and luminosities 
$\sqrt s=250$, 500 and 1000 GeV and a  luminosity of 250, 500 and 1000 fb$^{-1}$
\cite{Moortgat-Picka:2015yla}. The results are added to those obtained at the
LHC with  $\sqrt s=14$ TeV and 300 fb$^{-1}$ data.  There is a vast improvement
in the accuracy and this can be particularly seen in the $WW,ZZ$ and $b\bar b$
cases where a precision below 1\% can be achieved. 

At $e^+e^-$ colliders, the total decay width of the Higgs boson can be measured
in a model independent way since the  $ZZ$ coupling that enters the $H\to 
ZZ^*$  partial width and, hence, the measured branching ratio, can be determined
from the total cross section $\sigma(e^+e^- \to HZ)$.  The total width can also
be determined from the combined measurement of the cross section in the  $WW$
fusion process  process $e^+e^- \to W^*W^* \to H \nu \bar \nu$ at high energies 
($\sqrt s \gsim 500$ GeV) and the $H\to  WW^*$ branching ratio.  Alternatively,
$\Gamma_H$ can be directly determined by measuring the cross section of the
$\gamma \gamma \to H$ fusion process for single Higgs production at the $\gamma
\gamma$ option of the  $e^+e^-$ collider (see Appendix A3).   All these
processes allow an unambiguous indirect determination of the invisible Higgs
decay branching ratio at the level of BR($H \to {\rm inv}) < 5\%$ at $\sqrt
s=240$--250 GeV and  BR($H \to {\rm inv}) < 2.5\%$ if energies higher than 500
GeV are also possible \cite{Moortgat-Picka:2015yla,JapanInv}.

Finally, invisible Higgs decays can also be directly probed  in the $e^+ e^- \to
HZ \to f\bar f$ process thanks to the missing mass technique and measured with a
very good accuracy together with the Higgs mass and cross section. This is 
exemplified  in the right-hand side of Fig.~\ref{Fig:kappaILC} where the recoil
mass distribution of the $Z\to \ell \ell$ pair in the $e^ + e^- \to HZ$ process
at an energy of 250 GeV and a luminosity of 250 fb$^{-1}$; initial polarisation
for the electron and positron beams have been assumed \cite{Li:2012taa}. 

Recent studies similar to those that led to this figure \cite{JapanInv} show
that at this energy and luminosity, the missing mass technique allows to measure
the $e^ + e^- \to HZ$ cross section with an accuracy of about  2\%, the Higgs
mass with $\Delta M_H \approx 30$ MeV and, most important in the DM context,
limit the invisible  Higgs decay branching ratio to BR($H \to {\rm inv}) <1\%$
at the 95\%CL. In an earlier analysis,  performed at $\sqrt{s}=350$ GeV with 500
fb$^{-1}$ integrated luminosity, it was also shown that in the $e^+ e^- \! \to
\! HZ$ process, an accuracy of $\sim 10\%$ can be obtained on an invisible Higgs
decay with a branching ratio of 5\% and a $5\sigma$ signal can be observed for
an invisible branching ratio as low  as 2\% \cite{AguilarSaavedra:2001rg}.

To recapitulate, this discussion on the prospects for the measurement of the 
invisible Higgs decay branching ratio at future colliders can be summarized as
follows. While the present limits from ATLAS and CMS are BR($H \to {\rm inv})
<20\%$ at the 95\%CL, the  sensitivity could ultimately reach  the  10\% level
when the direct and indirect results of the two collaborations at $\sqrt s= 13$
TeV with the full collected set of data  will be combined or, in the worst case,
at the next LHC upgrade when 300 fb$^{-1}$ data will be available possibly at
$\sqrt s =14$ TeV. At the high--luminosity option of the LHC with about 3000
fb$^{-1}$ data at $\sqrt s =14$ TeV, the sensitivity could reach the 5\% level.
At a future $e^+e^-$ collider with an energy above $\sqrt s=240$ GeV and a
luminosity of a few 100 fb$^{-1}$,  an accuracy of about 1\%  could be reached
on the invisible Higgs branching ratio.

\begin{figure}[!h]
\begin{center}
\vspace*{-1mm}
\mbox{\hspace*{-5mm}
\includegraphics[width=7cm]{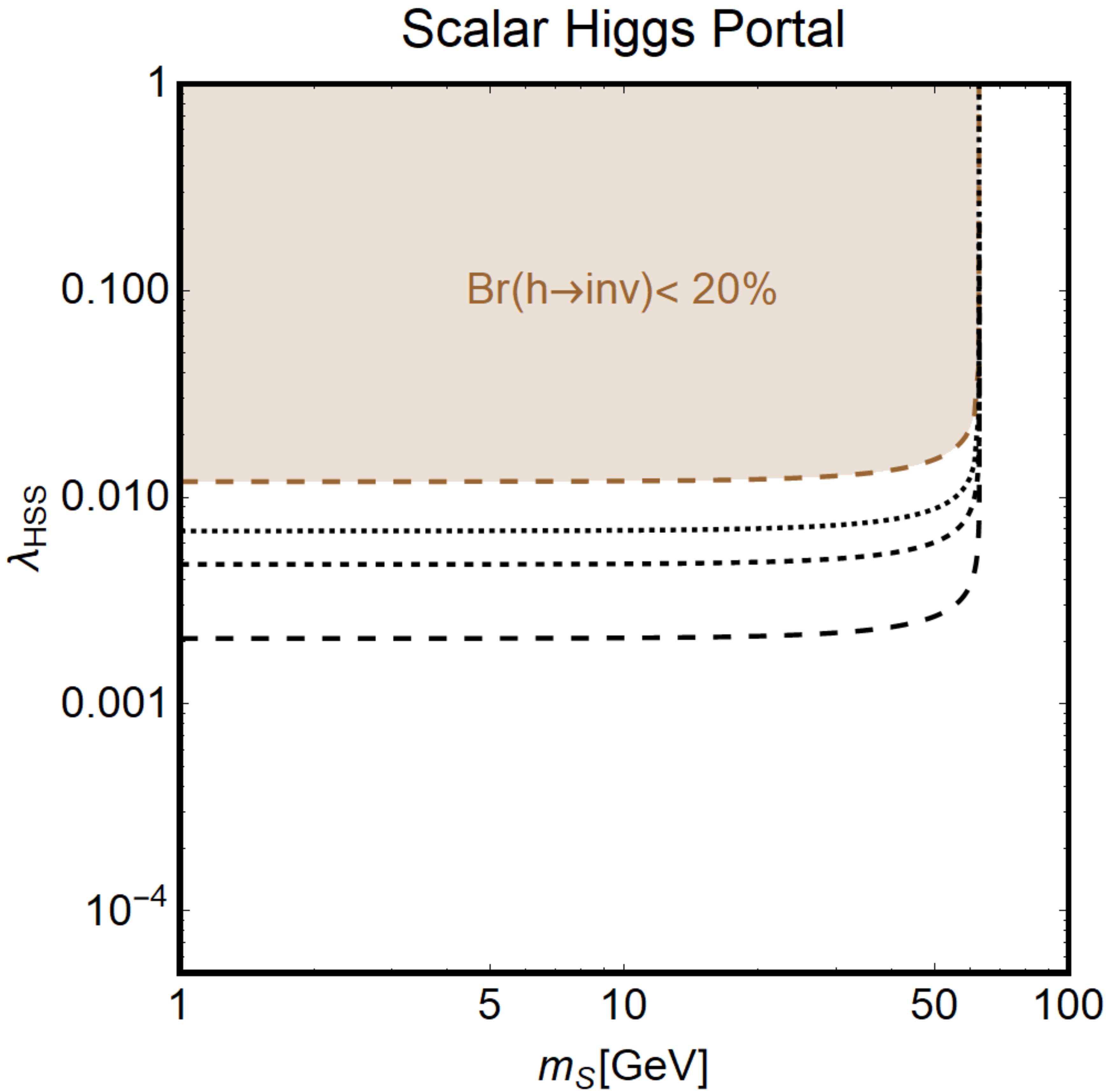}~~~~
\includegraphics[width=7cm]{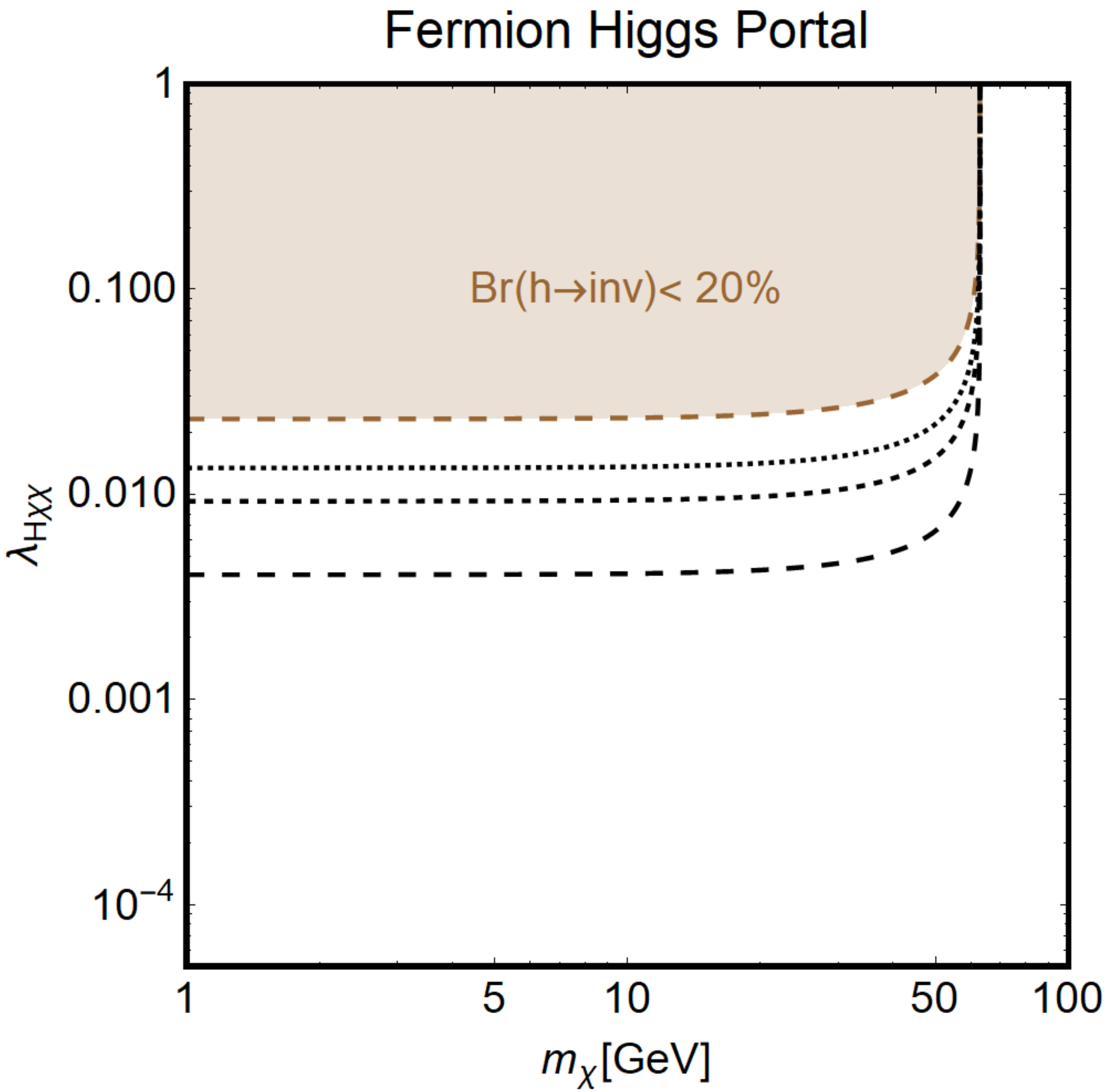}
}\\
\includegraphics[width=7cm]{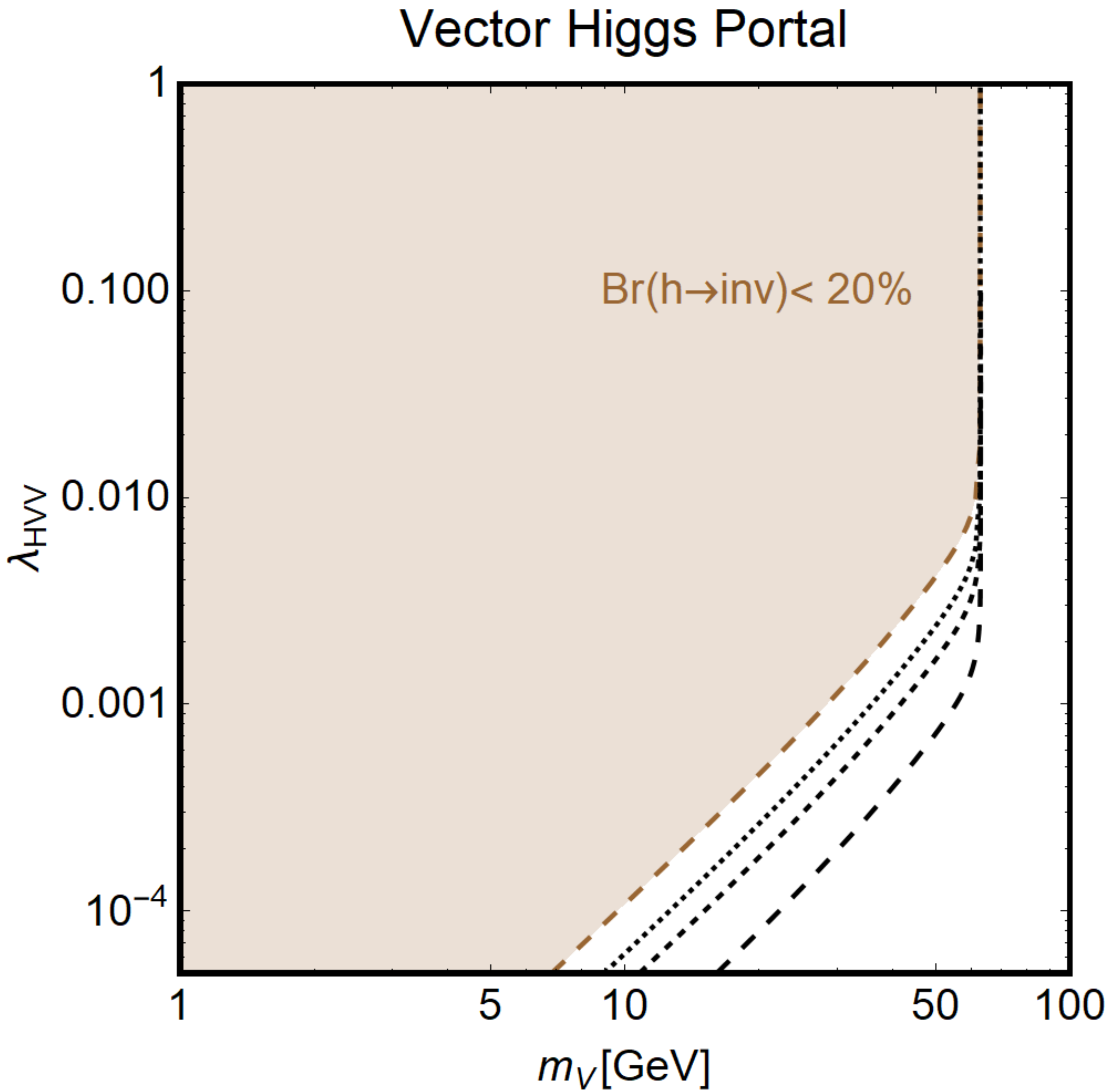}
\end{center}
\vspace*{-5mm}
\caption{Summary of constraints on the invisible Higgs branching ratio in the planes $[m_X,\lambda_{HXX}]$ for the Higgs--portal DM scenarios in the scalar (top left), fermionic  (top right) and vector (bottom panel) cases.  The magenta area  is the one excluded by the present limit BR$(H \to {\rm inv})<20\%$, while the full, dotted and dashed contour lines correspond to sensitivities on
the branching ratio of 10\%, 5\% and 1\%, respectively.} 
\label{fig:BR-DM}
\vspace*{-3mm}
\end{figure}

These bounds on the invisible Higgs branching fractions  will set tight
constraints on the masses and the couplings of the DM particles.  In the
effective approach that we are adopting for the Higgs--portal scenarios with DM
particles of spin--0, $\frac12$ or 1, recalling the expressions of the invisible
Higgs partial widths given in eq.~(\ref{GammaInv}), one obtains the countours in
the planes $[m_X, \lambda_{HXX}]$ shown in Fig.~\ref{fig:BR-DM}.  The colored
regions are those excluded by the bounds on  BR($H \to {\rm inv})$ from present
LHC data, while the other countours are for the sensitivities of  10\%, 5\% and
1\% expected for the branching ratio in the future. As can be seen, for $m_X
\lsim 62.5$ GeV, Higgs couplings approximately above $\lambda_{HXX} \approx
10^{-2}$ are already excluded in the scalar and fermion (assuming $\Lambda=1$
TeV) cases. In the vector case, even smaller couplings are excluded by the LHC
data,  in particular at low DM masses and for instance, one has  $\lambda_{HVV}
\lsim 10^{-4}$ for  $m_V \approx 10$ GeV. Future measurements would further
constrain these couplings, e.g. by  an order of magnitude  if BR($H \to {\rm
inv})<1\%$. 

\subsubsection{DM production through off--shell Higgs bosons}

In the previous discussion, we have assumed that the DM particles $X$  were
light enough, $m_X < \frac12 M_H$, so that the two--body  decays $H \to XX$ 
occur.   In turn, if the mass of the DM particle is larger than half the Higgs
mass, $m_X \geq \frac12 M_H$, there is no invisible two--body Higgs decay and
the detection of the DM  particles in collider experiments becomes much more
difficult. In fact, the only possible way  to observe the invisible states would
be through their pair production in the continuum  via the exchange of the Higgs
boson \cite{Djouadi:2011aa,Craig:2014lda,Baglio:2015wcg}. The latter needs to be
produced in association with visible  particles and, at hadron colliders, three
main processes are at hand similarly to single Higgs production:  

$i)$  double production in Higgs--strahlung $q\bar{q} \to  VH^* \to V XX$ with $V=W,Z$ bosons, 

$ii)$ vector boson fusion, $qq \to H^* qq  \to qq XX$, leading to two jets and missing energy,

$iii)$ the gluon fusion mechanism, $gg, qg \to jH^* \to jXX$, but in which at
least   an additional final state jet is emitted to render the process visible.

Here again, associated production with heavy quark pairs, $gg \rightarrow \bar t
t H$ and/or $gg \rightarrow \bar b b H$, have too low rates at the LHC to be
useful in this context (see however Ref.~\cite{Craig:2014lda}).  

Analytical expressions of the cross  sections for the three processes are given 
in Appendix A4. In the following, we simply present numerical results for the DM
pair production cross sections through Higgs splitting in the three possible
production processes listed above and in the three spin cases for the DM
particles using the effective field theory approach. The results will be  shown
for the c.m. energy $\sqrt s=14$ TeV which would be the  ultimate energy to be
reached at the HL--LHC \cite{ATLAS:2013hta,CMS:2013xfa,Cepeda:2019klc} and they
are compared to what can be obtained at $\sqrt s=100$ TeV, an energy to be
reached at the future hadron colliders foreseen at CERN and in China
\cite{Mangano:2017tke,Tang:2015qga}. Note that in all cases, one can implement
the most important radiative corrections to the processes, borrowing them from
what is known for the production of an on--shell Higgs
boson~\cite{Djouadi:1991tka,Dawson:1990zj,Spira:1995rr,Harlander:2002wh,Anastasiou:2002yz,Ravindran:2003um,Anastasiou:2015ema}.
These corrections are taken into account in our analysis as they significantly 
increase the tree--level cross sections. In all cases, we adopted the MSTW  PDFs
set for the parton distribution functions~\cite{Martin:2009iq}. 

\begin{figure}[!ht]
\vspace*{-1mm}
\centerline{
\includegraphics[scale=0.7]{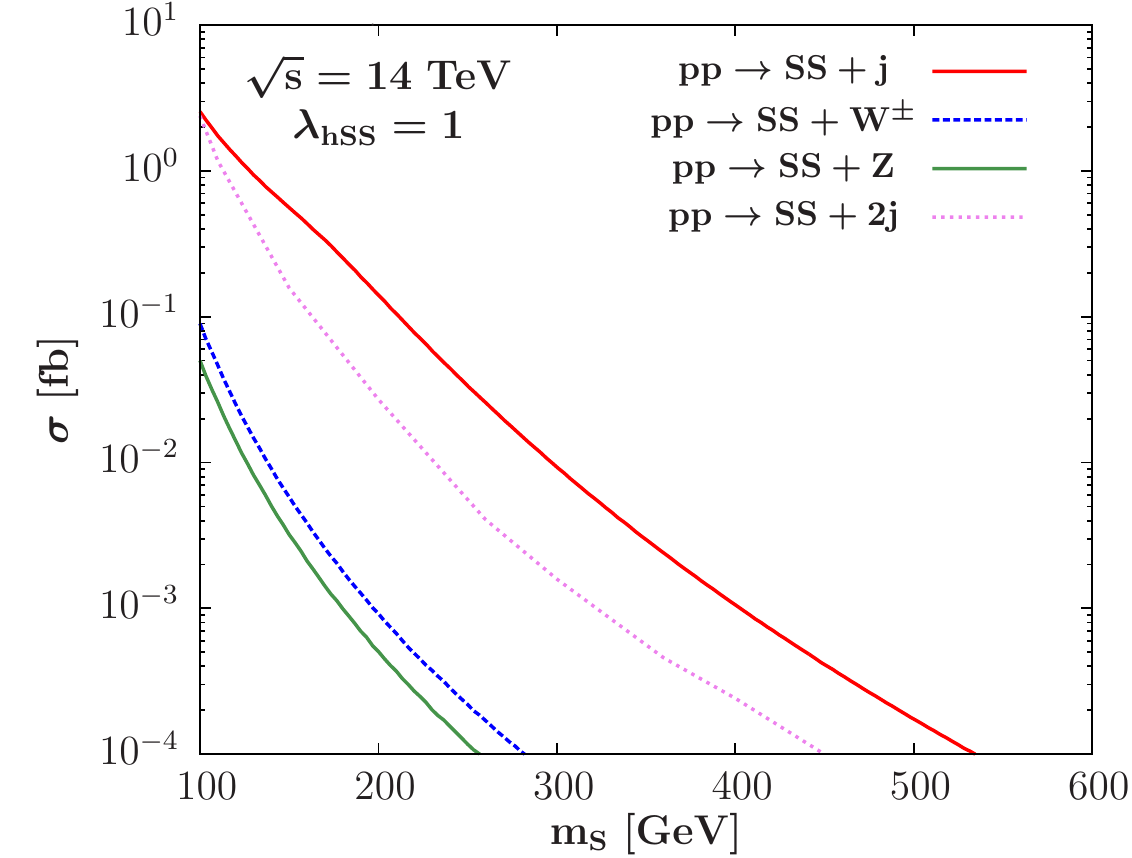}
\includegraphics[scale=0.7]{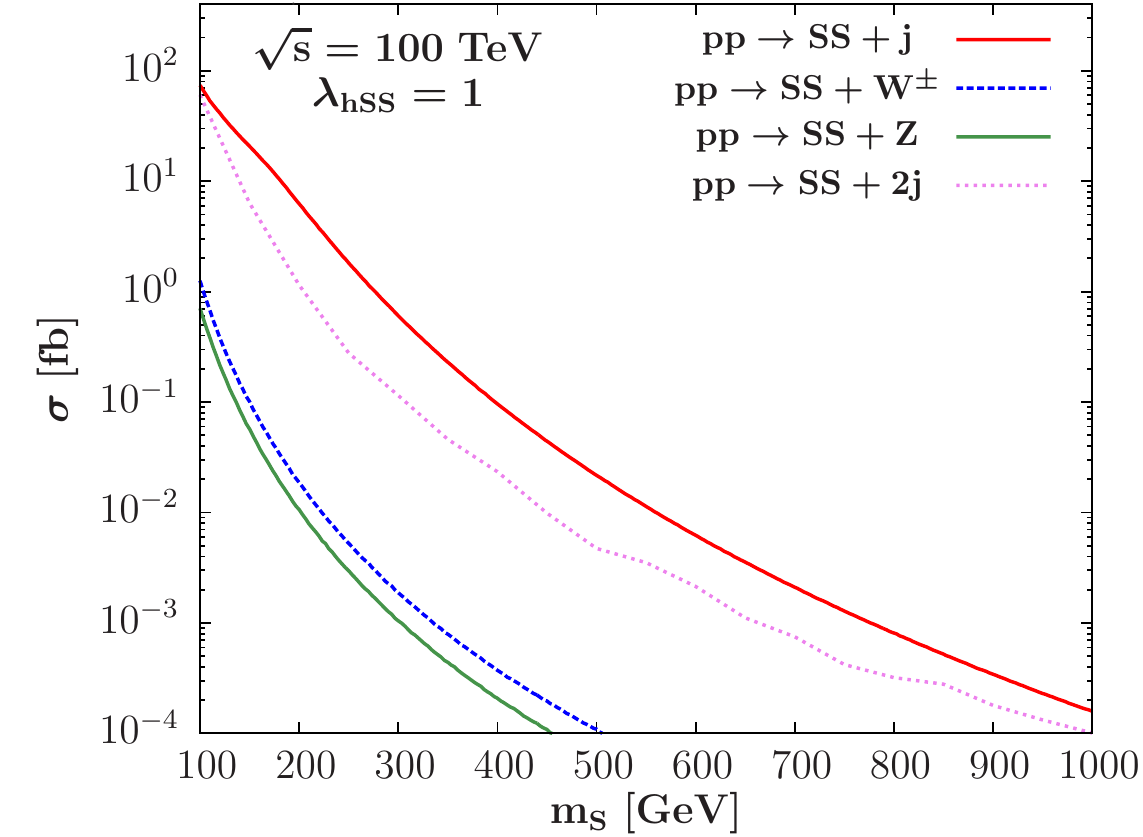} }
\vspace*{-1mm}
\caption[]{DM pair production cross sections in the continuum  at proton colliders with c.m. energies of $\sqrt{s}=14$ TeV (left) and 100 TeV (right) as a function of the scalar DM mass $m_S$. We assume scalar DM particles with $\lambda_{HSS}=1$ in the various processes.}
\label{prodDM-off1}
\vspace*{-.1mm}
\end{figure}

\begin{figure}[!ht]
\vspace*{-.1mm}
\centerline{
\includegraphics[scale=0.7]{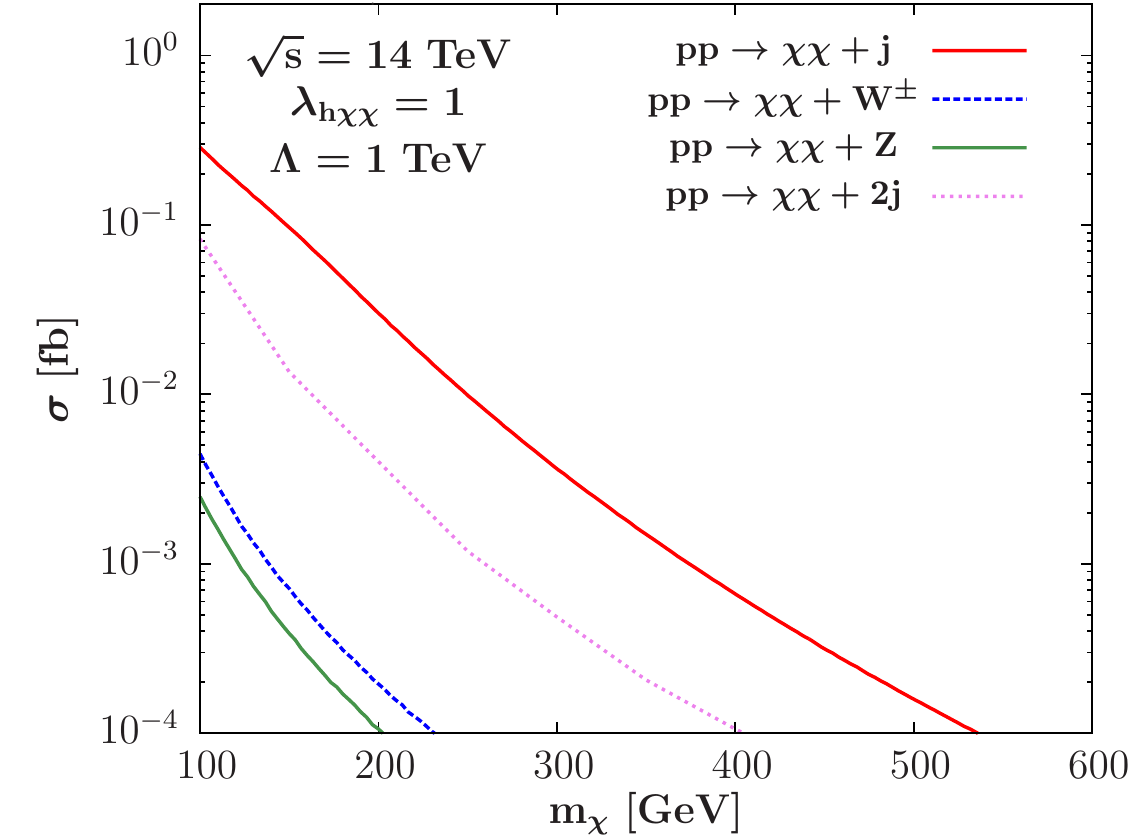}
\includegraphics[scale=0.7]{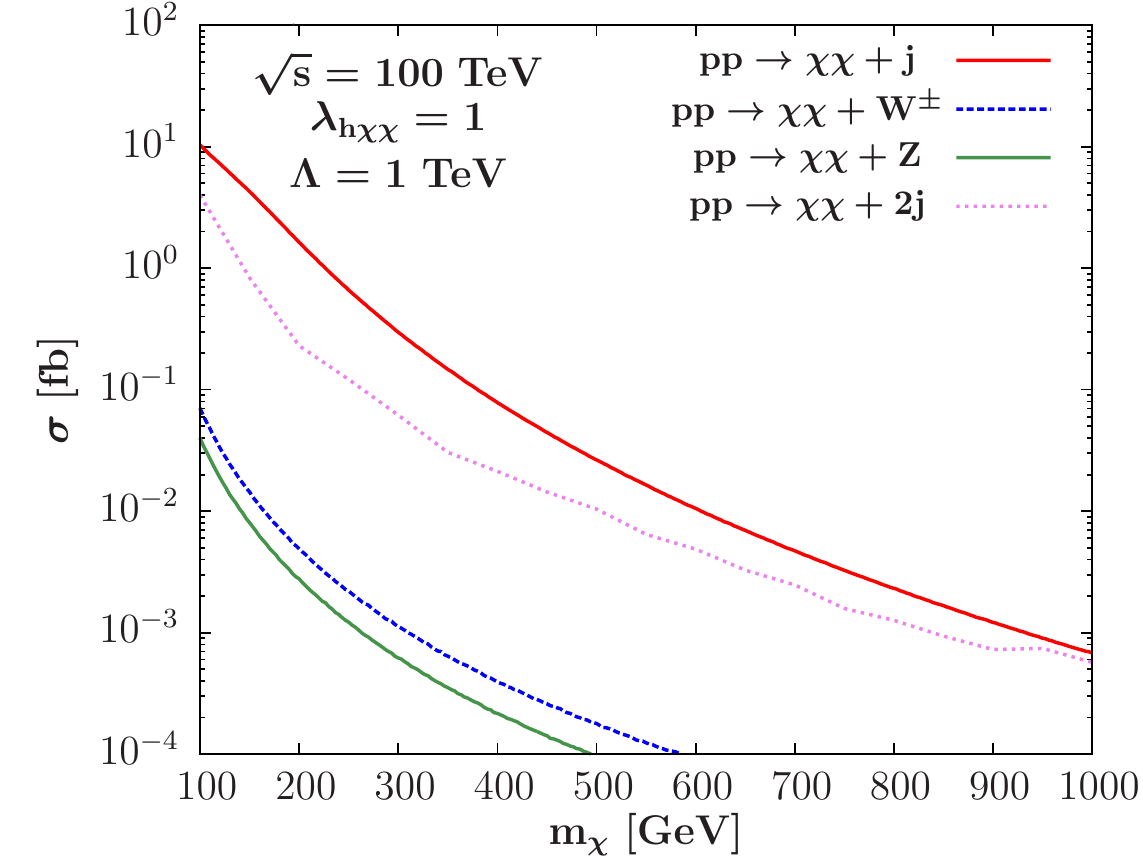} }
\vspace*{-1mm}
\caption[]{The same as in Fig.~\ref{prodDM-off1} but for fermionic DM  
with $\lambda_{H \chi\chi}=1$ and  $\Lambda=1$ TeV.}
\label{prodDM-off2}
\vspace*{-.1mm}
\end{figure}

\begin{figure}[!ht]
\vspace*{-.1mm}
\centerline{
\includegraphics[scale=0.7]{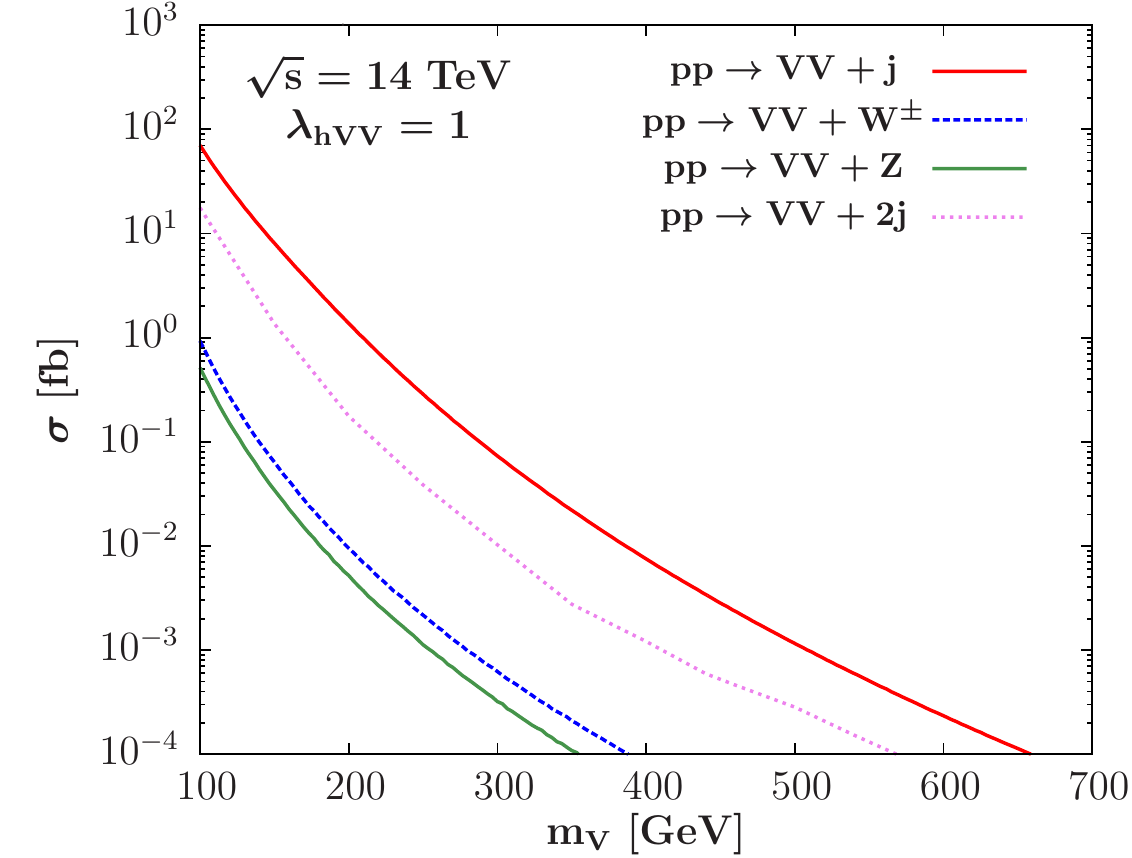}
\includegraphics[scale=0.7]{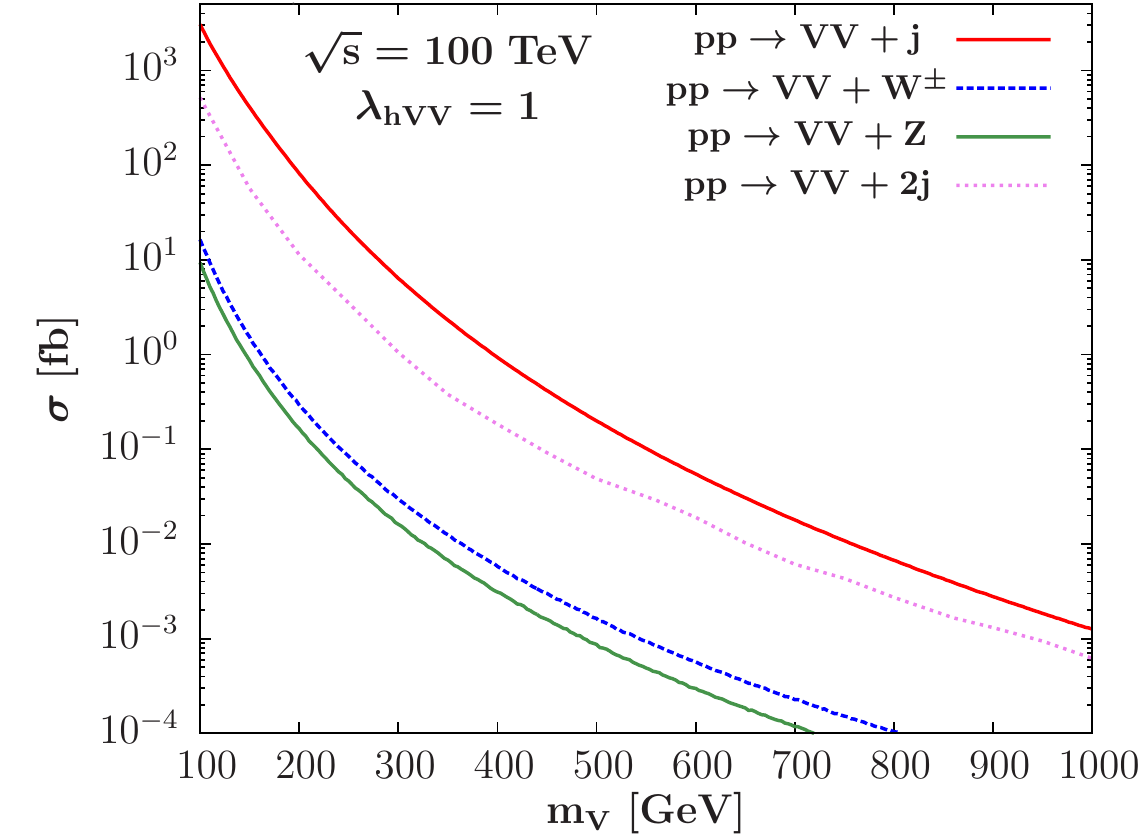} }
\vspace*{-1mm}
\caption[]{The same as in Fig.~\ref{prodDM-off1} but fore a vectorial DM state 
with $\lambda_{H VV}=1$.}
\label{prodDM-off3}
\vspace*{-.1mm}
\end{figure}

The DM pair production cross sections, which include the dominant QCD
corrections, are shown in Figs.~\ref{prodDM-off1}, \ref{prodDM-off2} and 
\ref{prodDM-off3} in the scalar, fermion and vector cases respectively, as
functions of the mass of the generic DM particle $X$~\cite{Baglio:2015wcg}. We
have set the DM couplings to the $H$ portal to $\lambda_{H XX}=1$ (in the
fermionic case, we also set $\Lambda=1$ TeV); for other Higgs--DM couplings one
simply has to multiply the rates by a factor  $\lambda_{HXX}^2$. The results are
shown for the c.m. energies $\sqrt s=14$ TeV (left) and $\sqrt s=100$ TeV
(right). 

For DM double production in the Higgs--strahlung process with either a $W$ or a
$Z$ boson,  the cross sections are extremely small at the  LHC even for light DM
states,  $m_X=100$ GeV.  For such particle masses, they do not exceed the fb
level in the spin--1 case and they are one and two orders of magnitude smaller
in, respectively,  the spin--0 and spin--$\frac12$ cases, with the rate for
$WXX$ being twice as large as the one for $ZXX$, as it is usually the case due
to the larger charged current  couplings compared to the neutral ones.  

For the vector boson fusion case in which the pair of escaping DM particle is
produced in association with two jets, the cross sections at the LHC are one
order of magnitude larger than in Higgs--strahlung for the three
spin--configurations and the hierarchy is the same: one order of magnitude 
larger for spin--1 DM particles than for spin--0 and than for spin $\frac12$
when the New Physics scale is assumed to be $\Lambda=1$ TeV.  

For  DM double production in association with one jet either from  gluon fusion
$gg\to XX g$ or $qg$ annihilation $gq \to q XX$,  the cross sections are a
factor 3 to 10  larger than in the vector boson fusion case, except at very low $m_X$ for the spin--0 case where they are approximately the same. 

Hence, when  the DM  particles are heavier than $\frac12 M_H$ and  the Higgs
boson is virtual in the process $H^* \to XX$, the production rates are rather
modest at the LHC and the present luminosity will not allow to probe a
significant portion of the  parameter space allowed of the Higgs--portal models.
To have more sensitivity to the DM particle masses, one needs a significantly
larger sample of produced Higgs bosons. This could occur first at the 
high--luminosity LHC option (HL--LHC) in which up to 3 ab$^{-1}$ of data could
be collected at a c.m. energy of $\sqrt s= 14$ TeV
\cite{ATLAS:2013hta,CMS:2013xfa,Cepeda:2019klc} or at higher energy pp colliders
such the ones with $\sqrt s=100$ TeV planed at CERN and in China
\cite{Mangano:2017tke,Tang:2015qga}.   

In the latter case, the production rates can be enhanced by several orders of
magnitude as can be seen from the right--hand side of the figures above.    For
light DM states, $m_X \approx 100$ GeV and in all spin configurations for the DM
particle, the cross sections are approximately 50 times larger at $\sqrt s=100$
TeV than at  $\sqrt s=14$ TeV in the ggF and VBF cases while in the VH process
the enhancement factor is only about a factor of 10. If, in addition, the
luminosity at these high--energy colliders is as high as the one planed for the
HL--LHC, i.e. at the level of a few   ab$^{-1}$ or even larger, one could have a
large enough number of events to probe the DM particles in these channels.

\begin{figure}[!h]
\vspace*{-1mm}
\centerline{
\includegraphics[scale=0.82]{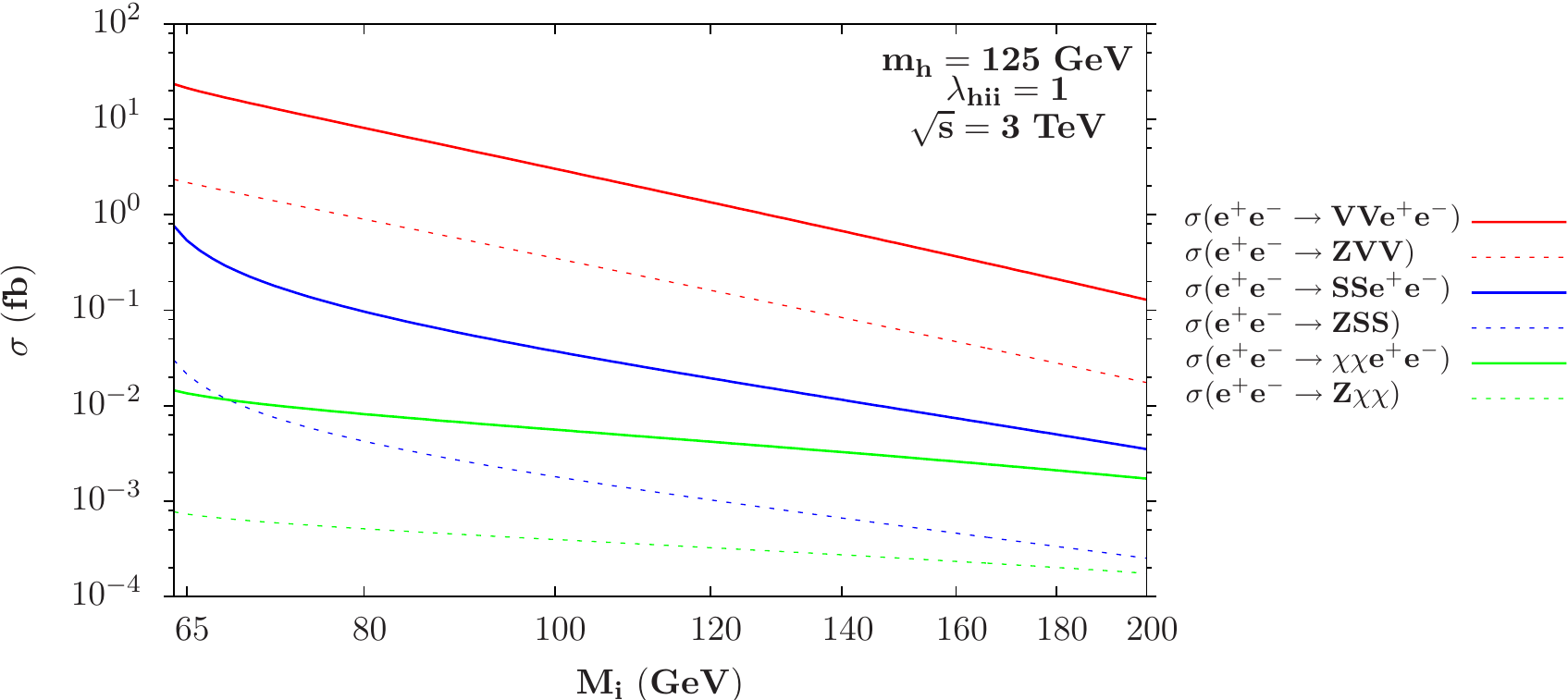}}
\vspace*{-1mm}
\caption{{Scalar, fermion and vector DM pair production cross sections in the processes  $e^+ e^- \to ZXX$  and  $ZZ \to XX$  with $\sqrt s=3$ TeV, as a function of their mass for $\lambda_{HXX}=1$.}} 
\label{Fig:CLIC}
\vspace*{-2mm}
\end{figure}

Finally, observing DM pair production in the continuum might be easier in the
cleaner environment of  $e^+e^-$ colliders
\cite{Djouadi:1994mr,Accomando:1997wt,AguilarSaavedra:2001rg}. The two most
important production  processes for a pair of DM particles are $e^+ e^- \to ZXX$
which is similar to VH in proton collisions and $e^+ e^- \to Z^* Z^* \to e^+
e^-XX$ which is similar to $ZZ$ fusion in VBF\footnote{The production rate in
$WW$ fusion,  $e^+ e^- \to W^* W^* \to \nu \bar \nu XX$, is one order of
magnitude larger than in $ZZ$ fusion but leads to a fully invisible signal
unless an additional  photon is  radiated.}.  Analytical, results for the cross
sections are again given in Appendix A4 and the results for the two processes
are shown in Fig.~\ref{Fig:CLIC}  at the c.m. energy $\sqrt s=3$ TeV expected
for the CERN CLIC machine, again in the case of a scalar, fermionic and
vectorial DM candidates \cite{Quevillon:2014owa} with couplings set to
$\lambda_{HXX}=1$. They are higher for $ZZ$ fusion than in Higgs--strahlung and
for the spins of the DM states, the largest cross sections are obtained for
vector, then fermion, then scalar states and differ by an order of magnitude in
each case.

\subsection{Constraints from astroparticle experiments}

\subsubsection{Astrophysical set-up}

We come now to the astroparticle aspects of the Higgs--portal to DM scenarios
and start by briefly summarizing the main features of WIMP phenomenology in this
context;  for a more extensive discussion  we refer for example to the reviews 
of Refs.~\cite{Arcadi:2017kky,Roszkowski:2017nbc}.

The most peculiar feature of a WIMP DM candidate is the mechanism that generates
its cosmological relic density which, under the assumption of a standard
cosmological evolution of the Universe, is described by a Boltzmann equation of
the form
\begin{equation}
\frac{dY_{\rm DM}}{dt}=\frac{ds}{dt}\frac{\langle \sigma v \rangle}{3 \mathcal{H}}Y_{\rm DM}^2 \left(1-\frac{Y_{\rm DM,eq}^2}{Y^2_{\rm DM}}\right) , 
\end{equation}
with $\mathcal{H}$ and $s$ being, respectively, the Hubble expansion parameter and the
entropy density which, in standard cosmology, obeys the conservation law
${ds}/{dt}=-3 \mathcal{H} s$. In the equation above, $Y_{\rm DM}$ is the DM yield or
comoving number density defined as $Y_{\rm DM}={n_{\rm DM}}/{s}$ with
$n_{\rm DM}$ being the number density. $Y_{\rm DM,eq}$ represents the DM yield
assuming a Maxwell--Boltzmann thermal distribution function giving 
\begin{equation}
n_{\rm DM,eq}=\frac{g_{\rm DM}}{2\pi^2}m_{\rm DM}^2 T K_2 
\left({m_{\rm DM}}/{T}\right) , 
\end{equation}
where $g_{\rm DM}$ stands for the DM internal degrees of freedom, $T$ for the temperature of the primordial thermal bath while $K_i$ is the modified Bessel's function of $i$th type. $\langle \sigma  v \rangle$ finally stands for the thermally averaged DM pair annihilation cross section and encodes the information from the particle physics framework in which the DM
is embedded.

After having been in thermal equilibrium at the early stages of the evolution of
the Universe, the DM decouples at a typical temperature $T_{\rm fo}\sim
\frac{1}{20}{m_{\rm DM}}$--$\frac{1}{30}{m_{\rm DM}}$, the freeze--out
temperature,  and  its final relic density can be approximately expressed
as~\cite{Gondolo:1990dk}:
\begin{equation}
\label{eq:relic_density}
\Omega_{\rm DM}h^2 \approx 8.76 \times 10^{-11}{\mbox{GeV}}^{-2} {\left[\int_{T_0}^{T_{f.o.}} g_{*}^{1/2} \langle \sigma v \rangle \frac{dT}{m_{\rm DM}}\right]}^{-1}, 
\end{equation} 
where $h \sim 0.7$ it the Hubble expansion rate at present times in units of 100 $(\mbox{km}/\mbox{s})/\mbox{Mpc}$. From this equation, it is clear that the requirement
of a correct DM relic density translates into the requirements of a specific
value of $\langle \sigma v \rangle$, found to be of the order of
$10^{-26}{\mbox{cm}}^3 {\mbox{s}}^{-1}$. The thermally averaged cross section is
related to the conventional annihilation one $\sigma$, when implicitly
assuming a sum over the final states that are kinematically allowed
\begin{equation}
\langle \sigma v \rangle=\frac{1}{8m_{\rm DM}^4 T K_2\left({m_{\rm DM}}/{T}\right)^2}\int_{4 m_{\rm DM}^2}^{\infty} ds \sqrt{s}(s-4 m_{\rm DM}^2) \sigma(s) K_1\left({\sqrt{s}}/{T}\right) . 
\label{eq:general_sigmav}
\end{equation} 

In specific physics scenarios, extra states very close in mass with the DM
particle might be present. In such a case, $\langle \sigma v \rangle$ should be
replaced by an effective cross section including co--annihilations, i.e.
annihilation processes involving these extra states~\cite{Edsjo:1997bg}
\begin{equation}
\langle \sigma_{\rm eff} v \rangle=\sum_{i,j} \langle \sigma_{ij} v_{ij} \rangle \frac{n_i}{n_{i,\rm eq}}\frac{n_j}{n_{j,\rm eq}}, 
\end{equation}
where the indices $i,j$ run over all the states that might affect the DM relic density.

The constraints from the DM relic density on the scenarios discussed in this
work have been determined by implementing the models in the numerical package
micrOMEGAs~\cite{Belanger:2001fz,Belanger:2007zz,Belanger:2008sj} which
determines with high accuracy the solution of the Boltzmann equation for the DM
particle as a function of the basic model parameters. In our discussion, this
numerical treatment will be nevertheless accompanied, whether possible, with
analytic estimates based on the so called velocity expansion, the elements of
which are given  in Appendix B.  

Indeed, the DM pair annihilation cross section can be decomposed into a
temperature (and hence time) independent contribution, dubbed $s$--wave term,
and temperature (time) dependent contribution, dubbed $p$--wave term, with the
temperature dependence encoded in the DM velocity,  $v_r \sim 0.3$ at the freeze out time, 
\beq 
\langle \sigma v \rangle \simeq a+b v_r^2 /2,
\eeq
Depending on the particle physics model the $s$--wave term can be the dominant
contribution to the annihilation cross section or, on the contrary, has a  null
value. $p$--wave dominated cross sections are in general more suppressed than
$s$--wave dominated one since $v_r \sim 0.3$ at freeze--out; consequently
stronger couplings between the DM and the SM states are needed to obtain the
measured relic density. As will be clarified below, whether the DM annihilation
cross sections is $s$--wave or $p$--wave dominated is of utmost importance to
asses whether the considered scenario can be tested through DM indirect
detection. A final important remark is that the velocity expansion is not valid
in some phenomenologically relevant scenarios, such as the presence of $s$--channel resonances, the opening of thresholds with new annihilation channels and co--annihilations~\cite{Griest:1990kh}.

Besides featuring the correct cosmological relic density, a viable WIMP DM
should also evade the present constraints from DM searches in astroparticle
physics experiments. The search strategies in this case are mainly subdivided
into two categories: direct detection and indirect detection. These searches can
be  complementary to the collider searches already discussed in the previous
subsections.

Direct detection (DD) strategies of DM particles are based on the possibility of
measuring the energy deposited in target detectors by scattering processes of DM
particles on their nuclei. The event rate can be generically written as
\begin{equation}
\label{eq:DDsignalRate}
\frac{dR}{dE}=\frac{N_T \rho_{\rm DM}}{m_{\rm DM} m_T}\int_{v_{\min}}^{v_{\rm max}} v f_E (v) \frac{d \sigma (v,E)}{dE}d^3 v , 
\end{equation}
where $E$ is the recoil energy associated to the scattering events, $m_T$ the mass of the target nucleus and $f_E$ is the distribution of the velocity of the WIMPs in the frame of the Earth, i.e. the probability of finding a WIMP with velocity $v$ at the time $t$. The integration interval is represented by $v_{\rm min}=\sqrt{m_T E/(2 \mu_T^2)}$, i.e. the minimum WIMP speed to induce a scattering process with recoil energy $E$, with $\mu_T= m_{\rm DM} m_T/(m_{\rm DM}+m_T)$ being the reduced DM--nucleus mass and $v_{\rm max}=v_{\rm esc}$ with $v_{\rm esc}$ being the DM escape velocity, i.e. the velocity above which the DM is no longer gravitationally bound to the Milky Way.

The DM differential scattering rate, ${d\sigma}/{dE}$ can be conventionally
decomposed into a spin independent (SI) and a spin dependent (SD)
component\footnote{This is actually a simplification which is reliable for the
type of models which will be discussed in our  work. For a more general
formalism,  we refer for example to
Refs.~\cite{Fitzpatrick:2012ix,Anand:2013yka,DelNobile:2013sia,Bishara:2017pfq}.}
\begin{equation}
\frac{d\sigma}{dE}=\frac{m_N}{2 \mu_T^2 v^2}\left(\sigma^{\rm SI}|F_{\rm SI}(q)|^2+ \sigma^{\rm SD}|F_{\rm SD}(q)|^2\right) , 
\end{equation}
where $F_{\rm SI},F_{\rm SD}$ are form factor functions of the momentum transfer
$q$, while $\sigma^{\rm SI}$ and $\sigma^{\rm SD}$ are the spin--independent and
spin--dependent  scattering cross sections of the DM on nucleons in the limit of
vanishing momentum transfer. Null results from direct detection experiments are
translated, by use of the expression above, into limits on the spin--independent
or spin--dependent cross sections on nucleons (customarily protons but limits
from the different type of nucleons can be eventually combined, as e.g. done in
Ref.~\cite{Tovey:2000mm}) as a function of the DM mass. At the moment, the most
severe constraints are imposed to spin--independent interactions as a result of
their coherent nature.

The  Higgs--portal scenarios that we discuss here lead mostly to 
spin--independent  interactions of the DM with nuclei. We will apply the
constraints from the current world leader experiment, 
XENON1T~\cite{Aprile:2018dbl}, possibly complemented at light DM masses by the
ones from the DarkSide--50 experiment \cite{Agnes:2014bvk,Agnes:2018oej}. It is
also worth mentioning that weaker but nevertheless relevant limits on
spin--independent  interactions of the DM have been also obtained by the
LUX~\cite{Akerib:2016vxi} and PandaX~\cite{Tan:2016zwf} experiments. We will
also investigate whether future Xenon based detectors
XENONnT~\cite{Aprile:2015uzo} (a similar sensitivity is expected for the LZ
experiment~\cite{Szydagis:2016few} as well) and DARWIN~\cite{Aalbers:2016jon}
can further probe the available parameter space of the DM state.

Indirect detection (ID) of DM particles consists into the search of the products
of DM annihilation at galactic or extragalactic scales, over the expected
background from known astrophysical sources. This can be done using earth based
telescopes such as HESS~\cite{Abramowski:2011hc,Abramowski:2014tra} and CTA
\cite{Acharya:2017ttl,Balazs:2017hxh}, or space detectors such as
AMS~\cite{Lu:2015pta} and Fermi--LAT~\cite{Abdo:2010ex}. Similarly to the case
of direct detection, the experimental signal, a differential flux in this case,
depends on a combination of astrophysical inputs, like the DM distribution in
the sources target of experimental searches, and particle physics inputs, such
as the DM annihilation cross section.  The absence of experimental signals can
be translated into  upper bounds on the DM annihilation rates into given final
states. 

The detection rate is, however,  sensitive to several astrophysical inputs, in
particular to the DM energy density $\rho(r)$ in the source, through the so called
$J$--factor.  Among the possible final states, the most compelling limits are
provided at the present time by searches of gamma--rays produced in the
interactions of the primary final state products of DM annihilation, e.g. during
the hadronization processes.

As both the relic density and the detection signal rates are mostly determined
by the particle physics inputs, some complementarity between them can be in
general established~\cite{Roszkowski:2016bhs}. In particular, experimental
exclusion limits can be converted into upper bounds on the DM annihilation cross
section responsible for the relic density. In the next subsections, we will
present and discuss the most relevant constraints from DM phenomenology and how
they complement information from collider searches, focusing again on the
simplest DM models with a SM--like Higgs sector. Since limits from DM indirect
detection are, in most scenarios, subdominant with respect to the ones from
direct detection, we will omit them  for simplicity   unless otherwise
specified.

\subsubsection{The DM cross sections}

In this subsection we will introduce some elements that will be helpful for the
understanding of the numerical results which will be described here. As already
pointed out, a reliable (with some notable exceptions) approximation of the DM
relic density is obtained by performing the velocity expansion of the thermally
averaged annihilation cross section. In Appendix B2, we present in
eqs.~(\ref{eq:Sc_in_SM})--(\ref{eq:Vec_in_SM}), the expansions of the
annihilation cross sections of scalar, fermionic and vector DM, retaining only
the leading order contributions.  The annihilation into $XX\to \bar f f$, $WW,ZZ$ final states for which the expressions are rather simple and the more 
complicated ones for the $XX \to HH$ final state are given. 

As can be seen, the cross sections for the different spin assignments of the DM
state feature similar dependence on the masses of the DM and of the Higgs
states. The annihilation cross sections of scalar and vector DM states are
$s$--wave dominated, while the ones of fermion DM are $p$--wave and proportional
to the squared DM velocity $v_r^2$. We thus expect that, for a given value of
the DM mass, the annihilation cross section of fermionic DM is more suppressed
with respect to the other two cases as one has $v_r^2 \sim 0.1$ at the typical
freeze--out temperature for a WIMP DM. Higher values of the coupling are then,
in general, required for fermionic DM to comply with the requirement of a
correct relic density. 

The velocity dependence has also important implications for indirect detection.
$s$--wave dominated cross sections have a very weak time dependence (as it is in
a subleading term) and indirect detection experiments are then capable of
probing, at least for masses below 100 GeV, the thermally favored values of the
annihilation cross sections. In the case of $p$--wave dominated cross sections,
the value of the velocity at present times is very different from that at
thermal freeze--out, the former being $v_r \sim  10^{-3}$. Fermionic DM
states with the correct relic density would then lie well below  the expected
sensitivity of indirect detection experiments.

Let us now move to DM scattering on nuclei. The involved processes   have a  characteristic energy scale of the order of 1 GeV and very low momentum exchange between the DM state and the nucleon, $q \sim {\cal O}(100\,\mbox{MeV})$. Furthermore, the  present time low DM velocity allows to consider this process in the non--relativistic limit. Given this, the scattering of DM states with nucleons can be described, at the microscopic level, starting from effective four field interactions between the DM and the SM quarks:
\begin{equation}
    \mathcal{L}=\frac{\lambda_{HSS} y_q}{M_H^2}S^2 \bar q q,\ \ \ \  \mathcal{L}=\frac{\lambda_{H\chi \chi} y_q}{M_H^2}\bar \chi \chi \bar q q,\ \ \ \  \mathcal{L}=\frac{\lambda_{HVV} y_q}{M_H^2}V^\mu V_\mu \bar q q , 
\end{equation}
where a sum over the six quarks is implicitly assumed with $y_q$ being their corresponding Yukawa couplings. From this, it is possible to obtain effective interactions between the DM particle and a nucleon $N=p,n$

\begin{equation}
\label{eq:eff_DD}
    \mathcal{L}=\frac{\lambda_{HSS} \lambda_N}{M_H^2}S^2 \bar N N,\ \ \ \  \mathcal{L}=\frac{\lambda_{H\chi \chi} \lambda_N}{M_H^2}\bar \chi \chi \bar N N,\ \ \ \  \mathcal{L}=\frac{\lambda_{HVV} \lambda_N}{M_H^2}V^\mu V_\mu \bar NN,
\end{equation}
where
\begin{equation}
    \lambda_N=m_N \sum_{q=u,d,s,c,b,t}\, {y_q f_q^N}/{m_q} . 
\end{equation}
The coefficients $f_q^N$ with $q=u,d,s$ represent the contributions of the light quarks to the mass of the nucleon, namely
\begin{equation}
    f_q^N \equiv {\langle N| m_q \bar q q |N \rangle}/{m_N} .
\end{equation}
The coefficients associated to the heavy $c,b,t$ quarks are, in turn, expressed in terms of a unique coefficient associated to the gluon, 
\beq
f_c^N=f_b^N=f_t^N={2}f_{TG}/27 ={2}(1-\sum_{q=u,d,s} f_q^N)/27. 
\eeq
This is because, at the typical energy scale of scattering processes, it is possible to integrate out the heavy quarks using the relation
\begin{equation}
    m_Q \bar Q Q=-\frac{\alpha_s}{12 \pi}G^{\mu \nu}_a G_{\mu \nu a},\,\,\,\,\, Q=c,b,t , 
\end{equation}
and
\begin{equation}
\label{eq:eff_GG_DD}
  \frac{\langle N| G^{\mu \nu}_a G_{\mu \nu a} | N \rangle}{m_N}=-\frac{8 \pi}{9 \alpha_s}f_{TG} . 
\end{equation}
The parameters $f^N_{u,d,s}$ can be determined from pion--nucleon scattering~\cite{Alarcon:2011zs,Crivellin:2013ipa,Hoferichter:2015dsa}. The numerical results presented in this work have been obtained by taking the   central values of the following measurements:
\begin{align}
    & f_u^p=(20.8 \pm 1.5) \times 10^{-3},\,\,\,\,\,\,f_u^n=(18.9 \pm 1.4) \times 10^{-3}, \nonumber\\
    & f_d^p=(41.1 \pm 2.8) \times 10^{-3},\,\,\,\,\,\,f_u^n=(45.1 \pm 2.7) \times 10^{-3}, \nonumber\\
    & f_s^p=f_s^n=0.043 \pm 0.011 \,,
\end{align}
which lead to $f_{TG}\approx 0.894$.
 From eq.~(\ref{eq:eff_GG_DD}) we immediately see that the DM scattering cross section will receive additional contributions in the presence of an additional effective coupling of the Higgs boson with gluons,  which could be induced for example by extra degrees of freedom,  of the form
\begin{equation}
    \mathcal{L}_{\rm eff}=\frac{k_g}{\Lambda}\frac{\alpha_s}{12 \pi}H G^{\mu \nu}_a G_{\mu \nu a} \, .
\end{equation}
with $\Lambda$ a suitably chosen scale of New Physics. In such a case one would have indeed
\begin{equation}
    \lambda_N=\sum_{q=u,d,s} \frac{m_N}{m_q} f_q^N y_q+\frac{2}{27}f_{TG}\left(\sum_{Q=c,b,t} \frac{m_N}{m_Q}-k_g \frac{m_N}{\Lambda}\right) \, .
\end{equation}

In the case where only the SM Higgs sector is assumed, $y_q=m_q/v$ so that $\lambda_N$ takes the very simple expression $\lambda_N={m_N}/{v}f_N$ with $f_N=\sum_{q=u,d,s} {6}f_{TG}/{27} \approx 0.3$.

Eq.~(\ref{eq:eff_DD}) correspond to spin--independent interactions which, using the expression for $\lambda_N$ just written above, gives rise to the following DM scattering cross section on nucleons
\begin{eqnarray}
&& \sigma^{\rm SI}_{SN} = \frac{\lambda_{HSS}^2}{16 \pi M_H^4} \frac{m_N^4  f_N^2}{ (m_S + m_N)^2} \;,  \nonumber\\
&& \sigma^{\rm SI}_{\chi N} = \frac{\lambda_{H\chi \chi}^2}{4 \pi \Lambda^2 M_H^4} \frac{m_N^4 m_\chi^2  f_N^2}{ (m_\chi + m_N)^2}, \nonumber\\
&& \sigma^{\rm SI}_{VN} = \frac{\lambda_{HVV}^2}{16 \pi M_H^4} \frac{m_N^4  f_N^2}{ (m_V + m_N)^2} \; . 
\end{eqnarray}

\subsubsection{Direct and indirect DM detection and Higgs physics}

In order to be viable, a WIMP Dark Matter candidate should have a relic density
(and thus an annihilation cross section at thermal freeze--out) compatible with
the experimental determination of $\Omega_{\rm DM} h ^2$ as well as rates for
direct and indirect detection below the present exclusion bounds. As can be seen
from the analytical expressions of the previous subsection, the relevant DM
interaction rates  for the effective Higgs--portal, depend simply on two
parameters, the DM mass and the coupling to the Higgs boson. Before presenting
the numerical analysis of the combination of these constraints, let us first see
whether the rates for DM direct and indirect detection can be related to the
outcome of searches of invisible Higgs decays at colliders. We anticipate that
this type of comparison will be subject to specific hypotheses that will be
more critically discussed in the next subsection.

One can notice that the partial Higgs decay width into the DM particles $X$, $\Gamma (H\to XX)$, and the spin--independent  $X$--proton elastic cross section $\sigma^{\rm SI}_{ X p}$ are both proportional to the coupling squared
$\lambda_{HXX}^2$. They can then be related and the ratio  $r_X= \Gamma (H
\to XX)/\sigma^{\rm SI}_{X p}$ depends only on the DM particle mass $m_X$ and
known SM parameters such as the Higgs mass $M_H=125$ GeV.  This allows to
relate the invisible Higgs branching fraction to the  direct detection cross
section in a very simple way: 
\beq 
{\rm BR}^{\rm inv}_X \equiv {\rm BR}(H \to XX) = \frac{ \Gamma (H \rightarrow XX)} {\Gamma^{\rm SM}_H
+ \Gamma (H \rightarrow XX)} =  \frac{\sigma^{\rm SI}_{X p}}{\Gamma^{\rm
SM}_H/r_X + \sigma^{\rm SI}_{X p}} 
\eeq
with $\Gamma^{\rm SM}_H=4.07$ MeV being the Higgs total decay width into all particles in the SM. From this relation, it is possible to determine for a given value of the DM mass, the maximal value of the invisible Higgs branching fraction compatible with present direct detection constraints. In the limit  $m_p \ll m_X \ll \frac12  M_H$, one can write the following simple approximate relations for the different spin assignments of the DM   
\begin{eqnarray}
\label{eq:Brsigma}
{\rm BR}^{\rm inv}_{S} &\simeq& { \left(\frac{\sigma^{\rm SI}_{S p}}{10^{-9} {\rm pb}}\right)} \bigg[ {400 \left(\frac{10\; {\rm GeV}}{m_S}\right)^2 + \left(\frac{ \sigma^{\rm SI}_{S p}}{10^{-9}{\rm pb}} \right)} \bigg]^{-1} \, ,  \nonumber \\
{\rm BR}^{\rm inv}_{V}&\simeq& { \left(\frac{ \sigma^{\rm SI}_{V p}}{10^{-9} {\rm pb}}\right)} \bigg[{4 \times 10^{-2} \left(\frac{m_V}{10\; {\rm GeV}}\right)^2 + \left( \frac{ \sigma^{\rm SI}_{V p}}{10^{-9}{\rm pb}} \right)}\bigg]^{-1} \, , \nonumber \\
{\rm BR}^{\rm inv}_{f}&\simeq& { \left(\frac{ \sigma^{\rm SI}_{f p}}{10^{-9} {\rm pb}}\right)} \bigg[ {3.5 + \left(\frac{ \sigma^{\rm SI}_{f p}}{10^{-9}{\rm pb}} \right)} \bigg]^{-1} \, .\label{Eq:brmax}
\end{eqnarray}

The relation between the invisible branching fractions and  the direct detection
cross sections strongly depends on the spinorial nature of the DM  particle; in
particular, the strongest (weakest) bound is obtained in the vectorial  (scalar)
case as will be seen shortly.

A correlation between DM observables and the invisible decay width of the Higgs
boson can be established also in the case of indirect detection. In this case,
the invisible Higgs branching fraction can be related to the DM
annihilation cross section responsible of the indirect signal,  the latter being
mainly due to a $\gamma$--ray continuum mostly originating from DM annihilation
into $\bar b b$ final states. By further simplifying the annihilation cross section into fermions given in eqs.~(\ref{eq:Sc_in_SM})--(\ref{eq:Vec_in_SM}) of Appendix B by taking the limit $m_{X} \ll M_H$ so that $\langle \sigma v_r \rangle (XX \rightarrow \bar b b) \propto {m_{X}^2 m_b^2}/{(v^2 M_H^4)}$, one obtains the following analytic expressions for the invisible Higgs branching ratios
\begin{align}  
& {\rm BR}^{\rm inv}_S \simeq {\left(\frac{\langle \sigma v_r
\rangle} {10^{-10}{\mbox{GeV}}^{-1}}\right)} \bigg[ {2.4 \times 10^{-2}+\left(
\frac{\langle \sigma v_r \rangle} {10^{-10}{\mbox{GeV}}^{-1}}\right)}
\bigg]^{-1} \, , \nonumber\\  
& {\rm BR}^{\rm inv}_f \simeq  {\left(\frac{\langle \sigma v_r \rangle}
{10^{-10}{\mbox{GeV}}^{-1}}\right)} \bigg[ {3.9 \times 10^{-11}{\left(\frac{m_\chi}
{10\,\mbox{GeV}}\right)}^2+\left(\frac{\langle \sigma v_r \rangle}
{10^{-10}{\mbox{GeV}}^{-1}}\right)} \bigg]^{-1},\,\nonumber\\
& {\rm BR}^{\rm inv}_V \simeq {\left(\frac{\langle
\sigma v_r \rangle} {10^{-10}{\mbox{GeV}}^{-1}}\right)} \bigg[ {1.3 \times
10^{-6}{\left(\frac{m_V} {10\,\mbox{GeV}}\right)}^4+\left(\frac{\langle \sigma
v_r \rangle} {10^{-10}{\mbox{GeV}}^{-1}}\right)} \bigg]^{-1} \, . 
\end{align} 

One can notice from these rates that the invisible branching fraction is
practically not constrained in the case of a fermionic DM. This is due to the
$p$--wave suppression of its annihilation cross section which makes it
practically not sensitive to  indirect detection experiments. One should  thus
expect that DM annihilation cross sections of the order of the thermally
favoured value, which coincides with the current experimental sensitivity at low
DM masses, imply a Higgs boson that is dominantly decaying into invisible
states.

\subsubsection{Numerical analysis}

We have now all the elements that allow to evaluate the constraints on the
various types of DM particles first from the correct relic density, assuming
standard thermal production, and then from direct detection experiments in
combination with present and eventually future constraints from the invisible
width of the  Higgs boson. For the latter, we have adopted the mass value
$M_H=125\,\mbox{GeV}$ and assumed SM--like couplings, apart from the additional
coupling with the DM particles.  For illustrating the invisible branching
ratios, we have assumed the present limit of 20\% from LHC data and the
values    10\%, 5\% and 1\% which would correspond to the ultimate constraints 
that could be obtained, respectively, at the LHC, the HL--LHC and at a future
$e^+e^-$ collider or a 100 TeV pp collider.  Analogous analyses related to the one
presented here can be found in
Refs.~\cite{Silveira:1985rk,McDonald:1993ex,Burgess:2000yq,Kim:2006af,Andreas:2010dz,Kanemura:2010sh,Lebedev:2011iq,Mambrini:2011ik,Djouadi:2011aa,LopezHonorez:2012kv,Djouadi:2012zc,Davoudiasl:2004be,Patt:2006fw,He:2008qm,He:2009yd,Barger:2010mc,Barger:2007im,Clark:2009dc,Lerner:2009xg,Goudelis:2009zz,Yaguna:2008hd,Cai:2011kb,Biswas:2011td,Farina:2011bh,Hambye:2008bq,Hambye:2009fg,Hisano:2010yh,Englert:2011yb,Englert:2011aa,Andreas:2008xy,Foot:1991bp,Melfo:2011ie,Raidal:2011xk,He:2011de,Mambrini:2011ri,Chu:2011be,Ghosh:2011qc,Greljo:2013wja,Kahlhoefer:2017dnp}.

\begin{figure}
\begin{center}
\vspace*{-3mm}
\subfloat{\includegraphics[width=0.38\linewidth]{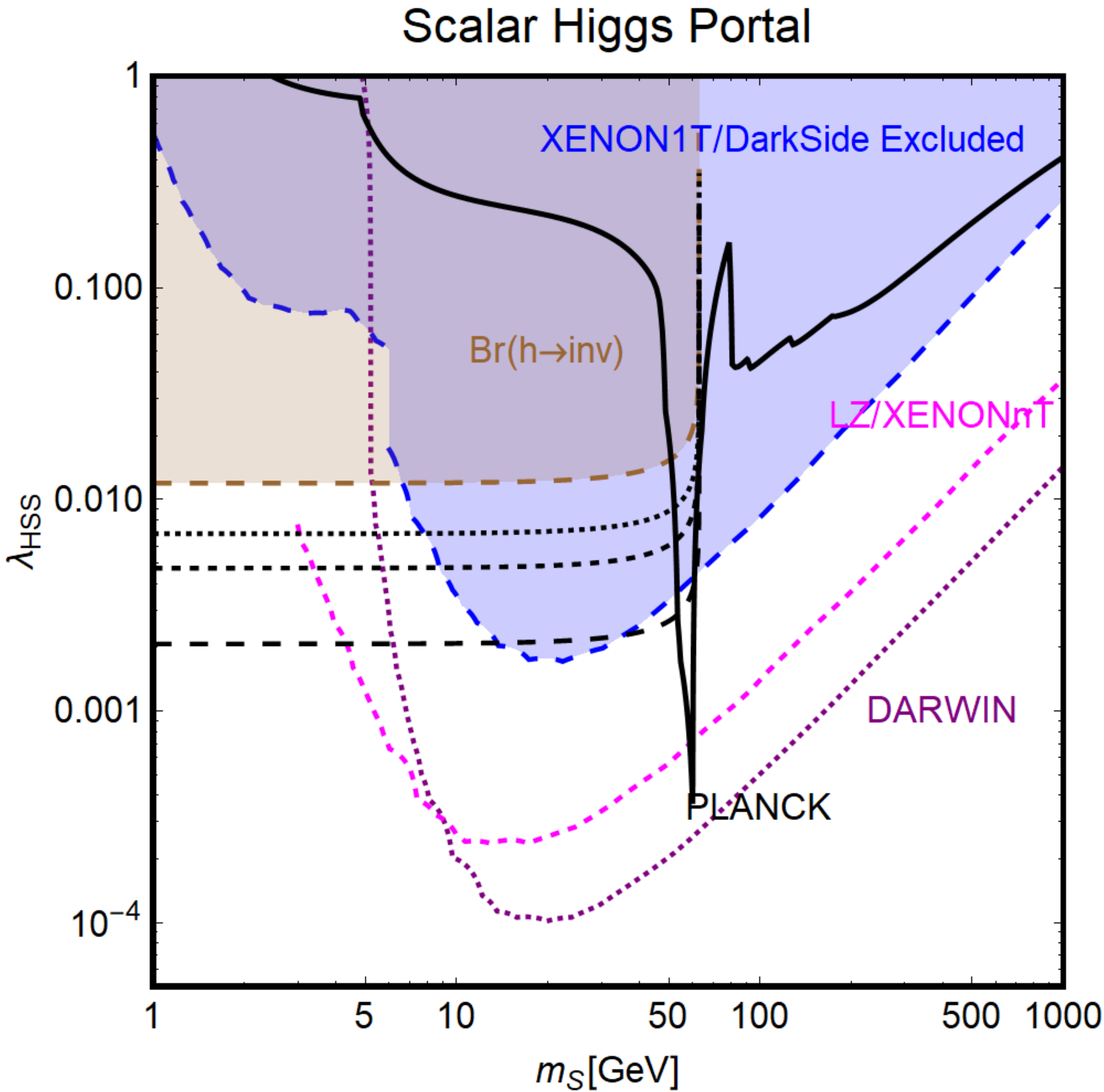}}~~~~
\subfloat{\includegraphics[width=0.38\linewidth]{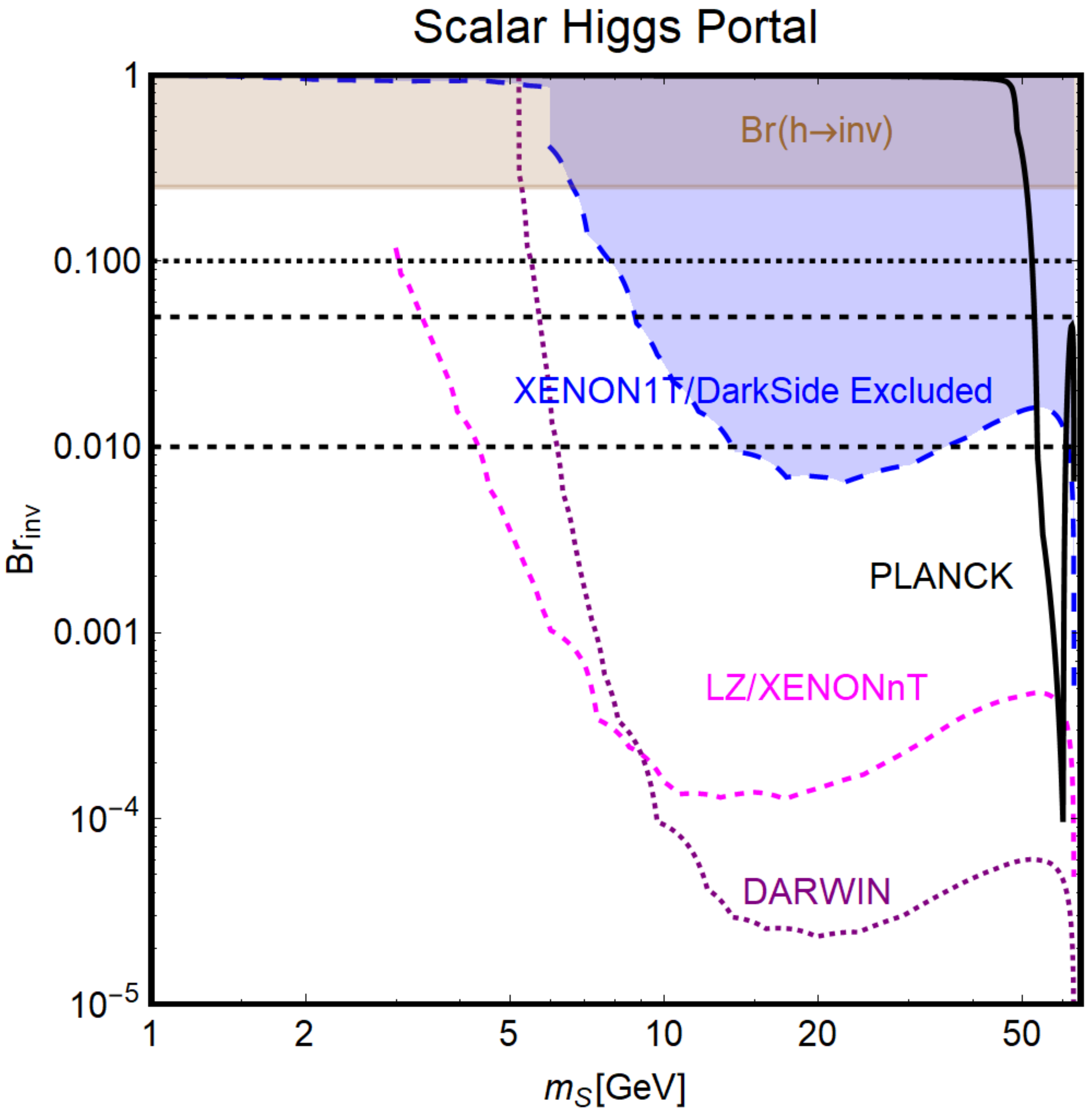}}\\[-2mm]
\subfloat{\includegraphics[width=0.38\linewidth]{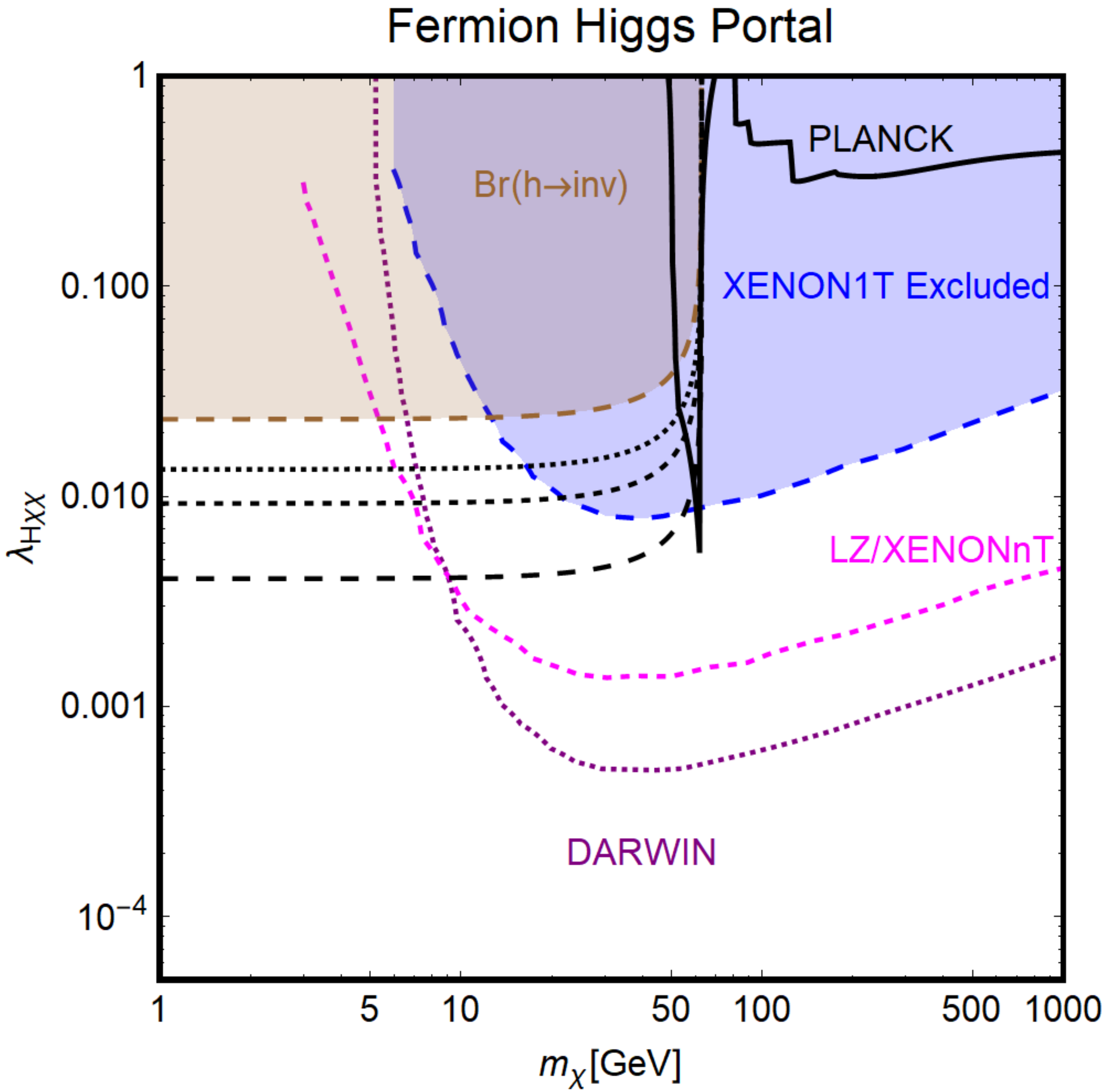}}~~~~
\subfloat{\includegraphics[width=0.38\linewidth]{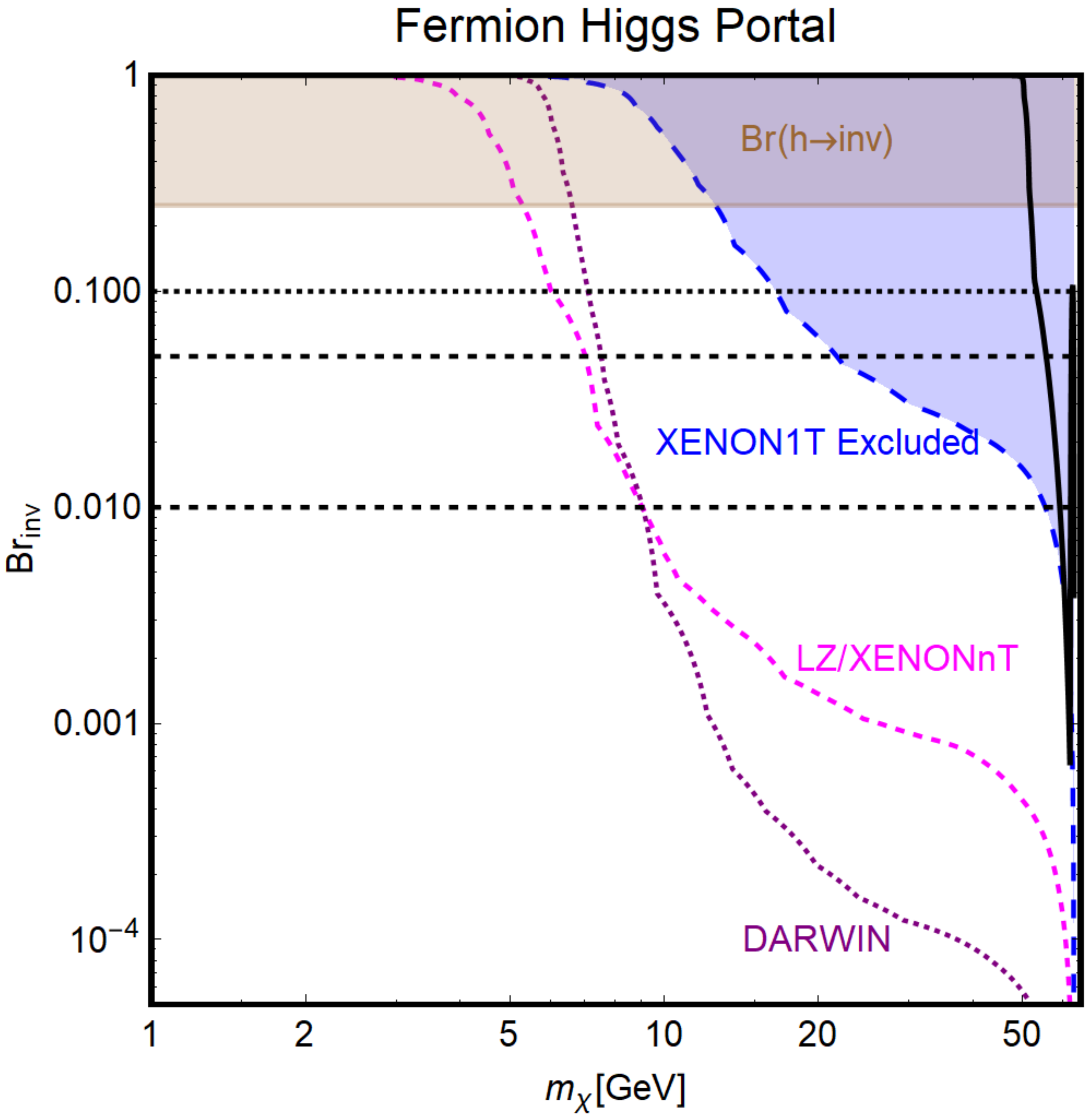}}\\[-2mm]
\subfloat{\includegraphics[width=0.38\linewidth]{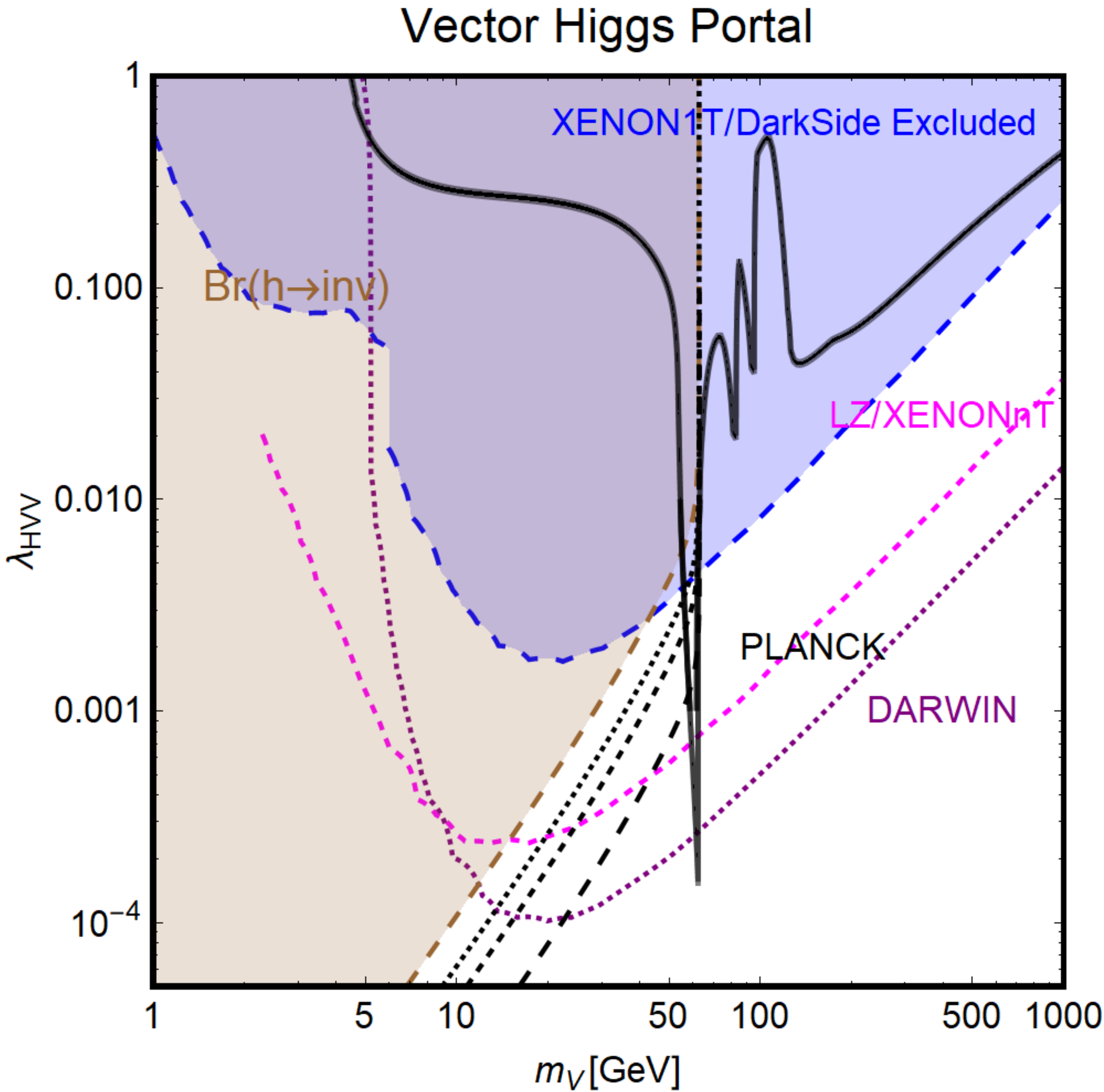}}~~~~
\subfloat{\includegraphics[width=0.38\linewidth]{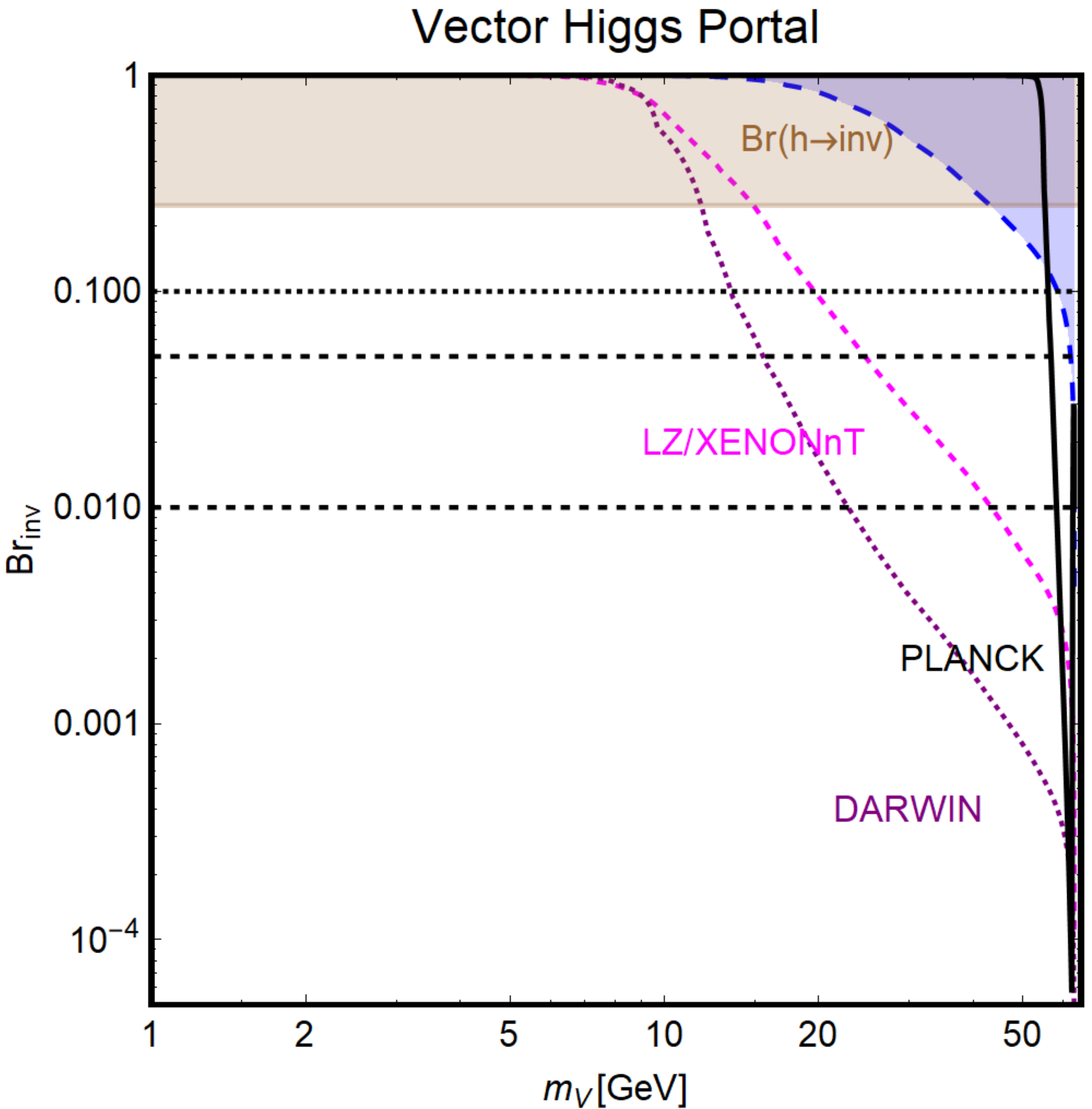}}
\end{center}
\vspace*{-3mm}
\caption{Summary of constraints in the planes $[m_X,\lambda_{HXX}]
$ (left panels) and $[m_X, {\rm BR}(H \to XX)]$ (right panels)  
for the Higgs--portal DM in the scalar (top), fermionic  (middle) and vector
(bottom) cases. The black contours correspond to the correct DM relic density. The blue and brown regions are excluded, respectively, by direct detection limits from XENON1T and the invisible Higgs decay width. The black
contour lines correspond to invisible Higgs branching ratios of $10\,\%$,
$5\,\%$ and $1\,\%$. The magenta and purple contours represent the sensitivity
reach of next generation direct detection experiments such as LZ/XENONnT and DARWIN.}
\label{fig:Hp}
\vspace*{-3mm}
\end{figure}

The outcome of our analysis is shown in Fig.~\ref{fig:Hp}. The pairs of panels
in the figure correspond, respectively, to scenarios of scalar ($S$, upper
panel), fermion ($\chi$, middle panel) and vector ($V$, lower panel) DM states. 
Left panels show results in the bidimensional plane [$m_{X}$, $\lambda_{HXX}$].
The black contours, labelled as PLANCK, correspond to the correct relic
density,  i.e one has to lie exactly in that line to have the correct DM
abundance: above the line the DM is underabundant and below overabundant. The
blue regions are excluded by DM direct detection: the regions $m_X \gtrsim
5\,\mbox{GeV}$ are excluded by XENON1T  and, in the case of scalar
and vector DM states, these exclusion bounds are complemented by a weaker
constraint from the DarkSide--50 experiment \cite{Agnes:2014bvk,Agnes:2018oej}
for $1\,\mbox{GeV} \leq m_X \leq 5\,\mbox{GeV}$. 

As will be discussed in more details in the next subsection, given the
dependence of the direct detection rate on the local DM density, in order to
draw exclusion regions like the ones shown in Fig.~\ref{fig:Hp}, one has to
implicitly assume that the DM features the correct relic density (e.g. through a
non--thermal mechanism and/or modifications of the cosmological history of the
Universe~\cite{Arcadi:2011ev,Gelmini:2006pw,Gelmini:2006pq,Baer:2014eja,Roszkowski:2017dou})
even outside the ``PLANCK'' isocontours. The brown region is excluded by the
current limit from invisible Higgs decays. The black lines correspond to BR$(H\!
\to \! {\rm inv})=10\%$, $5\%$ and $1\%$. Finally, the magenta and purple lines
are the sensitivity prospects of forthcoming DM direct detection experiments,
LZ/XENONnT (given the similarities in the expected sensitivities, the two are
represented by a unique line) and DARWIN. By sensitivity prospects, it is meant
that the regions above each line will be ruled out if the corresponding
experiment fails to see any signal. The same caveats mentioned for the
XENON1T/DarkSide--50 excluded region apply also for these curves. 

\begin{figure}[!h]
\hspace*{1.cm}
\includegraphics[width=0.7\linewidth]{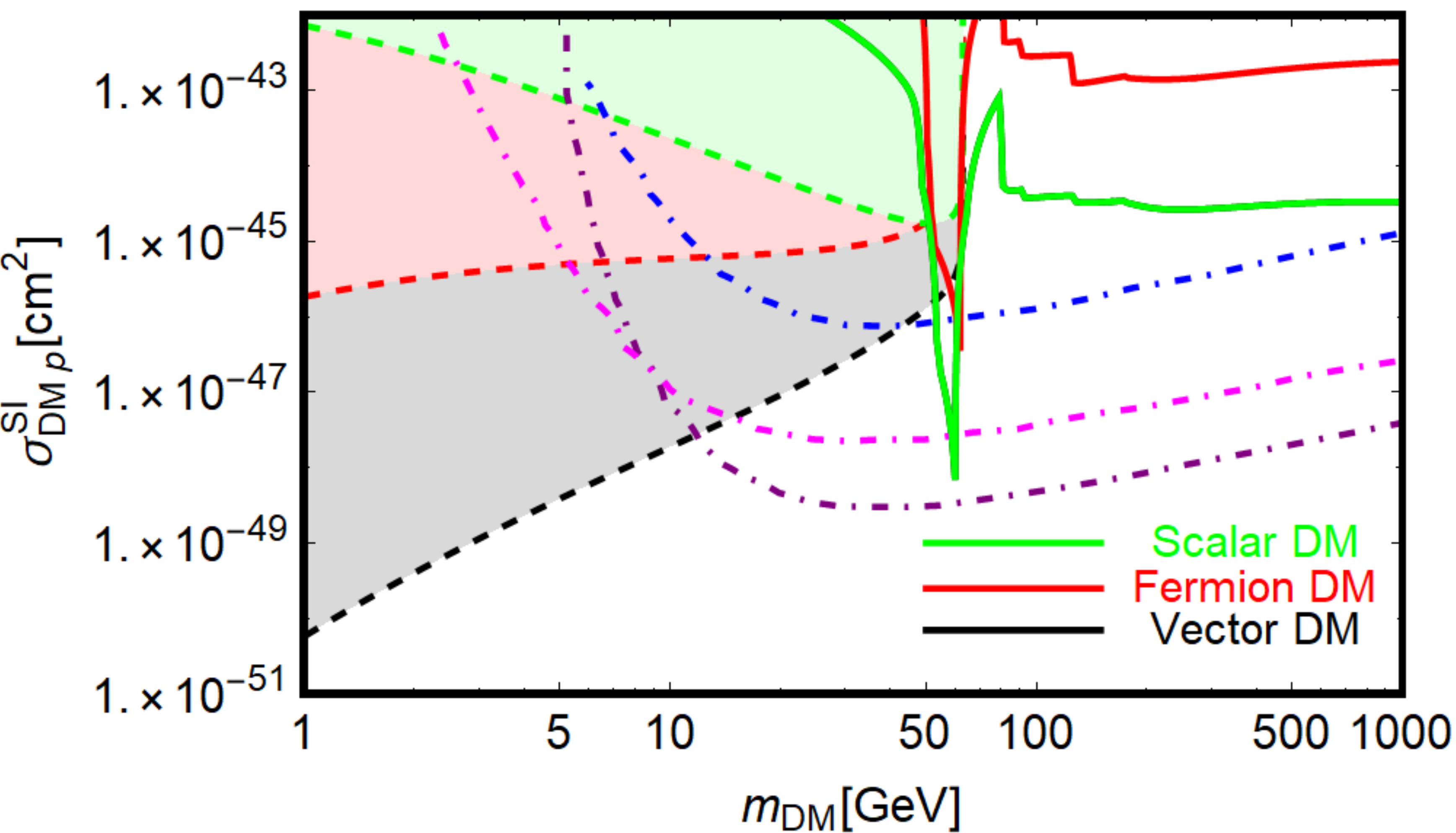}
\caption{Comparison of the LHC constraints from the invisible decays of a SM--like Higgs boson and limits/sensitivities from various direct DM detection experiments. The dot--dashed blue/magenta/purple curves are for the constraints from XENON1T and the prospects from XENONnT/LZ and DARWIN
respectively.}
\label{ATLAS-comparison}
\vspace*{-.1cm}
\end{figure}

The previous outcome  has been then re--cast, in the right panels of
Fig.~\ref{fig:Hp} in the bidimensional plane [$m_{X}$, BR$(H\rightarrow XX)$].
This type of comparisons between LHC Higgs results and astroparticle physics
experiments have also been made by the LHC experimental collaborations
themselves. We have, for instance,  reproduced and complemented in
Fig.~\ref{ATLAS-comparison} an  analysis performed by ATLAS  in
Ref.~\cite{Aad:2015pla}.

In the figure, the main constraints for the three DM spin assignments are shown
in the bidimensional plane $[m_{\rm DM},\sigma_{\rm DM p}^{\rm SI}]$. The dashed
curves correspond to the predicted DM--nucleon scattering cross section for a DM
coupling with the Higgs boson corresponding to BR$(H\! \rightarrow \! {\rm
inv})\!=\! 0.2$, i.e. the 95\%CL limit on the invisible Higgs branching ratio
derived using both the visible and invisible decay channels. The dot--dashed
blue/magenta/purple curves, represent, according to the previous color code, the
constraints from XENON1T and the prospects from XENONnT/LZ and DARWIN. The solid
lines represent the spin--independent cross section obtained by fixing the DM
Higgs coupling to the correct DM relic density (the curves for scalar and vector
DM overlap).

One sees that for Higgs masses below approximately 62 GeV, the LHC limits from
the invisible Higgs decays, when compared to those obtained  from direct
detection experiments, are in general weaker and less severe as one approaches
the $m_{\rm DM}=\frac12 M_H$ threshold.  An exception is the case of vector DM,
in which the limits from LHC are competitive with the ones from direct detection
in the low mass region, i.e. $m_{\rm DM} \lesssim 10\,\mbox{GeV}$, where the
latter are limited by the threshold in the detected recoil energies. The LHC
limits for vector DM will be instead superseded by next generation of direct
detection experiments.

In fact, for even smaller DM masses, $m_{\rm DM} \lesssim 5$--7 GeV, the
sensitivity of direct detection experiments is limited by the energy threshold
of the detectors (especially XENON--based ones) and the LHC plays a crucial role
in constraining this possibility.

\subsubsection{Uncertainties and caveats in the comparison of DM limits}
\label{sec:caveats}

As seen before, at least for DM masses above 10 GeV, direct detection
experiments are much more constraining than the LHC. Nevertheless, some care
should be  taken when comparing the outcome of the two types of experiments.
Indeed, limits from the invisible Higgs width or searches just require, to be
applied, that the particle in which the Higgs decays is stable at the detector
level  and appears as missing transverse energy. No assumptions concerning the
eventual relic density or other astrophysical properties are needed. This is
absolutely not the case for direct detection limits. As a matter of fact, the DM
scattering rate on a target detector depends  not only on the particle physics
input, represented by the DM scattering cross section, but also on the DM
velocity distribution $f(v)$ and its local density $\rho_{\rm DM}$. This last
quantity, in particular, serves as a normalization for the signal rate. As we
have already seen, experimental limits are customarily expressed in the
bidimensional plane $[m_X,\sigma_{X N}]$, fixing the assignment for
the astrophysical inputs according to the so--called Standard Halo Model
(SHM)~\cite{Drukier:1986tm}. In this model, the DM is represented in the
galactic frame, by an isotropic velocity distribution of the form
\begin{equation}  
f_{\rm gal}(v)=\left \{ \begin{array}{cc} N \exp\left(-|v|^2/v_c^2 \right)  & |v| \leq v_{\rm esc}  \\ 0     &  |v| \geq v_{\rm esc} \end{array} \right. \, , \end{equation}  
describing an isothermal sphere. $N$ is a normalization factor such that $\int
f_{\rm gal}(v)dv=1$ while $v_c$ is a circular velocity. The function $f_E$ defined in
eq.~(\ref{eq:DDsignalRate}) is the DM velocity distribution in the detector
frame and satisfies the relation $f_{E}(v)=f_{\rm
gal}(|\vec{v}+\vec{v}_s+\vec{v}_{e}(t)|)$ with $\vec{v}_s$ and $\vec{v}_e$
being, respectively, the Sun's velocity with respect to the center of the Galaxy
and the Earth's velocity with respect to the Sun. The local DM density is
determined from astrophysical observations either through local methods, i.e.
using kinematical data from nearby population of stars, or through global
methods, i.e. modelling the DM and baryon content of the Milky Way and using
kinematical data from the whole Galaxy;  see
Refs.~\cite{Read:2014qva,Catena:2009mf,Weber:2009pt,Salucci:2010qr,McMillan:2011wd,Garbari:2011dh,Iocco:2011jz,Bovy:2012tw,Zhang:2012rsb,Bovy:2013raa}
for more details. 

The SHM adopts for these three parameters  the fiducial values $\rho_{\rm
DM}=0.3\,\mbox{GeV}/{\mbox{cm}}^3$, $v_c=220$ (or 230) km/s and $v_{\rm
esc}=544$ km/s. These parameters are nevertheless subject to sizable
experimental uncertainties.  For example, one has  $\rho_{\rm DM} \in
[0.2-0.6]\,\mbox{GeV}/{\mbox{cm}}^3$ and $v_c$ can range from $220 \pm 20$ km/s
to $279 \pm 33$ Km/s~\cite{McMillan:2009yr}. These variations in the
astrophysical inputs translate into different predictions of the DM scattering
rate and, consequently, weaker or stronger limits with respect to the one
customarily quoted. One should also remark that the SHM provides a simplified
approximate description of the DM galactic distribution, challenged by the
results from the most recent DM hydrodynamical simulations, including the
effects of baryons~\cite{Bozorgnia:2016ogo,Bozorgnia:2017brl}, as well as by
observational evidences from the GAIA
collaboration~\cite{Necib:2018iwb,Necib:2018igl,Evans:2018bqy}. Notice that a
proper assessment of the astrophysical inputs is also crucial when the outcome
of different experiments is compared; this problem can nevertheless be
encompassed through  the so--called halo independent
methods~\cite{Fox:2010bz,McCabe:2011sr,DelNobile:2013cva,Ibarra:2017mzt,Gondolo:2017jro,Catena:2018ywo,Kahlhoefer:2018knc}.

\begin{figure}[!h]
\vspace*{-1mm}
    \centering
    \subfloat{\includegraphics[width=0.45\linewidth]{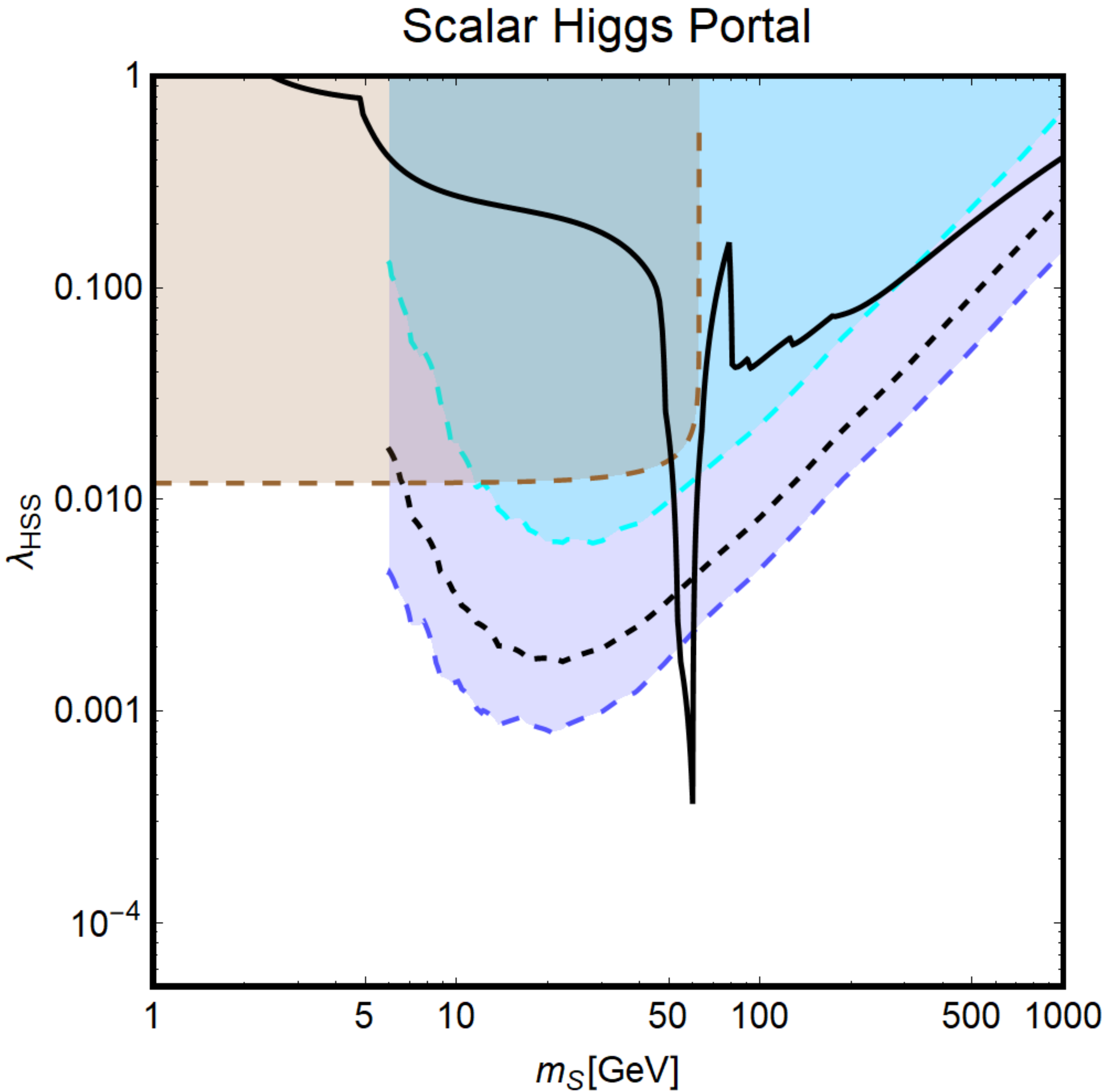}}
    \subfloat{\includegraphics[width=0.45\linewidth]{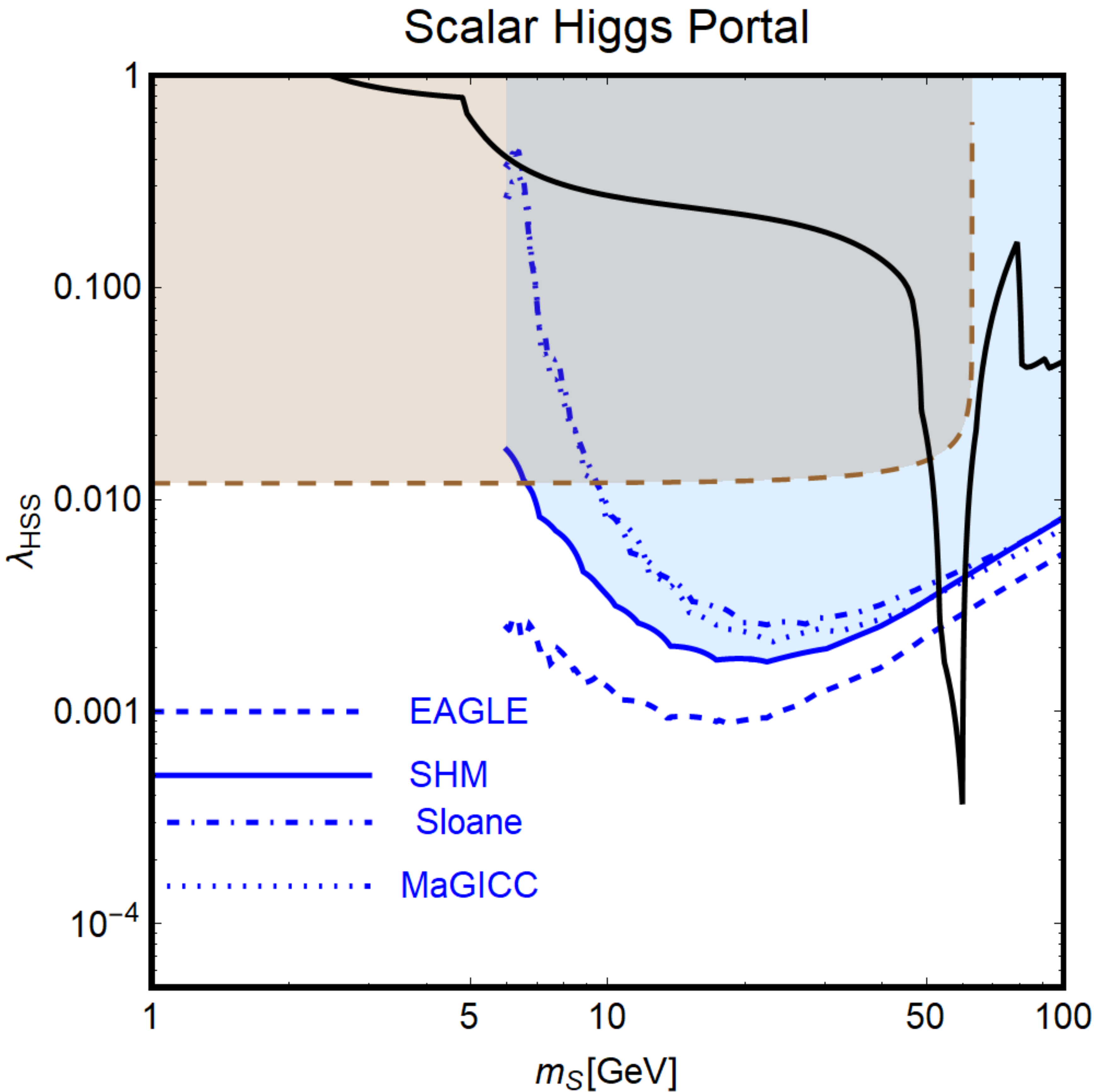}}
    \caption{Impact on direct detection constraints of deviations from the SHM. In the left panel, shown is the effect of varying the astrophysical inputs, as performed in Ref.~\cite{Benito:2016kyp}. The blue region between the dashed cyan and blue lines represent the variation of the XENON1T exclusion bound with the astrophysical inputs while the black solid lines represent the limit adopting the SHM. In the right panel, the conventional excluded region above the solid blue line is compared with cases in which the DM distribution adopted in the SHM is replaced by the outcome of the hydrodynamical simulations indicated in the plot~\cite{Bozorgnia:2017brl}. In both panels, together with the direct detection excluded regions, we show the (black) isocontour of the correct DM relic density according to the WIMP paradigm and the excluded regions by the invisible width of the Higgs boson.}
    \label{fig:Hpun}
\vspace*{-3mm}
\end{figure}

The effect on the limits from direct detection, focussing for definiteness on
the XENON1T experiment, when varying the astrophysical inputs in the signal rate
is shown in two examples in Fig.~\ref{fig:Hpun}. For simplicity we illustrate 
just the case of the scalar Higgs--portal. In the left panel, we follow the
analysis done in Ref.~\cite{Benito:2016kyp} in which a substantial variation of
the astrophysical inputs has been assumed in order to maximize the impact on the
direct detection limits (this reference considered the limits from the LUX
experiment but,  to a good approximation,  the results can be translated to the
case of XENON1T since the two detectors have the same material and a similar
design). In the right panel, we show, following the results presented in
Ref.~\cite{Bozorgnia:2017brl}, different exclusion lines from the XENON1T
experiment, obtained by replacing the DM velocity distribution of the SHM with
the one inferred from the result of some recent hydrodynamical simulations.
Here, we have focussed on the region $m_S < 100\,\mbox{GeV}$ where the impact of
the different DM distribution is mostly prominent. In both cases, we have
reported for comparison the isocontour of the correct DM relic density and the
excluded region by searches of the invisible branching fraction of the Higgs
boson. The impact of the astrophysical uncertainties is clear but no new viable
region for the DM relic density in the effective scalar Higgs--portal is opened.

A final important remark is that normalizing the DM scattering rate with the
experimental determination of $\rho_{\rm DM}$ (modulo the uncertainties),
implies the assumption that the scattering particle represents the total DM
component of the Universe and features the correct relic density. One could, in
principle, relax this assumption and consider that the WIMP Dark Matter
candidate represents only a fraction $f$ of the total DM component. In such a
case, the DM signal rate would be normalized by a factor $f=\Omega_{\rm
DM}/\Omega_{\rm DM}^{\rm  PLANCK}$ and the corresponding limits would be weaker
than the ones presented in the previous subsection. In such a scenario, limits
from searches of invisible Higgs decays can become competitive since they do not
depend on $f$.

\begin{figure}[!h]
    \centering
\subfloat{\includegraphics[width=0.4\linewidth]{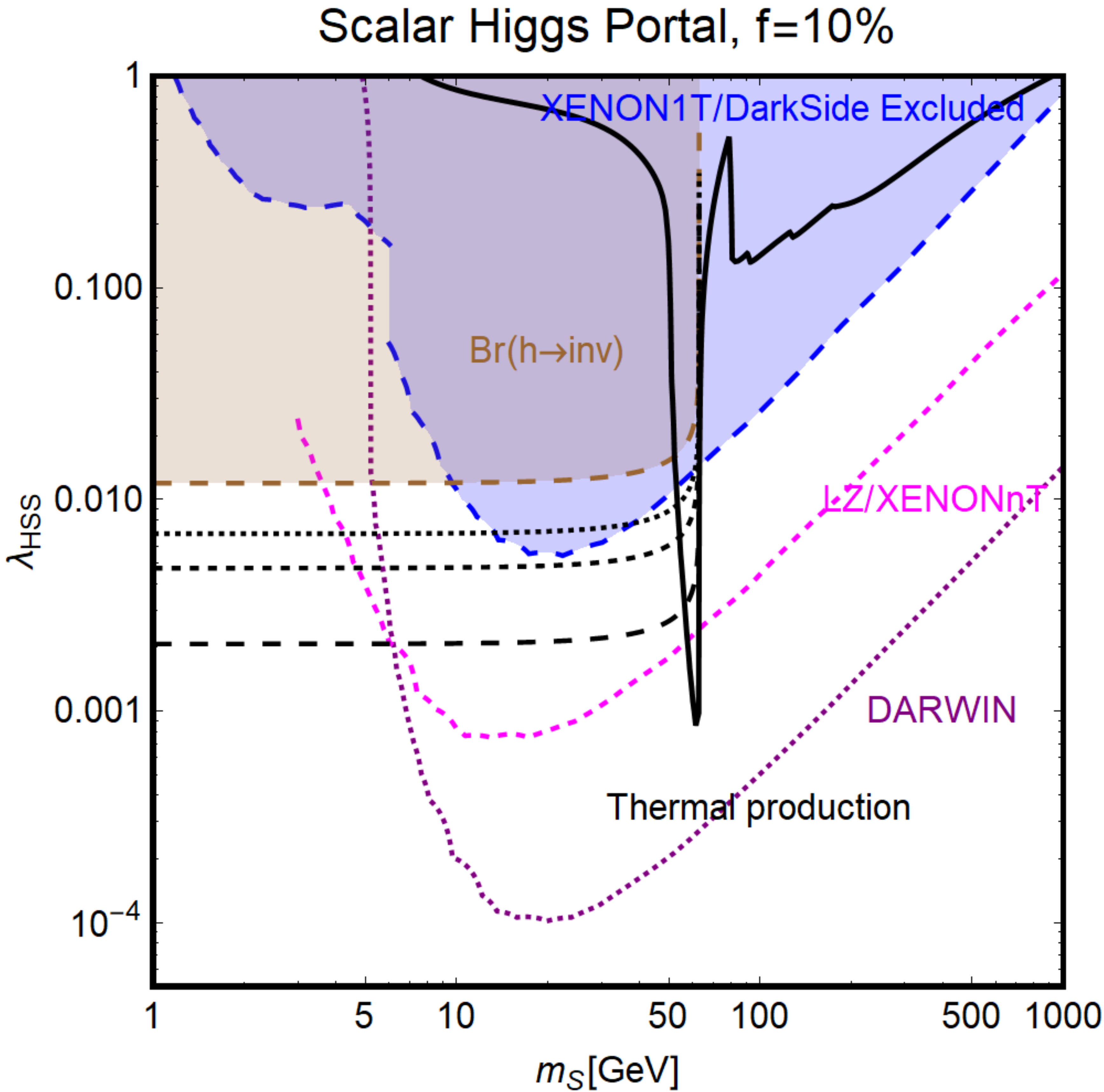}}~~~~
\subfloat{\includegraphics[width=0.4\linewidth]{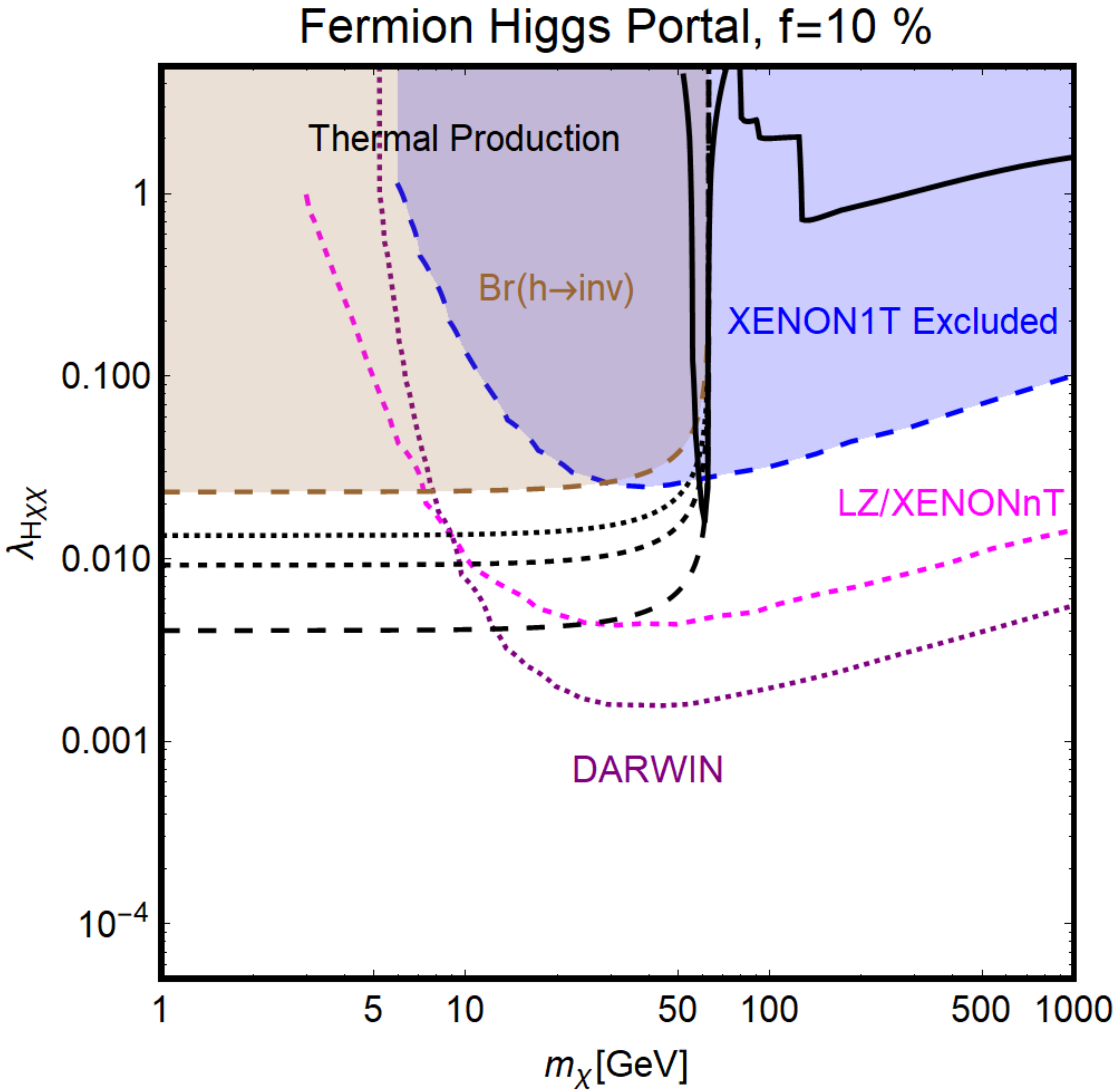}}\\
\subfloat{\includegraphics[width=0.4\linewidth]{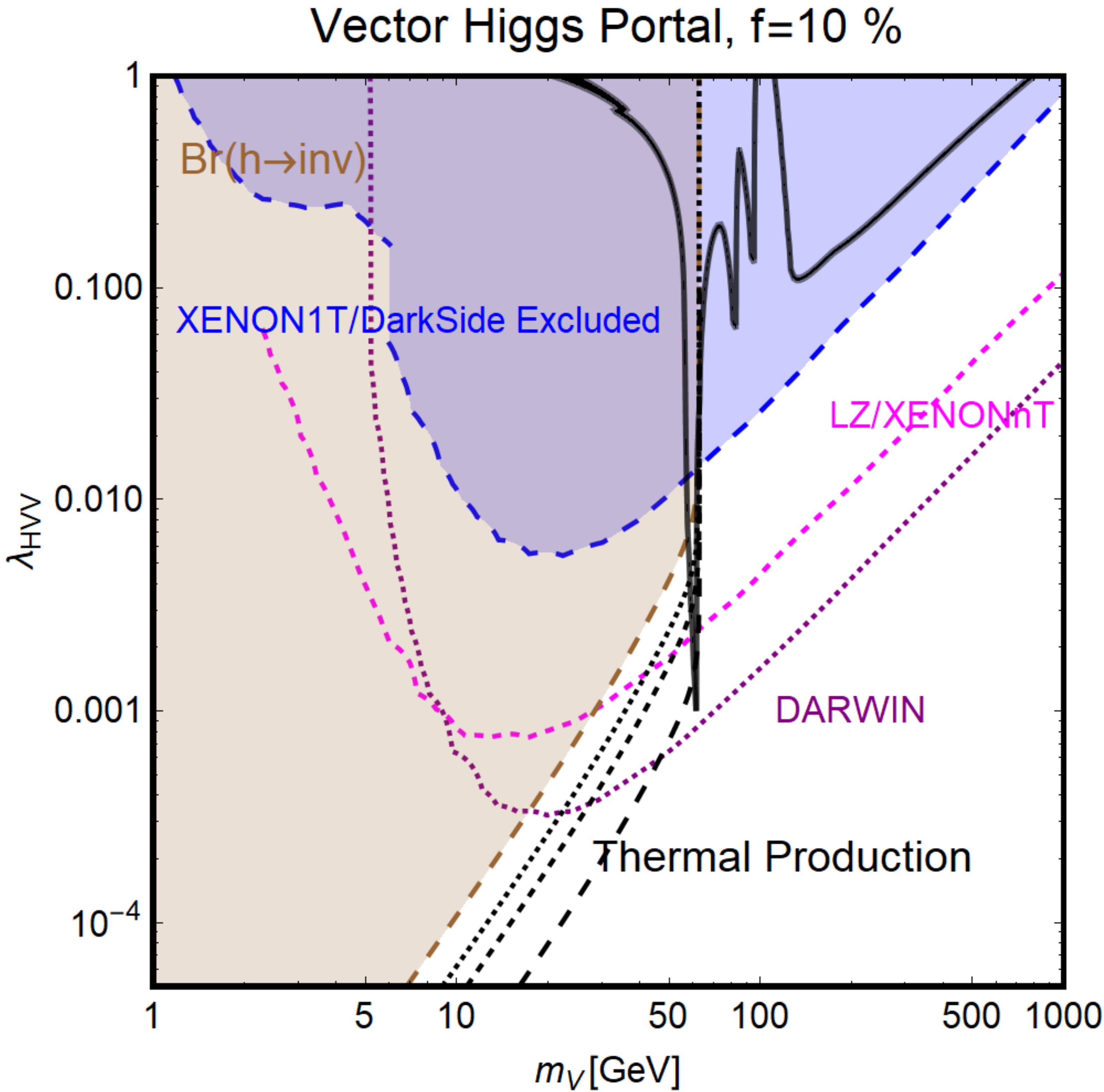}}
    \caption{The same as the left column of Fig.~\ref{fig:Hp} but assuming that the WIMP DM candidate contributes only a fraction $f=10\%$ of the total DM component of the Universe. In contrast to Fig.~\ref{fig:Hp}, the black isocontours have been labelled as ``thermal production'' since they do not correspond to the experimentally favored value of the relic density.}
    \label{fig:Hpf10}
\end{figure}

We have repeated the analysis shown in Fig.~\ref{fig:Hp} in the case of $f=10
\%$. By comparison with Fig.~\ref{fig:Hp} (left),  one sees that the direct
detection and relic density curves shifted in an analogous way towards higher
value of the DM coupling. This is to be expected since the annihilation (except
in the $HH$ channel) and scattering cross sections are proportional to
$\lambda_{HXX}$. More importantly, one notices that for $f=10\%$, the bounds
from the invisible Higgs width are comparable with the one from direct detection
in the cases of scalar and fermionic DM and even more competitive in the case of
vector DM. A similar reasoning as above can also be applied to the limits from
DM indirect detection. But as the latter will have only a marginal impact on the
results presented in this work, we will not discuss the effect of astrophysical
uncertainties (for a discussion see for example
Refs.~\cite{Calore:2015oya,Bernal:2016guq,Benito:2016kyp}).

In the next sections, we will discuss a series of more refined and extended DM
scenarios. While keeping in mind the caveats discussed in this subsection, we
will nevertheless assume for the results that will be presented there,  first
the SHM and second that the DM candidate represents the total DM component of
the Universe.

%% file: sec-NF.tex
\section{The SM Higgs and extended fermionic sectors}

\subsection{The physical landscape}

As it was mentioned in the previous section,  the considered effective model in
which the DM particle is a singlet fermion under the SU(2) gauge group  was,
contrary to the vector and scalar DM cases, clearly  non--renormalisable. In order to have a renormalizable interaction, at least two multiplets, whose hypercharge differ by a factor $1/2$, should be present in the theory, with the DM being a mixing of their electrically neutral component (we have two require, in addition, that the DM is the lightest among the new states) \cite{Freitas:2015hsa}. In this review we will stick on the minimal option, represented by the mixing between and electroweak isodoublet and an isosinglet. Less minimal possibilities have been extensively reviewed e.g. in \cite{Lopez-Honorez:2017ora}. Appropriate options for a renormalisable Higgs--portal with a
spin--$\frac{1}{2}$ DM, complying at the same time with LHC data, are the ones in which 
the fermions that are present belong to real representations,  i.e. have
Majorana mass terms, or form vector--like pairs. For instance, a simple
realization are the so called singlet--doublet models\footnote{More complete
realizations are represented by the minimal supersymmetric Standard Model (MSSM)
\cite{Martin:1997ns,Djouadi:1998di} and its next-to-minimal version (NMSSM)
\cite{Ellwanger:2009dp} to be discussed later.}
\cite{Cohen:2011ec,Cheung:2013dua,Calibbi:2015nha,Yaguna:2015mva}. Another
simple realization  of a renormalizable Higgs--DM interaction is when a complete
vector--like family of leptons (as well as quarks) is added to the SM fermionic
spectrum \cite{Fujikawa:1994we,Hambye:2008bq,Hambye:2009fg,Hisano:2010yh,Lebedev:2011iq,Angelescu:2015uiz,Angelescu:2016mhl}.  These are the two options that we will
discuss in this section, assuming that  the Higgs sector is still SM--like.  The
fermionic Higgs--portal discussed previously can be then interpreted as an
effective limit of such a  setup in which the extra fermionic fields, except
from the DM, are heavy and integrated out; see Ref.~\cite{Baum:2017enm} for a
concrete example.  

Nevertheless, the most straightforward and obvious possible extension in this
context would have been a fourth generation of fermions
\cite{Belotsky:2002ym,Kribs:2007nz,Feng:2010gw,Denner:2011vt}. The simplest
version, a fourth family that behaves exactly like the first three ones,  is in
fact completely and unambiguously ruled out by LHC data
\cite{Djouadi:2012ae,Kuflik:2012ai} but some variants have survived  until
recently \cite{Bao:2013zua,Lee:2011jk}. Such a scenario is worth discussing  as
it incorporates many ingredients that appear in other viable scenarios,  and we
thus start this subsection by briefly summarizing it.

\subsubsection{The possibility of a fourth generation}

In the extension of the SM with a fourth generation of fermions that we denote
by SM4, one simply needs to add to the SM fermionic pattern with three
generations, two quarks $u'$  and $d'$, a charged lepton $e'$ and a neutrino
$\nu'$ but with a right--handed component $\nu_R$ in such a way that it becomes massive \cite{Kribs:2007nz}
\begin{eqnarray}
{\rm SM4:} \quad \left( \begin{array}{c} \nu ' \\ e'^- \end{array} \right)_L \, , \ \  \nu'_R \, ,\ e'^-_R  \, , \ \  \left( \begin{array}{c} t' \\ b' \end{array} \right)_L,\ \ t'_R \, ,\ b'_R \; .
\end{eqnarray}
Being a right--handed  SM singlet, it is allowed to have a Majorana mass term, so that in general one can define two (Majorana) mass eigenstates $\nu_1$ and $\nu_2$ from the diagonalization of the mass matrix
\begin{equation}
    m_{\nu,4}=\left(\begin{array}{cc}
    0  & m_D  \\
    m_D & m_M 
    \end{array}
    \right) \, , 
\end{equation}
with $m_D=y_\nu v$ and $m_M$ being, respectively, the Dirac and Majorana masses. In the absence of a Majorana mass, the fields $\nu',\nu_R$ would form a Dirac state. As will be seen in the next subsection, a Dirac fermionic DM with non--zero hypercharge is extremely constrained as it has full couplings with the $Z$ boson. We will thus conservatively stick to the case in which the fourth generation neutrino DM is a Majorana fermion. In the mass basis, the coupling of the fourth generation neutrinos to the $H$ and $Z$ bosons are given by
\begin{align}
& \mathcal{L}_H=-\frac{m_{\nu_1}}{v}\frac{c_\theta}{s_\theta}\left[c_\theta s_\theta \bar \nu_1 \nu_1+c_\theta s_\theta \bar \nu_2 \nu_2-i (c_\theta^2-s_\theta^2) \bar \nu_1 \gamma^5 \nu_2\right]H \, , \nonumber\\
&\mathcal{L}_Z=\frac{g}{4 \cos\theta_W}\left[-c_\theta^2 \bar \nu_1 \gamma^\mu \gamma^5 \nu_1-s_\theta^2 \bar \nu_2 \gamma^\mu \gamma^5 \nu_2+2 i c_\theta s_\theta \bar \nu_1 \gamma^\mu \nu_2 \right]Z_{\mu}    \, , 
\end{align}
where we use the  abbreviations $s_\theta=\sin\theta, c_\theta=\cos\theta$ with $\theta$ being the mixing angle which diagonalizes the neutrino mass matrix.

In the general case in which the SM4 fermions mix with their SM light partners, 
two strong constraints should in principle apply on the new spectrum. First, the
fourth neutrino should be much heavier than those of the three first
generations, more specifically $m_{\nu_1} \gsim \frac12 M_Z$, as required by the invisible width
of the $Z$ boson measured at LEP1 and, in the case of the charged lepton,  the
LEP2 bound  $m_{e'}\! \gsim\! 100$ GeV should apply \cite{Tanabashi:2018oca}.
Second, direct LHC searches exclude too light fourth generation  quarks with
masses close to the unitarity bound, $m_{t'},  m_{b'} \! \lsim \! 600$ GeV
\cite{Chanowitz:1978mv,Aad:2012uu,Chatrchyan:2012fp}. However, if mixing between
SM4 and SM fermions is forbidden by some symmetry and if one assumes that the DM
neutrino interacts only with the Higgs boson, some of these constraints can be
evaded (see later for a summary of these constraints). 

Nevertheless, if the new fermions acquire masses through electroweak symmetry
breaking as in the SM, strong constraints can also be set on the SM4 scenario
from the measurements of the Higgs properties at the LHC. This is due to the
fact that in the loop induced $Hgg$ and $H\gamma\gamma$ couplings, see
Appendix A1, any heavy particle coupling to the Higgs proportionally
to its mass, as it should be also the case in SM4, will not decouple from  the
amplitudes and have a drastic impact.  In particular, for the  dominant $gg\to
H$ production process, the additional $t'$ and $b'$ contributions increase the
rate by an order of magnitude at leading order (LO). However, large ${\cal
O}(G_F m_{f'}^2)$ electroweak corrections affect these couplings
\cite{Djouadi:1994ge,Djouadi:1997rj,Passarino:2011kv,Denner:2011vt} resulting,
in the usual  SM4,  in a very strong suppression of the  $gg\to
H\!\to\! \gamma \gamma$ rate to the level where the channel becomes unobservable
at the LHC. 

\begin{figure}[!h]
\vspace*{-2cm}
\mbox{\hspace*{-5.4cm}
\resizebox{1.1\textwidth}{!}{\includegraphics{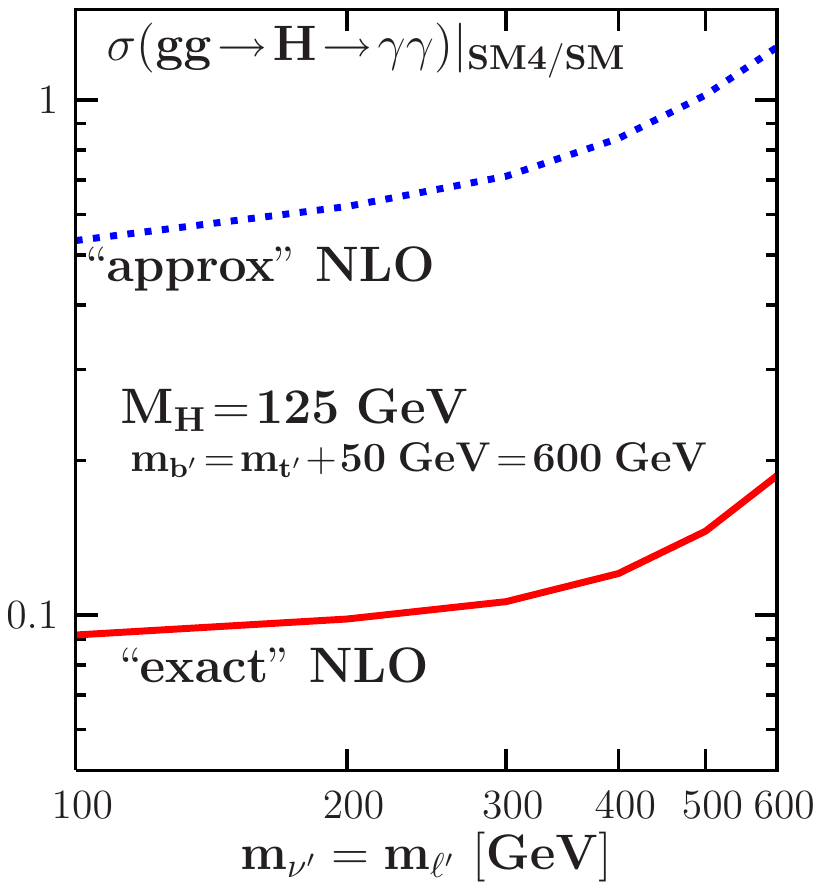}}\hspace*{-10cm}
\resizebox{1.1\textwidth}{!}{\includegraphics{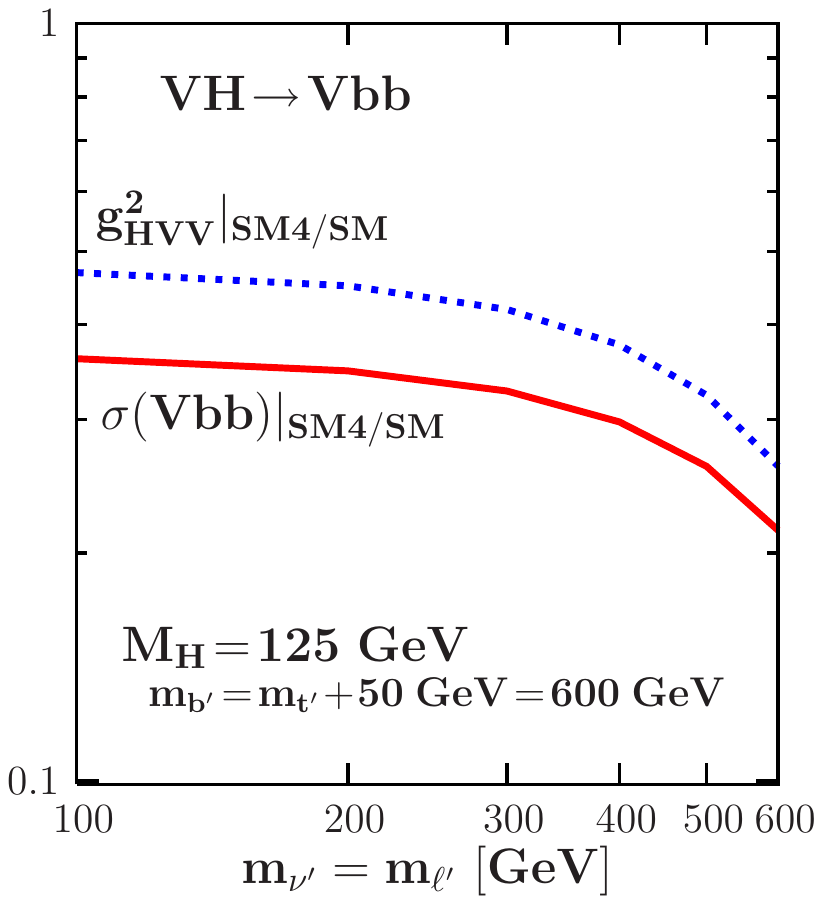}}
} 
\vspace*{-15.2cm} 
\caption[]{Left: $\sigma(gg\! \to\! H)\!\times\! {\rm BR}(  H\! \to\!
\gamma \gamma)|_{\rm SM4/SM}$ for a $125$ GeV Higgs as a function of
$m_{\nu'}\!=\!m_{\ell'}$ when the leading ${\cal O}(G_Fm_{f'}^2)$ 
corrections are  included naively  (``approx" NLO) or in a way that mimics the exact  NLO results (``exact" NLO). Right:  the $HVV$ coupling squared and the production times decay rate $\sigma(q\bar q\! \to\! VH)\!\times\! {\rm BR}(H\!\to\! b\bar b)$ in SM4, normalized to the SM values. The program HDECAY for SM4 has been used; from Ref.~\cite{Djouadi:2012ae}.} 
\label{fig:4th}
\vspace*{-.3cm}
\end{figure}

To illustrate this feature, using a version of  {\tt HDECAY} for SM4 which
includes these next--to--leading order (NLO) corrections, the  rate $\sigma(gg\!
\to\! H)\!\times\! {\rm BR}(  H\! \to\! \gamma \gamma)|_{\rm SM4/SM}$ for
$M_H\!=\!125$ GeV is shown as a function of the masses $m_{\nu'}\!=\!m_{e'}$
and  $m_{b'}\!=\!m_{t'}\!+ \!50\! = \!600\;$GeV in the left--hand panel of
Fig.~\ref{fig:4th}. One notices that it is a factor of 5 to 10 smaller than in
the SM despite of the increase of $\sigma(gg\!\to\!  H)$ by a factor of $\approx
9$  in SM4. In the right--hand side of the figure,  the ratio $\sigma(q\bar q\!
\to\! VH)\!\times\! {\rm BR}(  H\! \to\! b\bar b)|_{\rm SM4/SM}$ in SM4 shows
that the $Vb\bar b$ signal rate  would be reduced by a factor 3 to 5 depending
on $m_{\nu'}$. Looking at the Higgs signal strengths discussed in the previous
section and shown in Fig.~\ref{Fig:constraints}, it is clear that the
possibility of a fourth fermion family is excluded in such a case.   

Nevertheless, in Higgs--portal to DM models, some protecting symmetry like a
$\mathbb{Z}_2$ parity should be present to forbid or suppress the transitions
between the SM4 fermions and the ones of the first three generations. This will 
allow the lightest of them, the fourth neutrino, to be stable  and, as it should
also be massive, to be a good candidate for DM. In this case, the decay pattern
of the charged heavy fermions becomes rather complicated as will be discussed in
the next subsection, invalidating the usual experimental bounds on new quarks
and leptons decaying into light SM ones and gauge or Higgs bosons.  In
Refs.~\cite{Bao:2013zua,Lee:2011jk}, it was advocated that within such a
scenario, quark  masses $m_{t'}, m_{b'}$ as small as a hundred GeV are
possible and could modify significantly the phenomenology of the SM4 model. 

 Indeed, in the case where $m_{t'} \approx m_{b'} \approx 200$ GeV, the
suppression of the $H\to \gamma\gamma$ decay width, as a result of the negative
interference between the $W$ boson and the heavy quark loops, will not be drastic and a rate $\Gamma(H\to \gamma\gamma)|_{\rm SM4/SM} \sim 0.1$ would be
possible.  This has to be contrasted with the previous case in which the quarks
had to be as  heavy as 500 GeV and a two order of magnitude suppression of the
$H\to \gamma\gamma$ rate took place. As the contribution of the extra quarks in
gluon fusion lead to a cross section ratio $\sigma (gg \to  H)|_{\rm SM4/SM}
\sim 9$, one could ultimately arrange so that $\sigma(gg\! \to\! H)\!\times\!
{\rm BR}(  H\! \to\! \gamma \gamma)|_{\rm SM4/SM}$ is very close to the SM value
measured at the LHC. However, in order to fix the rates for the other detection
channels, in particular $WW$ and $ZZ$, the increase of the $gg\to H$ production
cross section should be compensated by a suppression of the $H\to WW$ and $H\to
ZZ$ branching ratios which result from the additional Higgs decays into the
invisible SM4 neutrino, $H \to \nu_1  \bar \nu_1$. One can then choose the
values of $m_{\nu_1}$ and $\theta$ which enter in the expression of this decay
width (both directly and in the phase space) in such a way that it suppresses by
an order of magnitude the branching fraction of all visible Higgs decays. 

\begin{figure}[!h]
\vspace*{-.2mm}
    \centering
    \includegraphics[width=0.52\linewidth]{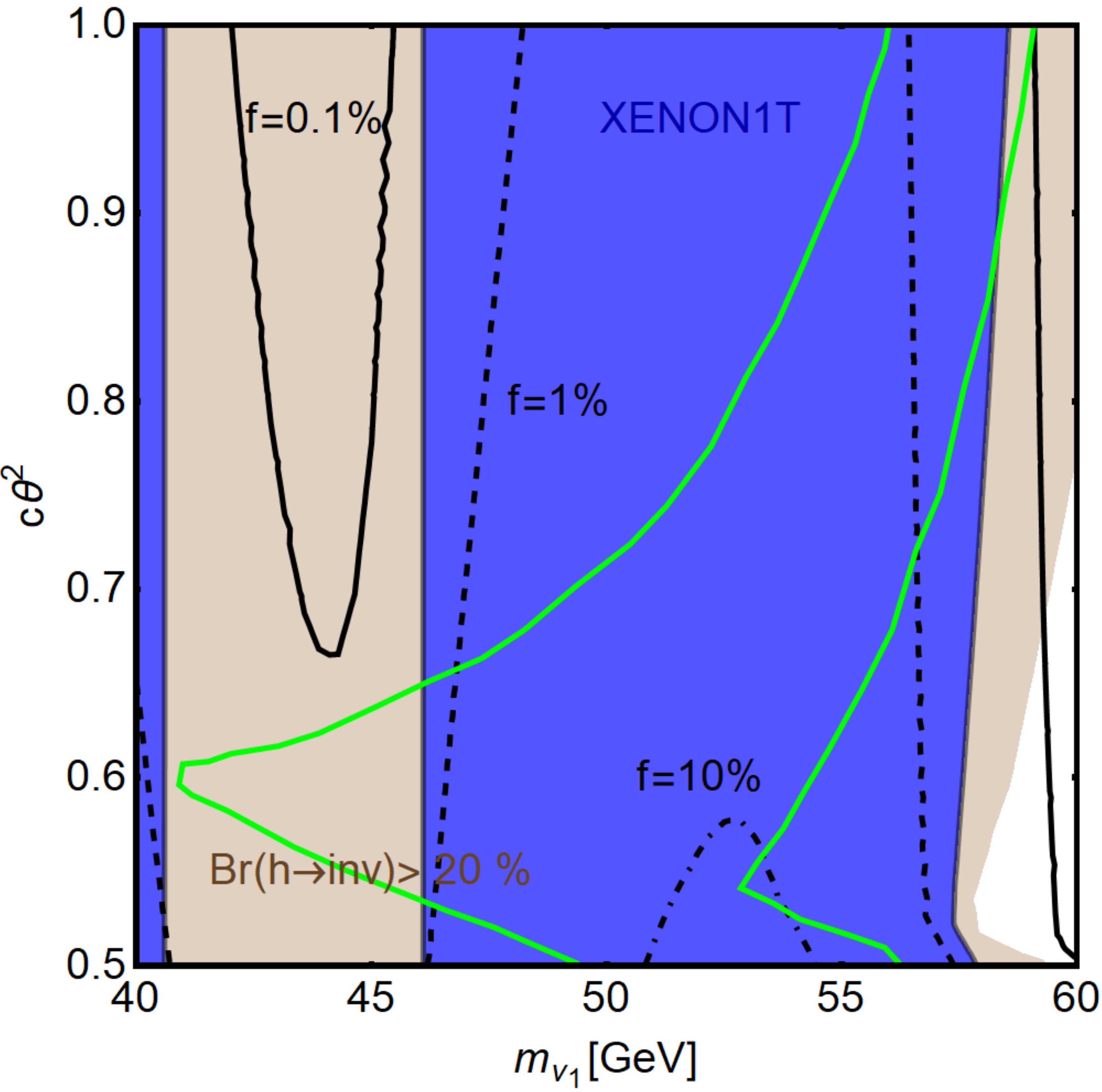}
    \caption{Summary of constraints on the SM4 in the bidimensional plane $[m_{\nu_1},\cos^2 \theta]$. The region enclosed in the green curves provides a fit of the Higgs signal strength into gauge bosons compatible with experimental data. The brown region corresponds to an invisible branching ratio of the Higgs above above $20\%$, while the blue region is excluded by DM direct detection. The solid/dashed/dot-dashed black represent isocontours corresponding to the values $f=0.1,1,10\%$ of the fraction of the DM relic density accounted by thermal production of the DM candidate.}
    \label{fig:pSM4}
\vspace*{-.2mm}
\end{figure}

Following the analysis presented in Ref.~\cite{Bao:2013zua}, we have delineated
the region of the $[m_{\nu_1}, \cos^2\theta]$ plane in which this occurs, namely
the region enclosed between the two green lines in Fig.~\ref{fig:pSM4}. In this
area,  the signal strengths for all Higgs decay channels measured in the
dominant gluon fusion mechanism,  in particular $\mu_{WW},  \mu_{ZZ}$ but also
$\mu_{\gamma\gamma}$ if the masses of the new quarks are also chosen
appropriately,  are compatible with the LHC measurements shown in
Fig.~\ref{Fig:constraints} \cite{Bao:2013zua,Lee:2011jk}.  The depicted scenario
is nevertheless very problematic as also shown in Fig.~\ref{fig:pSM4}. First, 
in order to obtain the correct fit of the signal strengths into gauge bosons, 
the SM Higgs boson should decay invisibly, i.e. into pairs of fourth
generation neutrinos, to a large extent. Indeed, the brown area depicted in
Fig.~\ref{fig:pSM4}  leads to an invisible Higgs branching that is larger than
20\%, the value excluded by LHC measurements and searches, and includes the area compatible with the signal strengths.  

In addition, the sizable couplings between of the DM with the Higgs and $Z$
bosons imply too strong annihilation processes for the DM state, so that it can
contribute at most to a fraction $f=10\%$ of the total DM component unless some
non--thermal production mechanism is assumed. For the same reason, very large 
scattering cross sections of the DM on nuclei are expected. As shown by the blue
region in Fig.~\ref{fig:pSM4}, even if one considers rescaling the direct
detection exclusion limits with a factor $f$ as discussed in the previous
section, sizable regions of the parameter space are ruled out. 

One should finally note that there are other Higgs production channels besides
gluon fusion and  some of them, such as VBF and HV, have been also probed by the
LHC experiments. In the scenario above, the Higgs decay signal strengths
measured in these two production channels would have been an order of magnitude
lower than in the SM and thus, in total contradiction with the $\mu$ values
shown in  Fig.~\ref{Fig:constraints} which are compatible with the SM at the
level of a few 10\%. This can be, in fact,  seen from the recent $5\sigma$
observation of the  Higgs boson in the $pp \to VH \to V b\bar b$ channel which
mainly proceeds via the process  $q\bar q \!\to \!VH$ which is not affected by
any loop, except for those appearing in the NLO radiative
corrections which tend to suppress the rate, see Fig.~\ref{fig:4th} (right), 
while $H\to b \bar b$ is affected by the same ingredients as the other visible
decays probed in gluon fusion. The signal strengths measured by the ATLAS and
CMS collaborations,  $\mu_{bb}\!= \!1.01\! \pm\! 0.20$ \cite{Aaboud:2018zhk} and
$\mu_{bb}= 1.04 \pm 0.20$ \cite{Sirunyan:2018kst}, leave no chance for this
scenario to occur.  

Hence, even in the less constrained case in which the SM4 fermions do not mix 
with the standard ones leading to a stable DM heavy neutrino, the possibility of
a fourth generation and,  more generally,  of  new fermions whose  masses
entirely originate from the Higgs sector of the SM,  is totally  excluded by the
LHC Higgs data. 

\subsubsection{Singlet--doublet DM model}

A most economical extension of the fermionic DM sector to accommodate
renormalizable interactions with the Higgs field is the so called
singlet--doublet model (SDM)~\cite{Cohen:2011ec,Cheung:2013dua,Calibbi:2015nha}
which, as mentioned earlier, has the additional advantage of representing a
simple limiting case of well studied scenarios such as the MSSM
\cite{Martin:1997ns,Djouadi:1998di} and its extensions such as the  NMSSM
~\cite{Ellwanger:2009dp}, hence providing a very useful benchmark for
supersymmetric scenarios\footnote{Notice, however, that in these two SUSY cases the DM state can have, besides a singlet (bino and/or singlino) and a doublet (higgsino) component, an SU(2) triplet (wino) component.}.  In this model, two additional doublet and a singlet fermionic fields are introduced 
\begin{equation}
L_L=\left( \begin{array}{c} N_L \\ E_L \end{array} \right), \,\,\,\,\,
L_R=\left( \begin{array}{c} -E_R \\ N_R \end{array} \right), \,\,\, N' \; , 
\end{equation}
which lead to different quantum assignments compared to those of the SM leptons
(instead, one should note  that the $L_L$ and $L_R$ fields have the same quantum
numbers as the higgsinos $\widetilde{H}_d$ and $\widetilde{H}_u$ of the MSSM as will be seen later). Their masses and interactions with the SM Higgs field $\Phi$ and its conjugate field $\widetilde{\Phi}$ are defined by the following Lagrangian
\begin{equation}
\label{eq:SD_lagrangian}
\mathcal{L}=-\frac{1}{2}M_N N^{'\,2}-M_L L_L L_R -y_1 L_L \Phi N^{'}-y_2 L_R \widetilde{\Phi} N^{'}+\mbox{h.c.},
\end{equation}
The  three lepton states are described by the following mass matrix
\begin{equation}
\label{eq:SD_mass_matrix}
M=\left(
\begin{array}{ccc}
M_N & {y_1 v}/{\sqrt{2}}~~ & {y_2 v}/{\sqrt{2}}~~ \\
{y_1 v}/{\sqrt{2}}~~ & 0 & M_L \\
{y_2 v}/{\sqrt{2}}~~ & M_L & 0
\end{array}
\right),
\end{equation}
which can be diagonalised with a unitary $3\!\times\! 3$ matrix $U$, leading to three Majorana mass eigenstates\footnote{A realization of the singlet--doubled model with a Majorana DM has been proposed in Ref.~\cite{Yaguna:2015mva}. It will not be reviewed here since its phenomenology features strong similarities with the one of the model proposed in the next subsection.} $N_i$ 
\begin{equation}
N_i=N^{'} U_{i1}+N_L U_{i2}+N_R U_{i3}\, ,
\end{equation}
the lightest of which, $N_1$, is assumed to be the DM candidate. The mass
spectrum of the new states is completed with an electrically charged Dirac
lepton $E^{\pm}$ with a mass $m_{E^{\pm}} \approx M_L$. 
Note that the Lagrangian of eq.~(\ref{eq:SD_lagrangian}) is defined under the
assumption that the new fermions are odd under a $\mathbb{Z}_2$ symmetry, with
the SM states being even, so that couplings through the Higgs boson  between 
new and SM fermions are forbidden.  Furthermore, given the transformation
properties of the fields $N^{'},\,L_L,\,L_R$,  under the discrete  $\mathbb{Z}_2$
symmetry, only one between the signs of the mass term $M_L$ and of the couplings
$y_1$ and $y_2$ is  physical. We will assume here a positive sign for $M_L$ and
leave free the signs of $y_1$ and $y_2$. 

In the physical basis, the interaction Lagrangian of the new fermions reads
\begin{align}
\label{eq:physical_SD}
\mathcal{L}&=\bar N_i \gamma^\mu \left(g_{Z N_i N_j}^V -g_{Z N_i N_j}^A \gamma_5\right) N_j Z_\mu+\bar E^{-}\gamma^\mu \left(g_{W^{\mp} E^{\pm} N_i}^V-g_{W^{\mp} E^{\pm} N_i}^A \gamma_5 \right) W^{-}_\mu N_i \nonumber\\
& -e \bar E^- \gamma^\mu E^- A_\mu -\frac{g}{2 c^2_W}(1-2 s^2_W) \bar E^- \gamma^\mu E^- Z_\mu + g_{HN _i N_j}H \bar N_i N_j +\mbox{h.c.} \; , 
\end{align}
with $g$ and $s_W,c_W$ as defined earlier.  The vector and axial--vector couplings of the  $N_i$ fields to the $Z$ boson are given by
\begin{eqnarray}
 g^{V/A}_{ZN_i N_j}=c_{ZN_i N_j} \mp c^{*}_{ZN_i N_j}
~~{\rm with}~~ c_{ZN_i N_j} =\frac{g}{4c_W}\left(U_{i3}U_{j3}^{*}- U_{i2}U_{j2}^{*}\right),
\end{eqnarray}
showing that, as expected for Majorana fermions, the coupling of the DM with the
$Z$ boson is only axial--vector like. The couplings between the neutral states $N_i$, their  charged partner and the $W^\pm$ bosons are instead given by
\begin{eqnarray}
g^{V/A}_{W^\mp E^{\pm} N_i}&&=\frac{g}{2\sqrt{2}}\left(U_{i3} \mp U_{i2}^{*}\right) .
\label{eq:Zpsi1}
\end{eqnarray}

The last term of the Lagrangian eq.~(\ref{eq:physical_SD}) represents the
coupling between a DM pair and the Higgs boson, already introduced in the simple
Higgs--portal model previously discussed. The couplings with the Higgs boson are explicitly given by
\begin{equation}
g_{H N_i N_j}=\frac{1}{\sqrt{2}}\left(y_1 U_{i2}^{*}U_{j1}^{*}+y_2 U_{j3}^{*}U_{i1}^{*}\right).
\label{eq:Hpsi1}
\end{equation}
\noindent
In our study, we will trade  the  parameters $y_1,y_2$ with the parameters $y,\theta$ so that \cite{Calibbi:2015nha}
\begin{equation}
y_1=y\cos\theta,\,\,\,\,\,y_2=y \sin\theta.
\end{equation} 
\noindent
Note however, the presence of the hypercharge and SU(2) components for the DM
particle, inherited from the mass mixing between the vector--like lepton, which  implies couplings also with the $Z$ and $W$ bosons, in contrast to the effective Higgs--portal model. This feature has a very strong impact on the phenomenology, as will be shown later.

As already mentioned, the Majorana singlet--doublet model can be interpreted as
a simplified version of a supersymmetric scenario. By performing the following
substitutions
\begin{eqnarray}
&& \sqrt{2} y\rightarrow {g \tan\theta_W},\,\,\,\,M_N \rightarrow M_1,\,\,\,\,M_L \rightarrow  -\mu,\nonumber \\
&&~~~~~~~\cos\theta \rightarrow -\cos\beta,\,\,\,\,\sin\theta \rightarrow \sin\beta,
\end{eqnarray}
it is straightforward to identify the SM singlet state $N^{'}$ with the bino (or
alternatively the singlino in the NMSSM) with the Majorana mass parameter $M_1$,
and the $L_{L,R}$, as already said, with the higgsinos, with a mass parameter
$\mu$. The angle $\beta$ is defined, as usual, such that $\tan\beta =v_2/v_1$
represents the ratio of the vacuum expectations values of the two Higgs doublets
in the MSSM/NMSSM. The singlet--doublet model can be then seen as a
supersymmetric model with mixed bino/higgsino DM in which all the scalar states,
apart from the SM--like Higgs boson are integrated out; see section 6. 

\subsubsection{Extensions with vector--like fermions}

The fermionic spectrum can be extended in a theoretically consistent  and
renormalisable way, while evading the stringent bounds from the 125 GeV Higgs
data and direct searches that apply in the fourth fermionic generation case, 
also with families of vector--like fermions (VLF). While less-economical with respect to the case of the singlet-doublet model discussed in the previous subsection, it is of great phenomenological interest since allows, as clarified below, for modification of the Higgs signals at colliders. We define a ``family'' of VLFs as a set of states composed by two ${\rm SU(2)_L}$ singlets and one ${\rm
SU(2)_L}$ doublet, all of them belonging to a same representation $R_c$ of SU(3)
and with a hypercharge which can be expressed in terms of a unique parameter
$Y$ as follows
\begin{eqnarray}
{\cal D}_{L,R}  \sim (R_c, 2, Y-1/2)~, \quad   
U^\prime_{L,R} \sim  (R_c, 1, Y)~,~ \quad 
D^\prime_{L,R} \sim  (R_c, 1, Y-1)~.
 \label{general-VL-family}
\end{eqnarray}
In this setup, a VLF family can be described, in terms of the SM Higgs field 
$\Phi$ and the decomposition for the SU(2) doublets ${\cal D}_{L,R}\equiv \begin{pmatrix} U & D \end{pmatrix}^T_{L,R}$, by the Lagrangian 
~\cite{Angelescu:2016mhl}
\begin{align}
-{\cal L}_{\rm VLF} & = y^{U_R} \overline{{\cal D}_L} \tilde{\Phi} U^\prime_R + y^{U_L} \overline{U^\prime_L} \tilde{\Phi}^\dagger {\cal D}_R +
y^{D_R} \overline{{\cal D}_L} \Phi D^\prime_R + y^{D_L} \overline{D^\prime_L} \Phi^\dagger {\cal D}_R  \nonumber \\
& +M_{UD} \overline{{\cal D}_L} {\cal D}_R 
+ M_U \overline{U^\prime_L} U^\prime_R +M_D \overline{D^\prime_L} D^\prime_R + {\rm h.c.}~.
\end{align}

According to the color and hypercharge assignments, the VLFs can share the same
quantum numbers as SM quarks and leptons; in this last case, however, neutral
singlets analogous to right--handed neutrinos would be present as well. This
would imply the presence of mixing between the vector--like and the SM fermions
with the same quantum numbers.  This mixing would allow for the decays of the
VLF fermions, including an eventual DM candidate, into SM states, originating a tension with requirement of stability at cosmological scales for the DM. We will thus
impose the discrete symmetry, dubbed $\mathbb{Z}_2^{\rm VLF}$, under which the
VLF and the SM fermions feature opposite charges, so that their couplings with
the Higgs boson and the subsequent mixing, will not allow for this possibility. 

The states that appear in eq.~(\ref{general-VL-family}) are in the
``interaction'' basis and, after electroweak symmetry breaking, their coupling
with the Higgs boson induces a mass mixing between the ``up'' ($U', U$) and
``down'' ($D', D$) states, having the same electric charges, $Q_U=Y$ and
$Q_D=(Y-1)$ respectively, described by the following transformation:
\begin{eqnarray}
U_L^{F} {\cal M}_{F} \left(U_R^{F} \right)^\dagger \!=\!
\begin{pmatrix}\! m_{F_1} & 0\\ 0 & m_{F_2} \!\end{pmatrix},\ 
U_L^{F} \!=\! \begin{pmatrix}\! \cos\theta_L^{F} & \sin\theta_L^{F}\! \\ \!-\!\sin\theta_L^{F} & \cos\theta_L^{F} \! \end{pmatrix},\ 
U_R^{F}\!=\! \begin{pmatrix} \! \cos\theta_R^{F} & \sin\theta_R^{F} \\ \!-\!\sin\theta_R^{F} & \cos\theta_R^{F} \! \end{pmatrix}~,
\end{eqnarray}
where the sub/superscripts $F=U,D$ distinguish between the two sectors.
Here, we will denote the lighter  mass eigenstate as $F_1$. The limit
where one of the singlets is decoupled, e.g. when $y_{U_R}=y_{U_L}=0$ and
$M_U\rightarrow \infty$, has been studied in detail in Ref.~\cite{Bizot:2015zaa}. The mass matrices $\mathcal{M}_{U,D}$ for the ``up'' and ``down'' states are defined as
\begin{eqnarray}
{\cal M }_U=  \begin{pmatrix} M_U & y^{U_L} v/ \sqrt{2} \\
y^{U_R} v/ \sqrt{2} & M_{UD} \end{pmatrix} ~,\quad
{\cal M}_D= \begin{pmatrix} M_D &  y^{D_L} v/ \sqrt{2} \\
y^{D_R} v/ \sqrt{2} & M_{UD} \end{pmatrix}~.
\label{SM-mass-matrices}
\end{eqnarray}  
In our analysis, we will consider the assignments $Y\!=\!0, R_c\!=\!1$ for the hypercharge $Y$ and the color index $R_c$ of the vector--like lepton families. In this case, the new fermions have the same quantum numbers as a generation of SM leptons plus additional right--handed neutrinos. This is why we will call them  ``VLLs" and we adopt for them the notation
\begin{eqnarray}
L_{L,R}=(1,2,-1/2),\,\,\,\,E^{'}_{L,R}=(1,1,-1),\,\,\,\,N_{L,R}^{'}=(1,1,0)\; .
\end{eqnarray} 
The vector--like lepton family features electrically neutral states,  the neutral components of the $L_{L,R}$ doublet as well as the ``right--handed'' vector--like neutrinos, the latter being complete SM singlets. It is then evident that the DM particle can be potentially embedded in this kind of construction. In this review we will consider the case that the mass eigenstate corresponding to the DM candidate is a Dirac fermion. Majorana DM can be nevertheless accommodated in this setup as shown, e.g., in~\cite{Joglekar:2012vc}.

For many reasons, such as chiral anomaly cancellation, VLLs should always 
be accompanied by vector--like quarks (VLQs)  in order to form a complete family. In our analysis, we will sometimes  consider such a
family of VLFs with the leptonic sector chosen as above and  for the quark
sector, we assume that the new quarks have the same quantum numbers as a
generation of SM quarks with $Y\!=\!\frac23$ and $R_c\!=\!3$, using the following
notation
\begin{eqnarray}
Q_{L,R}=(3,2,1/3),\ \ \ T^{'}_{L,R}=(3,1,2/3),\ \ \ B_{L,R}^{'}=(3,1,-1/3)\; .
\end{eqnarray}
We now illustrate in more detail the features of the family of vector--like leptons related to the DM heavy neutral lepton. For a single vector--like family, the Lagrangian with the particle content of eq.~(\ref{general-VL-family}) reads
\begin{align}
-{\cal L}_{\rm VLL}&= y_H^{N_R} \overline{L}_L \tilde{\Phi} N_R^\prime + y_H^{N_L} \overline{N}^\prime_L \tilde{\Phi}^\dagger L_R+y_H^{E_R} \overline{L}_L \Phi E^\prime_R + y_H^{E_L} \overline{E}^\prime_L \Phi^\dagger L_R  \notag \\
&+M_L \overline{L}_L L_R + M_N \overline{N}^\prime_L N^\prime_R +M_E \overline{E}^\prime_L E^\prime_R + \mathrm{h.c.}~.
\label{2HDM-VLLs}
\end{align}
After the breaking of the electroweak symmetry, the spectrum of the new fermions features two neutral states with masses $m_{N_1}, m_{N_2}$ and two electrically charged states with masses $m_{E_1}, m_{E_2}$ obtained by rotating the mass matrices
\begin{equation}
\mathcal{M}_N = \left( \begin{array}{cc} 
M_N & v' y_H^{N_L} \\  v' y_H^{N_R} & M_L  
\end{array}\right), \ \ 
 \mathcal{M}_L = \left( \begin{array}{cc} 
M_E & v' y_H^{E_L} \\  v' y_H^{E_R} & M_L 
\end{array}\right)\, ,  
\label{eq:vll_massmat}
\end{equation}
with pairs of unitary matrices $U_{L,R}^{F}$, $F=N,E$ of angles $\theta^F_{L,R}$
\begin{eqnarray}
U_L^N \cdot \mathcal{M}_N \cdot \left( U_R^N \right)^{\dagger} = \mathrm{diag} (m_{N_1}, m_{N_2}), \quad
U_L^E \cdot \mathcal{M}_E \cdot \left( U_R^E \right)^{\dagger} = \mathrm{diag} (m_{E_1}, m_{E_2})\; .
\end{eqnarray}
The two mixing angles $\theta_{L/R}^N$ are given by
\begin{eqnarray}
\label{eq:LRangles}
\tan 2 \theta_{L/R}^N =\frac{2 \sqrt{2} v \left(M_{L/N} y_H^{N_L}+ M_{N/L} y_H^{N_R}\right)}{2 M_L^2-2 M_N^2 \mp v^2 \left(|y_H^{N_L}|^2-|y_H^{N_R}|^2\right)}\; , 
\end{eqnarray}
while $\theta_{L,R}^E$ are obtained by replacing $M_N \rightarrow M_E$ and
$y_H^{N_{L,R}} \rightarrow y_H^{E_{L,R}}$ in the expressions above.   In this
scenario, the DM candidate is the state $N_1$ if it is lighter than the charged
fermions $E_{1,2}$. In the mass basis, the DM interaction Lagrangian can be
written as
\begin{eqnarray}
\label{eq:hVLL_DM_simp}
\mathcal{L} &=&  \! \sum_{i,j=1,2} \bar N_i \gamma^\mu \left(y_{ZN_i N_j}^V\!-\!y_{ZN_i N_j}^A \gamma_5\right)N_j Z_\mu  +  \bar E_j \gamma^\mu \left(y_{W N_i E_j}^V\!-\!y_{WN_i E_j}^A \gamma_5\right)N_i W^{-}_\mu +\mbox{h.c.} \nonumber \\
&&\! +\frac{1}{\sqrt{2}} \sum_{i,j=1,2} y_{H N_i N_j} \bar N_i N_j H + 
\frac{1}{\sqrt{2}} \sum_{i,j=1,2} y_{H E_i E_j} \bar E_i E_j H \, , 
 \end{eqnarray}
where the couplings, using the abbreviations $\cos\theta=c_\theta$ etc..., are
\begin{align}
\label{eq:hcoupling}
y_{H N_i N_j}=\frac{1}{\sqrt {2}} \left[
\begin{array}{cc}
c_{\theta_N^L} s_{\theta_N^R} y_H^{N_L}+ c_{\theta_N^R} s_{\theta_N^L} y_H^{N_R} &c_{\theta_N^L} c_{\theta_N^R} y_H^{N_L}- s_{\theta_N^R} s_{\theta_N^L}y_H^{N_R} \\
-s_{\theta_N^L} s_{\theta_N^R} y_H^{N_L}+ c_{\theta_N^R} c_{\theta_N^L} y_H^{N_R} &  -s_{\theta_N^L} c_{\theta_N^R} y_H^{N_L}+ s_{\theta_N^R} c_{ \theta_N^L} y_H^{N_R}  
\end{array}  \right] ~,  \nonumber
\end{align}\vspace*{-6mm} 
\begin{align}
y^{V/A}_{Z N_i N_j}=\frac{g}{4 \cos\theta_W}\left[
\begin{array}{cc}
s^2_{\theta_L^N} \pm s^2_{\theta_R^N}      & s_{\theta_L^N} c_{\theta_L^N} \pm s_{\theta_R^N} c_{\theta_R^N} \\
s_{\theta_L^N} c_{\theta_L^N} \pm s_{\theta_R^N} c_{\theta_R^N}     &  c^2_{\theta_L^N} \pm c^2_{\theta_R^N}
\end{array} \right] ~, \nonumber
\end{align}\vspace*{-6mm}
\begin{align}
y^{V/A}_{W N_1 E_1}=\frac{g}{2 \sqrt{2}}\left[ 
\begin{array}{cc}
s_{\theta_L^N} s_{\theta_L^E} \pm s_{\theta_R^N} s_{\theta_R^E}  & s_{ \theta_L^N} c_{\theta_L^E} \pm s_{\theta_R^N} c_{\theta_R^E} \\
c_{\theta_L^N} s_{\theta_L^E} \pm c_{\theta_R^N} s_{\theta_R^E} & c_{\theta_L^N} c_{\theta_L^E} \pm c_{\theta_R^N} c_{\theta_R^E}
\end{array}
\right]~ . 
\end{align}
The couplings $y_{HE_i E_j}$ are obtained from $y_{HN_i N_j}$ by the
exchange $N \leftrightarrow E$ in the labels.  

\subsubsection{Theoretical constraints}

To have a theoretically consistent picture, the couplings of the new fermions in
extensions of the SM should obey severe requirements. First, these fermions
affect the running of the SM gauge couplings, potentially leading to Landau
poles at low energies. The Yukawa couplings of these states are also subject to
evolution with energy and could enter a non--perturbative regime if they are set
to too high values at the electroweak  scale. More important, the new
Yukawa couplings also affect the renormalisation group evolution of the Higgs
quartic coupling,  possibly rendering it negative and hence destabilizing the
potential in a way such that electroweak symmetry breaking does not occur.  

We illustrate in this subsection, the impact of these renormalisation group
constraints taking the case of vector--like fermions as an example; we will 
restrict to the leading order, which makes that our discussion can be viewed to
be only qualitative since it is essentially based on these one--loop
ingredients. 

The one--loop $\beta$ functions for the gauge couplings $g_i$ with $g_1=g',
g_2=g$ and $g_3=g_s$ in the presence of $N_{\rm VLL}$ and $N_{\rm VLQ}$ vector--like leptons and quarks can be written as
\begin{align} 
& \beta_{g_1}=\frac{41}{6}+\frac{4}{3}N_{\rm VLL} \left(Y_E^2+2
Y_L^2\right)+4 N_{\rm VLQ} \left(Y_B^2+Y_T^2+2 Y_Q^2\right) \, , \nonumber\\ 
& \beta_{g_2}=-\frac{19}{6}+\frac{2}{3}N_{\rm VLL} +2 N_{\rm VLQ} \ , \quad \quad  \beta_{g_3}=-7+\frac{8}{3}N_{\rm VLQ} \ . 
\label{eq:gi-RGE} 
\end{align}
From these equations, one can see that the $\beta_{g_i}$ functions get positive contributions that depend on the multiplicity of the VLF families and, in the case of the U(1) coupling $g_1$, on the hypercharge of the new states. Too large positive contributions would lead to Landau poles, i.e. $\alpha_i (\mu)  = g_i^2(\mu)/(4\pi) \geq 1$, at some scale $\mu$. In this work, we will thus
consider,  also for the sake of minimality, the case of only one vector lepton
family $N_{\rm VLL}=1$ with and without a vector--like family, $N_{\rm
VLQ}=0$ or 1.  

\begin{figure}[!h]
\begin{center}
\includegraphics[width=0.5\linewidth]{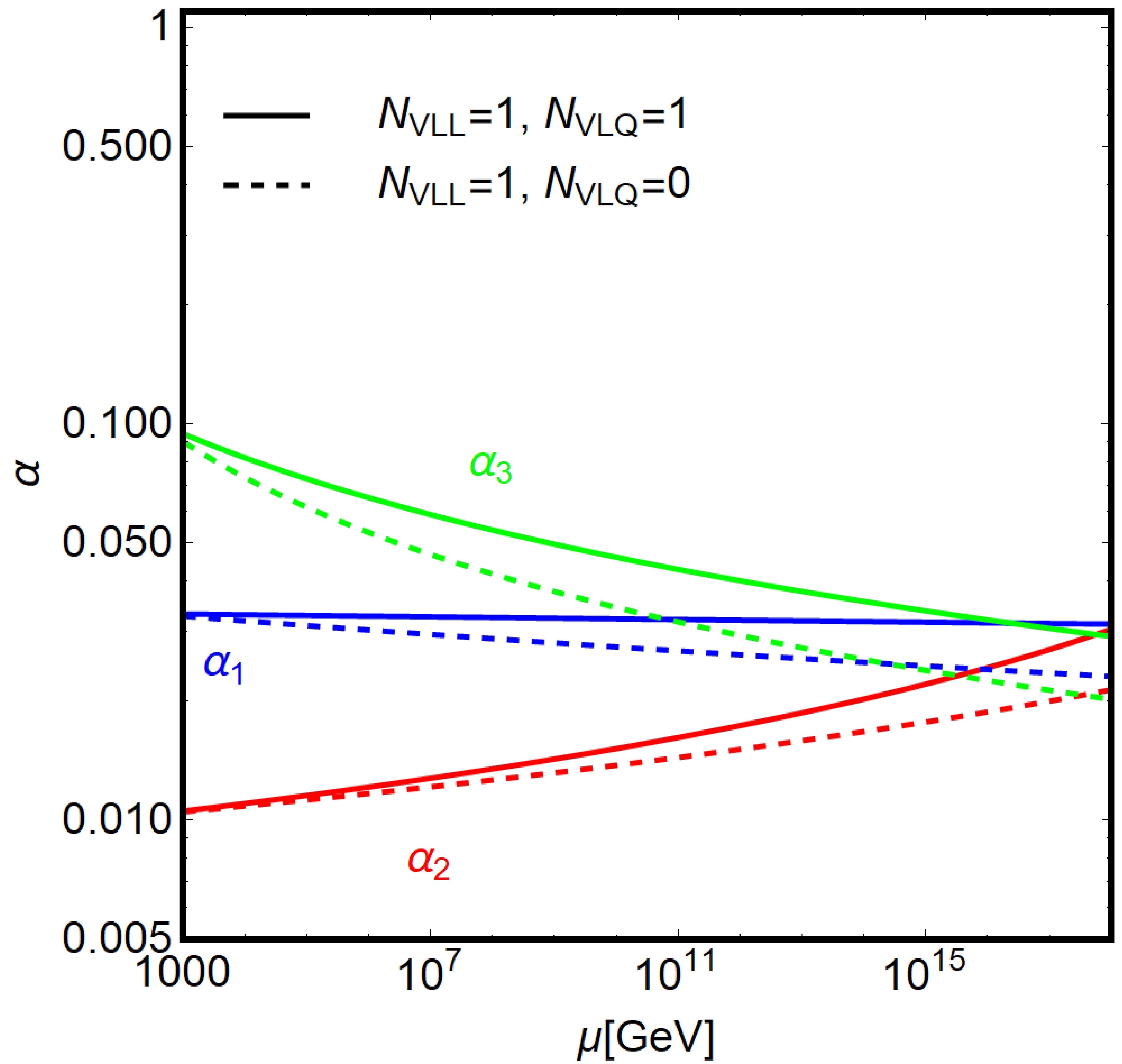}
\end{center}
\vspace*{-3mm}
\caption{Renormalisation group evolution of the SM gauge couplings $\alpha_i=g_i^2/(4\pi)$, as a function of the energy scale $\mu$, for different vector--fermion contents: $N_{\rm VLL}=N_{\rm VLQ}=1$ (solid lines) and  $N_{\rm VLL}=1, N_{\rm VLQ}=0$ (dashed lines).}
\label{fig:alphaRGE}
\vspace*{-2mm}
\end{figure}

The evolution of these couplings is shown in Fig.~\ref{fig:alphaRGE} as a
function of the energy scale and, as one can see,  the running is not affected
to a pathological extent by the presence of a sequential family of vector--like
leptons or even by a full family of vector--like fermions, so that the gauge
couplings remain perturbative up to $M_{\rm Planck}$. 

For the self--coupling $\lambda$ and the Yukawa couplings $y_H^F$, 
the situation is quite different. Their $\beta$ functions are given by 
\begin{align}
 \beta_{y_H^{F}}& =\frac{y_{H}^{F}}{16 \pi^2}\left[3 |y_{H}^{F}|^2+ 2\left(3 N_{\rm VLQ} X_H^{\rm VLQ} +N_{\rm VLL} X_H^{\rm VLL} \right) 
- \delta_H^F \right] \, , \nonumber \\
 \beta_{\lambda}&=\frac{1}{16 \pi^2} \left[ 3 \lambda^2-48 \left(N_{\rm VLL} (|y_H^E|^4+2 |y_H^{NE}|^4)+3 N_{\rm VLQ} (|y_H^B|^4+|y_H^T|^4+2|y_H^{BT}|^4)\right)\right. \nonumber\\
& \left.+8 \lambda \left(N_{\rm VLL} (|y_H^E|^2+2 |y_H^{NE}|^2)+3 N_{\rm VLQ} (|y_H^B|^2+|y_H^T|^2+2|y_H^{TB}|^2)~\right)\right]~, 
\label{eq:singletRGE}
\end{align}
where we have used the abbreviations,
\bea
X_H^{\rm VLQ}= |y_H^{B}|^2+|y_H^{T}|^2+2 |y_H^{TB}|^2 ~, ~~~
X_H^{\rm VLL}= |y_H^E|^2+|y_H^N|^2+|y_H^{NE}|^2) , \nonumber \\
\delta_H^{TB}= -8 g_3^2 -\frac{9}{4}g_2^2-\frac{1}{3}g_1^2 , \
\delta_H^{T}= -8 g_3^2 -\frac{8}{3}g_1^2 , \
\delta_H^{B}= -8 g_3^2 -\frac{2}{3}g_1^2 , \ \nonumber \\
\delta_H^{NE}= -\frac{9}{4}g_2^2-\frac{3}{2}g_1^2 , \
\delta_H^{E}= -6g_1^2 , \
\delta_H^{N}= 0\ . \hspace*{2cm}
\eea

As can be seen, the $\beta$ function of the Yukawa couplings of the new fermions
are proportional to the couplings themselves to the third power, and they vary
significantly with the energy scale in contrast to the gauge couplings.
Consequently, the Yukawa couplings $y_H^F$ of the new fermions can become
non--perturbative at relatively low energy scales, if they are $\gtrsim {\cal
O}(1)$ at the weak scale. More important, this potentially strong variation with
energy would dramatically affect the running of the quartic Higgs coupling
$\lambda$, since its $\beta$ function would receive two negative contributions
proportional to $\lambda_H (y_H^F)^2$ and $(y_H^F)^4$. A too steep increase of
the Yukawa couplings would then drive the quartic coupling to negative values,
destabilizing the scalar potential.

As was discussed in Refs.~\cite{Joglekar:2012vc,Angelescu:2016mhl} e.g., it is
possible to obtain constraints on the size of the Yukawa couplings $y_H^F$  by
solving the RGEs for the quartic Higgs coupling in combination with the ones of
the new fermions, the top Yukawa and the SM gauge couplings. 

\begin{figure}[!h]
\vspace*{-2mm}
    \centering
    \subfloat{\includegraphics[width=0.47\linewidth]{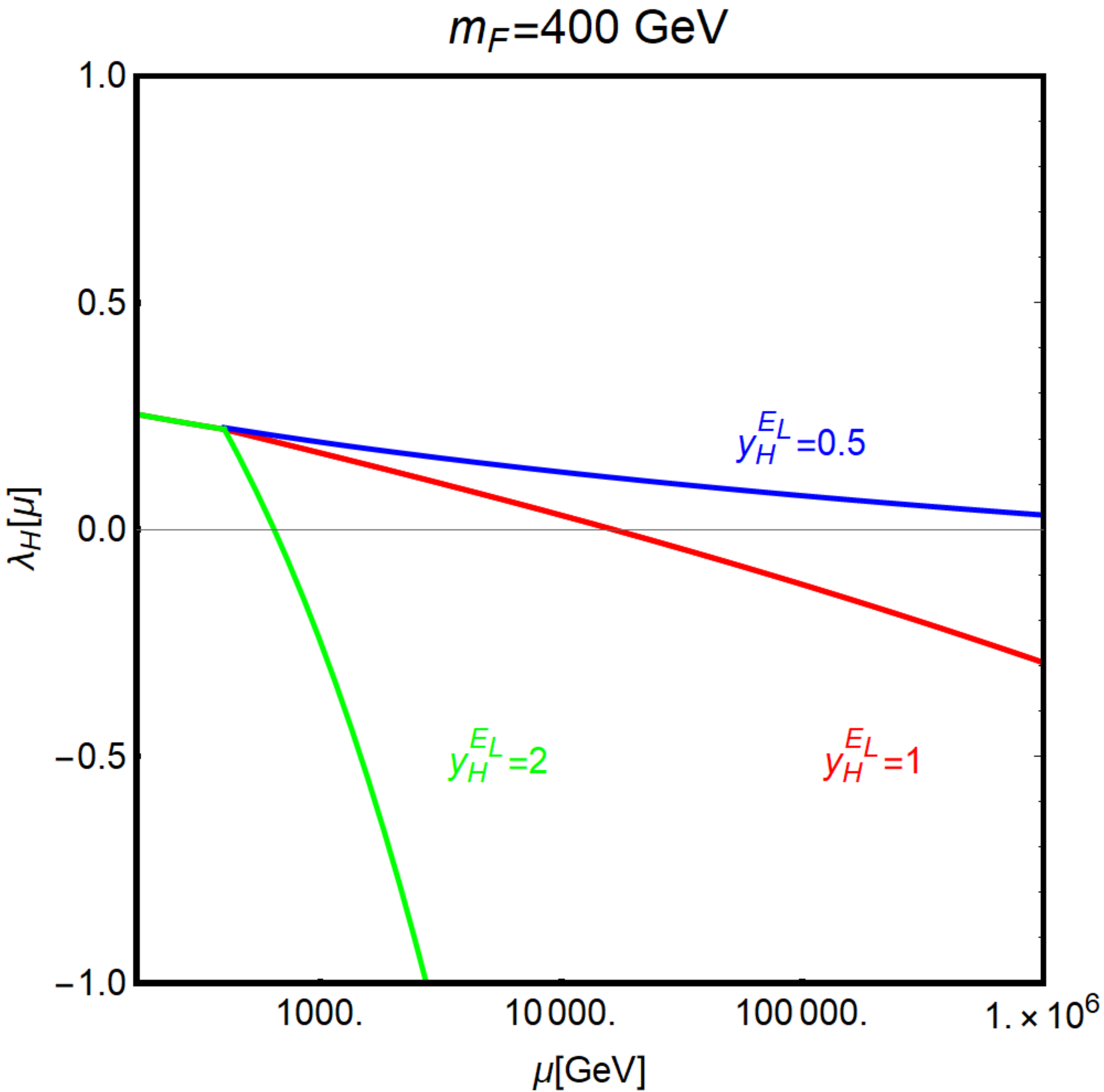}}
~~~\subfloat{\includegraphics[width=0.46\linewidth]{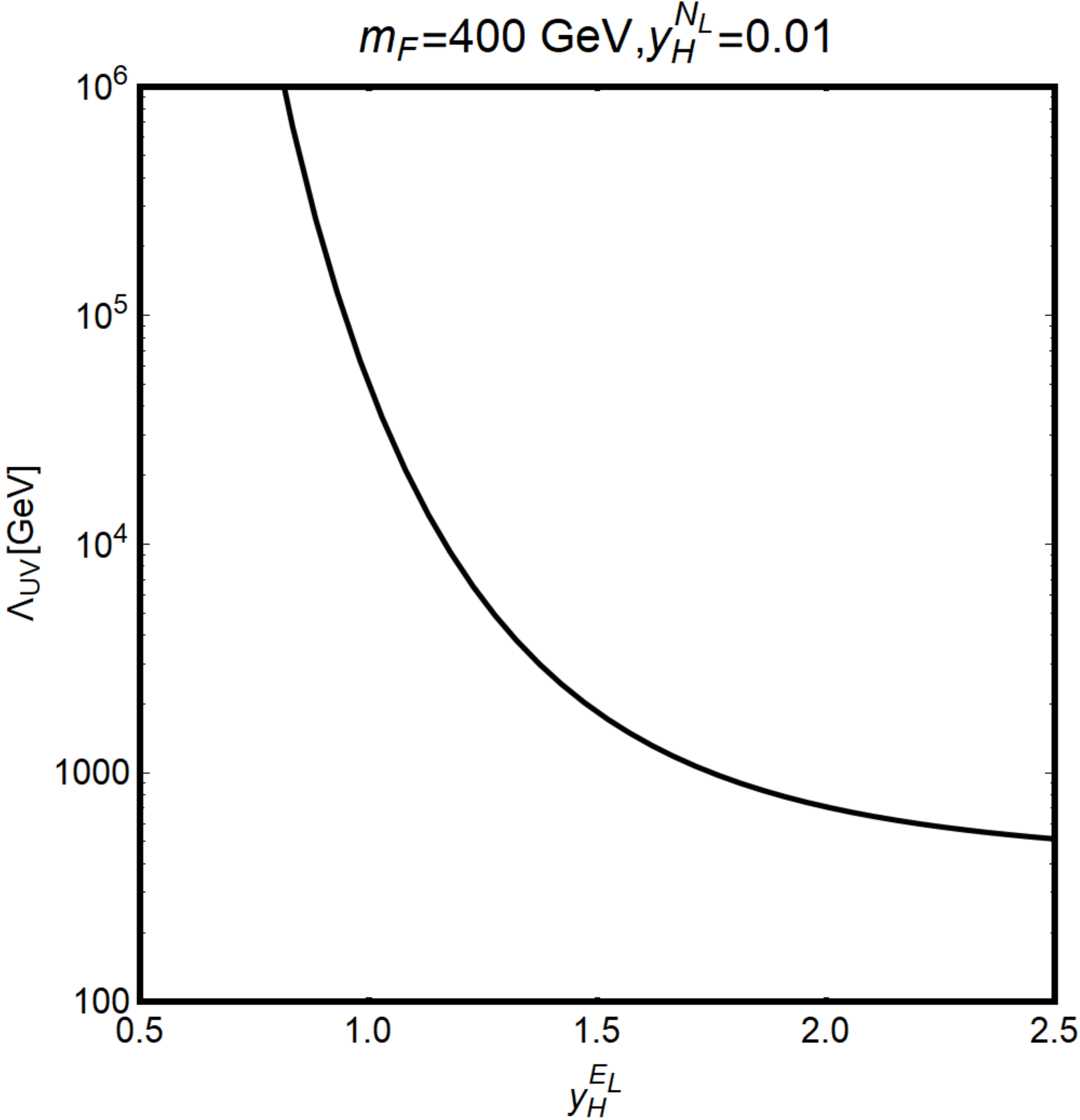}}
\vspace*{-2mm}
    \caption{Left panel: evolution of the Higgs quartic coupling as a function of the energy scale, assuming the extension of the SM with a family of vector--like fermions with common mass $m_F=400\,\mbox{GeV}$; the blue, red and green lines refer to, respectively, the values $y_H^{E_L}=0.5,1,2$, while the assignments of the other couplings has been fixed to $y_H^{N_L}=0.01,y_H^{N_R}= y_H^{E_R}=0$. Right panel: the value of the stability scale $\Lambda_{\rm UV}$ as defined in the text as a function of $y_H^{E_L}$; the other model parameters have been fixed as in the left panel.}
    \label{fig:VhLRGE}
\vspace*{-4mm}
\end{figure}

We illustrate through an example, the results of such an analysis in
Fig.~\ref{fig:VhLRGE}. For simplicity we have considered the case $N_{\rm VLL}=1,N_{\rm VLQ}=0$. The left panel of the figure shows the value of the
self-coupling $\lambda$ as a function of the energy scale $\mu$, in the case
where the SM is augmented by a full family of vector--like fermions which, for
simplicity, have been assumed to have the same mass $m_F=400$ GeV. The lepton 
Yukawa couplings have been set to $y_H^{E_L}=0.5,1,2$, while keeping
$y_H^{N_L}=0.01$ and $y_H^{N_R}=y_H^{E_R}=0$; these new Yukawa couplings have
been assumed to be zero below the scale $m_F$. As is clear from the figure, too
high initial values of the coupling $y_H^{E_L}$, namely $y_H^{E_L}=2$ in the
specific case considered here, would make that the Higgs quartic coupling
becomes negative immediately after the energy threshold $m_F$, hence
destabilizing the scalar potential.  

A slightly more refined analysis, which gives an idea of the size of the Yukawa
couplings as well as  the energy scale for which the scalar Higgs potential
becomes instable, is summarized in the right--hand side of
Fig.~\ref{fig:VhLRGE}.  Shown as a function of the coupling $y_H^{E_L}$ and for
the same values of the model parameters considered in the left panel of the
figure, the stability scale $\Lambda_{\rm UV}$. The latter corresponds to the
scale at which the negative runaway of the effective potential takes place
which, following an analysis performed in Ref.~\cite{Blum:2015rpa}, can be 
defined  by the condition $\lambda_H (\Lambda_{\rm UV})=-0.07$ which roughly
indicates the onset of vacuum instability. As can be seen from the figure, 
values $y_H^{E_L} \lesssim 1$ are needed in order that the instability scale
$\Lambda_{\rm UV}$ lies significantly above the TeV scale. 

Before closing this subsection, let us again emphasize the fact that one should
not interpret the results from these renormalisation group analyses as strict
constraints on the model parameters, but simply as an indication that the theory
requires a further ultraviolet completion, through the introduction of suitable additional degrees of freedom at the scale $\Lambda_{\rm UV}$. It is sufficient to ensure that the scale $\Lambda_{\rm UV}$ is sufficiently above the energy scales relevant for collider and astroparticle phenomenology, which
will be discussed later.

\subsection{Constraints on the new leptons and expectations at colliders}

We now turn to the phenomenology of the new leptons (and their possible partner
quarks) that accompany the DM particle in these models and summarize the various
experimental constraints to which they are subject. The production of the new
states at colliders is described at the end of the section. For additional
discussions, eventually outside of the DM context, see for instance
Refs.~\cite{Aguilar-Saavedra:2013qpa,Ellis:2014dza,Djouadi:2016eyy,Panella:2001wq,Han:2006ip,delAguila:2007qnc,delAguila:2008hw,delAguila:2008cj,Deppisch:2015qwa,Djouadi:1993pe,Azuelos:1993qu,Buchmuller:1991tu}. 

\subsubsection{Constraints on masses and couplings}

Let us start by describing the couplings of these particles to the SM ones in a
slightly more general form that suits at the same time the two cases that are
under discussion here, namely singlet--doublet and vector--like leptons. Except
for the singlet DM particle which should have no electromagnetic nor weak
charges, the other singlet--doublet like or vector--like fermions couple to,
besides the Higgs boson, the photon when electrically charged and the
electroweak gauge bosons $W/Z$ with typical electroweak strength. These
couplings allow for the pair production processes of these particles at
colliders.  For a generic lepton $L$ which could be either $E$ or $N',N$  with
$N$ being the lighter DM state (from now on, we omit the subscripts for
simplicity) with electric charge $e_L=-1$ or $0$ in units of the proton charge
and left-- and right--handed isospin assignments $I^{L}_{3L}$ and $I_{3R}^L$,
the vector and axial--vector couplings to gauge bosons are given by
eqs.~(\ref{Zffcouplings}) and  (\ref{Wffcouplings}) in the $Z$ and $W$ cases,
respectively.   Hence, in the vector--like case, the axial--vector couplings to
the  $Z$ boson are zero by construction; in addition, the fermion couplings to
the $W$ boson are  twice as large as the ones of SM fermions. 

Because of the $\mathbb{Z}_2$ symmetry under which the new fermions are odd
while the SM ones are even, there should be  no mixing between the two types of
fermions. The heavier states should then decay into  lighter companions and
gauge or Higgs bosons, $W$ bosons for charged decays and $Z,H$ bosons for
neutral decays. At the end of the chain, there must be the lightest odd particle
which is our DM candidate, namely one of the additional neutrinos. In this case,
the signatures will consist into  missing energy and Higgs or gauge bosons, the
so--called mono--Higgs or mono--$Z,W$ signatures. If the mass splitting between
the parent and daughter new particles in the decay is small (as is required by
the electroweak precision data to be discussed later),  the intermediate bosons
will be off mass shell and will decay into a pair of almost massless fermions,
$E \to NW^* \to N f\bar f'$ or $N' \to NZ^* \to N \bar f f$ for example (because
the $H$ couplings to light fermions is very small,  the intermediate states are
in general the weak  bosons).  The smaller is the mass difference between the
heavier new leptons and the DM particle,  the softer are the final state
fermions so that the signatures will be rather difficult to detect, in
particular at hadron colliders.  

In the case of the heavy quarks that appear when one considers a full 
vector--like fermion family, some amount of fermion mixing should be possible in
particular if baryon number is conserved. This will allow the new quarks, or at
least the lighter one since here also there might be decays of heavy to lighter
new quarks $Q' \to QV^* \to Qf\bar f$, to decay into the SM ones $q$ and some
gauge or Higgs bosons, $Q \to qV, qH$. However, the mixing  angle should be very
small and the lifetime of the new fermions could be very long making that they
decay outside the detectors of collider experiments which, in practice, make
them almost stable  in this context and hence not straightforward to detect. 

The present experimental constraints on the masses of the new heavy leptons and
quarks depends on whether they are considered as stable or not and we summarize
those quoted by the Particle Data Group \cite{Tanabashi:2018oca} in the
following. In the case of a Dirac DM neutrino which should be stable, a limit of
$m_N >45$ GeV has been set from the accurately measured invisible decay of the
$Z$ boson at LEP1; if the neutrino is of Majorana type\footnote{In the case of
Majoranas, the couplings to the $Z$ boson are only axial--vector like  and the
partial widths are suppressed  by three powers of the velocity $\beta_N = \sqrt
{1- 4m_N^2/M_Z^2}$ compared to only one power in the Dirac case where the
couplings are  vector--like; the partial width is thus much more suppressed near
the phase--space boundary, $M_Z \sim 2m_N$.} the limit is slightly weaker, $m_N
>39.5$ GeV.  In the case of the charged leptons, there are bounds from searches
at LEP2 with a c.m. energy beyond $\sqrt s=200$ GeV
\cite{Achard:2001qw,Tanabashi:2018oca}: $m_E > 102.6$ GeV for a  stable lepton
and  $m_E >  100.8$ GeV if it decays into a light neutrino and a $W$ boson. For
heavy neutral leptons that are not stable, bounds from LEP2 searches also apply
and give $m_{N'}> 90.3$ GeV for a Dirac and $m_{N'}>80.5$ GeV for a Majorana
state \cite{Achard:2001qw,Tanabashi:2018oca}. 

In the case of quasi-stable quarks, the only bound that is quoted is the one on 
a $b'$--like quark, $m_{b'}=190$ GeV from searches at the Tevatron, besides the
one from $Z$ decays at LEP1,  $m_{b'}=46$ GeV \cite{Tanabashi:2018oca}. If the
heavy quarks decay  visibly into light SM quarks and gauge or Higgs bosons,
bounds from negative searches at the LHC are much more severe. This is
particularly true for the heavy partners of the top and bottom quarks for which 
bounds of about  $m_{Q} \gsim 1$ TeV are set on their masses at LHC with $\sqrt
s=13$ TeV depending on the isospin and branching fractions, and above $m_{Q}
\gsim 750$ GeV in essentially all cases  \cite{Tanabashi:2018oca}.

Coming back to the new leptons and the searches that have been performed at the
LHC, the charged ones under the electroweak group can be pair produced in proton
collisions through the Drell--Yan processes of the type (details will be given
later
on)~\cite{Drell:1970wh}
\begin{equation}
p p \to q\bar q \to E^+ E^-  \, , \ N' N' \, , \ \ E^\pm N' \, , 
\label{Drell-Yan}
\end{equation}
with the charged fermions subsequently decaying into a $W$ (either on-- or 
off--shell) and a neutral lepton (typically the lightest one, i.e. the DM
candidate) while the heavy neutral fermions feature different possible decay
channels, i.e. into a lighter one and a $Z$ (again either on-- or off--shell) or
possibly $H$ boson or the charged fermion $E^{\pm}$ and a $W$ boson. The relative
strength of the possible signals depends on the amount of hypercharge and
doublet components of the new lepton fields, set by the elements of the mixing
matrix $U$ or by the angles $\theta_{L,R}^{N,E}$ in the two cases that we are
interested in. The cleanest signature is in general represented by events with
missing energy accompanied by multileptons. 

A detailed classification of all the possible event topologies has been
presented for instance in Ref.~\cite{Calibbi:2015nha} in the case of the
singlet--doublet model and we refer to it for details. In this case, among all
these processes,  the strongest constraints come from the charged current
channel $pp \rightarrow E^\pm N' \rightarrow W^{\pm}Z NN$ which leads to  a
three leptons plus missing energy signature. The limits obtained by   CMS and
ATLAS in various searches that mimic this topology, see for instance
Refs.~\cite{Aaboud:2018jiw,Sirunyan:2018ubx} for the most recent ones,  should
be appropriately recast in order to be applied for the scenario under
consideration. Such recasting has been performed, for example, in
Ref.~\cite{Calibbi:2014lga} for analyses of the $\sqrt{s}=8\,\mbox{TeV}$ set of
data \cite{Khachatryan:2014qwa,Aad:2014nua}. It has been found that these
searches constrain the regions $m_{N'},m_{E^\pm} \lesssim 270\,\mbox{GeV}$ for
the heavier short lived leptons and $m_{N}\lesssim 75\,\mbox{GeV}$ for the
stable DM lepton. One should however note that these constraints mostly apply to
the case $M_N < M_L$ and in the opposite case, the lightest neutral leptons are
mostly doublet--like and are very close in mass with the charged state $E^\pm$.
This type of configuration is more complicated to probe at hadron colliders
since it would correspond to the production of long--lived particles, leading to
displaced vertices or particles being eventually stable at the detector level.

\subsubsection{Constraints from electroweak observables} 

Besides these direct collider constraints, there are also indirect ones as these new fermions interact with the $W,Z$ gauge bosons and thus affect
electroweak precision observables (EWPO) 
\cite{He:2001tp,Barbieri:2006bg,Enberg:2007rp,DEramo:2007anh,Joglekar:2012vc}. 
Depending on their masses and their gauge and Yukawa couplings, they could
induce large deviations with respect to the SM predictions. It turns out that,
to a good approximation, the contributions induced by the new fermions  can be
mapped  into corrections to the $\rho$ parameter \cite{Veltman:1977kh} which
historically was used to measure the strengths of the ratio of the neutral to the charged currents at zero--momentum transfer, $q^2=0$. 
In the SM, this parameter is equal to unity at tree--level as a result of an SU(2) custodial symmetry but it receives higher--order corrections, via the
$W$ and $Z$ boson self--energies $\Pi_{WW}, \Pi_{ZZ}$, parameterized by 
\beq
\rho = \frac{ 1}{1 -\Delta \rho} \ \ , \ \ 
\Delta \rho = \frac{\Pi_{WW}(0)} {M_W^2}  - \frac{\Pi_{ZZ}(0)} {M_Z^2} \, .
\label{rho-def}
\eeq
The contribution of two particles $A,B$ with masses $m_A,m_B$ to the $W,Z$ boson self-energies and hence to the $\rho$  parameter (factorizing out complicated coefficients of the couplings to gauge bosons and ignoring higher orders for simplicity)  read 
\beq
\Delta \rho \propto  \frac{G_F}{8\pi^2\sqrt 2} f(m^2_A,m^2_B)~~{\rm with}~~
f(x,y) = x+y- \frac{2 x y}{x-y} \log \frac{x}{y}\, .
\label{eq:f-deltarho}
\eeq
The function $f$ vanishes if the $A$ and $B$ particles are degenerate in mass
$f(m_A^2,m_A^2)=0$ while, in the limit of a large mass splitting, one has  
$f(m_A^2,0)=m_A^2$ instead. 

Hence, in the case where the members of an SU(2) doublet have masses that are
quite different, contributions that are quadratic  in the mass of the heaviest
particle appear. As the $\rho$ parameter has been measured with a precision of a
fraction of a permile and found to agree with the SM expectation
\cite{Tanabashi:2018oca}, it sets a very strong constraint on the mass splitting
between the two leptons that belong to the same SU(2) isodoublets. 

More precisely,  the new states will not only contribute to the $\rho$ parameter
above but more generally to the Peskin--Takeushi $S,T,U$ parameters 
\cite{Peskin:1991sw}. While  the leading New Physics contributions i.e. $\Delta
\rho$ is denoted by $T$, namely $T \propto \Delta \rho - \Delta
\rho|_{\rm SM}$, $S$ parametrizes the New Physics contributions to neutral current processes at different energy scales  while the $U$ parameter describes  new  charged current contributions to the $W$ boson mass. A global fit to all
electroweak precision observables available today has been made leading to the
following $\chi^2$~\cite{Baak:2011ze,Baak:2014ora}
\begin{equation}
\label{eq:chi2}
\chi^2=\sum_{i,j}(x_i-x_i^{\rm SM}){\left(\sigma_i V_{ij} \sigma_j\right)}^{-1}(x_j-x_j^{\rm SM}) \, , 
\end{equation}
determined as functions of the deviation of the $x=(S,T,U)$ parameters with respect to their SM corresponding values
\begin{equation}
 x^{\rm SM}= (S,T,U)^{\rm SM} =  (0.05,0.09,0.01),
\end{equation}
with the standard deviations and the covariance matrix given by 
\begin{equation}
\sigma=(0.11,0.13,0.11) \ , \quad  V=\left(
\begin{array}{ccc} 1 & 0.9 & -0.59 \\ 0.9 & 1 & -0.83 \\ -0.59 & -0.83 & 1
\end{array} \right) \, .
\label{eq:covariance}
\end{equation}

We have calculated these radiative corrections adapting the complete and general expressions quoted in Ref.~\cite{Joglekar:2012vc} to our two singlet--doublet and vector--like lepton cases. 

\begin{figure}[!h]
\vspace*{-1mm}
\begin{center}
\includegraphics[width=0.51\linewidth]{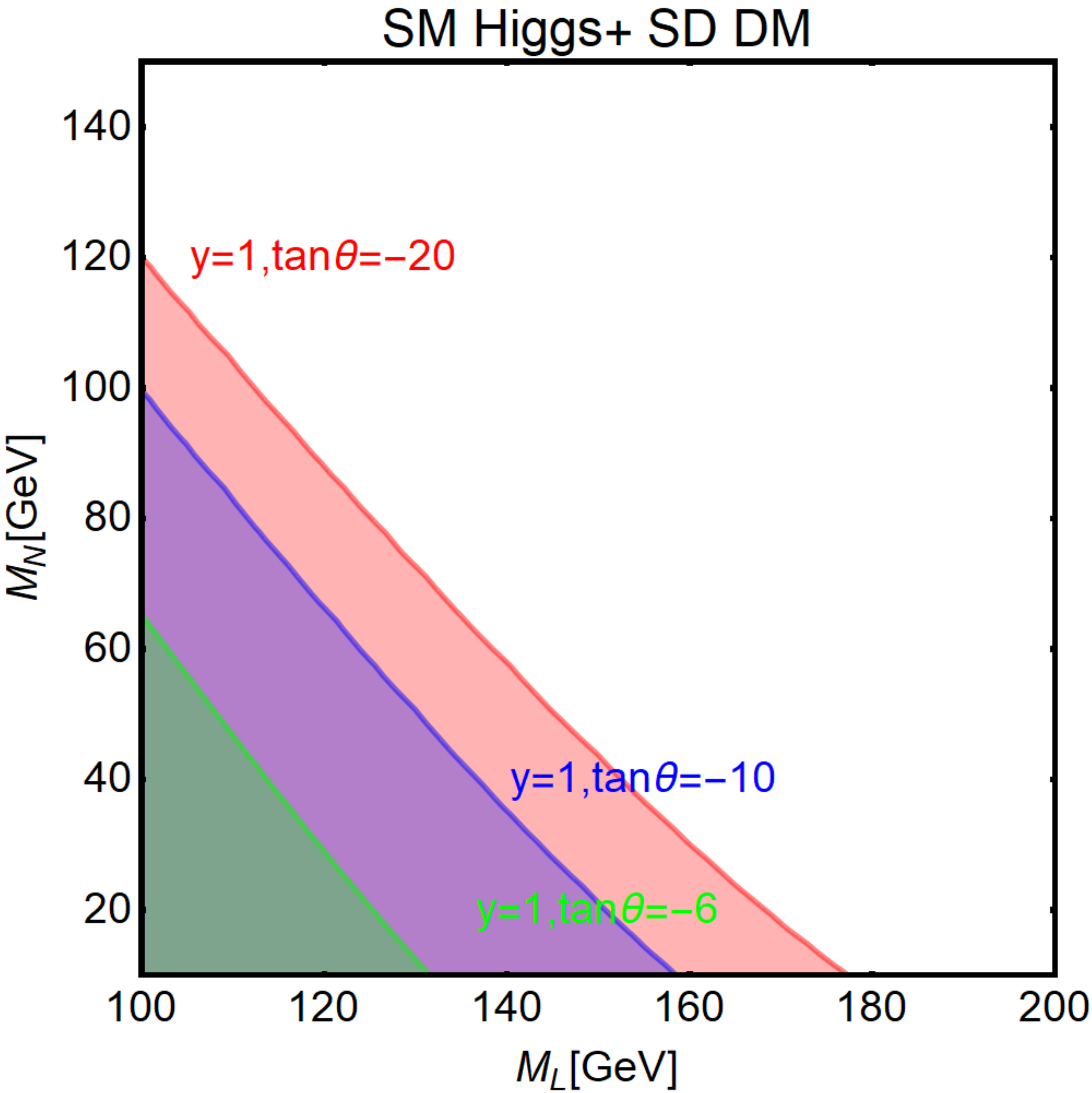}
\end{center}
\vspace*{-4mm}
\caption{Excluded regions by electroweak  precision observables  in the plane $[M_L,M_N]$ in the singlet--doublet DM model for three assignments of the $(y,\tan\theta)$ parameters.}
\label{fig:sdSTU}
\vspace*{-1mm}
\end{figure}

The regions excluded by precision observables in the singlet--doublet scenario
are shown in Fig.~\ref{fig:sdSTU} in the $[M_L,M_N]$ plane for the value $y=1$
and three assignments  for $\tan\theta$. As discussed in
Refs.~\cite{Cynolter:2008ea,Calibbi:2015nha}, these bounds are not very effective in this scenario as they essentially
affect only the $T$ parameter. As can be seen from Fig.~\ref{fig:sdSTU}, only a
portion of parameter space at small values of $M_N,M_L$ is excluded. This is
because, in this region, $M_L \lesssim y_{1,2} v$, implying sizable mass
splitting between the SU(2) doublet states. The different extension of the
excluded regions with $\tan\theta$ is explained by the fact that the
contribution of the new fermions to the $T$ parameter is proportional to
$(y_1^2-y^2_2)^2 \propto y^4 (1-\tan^2\theta)^2$. Consequently, the excluded
regions grow with $\tan\theta$ while, in turn, we have $\Delta T\!=\!0$
for $\tan\theta\!=\!1$. As will be justified later, we have focused in
Fig.~\ref{fig:sdSTU} on the case where  $\tan\theta<0$ as it is more interesting
for DM phenomenology in the astroparticle physics context.

\begin{figure}[!h]
\begin{center}
\subfloat{\includegraphics[width=0.49\linewidth]{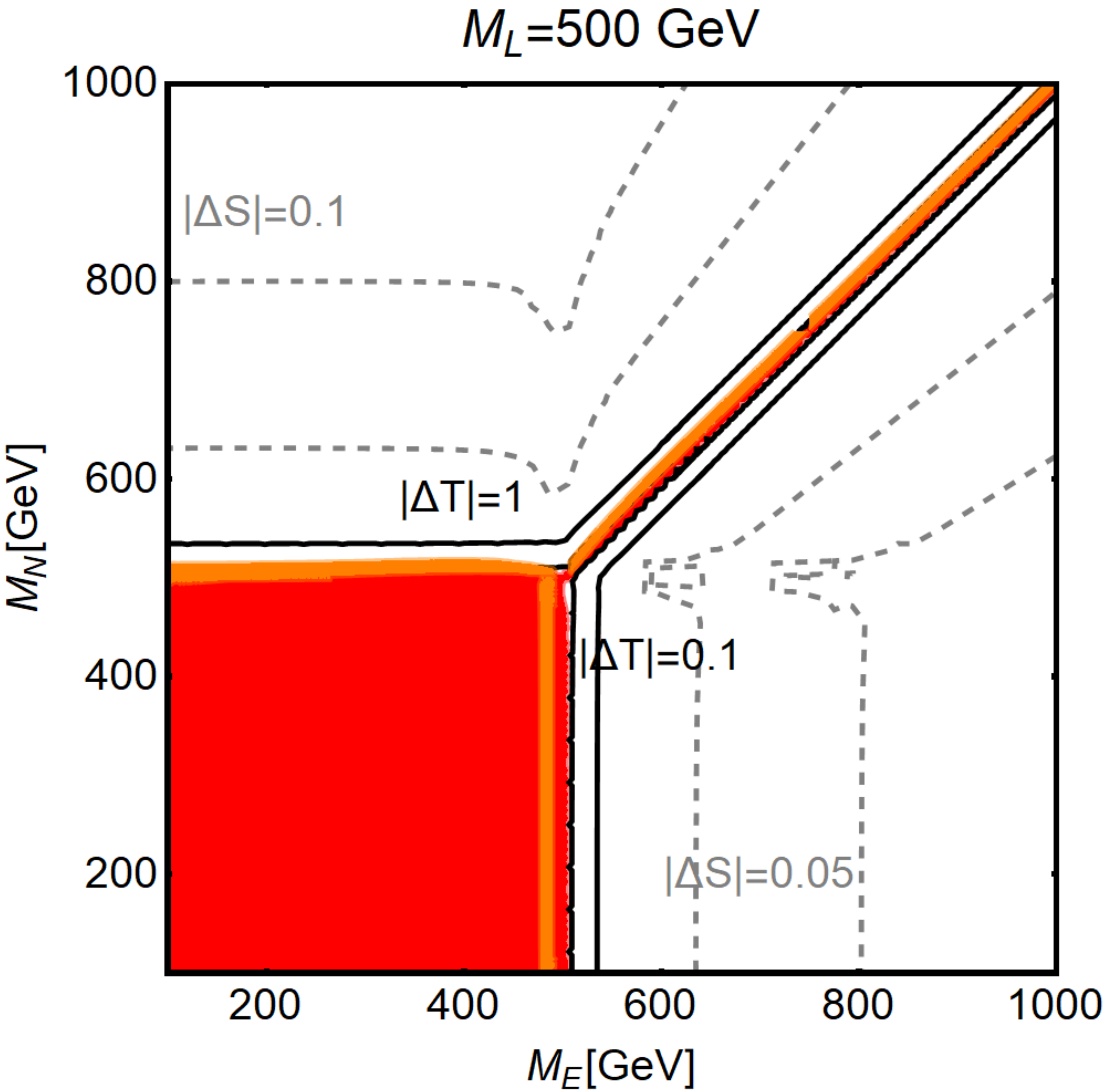}}~
\subfloat{\includegraphics[width=0.47\linewidth]{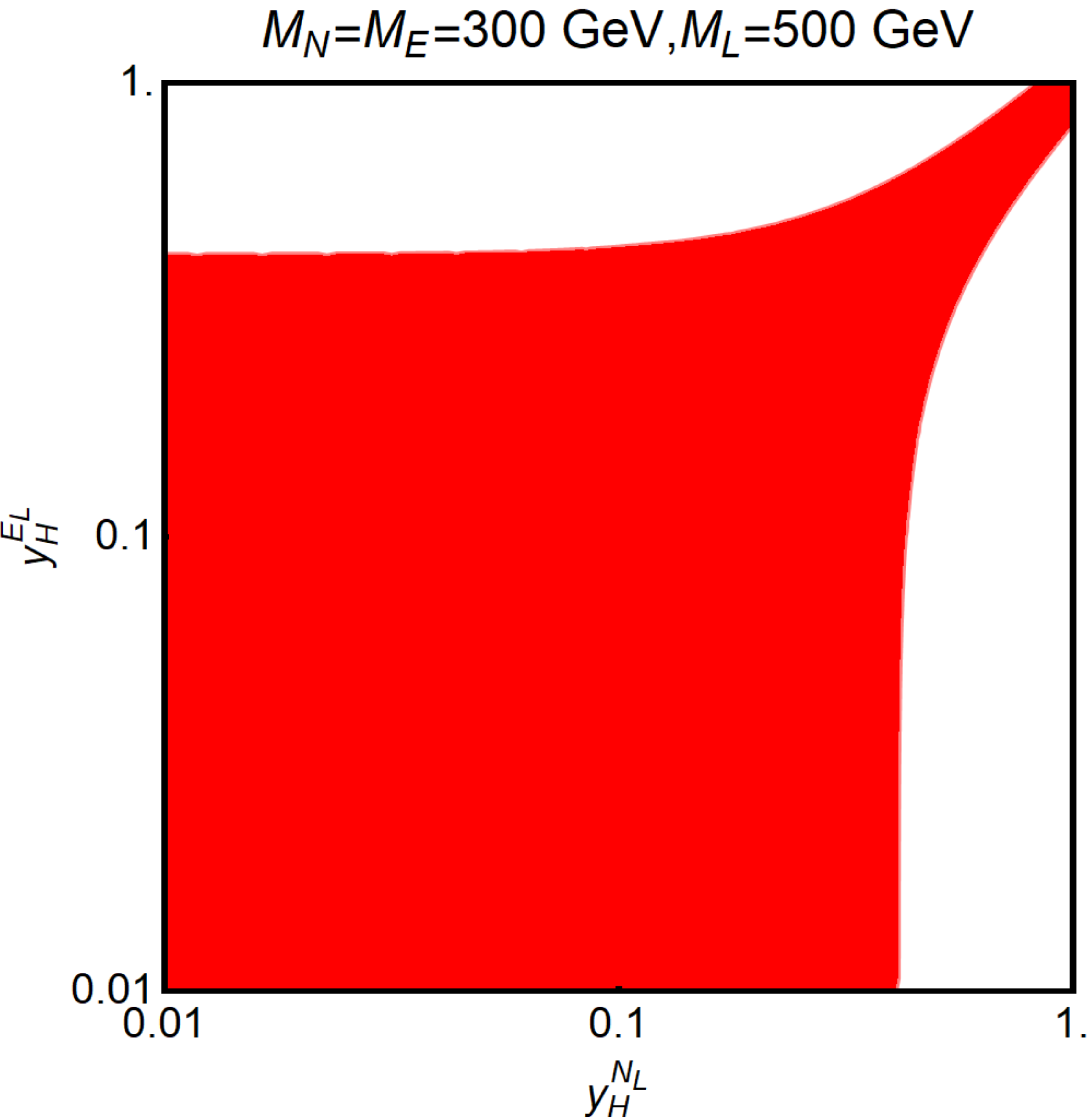}}
\end{center}
\vspace*{-3mm}
\caption{Allowed regions by electroweak precision data for the extension of the SM with a vector--like family. Left: constraints in the bidimensional plane $[M_E,M_N]$ for $M_L=500$ GeV with the orange and the orange+red areas standing for the allowed ones for Yukawa coupling assignments $(y_H^{N_L},y_H^{E_L}) =(0.01,0.5)$ and $(0.1,0.1)$;  isocontours for $\Delta S$ and $\Delta T$ with values reported on the plots are also shown.  Right: the constraints in the $[y_H^{N_L},y_H^{E_L}]$ plane for $M_N=M_E=300\,\mbox{GeV}$ and $M_L=500\,\mbox{GeV}$.}
\label{fig:VhLSTU1}
\vspace*{-3mm}
\end{figure}

The case of a full vector family is more complicated as a result of the larger
number of free parameters, i.e. three masses, $M_N,M_E,M_L$, and four Yukawa
couplings $y_H^{N_{L,R}}$ and $y_H^{E_{L,R}}$. In this scenario,  both the $S$
and $T$ parameters receive substantial contributions that can be  fully
described only if an extensive numerical analysis based on eq.~(\ref{eq:chi2})
is performed. In the left--hand side of   Fig.~\ref{fig:VhLSTU1}, we
nevertheless show two examples of the constraints that can be set on the
$[M_N,M_E]$ plane for $M_L=500\,\mbox{GeV}$, $y_H^{N_R}=y_H^{E_R}=0$ and two
assignments of the couple $(y_H^{N_L},y_H^{E_L})$, namely  $(0.1,0.1)$ and
$(0.01,0.5)$.

We have marked in red or orange the regions of the parameter space that lead to values of the precision observables that are less than three standard deviations from the  experimental measurements and, in order to facilitate the understanding of the figure, we have also drawn isocontours of the deviations $\Delta S$ and $\Delta T$ from the SM expectation. As it can be seen, the allowed regions are mostly determined by the constraint on the $T$ parameter which can be evaded only by imposing, at least partially, a custodial symmetry, 
\begin{equation}
    M_N=M_E,\,\,\,\,\,y_H^{N_L}=y_H^{E_L},\,\,\,\,y_H^{N_R}=y_H^{E_R} \, .
\end{equation}
This requirement is particularly severe at high values of the Yukawa couplings,
$y_H \gtrsim 0.1$. This is exemplified in the right--hand side of
Fig.~\ref{fig:VhLSTU1}, where we show the same electroweak constraints but in
the plane $[y_H^{N_L},y_H^{E_L}]$ again for $M_L=500$ GeV and fixing the other
lepton masses to $M_N=M_E=300$ GeV. Large Yukawa couplings close to unity  are
allowed only if $y_H^{N_L}=y_H^{E_L}$. Notice that in our analysis, we have from
the start assumed $y_H^{N_R}=y_H^{E_R} =0$.  Besides reducing the number of free
parameters, this choice ensures to comply with collider bounds on the Higgs
signal strengths,  as will be clarified below. 

\subsubsection{Constraints from the Higgs sector}

New fermions coupled with the Higgs boson and, at least partially, charged under
the SM gauge group, can generate possibly large modifications of the Higgs
couplings to SM fermions and gauge bosons which are tightly constrained by the
measurement of the Higgs signal strengths at the LHC. Among the two SM 
extensions considered above, only the one with a full family of vector like
fermions is phenomenologically relevant in this regard. In this last case, given
the absence of mixing between the SM and the new fermions, only the couplings of
the Higgs with the SM gauge bosons are affected. If the SM is extended by only
vector--like leptons, the effective coupling of the Higgs boson with two photons
is affected in the most significant way. In the case where vector--like quarks
are also present, modifications of the effective coupling with gluons are
relevant as well. We will thus focus our discussion on these two couplings.  

\begin{figure}[!h]
\vspace*{-.3mm}
    \centering
    \subfloat{\includegraphics[width=0.47\linewidth]{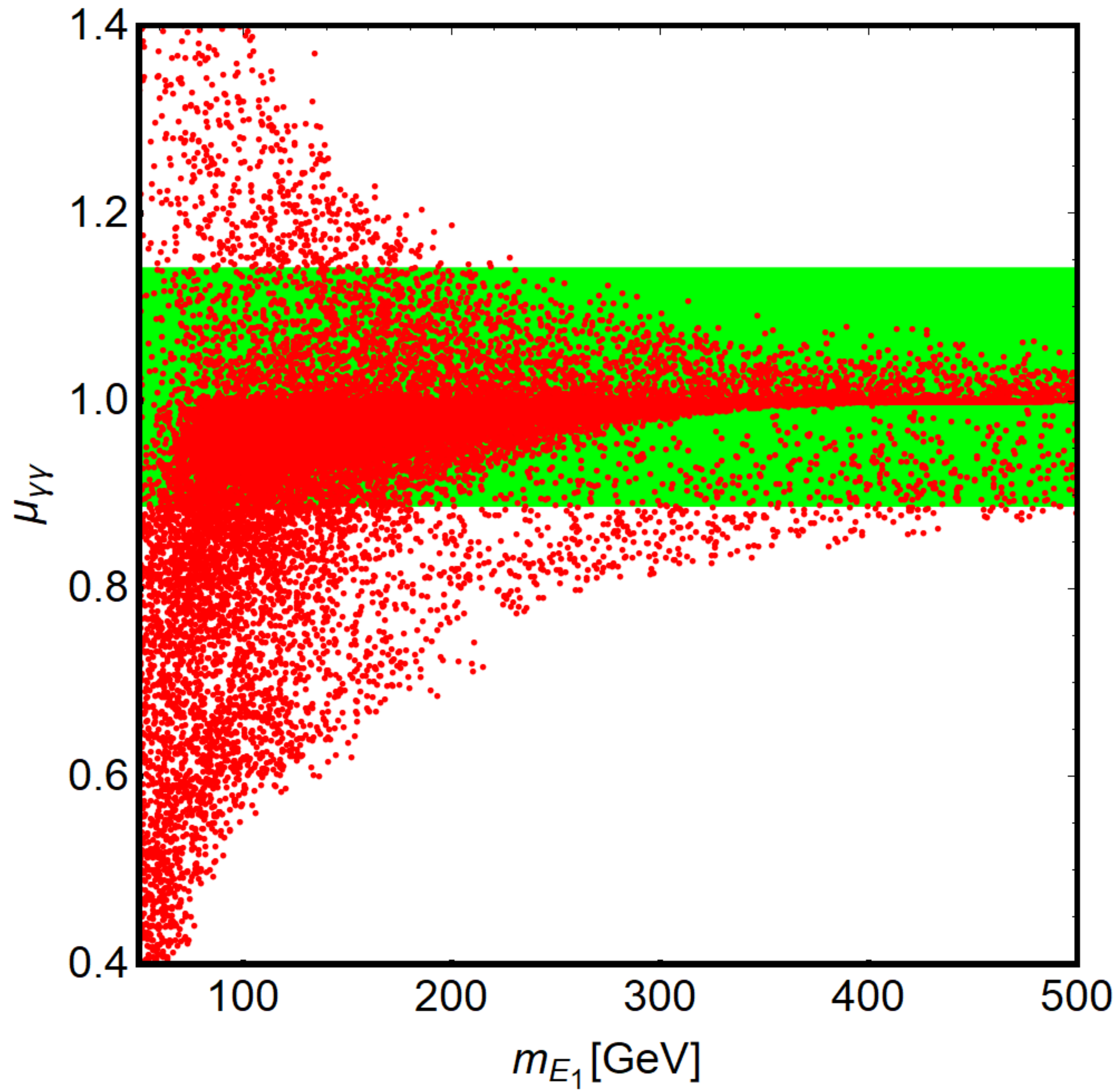}}~
    \subfloat{\includegraphics[width=0.47\linewidth]{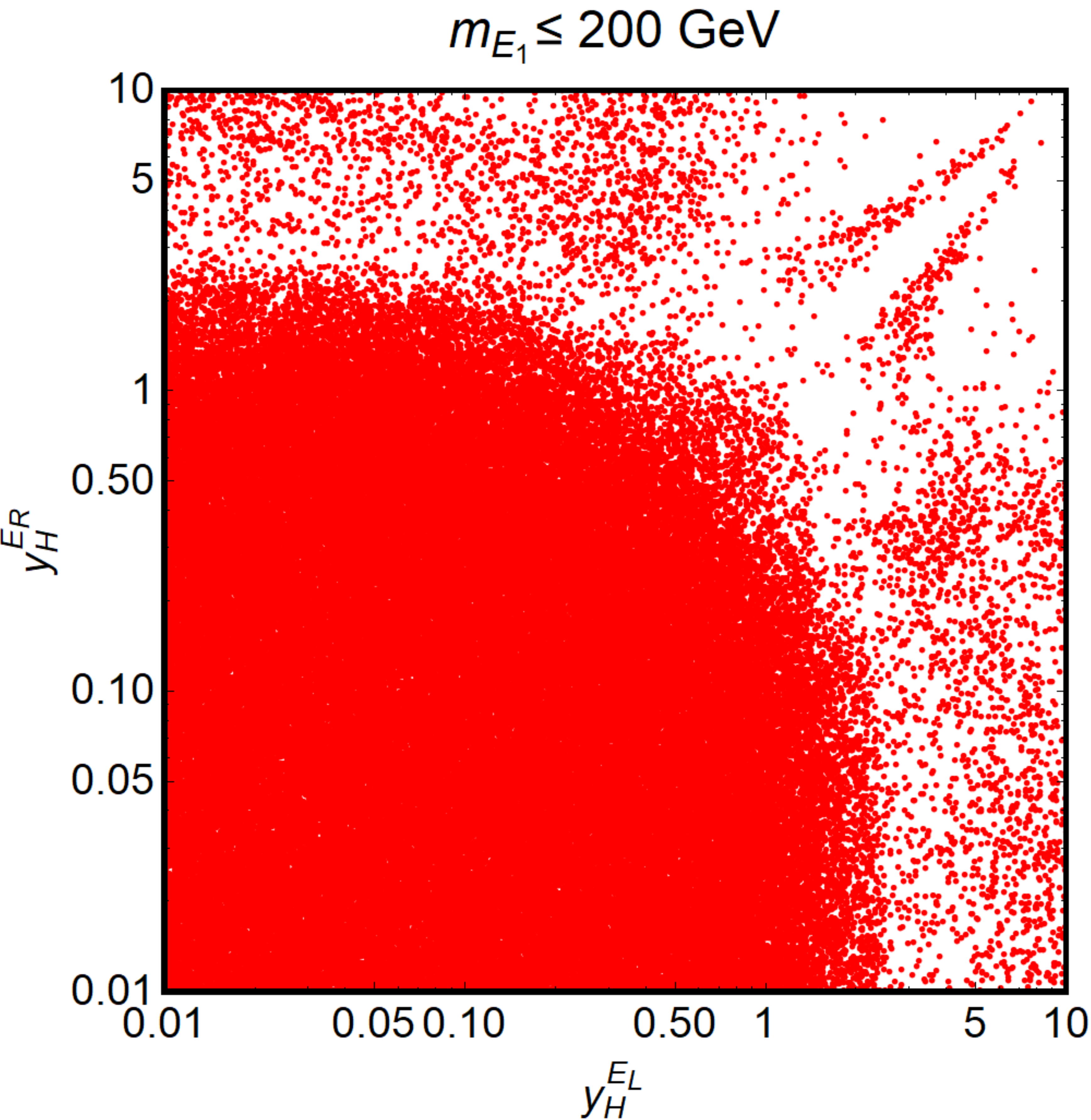}}
    \caption{Left: the Higgs diphoton signal strength $\mu_{\gamma \gamma}$ as a function of the mass of the lightest charged vector--like lepton when a scan over the other parameters of the model is performed; the green region represents the 68\%CL boundary allowed by experimental fits of $\mu_{\gamma \gamma}$. Right: model points with $\mu_{\gamma \gamma}$ complying with present constraints in the $[y_H^{E_L},y_H^{E_R}]$ plane with only $m_{E_1}\leq 200\,\mbox{GeV}$ values considered.}
    \label{fig:SMmu}
\vspace*{-1mm}
\end{figure}

In presence of new leptons, the decay amplitude of the Higgs into diphotons is given by
\begin{equation}
    \mathcal{A}^{H\gamma \gamma}=\mathcal{A}^{H \gamma \gamma}_{\rm SM}+\mathcal{A}^{H \gamma \gamma}_{\rm NP} \, , 
\end{equation}
where $\mathcal{A}_{\rm SM}^{h \gamma \gamma},\mathcal{A}_{\rm NP}^{h \gamma \gamma}$ are the SM and New Physics contributions, respectively. The latter can be schematically written as
\begin{equation}
{\cal A}^{H\gamma\gamma}_{\rm NP} =A_{1/2}^H(0) \sum\limits_F {\rm sign} (f_F)\left|\frac{y_H^{E_L} v}{m_{E_1}} \frac{y_H^{E_R} v}{m_{E_2}} \right| \, , 
\label{SM-hpp1}
\end{equation}
where $A^H_{1/2}$ is the loop function for spin--$\frac12$ particles 
given in Appendix A1,  for which we have take for simplicity  the asymptotic limit $A^H_{1/2}(0)=4/3$,  which is a good approximation for $M_F \gsim 100$ GeV as it should be the case here. Notice that the sign of $\mathcal{A}_{\rm NP}^{H \gamma \gamma}$ relative to $\mathcal{A}_{\rm SM}^{H \gamma \gamma}$ is not fixed but determined by the sign of the function $f_F$ defined as
\begin{equation}
f_F(y_H^{E_L},y_H^{E_R},M_E,M_L)= \frac{-4 y_H^{E_L} y_H^{E_R} v}{2 M_E M_L - y_H^{E_L} y_H^{E_R} v^2} \, . 
\label{prescription}
\end{equation}
In absence of modification of the effective Higgs coupling to gluons, the Higgs signal strength can be written in this case as
\begin{equation}
    \mu_{\gamma \gamma}=\frac{\sigma_{\rm NP}^H \times {\rm BR}(H \rightarrow \gamma \gamma)_{\rm NP} } {\sigma_{\rm SM}^H \times {\rm BR}(H \rightarrow \gamma \gamma)_{\rm SM}} \simeq \frac{|\mathcal{A}_{\rm SM}^{H\gamma \gamma}+\mathcal{A}_{\rm NP}^{H\gamma \gamma}|^2}{|\mathcal{A}_{\rm SM}^{H \gamma \gamma}|^2} \, . 
\end{equation}
As it should be clear from eq.~(\ref{SM-hpp1}), the New Physics contribution to the Higgs decay amplitude into diphotons is proportional to the product $y_H^{E_L}y_H^{E_R}$ and, consequently, vanishes if one of these couplings is zero. The contribution also decreases with the masses of the charged vector leptons. In order to asses the impact of the contribution of the new sector to the diphoton Higgs signal strength $\mu_{\gamma \gamma}$, we have performed a scan of the parameters $M_E,M_L,y_H^{E_L},y_H^{E_R}$ over the following ranges
\begin{equation}
    M_{E,L} \in \left[100,500\right]\,\mbox{GeV},\,\,\,\,\,y_H^{E_{L,R}} \in \left[10^{-3},10\right] \, . 
\end{equation}
The result have been confronted to the experimental determination  $\mu_{\gamma \gamma} = 1.14 ^{+0.19}_{-0.18}$ and shown in the left panel of Fig.~\ref{fig:SMmu} as a function of $M_{E_1}$. One can see that $\mu_{\gamma \gamma}$ deviates with respect to the measured value mostly for masses of the lightest charged vector lepton $m_{E_1}$ below 200 GeV. New Physics effects tend instead to rapidly decouple at higher masses.

We have then focussed on the range $m_{E_1} \leq 200\,\mbox{GeV}$ and we show in
the right panel of Fig.~\ref{fig:SMmu} the points of the $[y_H^{E_L},y_H^{E_R}]$
plane which lead to a $\mu_{\gamma \gamma}$ value compatible with the LHC
constraint.  One sees that most of the viable model points can have
$y_H^{E_{L,R}}$ up to 3, provided that the other Yukawa coupling is very small,
below $10^{-2}$. Nevertheless, there is a limited set of solutions with even
higher values of the Yukawa couplings: it is indeed possible to have a viable
$\mu_{\gamma \gamma}$ for high Yukawa couplings in the case in which a
cancellation occurs between the SM and new contributions to the Higgs diphoton
rate, such that $\mathcal{A}_{\rm NP}^{H \gamma \gamma} \simeq -2
\mathcal{A}_{\rm NP}^{H \gamma \gamma}$. By using eq.~(\ref{SM-hpp1}) and
assuming for simplicity $y_H^{E_L}=y_H^{E_R}=y_H$ and $m_{E_1}=m_{E_2}$, the
latter requirement can be translated into the following simple equation
\cite{Angelescu:2015uiz}
\begin{equation} 
{\cal A}^{H\gamma\gamma}_{\rm NP}\simeq \frac{4}{3} \left(\frac{y_H v}{m_E}
\right)^2  \simeq -2 {\cal A}^{H \gamma\gamma}_{\rm SM} \simeq 13~.
\label{cancellation-SM2}  
\end{equation}

\subsubsection{Prospects for heavy leptons at colliders}

We now discuss the prospects for producing the new leptons at high-energy
colliders (some elements have been touched upon in the previous section). First,
because they couple to gauge bosons with full strength, heavy non--singlet
leptons can be pair produced in proton--proton collisions
\cite{Panella:2001wq,Han:2006ip,delAguila:2007qnc,delAguila:2008hw,delAguila:2008cj,Deppisch:2015qwa} in the  Drell--Yan process $q \bar q \to V^* \to L\bar L$,
eq.~(\ref{Drell-Yan}). The cross
section will only depend on the $L$ electric charge and weak isospin. In the
case of an electrically--charged $E$ state, both the $\gamma$ and $Z$ boson
channels and their interference have to be included, while only the channel with
$Z$ boson exchange has to be considered for a neutral lepton $N$ (which stands now generically for all heavy neutral leptons) with
electroweak couplings.  The cross sections for pairs of charged or neutral leptons $L=E,N$ with velocities  $\beta_L= (1-4m_L^2/ \hat s)^{1/2}$ at 
partonic c.m. energy $\hat s$ simply read
\beq
\hat \sigma (q\bar q \to L \bar L) = \frac{2 \pi \alpha^2}{9 \hat s} \beta_L (3-\beta_L^2)  \left[e_q^2 e_L^2 + \frac{2 e_q e_L v_q v_L}{1-M_Z^2/\hat s} + \frac{(a_q^2+ v_q^2) (a_L^2+ v_L^2)}{(1-M_Z^2/ \hat s)^2} \right] \, ,
\eeq
where the electron and heavy lepton couplings are given in eq.~(\ref{Zffcouplings}).  In addition, one could produce pairs of charged and neutral leptons via $W$ boson exchange, $q\bar q ' \to W^{*\pm} \to E^\pm N$. For comparable masses, $m_E\! \approx\! m_N\! = \! m_L$,  the partonic cross section  is given by 
\beq
\hat \sigma (q\bar q \to E^- \bar N + E^+ N ) = \frac{4 \pi \alpha^2}{9 \hat s} \frac{\beta_L (3-\beta_L^2) }{(1-M_W^2/ \hat s)^2} \times \frac{1}{8 s_W^4}\, .
\eeq

\begin{figure}[!h]
\vspace*{-3.4cm}
\centerline{
\includegraphics[width=17.5cm]{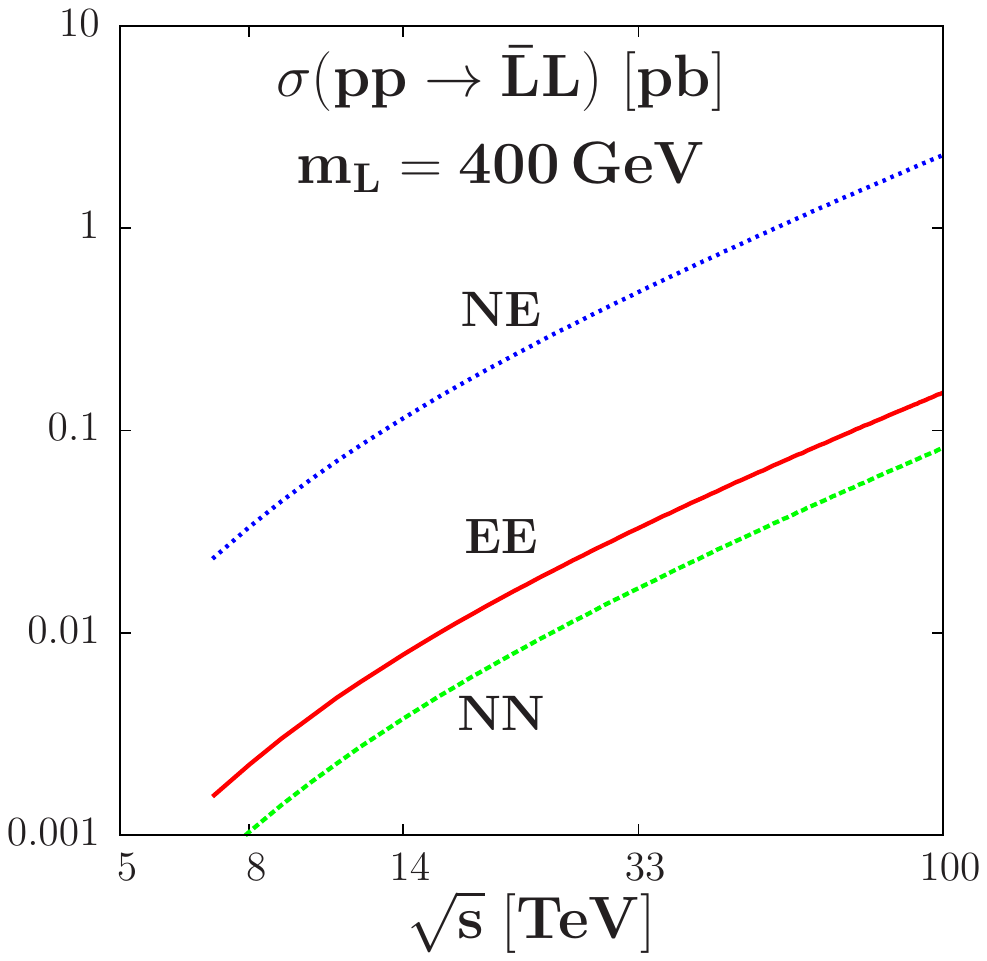}
}
\vspace*{-14.cm}
\caption{Cross sections for the pair production of heavy leptons in $pp$ collisions as a function of $\sqrt s$ for a mass $m_L=400$ GeV \cite{Djouadi:2016eyy}.}
\label{Fig:pp-VLL}
\vspace*{-1mm}
\end{figure}

The cross sections for producing pairs of charged and neutral vector leptons
with masses $m_L=400$ GeV are shown in Fig.~\ref{Fig:pp-VLL} at a proton
collider as a function of the total energy $\sqrt s$ \cite{Djouadi:2016eyy}.  The heavy leptons considered have the assignments  for electric charge and weak isospin $E (-1,-\frac12)$ and $N (0, +\frac12)$. One notices that the rates are much smaller for the neutral processes with $(\gamma)Z$ exchanges than for the charged one with $W$  exchange: the cross sections  for $EE,NN$ production are comparable and are only at the  fb level at RunI LHC while they are a
factor 20 larger for $NE$ production. The latter  process is thus the best probe
of heavy leptons in pair production. Note also that  the rates increase by
two orders of magnitude when moving from a $\sqrt s\!=\!8$ TeV to a $\sqrt s\!=\! 100$ TeV collider.

\begin{figure}[!h]
\vspace*{-2.6cm}
\centerline{\hspace*{-1.5cm}
\includegraphics[width=17cm]{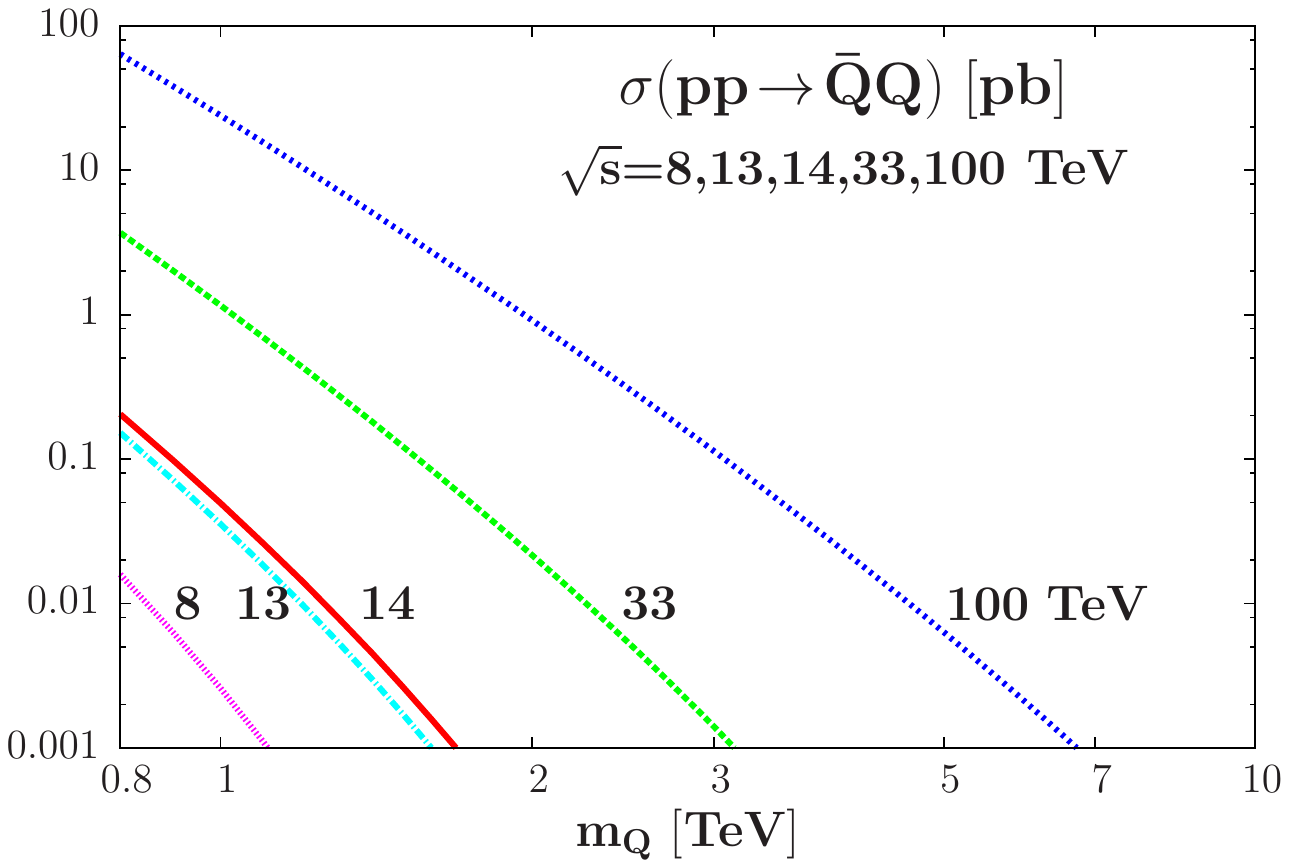} 
}
\vspace*{-14.3cm}
\caption{The production cross sections in $pp$ collisions of vector--like quark pairs  as functions of the  mass for several collider energies \cite{Djouadi:2016eyy}.}
\label{Fig:pp-VQQ}
\vspace*{-2mm}
\end{figure}

In the case of an entire vector--like family, some quarks should also accompany
these leptons and their production at hadron colliders is more favorable.
Indeed, as they couple to gluons like SM quarks, they can be pair produced in
the strong interaction process $pp \to Q \bar Q$ with rates that depend only on
the mass $m_Q$ and the strong coupling constant $\alpha_s$ (single production
with a SM quark is suppressed by the tiny or null mixing angle)
\cite{Aguilar-Saavedra:2013qpa,Ellis:2014dza,Djouadi:2016eyy}.  The total
hadronic cross section, i.e., after folding with the parton luminosities (which
are taken here to be those of the MSTW2008 fit~\cite{Martin:2009iq}),  is shown
in Fig.~\ref{Fig:pp-VQQ} as a function of the heavy quark mass for several
center of mass energies \cite{Djouadi:2016eyy}. For $m_Q=1$ TeV,  the cross
section is at the few fb level at $\sqrt s=8$ TeV and increases by more than 
one order of magnitude at $\sqrt s=13$ or 14 TeV and four orders of magnitude
at  $\sqrt s=100$ TeV. For higher quark masses, the increase of the rate with
energy is even steeper, highlighting the advantage of a higher energy proton
collider in this context.

\begin{figure}[!h]
\vspace*{-2.4cm}
\centerline{ \includegraphics[scale=0.83]{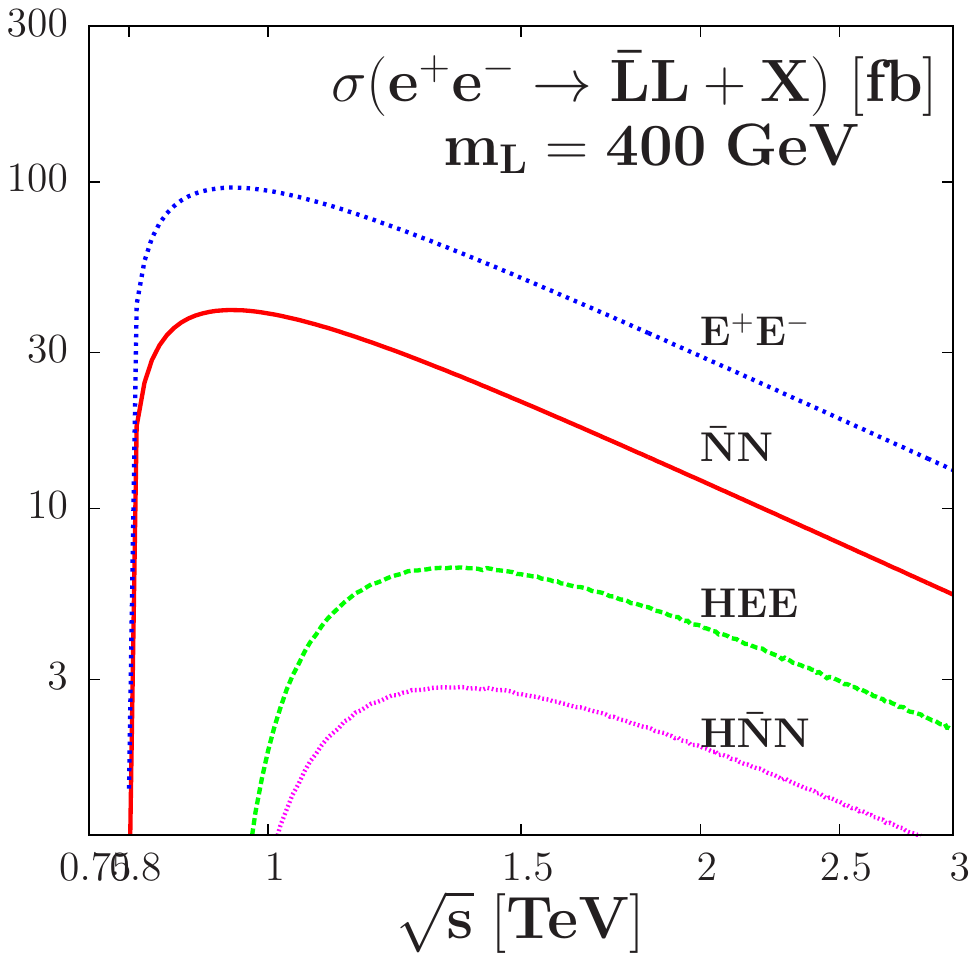} }
\vspace*{-13.9cm}
\caption{The cross sections for the pair production of heavy vector--like charged and neutral leptons at an $e^+ e^-$ collider as a function of the c.m. energy $\sqrt s$ for $m_E\! =\! m_N \!= \! 400$ GeV.  Also shown are the cross sections for associated production of the charged and neutral leptons with the SM Higgs boson, $e^+ e^- \to H \bar LL$, for masses $m_L=400$ GeV and Yukawa couplings $y_L=m_L/v$ again as a function of $\sqrt s$.}
\label{Fig:eeVLL}
\vspace*{-2mm}
\end{figure}

Before closing this section, we should briefly comment on the production of the heavy leptons at future high--energy $e^+e^-$ colliders. There is first lepton pair production just like in the Drell--Yan process at hadron colliders,   $e^+ e^- \to \bar L L$, which is kinematically possible for masses $m_{L} \lsim \frac12 \sqrt s$. As in pp collisions, the total pair production cross section is given in terms of the usual couplings to the photon and $Z$ boson by \cite{Djouadi:1993pe,Azuelos:1993qu,Buchmuller:1991tu}
\beq
\sigma (e^+ e^- \to L \bar L) = \frac{4\pi \alpha^2}{3s}  \frac{\beta_L (3-\beta_L^2)}{2}  \left[e_e^2 e_L^2 + \frac{2 e_ee_L v_e v_L}{1-M_Z^2/s} + \frac{(a_e^2+ v_e^2) (a_L^2+ v_L^2)}{(1-M_Z^2/s)^2} \right] \, .
\label{eq:eeLL}
\eeq
The cross sections for  pair producing  $E,N$ that are part of the same
isodoublet are shown in Fig.~\ref{Fig:eeVLL} as functions of the energy
$\sqrt s$ for the masses $m_E = m_N = 400$ GeV. Again, the rate is much smaller
for  the neutral $N$ compared to the charged $E$ leptons,  as the process 
proceeds  only through $Z$ boson exchange for the former but the rates are
still large, above 10 fb in the chosen example, to easily detect the particles
in the clean $e^+ e^-$ environment. Of course, the production rates are higher
for lighter leptons, i.e with masses that are closer to the present limit of
$m_L \approx 100$ GeV.    

Another process of interest in $e^+ e^-$ collisions  is  associated production
of the lepton pair with the SM Higgs boson, $e^+ e^- \to H \bar L L$. This
process is similar to associated Higgs production with top quark pairs, $e^+ e^-
\to  H t\bar t$  \cite{Djouadi:1991tk,Djouadi:1992gp}, which has been discussed
in the context of the SM. Compared to the latter,  the cross sections  should be
smaller as a result of the absence of the color factor and the presumably
smaller Yukawa couplings but this might be partly compensated by a more
favorable phase space, since one can consider the associated production with the
lightest neutral fermion, i.e. the DM particle, which has a mass as low as
$\approx 70$ GeV from present constraints.  The signature in this case would be
a mono--Higgs topology since the neutral heavy leptons are invisible. The 
charged leptons which are  expected to be heavier, might have larger rates as a
result of the additional photon exchange contribution and would lead to events
with Higgs bosons, light SM fermions from the $E$ decays into on-- or off--shell
$W$ bosons and missing energy.  

Also in Fig.~\ref{Fig:eeVLL}, we have displayed the cross sections for the 
process $e^+ e^- \to H \bar LL$ with the leptons being $L=E,N$ with masses
$m_L=400$ GeV and with Yukawa couplings that are similar to the SM--like ones, 
$y_L=m_L/v$. For this set of parameters, the rates are still  significant, being
only one order of magnitude smaller than those for lepton pair production and follow
exactly the same  trend with phase--space.    For $HLL$ couplings compatible
with LHC Higgs data, the rates should be much smaller though.

Finally, concerning the eventual heavy quark partners that would appear when a
full vector--like fermion family is considered, one would need a very high
energy  collider such as the CLIC machine in view of the bounds on the masses of
these particles (in particular if they decay inside the detectors into light
quarks and gauge or Higgs bosons). Even in that case, the cross sections are not
that large as the production occurs through $s$--channel $\gamma,Z$ boson
exchange which, as in the case of leptons eq.~(\ref{eq:eeLL}), are suppressed
like $1/s$ at high center of mass energies. The best place for these hadronic
states is thus definitely at proton machines as shown before.   

\subsection{Constraints from astroparticle physics}

\subsubsection{Constraints in the singlet--doublet model}

For what concerns the DM phenomenology from the astroparticle physics
perspective, the singlet--doublet scenario presents sizeable differences with
respect to the effective fermionic portal discussed in the previous section, a
feature primarily due to the DM couplings with the  gauge bosons which are
absent in the latter scenario. As the DM state is a Majorana fermion, no new
spin--independent interactions are present since they  require a vectorial
coupling between the DM and the $Z$ boson which is automatically 
zero in the Majorana case~\cite{Arcadi:2014lta}. As a consequence, the spin--independent scattering cross section of the DM on protons is again given by an expression of the type
\begin{equation}
\sigma_{N_1 p}^{\rm SI}=\frac{\mu_{N_1 p}^2}{\pi M_H^4}|g_{HN_1 N_1}|^2\frac{m_p^2}{v^2}
{\left[f_p \frac{Z}{A}+f_n \left(1-\frac{Z}{A}\right)\right]}^2,
\end{equation}
where the Higgs to DM coupling, eq.~(\ref{eq:Hpsi1}), is explicitly given by
\begin{equation} 
g_{HN_1 N_1}=-\frac{y^2 v \left(m_{N_1} +M_L \sin 2\theta\right)}{M_L^2
+ 2 M_L m_{N_1}-3 m_{N_1}^2  +y^2 v^2/2} 
\end{equation}
and, as can be seen, it has a rather complicated structure and depends on many
different parameters. In particular, it becomes zero in the case where the
condition $m_{N_1}+M_L \sin2\theta=0$ is met. Spin--independent interactions
would then encounter a so--called ``blind
spot''~\cite{Choudhury:2015lha,Choudhury:2017lxb} since there will be no Higgs exchange channel there.

Contrary to the case of the effective Higgs--portal model, spin--dependent 
interactions are present in the singlet--doublet model as a result of the axial
coupling of the DM with the $Z$ boson. The corresponding cross section is given
by
\begin{equation}
\sigma_{N_1 p}^{\rm SD}=\frac{\mu_{N_1 p}^2}{\pi M_Z^4}|g_{ZN_1 N_1}^A|^2 {\left[A_u^{Z} \Delta_u^p+ A_d^Z \left(\Delta_d^p+\Delta_s^p\right)\right]}^2 \, ,
\end{equation}
and it can also vanish if the coupling $g_{ZN_1 N_1}^A=0$ which, as  can be seen from eq.~(\ref{eq:Zpsi1}), occurs for the configuration $|U_{12}|^2=|U_{13}|^2$.

\begin{figure}[!h]
\vspace*{-3mm}
\begin{center}
\subfloat{\includegraphics[width=0.45\linewidth]{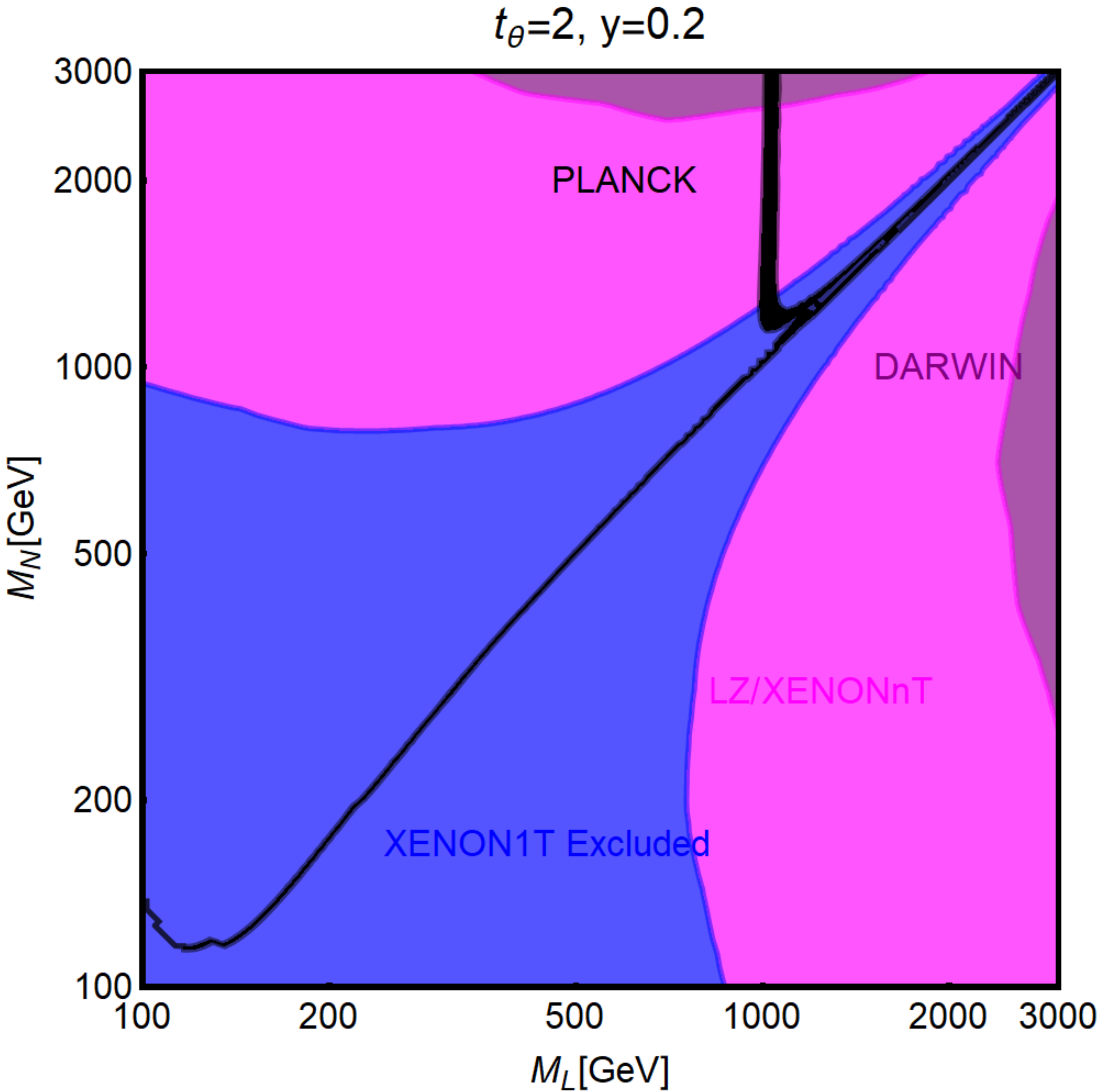}}~~~~
\subfloat{\includegraphics[width=0.45\linewidth]{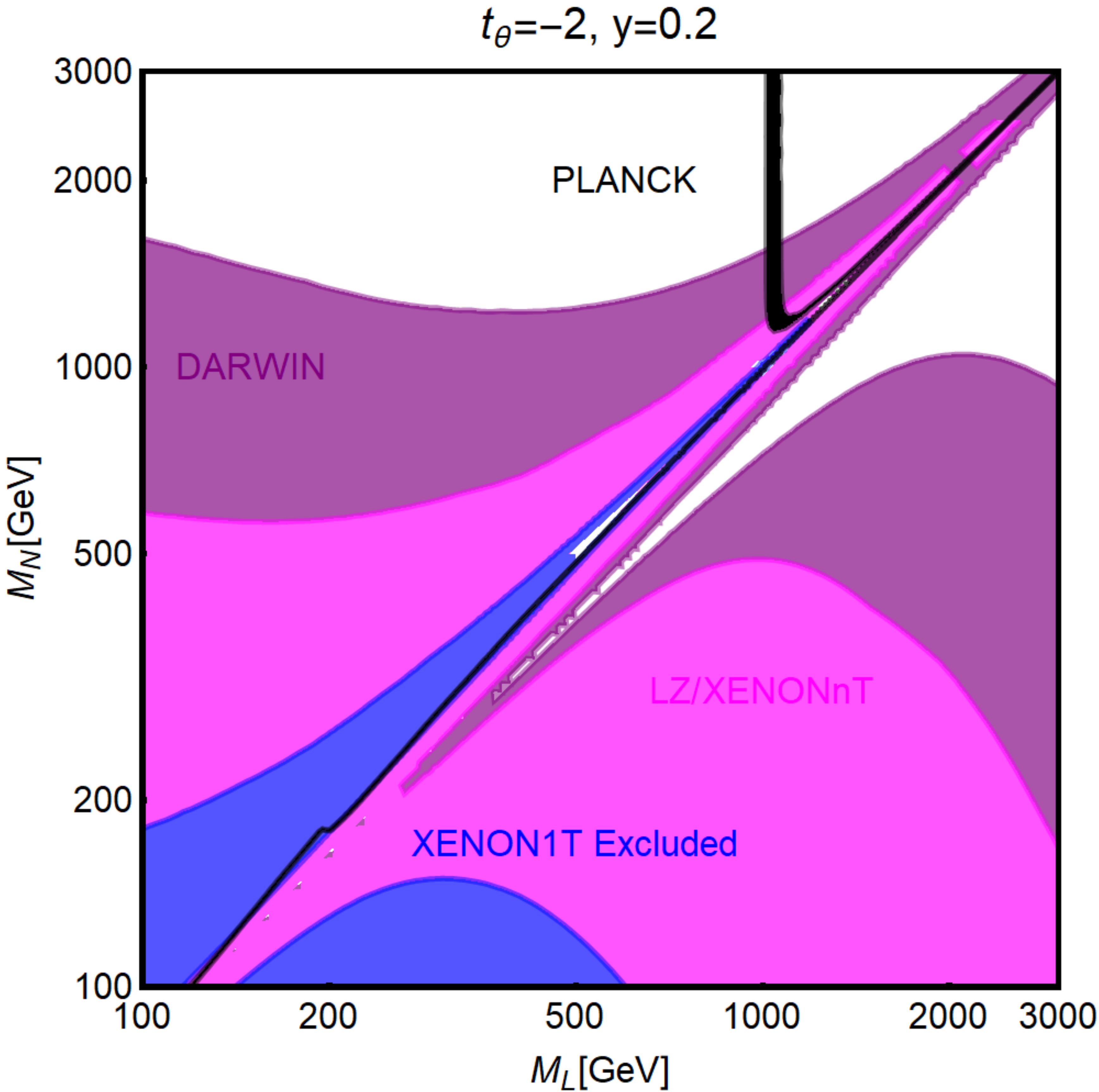}}\\
\subfloat{\includegraphics[width=0.45\linewidth]{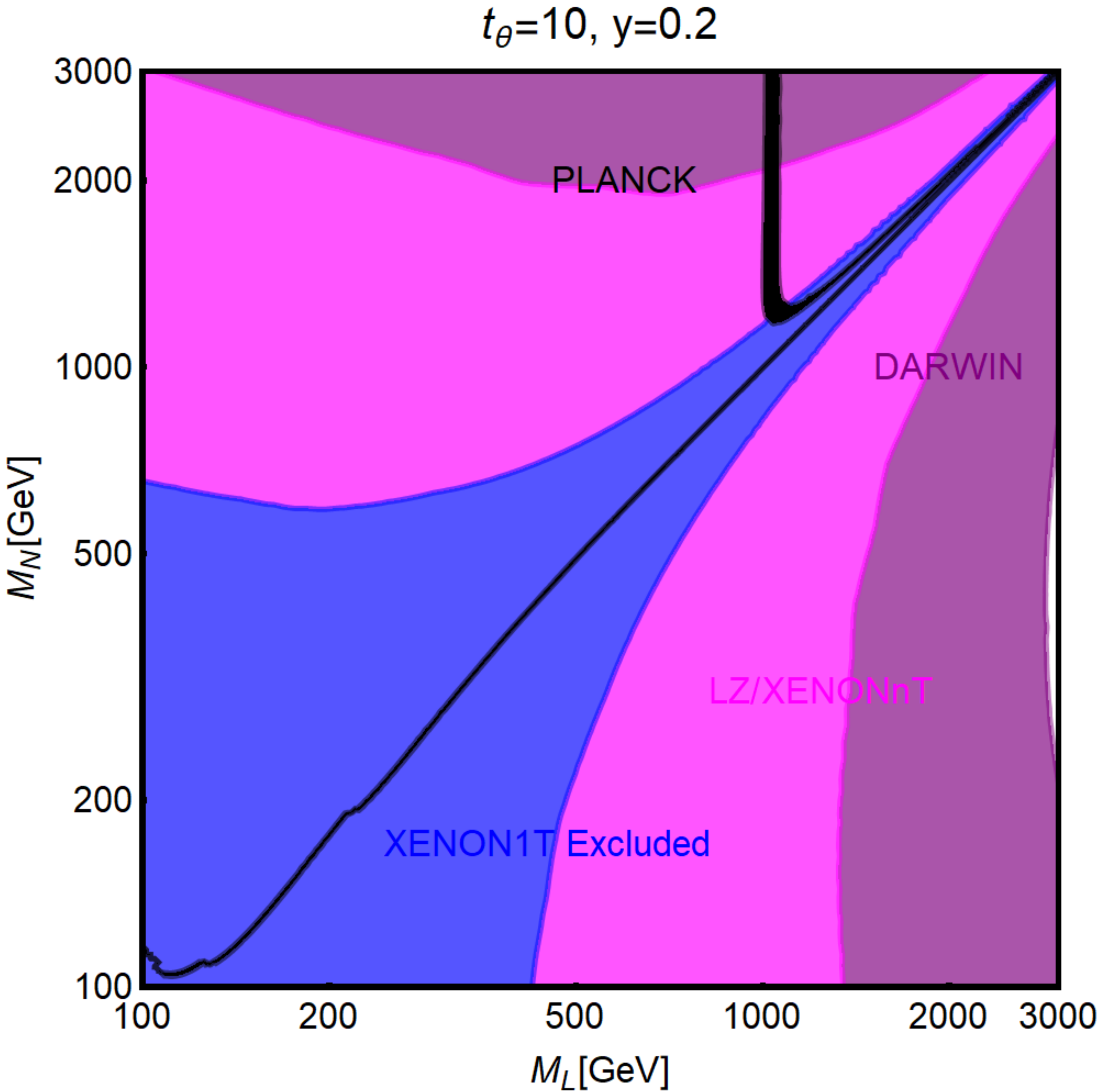}}~~~~
\subfloat{\includegraphics[width=0.45\linewidth]{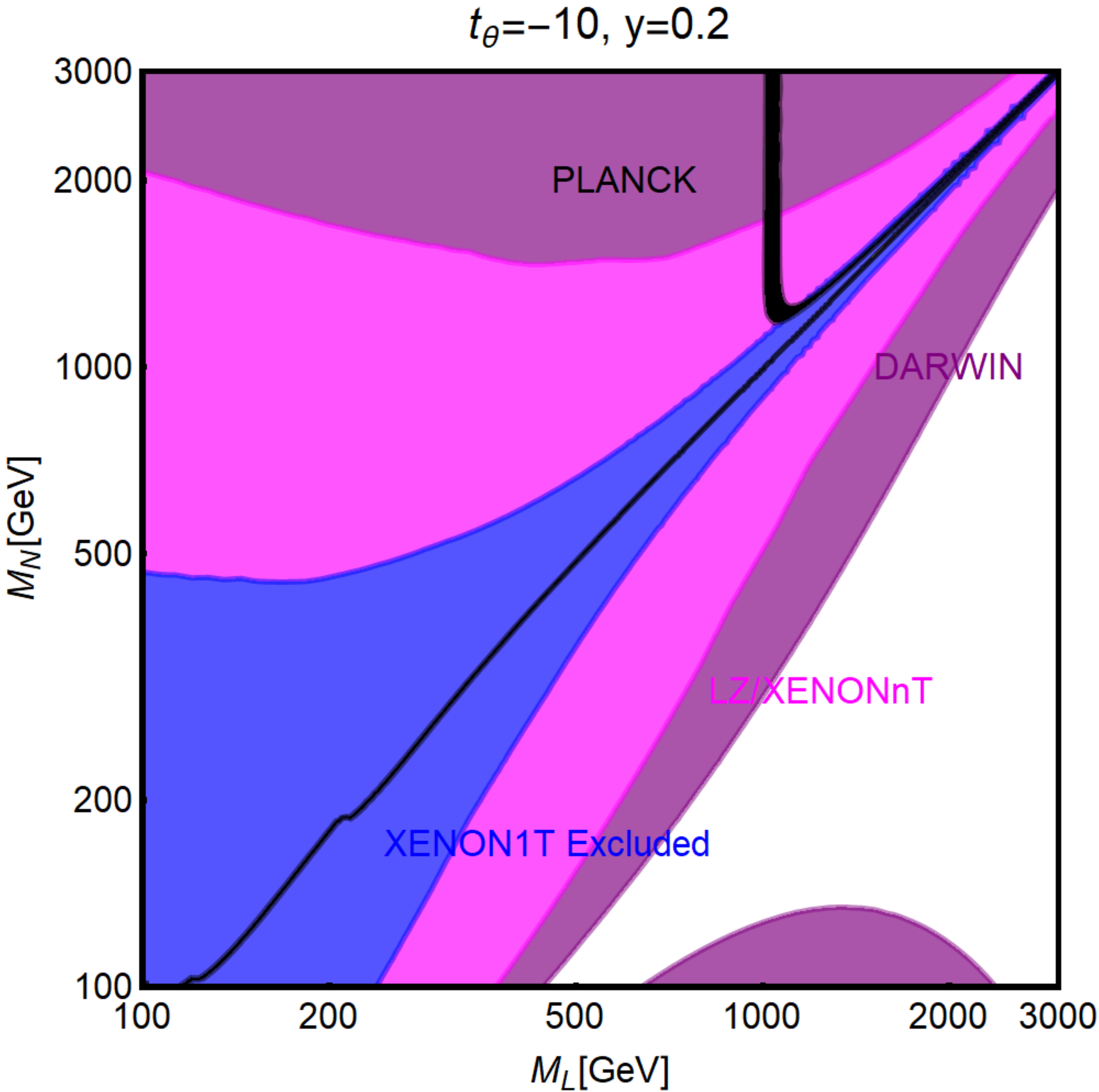}}
\end{center}
\vspace*{-5mm}
\caption{Summary of the constrains for the singlet--doublet model with  a
Majorana fermion DM, in the plane  $[M_L,M_N]$ for $y=0.2$ and $\tan\theta=2$
(top left),  $\tan\theta=-2$ (top right),  $\tan \theta=10$ (bottom left), 
$\tan\theta=-10$ (bottom right). The black isocontours correspond to the correct DM relic density. The colored blue, magenta and purple regions represent the current constraints and future sensitivities of direct detection
experiments.} \label{fig:SD_y02}
\vspace*{-3mm}
\end{figure}

\begin{figure}[!h]
\begin{center}
\vspace*{-1mm}
\subfloat{\includegraphics[width=0.46\linewidth]{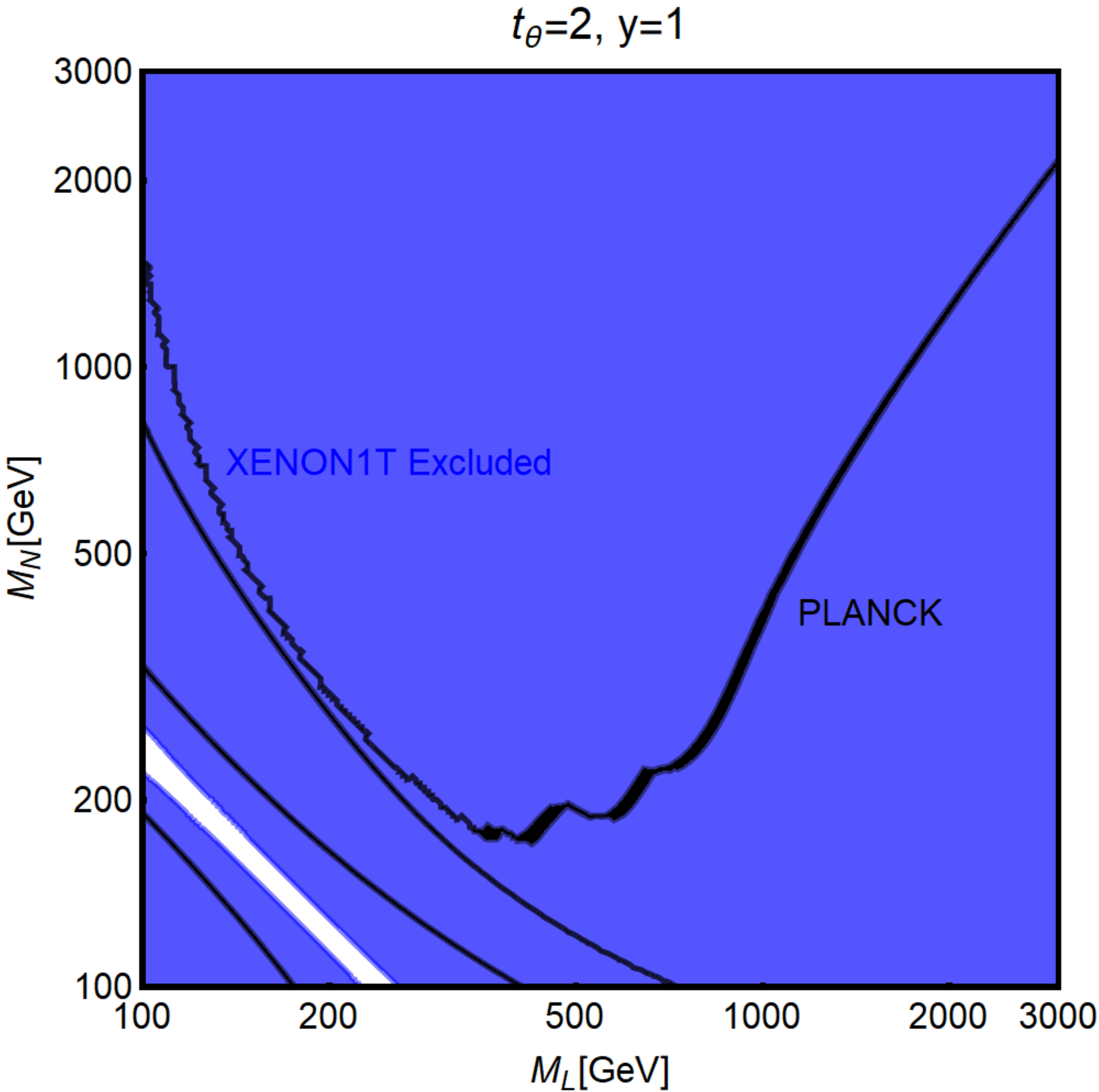}}~~~~
\subfloat{\includegraphics[width=0.46\linewidth]{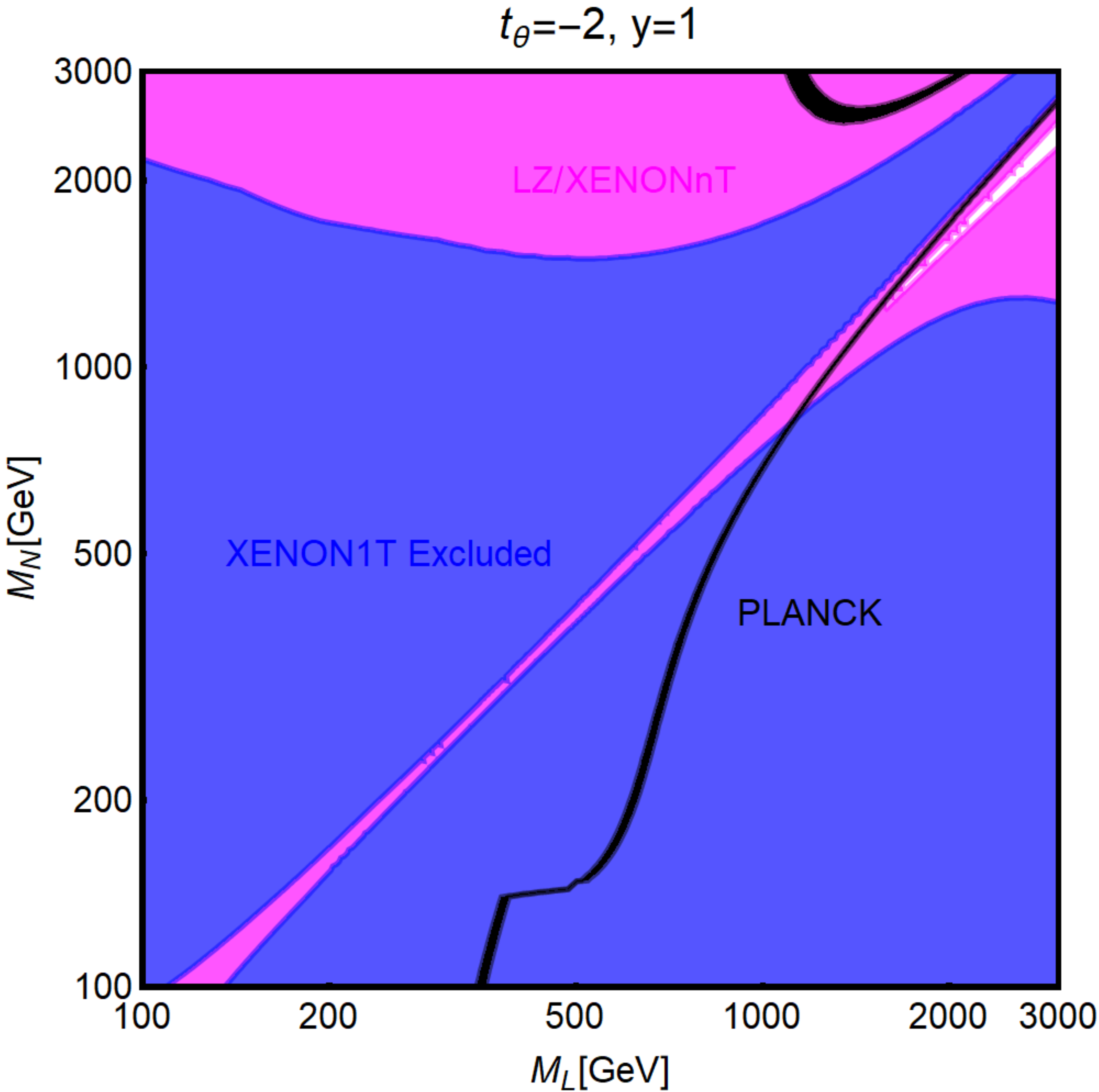}}\\
\subfloat{\includegraphics[width=0.46\linewidth]{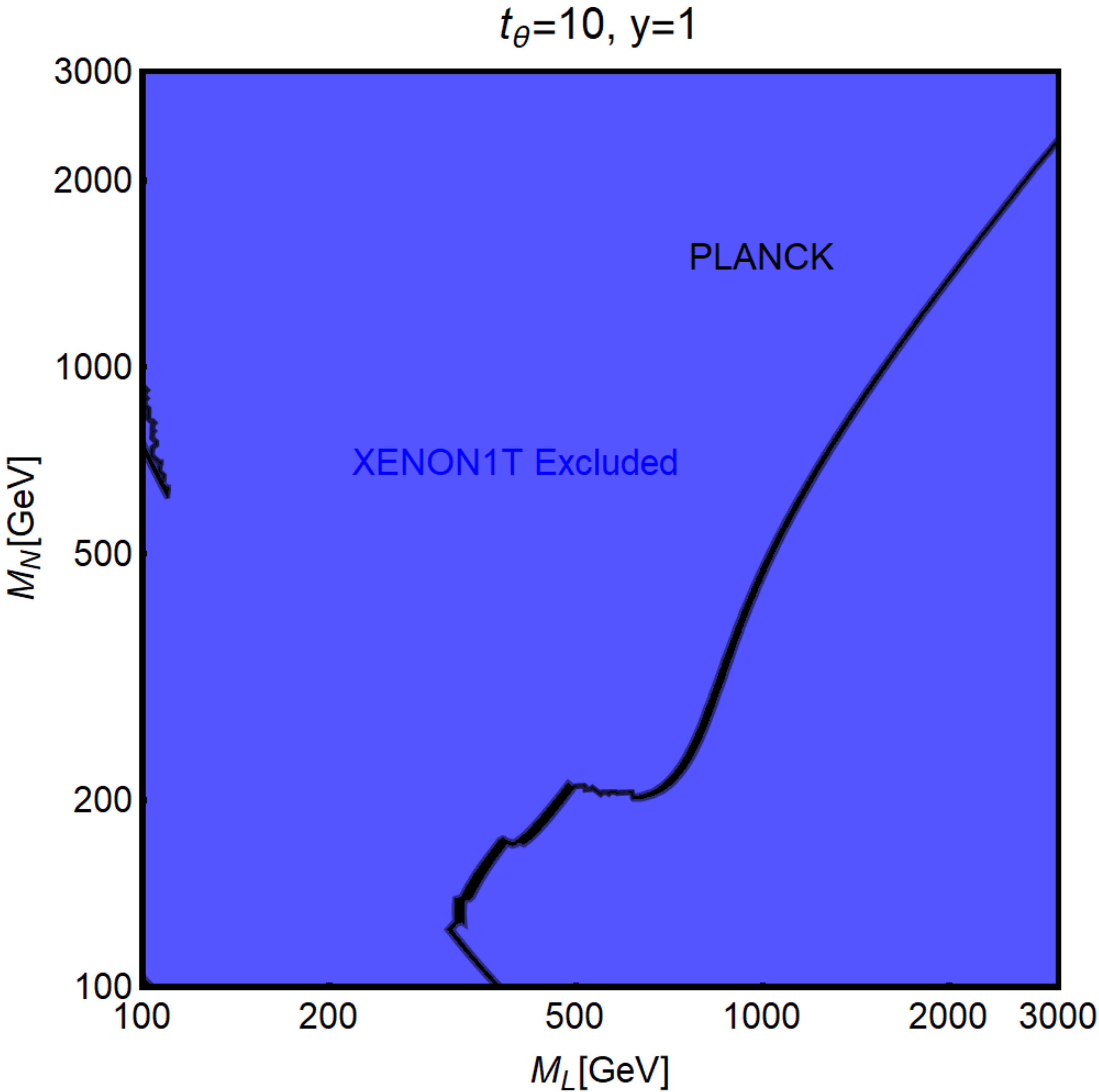}}~~~~
\subfloat{\includegraphics[width=0.46\linewidth]{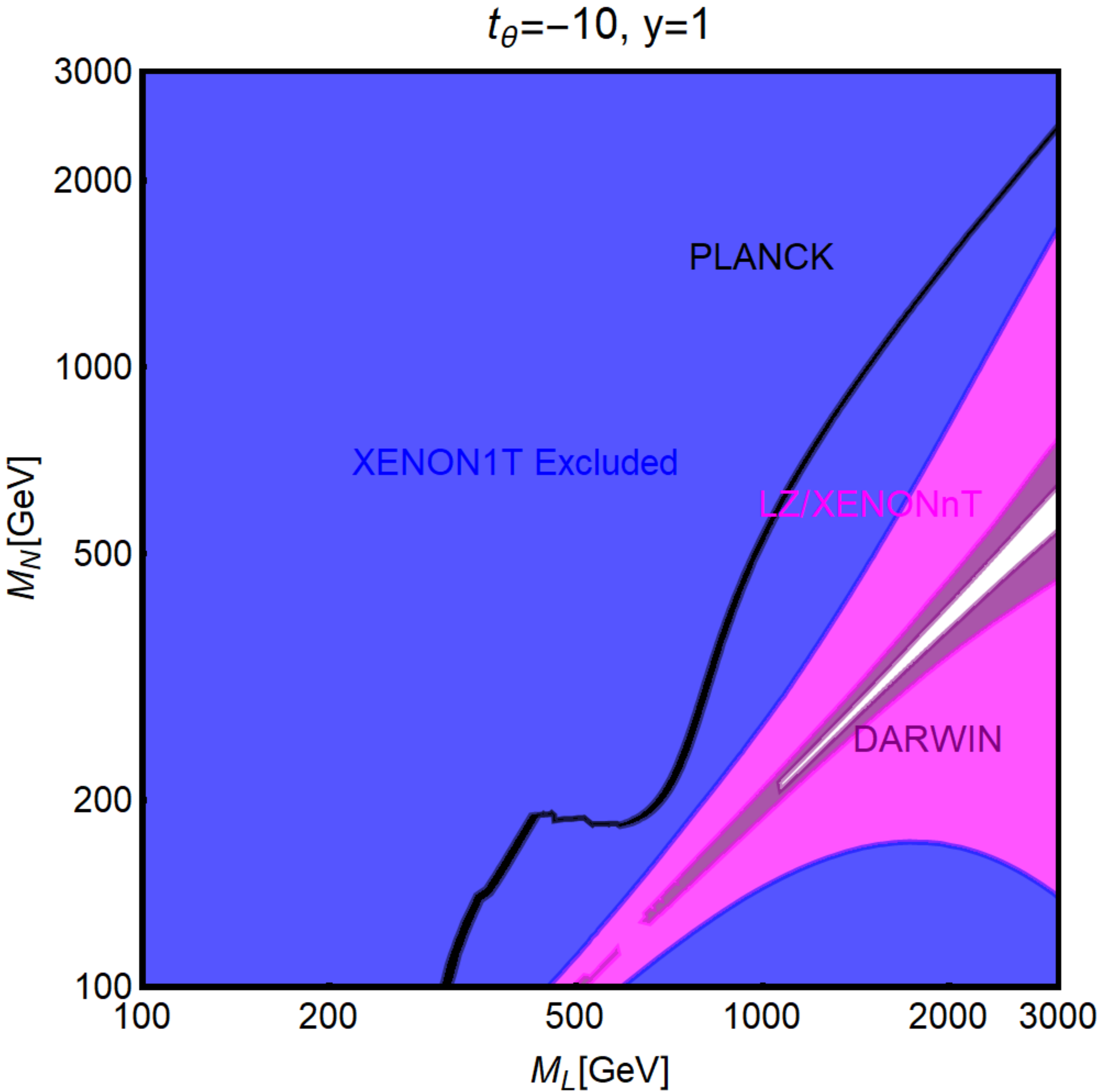}}
\end{center}
\vspace*{-3mm}
\caption{ The same as in Fig.~\ref{fig:SD_y02} but for the assignment $y=1$.}
\label{fig:SD_y15}
\vspace*{-1mm}
\end{figure}

Concerning the relic density, the main DM annihilation channels are again into
SM fermions pairs as well as $WW, ZZ,ZH$ final states, induced by $s$--channel
Higgs exchange but also  $Z$ exchange. Moreover, annihilation processes into
gauge boson final states can be mediated by $t$--channel exchange of the new
fermions. Approximate expressions for the corresponding cross sections are given
in Appendix \ref{App-DM-NF-SD}. 

In contrast to the effective Higgs--portal model, the annihilation cross section
into pairs of SM states features an $s$--wave contribution. The latter is
nevertheless helicity suppressed, being proportional to the mass square $m_f^2$
of the final state fermion, implying that the cross section is dominated by its
$p$--wave term. On the contrary, the additional $t$--channel diagrams are
responsible for unsuppressed $s$--wave contributions to the $WW$ and $ZZ$
annihilation cross sections. The coupling of the DM with the $Z$ boson gives
finally rise to the additional final state $ZH$, with respect to the effective
Higgs--portal  model of the previous section. Given the presence of $s$--wave
dominated cross sections, the singlet--doublet model is potentially testable
also through indirect detection. The corresponding limits are, however, not
competitive with the ones coming from direct detection and, hence, will not be
explicitly reported here. For more details, see eventually
Refs.~\cite{Calibbi:2015nha,Arcadi:2017kky}. 

Given the small number of free parameters in the model, a nice illustration of
the DM phenomenology can be achieved by varying the two masses $[M_L,M_N]$ while
keeping fixed the parameters $y$ and $\theta$. The outcome of such an 
analysis is shown in Figs.~\ref{fig:SD_y02} and~\ref{fig:SD_y15}.

In Fig.~\ref{fig:SD_y02}, we have considered an ``MSSM--like'' assignment of the
$y$ coupling, i.e. $y\sim {g \tan\theta_W}/{\sqrt{2}}\sim 0.2$, accompanied by
four choices of $\tan\theta$, $\pm 2$ and $\pm 10$. For these  couplings, the
correct relic density, indicated by the black isocontours in the figure, is
achieved in proximity of the diagonal $M_L \sim M_N$, representing the so called
``well tempered'' DM regime~\cite{Banerjee:2016hsk,Bharucha:2017ltz}, until it
becomes saturated for $M_L \sim 1.1\,\mbox{TeV}$ by a mostly doublet--like DM.
Notice that in our analysis, we have neglected the Sommerfeld
enhancement~\cite{Cirelli:2007xd,Cirelli:2015bda,Garcia-Cely:2015quu,Lopez-Honorez:2017ora}
which would be responsible for a slight shift of the saturation of the $L$ mass
value to $M_L \sim 1.4\; \mbox{TeV}$. Since, contrary to the MSSM, $y$ is a free
parameter, we have repeated in Fig.~\ref{fig:SD_y15} the same analysis but
taking $y=1$, to highlight  the possibility of achieving the correct relic
density for a mostly singlet--like DM state.

All the benchmarks presented are, nevertheless, strongly affected by the bounds
from DM direct detection. The only surviving regions appear to be those related
to the mostly doublet regime at $y=0.2$ and some some limited regions of the
parameter space (see e.g. the top right panel of Fig.~\ref{fig:SD_y15}) where
direct detection constraints are weaker thanks to the occurrence of the blind
spots which for our sign convention arise only for negative $\tan\theta$. The
singlet--doublet model will be ruled out for DM masses up to a few TeV in the absence of signals at the next direct detection experiments.

\begin{figure}[!h]
\vspace*{-2mm}
\begin{center}
\subfloat{\includegraphics[width=0.46\linewidth]{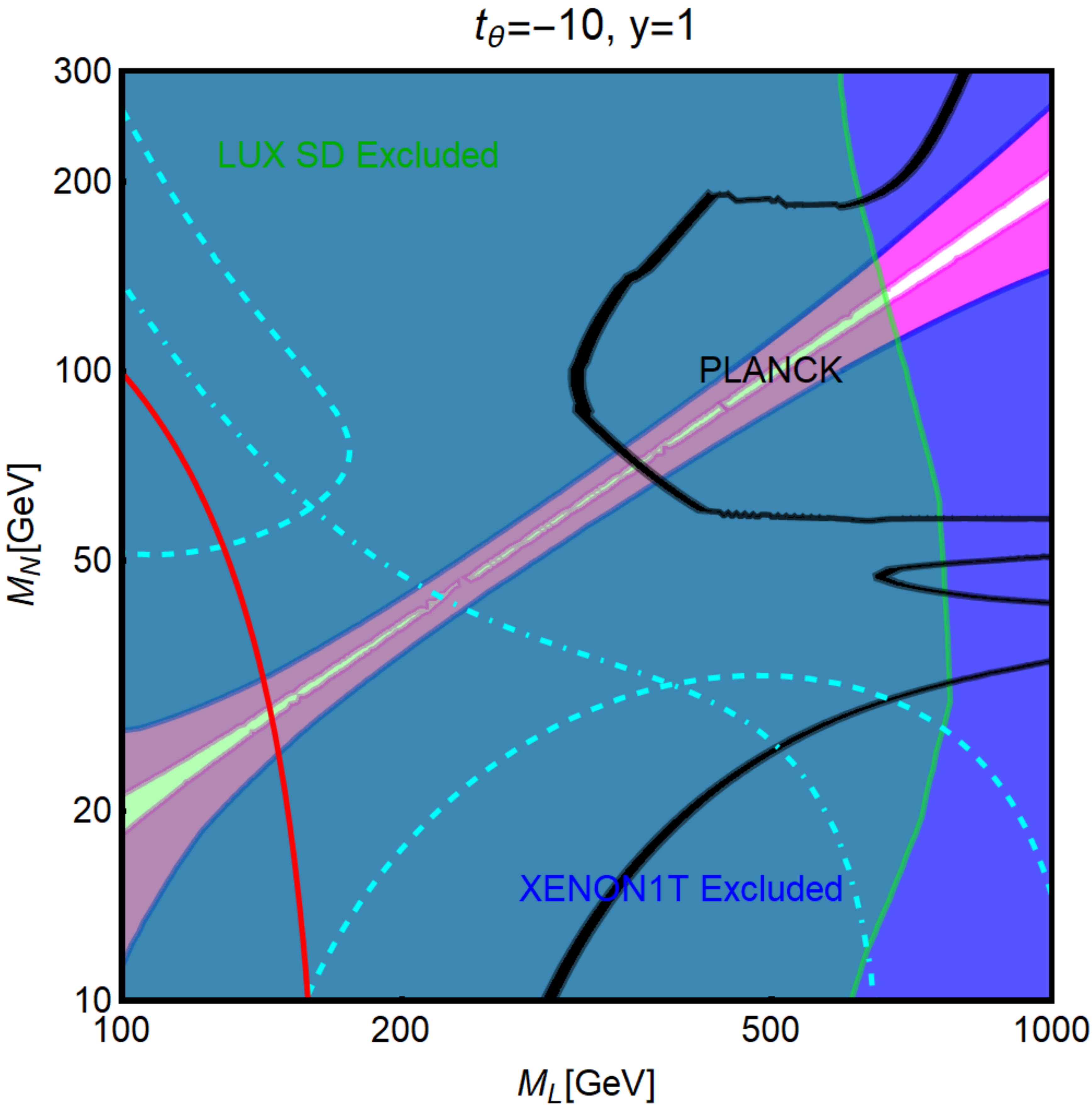}}~~~~
\subfloat{\includegraphics[width=0.46\linewidth]{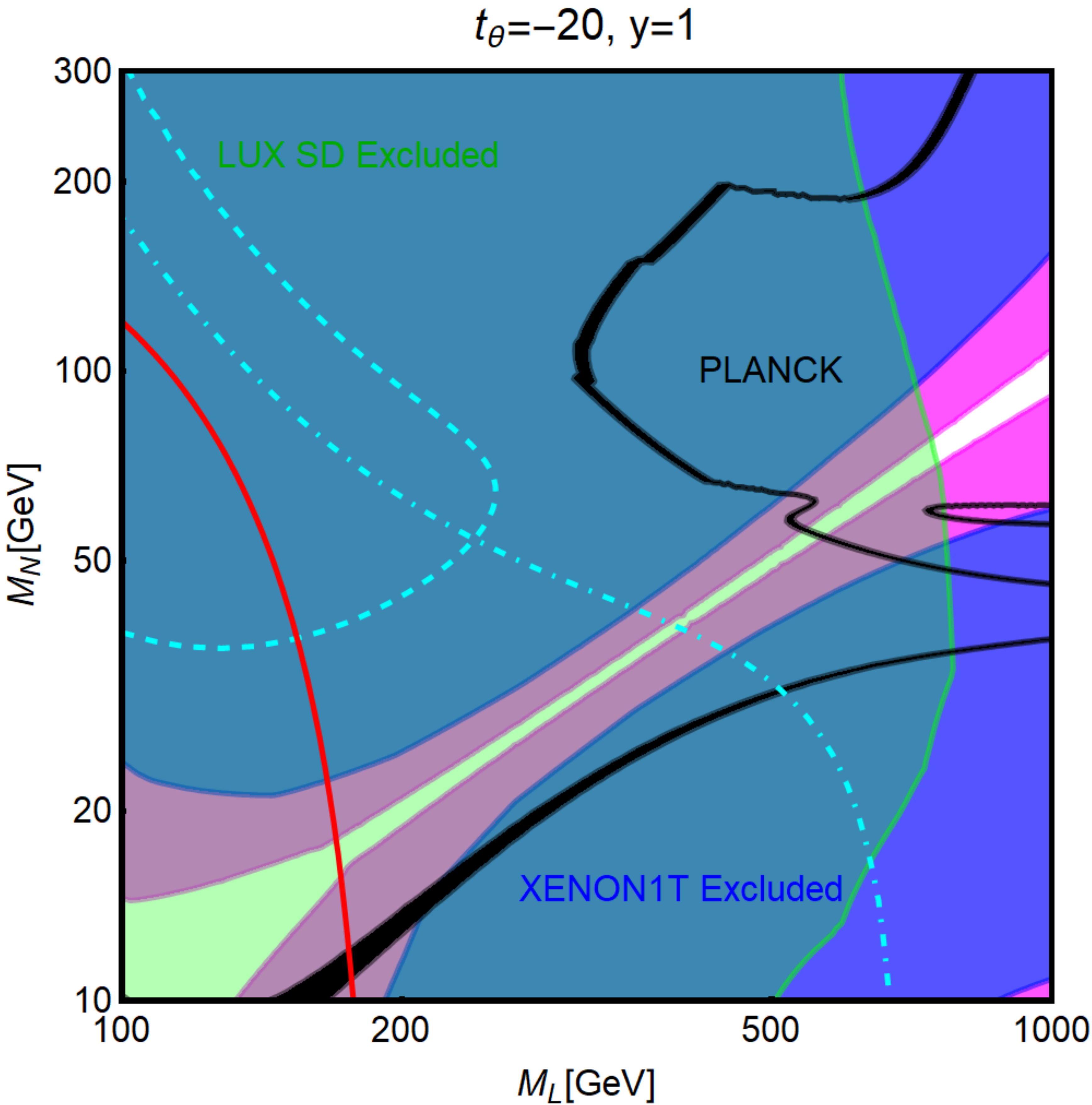}}
\end{center}
\vspace*{-2mm}
\caption{Summary of constraints for the singlet--doublet (Majorana) model with a
SM Higgs sector with $\tan\theta= -10$ (left panel) and $-20$ (right panel); the black isocontours correspond to the correct DM relic density. The blue and the green regions represent current exclusion bounds from XENON1T and  LUX respectively. The magenta region is the projected exclusion on spin--independent interactions by XENONnT/LZ. The regions enclosed in the dashed (dot-dashed) curves are excluded by limits on the invisible width of the $Z\,(H)$ boson while the red region is excluded by electroweak precision data.}
\label{fig:SD_light}
\vspace*{-2mm}
\end{figure}

In the analysis shown in Figs.~\ref{fig:SD_y02} and \ref{fig:SD_y15}, we have
focused on the region $M_{N,L} \geq 100\,\mbox{GeV}$, where the considered
scenario is expected to feature the largest differences with respect to the
effective Higgs--portal and in order to comply with the bounds from LEP on the
mass of the new charged fermion. A focus on lighter DM has been instead made in
Fig.~\ref{fig:SD_light}. The two benchmark considered in the figure are
characterized by two high negative values of $\tan\theta$, namely $-10$ and
$-20$, for which  the blind spot in direct detection is achieved for a high
value of $M_L/M_N$, corresponding to a mostly singlet--like DM state (we recall
that a mostly SU(2) doublet DM would be depleted too efficiently in the early
Universe \cite{Cirelli:2007xd}). 

As it is clear from the figure, in the low mass
regime the relic density isocontours show two ``cusps'', corresponding to
$s$--channel resonances for $m_{N_1} \sim \frac12 M_Z$ and $m_{N_1} \sim \frac12
M_H$. This result is in contrast with the effective Higgs--portal where only the
``pole'' at  $ \frac12 M_H$ is present. Despite the presence of sizable portions
of parameter space in which the DM spin--independent cross section is suppressed
by the occurrence of blind-spots, the considered benchmarks are still in strong
tension with DM direct detection as a result of the complementary constraint
(green region in the plot) from spin--dependent interactions, as given by the LUX collaboration \cite{Akerib:2017kat} \footnote{In the final stage of the work, the XENON1T collaboration released slightly stronger exclusion bounds on  spin--dependent interactions \cite{Aprile:2019dbj}.}.
Fig.~\ref{fig:SD_light} shows also the impact of the constraints from the
invisible decay widths of the $Z$ and Higgs bosons which, similarly to the case
of the effective Higgs--portal discussed previously, appear to be in general
less competitive than the bounds from DM direct detection.

\subsubsection{Constraints on the vector--like lepton DM}

We now turn to the case of vector--like leptons. The DM being a Dirac fermion,
some differences with respect to the singlet--doublet model emerge. Concerning
the relic density, the most relevant annihilation channel is now represented by
the one into $\bar f f$,  the corresponding cross section being not anymore
helicity suppressed, thanks to the presence of a vectorial coupling between the
DM and the $Z$ boson.  The expressions for the annihilation cross sections are
given in Appendix \ref{App-DM-NF-VLL}. 

The Dirac nature of the DM has an even more significant impact on DM direct detection since the vectorial coupling of the DM with the $Z$ boson is responsible of the dominant contribution to the DM spin--independent cross section which reads 
\begin{align}
& \sigma_{N_1 p, Z}^{\rm SI}=\frac{\mu_{N_1}^2}{\pi}\frac{1}{M_Z^4}|y_{V,Z N_1 N_1}|^2 {\left[\left(1+\frac{Z}{A}\right)V_u + \left(2-\frac{Z}{A}\right)V_d\right]}^2\nonumber\\
& \approx 2 \times 10^{-39}\,{\mbox{cm}}^2 {\left( \sin^2 \theta_L^N+\sin^2 \theta_R^N\right)}^2\, .
\end{align}

\begin{figure}[!h]
\vspace*{-2mm}
    \centering
    \includegraphics[width=0.55\linewidth]{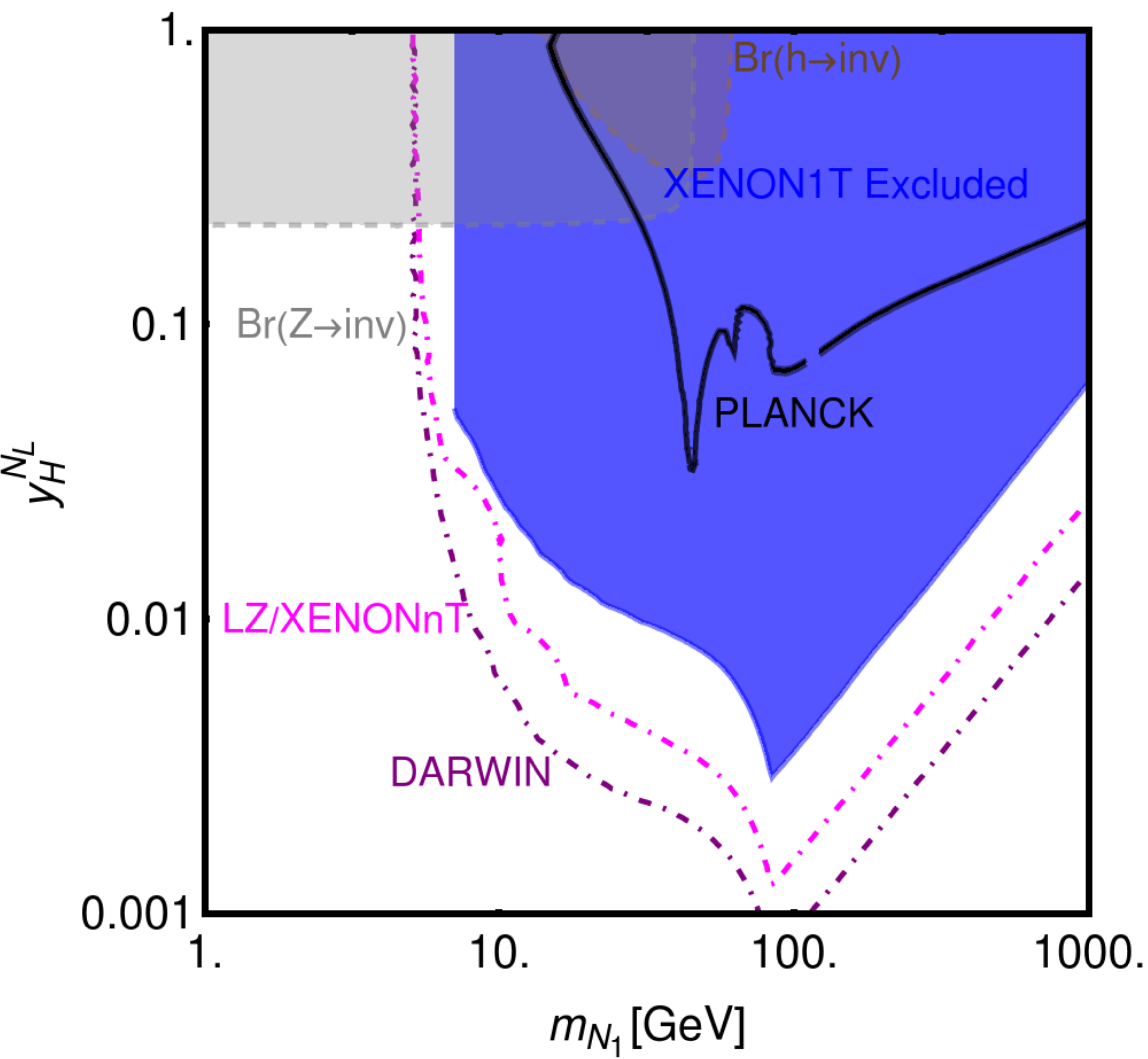}
\vspace*{-2mm}
    \caption{Summary of constraints, in the plane $[m_{N_1},y_H^{N_L}]$, for the vector--like lepton DM model. Concerning the other model parameters we have considered the assignments $y_H^{E_L}=y_H^{N_L}$, $y_H^{E_R}=y_H^{N_R}=0$ and $M_L=M_E=1.2 M_N$ (except for the region $m_{N_1}<100\,\mbox{GeV}$ were we have fixed $M_L=M_E=100\,\mbox{GeV}$). The black isocontours represents the correct relic density according to the WIMP paradigm. The blue region is excluded by current limits from XENON1T. The regions enclosed by the dot-dashed magenta and purple lines will be excluded in absence of signals at LZ/XENONnT and DARWIN, respectively. The brown and cyan regions are excluded by the invisible Higgs and $Z$ boson decay widths.}
    \label{fig:pVhlportal}
\vspace*{-.2mm}
\end{figure}

We compare in Fig.~\ref{fig:pVhlportal} the combined DM constraints on this model with the case of the effective fermionic Higgs--portal by considering the bidimensional plane $[m_{N_1},y_H^{N_L}]$ and assigning the other parameters of the model so that the DM is mostly singlet--like.

As evident from Fig.~\ref{fig:pVhlportal}, this scenario is to a large extent already ruled out by XENON1T limits, and thus even more disfavored than the effective Higgs--portal. This is due to the fact that the most relevant interactions for both relic density and direct detection are actually the ones mediated by the $Z$ boson, hence leading to results similar to the ones of  the so--called $Z$--portal model~\cite{Arcadi:2014lta}. We also note that the limit from the Higgs invisible width is superseded by the one on the invisible $Z$ width and it results is thus not competitive.

The vector--like model defined by the Lagrangian of eq.~(\ref{2HDM-VLLs}) can nevertheless give results that are substantially different from those of the simple effective Higgs--portal,  given the presence of seven free parameters, namely $M_N,M_L,M_E,y_H^{N_{L,R}}, y_H^{E_{L,R}}$. In order to properly account for these additional effects, we have performed a parameter scan over the ranges
\begin{align}
M_{N,E,L} \in \left[100,1500\right]\,\mbox{GeV} \ , \ \ \ 
y_H^{N_{L,R}} \in \left[10^{-6},1\right] \ , \ \ \ 
y_H^{E_{L}} \in \left[ 10^{-6},1\right] \ , 
\label{eq:scanpara}
\end{align}
where we have reduced the number of inputs to only five by imposing the equality  $y_H^{E_R}=0$. With such an assumption,  as already pointed out, it is possible to automatically fulfill the requirement that the signal strength for
the Higgs decays into diphotons coincides with the SM prediction~\cite{Angelescu:2016mhl}.  For each model point, we have required the compatibility with the limits from electroweak precision observables as well as with the correct DM relic density.

\begin{figure}[!h]
\begin{center}
\vspace*{.2mm}
\vspace*{-.2cm}
\subfloat{\includegraphics[width=0.48\linewidth]{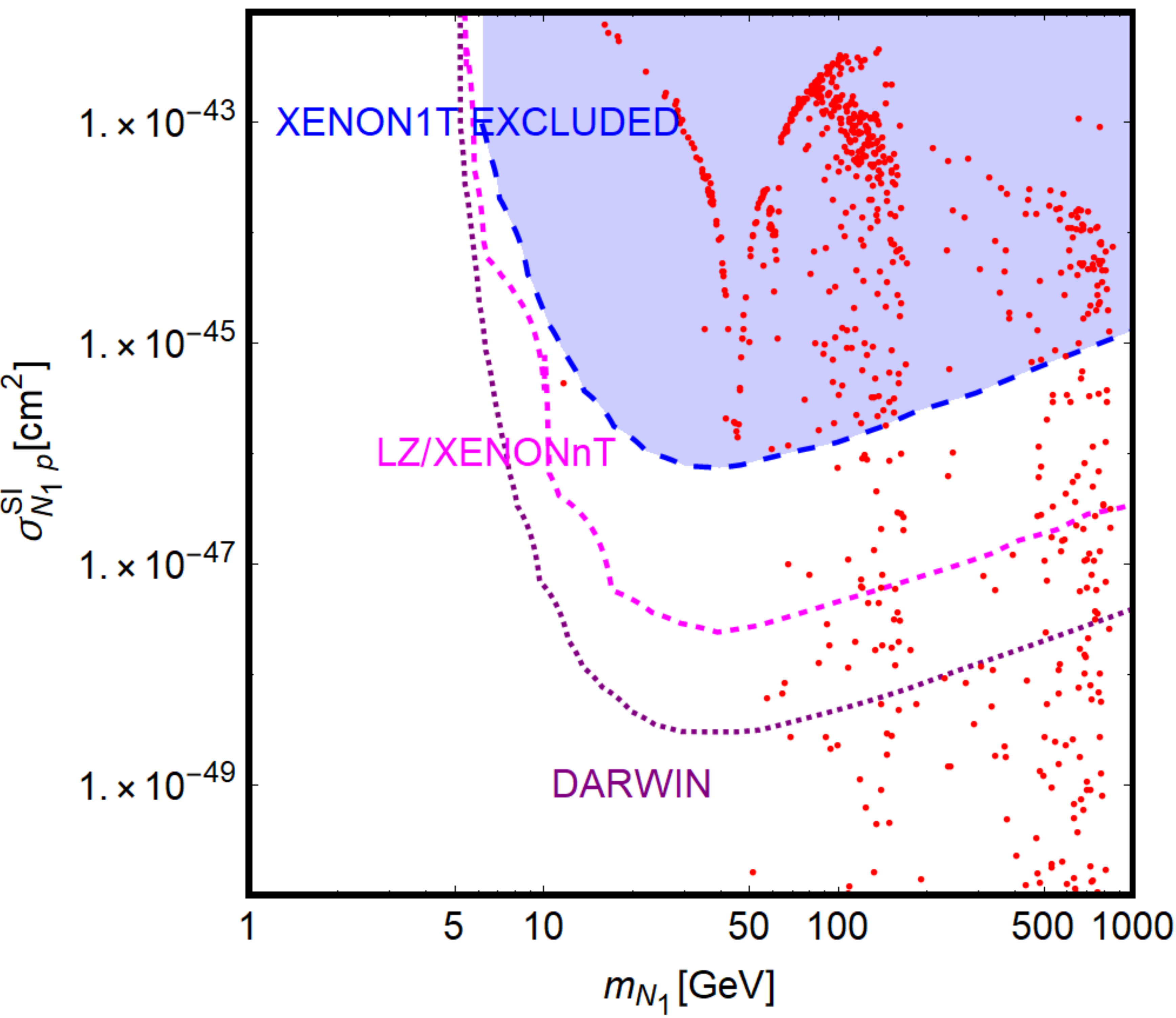}}~
\subfloat{\includegraphics[width=0.47\linewidth]{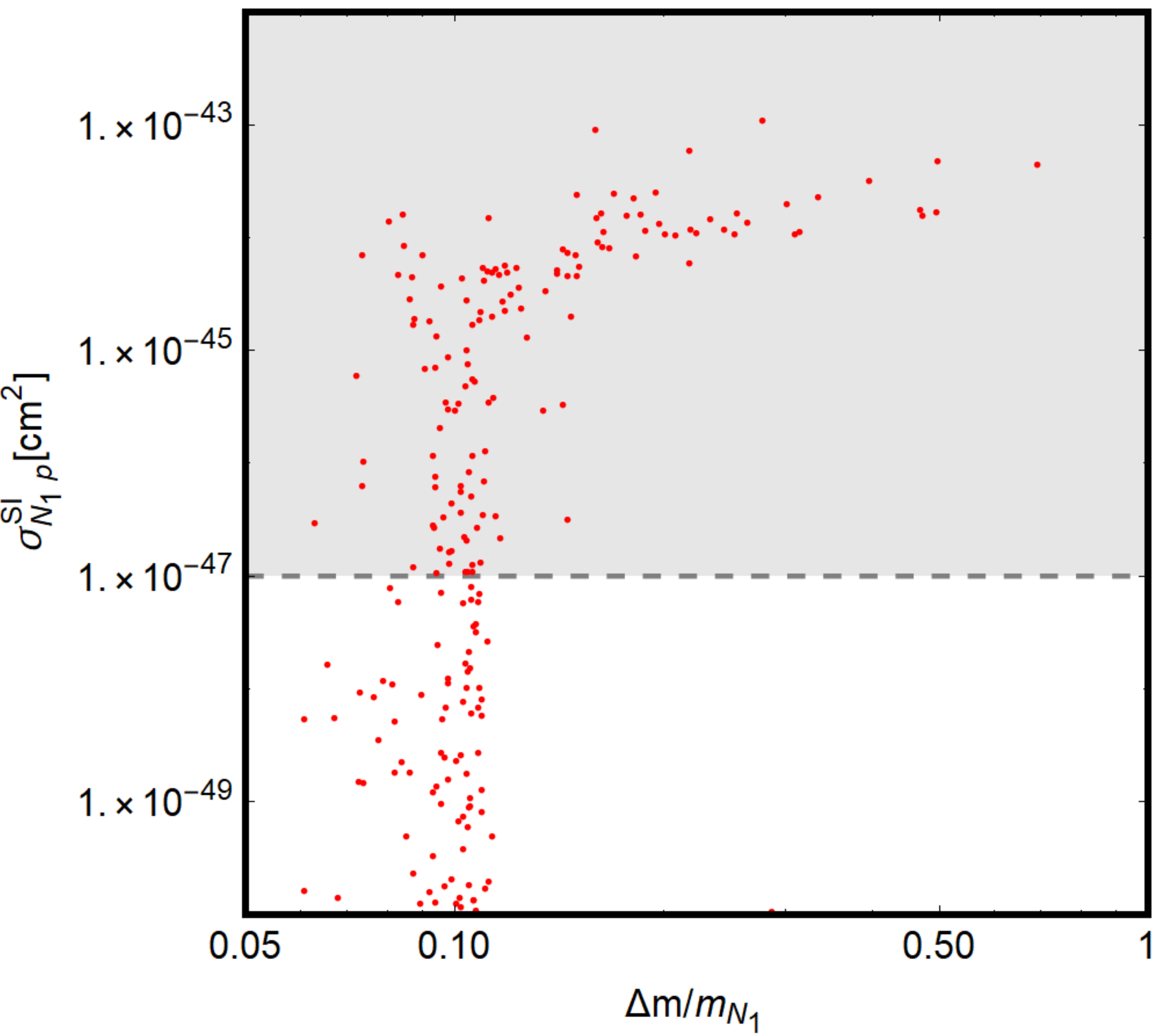}}
\end{center}
\vspace*{-4mm}
\caption{Model points (in red) satisfying the correct DM relic density in the bidimensional plane $[m_{N_1},\sigma_{N_1 p}^{\rm SI}]$ when performing a scan 
of the parameter space in the ranges given in  eq.~(\ref{eq:scanpara}). The blue region is excluded by XENON1T limits while the magenta and purple regions correspond to the reach of LZ/XENONnT and DARWIN, respectively.}
\label{fig:VLLh1}
\vspace*{-3mm}
\end{figure}

The model points successfully passing this constraint are reported in the left
panel of Fig.~\ref{fig:VLLh1} in the bidimensional plane $[m_{N_1},\sigma_{N_1
p}^{\rm SI}]$, and compared with bounds and prospects from DM direct detection
experiments. As can be seen, viable solutions, even in the absence of signals at
the future DARWIN experiment, appear for $m_{N_1}\gtrsim 100\,\mbox{GeV}$. 

As
clarified by the right panel of Fig.~\ref{fig:VLLh1}, these viable model points
correspond to configurations in which the next--to--lightest electrically
neutral vector--like lepton and the lightest electrically charged one, are very
close in mass to the DM, namely $\Delta m/m_{N_1} \lesssim 10\%$. In this case
the correct DM relic density is determined by coannihilation processes while the
coupling $y_H^{N_{L,R}}$ can assume small enough values to pass the constraints
from direct detection. For DM masses below 100 GeV, coannihilation processes are
not relevant since the masses of the charged vector leptons are limited by the
LEP2 bound (the mass of the next--to--lightest vector--neutrino cannot be
similarly low since it is related to the doublet mass parameter $M_L$). In this
region, the result of the scan mostly coincides with the one of
Fig.~\ref{fig:pVhlportal}, so that no viable model points are left by the
constraints  from direct detection.

It is then clear from the discussion in this section and the previous one, that
it is extremely difficult to accommodate a viable DM phenomenology within the
effective SM Higgs--portal as well as in its minimal extensions with additional
matter particles. This is mostly a  consequence of the fact that the correct
relic density requires too strong interactions of the DM with the Higgs (and
possibly gauge) boson, already excluded by the direct searches of DM. We will
investigate in the next section whether this tension can be relaxed by
considering a richer Higgs sector.

%% file: sec-Sing.tex
\section{Singlet extensions of the  Higgs sector}

In this section and the following ones, we will investigate the possibility of
enlarging the Higgs  sector of the theory, limited until now to a unique scalar
doublet, in combination with the already considered extensions in the fermionic
matter sector  to incorporate a DM candidate.  There are many different
possibilities for such an extension and we therefore consider only some specific
scenarios that should, nevertheless, be representative of the richer collider
and astroparticle physics  phenomenology that is induced. In this section, we
will review a set of models in which the scalar sector is enlarged by singlets
under the SM gauge group while the next section will be devoted to two Higgs
doublet models (2HDM).

As straightforward extension of the SM Higgs sector consists into the
introduction of a real singlet coupled with the Higgs doublet and with a
non--zero vacuum expectation value. This implies mixing between the SM--like
Higgs and the new singlet state which, consequently, makes that the latter
acquires a coupling with the SM particles. The new field can be also coupled
with pairs of a further SM singlet, the DM candidate, which can be of spin--0, 1
or $\frac12$ and  studied in the effective approach that we introduced in
section 2. Notice, however,  that this is one possible way to realize in a
renormalizable way the fermionic Higgs--portal. While these models can be
reliably studied in a model--independent way by considering all parameters of
the DM sector as free, we will also investigate more concrete realizations in
which the mass of the DM originates from the vev of the new state. 

In a second step, we will consider the somehow orthogonal scenario in which the
new scalar resonance is not coupled with the SM--like Higgs boson and does not
acquires a vev. In such a framework,  one can then consider also the case of a
pseudoscalar resonance in addition to a scalar one. The new scalar or
pseudoscalar particles,  together with the SM Higgs boson, will serve as a
double portal to the DM states. Apart from a brief discussion of a top--philic
resonance coupled minimally to the DM, we will consider in some detail the
scenario in which a full sequential family of vector--like fermions is added to
the SM spectrum, in which the corresponding lightest neutrino is the DM
candidate. The scalar resonances will only couple to these new fermions at
tree--level but a radiative coupling to the SM gauge bosons will also be
generated  through the new states. This will have an important impact on the
phenomenology, in particular for the detection of the DM particle. 

Finally, we will consider a further refinement of the latter scenario in which
one assumes the simultaneous presence of a new scalar and a pseudoscalar state,
taken to be the two components of a complex field, coupled with a family of
vector--like fermions whose masses are generated by the spontaneous breaking of
a global symmetry. A peculiar feature of such a scenario is that the
pseudoscalar state can be chosen to be much lighter than the DM particle. This
would allow to enhance the DM annihilation cross section without strengthening
the bounds from direct detection in astrophysics experiments.

The section is organised in a way analogous to the previous one. We first
illustrate the main generalities of the proposed models, including  the various
theoretical and experimental constraints to which they are subject. We then
discuss their collider phenomenology and, finally, combine this information with
astroparticle bounds related to the DM particle.

We should remark that in this section and, in fact, to the end of this review,
the SM Higgs boson will be relabelled as $h$ as we will reserve the label $H$ to
the extra heavy CP--even scalar state which is present in the extended Higgs
sectors.

\subsection{Models with additional Higgs singlets} 

\subsubsection{A heavy Higgs--like scalar boson}

The simplest extension of the SM Higgs sector that one could think of  would be the addition of a scalar field $\phi$ that develops a vev and mixes with the 
SM--like Higgs doublet $\Phi$~\cite{Schabinger:2005ei,Patt:2006fw,OConnell:2006rsp,Barger:2007im,Profumo:2007wc,Baek:2011aa,Bertolini:2012gu,Robens:2015gla,Godunov:2015nea,Falkowski:2015iwa}. This scenario can be described through the following potential 
\begin{equation}
V(\Phi, \phi)= \lambda ( \Phi^\dagger \Phi)^2 + \mu^2 \Phi^\dagger \Phi+ 
{\lambda_{hH}} \Phi^\dagger \Phi {\phi}^2+{\lambda_\phi} {\phi}^4+\mu^2_\phi {\phi}^2,
\label{Vscalar2}
\end{equation}
in which $\Phi$ and $\phi$ represent, respectively, the SM Higgs doublet and the
new singlet field. Notice that the parameter with dimension of mass $\mu_\phi$
is such that $\mu^2_\phi  <0$ so that the field $\phi$ develops a non zero vev,
denoted $v_\phi$. The scalar potential would in general allow terms containing
an odd number of scalar fields but we have implicitly assumed the existence of a
$\mathbb{Z}_2$ symmetry forbidding these terms.  After electroweak symmetry
breaking, the real part of the field $\phi$, which is decomposed as
$\phi=(v_\phi+\rho)/\sqrt{2}$, mixes with the real part of the neutral component
of the SM doublet $\Phi$ giving the two mass eigenstates
\begin{eqnarray}
\label{eq:rotation}
\left( \begin{array}{c} h \\ H \end{array} \right)= \Re_\theta 
\left( \begin{array}{c} {\rm Re} (\Phi^0) \\ \rho \end{array} \right),
~~{\rm with}~~\Re_\theta \equiv  \left( \begin{array}{cc}
\cos\theta & \sin\theta \\ -\sin\theta & \cos\theta \end{array}
\right) . 
\end{eqnarray}

In the equation above, the mixing angle $\theta$ is defined in terms of the SM and new vevs, respectively, $v$ and $v_\phi$, by 
\begin{equation}
\label{eq:mixinghS}
\tan 2\theta=\frac{\lambda_{hH} v v_\phi}{\lambda_\phi v_\phi^2-\lambda v}
~~{\rm with}~~\frac{\pi}{8} \leq \pm \lambda_{hH}  \pm \frac{\pi}{8} \leq 
\frac{3\pi}{8}.
\end{equation}
The two physical masses, are given by:
\bea
M_{h/H}^2 = (\lambda v + \lambda_\phi v_\phi) \pm |\lambda v^2 - \lambda_\phi 
v_\phi^2| \sqrt{1+ \tan^22\theta} \, .
\eea
For our analysis, we will always identify the $h$ eigenstate with the SM--like Higgs boson, so that $M_h=125\,\mbox{GeV}$ and further assume $M_H> M_h$.

The phenomenology of the model can be thus described by three parameters in addition to the SM---like ones $v$ and $\lambda$: $\lambda_\phi, v_\phi,\,\lambda_{hH}$ or equivalently in terms of  $M_H,  \lambda_{hH}$ and $\sin\theta\equiv s_\theta$ using the  abbreviation $\Delta M^2 = M^2_H-M^2_h$
and the following relations \cite{Falkowski:2015iwa}
\bea
\label{eq:hSrelation}
\lambda = \frac{M^2_h}{2 v^2} + \frac{\Delta M^2 s^2_\theta}{2v^2},
\,\, \lambda_\phi = \frac{2\lambda^2_{hH} v^2}{s^2_{2\theta} \Delta M^2}
\left(\frac{M^2_H}{\Delta M^2}-s^2_\theta\right), \, \,  v_\phi = -  \frac{\Delta M^2s_{2\theta}}{2\lambda_{hH}v} . 
\eea
The Higgs mixing makes that the two mass eigenstates $h$ and $H$ will share the couplings of the SM Higgs to fermions and gauge bosons and the relevant Lagrangian can be written as
\begin{equation}
\mathcal{L}^{hH}_{\rm SM}=({h c_\theta - H s_\theta}) \left[\frac{2 M_W^2}{v} W^{+}_\mu W^{\mu -}+ \frac{M_Z^2}{v} Z^\mu Z_\mu-\sum_f \frac{m_f}{v} \bar f f \right] . 
\end{equation}
The trilinear Higgs couplings are slightly more complicated than in the SM, being
\begin{eqnarray}
\mathcal{L}_{\rm scal}^{hH} = -\frac{v}{2} \bigg[ \kappa_{hhh}~h^3+ \kappa_{hhH} s_\theta~h^2 H+ \kappa_{hHH} c_\theta ~h H^2 + \kappa_{HHH}~ H^3 \bigg],
\end{eqnarray}
with the reduced couplings $\kappa_{hhh}, \kappa_{hhH}$ etc... given by 
\bea
\kappa_{hhh}= \frac{M^2_h}{v^2 c_\theta} \left(c^4_\theta - s^2_\theta \frac{\lambda_{hH}v^2}{\Delta M^2} \right), \,    \,
\kappa_{hhH}= \frac{2M_h^2+M_H^2}{v^2}\left ( c^2_\theta+\frac{\lambda_{hH} v^2}{\Delta M^2}\right) , \nonumber \\
\kappa_{HHH}= \frac{M^2_h}{v^2 s_\theta} \left(s^4_\theta + c^2_\theta \frac{\lambda_{hH}v^2}{\Delta M^2} \right), \,    \,
\kappa_{HHh}= \frac{2M_h^2+M_H^2}{v^2}\left ( s^2_\theta+\frac{\lambda_{hH} v^2}{\Delta M^2}\right). 
\eea
The parameters  $\sin\theta, M_H$ and $\lambda_{hH}$ are subject to experimental
and theoretical constraints to be summarised shortly. We will  slightly
anticipate this discussion and note that most constraints become increasingly
stringent with a lighter $H$ state. For this reason, as already mentioned, we will assume $M_H> M_h$ in most of our study.    

Turning to the DM sector and analogously to the minimal case discussed in
section 2, it will consist into a scalar $S$, a fermion $\chi$ that we assume
here  to be of the Dirac type only, or a vector $V$, the interactions of which 
are described again in an effective and model--independent approach. The Higgs
sector discussed above will act as a two--portal scenario for these DM particles. We present below the Lagrangians linking the two sectors. 

In the \underline{scalar DM case},  the original Lagrangian to be added to the one involving only the SM and the two Higgs fields and using the notations of section 2 whenever possible, is    
\begin{equation}
\mathcal{L}_{S}=\lambda_{\Phi}^S |S|^2  \Phi^\dagger \Phi
               +\lambda_{\phi}^S |S|^2  \phi^2 , \end{equation}
which leads, after electroweak symmetry breaking, to the following effective Lagrangian
\begin{equation}
\mathcal{L}_S=g_{hSS}|S|^2 h +g_{SSH}|S|^2 H+g_{SS hh}|S|^2 h^2
 +g_{SS hH}|S|^2 h H+g_{SSHH}|S|^2 H^2,
\end{equation}
where, using the relations of eq.~(\ref{eq:hSrelation}),  the various couplings are given by
\begin{align}
\label{eq:bsmhSDMS}
 g_{SSh}= \lambda_\Phi^S v c_\theta - \lambda_\phi^S s^2_\theta c_\theta {\Delta M^2}/({\lambda_{hH} v}) , \ \ 
 g_{SS H}= -\lambda_\Phi^S v s_\theta - \lambda_\phi^S c^2_\theta 
s_\theta {\Delta M^2}/({\lambda_{hH}v }), \nonumber\\
 g_{SS h h}= \lambda_\Phi^S c^2_\theta + \lambda_\phi^S s^2_\theta ,\ \  
g_{SSh H}=2 s_\theta c_\theta (  \lambda_\phi^S -\lambda_\Phi^S ), \ \
 g_{S S HH }= \lambda_\Phi^S s^2_\theta + \lambda_\phi^S c^2_\theta . 
~~~~~~
\end{align}
From the equations above, one can see that without loss of generality, one of  the two couplings $\lambda_\Phi^S$ and $\lambda_\phi^S$ can be set to zero and
we assume in the remaining discussion $\lambda_\Phi^S=0$. With this assumption, the model will have five free parameters: $\lambda^S_\phi, \lambda_{hH}, s_\theta, m_S$ and $M_H$.

In the \underline{Dirac fermion DM case}, we will assume the following Yukawa 
Lagrangian for the field $\phi$ (for the field  $\Phi$, the Lagrangian is still as in section 2)
\begin{equation}
\mathcal{L}_\chi=y_\chi \bar \chi \chi \phi \, ,
\end{equation}
so that the DM mass is dynamically generated by the vev of the field $\phi$,
leading to the fact that the mass and the coupling of the DM fermion are not
independent but related as $y_\chi \propto m_\chi/v_\phi$. This choice will
allow to reduce the number of free parameters compared to the scenario of a scalar DM state. The effective couplings of the DM fermion with the $h$ and $H$ fields can be straightforwardly derived by using eqs.~(\ref{eq:rotation}) and
(\ref{eq:hSrelation}).    

In the \underline{case of a vector DM particle}
\cite{Hambye:2008bq,Lebedev:2011iq,Baek:2012se,Farzan:2012hh,Arcadi:2016qoz}, a
dynamical generation of the DM mass can also be envisaged. This occurs, for
instance, when  one  identifies the DM state with the stable gauge boson of a
U(1) dark gauge group spontaneously broken by the vev of the  complex field
$\phi$. The interaction between the gauge boson and the DM fields are then
embedded in the covariant derivative \cite{Arcadi:2016qoz}
\bea
(D_\mu \Phi)^{*}D^\mu \Phi~~{\rm with}~D_\mu= \partial_\mu -\frac12 i \eta_V^H V_\mu , 
\eea
which, after symmetry breaking and performing the usual field shift $\phi \to (v_\phi+H)/\sqrt 2$, leads to the following  Lagrangian for the DM state
\begin{equation}
\mathcal{L}_{V}=\frac{1}{2}{\eta_V}m_V V^\mu V_\mu H+ \frac{1}{8} H^2 V^\mu V_\mu + \frac{1}{2} m^2_V V^\mu V_\mu , 
\end{equation}
where one gets $m_V=\frac{1}{2} \eta_V v_\phi$ for the $V$ boson mass and the
coupling $\eta_V$ then represents a  gauge coupling. Hence, as in the case of a
fermion DM,  the effective set--up is minimal compared to the SM case  and only a small number of extra parameters is needed. 

To complete the theoretical aspects needed to describe this scenario in which 
the two Higgs states $h$ and $H$ will serve as a portal to the spin--0, $\frac12$ or 1 DM states in an effective approach, let us briefly summarize the constraints that one can impose on the model. 

First, a combination of the two Higgs states needs to satisfy the constraints
that usually apply for a SM--like heavy Higgs boson, namely from the unitarity
in the scattering amplitudes of massive gauge bosons at high--energy  and from electroweak precision observables, i.e. the contributions  to the $\rho$ parameter. Naively, one should have \cite{Englert:2011yb}: 
\bea
M_H^2 s^2_\theta + M_h^2 c^2_\theta \lsim 4\sqrt 2 \pi/3 G_F \approx (700~{\rm GeV})^2 , 
\eea
from perturbative~unitarity and, from the electroweak precision observables or  
$\Delta \rho$: 
\bea
\log M_H^2 s^2_\theta + \log M_h^2 c^2_\theta \lsim \log (v/\sqrt 2)^2 = \log (175~{\rm GeV})^ 2 \, .
\eea
But in fact, the constraints from the requirement that the three couplings of
the scalar potential, namely $\lambda, \lambda_\phi$ and  $\lambda_{hH}$, remain
perturbative up to the Planck scale are much stronger.  There are also
constraints from the requirement of the stability of the electroweak vacuum as
well as a positive Higgs mass spectrum and, actually, one needs   $4 \lambda
\lambda_\phi > \lambda^2_{hH}$ to have $M_h,M_H>$ and $\lambda \lambda_\phi >0$
for the scalar potential to be  bounded from below. 

These constraints can be determined by the renormalisation group evolution of the quartic couplings $\lambda,\lambda_{hH},\lambda_\phi$ described by the following equations (we again limit ourselves to one--loop $\beta$ functions):
\bea
16\pi^2 {d \lambda}/{dt}&\!=\!&  24 \lambda^2 \!-\! 6 y_t^4 \!+ \! \frac38 \left( 2 g^4 \!+\! (g^2 + g^{\prime 2})^2 \right) \!+\! (\!-\!9 g^2 \!-\! 3 g^{\prime 2}\!+\! 12 y_t^2) \lambda \!+\! \frac12 \lambda_{hH}^2\;, \nonumber \\
16\pi^2 d \lambda_{hH} /{dt}  &\!=\!&  4 \lambda_{hH}^2 + 12 \lambda \lambda_{hH} -\frac32 (3 g^2 + g^{\prime 2}) \lambda_{hH} + 6 y_t^2 \lambda_{hH} + 6 \lambda_\phi \lambda_{hH} \;, \nonumber  \\
16\pi^2d \lambda_{\phi} /{dt}&\!=\!&2\lambda_{hH}^2 +18\lambda_\phi^2\;. 
\eea

The vacuum stability conditions depend on the sign of $\lambda_{hH}$ and there
are then two possibilities \cite{Falkowski:2015iwa,EliasMiro:2012ay}.  For
$\lambda_{hH}\! >\! 0$, the condition $\lambda \!> \!\lambda_{hH}^2/(4
\lambda_\phi)$ imposed at the weak scale so that $v$ and $v_\phi$ are minima of
the scalar potential may not hold at high energies. If all  quartic couplings
$\lambda_i\!> \! 0$, the potential is positive definite and there are no 
runaway directions. For $\lambda_{hH} \!<\! 0$, the potential has a runaway
direction at large field values unless $\lambda \!>
\!\lambda_{hH}^2/(4\lambda_\phi)$. This condition and $\lambda_\phi>0$ should be
valid at all scales. Note again that $\lambda, \lambda_\phi$ can be  expressed
in terms  of $M_H, \sin\theta$  and $\lambda_{Hh}$  which can be used as inputs
instead.

\begin{figure}[!ht] 
\vspace*{-1mm}
\centerline{
\includegraphics[width=0.45\textwidth]{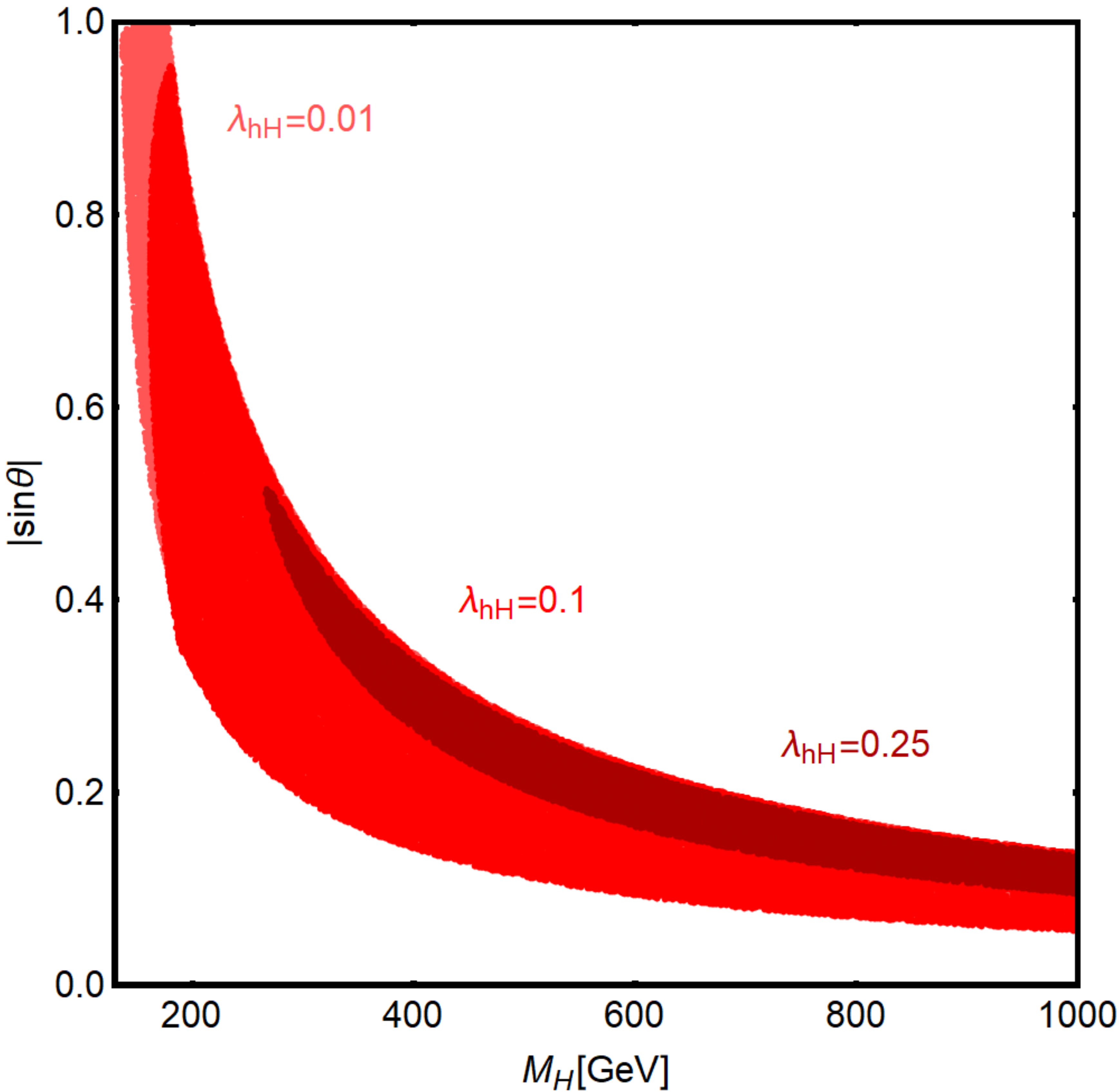}~~~~
\includegraphics[width=0.45\textwidth]{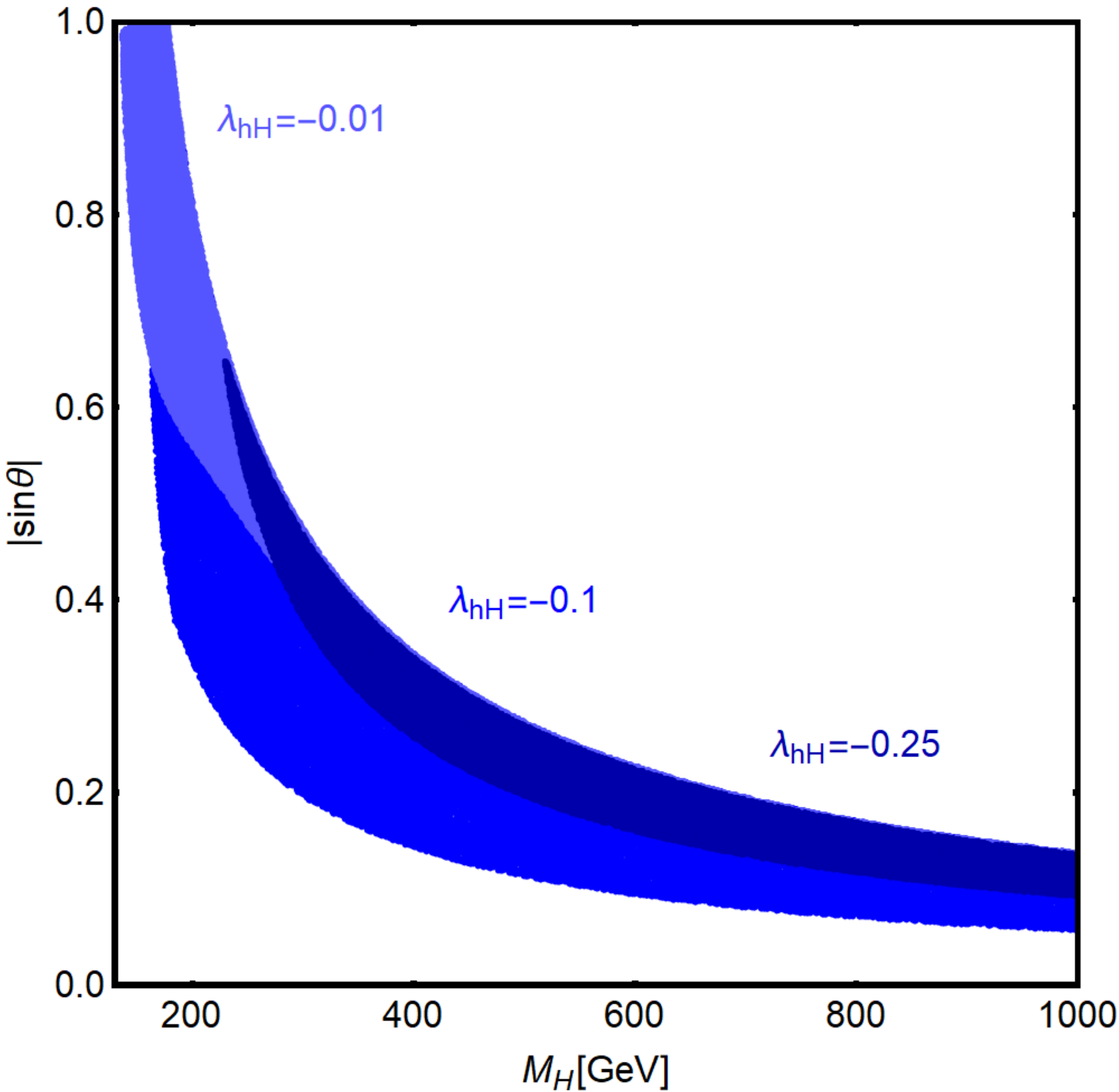} }
\caption{ \label{fig:lhs-stability}
Regions of the parameter space (shaded) where the couplings remain perturbative and the electroweak vacuum stable up to the Planck scale for several $\lambda_{hH}$ values at the weak scale; in the left (right) panels these are: $\lambda_{hH}= +0.01 (-0.01)$ (lightest red/blue), $\lambda_{gH}=+0.1(-0.1)$ (red/blue) and $\lambda_{hH}=+0.25(-0.25)$ (darkest red/blue).}
\vspace*{-2mm}
\end{figure}

Closely following  an analysis performed in Ref.~\cite{Falkowski:2015iwa}, we
have delineated the areas of parameter space that are still allowed by the
constraints  that the couplings remain perturbative $\lambda_i< 4 \pi$ up to the
scale $M_{\rm Planck}$ and the electroweak vacuum  remains stable too. The
results are shown in Figure~\ref{fig:lhs-stability} in the  plane $[M_{H}, 
|\sin \theta|]$ for three positive (left) and negative (right) values of 
$\lambda_{Hh}$.  The allowed regions are in red or blue and one can see that
they are smaller for larger values of $\lambda_{Hh}$ and, in particular  for
$\lambda_{Hh} \approx \pm 0.25$, they become extremely narrow. In all cases, the
smaller is the mixing angle and the heavier should be the $H$ state. In fact, if
$M_H <M_h$, the scalar coupling $\lambda$ at the weak scale should be smaller 
than in the SM and hence, the vacuum cannot remain stable up to the scale
$M_{\rm Planck}$. The possibility of a light $H$ state is thus also very
unlikely from this perspective. 

Furthermore, there are two additional constraints on this scenario from the LHC
Higgs data. The first one is due to the measurement of the signal strengths of
the 125 GeV Higgs boson: since the $H$ and $h$ states share the couplings of the
SM Higgs to gauge bosons and fermions, the signal strengths of the  $h$ boson
will come with a factor of $\cos^2\theta$ for all production cross sections (but
not for the branching ratios in the decays to the various final states) and will
be strongly constrained by data. Another complementary constraint will come from
the direct searches at the LHC of heavy Higgs bosons decaying into $WW,ZZ,\gamma
\gamma$ and other final states. Both constraints will be discussed in the next
subsection. 

\subsubsection{Singlet scalar or pseudoscalar states}

In the extension of the Higgs sector discussed in the previous subsection, the
new singlet scalar field had a non--vanishing vev and the corresponding Higgs
state had a sizeable mixing with the SM--like Higgs boson, $\sin\theta \neq 0$.
An almost opposite option would be that the singlet scalar field does not
develop a vev  and has no mixing with the SM Higgs, $\sin\theta\! \to \! 0$. In
this case, the possibility that the new state is a pseudoscalar instead of a
scalar can also be considered since, if CP symmetry is conserved, a pseudoscalar
will automatically have no mixing with the SM Higgs boson. The scalar sector of
the theory is then still described by the potential of eq.~(\ref{Vscalar2}) but
with the mass term $\mu_\phi^2$ being positive and, to forbid mixing, a very
small or vanishing value for the parameter $\lambda_{hH}$.  

In the absence of such a mixing with the SM--like Higgs doublet, one should find
alternative ways to generate a coupling of the new field with the SM fermions
and gauge bosons and with the DM particle. We will discuss two possibilities
which, despite of the fact that they are both effective and non--renormalisable,
lead to different and interesting phenomenology and experimental signatures for
the DM particle.

A minimal option in this context is represented by the introduction of an effective interaction of the scalar or pseudoscalar Higgs state, that we will denote respectively by $\phi = H$ and $A$, with the top quark neglecting all other fermion masses, 
\begin{eqnarray}
\mathcal{L}_{\rm Yuk} \supset - g_{Htt} \; \bar{t}t H \quad {\rm or} \quad i g_{Att} \; \bar{t} \gamma_5 t A \, ,
\label{SP}
\end{eqnarray}
which can be generated via a dimension--5 or higher operator for instance~\cite{Alanne:2017ymh}. 
Using the SM--Higgs Yukawa coupling to fermions as a reference, one  can express these new couplings of $\phi = H,A$ to the top quark as
\begin{eqnarray} 
g_{\phi t t}= \frac{m_t}{v} \times \hat g_{\phi t t}    , 
\label{eq:yukawa}
\end{eqnarray} 
with  $v$ the SM vev; in this case,  the coupling of the SM--like Higgs boson
$g_{h t t}= {m_t}/{v}$ would  correspond to $\hat g_{h t t}\! =\! 1$. These
new interactions,  even if one has $\sin \theta \rightarrow 0$, induce 
couplings of the $\phi$ state to massive gauge bosons, gluons and photons via quantum corrections and more precisely here, triangular diagrams involving the contributions of the top quark. 

As for the DM sector for which these singlet--Higgs states would serve as
portals,  the simplest model that we use as a benchmark in when a neutral and colorless DM particle $N$  of mass $m_N$ is present and would couple to the $H/A$ bosons exactly like the top quark
\begin{eqnarray} 
g_{\phi N N}= \frac{m_N}{v} \times \hat g_{\phi  NN} ,    
\label{eq:yukawa2}
\end{eqnarray} 
leading to two free parameters, $m_N$ and  $\hat g_{\phi NN}$, in addition to those of the Higgs sector.

An alternative possibility would be to have only radiatively induced couplings between the new singlet fields and the SM particles. This can be achieved by the introduction of a full family of vector--like fermions, including as well a vector--like neutrino which, as already discussed in 3.1.3, would correspond to the DM particle.  Their couplings with the new mediators are described by the following Lagrangian
\begin{equation}
-{\cal L}_{\phi F }=  \varepsilon_\phi ~\phi\left(y_\phi^{UD}  \overline{{\cal D}_L} {\cal D}_R 
+ y_\phi^U   \overline{U^\prime_L} U^\prime_R 
+ y_\phi^D  \overline{D^\prime_L} D^\prime_R\right) + {\rm h.c.} ~,
~~{\rm with}~U\!=\!N,T~;~D\!=\!E,B , 
\label{Larangian-S+VL}
\end{equation}
where $\varepsilon_\phi= 1, i$ for the $\phi=H,A$ possibilities. The Yukawa coupling matrices read
\begin{equation}
Y_{Q}^h = \frac{1}{\sqrt{2}}
\begin{pmatrix} 0& y_\phi^{Q_L}  \\ y_\phi^{Q_R}  & 0 \end{pmatrix} ~,
\quad Y_{Q}^H=\frac{1}{\sqrt{2}}
\begin{pmatrix} -y^Q_\phi & 0 \\ 0 & -y^{UD}_\phi \end{pmatrix}
=- Y_{Q}^A . 
\label{singlet-Yukawa-couplings}
\end{equation}

In general, in order to compute the new fermion loop contributions, one has to
account for the mixing induced by the Yukawa couplings with the SM Higgs
boson $y_h^{Q_L}$ and $y_h^{Q_R}$,  as can be seen in eq.~(\ref{SM-mass-matrices}). One  can nevertheless consider a rather simplified
scenario in which these Yukawa couplings are set to zero. In this way, there is
no mixing between the vector--like fermions and, hence, automatically no
corrections at leading order from these new fermions to the SM Higgs couplings
and to electroweak precision observables.

Similarly to what occurs in the case of couplings with the top quark, effective interactions between the $H/A$ states and the SM gauge bosons are induced by triangle diagrams with vector--like fermions with the appropriate quantum numbers running in the loop. The interactions can be described through the following effective Lagrangian~\cite{Chu:2012qy,Mambrini:2015wyu,Backovic:2015fnp,DEramo:2016aee}
\begin{align}
& \mathcal{L}_{H}=c_{\rm gg}^H H G_{\mu \nu}G^{\mu \nu}+c_{\rm WW}^H H W_{\mu \nu}W^{\mu \nu}+c_{\rm ZZ}^H H Z_{\mu \nu}Z^{\mu \nu}+c_{\rm Z\gamma}^H H F_{\mu \nu}Z^{\mu \nu}+c_{\rm \gamma \gamma}^H H F_{\mu \nu}F^{\mu \nu} , \nonumber\\
& \mathcal{L}_{A}=c_{\rm gg}^A A G_{\mu \nu}\tilde{G}^{\mu \nu}+c_{\rm WW}^A A W_{\mu \nu}\tilde{W}^{\mu \nu}+c_{\rm ZZ}^A A Z_{\mu \nu}\tilde{Z}^{\mu \nu}+c_{\rm Z\gamma}^A A F_{\mu \nu}\tilde{Z}^{\mu \nu}+c_{\rm \gamma \gamma}^A A F_{\mu \nu}\tilde{F}^{\mu \nu} ,
\label{eq:phi-couplings}
\end{align}
with $F_{\mu \nu}\!=\! (\partial_\mu A_\nu \!-\! \partial_\nu A_\mu)$ being the field strength of the photon field $A_\mu$, $\tilde{F}_{\mu\nu}=\epsilon_{\mu \nu \rho \sigma} F^{\rho \sigma}$ and likewise for the SU(3) and SU(2) gauge fields. Using the form--factors $A_{1/2}^{\phi}(\tau_F)$ given in Appendix A1 for the scalar and pseudoscalar $\phi=H,A$ cases, with the fermions $F$ running  in the triangular loops having mass variables $\tau_F= {M_\phi^2}/{4 m_F^2}$, the coefficients $c^{H,A}_{ij}$  in these Lagrangians are given by \cite{Altmannshofer:2015xfo,Bae:2016xni}:
\begin{align}
& c_{gg}^{\phi=H,A}=\frac{\alpha_s}{8 \pi} \sum_{F=B,T,Q} \eta_F \frac{y_\phi^F}{M_F} A_{1/2}^\phi(\tau_F), \nonumber\\
& c_{\gamma \gamma}^{\phi=H,A}=\frac{\alpha}{8 \pi} \sum_{F=E,L,B,T,Q} N_c^F Q_F^2 \frac{y_\phi^F}{M_F} A_{1/2}^\phi(\tau_F) , \nonumber\\
& c_{ZZ}^{\phi=H,A}=\frac{\alpha}{8 \pi} \sum_{F=E,L,B,T,Q}N_c^F \left(\frac{s_W^2}{c_W^2} Y_F^2  \eta_F +\frac{c_W^2}{s_W^2} \xi_F \right) \frac{y_\phi^F}{M_F} A_{1/2}^\phi(\tau_F),\nonumber\\
& c_{WW}^{\phi=H,A}=\frac{\alpha}{2 \pi s_W^2} \sum_{F=L,Q} N_c^F \frac{y_\phi^F}{M_F} A_{1/2}^\phi(\tau_F) , \nonumber\\
& c_{Z\gamma}^{\phi=H,A}=\frac{\alpha}{4 \pi} \sum_{F=E,L,B,T,Q} N_c^F \left(\frac{c_W}{s_W} \xi_F -\frac{s_W}{c_W} Y_F^2  \eta_F\right) \frac{y_\phi^F}{M_F} A_{1/2}^\phi(\tau_F) \, , 
\label{eq:phi-couplings_2}
\end{align}
where in the case of the $\phi gg$ coupling, $\eta_F= \frac12(1)$ for a singlet (doublet) while in the case of the $\phi ZZ$ coupling, $\eta_F=1(2),\xi_F=0 (\frac12)$ for a singlet (doublet).

Besides their quantum numbers under the SM gauge group, the effective couplings
depend on the masses and Yukawa coupling of the new vector fermions. The latter
should, however, comply with the severe constraints from their renormalisation
group evolution. Having the coupling between the new fermions and the SM--like
Higgs boson set to zero, these effects affect mainly the quartic coupling
$\lambda_{H,A}$ of the new scalar singlets. Similarly to the quartic coupling of
the SM--like Higgs boson and, as discussed in section 3.1.4,  we have to
require that $\lambda_{H,A}$ remains positive at least up to the energy scales
that are relevant for their collider and DM phenomenology.

All the renormalisation group elements that allow to understand, at least
qualitatively  since they are based  one--loop $\beta$ functions only,   how the
evolution affects the couplings, have been discussed in the previous section and
given in eqs.~(\ref{eq:singletRGE}); one simply has to replace $H$ by $\phi$. 
In order to have an insight on the impact of the running of the new Yukawa
couplings in this model, we have solved the sets of these equations  in
combination with the evolution of the SM gauge couplings given in
eq.~(\ref{eq:gi-RGE}), for both the scalar and pseudoscalar singlet cases, for
some chosen weak scale values of the Yukawa couplings of the new fermions and of
the quartic coupling $\lambda_{\phi}$. For a numerical illustration, we have set
for simplicity the vector--like lepton and quark  masses to common values of,
respectively, $m_{L}=400\,\mbox{GeV}$ and $m_{Q}=1\,\mbox{TeV}$ while the mass
of the $\phi$ state is set at $M_\phi=750$ GeV. 

\begin{figure}[!h]
\begin{center}
\hspace*{-5mm}
\subfloat{\includegraphics[width=0.52\linewidth]{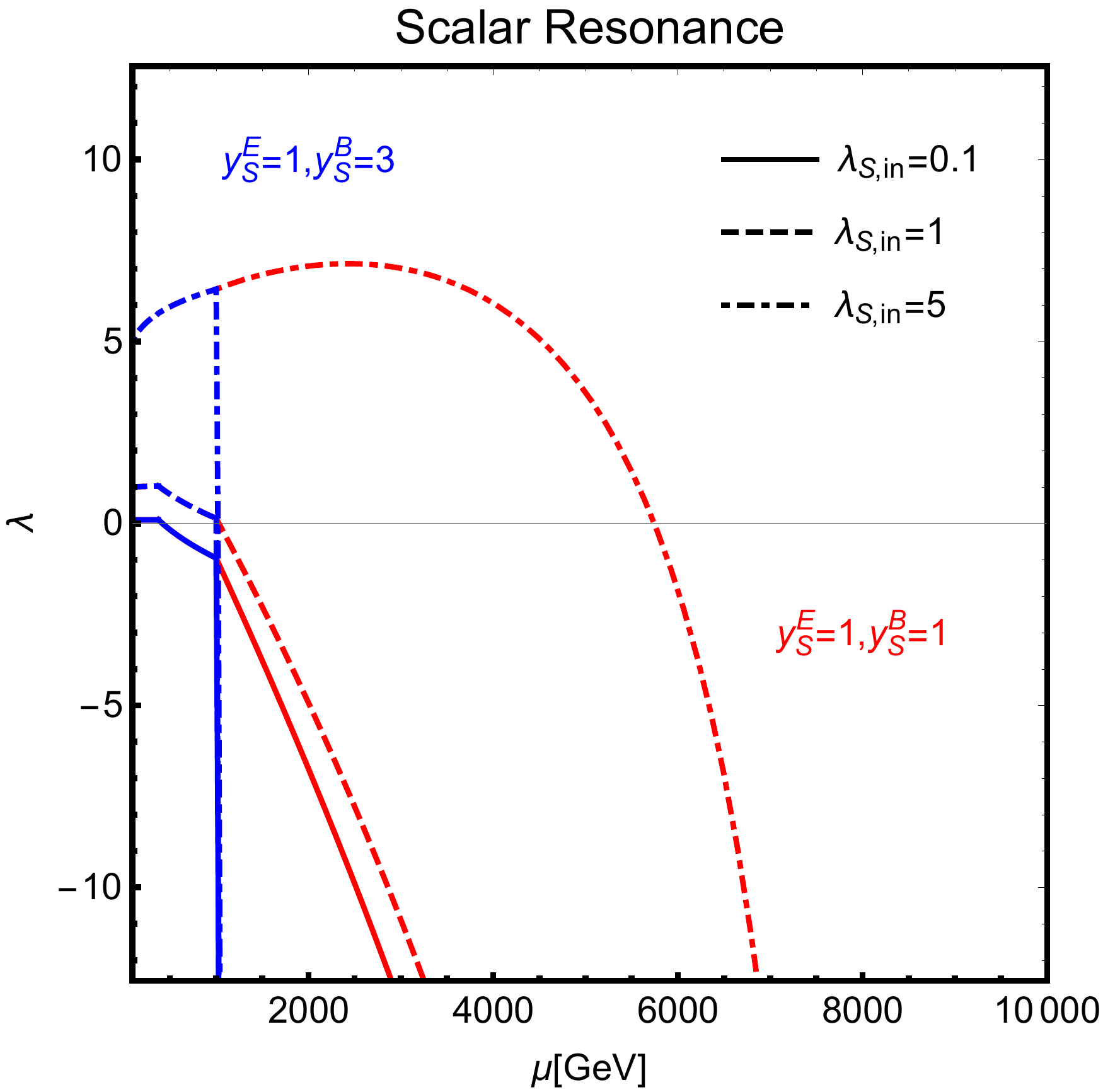}}
\hspace*{-3mm}
\subfloat{\includegraphics[width=0.52\linewidth]{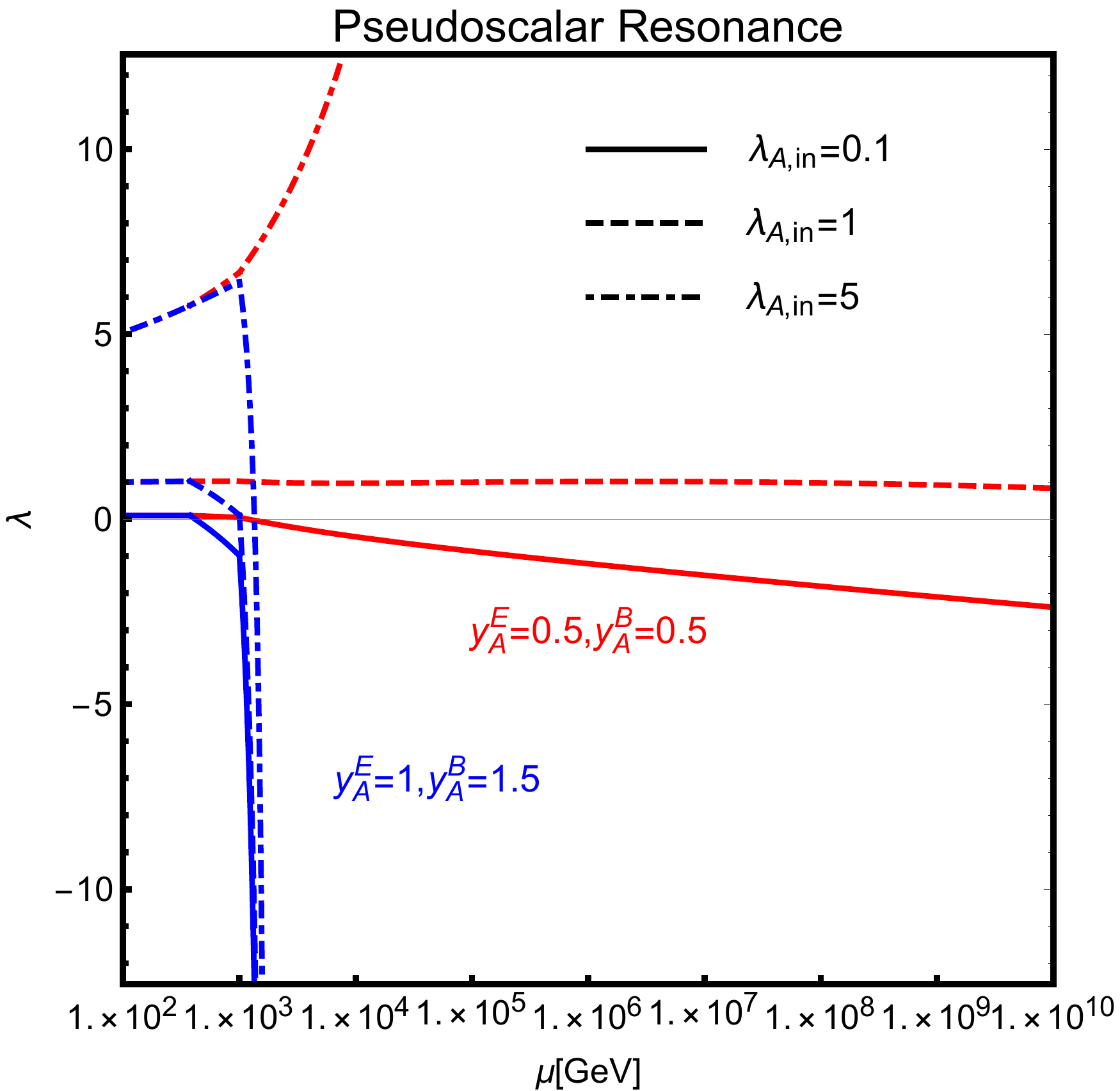}}
\end{center}
\vspace*{-3mm}
\caption{Evolution of the quartic coupling of singlet scalar (left) or pseudoscalar (right) resonances of  mass $M_\phi=750$ GeV for some assignments of the model parameters and three values of $\lambda_\phi =0.1,1$ and 5. For all models, we have assumed $y_\phi^E=y_\phi^L$ and $y_\phi^B=y_\phi^T=y_\phi^Q$
and common masses of $m_{\rm VLL}=400$ GeV and $m_{\rm VLQ}=1$ TeV for the 
vector--fermions. }
\label{fig:RGElambdasinglet}
\vspace*{-3mm}
\end{figure}

Our results are shown in Fig.~\ref{fig:RGElambdasinglet} for the cases of a
scalar (left panel) and a pseudoscalar (right panel) singlet for some initial
conditions for their couplings with the vector fermions  and for the quartic
coupling $\lambda_\phi$. The plots display the evolution of the quartic
coupling, which is the most severely affected by the effects of running. In both
panels, the starting values of $\lambda_\phi$ are the same, namely
$\lambda_\phi= 0.1,1$ and 5. 

The behaviour of the curves can be explained as follows. For energies $\mu \leq
m_{F}$, the $\beta$ function of the quartic couplings is positive and dominated
by the term $\propto \lambda_{\phi}^2$ (this also explains the dependence on the
initial conditions). Above the energy threshold corresponding to the masses of
the vector--like fermion,  the $\beta$ function is affected by the negative
contribution related to the corresponding Yukawa couplings, hence becoming a
decreasing function eventually acquiring negative values. As should be clear
from the figure, for Yukawa couplings equal or greater than unity, the drop of
the quartic couplings is very sharp so that they become negative already at an
energy scales of a few TeV.  

We note nevertheless that since we are having a theoretical bottom--up approach,
a negative value for the quartic coupling $\lambda_\phi$ at some energy scale
should not be interpreted as an exclusion constraint but simply as the
requirement of the completion of the theory with new degrees of freedom. Hence,
the scenario under consideration can be seen to be still valid, as long as the
pathology associated to the running effects do not appear at the energy scales
relevant for collider and DM phenomenology. Given this, from the outcome of
Fig.~\ref{fig:RGElambdasinglet}, we can infer a bound $y_F^\phi \lsim 1$ on the
couplings of the new fermions.

\subsubsection{The scalar and pseudoscalar double portal}

In the previous discussion, we have considered individually the extension of the
SM Higgs sector with a new scalar or a pseudoscalar singlet. From a Dark Matter 
perspective, it is nevertheless interesting to also consider the scenario of a
combined scalar--pseudoscalar portal. This is particularly true in the case
where one of the scalar mediators is significantly lighter than the DM particle,
which leads to some striking phenomenological features. This is the possibility
that we will briefly comment upon here.  

A double scalar portal scenario can be realized for example by introducing a
complex field $\phi$ that can be decomposed into a scalar and pseudoscalar
components as $\phi \to \frac{1}{\sqrt{2}}(H+ia)$. Here, we use the label $a$ to
highlight the difference with the discussion of the prevision subsection and the fact that the pseudoscalar state is much lighter than the scalar one, $M_H \ll M_a$. The model can be described by the following Lagrangian
\beq
\mathcal{L}_\phi =\partial_\mu \phi \partial^\mu \phi^{*}+\mu_\phi^2
|\phi|^2-\lambda_\phi |\phi|^4 +\frac12 \epsilon_\phi^2 (\phi^2 +
\mbox{h.c.} ) \label{eq:lag_dyn_a} 
\eeq 
The scalar field $\phi$ is charged under a global U(1) symmetry spontaneously broken by a vev $v_\phi$, with $v^2_{\phi}=\mu_{\phi}^2/\lambda_{\phi}$, acquired by its scalar component which generates the mass term $M_H=\sqrt{2\lambda_\phi} v_\phi$ for the state $H$. The pseudoscalar component acquires a mass from the explicit mass term $\epsilon_{\phi}$, assumed to be such that $\epsilon_{\phi} \ll \mu_{\phi}$, so that $M_a=\sqrt{2} \epsilon_\phi$. In this construction, $a$ can be identified as a pseudo--Goldstone boson associated to the U(1) symmetry \cite{Mambrini:2015nza}.  
One can then again consider the case in which the field $\phi$ is coupled only with the SM gauge bosons, through the introduction of a sequential family of vector--like fermions \cite{Arcadi:2016dbl,Arcadi:2016acg}. The Lagrangian can be then written as: 
\begin{align}
-\mathcal{L}&= \frac12 M_H^2 H^2+ \frac12 M_a^2 a^2+ 
\sqrt{ \frac{\lambda_\phi} {2}}M_H H a^2 +\sqrt{\frac{\lambda_\phi}{2}}M_H
H^3+\frac14 \lambda_\phi ( H^2+a^2 )^2 \nonumber \\ 
& +\sum_F m_F \bar F F+ \frac{y_F}{\sqrt{2}} H \bar F F
 + i\frac{y_F}{\sqrt{2}}a \bar F \gamma_5 F \nonumber\\ 
& +c_{\rm gg}^H H G_{\mu \nu}G^{\mu \nu}+c_{\rm WW}^H H W_{\mu \nu}W^{\mu \nu}  +c_{\rm ZZ}^H H Z_{\mu \nu}Z^{\mu \nu}+c_{\rm Z\gamma}^H H F_{\mu \nu}Z^{\mu \nu} +c_{\rm \gamma \gamma}^H H F_{\mu \nu}F^{\mu \nu} \nonumber \\ 
& +c_{\rm gg}^a a G_{\mu \nu}\tilde{G}^{\mu \nu}+c_{\rm WW}^a a
W_{\mu \nu}\tilde{W}^{\mu \nu}+c_{\rm ZZ}^a a Z_{\mu \nu}\tilde{Z}^{\mu
\nu}+c_{\rm Z\gamma}^a a F_{\mu \nu}\tilde{Z}^{\mu \nu}+c_{\rm \gamma \gamma}^a
a F_{\mu \nu}\tilde{F}^{\mu \nu} .  
\end{align}
In this setup, it is interesting to further assume that the masses of the vector fermions also arise from the breaking of the extra U(1) symmetry so that one can write 
\begin{equation} 
y_F=\sqrt{2}\frac{m_F}{v_\phi}=2 \sqrt{\lambda_\phi}\frac{m_F}{m_H} . 
\end{equation}

This choice renders the model under consideration very predictive since there is
only one fundamental new coupling, namely $\lambda_\phi$. The Lagrangian
eq.(~\ref{eq:lag_dyn_a}) represents a first simple but theoretically consistent
realization of a scenario with a light pseudoscalar mediator. This type of model
is very interesting for several different reasons. First of all, as will be
illustrated in the following, it leads to many interesting and not yet fully
explored collider signatures. A light pseudoscalar mediator is also very
appealing for DM phenomenology as it can, indeed, sensitively impact the DM
annihilation processes while affecting direct detection prospects and
constraints only to a  marginal extent. 

We close this section with some remarks on eventual limits from renormalization
group evolution for the scenario with a full family of VLFs. The relevant 
equations \cite{Arcadi:2016acg} are totally analogous to the ones considered in
the SM Higgs case  and will be hence not rewritten here. Given the
relation between the Yukawa couplings and the quartic couplings $\lambda_\phi$,
it is possible to relate the scale $\Lambda_{\rm NP}$  at which the quartic
coupling $\lambda_\phi$ becomes negative, and the value $\lambda_\phi$ of the quartic coupling at the electroweak scale. To show this relation, we have made a similar study as the one done in Ref.~\cite{Arcadi:2016acg} and performed a scan over the $\lambda_\phi,M_H,M_a$ parameters over the following ranges
\begin{align}
    & M_H \in \left[200,2000\right],\,\mbox{GeV} \, , \ \
     M_a \in \left[0.2,2\right],\,\mbox{GeV} \, , \ \ 
     \lambda_\phi \in \left[10^{-4},4\pi\right] \, , 
\end{align}
assuming that all the vector fermions have the same mass $m_F$ of which four different values, namely $m_F=0.5,1,2$ and 5 TeV, have been considered. As additional requirements, we have imposed $y_F < 4 \pi$, $M_H< m_F$ and that the ratio $M_a/M_H$ respects the conditions for collimated photons (to be discussed later). For each model point, we have solved the renormalisation group equations determining the scale $\Lambda_{\rm NP}$ at which $\lambda_\phi <0$. 

\begin{figure}[!h]
\vspace*{2mm}
    \centering
    \includegraphics[width=0.55\linewidth]{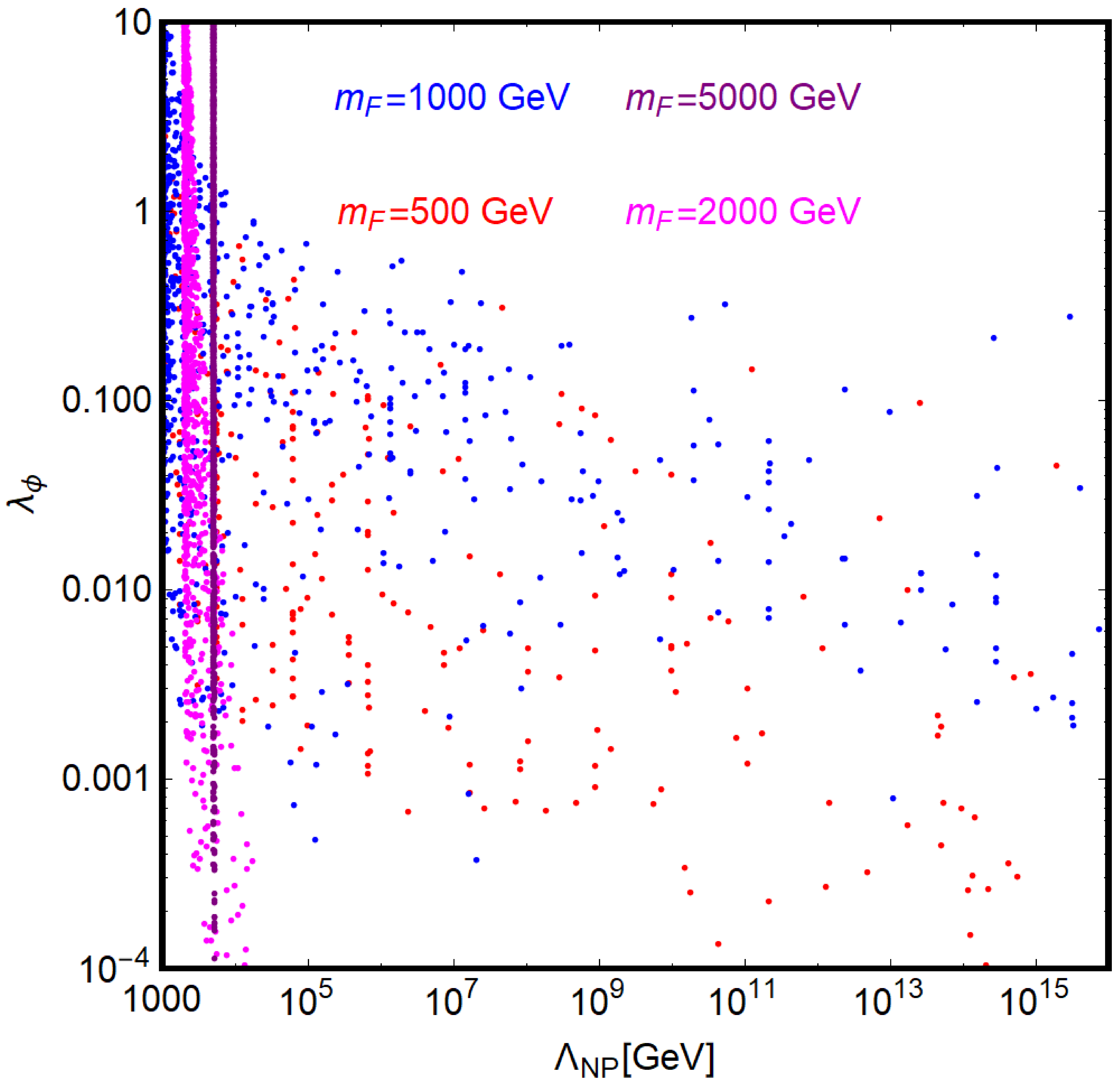}
\vspace*{2mm}
    \caption{Model points in the bidimensional plane $[\lambda_\phi,\Lambda_{\rm NP}]$ for a scalar plus a light pseudoscalar scenario with a vector fermion family which satisfy the theoretical constraints discussed in the text for several values of the common fermion mass $m_F$.}
    \label{fig:lam_dyn_RGE}
\end{figure}

The results of this analysis are shown in Fig.~\ref{fig:lam_dyn_RGE} in the
bidimensional plane $[\Lambda_{\rm NP},\lambda_{\phi}(M_Z)]$. The model points
have been marked with different colors according to the value of $m_F$. As can
be seen, for $m_F=0.5,1\,\mbox{TeV}$ very high values of $\Lambda_{\rm NP}$ can
be achieved, provided that the starting value of $\lambda_\phi$ is below 1.
Higher values of $m_F$ correspond, instead, to too high initial values of $y_F$,
which drive $\lambda_\phi$ to negative values already at the energy threshold
corresponding to the mass $m_F$ of the vector--fermions.

\newpage

\subsection{Constraints and expectations at colliders}

\subsubsection{The scalar Higgs with mixing}

We discuss now the possible collider constraints and the prospects for
observation in the scenarios with new scalar and pseudoscalar resonances just 
introduced before. In the case of a heavy scalar state which mixes with the
SM--like Higgs boson, a first constraint comes from the LHC data on the later.
Indeed, Higgs mixing will make that the couplings of the observed 125 GeV $h$
boson to fermions and gauge bosons, compared to those expected in the SM,  will
be multiplied by $\cos\theta$ and those of $H$ multiplied by $\sin\theta$. This results into $h$ production cross sections and decay branching ratios of
\bea
\begin{array}{c}
\sigma(h) = \cos^2 \theta \times \sigma(H_{\rm SM}) \\
{\rm BR} (h \! \to \! XX) \! =\!  {\rm BR} ( H_{\rm SM} \! \to \! XX) 
\end{array} \ \bigg\} \ 
\Rightarrow  \mu_{XX} = \cos^2 \theta \times \mu_{XX}|_{\rm SM}\; .
\eea
As a matter of fact, while the cross sections are suppressed by mixing, the
decay branching  ratios are not, as the factor $\cos\theta$ drops out in the
ratio of partial to total decay widths. The signal strength is hence only
suppressed by $\cos^2\theta$ compared to the SM expectation. Using the combined
total signal strengths of the 125 GeV Higgs as determined at RunI by ATLAS and CMS with all production and decay channels combined, eq.~(\ref{eq:mutotLHC}), one finds  
\beq 
\mu_{\rm tot}\leq 0.89~{\rm at}~95\%{\rm CL}\Rightarrow \sin^2\theta \leq 0.11 . \eeq
A first implication of such a result is that, as can be seen from an inspection of Figure~\ref{fig:lhs-stability}, the mass of the heavier $H$ state should be larger than $M_H \gsim 200\,(400)$ GeV for $|\lambda_{hH}| \simeq 0.1 \,(0.25)$, which justifies  our initial choice $M_H >125$ GeV. 

There are also constraints on $M_{H}$ and  $|\sin \theta|$ from the direct
searches of heavy Higgs bosons that have been performed by the ATLAS and CMS
collaborations in many channels such as $H\to WW,ZZ,\gamma\gamma$ as well as in 
$hh$ and $t\bar t$ final states. To discuss these, let us first summarize the
production rates and decay branching ratios of the $H$ state which, once the
mass $M_H$ and the mixing angle $\theta$ are known, are almost completely fixed.  
If $M_H$ is close to 125 GeV, the $H$ decay modes are similar to that of the
observed  $h$ boson if the invisible decays are assumed to be small or absent.
The main standard decays will be into $b\bar{b}$ followed by the decays into
$c\bar{c}$, $\tau^+\tau^-$ and $gg$ with branching ratios of the order of a few
percent. The $\gamma\gamma$ and $Z \gamma$ loop decay modes have small rates, a
few permile. The $H$ state will also decay into $WW$ and $ZZ$ pairs, one of the
gauge bosons being virtual. The former has a significant  branching ratio  at
$M_H \gsim 140$ GeV and becomes the dominant mode; in fact, in the mass range
$M_H=160$--180 GeV, it is the only relevant decay. Above $M_H \approx 180$ GeV, the $H$ state will mainly decay into real vector bosons, with fractions of  $\frac23$ for $WW$ and  $\frac13$  for $ZZ$ decays sufficiently above the thresholds. The opening of the $t\bar{t}$ decay channel for $M_H\gsim 350$ GeV does not alter  this pattern much as the branching ratio for this decay does not exceed the value of $\approx 20\%$  reached at $M_H \approx 400$ GeV and decreases with $M_H$ (the $t\bar t$ partial width is proportional to $M_H$
while it grows with  $M_H^3$ for the decays into $W,Z$ bosons as a result of
their longitudinal components). 

\begin{figure}[!h]
\vspace*{-3cm}
\centerline{ \includegraphics[scale=0.8]{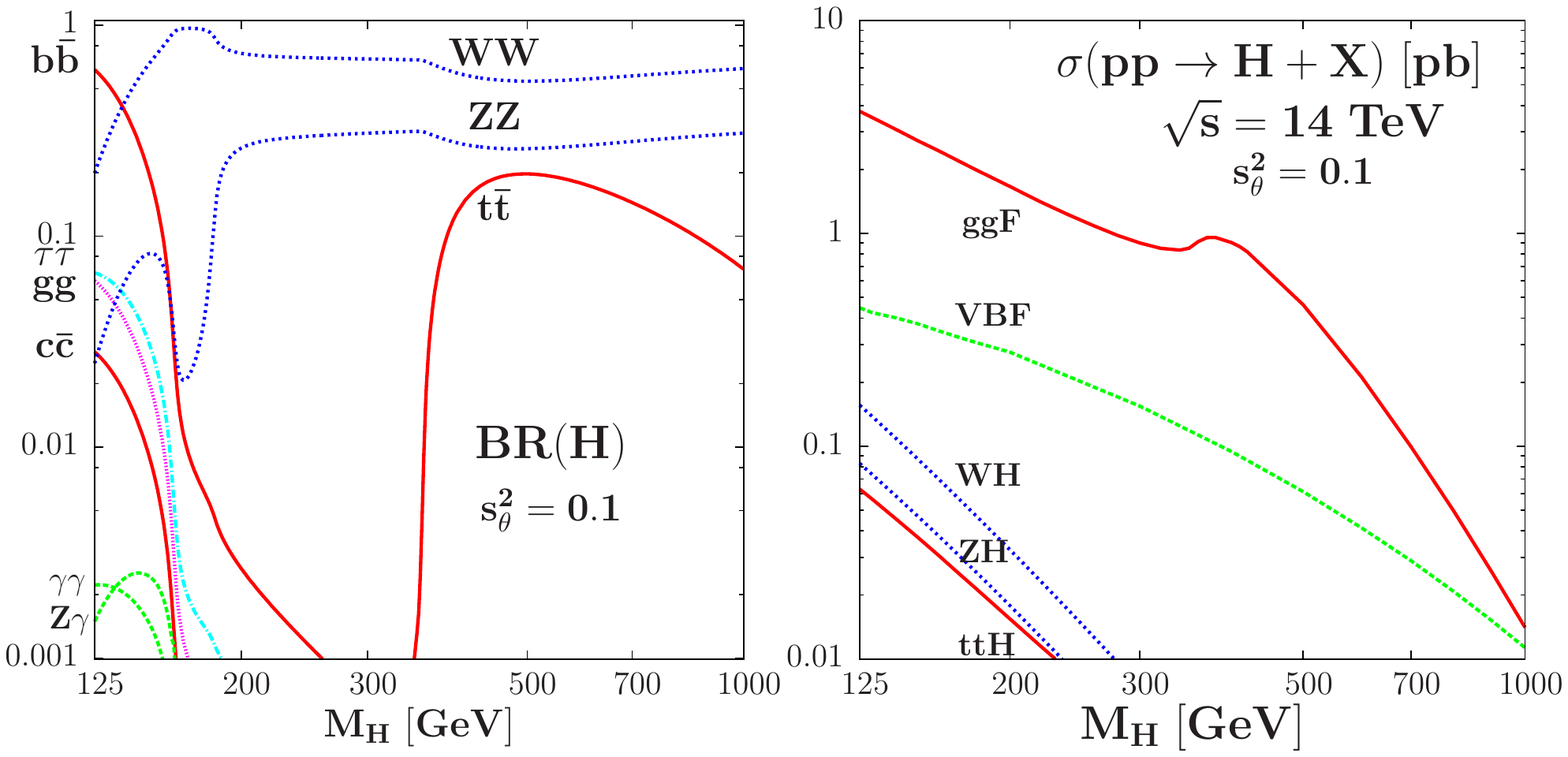} }
\vspace*{-13cm}
\caption{The decay branching fractions (left) and the cross sections at the LHC in the main production channels (right) for a heavy CP--even Higgs boson $H$ 
as a function of its mass and for a mixing angle $\sin^2\theta=0.1$.} 
\label{Fig:Hmixrates}
\vspace*{-1mm}
\end{figure}

The branching ratios, again obtained with the code {\tt HDECAY} 
\cite{Djouadi:1997yw,Djouadi:2006bz,Djouadi:2018xqq} adapted to this scenario,
are summarized in Fig.~\ref{Fig:Hmixrates} (left) as a function of $M_H$.  If
$M_H \lsim 180$ GeV, the $H$ state is very narrow with a total width of
$\Gamma_H \lsim 100$ MeV for $\sin^2\theta=0.1$ for instance, but the width
rapidly increases, reaching 50 GeV at $M_H=1$ TeV for the same mixing angle as a
result of the $M_H^3$ dependence of the $H\to WW,ZZ$ partial decay widths.
Nevertheless, thanks to the small mixing angle, the state does not become too
wide as it would have been the case of a SM--like Higgs for which one would have
$\Gamma_H \approx \frac12 M_H$ if $M_H \approx 1$ TeV).  

For the single production of the $H$ state at hadron colliders, the mechanisms
are again the same as for the SM--like Higgs boson.  There is first the $gg \to
H$ process  which proceeds through top (and to a lesser extent bottom quark)
loops and  is dominant at masses not too close to $M_H \approx 1$ TeV,  vector  
boson fusion $qq \to Hqq$ which has a one order of magnitude smaller rate than 
gluon--fusion for Higgs masses below 500 GeV but dominates for masses above 1
TeV,  and then come the Higgs strahlung processes $q\bar q\to VH$ with $V=W,Z$
and the associated production with top quark pairs $pp \to t\bar t H$ which have
reasonable production rates only for $M_H \lsim 200$--300 GeV for  $\sin^2\theta
\lsim 0.1$.  

The total cross sections at the LHC with $\sqrt s=14$ TeV, obtained with the 
programs of Ref.~\cite{Michael-web} which include the important higher order
corrections, are displayed in Fig.~\ref{Fig:Hmixrates} (right) as a function of
$M_H$ and again for a mixing angle $\sin^2\theta=0.1$. At low $M_H$ values, the
production rates are one order of magnitude smaller than for the production of
the 125 GeV Higgs state  and they  become increasingly smaller for a heavier $H$
state. Nevertheless, for not to small mixing angles,  they are  substantial
enough for the several searches that have been conducted at the LHC to be rather
constraining as is summarized below.  

\begin{figure}[!h]
\vspace*{-.2cm}
\centerline{\hspace*{-2mm}
\includegraphics[scale=0.44]{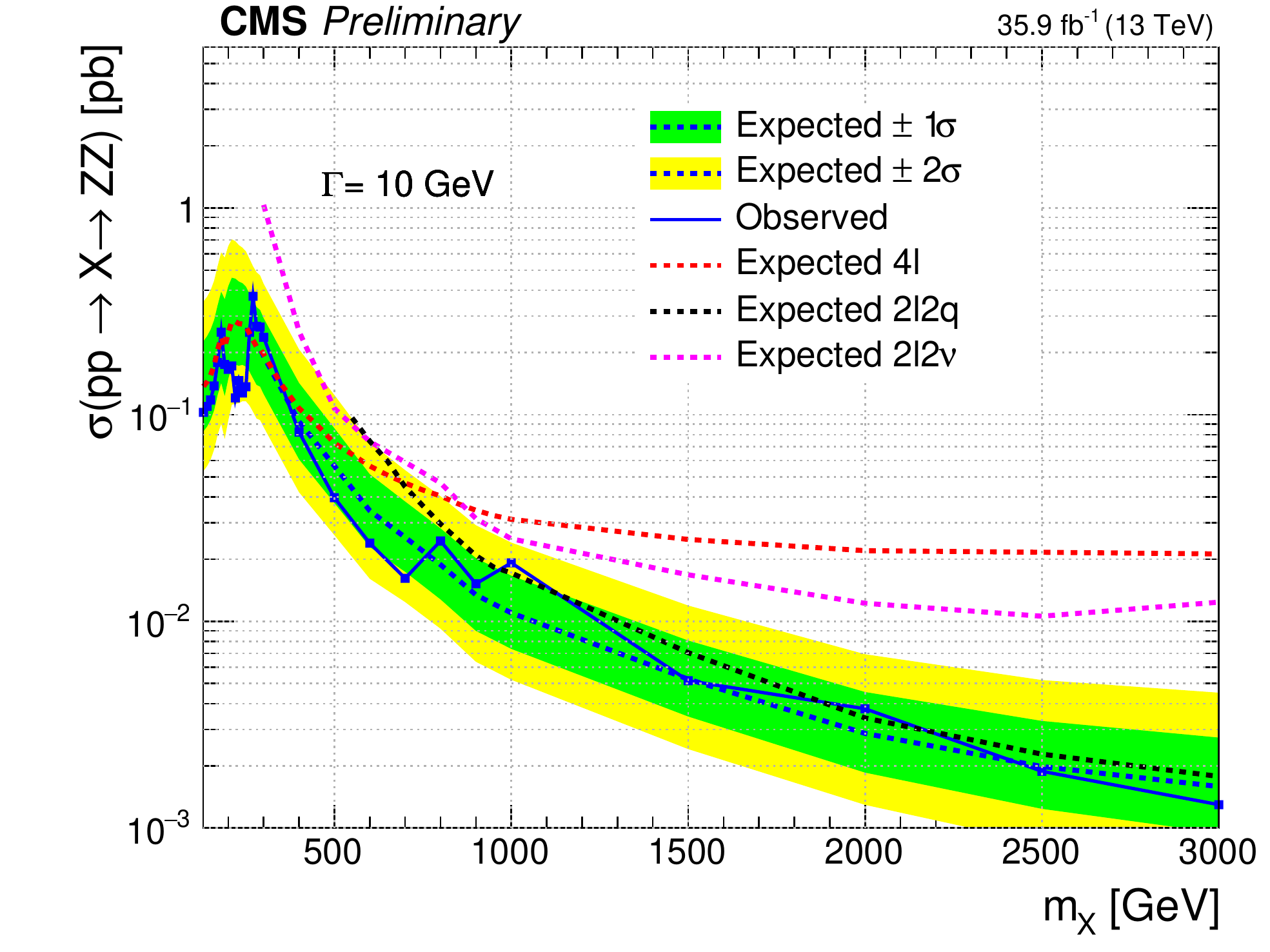}\hspace*{-2mm} 
            \includegraphics[scale=0.42]{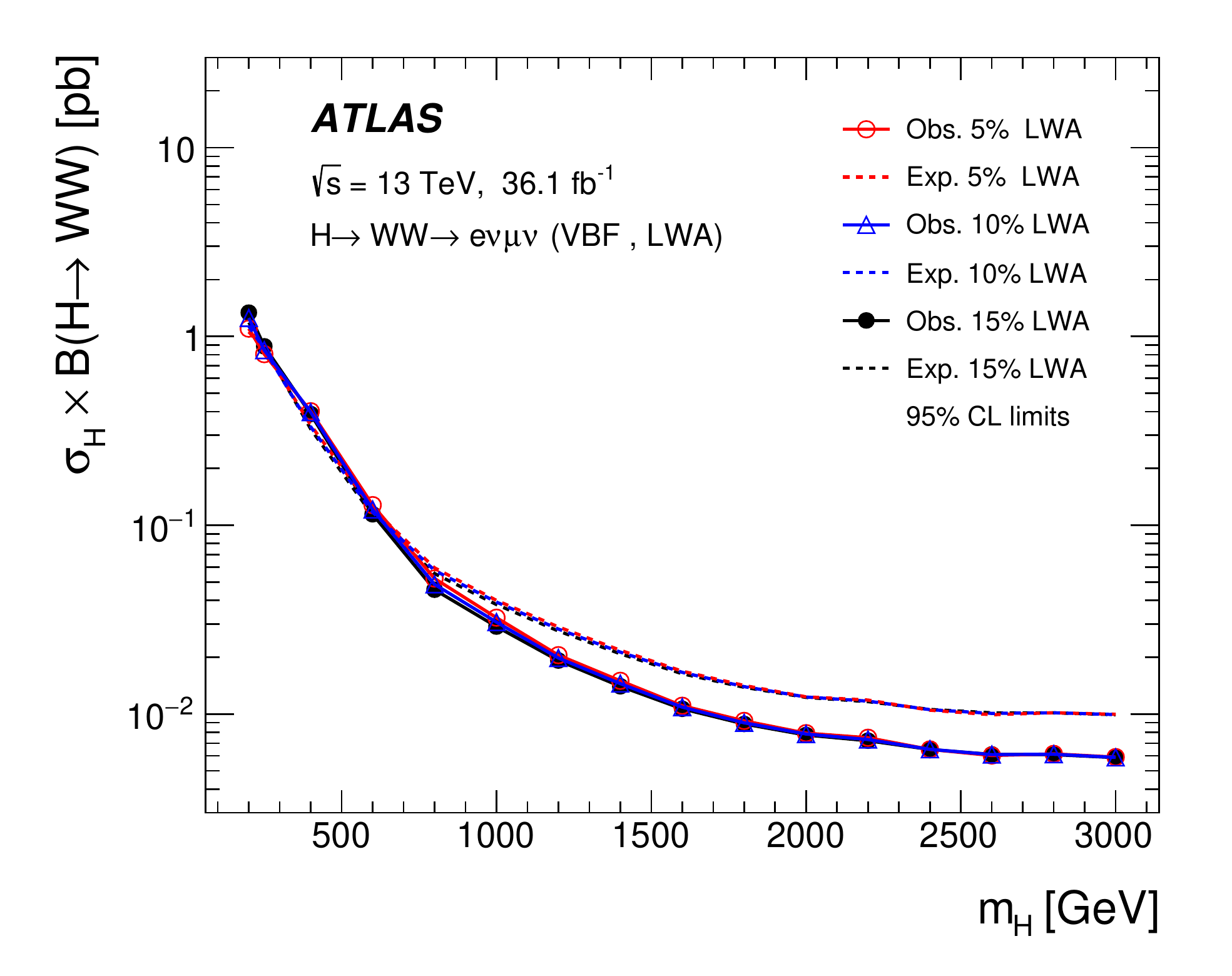} } 
\vspace*{-.1cm}
\caption{Expected and observed upper 95\%CL limits on the cross section times 
branching ratio as a function of the mass from two searches of a heavy Higgs boson at $\sqrt s=13$ TeV with about 36 fb$^{-1}$. Left: a CMS analysis in the  channel $pp\to H\to ZZ \to 4\ell, 2\ell 2q, 2\ell 2\nu$ separately and their combination for a width $\Gamma_H= 10$ GeV \cite{Sirunyan:2018qlb}. Right: an ATLAS analysis in the channel $gg\to H \to WW \to e\nu \mu \nu$ for a 
width of 5, 10 and 15\% of the Higgs mass \cite{Aaboud:2017gsl}.}
\label{Fig:pp-Hmix-VV}
\vspace*{-1mm}
\end{figure}

For $M_H \! \lsim \!  200$ GeV, the most promising searches for the $H$ boson
will be in the channels $gg \!\to \! H$ and $qq \! \to \! qqH$ with $H \! \to \!
WW$ and $H \! \to \! ZZ$ since  these decays are by far dominant if not
exclusive with rates of  respectively $\frac23$ and $\frac13$. In
Fig.~\ref{Fig:pp-Hmix-VV}, shown are the expected and observed 95\%CL upper
limits of the production cross section times the decay branching ratio as a
function of $M_H$ in these two channels.  The left panel shows a CMS analysis at
$\sqrt s=13$ TeV and 36 fb$^{-1}$ data of the  $H\to ZZ \to 4\ell, 2\ell 2q, 
2\ell 2\nu$ final states and their combination when the $H$ boson is produced
in  the ggF and VBF (with a small contribution of HV) processes and has a total
width of 10 GeV \cite{Sirunyan:2018qlb}. Cross sections at the 100 fb level are
excluded at low masses, meaning that $\sin^2\theta$ values smaller than 0.1 and
even 0.01 are excluded for $M_H \lsim 500$ GeV,  and the limit extends to 1 fb
at masses of 3 TeV.  In the right panel of Fig.~\ref{Fig:pp-Hmix-VV}, we show a
similar analysis performed by ATLAS \cite{Aaboud:2017gsl} at the same energy and
with a similar data sample but for the  process $qq \to qqH$ with $H \to WW \to
e\nu \mu \nu$. Here, the total Higgs width has been chosen to be 5, 10 and 15\%
of the Higgs mass. Once all leptonic channels have been added and when ggF
production is also included, the exclusion limits become comparable to those
derived in the $H\to ZZ$ mode.  

\begin{figure}[!h]
\vspace*{-.1cm}
\centerline{~~~\includegraphics[scale=0.36]{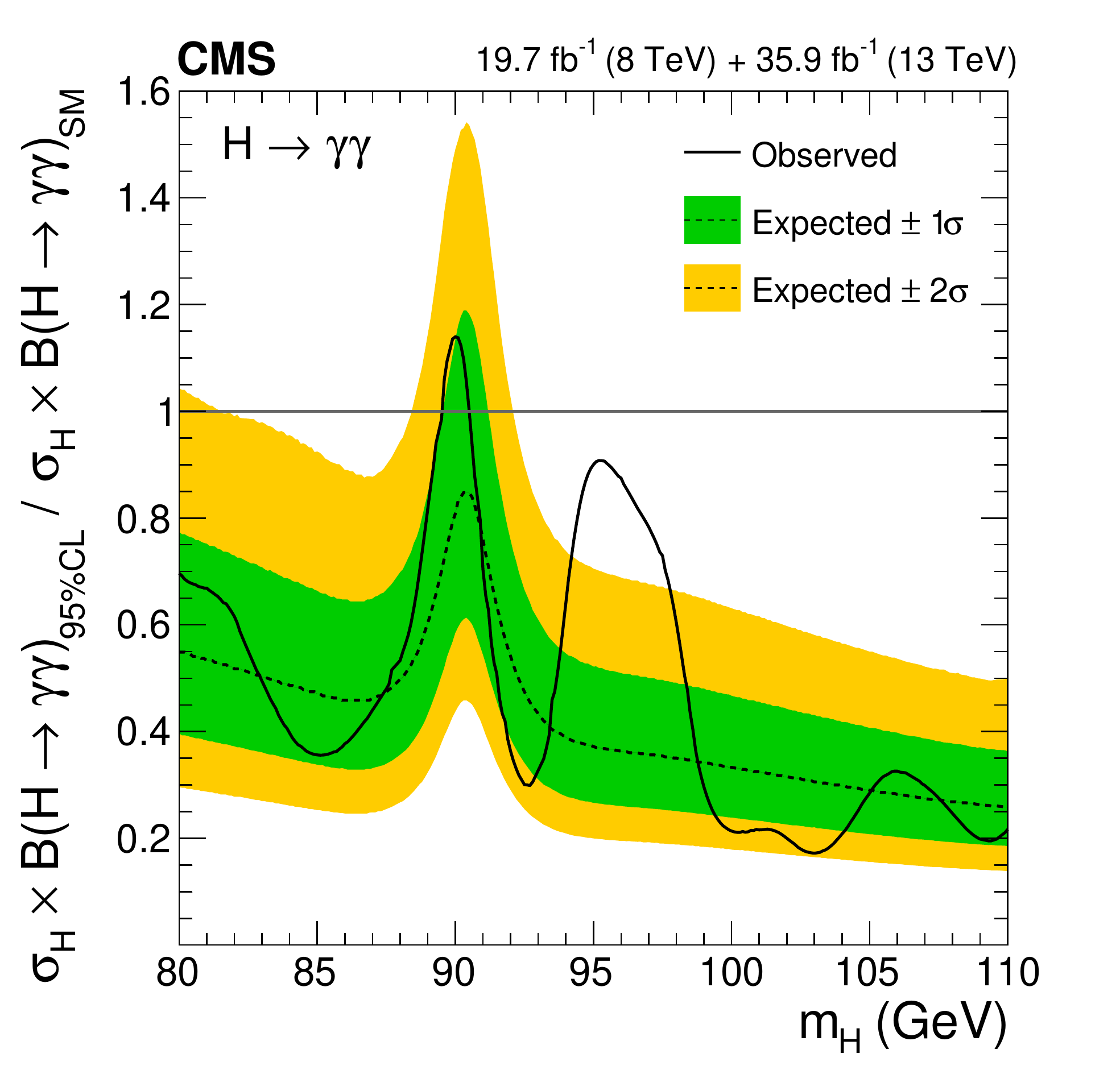}\hspace*{-8mm}
             \includegraphics[scale=0.48]{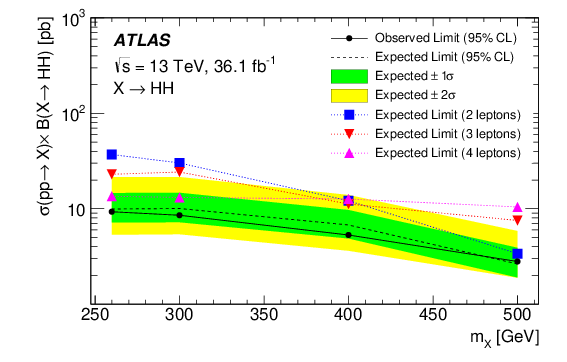} }
\vspace*{-.1cm}
\caption{Expected and observed 95\%CL exclusion limits on the product of the cross section and branching fraction as a function of the mass in two channels. Left: a CMS search at $\sqrt s=8$ and 13 TeV for the final state $H \to \gamma\gamma$ in the low mass range \cite{Sirunyan:2018aui}. Right: an ATLAS search in the channel $pp\to H\to hh$  at $\sqrt s=13$ TeV and 36 fb$^{-1}$ data in the 2, 3 and 4 lepton modes; shown are the individual limits and their combination \cite{Aaboud:2018ksn}.}
\label{Fig:pp-Hmix-others}
\vspace*{-2mm}
\end{figure}

Despite of the fact that the possibility $M_H \lsim 125$ GeV is highly
disfavored, it is nevertheless interesting to verify it experimentally,
especially that some excesses of events compatible with a Higgs mass slightly
below 100 GeV have appeared in the past, in particular at LEP2. Such low masses
cannot be probed using the $H\to WW^*$ and $ZZ^*$ modes which are too suppressed
by the Higgs virtuality, and the by far dominant $H \to b\bar b$ mode is of
little use since the production processes $q\bar q \to HW,HZ$ have too low rates
being damped by the small mixing. The most efficient channel is then the $H\to
\gamma\gamma$ mode which has a branching ratio at the permile level but one can
use all Higgs production mechanisms. An analysis of the CMS collaboration  in
this channel with the full set of RunI and 36 fb$^{-1}$ of RunII data 
\cite{Sirunyan:2018aui} is  given in the left panel of Fig.~\ref{Fig:pp-Hmix-others}. Shown again are the expected and observed 95\%CL
exclusion limits on the product of the $H$ production cross section times the
photonic branching fraction, compared to the SM value as a function of  the mass
in the range $M_H=80$--110 GeV. A local excess of approximately $2.8\sigma$ (but
only a $1.3 \sigma$ global excess mainly coming from the 13 TeV data) has been 
observed for a Higgs mass of approximately 95 GeV. This is rather close to the
value $M_H=98$ GeV for which a $2.3 \sigma$ local excess has been observed at
LEP in  the process $e^+e^- \to ZH \to Zb\bar b$ \cite{Barate:2003sz}. 

We come now to a channel that does not occur in the SM Higgs case, the  resonant
$pp\to H \to hh$ mode. In principle the $H\to hh$  branching ratio depends on
the coupling $\kappa_{Hhh}$ which is important at high $M_H$; but for these high
mass values, the partial widths for Higgs decays into $WW,ZZ$ bosons which grow
like  $M_H^3$, while it goes like $M_H$ for the former mode, are by far larger.
The mode $H\to hh$ can thus be important only for masses $M_H \lsim 500$ GeV. A
search for resonant $hh$ production in the topology $pp\to H\to hh \to WW^* WW^*
\to $ leptons has been performed by ATLAS at RunII with 36 fb$^{-1}$ data 
\cite{Aaboud:2018ksn} and the results, for $\sigma(pp \to H) \times {\rm
BR}(H\to hh)$ as a function of $M_H$ in the range 260--500 GeV, are shown in the
right--hand side of Fig.~\ref{Fig:pp-Hmix-others}. The individual limits in the
two, three and four lepton topologies and their combination are given. No excess
has been observed and a combined observed 95\%CL limit of  9.3 fb to 2.8 fb has
been set on the production times decay rate  for the two extreme Higgs mass
values, respectively 260 and 500 GeV. This corresponds to an observed 95\%CL
upper limit of 160 times the SM rate for non--resonant $hh$ production.

We finally come to the direct searches for invisible $H$ decays into DM
particles. The analytical expressions of the partial widths for the decays $H\to
XX$ where $X$ is the DM particle which can be a spin--0 $S$, a spin--$\frac12$
fermion $\chi$ or a spin--1 $V$ state, are exactly the same  as those given in
eq.~(\ref{GammaInv}) of section 2 for the SM Higgs boson but with $v \to
v_\phi$.  Only the Higgs to DM couplings in the case of fermion and vector DM
states have to be adapted:  $\lambda_{Hff}/\Lambda \to  m_f/v_\phi$  and
$\lambda_{HVV} = m_V^2/(v_\phi v)=2 \eta_V m_V/v$ since in our scenario the fermionic and
vector DM masses are dynamically generated by the vev of the additional singlet
field so that in these cases, only one additional free parameter is introduced
by the DM sector, namely the DM mass. In the scalar case, the Higgs--DM coupling
is given in eq.~(\ref{eq:bsmhSDMS}) and tends to zero if the mixing angle is
very small, $\sin\theta \to 0$. In any case,  for large $M_H$ values, they grow
like $1/M_H$ for a scalar DM, like $M_H$ in the fermionic case and $M_H^3$ for a
vector DM so that only in the latter  case that the invisible decay could
compete with the largely dominating  $H\to WW,ZZ$ modes that also grow like
$M_H^3 \! \times \! \sin^2\theta$. 

\begin{figure}[!h]
\vspace*{-.01cm}
\centerline{ \includegraphics[scale=0.57]{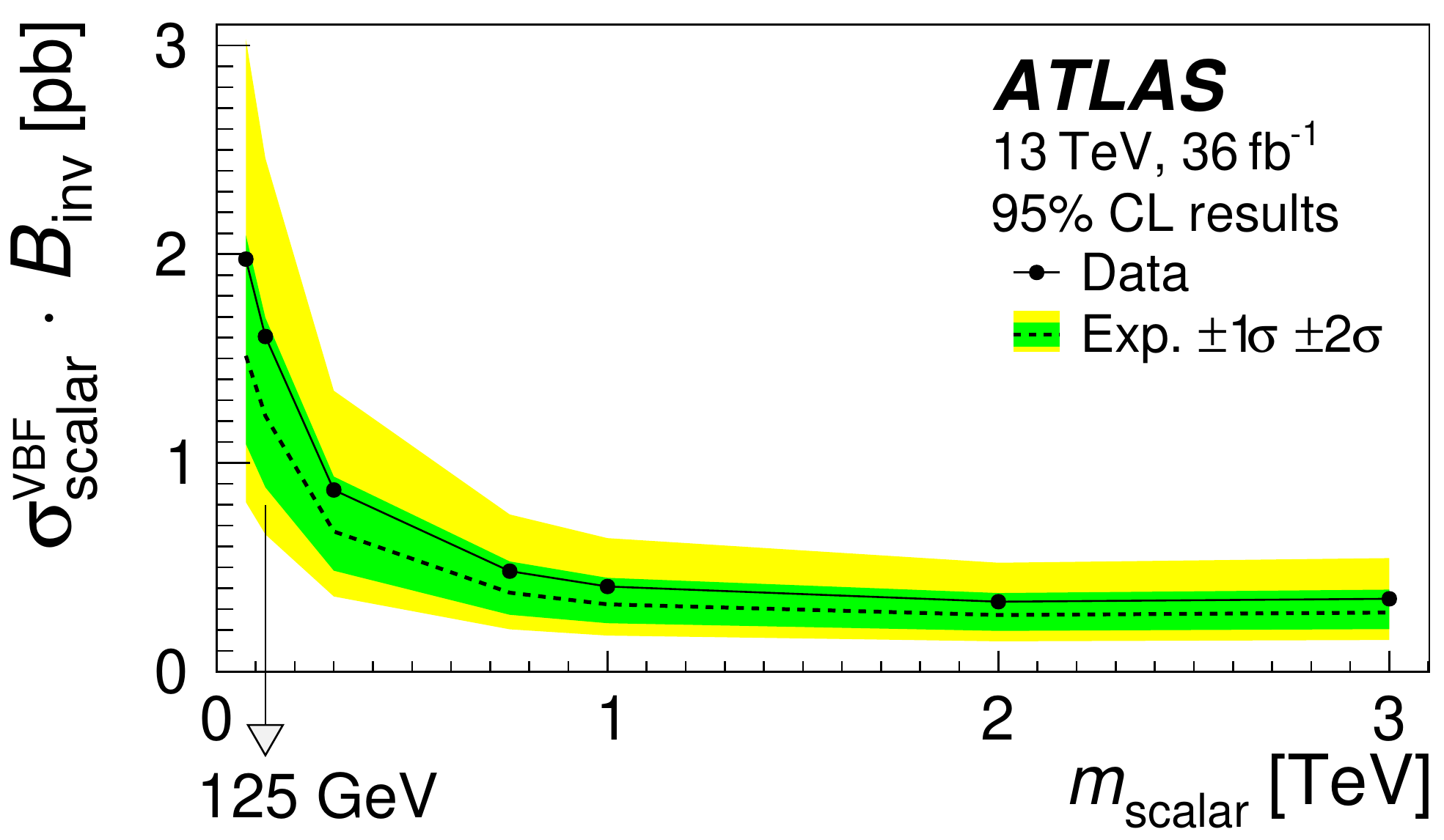}
}
\vspace*{-.01cm}
\caption{Left: The 95\%CL limit on the VBF cross section times the branching fraction to invisible decays of a heavy mixed Higgs boson as a function of its 
mass from an ATLAS analysis at $\sqrt s=13$ TeV and 36 fb$^{-1}$ data; from   Ref.~\cite{Aaboud:2018sfi}. 
}
\label{Fig:Hmix-BRinv}
\vspace*{-2mm}
\end{figure}

Direct searches for invisible decays of a heavy Higgs boson have been performed
at the LHC in various channels. In the left--hand side of
Fig.~\ref{Fig:Hmix-BRinv} we display an example of a search performed by ATLAS
in the VBF production mode at $\sqrt s=13$ and 36 fb$^{-1}$ data. The 95\%CL
exclusion limits on the production cross section $\sigma (qq \to qqH)$ times the
invisible branching  fraction BR($H \to XX)$ for decays into DM particles is
shown as a function of $M_H$. As can be seen, it ranges from 1pb  at $M_H=300$
GeV to 0.3 pb at $M_H=1$--3 TeV and, when confronted with
Fig.~\ref{Fig:Hmixrates} of the total cross section including the one for VBF
for $\sin^2\theta=0.1$, one sees that the search is not yet constraining for
values of the mixing angle allowed by indirect constraints.  A similar analysis
has been conducted by the CMS collaboration using  RunI data and the outcome has
been shown in Fig.~\ref{inv-exp}.

\begin{figure}[!h]
\begin{tabular}{ll}
\vspace*{-.1cm}
\begin{minipage}{8cm}
\hspace*{-5mm}
\centerline{ \includegraphics[scale=0.71]{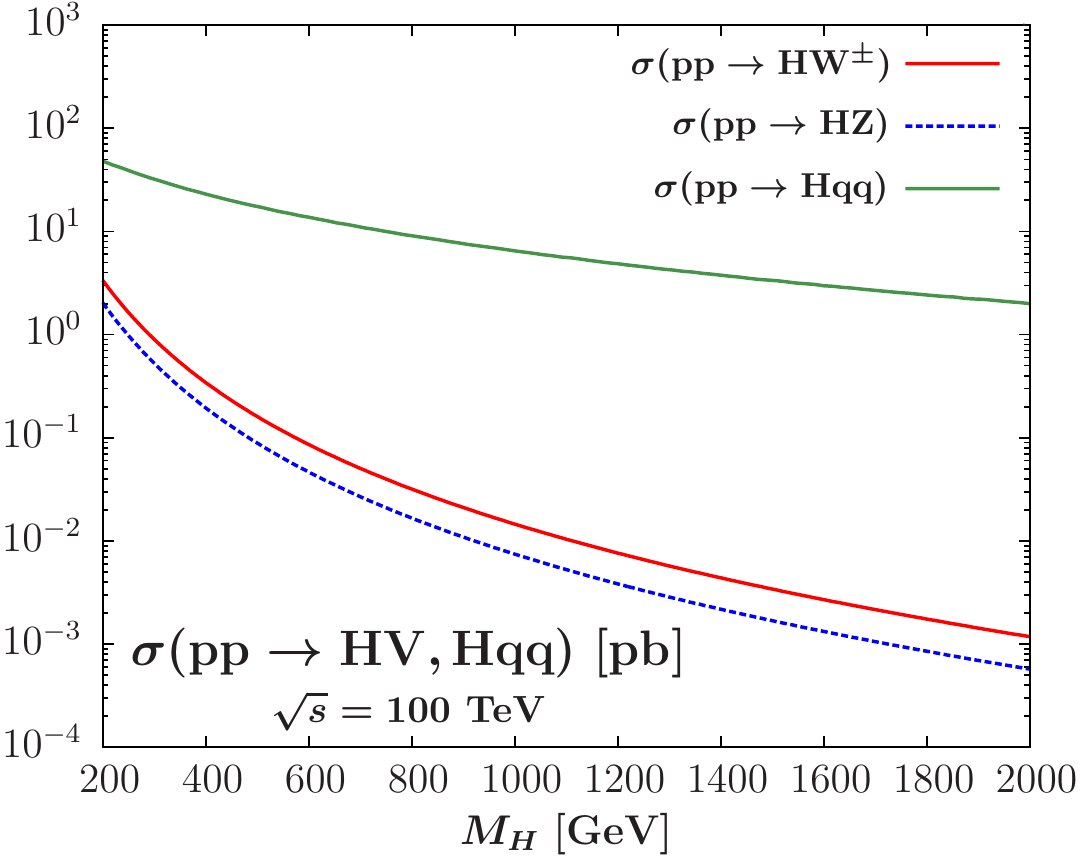}}
\end{minipage}
& \hspace*{-1.cm} 
\begin{minipage}{8cm}
\vspace*{-3.3cm}
\centerline{\includegraphics[scale=0.62]{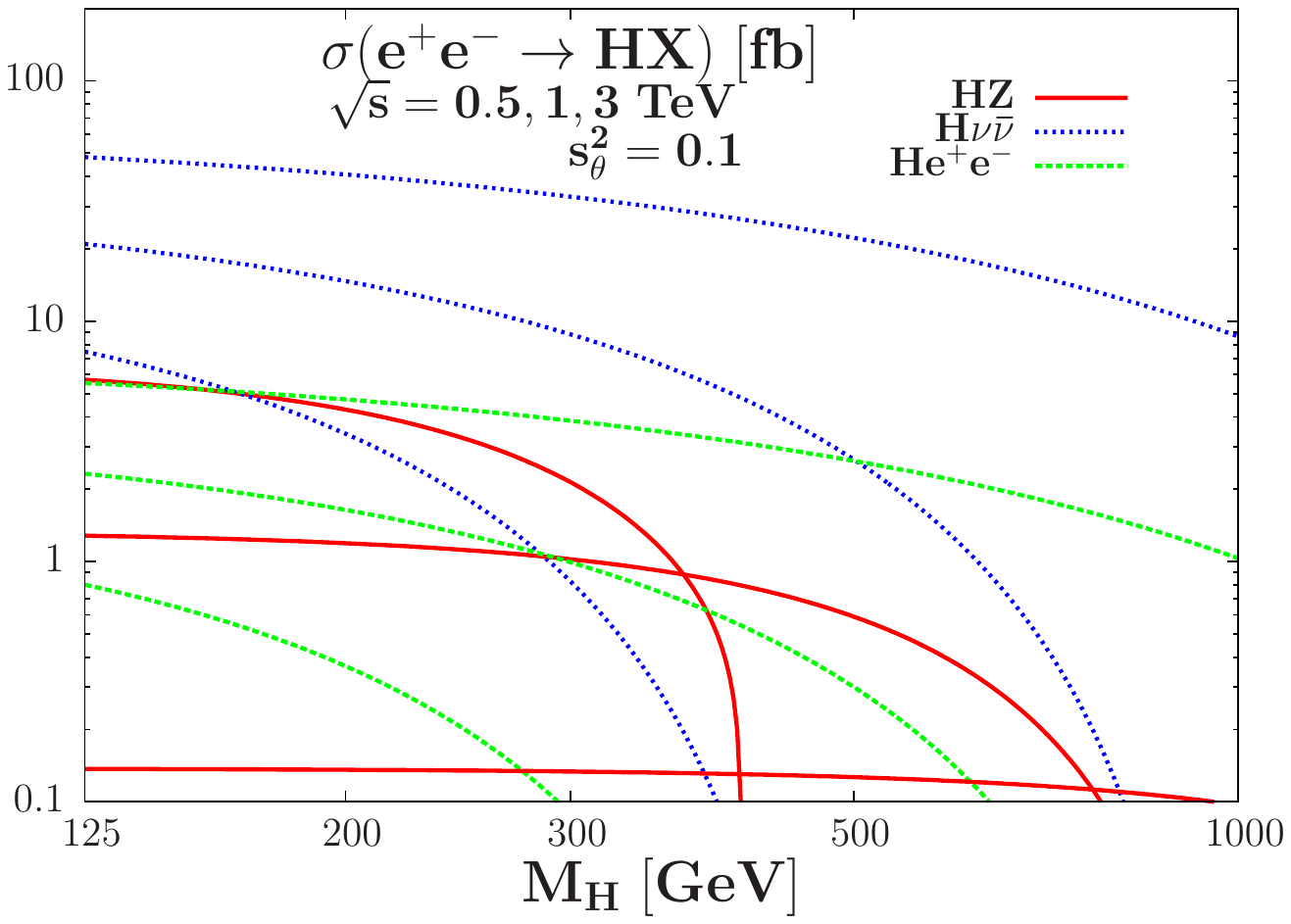} }
\vspace*{-11.1cm}
\end{minipage}
\end{tabular}
\vspace*{-.2cm}
\caption{Left: the cross section for Higgs production at a 100 TeV pp collider 
in the VBF and HV processes as a function of $M_H$ and no mixing angle suppression.  Right: cross sections for $H$ production in $e^+ e^-$ collisions in the main channels  as a function of $M_H$ and a mixing $s^2_\theta=0.1$ at three center of mass energies $\sqrt s=0.5, 1$ and 3 TeV.} 
\label{Hmix-prosps}
\vspace*{-2mm}
\end{figure}

These limits, if no signal is observed, will certainly be improved at the
high--luminosity option of the LHC with $\sqrt s\!=\! 14$ TeV and 3 ab$^{-1}$
data.  A vast improvement of the sensitivity could be achieved at higher
energy  pp colliders and, in particular, at 100 TeV where a two orders of
magnitude increase of the rates in the main production channels is
expected for a mass $M_H=1$ TeV, allowing to probe very small values of the
mixing angle.  An example of cross sections at $\sqrt s=100$ TeV is shown in the
left--hand side of Fig.~\ref{Hmix-prosps} for the VBF and HV channels as a
function of $M_H$  when no mixing suppression is present, $\sin^2\theta =1$.  

Finally, some remarks owe to be devoted to the prospects for Higgs production at
future $e^+e^-$ colliders. A heavy Higgs state can be produced in the  usual
$e^+e^- \to HZ$ channel discussed in section 2, but also in the $WW$ and $ZZ$
fusion processes,  $e^+e^- \to H \nu\bar \nu$ and $e^+e^- \to H e^+e^-$. In
fact, these processes are more interesting at high energies as  the cross
section rise like  $\log s/M_H^2$ in contrast to the Higgs--strahlung process for
which the rates drop as $1/s$. The cross sections for these production modes are
shown in the right--hand side of Fig.~\ref{Hmix-prosps} as a function of $M_H$
and for a mixing angle $\sin^2\theta=0.1$  at three cm. energies, $\sqrt s=0.5,
1$ and 3 TeV. As can be seen, for relatively small $M_H/\sqrt s$ values, the 
$e^+e^- \to H \nu\bar \nu$ process is by far dominant with extremely large cross
sections  while the mode $e^+e^- \to H e^+e^-$ has an order of magnitude lower
rate. The $e^+e^- \to HZ$ mode is only interesting at low $M_H$ and $\sqrt s$
values but allows many interesting measurements in case of discovery, as it has
been discussed in the SM--Higgs case in section 2.

\subsubsection{Singlet scalar or pseudoscalar resonances}

Let us now consider the extension of the SM Higgs sector in which the  $\phi$
resonance is an isospin singlet scalar or pseudoscalar state that does not mix
with the SM Higgs boson\footnote{These singlet models gained some popularity
when a significant excess of diphoton events (which turned to be a statistical
fluctuation)  was observed at the end of RunI by both ATLAS and CMS at an
invariant mass of about 750 GeV; we will thus use such a mass value as an
example in several instances.}. In this case, the DM  particle will be assumed to be  fermionic: either a singlet heavy neutrino  or  the electrically neutral
member of a full vector--like family of quark and leptons. At tree--level, the
$\phi$ state couples only to the top quark in the first case or the VLFs in the
second one, while the $\phi$ couplings to gauge bosons are generated through the
exchange of these heavy fermions.  We discuss below two important ones: the
$\phi gg$ coupling which allows the production of the $\phi$ state at proton
colliders in the by far dominant gluon fusion mechanism $gg \to \phi$ and  the
$\phi \gamma \gamma$ coupling which allows its detections in the cleanest
possible decay channel $\phi \to \gamma \gamma$ both at colliders and in
astroparticle physics experiments. 

Considering the effective Lagrangian eq.~(\ref{eq:phi-couplings}) in the scalar
and pseudoscalar cases,  the partial widths of the $\phi=H/A$ resonance decays into two gluons and two photons read 
\beq 
\Gamma (\phi \to \gamma \gamma) = \frac{(c^\phi_{\gamma\gamma})^2}{64 \pi^2} M_\phi^3    \, , \ \ \  
\Gamma (\phi \to g g) = \frac{(c^\phi_{gg})^2}{8 \pi^2} M_\phi^3\, . 
\eeq
If only these two decays are present, or when the gluonic decay is by far dominant compared to the electroweak ones, one would have a branching ratio for the photonic decay 
\beq 
{\rm BR}(\phi  \to \gamma \gamma) = \frac{ \Gamma (\phi \to \gamma \gamma) }
{\Gamma (\phi \to \gamma \gamma) + \Gamma (\phi \to g g) } \approx 
\frac{ \Gamma (\phi \to \gamma \gamma) }{ \Gamma (\phi \to g g)  } \approx 
\frac18 \frac{ (c^\phi_{\gamma\gamma})^2  }{ (c^\phi_{gg})^2 } \, ,
\eeq
leading to  ${\rm BR}(\phi  \to \gamma \gamma) \approx 10^{-1}$ if $c^\phi_{ 
\gamma\gamma} \approx  c^\phi_{gg}$. However, $c^\phi_{gg}$ is in principle an 
order of magnitude larger than $c^\phi_{ \gamma\gamma}$  as it involves the
strong interaction coupling instead of the electromagnetic one,
eq.~(\ref{eq:phi-couplings_2}). Note also that, in general, decays into $WW,ZZ$ 
and $Z\gamma$ final states also occur through similar effective couplings given
by eq.~(\ref{eq:phi-couplings}) and we will use later $c^\phi_{WW} =c_2$ and 
$c^\phi_{BB} =c_1$ for the resonance couplings to the SU(2) and U(1) fields. 

\begin{figure}[!h]
\vspace*{-2.1cm}
\centerline{
\includegraphics[scale=0.82]{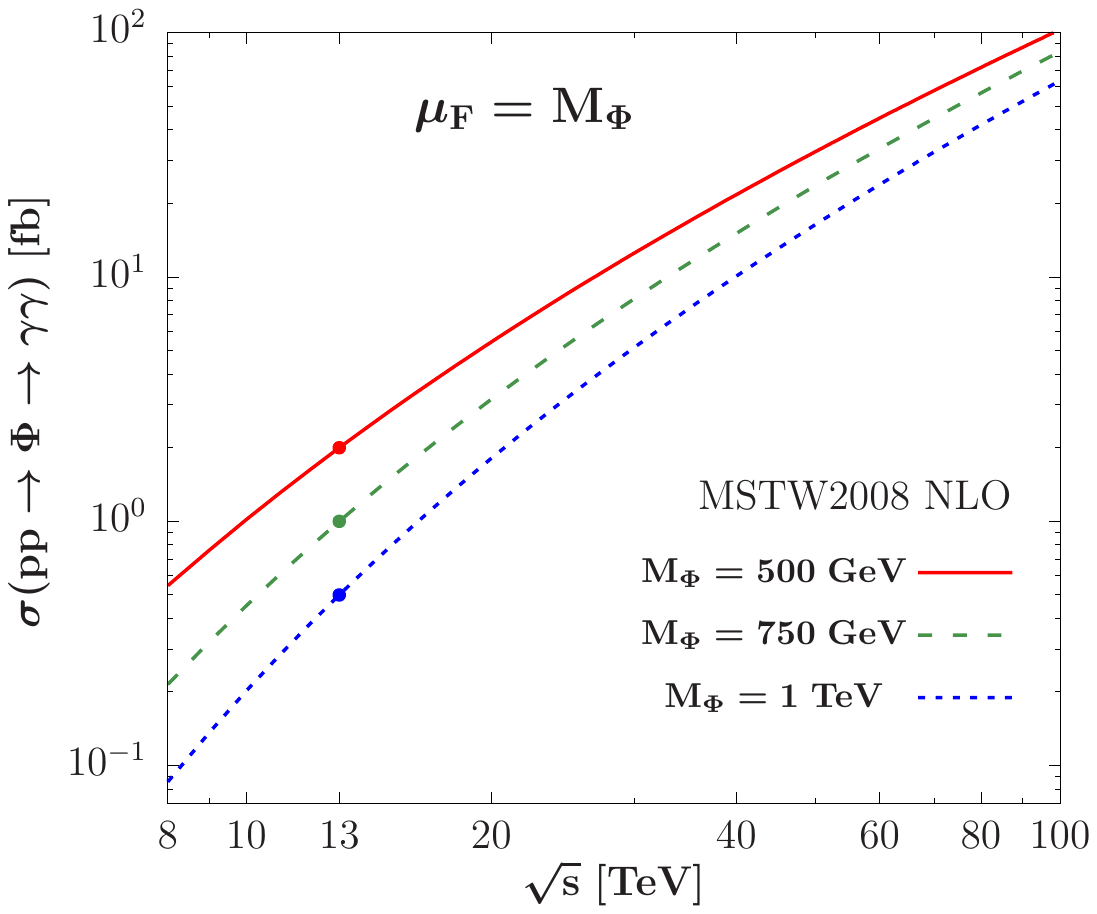} 
}
\vspace*{-14.6cm}
\caption{Cross sections for producing a singlet $\phi$  with a mass
$500,750,1000$~GeV decaying into two photons in $gg$ fusion at a $pp$ collider
as a function of the energy; from  Ref.~\cite{Djouadi:2016eyy}.}
\label{fig:ggPhi}
\end{figure}

For the production of the $\phi$ resonance at $pp$ colliders, one should focus  
on the gluon fusion mechanism $gg\to \phi$ as additional processes likes
Higgs--strahlung $q\bar q\to \phi W, \phi Z$ (as well as $\phi \gamma$) and
vector boson fusion $qq \to \phi qq$ involve the electroweak $\phi VV$ couplings
and will have much smaller cross sections in principle. At leading--order, the
cross section $\sigma(gg\to \phi)$ of the partonic subprocess is proportional to
the  $\phi \to gg$ partial width: \begin{eqnarray}   \sigma(pp \to \phi) =
\frac{1}{M_\phi s} C_{gg} \Gamma(\phi \to gg) \, :~ C_{gg} =  \frac{\pi^2}{8}
\int_{M_\phi^2/s}^{1} \frac{dx}{x} g(x)g(\frac{M_\phi^2}{s x} ) \, ,  
\end{eqnarray}   where $g(x)$ is the gluon PDF inside the proton at a
factorization scale $\mu_F$. Assuming that the $\phi$ state will be detected
through its clean  photonic decay mode, the $gg \to \phi \to \gamma \gamma$
production cross section times branching ratio at different c.m. energies of the
$pp$ collider can be obtained directly from a value of its rate  at a given
energy simply by rescaling the $gg$ luminosity. Assuming, for instance,  this
rate to be $\sigma \times {\rm BR}=1$~fb at $\sqrt s=13$ TeV for a resonance
with a mass of $M_\phi=750$ GeV, it is shown in  Fig.~\ref{fig:ggPhi} for mass
values of  $M_\phi=500,750$ and $1000$ GeV as a function of the collider energy
using the  MSTW2008 NLO PDFs ~\cite{Martin:2009iq} for the choice  of the
factorization scale $\mu_F= M_\phi$.  One notices that for these $M_\phi$
values, the cross sections grow by a modest factor $\simeq 1.2$ from 13 to
14~TeV, but by larger factors $\sim 10 \, (84)$ at $\sqrt s= 33\,  (100)$ TeV
which correspond to the energies considered for the HE--LHC~\cite{Baur:2002ka},
SPPC or FCC--hh
~\cite{CEPC-SPPCStudyGroup:2015csa,Contino:2016spe,Mangano:2017tke}. The
uncertainty associated with the variation in $\mu_F$ in the range $\mu_F = 2
M_\phi$ and $\mu_F = \frac12 M_\phi$ is about $20$\% at 100~TeV and there is an
additional uncertainty of about $20$\% again associated with different choices
of the PDFS that are recommended by the LHC Higgs working group
\cite{Dittmaier:2012vm}. 

\begin{figure}[!h]
\vspace*{-.2cm}
\centerline{
\includegraphics[scale=0.43]{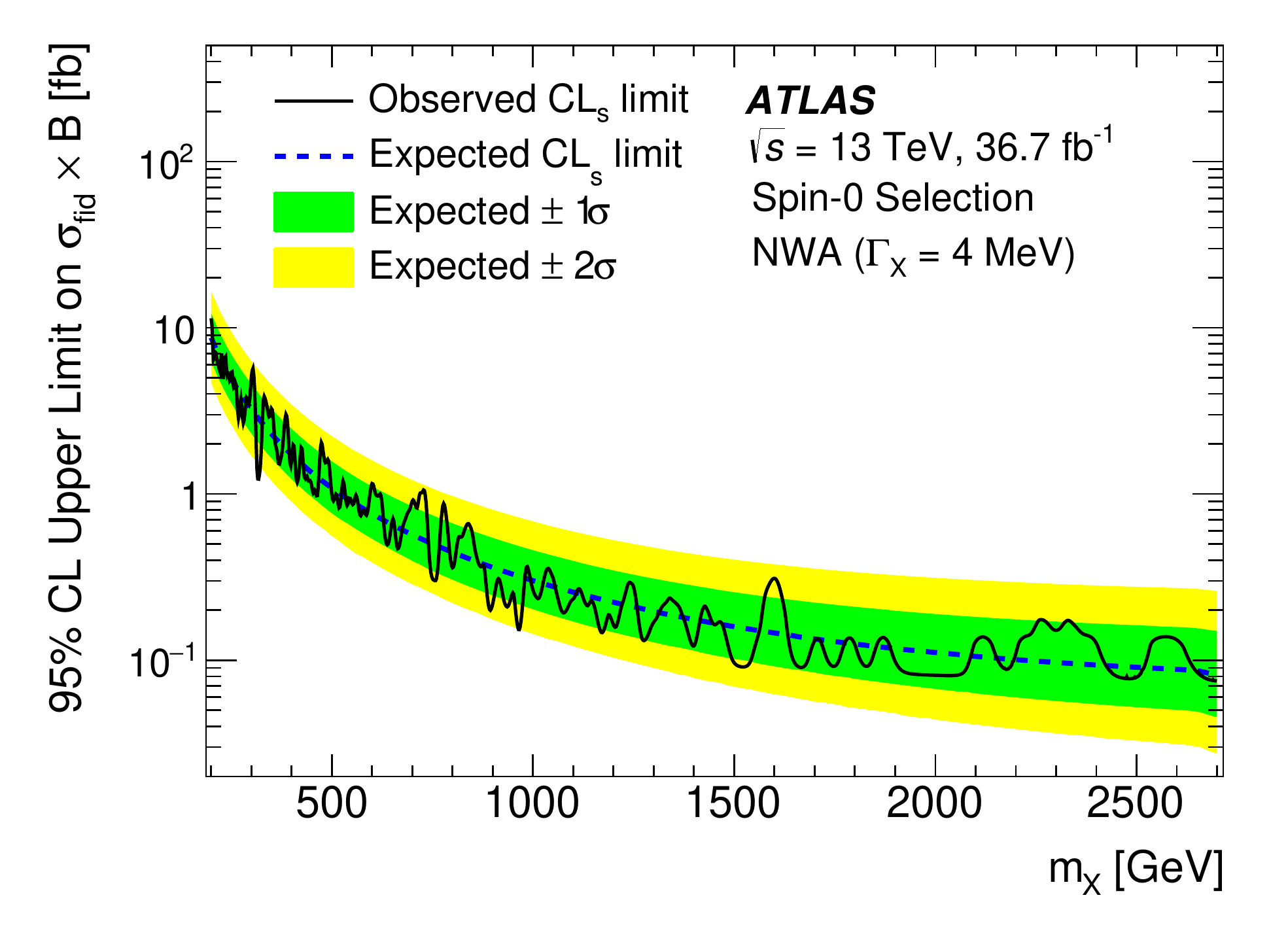}\hspace*{-.4cm}
\includegraphics[scale=0.43]{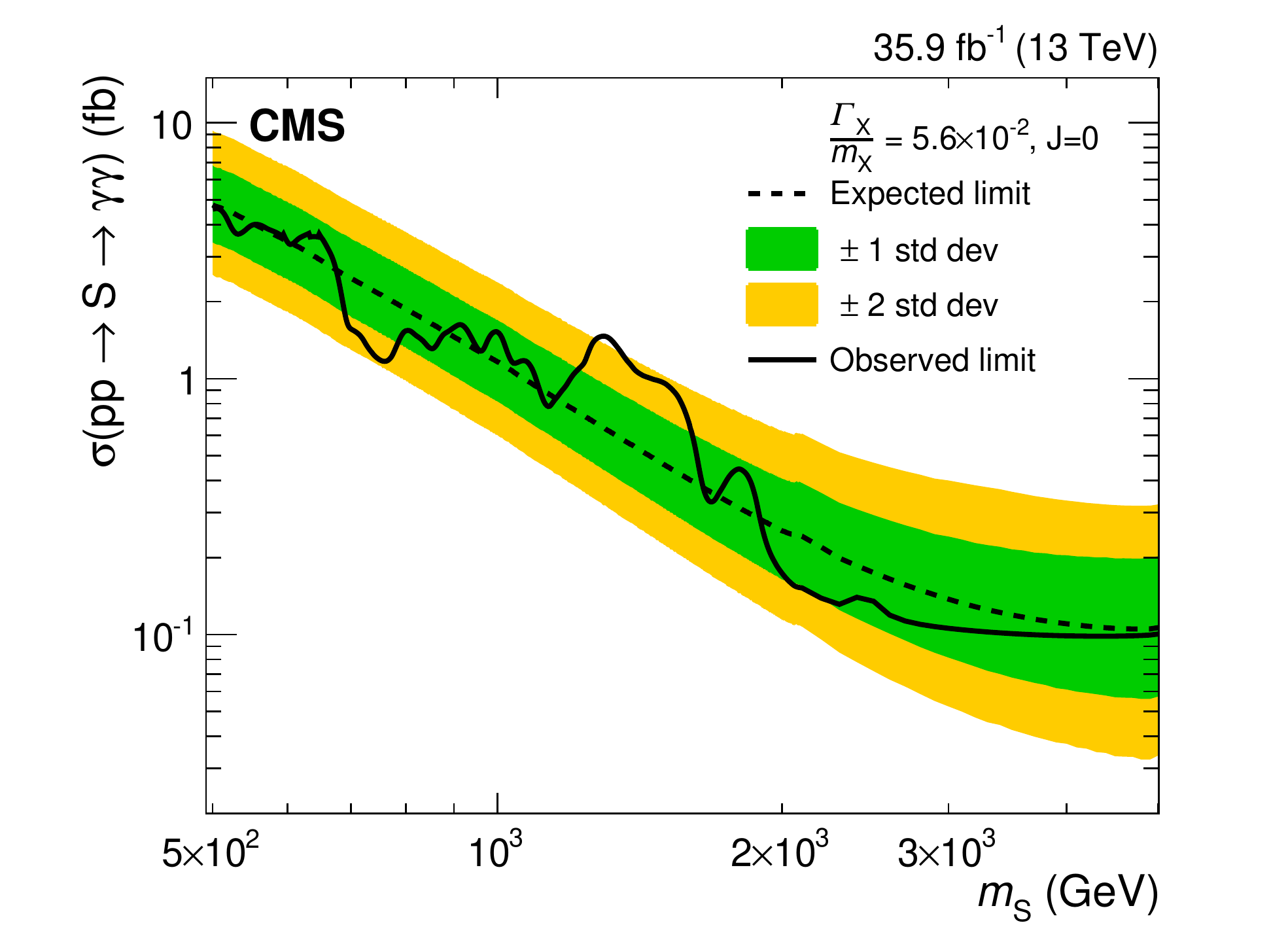}
}
\vspace*{-.3cm}
\caption{95\%CL upper limits on the gluon--gluon fusion cross section times the two photon branching ratio at $\sqrt s = 13$ TeV with about 36 fb$^{-1}$ data for a spin--zero resonance as a function of its mass. The ATLAS analysis (left) assumes a width $\Gamma_\phi = 4$ MeV \cite{Aaboud:2017yyg} and the CMS one (right) considers a width to mass ratio of  $\Gamma_\phi/M_ \phi=5.6 \times 10^{-2}$ \cite{Sirunyan:2018wnk}. }
\label{fig:Hsing-ggpp}
\vspace*{-.3cm}
\end{figure}

Diphoton resonances have been searched for by the ATLAS in CMS teams and, except
for the notorious bump at an invariant mass of about 750 GeV observed at RunI
which faded away with more statistics, no significant excess has been observed,
setting strong limits on heavy scalar or pseudoscalar states (as well as higher
spin resonances such as spin--one $Z'$ bosons and spin--two Kaluza--Klein
gravitons). Example of analyses are displayed in Fig.~\ref{fig:Hsing-ggpp} from
searches by ATLAS (left) and CMS (right) at $\sqrt s=13$ TeV and about 36
fb$^{-1}$ data. The expected and observed exclusion limits and their $1\sigma$
and $2\sigma$ bands for spin--0 resonances produced in gluon--gluon fusion and
decaying in two photons are shown as a function of the mass. In the  ATLAS case,
very narrow resonance with a constant width of 4 MeV is assumed, approximately
corresponding to a $\phi$ state coupling only with top quarks with a not too
large Yukawa. In the CMS case, the width to mass ratio is taken to be equal to
5.6\%, which can be reached e.g. if the resonance decays into top quarks with
a large Yukawa coupling. For a 1 TeV resonance, cross sections times branching
ratios below $\approx 0.3$ fb for a narrow width and  $\approx 1$ fb for a large
width have been excluded.

Nevertheless, these types of searches involve many assumptions and have several
caveats. A first one is that  the resonance signal and the QCD background cannot
be separated and in fact interfere. Indeed, at leading  order,  the  process $gg
\to \gamma\gamma$  receives  contributions from both the $gg \! \to \! \phi \! =
\! H,A \! \to \!  \gamma\gamma$ signal  and the  $gg\to \gamma\gamma$ continuum
background consisting of a box diagram in which the two photons are radiated
from the internal quark lines. Both types of diagrams lead to contributions that
have  an imaginary part: in the signal if there is a fermion in the triangular
$gg\phi$ and $\phi \gamma\gamma$ loops that has a mass $m_F \leq \frac12 M_\phi$
(this would be e.g. the case of the top quark if it couples to a $\phi$ state
with a mass above 350 GeV) and in the case of the background as the main
contribution in the box diagram will be due to light quarks. Furthermore, the
interference will not only affect the signal to background ratio, but it will
also  significantly change the line--shape  of the $\phi$ resonance. 

\begin{figure}[!h]
\vspace*{-.2cm}
\centerline{\hspace*{-5mm}
\includegraphics[scale=0.32]{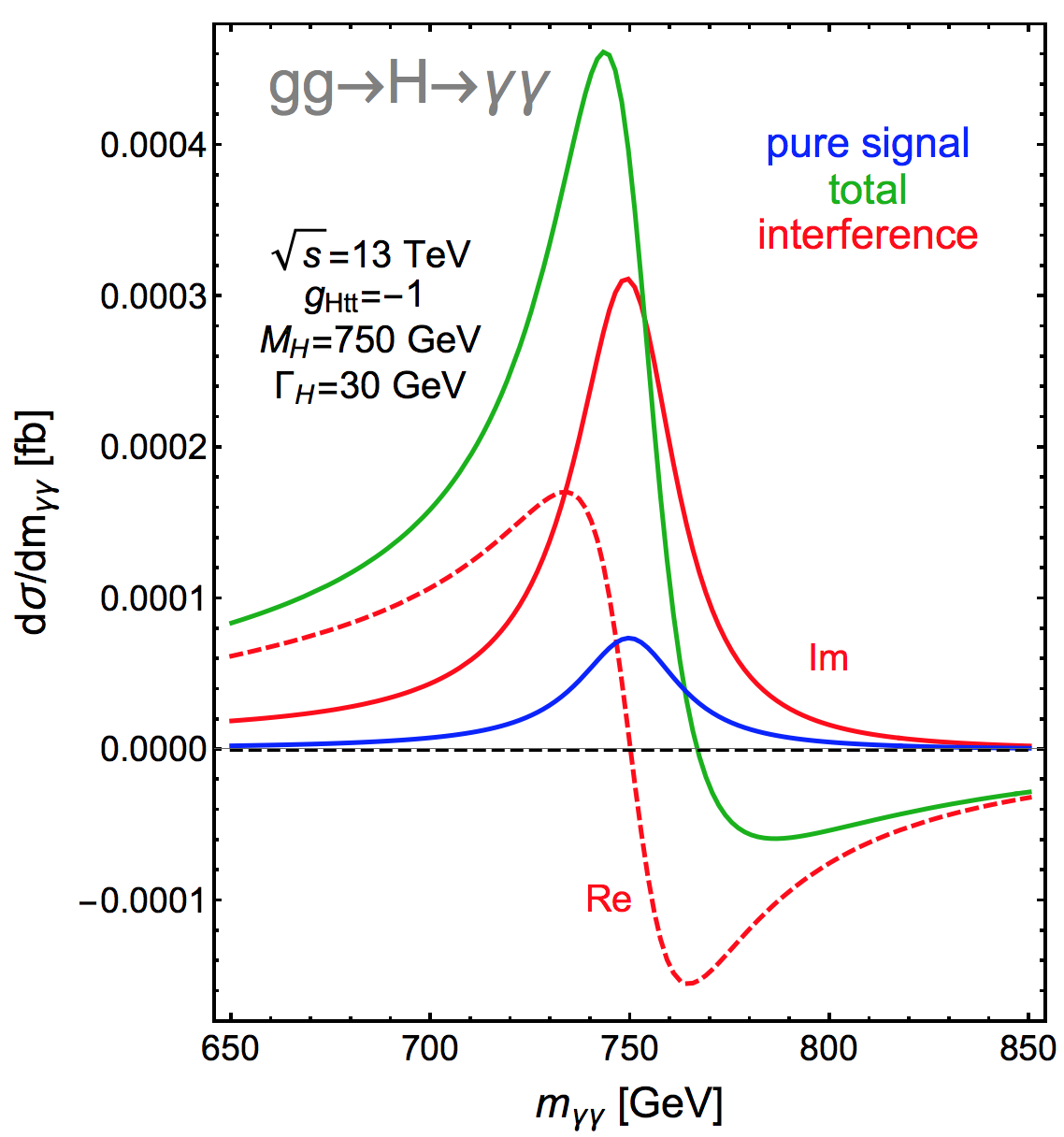}~~~
\includegraphics[scale=0.16]{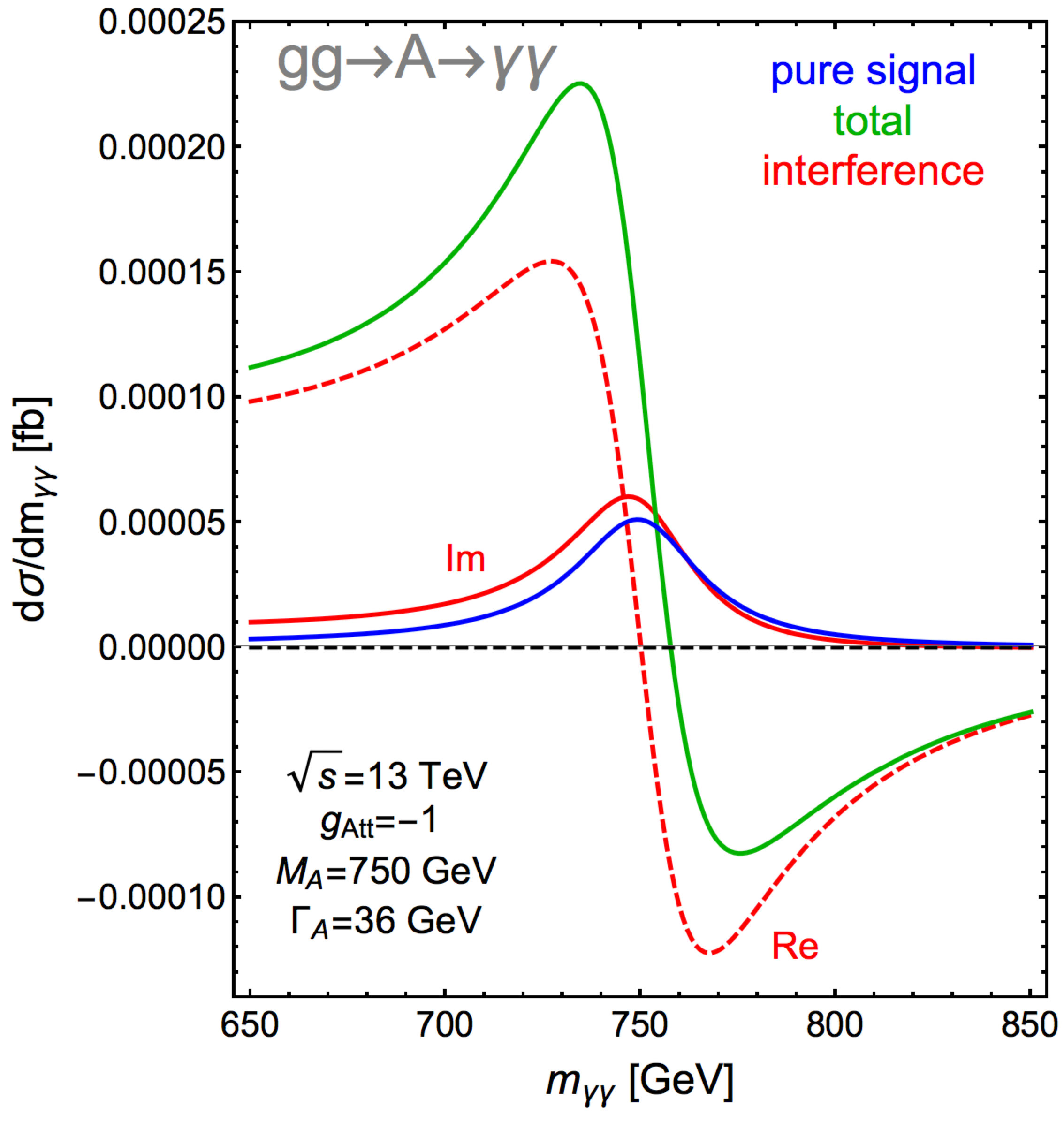}}
\centerline{
\includegraphics[scale=0.30]{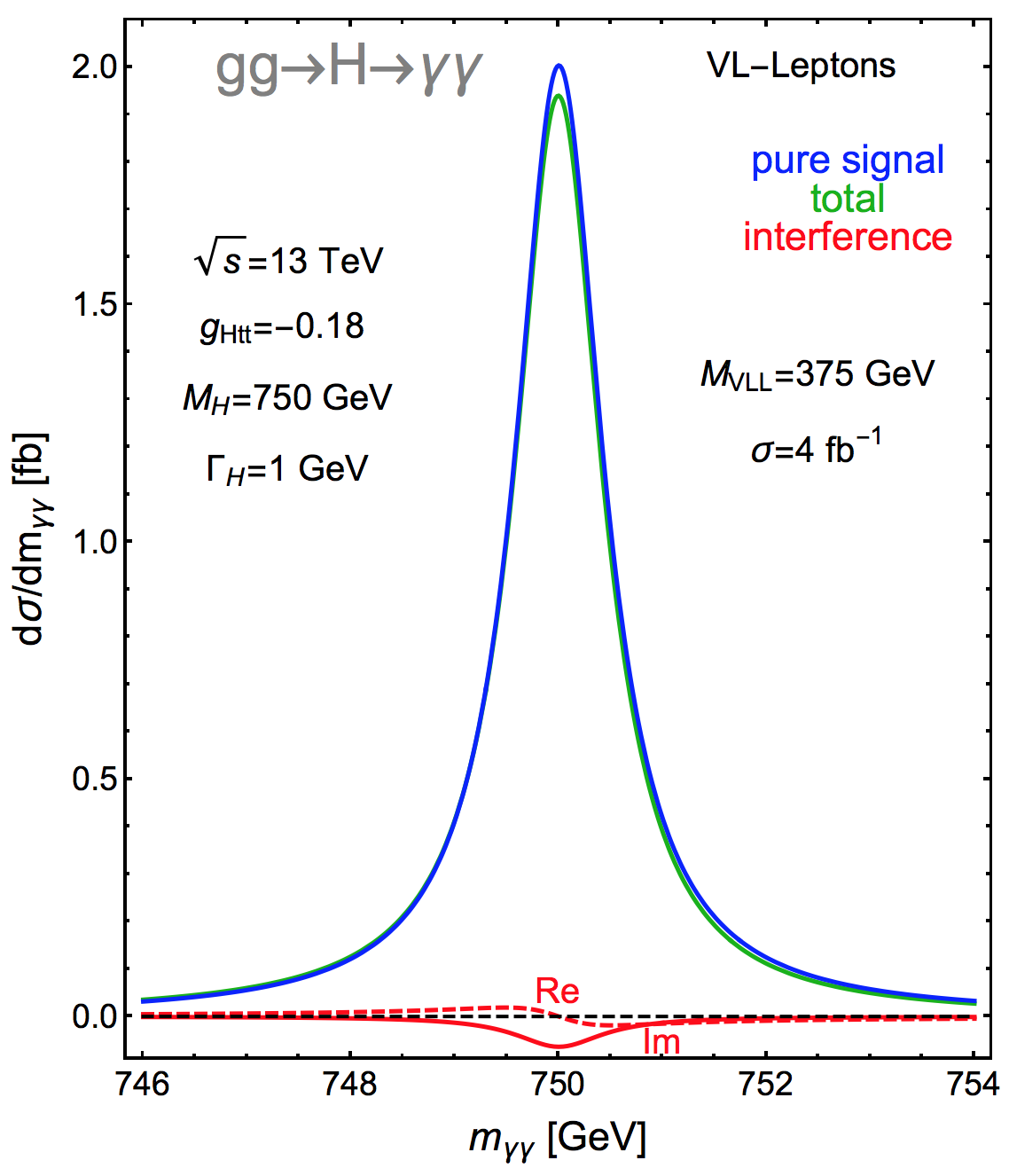}~~~~~~~
\includegraphics[scale=0.14]{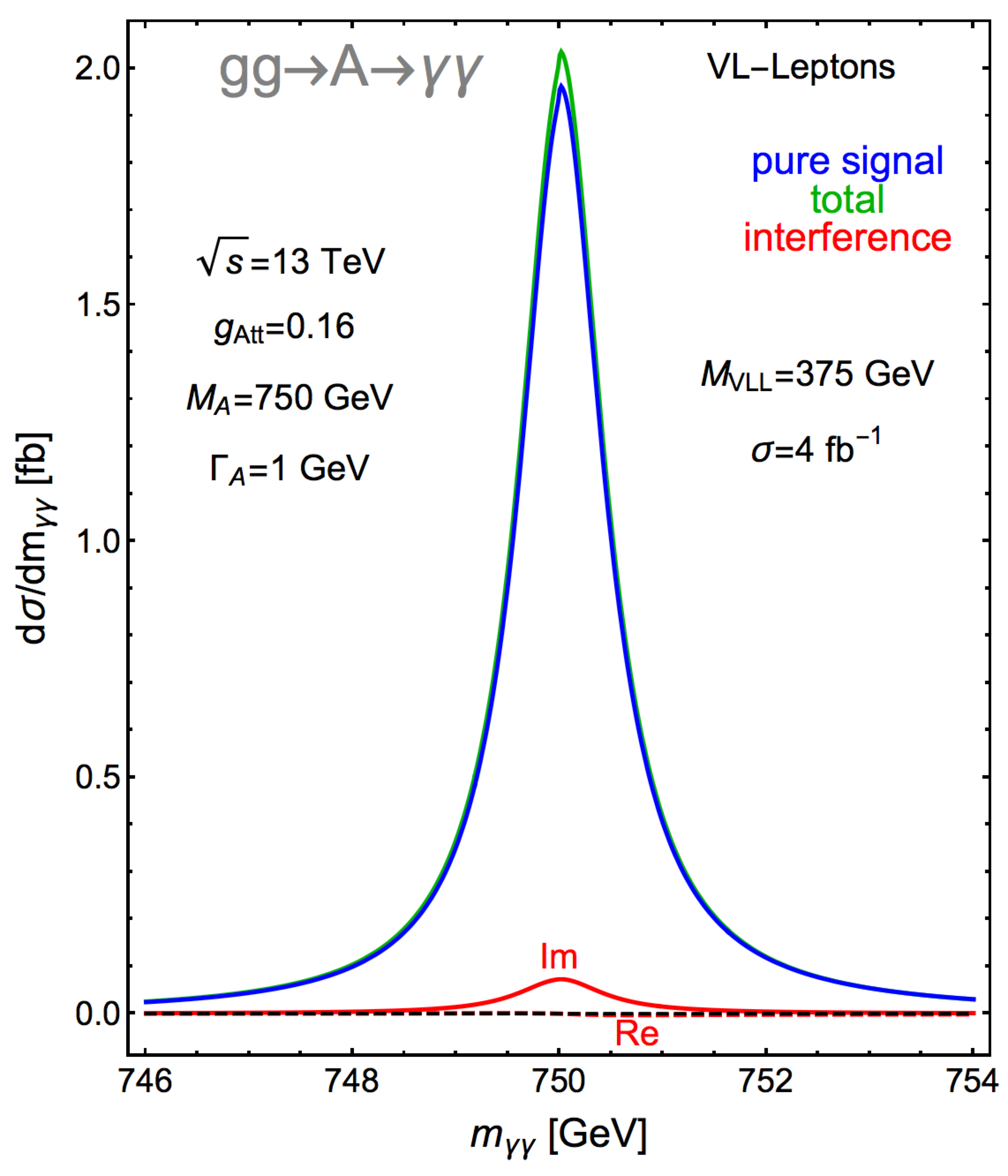}
}
\vspace*{-.1cm}
\caption{Upper panels: the contributions to the line--shapes of a scalar $H$ and a pseudoscalar $A$ state with mass $M_H=M_A=750$~GeV and total widths $\Gamma_H \approx 30$~GeV  and $\Gamma_A \approx 36$~GeV in the process $gg \to \phi \to \gamma \gamma$ where only the top quark with a large Yukawa coupling
contribute. Lower panels:  the line--shapes for both $H$ and $A$ contributions in a scenario with 750 GeV mass and  $\Gamma_H=\Gamma_A=1$ GeV with vector--like leptons in the loops that lead to a large rate of $\sigma=4$ fb. Shown are the rates with the pure signal only, the interference  and the total rate including the interference. From Ref.~\cite{Djouadi:2016ack}.}
\label{Hsing_gg_inter}
\vspace*{-2mm}
\end{figure}

This is illustrated in Fig.~\ref{Hsing_gg_inter} which shows contributions to
the line--shape of a scalar $H$ and pseudoscalar $A$  states of mass $M_\phi=
750$ GeV to be observed in the $gg\to H/A \to \gamma \gamma$ process. We have
assumed two scenarios: one in which the resonances interact only with the top
quark with a coupling $g_{\phi tt}=1$ leading  to total widths $\Gamma_\phi
\approx \Gamma (\phi \to t \bar t)$ of 30 (36) GeV in the  (pseudo)scalar case.
In a second scenario, we have assumed that vector--like leptons with masses
$m_{\rm VLL}=375$ GeV are running in the loops  with sufficiently large enough
Yukawa couplings to give a cross section times branching ratio of 4 MeV at
$\sqrt s=13$ TeV (which is by now excluded). However, the top quark Yukawa
coupling is so tiny $g_{H tt}\!=\!-0.16$ or $g_{A tt}\!\!=-0.18$ that one has a
small resonance width,  $\Gamma_\phi\! \approx \!\Gamma (\phi \!\to \! t \bar
t)=1$ GeV. 

In all cases, the $\phi$ line--shapes are shown without (blue lines) and with
interference (green lines); the contributions of interferences in the real and
imaginary parts of the  amplitude are also shown (dashed and solid red lines).
When only the top quark is present, the imaginary part of the interference is
important and  leads to a total cross section much larger than the one with pure
signal. The real part is also large and changes sign at the nominal Higgs mass.
The overall combination not only changes the rate but also the shape as it
exhibits a peak slightly below the nominal mass and a more modest dip just above
it. In the case where vector-- leptons are considered in the amplitudes with
large Yukawa couplings, since their contributions are mostly real and the
imaginary part small, they lead to a tiny difference between the  pure signal
and total cross section including interference. 

A second caveat is that when the resonances couple to the top quark,  they will
decay at the two--body level into theses states hence strongly suppressing the
loop induced two--photon modes that are searched for. In fact, the $\phi \to
t\bar t$ decays represent more an opportunity than a nuisance in these
resonance  searches as top quarks in the final states are easy to detect
especially when they are boosted, allowing to use jet substructure techniques
that nowadays became  very efficient. Furthermore,  there are interference
effects that allow to probe small signals in an easier way and, once observed,
to collect more information on the new states. Indeed, as the Higgs resonance is
produced in the gluon  fusion process, the signal amplitude for $gg \to \phi \to
t\bar t$ will interfere with the QCD process  $gg\to t\bar t$ which is the main
background at high energies and is very large as it occurs already at
tree--level: at  $\sqrt s = 13$ TeV,  $\sigma(pp \to t\bar t)= 820$ pb for $m_t=
173$  GeV and is dominated by the  $gg \to t\bar t$ initiated process as the
$q\bar q \to t\bar t$ part represents only 15\% of the total rate. 

\begin{figure}[!h]
\vspace*{-.4cm}
\centerline{
\includegraphics[scale=0.32]{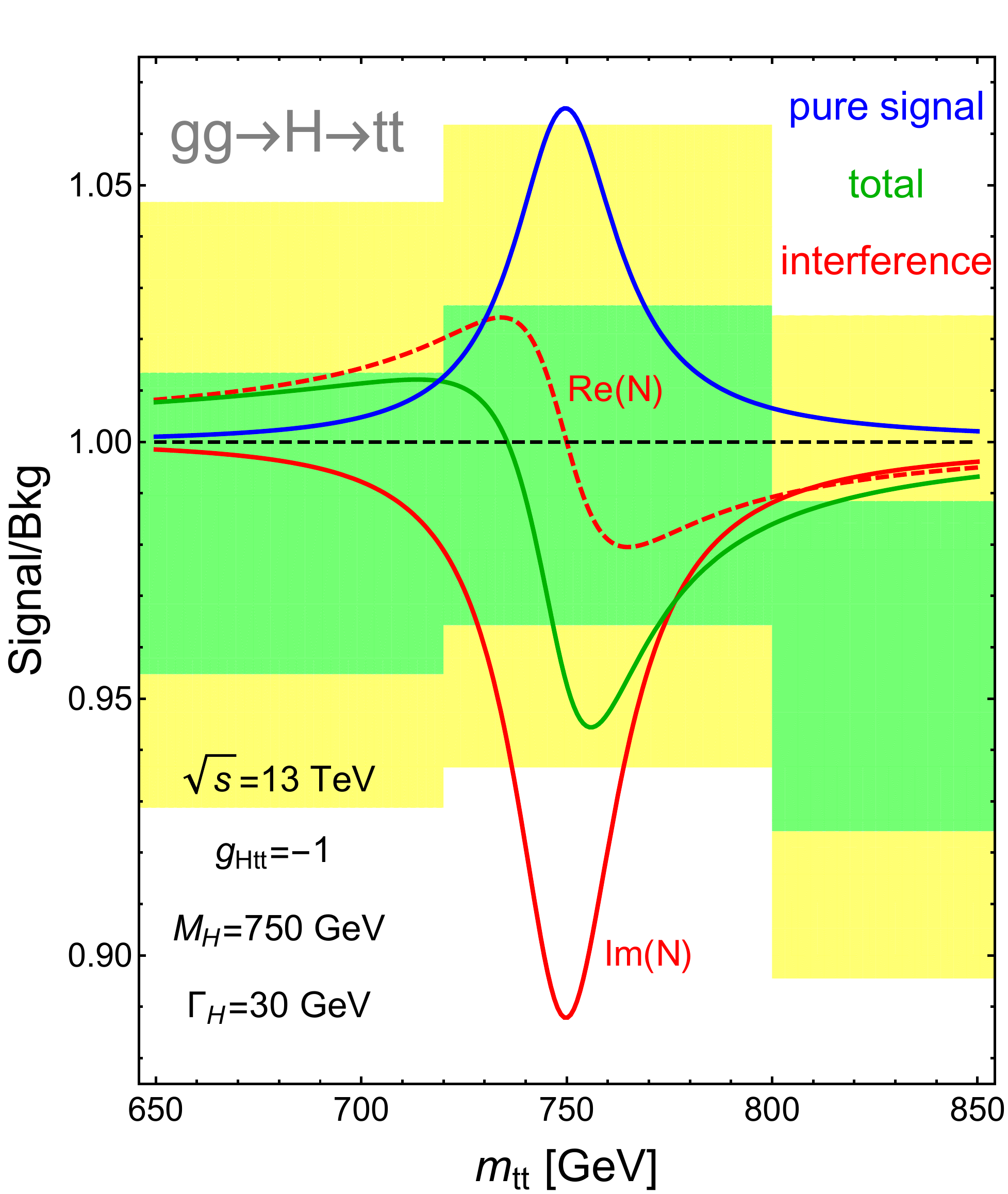}~~~~
\includegraphics[scale=0.32]{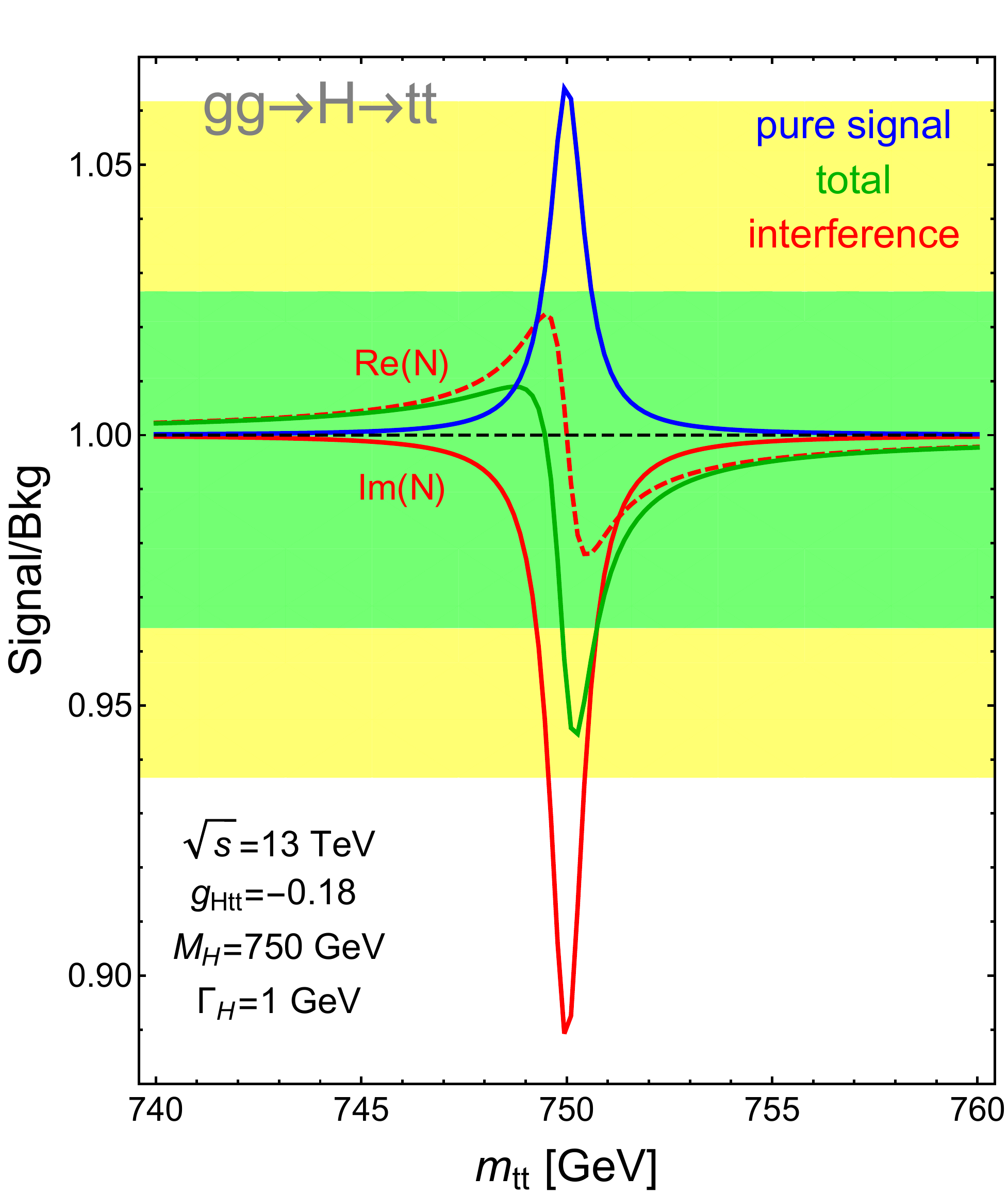}}
\vspace*{-.2cm}
\caption{The contributions to the line--shapes of a scalar $H$  
state with mass 750~GeV and total widths of $\Gamma_H = 30$~GeV (left) and
$\Gamma_H=1$ GeV (right) in the process $gg \to H \to t\bar t $; shown are the rates with the pure signal only, the interference  and the total rate including the interference \cite{Djouadi:2016ack}. }
\label{Hsing_tt_inter1}
\vspace*{-2mm}
\end{figure}

The effects of these interferences on the signal plus background normalized to
the background alone, are shown in Fig.~\ref{Hsing_tt_inter1} for the
distribution of the invariant mass of the $t\bar t$ system in exactly the same
two scenarios as for the $\gamma\gamma$ final states: $H$ and $A$  resonances
with masses of 750 GeV and either broad widths, $\Gamma_H=30$ GeV  and
$\Gamma_A=36$ GeV, or narrow ones, $\Gamma_H=\Gamma_A=1$ GeV (but the  VLLs not
affecting the rates).  However, we present the results only for the scalar $H$
case as those for $A$ are similar. The ATLAS RunI data~\cite{Aaboud:2017hnm} in
this channel but without the interferences, which are more constraining than
those of CMS, are included in the plots as ``Brazil" $1\sigma$ and $2\sigma$
green and yellow bands. One sees that the interference has a very important
impact. Its real part changes sign across the nominal $H$ mass, whereas its
imaginary part (due to the top quark loop in $gg \to H$) is larger in magnitude
and always negative. Hence, the combined effect is negative and overwhelms the
putative peak resulting finally in a dip in the $m_{t {\bar t}}$ distribution.
One notes that the dip is not symmetric about $M_H= 750$~GeV and a greater
sensitivity to interference effects could be obtained by comparing off--centre
bins. Nevertheless, the dip structure in the $\Gamma_ H = 1$~GeV  case is
unlikely to be observed in view of the resolution in $m_{t {\bar t}}$.

\begin{figure}[!h]
\vspace*{-.4cm}
\centerline{
\includegraphics[scale=0.65]{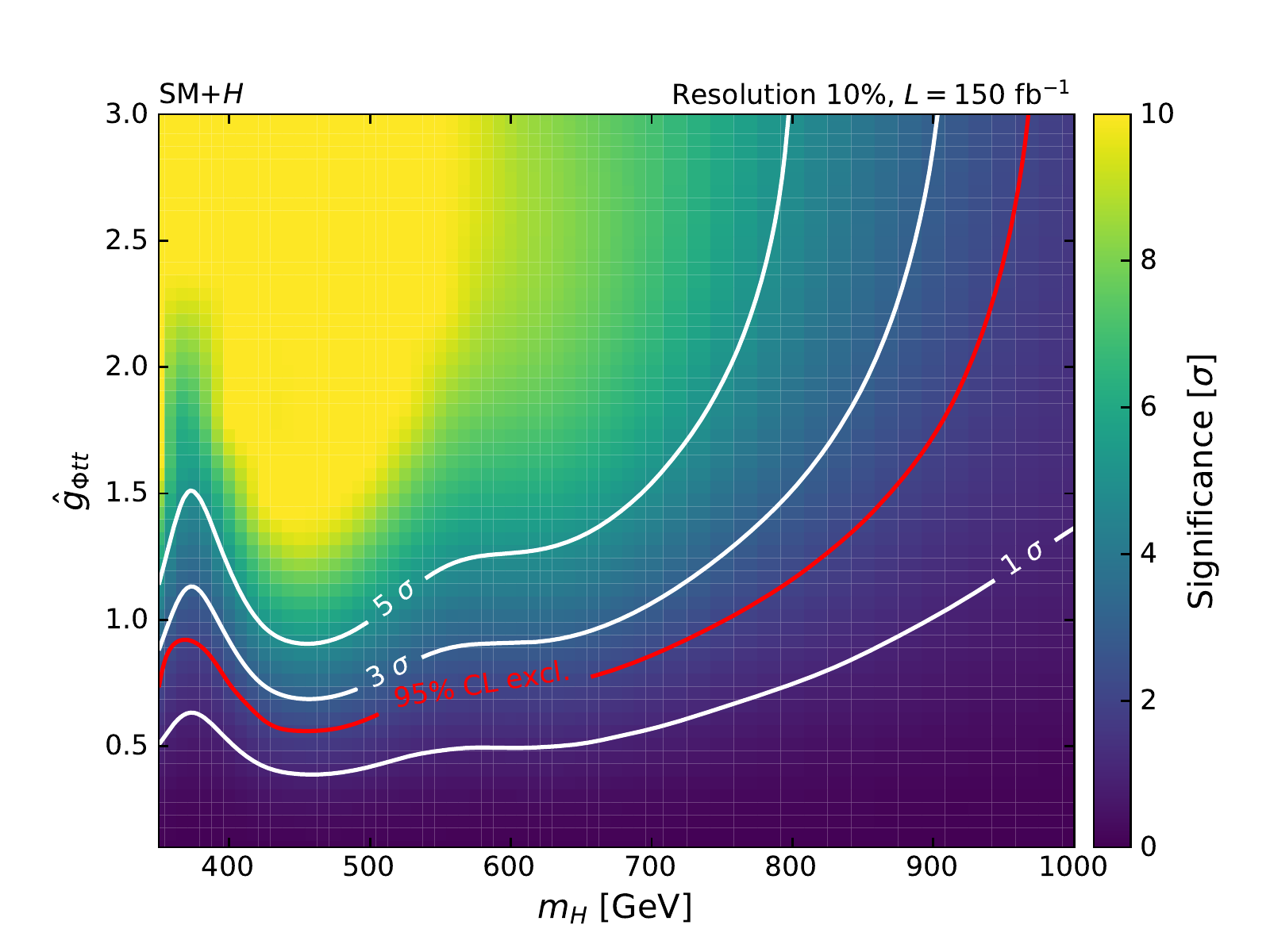} }
\vspace*{-.2cm}
\caption{Expected significance and exclusion potential 
in the plane $[M_H, \hat g_{Htt}]$ in the search for a scalar resonance in the channel $gg \to H\to t\bar t \to \ell+$jets at LHC with $\sqrt s\!=\! 13$
TeV for 150 fb$^{-1}$ data and a mass resolution $m_{t\bar t}=10\%$; from  Ref.~\cite{Djouadi:2019cbm}.}
\label{Hsing_tt_inter2}
\vspace*{-2mm}
\end{figure}

A simulation has been performed in Ref.~\cite{Djouadi:2019cbm} on the expected
sensitivity and the exclusion potential for scalar and pseudoscalar resonances
in the search channel $gg\to \phi \to t\bar t$ at the LHC that leads to lepton
plus jets final states.  Assuming an  experimental mass resolution of 10\% on
the $t\bar t$ system,  these are displayed in the case of an $H$ state in 
Fig.~\ref{Hsing_tt_inter2} in the parameter plane $[M_H, \hat g_{Htt}]$ when 
only top quark loops are present in the $ggH$ triangle amplitude with Yukawa
couplings between $\hat g_{Htt}=0.3$ and 3; an energy  $\sqrt s=13$ TeV and a
luminosity of 150 fb$^{-1}$, equivalent to what has been collected at RunII by
ATLAS and CMS, have been assumed. One sees that for $\hat g_{Htt} \approx 1$, a
$5\sigma$ discovery can be made up to $M_H \approx 500$ GeV and a $2\sigma$
sensitivity is achieved up to $M_H \approx 800$ GeV. Of course, this very
impressive sensitivity drops with smaller Yukawa couplings and a worse
resolution on $m_{t\bar t}$ (for the same $M_H$, only values $\hat g_{Htt}
\approx 2$ are probed if $m_{t\bar t}\!=\!20\%$) but it significantly increases
with the luminosity and at  HL--LHC with 3 ab$^{-1}$ data, a $5\sigma$ discovery
can be made up to $M_H=850$ GeV and a $2\sigma$  sensitivity up to $M_H=1$ TeV,
again for $\hat g_{Htt} \approx 1$. Note that the sensitivity in the
pseudoscalar $A$ case is slightly better than for $H$ with the same couplings
and experimental set--up, as the production rates are  higher.

\begin{figure}[!h]
\vspace*{-.02cm}
\centerline{
\includegraphics[scale=0.4]{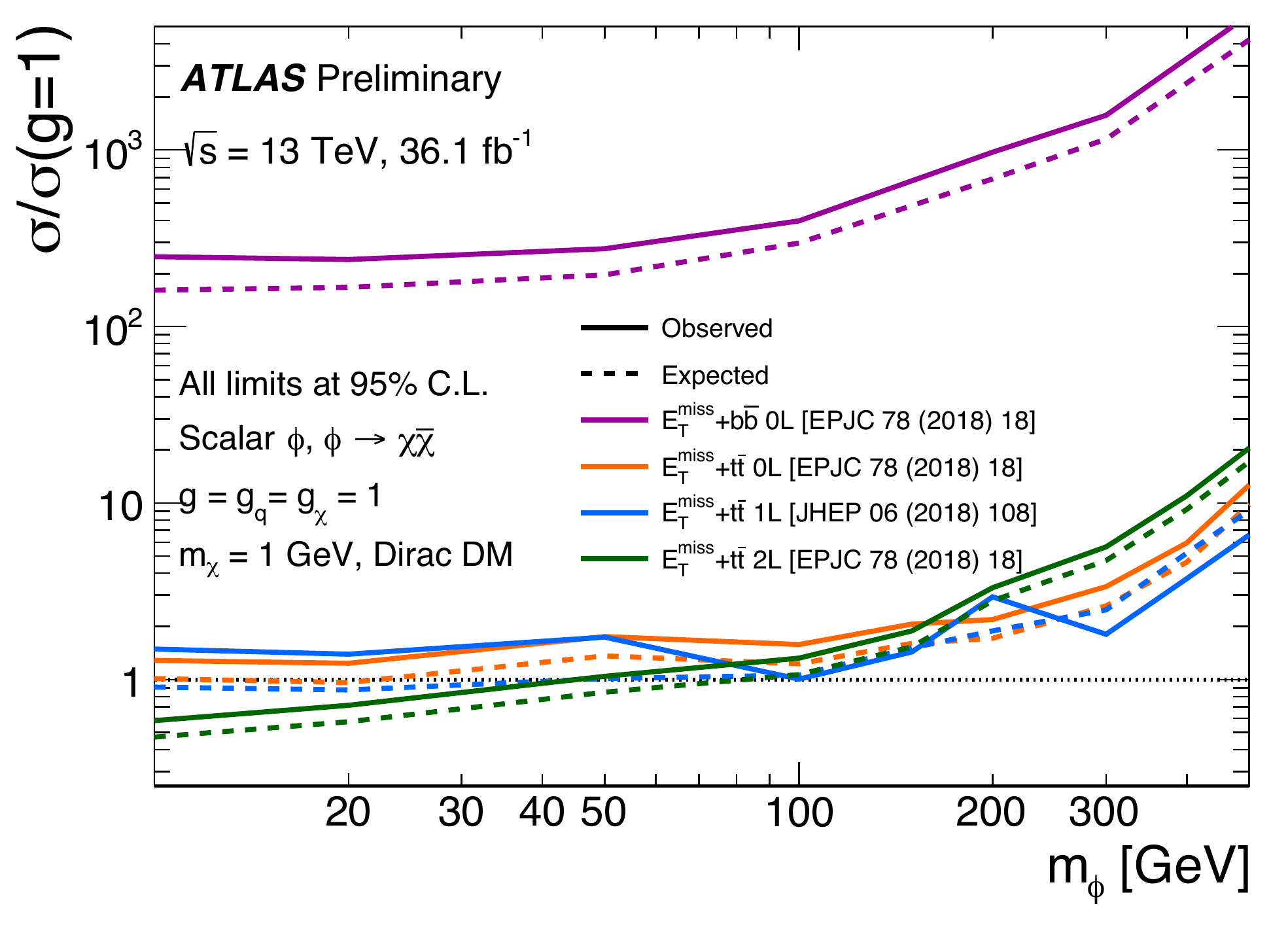}~~
\includegraphics[scale=0.4]{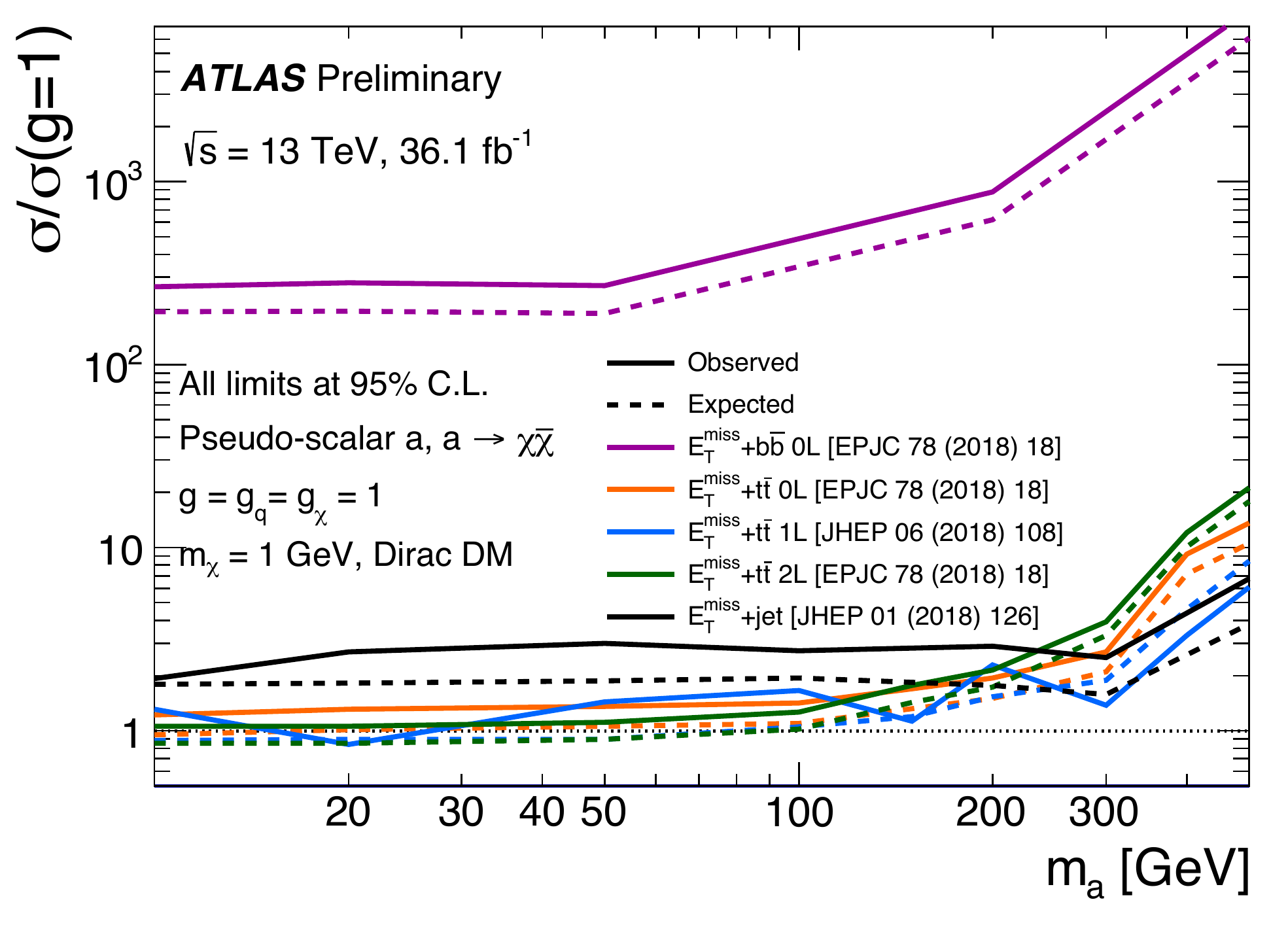}
}
\vspace*{-.1cm}
\caption{95\%CL exclusion and expected limits for scalar (left) or pseudoscalar
(right) DM mediator states as a function of their masses for a DM fermion with
mass of 1 GeV. The limits are given on the rates normalized to the case
where $g_{\phi tt}= g_{\phi NN} =1$; from \cite{ATLAS:2018vvx}.}
 \label{fig:Hsing-DM}
\vspace*{-3mm}
\end{figure}

Finally, a third caveat with the searches for spin--0 resonances in the diphoton
channel, but which in our context is turned into an advantage, is that decays
into new particles including the DM state can occur making the branching ratio
for the $\gamma\gamma$ mode marginal. Such searches have been performed at the
LHC in various final states and an example of an ATLAS analysis of a light
fermionic DM produced in association with $b\bar b$ and $t\bar t$ pairs  at
RunII with 36 fb$^{-1}$ data \cite{ATLAS:2018vvx} is  shown in 
Fig.~\ref{fig:Hsing-DM} where the 95\%CL exclusion limits for a scalar (left)
and pseudoscalar (right) state are shown as a function of $M_\phi$ below the
$t\bar t$ threshold. They are normalized to the rates calculated for unit Yukawa
couplings to quarks and to a Dirac fermion DM with a 1 GeV mass.  Bounds of
order unity are set on the ratio of cross sections in associated $\phi t\bar t$
production for not too large masses, while the bounds from $\phi b\bar b$
production are two orders of magnitude weaker. Hence, DM states with smaller and
more realistic $\phi NN$ couplings are still allowed by these searches. 

A similar search has been recently performed by CMS also at RunII with about 36
fb$^{-1}$ \cite{Sirunyan:2019gfm}, again assuming a very light DM particle with
a mass set to 1 GeV, and a scalar or a pseudoscalar Higgs mediator in the
simplified case where they couple only to top quarks with unit Yukawa couplings.
Both the associated production with top quark pairs $pp\to t\bar t \phi$ and the
production with a single top quark, $pp \to t\phi, \bar t \phi$ are considered
and except for high mediator masses where the phase--space is in favor of the
single top channel, the bulk of the cross section is generated by the $t\bar t
\phi$ process.  As no deviations from SM predictions have been observed, $H$
masses below 290 GeV and $A$ masses below 300 GeV have been excluded at the
95\%CL. 

\begin{figure}[!h]
\vspace*{-.02cm}
\centerline{
\includegraphics[width=0.52\textwidth]{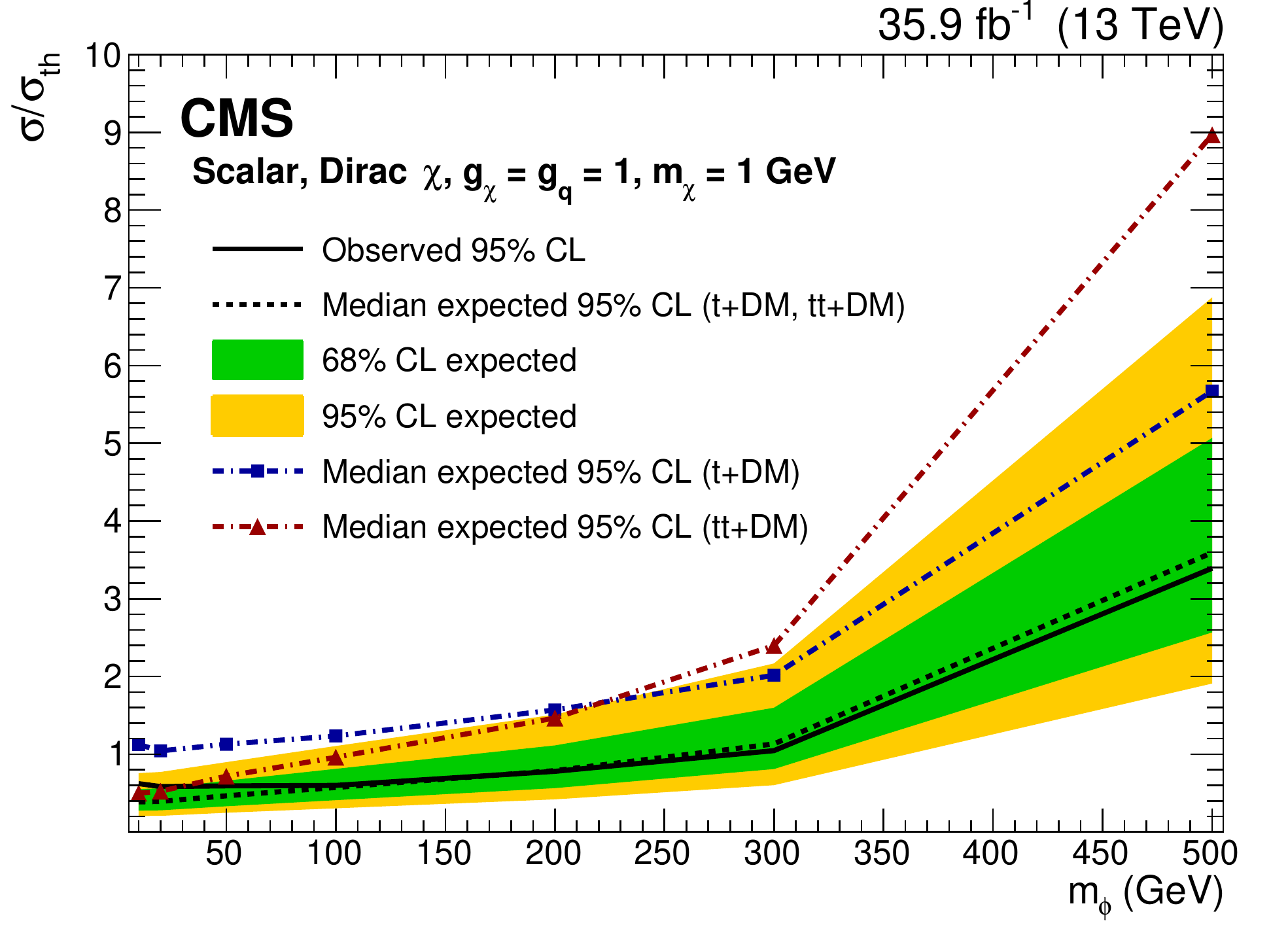}~
\includegraphics[width=0.52\textwidth]{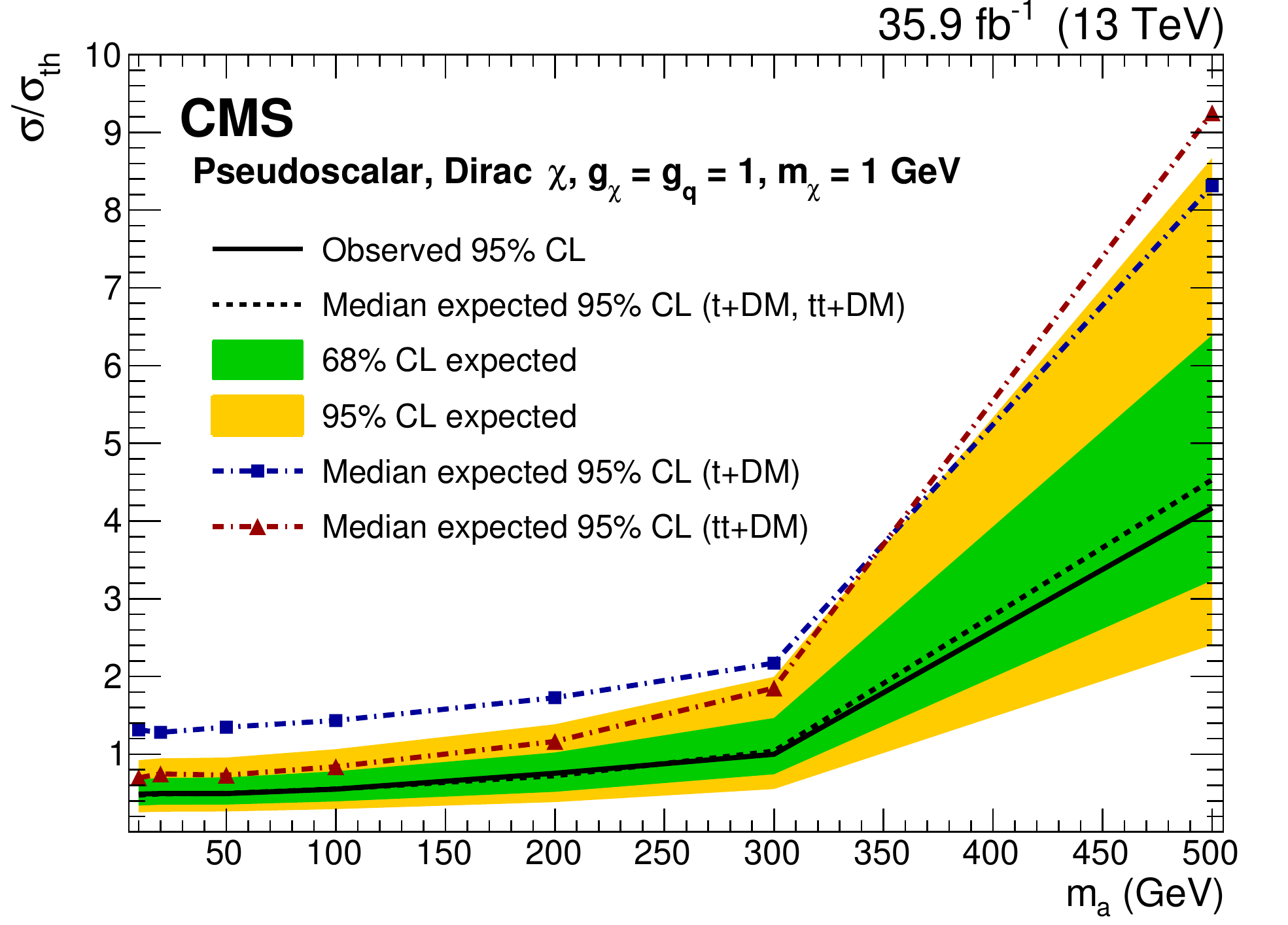}
}
\vspace*{-.1cm}
\caption{Expected (black dashed line) and observed (black solid line) 95\%CL limits on DM production cross sections relative to the theory predictions for a scalar (left) and a pseudoscalar (right) mediator in associated production with top quarks. The expected limit from the  associated $t\phi$ and $t\bar t \phi$ channels  alone are shown by the blue dash--dotted and  red dash--dotted lines respectively. From \cite{Sirunyan:2019gfm}.}
\label{fig:ttppfi-CMS}
\vspace*{-.3cm}
\end{figure}

\begin{figure}[!h]
\vspace*{-3cm}
\begin{tabular}{ll}
\hspace*{-1.cm}
\begin{minipage}{8cm}
\centerline{\includegraphics[scale=0.9]{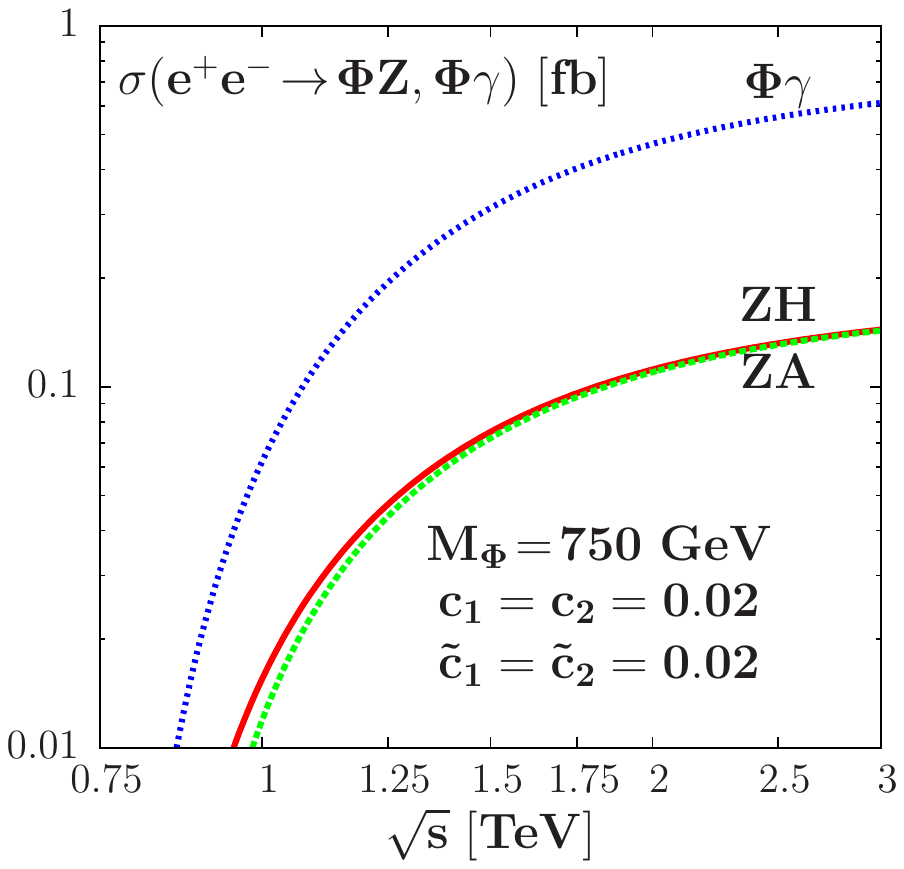} } 
\end{minipage}
& \hspace*{-.3cm} 
\begin{minipage}{8cm}
\vspace*{-2.2cm}
\centerline{\includegraphics[scale=0.8]{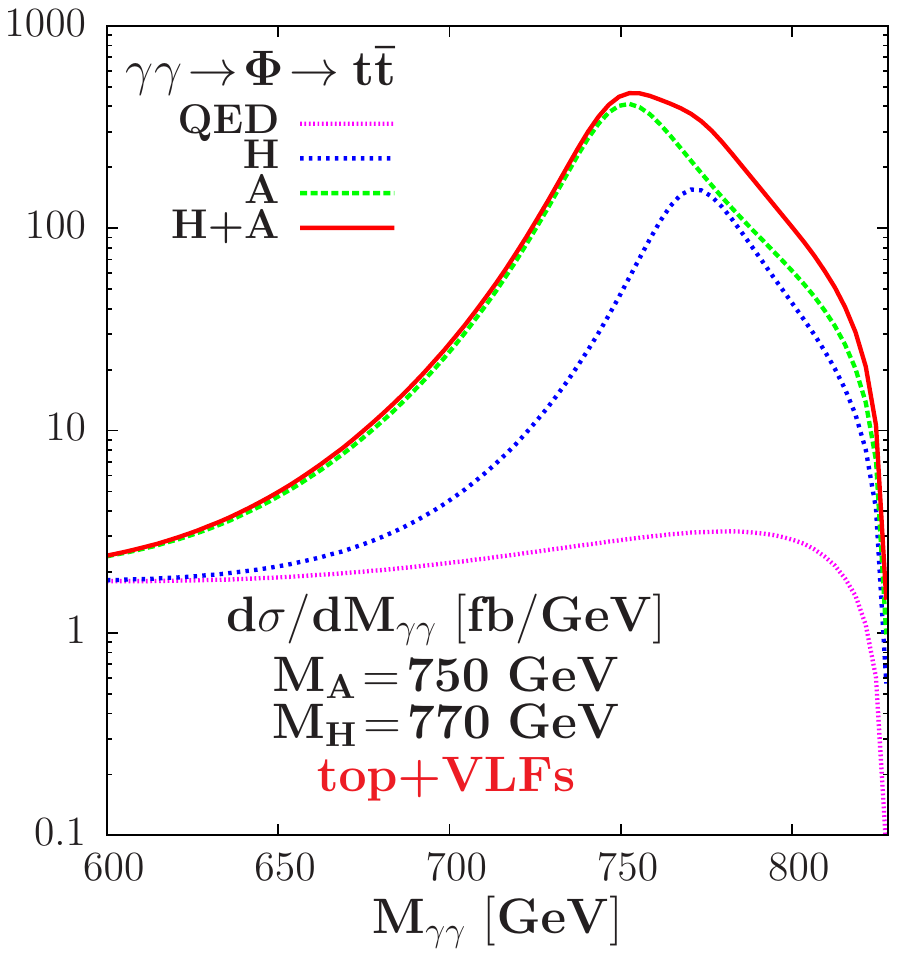} }
\end{minipage}
\end{tabular}
\vspace*{-16cm}
\caption{Left: cross sections for the $e^+e^- \to \phi Z, \phi \gamma$ processes with $\phi=H,A$ as functions of the total energy $\sqrt s$ for masses $M_\phi=750$ GeV and induced couplings to gauge bosons, $c_1=c_2=0.02= \tilde c_1= \tilde c_2$.  Right: invariant mass distribution ${\rm d\sigma/d}M_{\gamma \gamma}$ in fb/GeV for the process $\gamma\gamma \to t\bar t$ in the photon  mode of a linear $e^+e^-$ collider; shown are the QED background, the separate and combined contributions of the $H$ and $A$ states, and  the full  QED$\; +H+A$ contributions; the values $M_A=750$ GeV, $M_H=770$ GeV, $\Gamma_A=35$ GeV and  $\Gamma_H=32$ GeV are assumed. From
Ref.~\cite{Djouadi:2016eyy}.} 
\label{Hsing-ee}
\vspace*{-2mm}
\end{figure}

Before closing this section, let us note that the singlet (pseudo)scalar states
can also be produced in $e^+e^-$ collisions, in association either with a photon
or a $Z$ boson. The cross sections for the processes $e^+e^- \to \phi Z, \phi
\gamma$ for $M_\phi=750$ GeV are shown in the left panel of Fig.~\ref{Hsing-ee} 
as a function of the energy  for induced $\phi$ couplings to the  U(1) and SU(2)
gauge fields in units of $e/v$, $c_1=c_2=0.02= \tilde c_1= \tilde c_2$. For such
tiny couplings, the rates in the $\phi \gamma$ mode are smaller than 1 fb even
at $\sqrt s \!=\! 3$ TeV and those in $Z\phi$ production are even a factor of
about five lower. High luminosities are thus necessary in order to probe these 
spin--0 states. Note that one can also produce them in $WW$ and $ZZ$ fusion,
$e^+ e^- \! \to \! \phi \nu \bar \nu$ and $e^+ e^- \! \to \! e^+ e^- \phi$ but
the rates are even smaller: for the same $\phi$ mass and couplings, they are
respectively, one and two orders of magnitude lower than the rates in $e^+e^-
\to \phi \gamma$ \cite{Djouadi:2016eyy}.

The best probe of these $\phi$ resonances is presumably their production via the
$\gamma \gamma$ option of future linear $e^+ e^-$ colliders  constructed using
Compton back--scattering from laser light
\cite{Ginzburg:1982yr,Badelek:2001xb,Godbole:2002qu} leading to photon beams
that carry a large fraction of the energy and luminosity of the parent $e^+/e^-$
beams. The advantage of  such a collider is that it provides a direct access to
the state in single production, $\gamma\gamma \to \phi$,  and gives the
opportunity to probe its CP properties. The cross sections for $\gamma\gamma \to
\phi=H,A$ and $H\!+\!A$ production with subsequent decays into $t\bar t$ final
states are presented in the right--hand side of Fig.~\ref{Hsing-ee} as a
function of the $\gamma\gamma$ energy. The laser energy and the helicities of
$e^{-}, e^{+}$  beams and those of the lasers have been chosen to make the $J_{Z
} =0$ $\phi$ contribution dominant. The latter is calculated in the case where
one  includes in the loop the top quark with SM--like couplings as well as
additional vector--like fermions that increase the $H (A)\to \gamma\gamma$ 
amplitude by a factors $10\, (15)$. The masses of the resonances are assumed to
be $M_H=770$ GeV and $M_A=750$ GeV and their total widths $\Gamma_A=35$ GeV and 
$\Gamma_H=32$ GeV. Shown are the pure continuum QED contribution $\gamma\gamma
\to t\bar t$, the additional separate contributions due to $s$--channel
exchanges of the $H$ and $A$ states, and  the full set of contributions
QED+$H$+$A$. For such large loop contributions, the signals stand clearly above
the QED backgrounds.

\newpage

\subsubsection{The scalar and pseudoscalar portals}

If a scalar and a pseudoscalar resonances are  simultaneously present,  two
cases are worth discussing. A first one, which is interesting from the collider
physics point of view, is when they are degenerate in mass. A second scenario 
which is interesting from the astroparticle physics point of view is when the
pseudoscalar resonance is much lighter than the scalar one and, in fact, even
lighter that the SM--like Higgs state. We briefly summarize the main features of
these two scenarios and the constraints to which they are subject.  

When both the $H$ and $A$ resonances are present with masses that are
significantly different, more precisely $|M_H-M_A|$ is larger than the
experimental resolution  so that the two states can be disentangled, all the
discussions of the previous subsections  hold as one just needs to search or
study these two states independently from one another. If the two masses are
almost equal, namely $|M_H- M_A|$ is smaller than the experimental resolution,
one simply needs to arrange that the signals cross sections for $H$ and $A$
production are added and the branching ratios of the two states weighted. There
is however a notable exception to this state of affairs: when $H$ and $A$ are
produced in the same process and decay into the same final states, the
amplitudes will interfere and might not only change the signal rates but also
also the distributions and shape of the signal to background. We have already
encountered two cases before in which such interference effects are important:
$\phi$ production in $gg$ fusion and subsequent decays into diphotons or $t \bar
t$ pairs. 

\begin{figure}[!h]
\vspace*{-.2cm}
\centerline{
\includegraphics[scale=0.35]{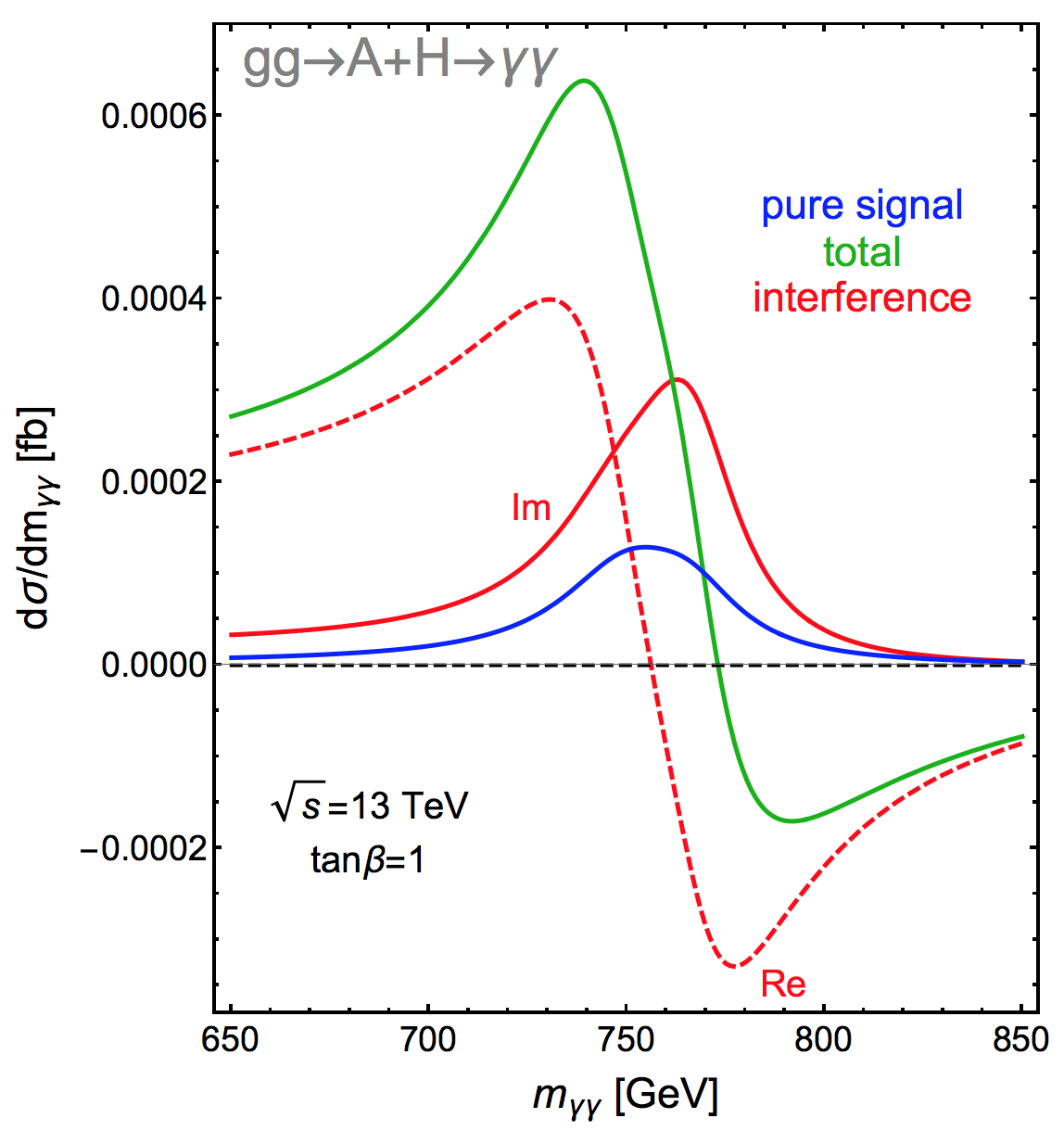}~~~~
\includegraphics[scale=0.33]{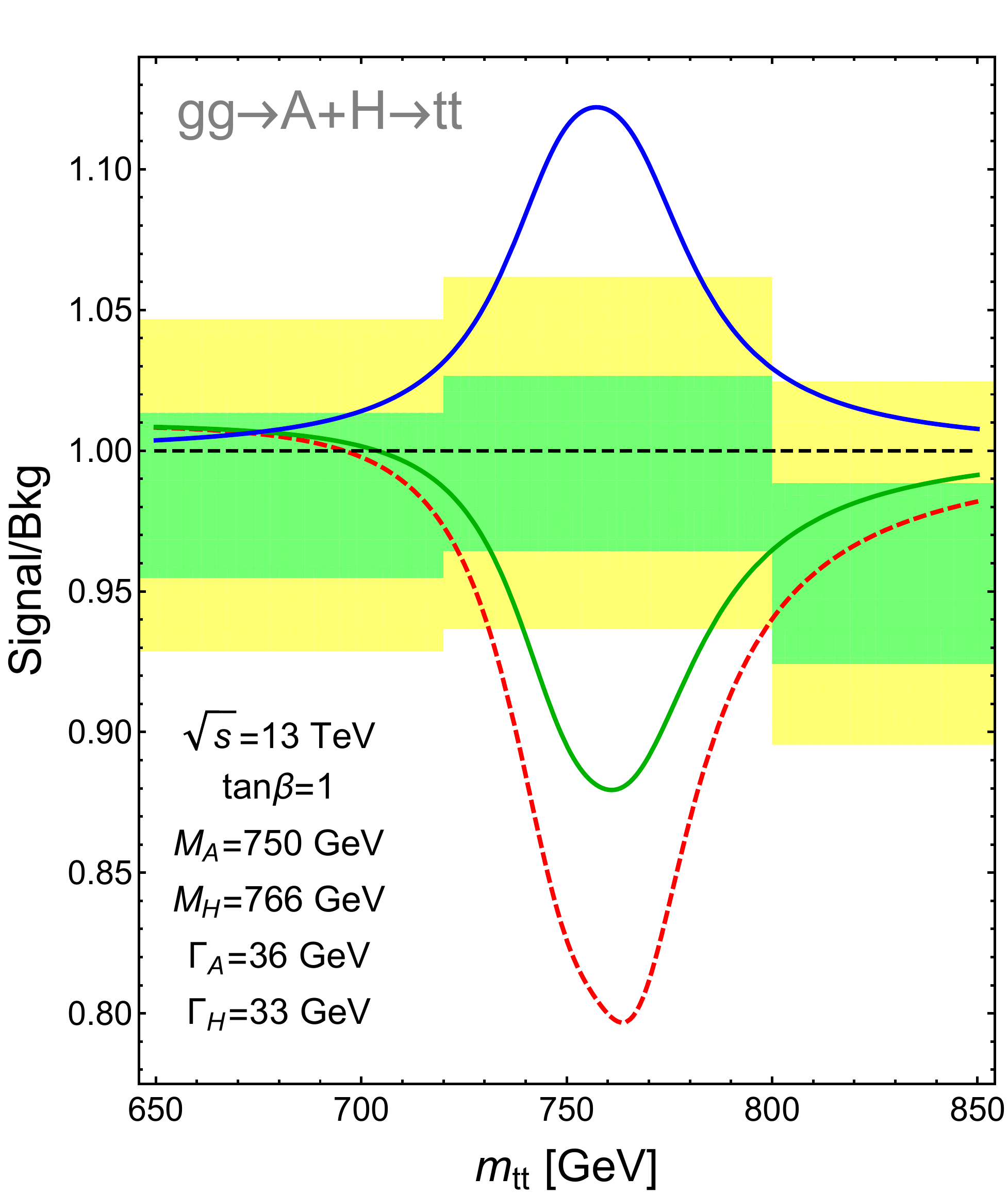}
}
\vspace*{-.1cm}
\caption{The contributions to the line--shapes of the sum of a scalar $H$ and pseudoscalar $A$ states with masses and total decay widths as depicted in the figures, in the processes $gg\! \to\! H\!+\!A \! \to \! \gamma\gamma$ (left) and $t\bar t $ (right); shown are the rates with pure signal, interference with the backgrounds and the total rate including interference \cite{Djouadi:2016ack}.}
\label{Hsing_H+A_inter1}
\vspace*{-3mm}
\end{figure}

\begin{figure}[!h]
\vspace*{-.1cm}
\centerline{
\includegraphics[scale=0.64]{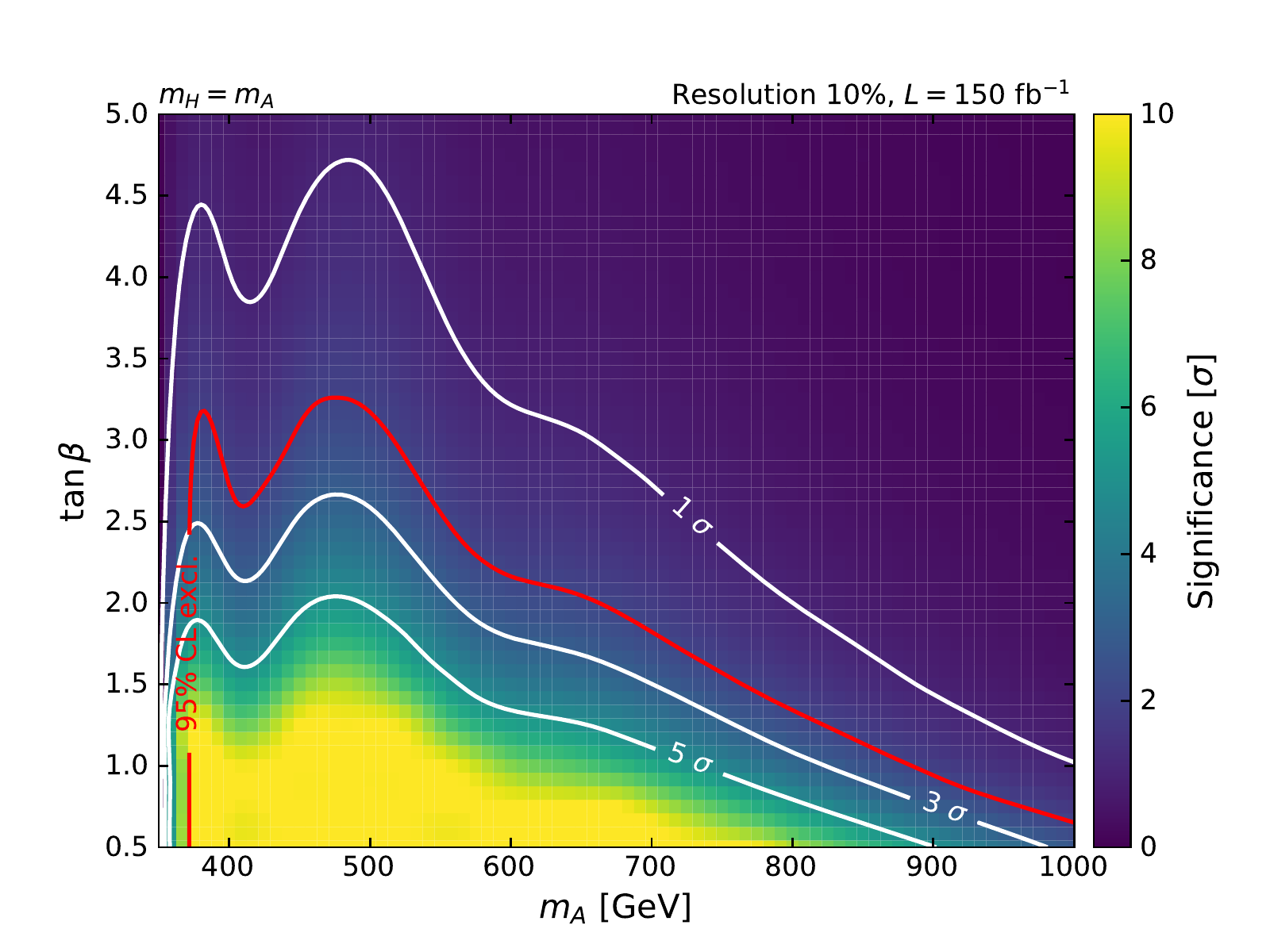}
}
\vspace*{-.3cm}
\caption{The expected significance and exclusion potential  in the plane $[M_H=M_A, \tan \beta= \hat g_{\phi tt}]$ in the searches in the channels $gg \to H+A\to t\bar t \to \ell+$jets at LHC with $\sqrt s\!=\! 13$ TeV for 150 fb$^{-1}$ data and a resolution $M_{t\bar t}=10\%$; from Ref.~\cite{Djouadi:2019cbm}.}
\label{Hsing_H+A_inter2}
\vspace*{-2mm}
\end{figure}

In Fig.~\ref{Hsing_H+A_inter1}, we repeat the analyses done before separately
for $H$ and $A$ for the case of a simultaneous $H+A$ signal in the processes
$gg\to H+A \to \gamma\gamma$ (left) and $gg\to H+A\to t\bar t$ (right)
highlighting the effects of interferences. As previously, we have chosen a
scenario with $M_H=766$ GeV, $M_A=750$ GeV, $\Gamma_H \approx 33$~GeV and
$\Gamma_A=36$ GeV and we include only the contribution of top quarks with
SM--like Yukawa couplings; $\tan\beta=1$ means here that  $\hat g_{\phi tt}=1$.
The previous features for $H$ and $A$ are amplified in this combined case with
not only a higher signal rate but also an  interference that has different
structure from the pure $H$ or $A$ cases. The sensitivity or discovery
potentials displayed in Fig.~\ref{Hsing_H+A_inter2}  are also higher compared to
the previous cases, and for $ \tan\beta=g_{\phi tt}=1$ for instance, a
$2\, (5)\sigma$ signal can be observed for $M_H=M_A=  0.9\, (0.7)$ TeV. 

The other scenario which leads to an interesting phenomenology in the context of
a simultaneous presence of a scalar and a pseudoscalar resonances, is when the
latter is very light $M_a \ll M_H$ and even $M_a \ll M_h$.   

If the $a$ state has very small couplings to SM fermions, its only possible
decays would be the $a\to gg$ and $a\to \gamma\gamma$ modes induced by the loops
involving the heavy VLFs, with a branching ratio of the latter being of the
order 1\% to 10\% depending on the relative magnitude of the couplings
$c_{gg}^a$ and $c_{\gamma\gamma}^a$. The only process which could allow for the
detection of the  light $a$ boson would be thus $pp \to h\to aa$ with $a \to
\gamma\gamma$  since the $a\to gg$ mode would have a too large background as the
jets are at a small invariant mass. 

The cross section for the $pp \rightarrow h \rightarrow 4\gamma$ process can be written, assuming for simplicity a common mass $m_F$ for all the vector--like fermions, as:
\begin{align}
\label{eq:4gamma}
& \sigma_{4\gamma}=\frac{\pi^2}{8 M_h s}\Gamma(h \rightarrow gg)\; {\rm BR}(h \rightarrow aa) \; \left[ {\rm BR}(a \to \gamma \gamma)\right]^2 c^{\phi=h}_{\rm gg}(M_h/\sqrt{s}) \nonumber\\
&    \simeq\left \{    \begin{array}{cc}
0.82 \,\mbox{fb}\frac{\Gamma_h/M_h}{10^{-4}}\left(\frac{c^{\phi=h}_{\rm gg}(M_h/\sqrt{s})}{1000}\right) \simeq 0.16 \,\mbox{pb} \lambda_\phi \left(\frac{c^{\phi=h}_{\rm gg}(M_h/\sqrt{s})}{1000}\right)      
& {\rm for}~M_a \lesssim 3 \pi^0 \; , \\
0.32 \,\mbox{fb}\frac{\Gamma_h/M_h}{0.1}\left(\frac{c^{\phi=h}_{\rm gg}(M_h/\sqrt{s})}{1000}\right) \simeq 0.06 \,\mbox{fb} \lambda_\phi \left(\frac{c_{\rm gg}^{\phi=h}(M_h/\sqrt{s})}{1000}\right)        
&  {\rm for}~M_a \gtrsim 3 \pi^0 \; ,    
\end{array}    \right.
\end{align}
where $c^{\phi=h}_{gg}$ is the form factor parameterizing the $\phi gg$ loop
amplitude generated by the vector--like quarks given in
eq.~(\ref{eq:phi-couplings_2}). We have distinguished the regime $M_a < 3
m_{\pi^0}$ in which no hadronic final states are kinematically accessible, hence
automatically implying BR$(a \rightarrow \gamma \gamma)=1$, and the regime $M_a > 3 m_{\pi^0}$ in which the $4\gamma$ cross section is drastically reduced by a
factor BR$^2(a \rightarrow \gamma \gamma)={81\alpha^4}/(4\alpha_{\rm s}^4)$ as the $a\to gg$ mode is then present. For simplicity we have assumed, for this estimates, a common mass value of the fermions composing the vector--like family.

\begin{figure}[!h]
\vspace*{-.5cm}
\centerline{
\includegraphics[scale=0.4]{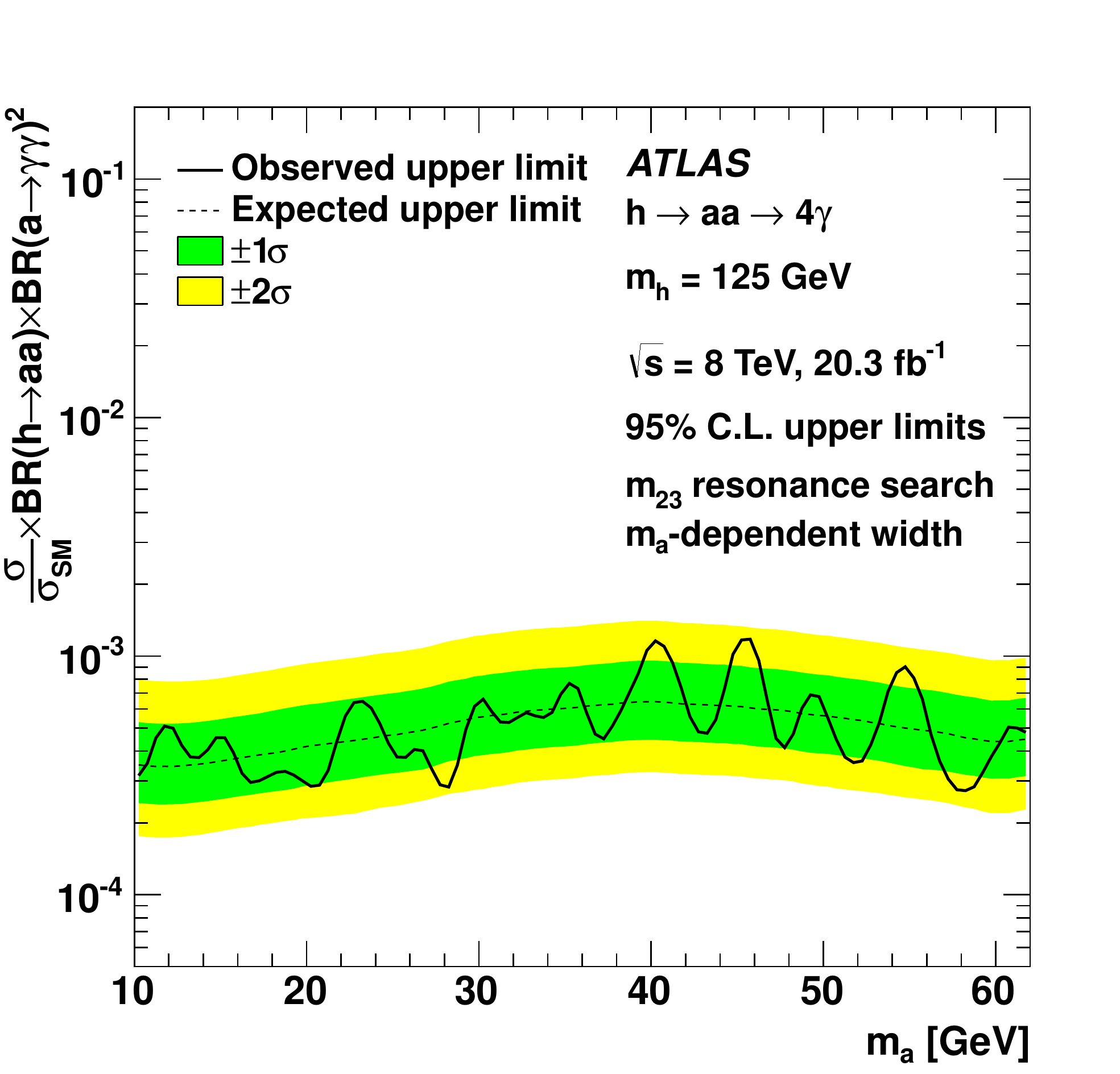}~ 
\includegraphics[scale=0.4]{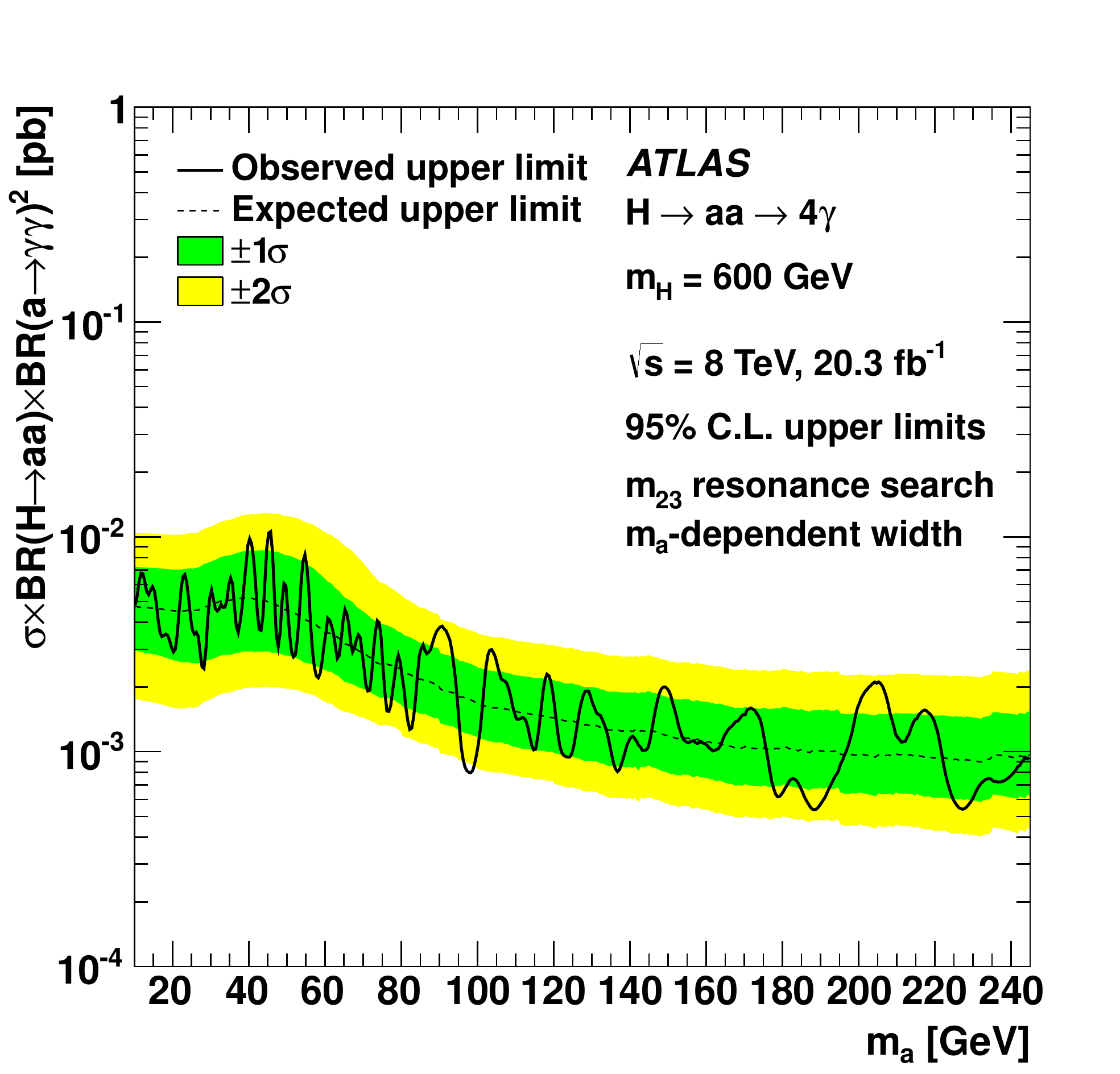}
}
\vspace*{-1mm}
\caption{95\%CL expected and observed upper limits on the $\sigma (\phi)/ \sigma_{\rm SM} \times {\rm BR} (\phi \to aa) \times  {\rm BR}^2 (a \to \gamma\gamma)$ rate, with $\phi=h$ with $M_h=125$ GeV (left) and $\phi=H$ with $M_H=600$ GeV (right),  in the search for a light pseudoscalar decaying into two photons at the LHC with $\sqrt s=8$ TeV and 20 fb$^{-1}$ data \cite{Aad:2015bua}. }
\label{Hsing-4gamma}
\vspace*{-.3cm}
\end{figure}

Multi--photon final states have been searched for at the LHC and the example of
an ATLAS analysis at $\sqrt s=8$ TeV with 20 fb$^{-1}$ data \cite{Aad:2015bua} 
is shown in the left--hand side of Fig.~\ref{Hsing-4gamma} for a resonance
decaying into four photons.  The search is performed separately for three
two--photon mass spectra defined by the three possible pairings for the photons
ordered by  $p_T$, from highest to lowest, $m_{12},m_{13}$ and $m_{23}$. For the
SM Higgs boson decay $h\to aa$, the mass range 10~GeV$\leq M_a\leq \frac12M_h
\simeq 62$ GeV has been considered for the pseudoscalar $a$ state.   As can be
seen, the cross section $\sigma(h)$ and BR$(h\to aa)$ multiplied by   BR$^2(a\to
\gamma\gamma)$ is constrained to be less than a fraction of a permile. 

These searches  can be extended to address heavy singlet $H$ production and
decay into $aa$ leading to the same four photon final states. The outcome is
displayed in Fig.~\ref{Hsing-4gamma} (right) where the rate is shown for a 600
GeV $H$ resonance decaying into $aa$  states with a mass in the range 10--250
GeV.  Here, the limits are at least one order of magnitude weaker as one has to
account for the suppressed rate $\sigma(pp\to H)$ compared to the SM--like $h$
boson.  

One should nevertheless notice that in the case of a very light pseudoscalar, $M_a \lesssim O(1\,\mbox{GeV})$, the photon pair emitted by the decay of this state result strongly collimated so that it is misidentified as a single photon~\cite{Agrawal:2015dbf,Dasgupta:2016wxw,Aparicio:2016iwr,Knapen:2015dap}. By requiring that the opening angle $\Delta \phi \sim 2/\gamma \sim 4 M_a/M_H$ of the emitted photons is below the energy resolution of the calorimeter, $O(20 \,\mbox{mrad})$ one obtains the following condition:
\begin{equation}
\label{eq:collimated}
    M_a \lesssim 2.5\mbox{GeV}\left(\frac{M_H}{1\,\mbox{TeV}}\right)
\end{equation}
for a collimated photon pair. If this condition is realized, eq.~\ref{eq:4gamma} represents a diphoton signal and then should be added to the diphoton cross-section from direct decay of the heavy resonance $H$. 

\subsection{Astroparticle constraints}

\subsubsection{The mixed Higgs case}

We turn now to the discussion of the astroparticle constraints on the different realizations of extensions of the SM Higgs sector with extra singlet states.
As already mentioned, the case of a real scalar with mixing with the SM Higgs boson represents from the DM perspective a double Higgs--portal. The DM relic density is then determined by annihilation processes into pairs of SM fermions and gauge bosons, via $s$--channel exchange of the two scalar mediators, as well as annihilations into $hh$, $hH$ and $HH$, if kinematically allowed. Despite analytical estimates for the $s$--channel cross sections can be straightforwardly derived from the case of the effective Higgs--portal, we have nevertheless reported them explicitly in Appendix B in order to pinpoint the dependence on the angle $\theta$. 

For what concerns direct detection of the DM, they are due to the spin--independent interactions mediated by $t$--channel exchanges of the $h/H$ states. The dependence of the cross sections on the parameters of the theory is exemplified by the following expressions for the three assignments of the DM spin:
\begin{align}
\label{eq:sigma_mixed}
& \sigma_{Sp}^{\rm SI}=\frac{\mu_{Sp}^2 (\lambda_\phi^S)^2}{4 \pi m_S^2}\sin\theta^2 \cos\theta^2{\left(\frac{1}{M_h^2}-\frac{1}{M_H^2}\right)}^2 {\left[\frac{Z}{A}f_p+\frac{A-Z}{A}f_n\right]}^2 , \nonumber\\
& \sigma_{\chi p}^{\rm SI}=\frac{m_\chi^2}{\pi v_\phi^2}\sin^2 \theta \cos^2 \theta {\left(\frac{1}{M_h^2}-\frac{1}{M_H^2}\right)}^2 {\left[\frac{Z}{A}f_p+\frac{A-Z}{A}f_n\right]}^2 , \nonumber\\
&    \sigma_{Vp}^{\rm SI}=\frac{(\eta_V^H)^2 \mu_{Vp}^2}{4\pi}\sin^2 \theta \cos^2 \theta {\left(\frac{1}{M_h^2}-\frac{1}{M_H^2}\right)}^2 {\left[\frac{Z}{A}f_p+\frac{A-Z}{A}f_n\right]}^2 . 
\end{align}
Corresponding limits from indirect detection are not competitive with the ones
from direct detection and will not be discussed explicitly here. Also,
fermionic DM states cannot be probed through indirect detection since they
annihilate only through $p$--wave processes.

Again in this mixed Higgs case, one observes a strong correlations between the
DM annihilation rate and the spin--independent cross sections, as witnessed by
the common $\sin^2 \theta \cos^2 \theta$ factor in the expressions of eq.~\ref{eq:sigma_mixed}. It is therefore possible to have a reliable insight on the phenomenology of
the DM state $X$ by focusing on the bidimensional plane $[M_H,m_X]$ while
setting the couplings and the mixing  parameter $\sin\theta$ to  ${\cal O}(1)$
values  when possible, or to the highest allowed values by complementary
constraints from colliders searches and the constraints on the scalar potential.
Indeed, lowering the DM couplings to comply with bounds from direct detection
would imply a comparatively increased difficulty in achieving the correct relic
density through the WIMP paradigm. 

\begin{figure}[!h]
\vspace*{-3mm}
\begin{center}
\mbox{\hspace*{-3mm}
\subfloat{\includegraphics[width=0.43\linewidth]{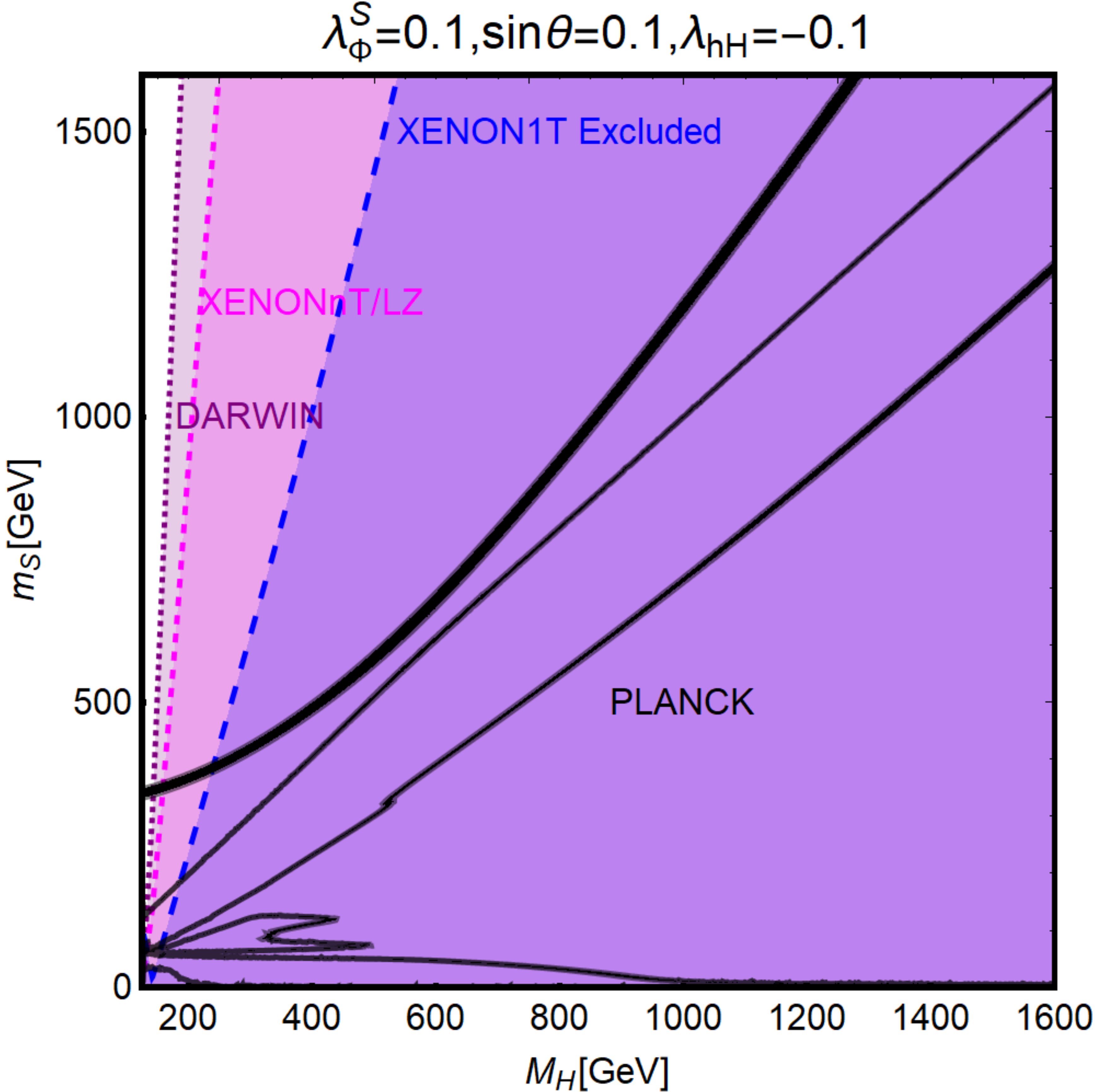}}~~~
\subfloat{\includegraphics[width=0.43\linewidth]{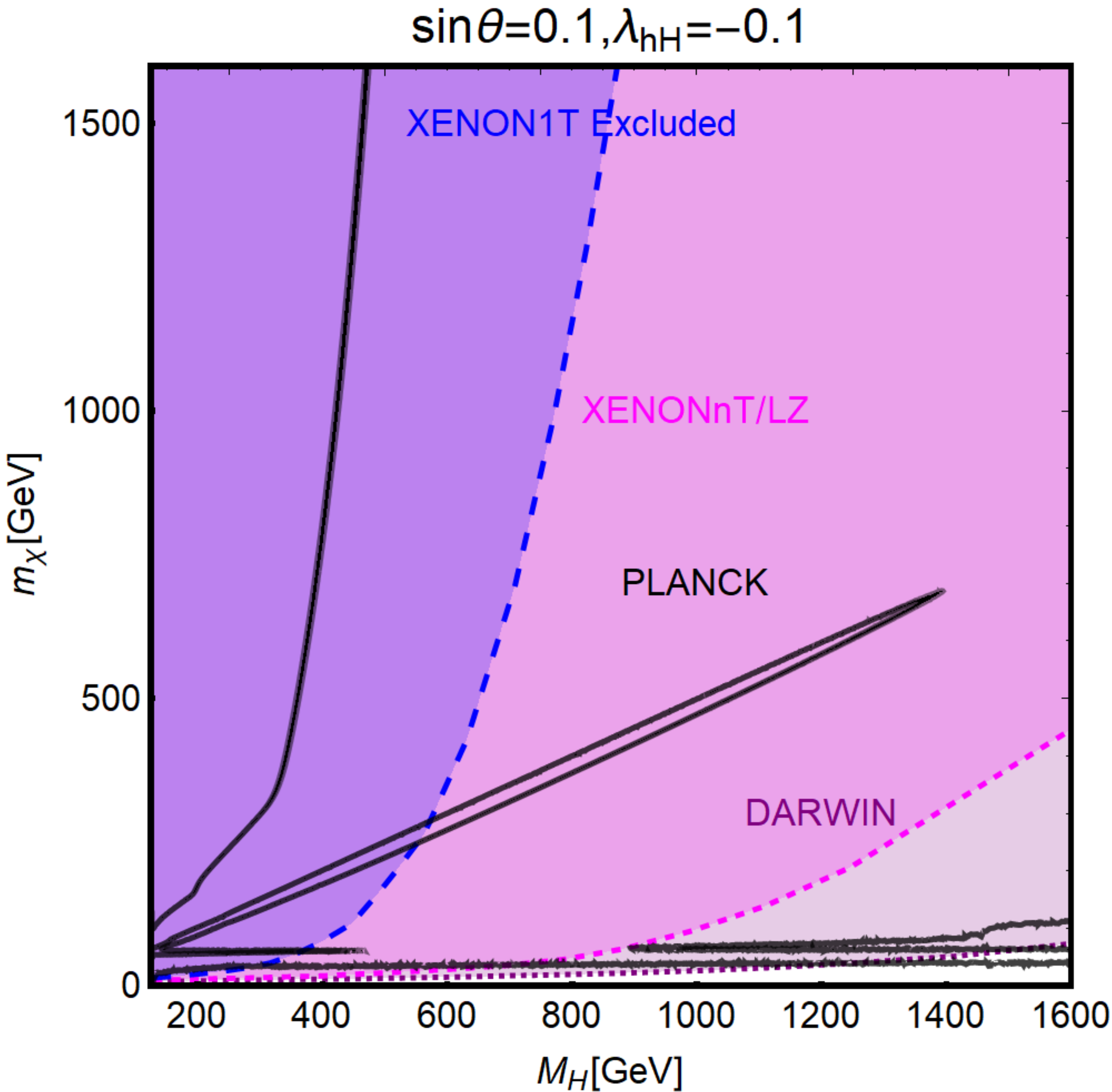}}
}\\[-2mm]
\subfloat{\includegraphics[width=0.43\linewidth]{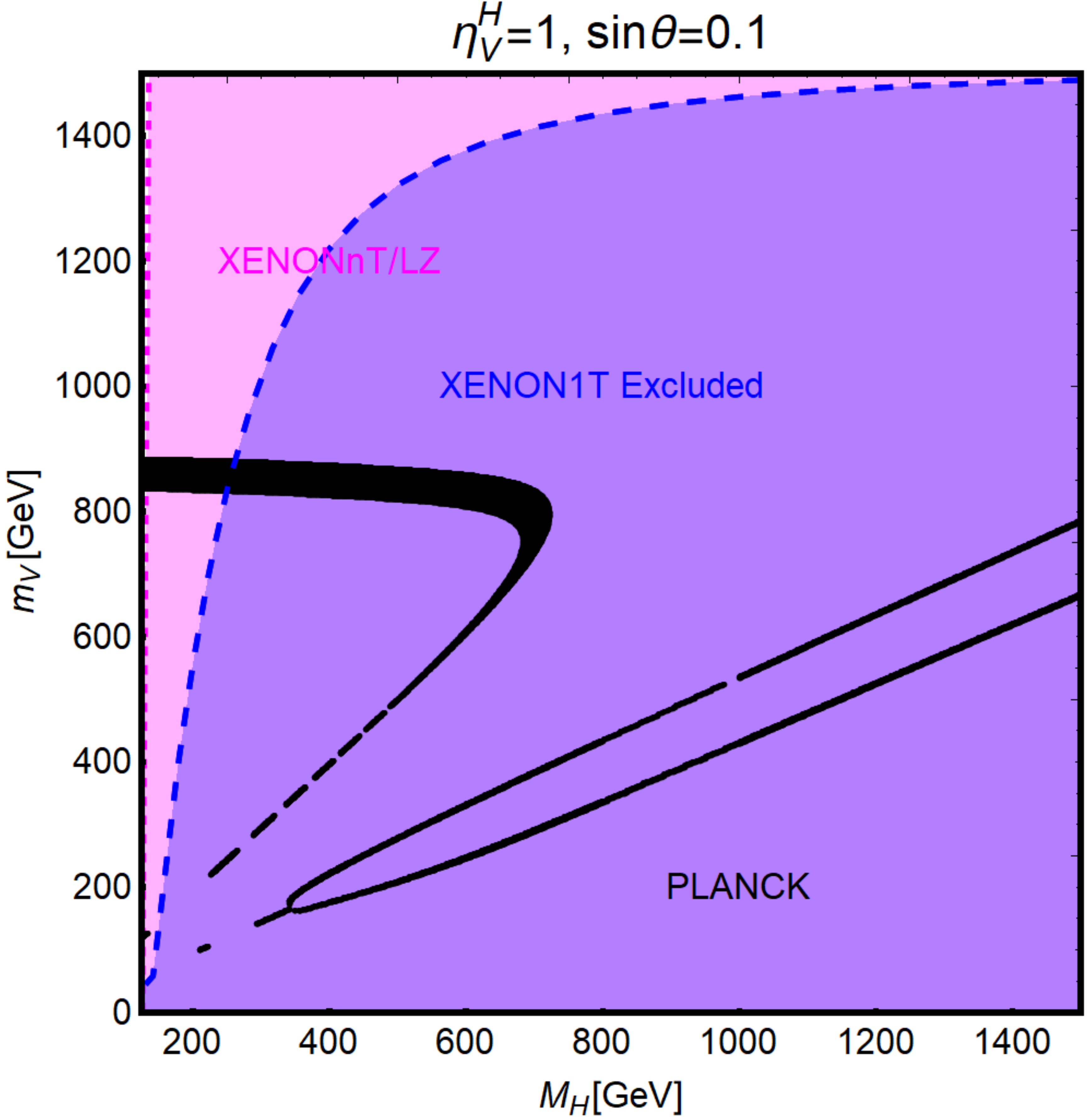}}
\end{center}
\vspace*{-3mm}
\caption{Summary of the DM constraints in the case of a heavy scalar Higgs boson mixing with the SM--like one in the plane $[M_H, m_X]$ where $X$ is a spin--0 (upper left), spin--$\frac12$ (upper right) and spin--1 state (bottom). The black contours represent the correct relic density according to PLANCK, the blue (magenta /purple) represent the current (projected) exclusion by XENON1T (XENONnT/LZ/DARWIN). The curves have been obtained for fixed assignments of the couplings reported on top of the different panels (see main text for the corresponding definitions). In all considered models, we have adopted the value $\sin\theta=0.1$ for the Higgs mixing parameter.}
\label{fig:DMHmix}
\vspace*{-3mm}
\end{figure}

The constraints from astrophysical experiments on the DM particles are
summarized in Fig.~\ref{fig:DMHmix} for the cases of scalar, fermionic and
vector DM.  Isocontours of the correct DM relic density (black lines) according
the WIMP paradigm  have been reported in the bidimensional plane
$[M_H,m_{S,\chi,V}]$. In order for the model to be viable, these contours should
lie, at least partially, outside the blue regions corresponding to the current
exclusion limits from the  XENON1T experiments. The magenta and purple regions
represent, as usual, the coverage expected in the next future XENONnT/LZ and
DARWIN experiments.

In agreement with the discussion above, the other parameters of the Higgs sector
in addition to $M_H$ have been chosen to be $\lambda_{hH}=-0.1$ and
$\sin\theta=0.1$, very close to the experimental sensitivity or limits. DM
couplings have been set to $\lambda_\phi^S=0.1$ and $\eta_V^H=1$ in the scalar
and vectorial DM cases respectively, while the coupling of the fermionic DM is
not a free parameter, being determined by $m_\chi$ and $v_\phi$. 

It is clear from the figures that the double portal model with a SM--like Higgs
plus a real singlet resonance is also strongly constrained by DM direct
detection searches. In the case of scalar and vectorial DM states,  the only
regions which could be still viable  correspond to the ones in which  $m_{S},
m_V  \gsim  M_H$ (called the secluded regime, see for instance 
Ref.~\cite{Arcadi:2016qoz} for a detailed discussion of this regime) where the
DM relic density is mostly due to the annihilation into $HH$ pairs, whose rate
is not correlated with the one of direct detection since it is not proportional
to $\sin^2 \theta$. In the case of a fermionic DM, the only viable region of
the  parameter space corresponds to the $s$--channel ``pole'' $m_\chi \simeq
\frac12 M_H$. The remaining allowed areas of the considered scenarios would be
ruled out in the absence of a signal at the future direct detection experiments.

\subsubsection{The singlet resonance case}

Let us now discuss the case in which the scalar sector of the theory is extended
by a scalar or a pseudoscalar resonance that does not mix with the SM--like
Higgs boson. As already pointed out, we will focus in this setup on fermionic
Dark Matter and consider two different scenarios. The first one is the case in
which the scalar/pseudoscalar resonance is coupled only with top quarks and the
DM particle $N$ (a similar scenario, limited to the case of a scalar mediator
only,  has been studied in great detail in Ref.~\cite{Arina:2016cqj}). The DM
relic density is determined, for $m_N \geq m_t$, by annihilation processes into
$t\bar t$ pairs occurring through a $p$--wave and an $s$--wave cross section in
the cases of scalar and pseudoscalar resonances, respectively,  with analytic
approximation that are totally analogous to the ones obtained in the previous
subsection and we do not reproduce them here. For $m_N \leq m_t$, the relic density is determined by annihilation processes into gluon pairs generated
at the one loop level (see below for further details). 

Concerning direct detection, in the case of the scalar resonance, a spin--independent  cross section is generated by $t$--channel exchange of the $H$ state, of the form 
\begin{equation} 
\sigma_{Np}^{\rm SI}=\frac{4 \mu_{Np}^2}{729 \pi
M_H^4}g_{HN\bar N}^2 g_{H t\bar t}^2 \frac{m_p^2}{m_t^2}|f_{TG}|^2 \, . 
\end{equation} 
In the case of a pseudoscalar mediator, its $t$--channel exchange leads to a
momentum suppressed cross section~\cite{Arina:2014yna,Dolan:2014ska} that is
very far from experimental sensitivity~\cite{Dolan:2014ska}. An unsuppressed
spin-independent cross section would instead arise at the one--loop
level~\cite{Freytsis:2010ne,Ipek:2014gua}. In the absence of coupling between
the pseudoscalar and the SM--like Higgs state, the corresponding cross section
is relevant only for $M_A \lesssim
10\,\mbox{GeV}$~\cite{Arcadi:2017wqi,Sanderson:2018lmj,Abe:2018emu,Ertas:2019dew}. We will
refrain from considering these low mass values in the present analysis.   Given
the simplicity of the possible models that can be considered, it is still
possible to see the main features of DM phenomenology from the astrophysical
perspective by performing an analysis in the bidimensional plane $[M_{H,A},m_N]$
for fixed values of the resonance couplings.

\begin{figure}[!h]
\vspace*{3mm}
    \centerline{
\subfloat{\includegraphics[width=0.48\linewidth]{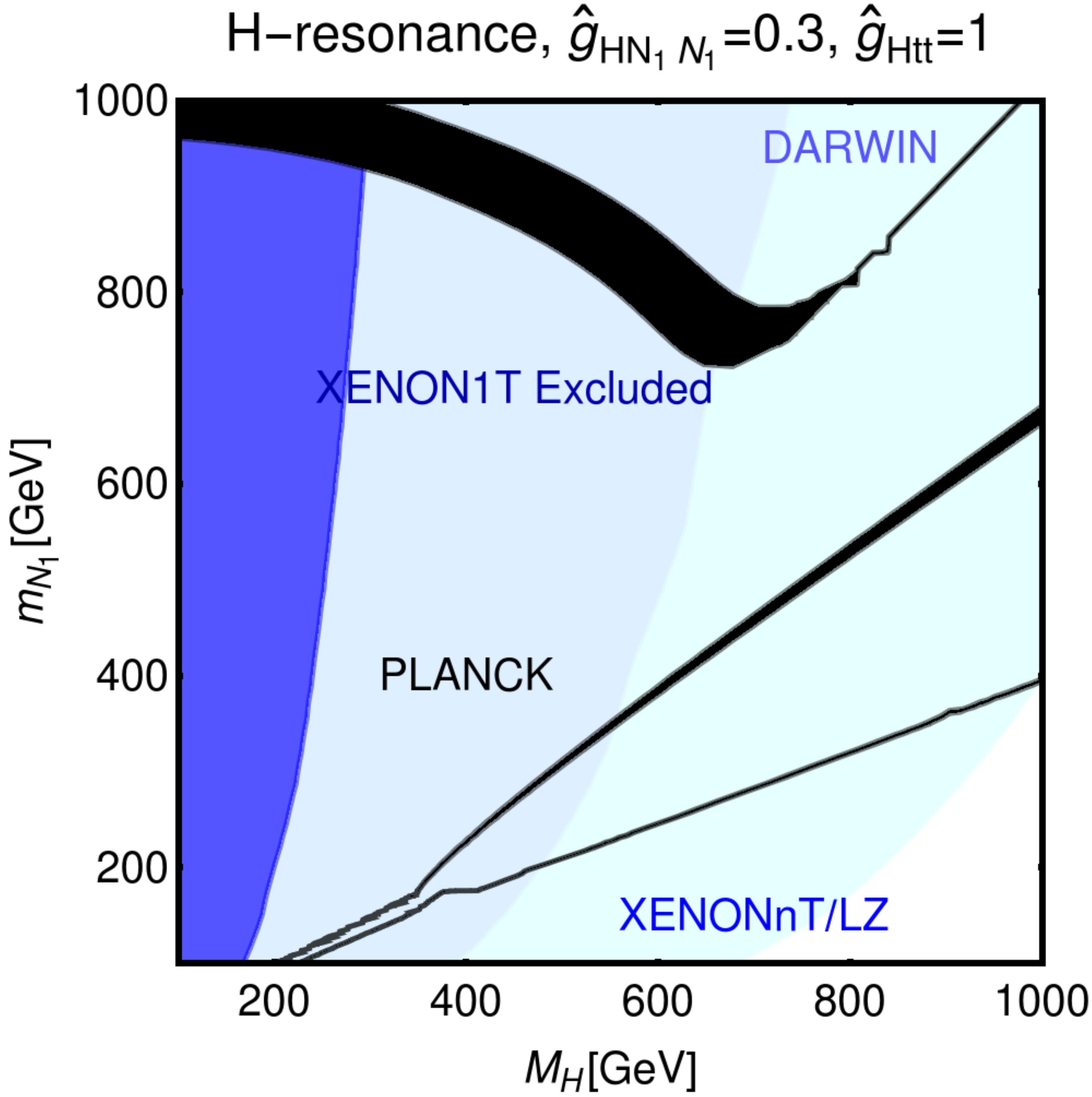}}~
    \subfloat{\includegraphics[width=0.48\linewidth]{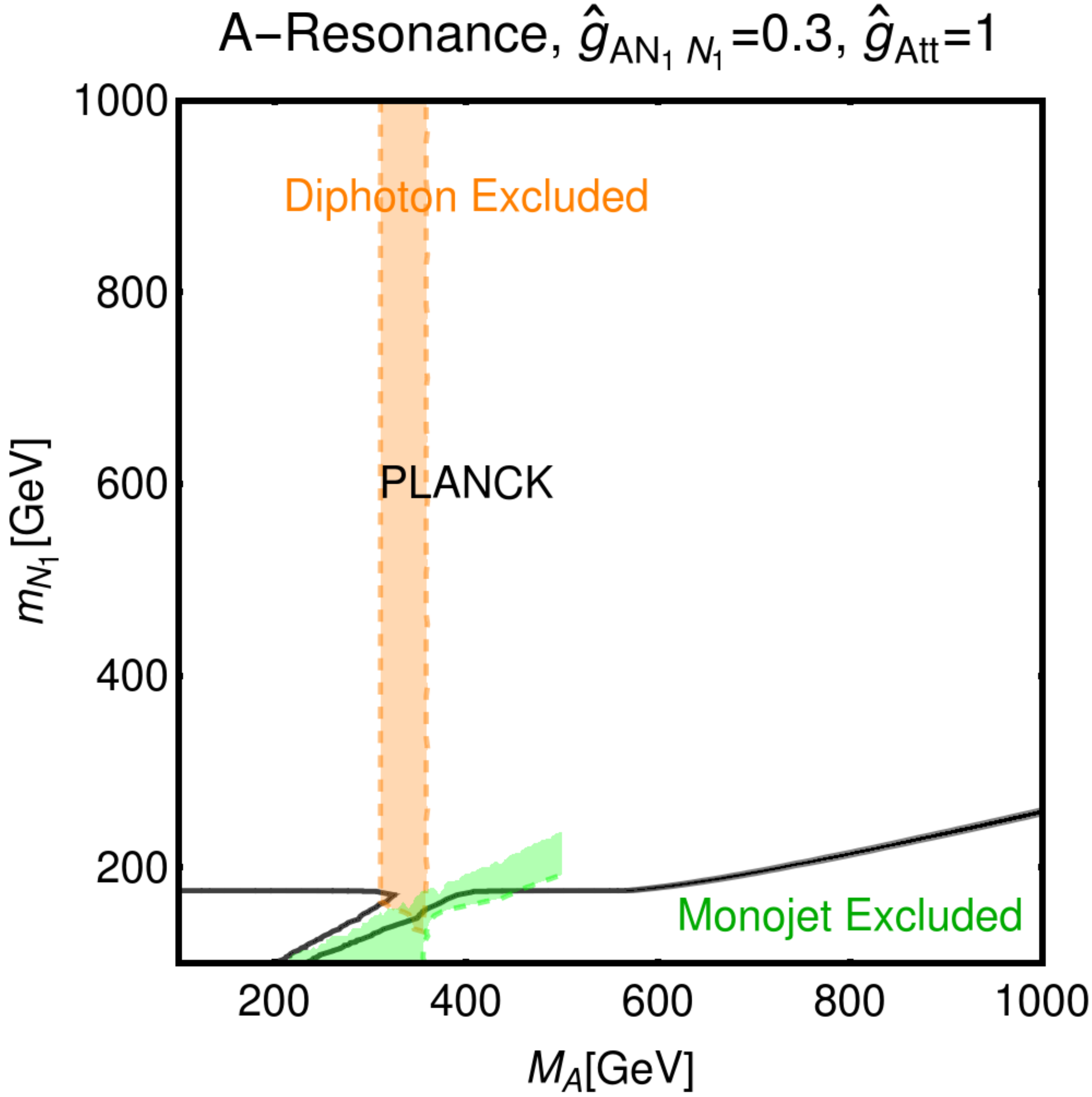}}
}
\vspace*{-1mm}
    \caption{Summary of constraints for the scenario of singlet scalar (left panel) and pseudoscalar (right panel) resonances coupled with the top-quark in the bidimensional plane $[M_{A,H},m_N]$ for a fixed assignment of the couplings, reported on top of the panels. In both plots, the black isocontours represent the correct DM relic density as measured by PLANCK. In the case of the scalar resonance, the current exclusion from XENON1T (blue region) and projected sensitivities from XENONnT/LZ (light blue) and DARWIN (cyan) are shown. In the case of a pseudoscalar resonance, there are instead marginal limits from searches of diphoton resonances (orange) and monojets (green). }
    \label{fig:singlet_tty}
\vspace*{-1mm}
\end{figure}

The outcome of such an analysis is displayed in Fig.~\ref{fig:singlet_tty} for
the scalar and pseudoscalar resonances in, respectively the left and right
panels of the figure. In the scalar case, it is clear that limits from direct
detection can be relaxed by taking a sufficiently high value of the mass of the
mediator (we also notice that the cross-section arises from top--quark loops
while in the case of the SM--like Higgs--portal, contributions of equal size are
also due to charm and bottom loops). Nevertheless, this would not be anymore the
case in the absence of signals at the future detectors. While being not affected
by direct detection constraints, the case of a pseudoscalar resonance is,
instead, marginally affected by collider constraints. These are related to the
production of the resonance in the gluon--gluon fusion process that is mediated
by top quark loops and subsequently decaying into diphotons also  generated   by
top--quark loop contributions (the excluded region by present constraints
\cite{Aaboud:2017yyg} is marked in orange) or into pairs of DM particles
accompanied by initial state radiation with a monojet signature (for which, the
excluded region \cite{Aaboud:2017phn,Sirunyan:2017hci} is marked in green).

In the second scenario that we consider, the singlet scalar/pseudoscalar
resonances feature no direct coupling with the SM states. They couple instead to
a sequential family of vector--like fermions to which the DM belongs which, in
turn, have suppressed or even vanishing Yukawa couplings with the SM--like Higgs
doublet. In this setup, the DM relic density is due to annihilation into pairs
of SM gauge bosons, mediated by the $s$--channel $H/A$ exchange through the
effective one--loop induced couplings given in
eqs.~(\ref{eq:phi-couplings})--(\ref{eq:phi-couplings_2}), as well as by
annihilations into $HH$ or $AA$  when kinematically possible. 

The annihilation cross sections, which are again given in the Appendix,  are
$p$--wave dominated in the case of the  $H$ and $s$--wave dominated in the case
of the $A$ resonances. The latter possibility  is thus potentially testable
through DM indirect detection. In this case, the main signature would be given by
gamma--ray lines which is constrained by the negative results of the searches
performed by the  FERMI  \cite{Ackermann:2015lka} and
HESS~\cite{Abramowski:2013ax} experiments.

In the case of the scalar resonance, its effective coupling with the gluons allows for the presence of a sizable spin--independent cross section of the form~\cite{DelNobile:2013sia}
\begin{equation}
\label{eq:SIgg}
    \sigma_{N_1 p}^{\rm SI}=\frac{64 \mu_{N_1 p}^2}{81 \pi M_H^4}g_{H N_1\bar N_1}^2 \frac{{(\bar c^H_{gg})}^2}{\bar \alpha_s^2} m_p^2 |f_{TG}|^2 \, , 
\end{equation}
where $\bar \alpha_s=\alpha_s(\mu_N=1\,\mbox{GeV})$ and $\bar c^H_{gg}$ is the value of the effective coupling between the $H$ state and gluons computed including  renormalisation group effects at the typical scale $\mu_N=1\,\mbox{GeV}$ of the DM scattering on nucleons \cite{DEramo:2016aee}. A spin--independent  cross section is also induced radiatively by the effective couplings of $H$ with the electroweak gauge bosons~\cite{Ovanesyan:2014fha} but  it is strongly suppressed and far from the current and next future experimental sensitivity and has thus not been included in our analysis. No relevant effects from direct detection are, instead, expected in the case of a pseudoscalar resonance.

\begin{figure}[!h]
\vspace*{-.4mm}
\centerline{
\subfloat{\includegraphics[width=0.48\linewidth]{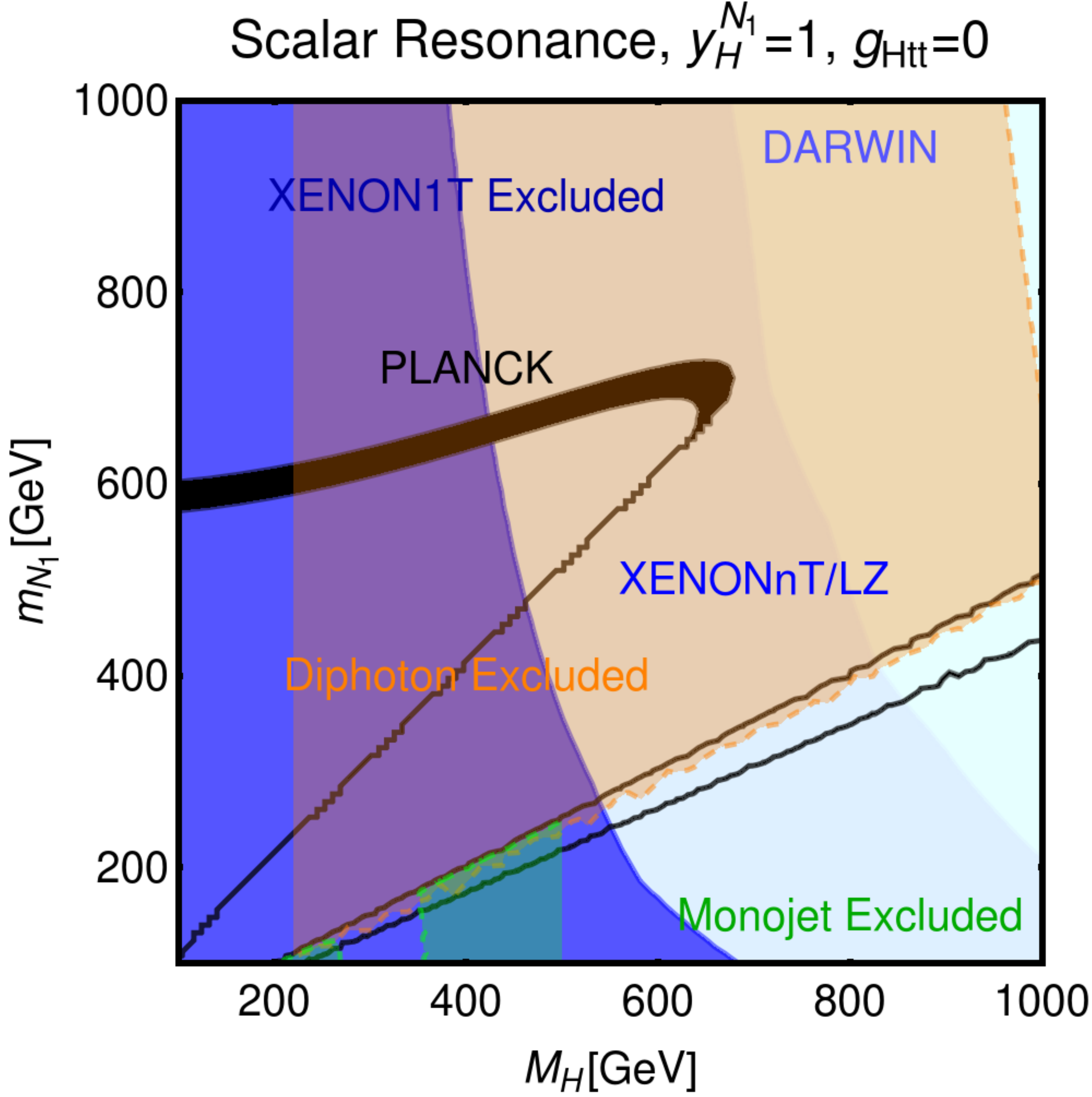}}~
\subfloat{\includegraphics[width=0.48\linewidth]{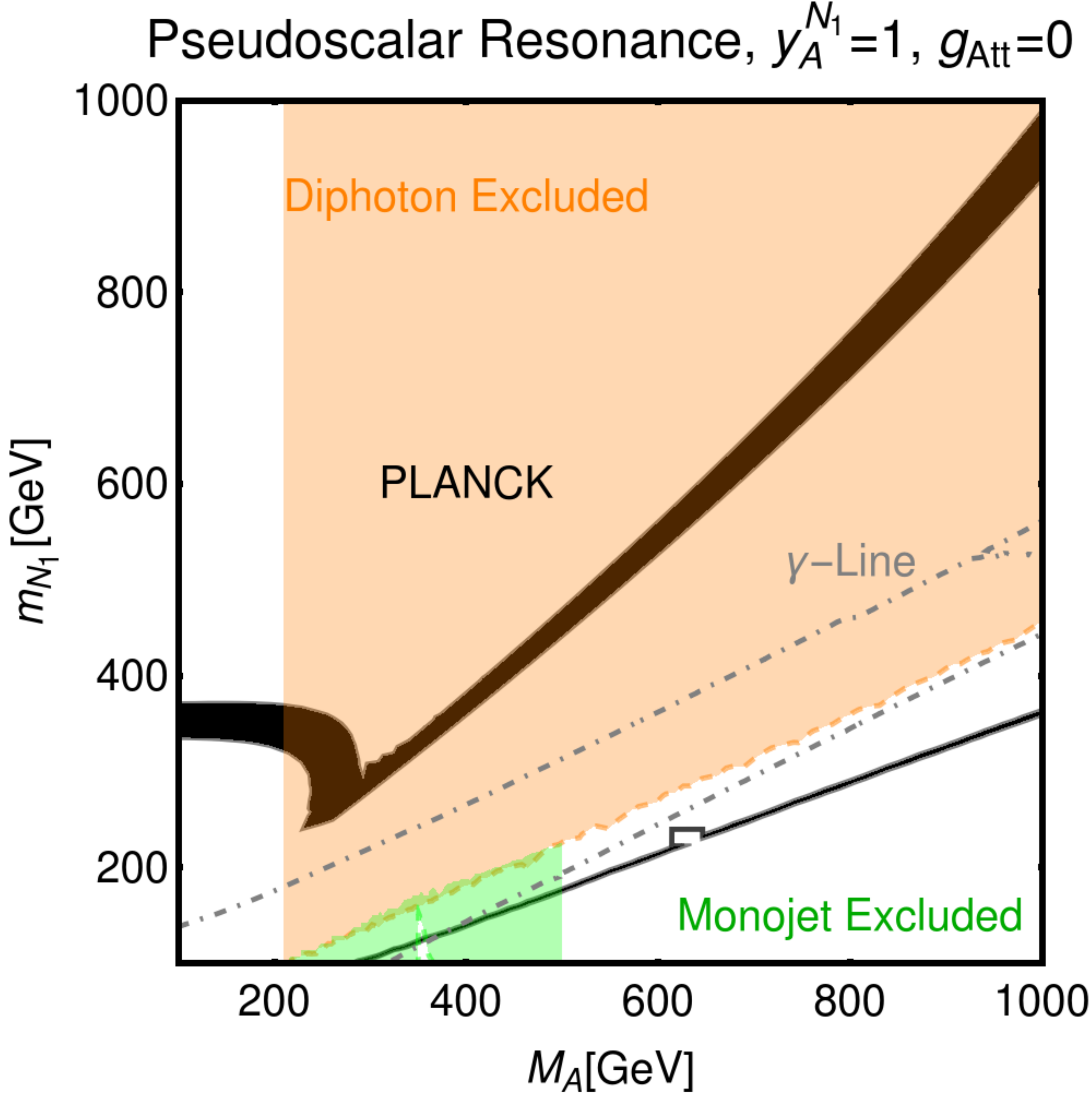}}
}
\vspace*{-1mm}
\caption{Summary of constraints for the models with a singlet scalar (left) and pseudoscalar (right) coupled with a sequential family of vector--fermions. In both panels, the black contours represent the correct DM relic density, while the orange and green regions represent the exclusions from the most recent LHC searches of diphoton resonances~\cite{Aaboud:2017yyg} and monojet events~\cite{Aaboud:2017phn,Sirunyan:2017hci}. In the scalar case, limits/prospects (blue/light blue/cyan regions) have been considered as well
while  the pseudoscalar case features instead the additional exclusion (between the dot--dashed gray lines) from searches of $\gamma$--ray lines.}
    \label{fig:singlet_VLF}
\vspace*{-2mm}
\end{figure}

As already mentioned, the case of singlet resonances coupled with sequential vector--like fermions features a high complementarity between DM phenomenology and collider searches, and all the relevant rates depend on the effective $c_{ii}^{\phi=H,A}$ couplings. An example of this complementarity is highlighted in Fig.~\ref{fig:singlet_VLF} in which we show  the combined constraints on the two types of searches for the scenarios of a scalar and a pseudoscalar resonance, in the bidimensional plane $[M_{\phi},m_{N_1}]$. For definiteness, we have fixed to 1 TeV the mass of the vector--quarks while the mass of the charged vector--leptons has been set to twice the DM mass. All the couplings between the vector fermions and the $H,A$ states have been set to unity. In the case of a scalar resonance, the correct relic density can be obtained, besides the ``secluded'' regime, only around the $s$--channel pole $m_{N_1} \sim \frac12 M_H$. This is due to the velocity suppression of the annihilation cross sections into SM particles. 

As can be seen from the left panel of the figure, this scenario is excluded by
the combined constraints from searches of diphoton resonances and DM direct
detection. Due to the absence of the direct detection constraints, a viable
region of the parameter space is present for the case of a pseudoscalar
resonance for $M_A < 200\,\mbox{GeV}$ and $m_{N_1} \simeq 400\,\mbox{GeV}$,
where there is no sensitivity from searches of diphoton resonances. We have
verified that in this region the correct cosmological relic density is mostly
determined by annihilations of the DM particle into two gluons.

\subsubsection{Scalar plus light pseudoscalar resonance}

We finally come to the simultaneous singlet scalar+pseudoscalar portal scenario.
While sharing many features with the models discussed in the previous
subsection, the presence of a very light pseudoscalar, in combination with the
scalar mediator, has a profound phenomenological impact. Indeed, it first guarantees the presence of the DM annihilation
channel into $aa$ final states and, more important, it introduces a new
annihilation channel into $Ha$ final states with an efficient $s$--wave
annihilation cross section. These processes being determined only by the
couplings to the new particle sector, the complex scalars and extra fermions,
the correlation between the relic density and the DM detection constraints is
weaker than in the previous scenarios.

As already pointed, we will focus on the scenario in which the connection between the complex field and the SM sector is provided by a full sequential family of vector like fermions. In such a case, the presence of only radiatively induced couplings leads to a very strong correlation between Dark
Matter searches and more general searches of New Physics at colliders. We also
remark that, under the assumption of dynamical generation of the VLF masses, the
effective couplings of the $H,a$ states with the SM gauge bosons are determined
by the masses of the VLFs, $M_H$ and a single coupling $\lambda_\phi$.   

\begin{figure}[!h]
\vspace*{1mm}
\centerline{
\hspace*{-2mm}
\subfloat{\includegraphics[width=0.52\linewidth]{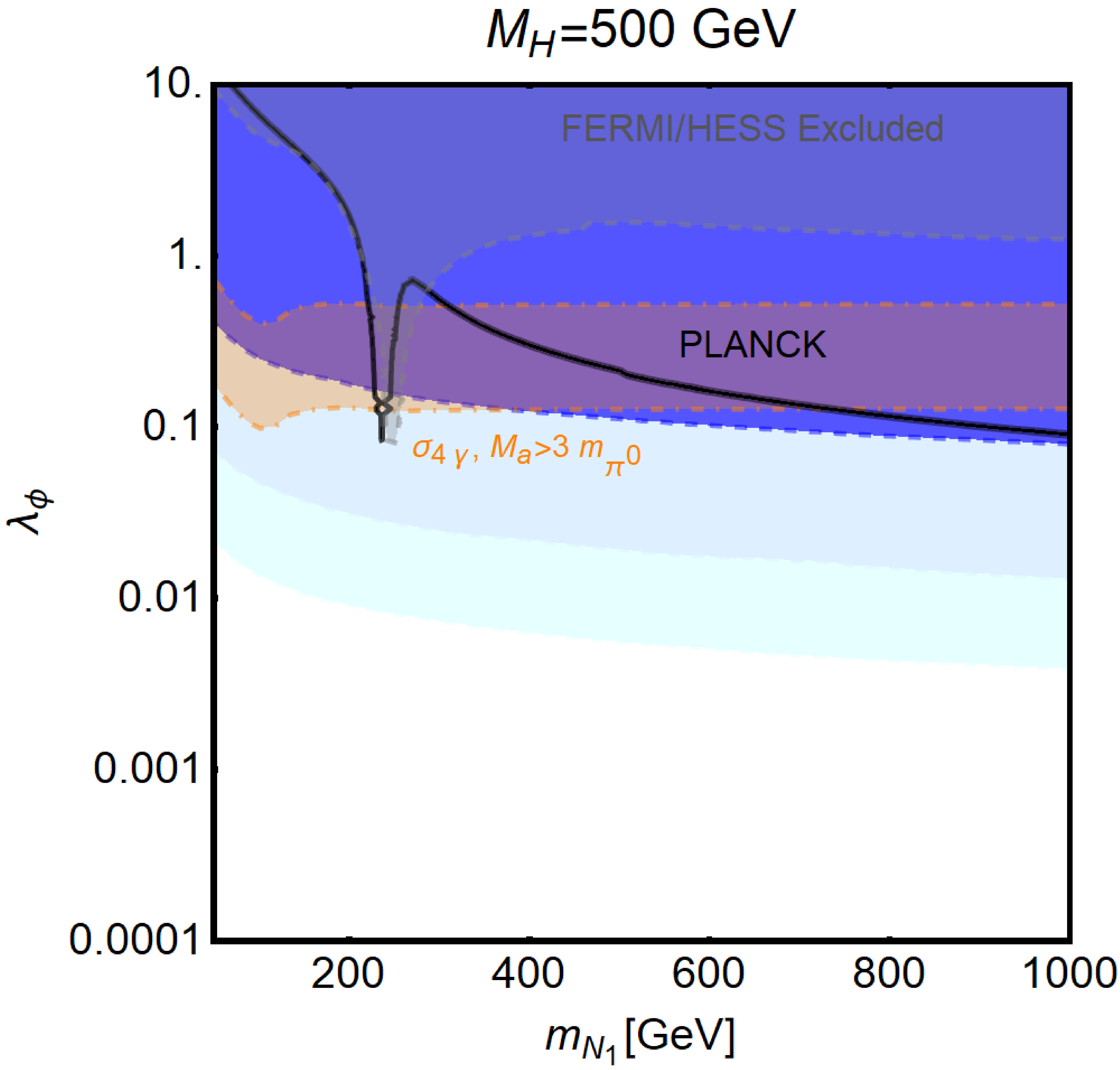}}
\hspace*{-2mm}
\subfloat{\includegraphics[width=0.52\linewidth]{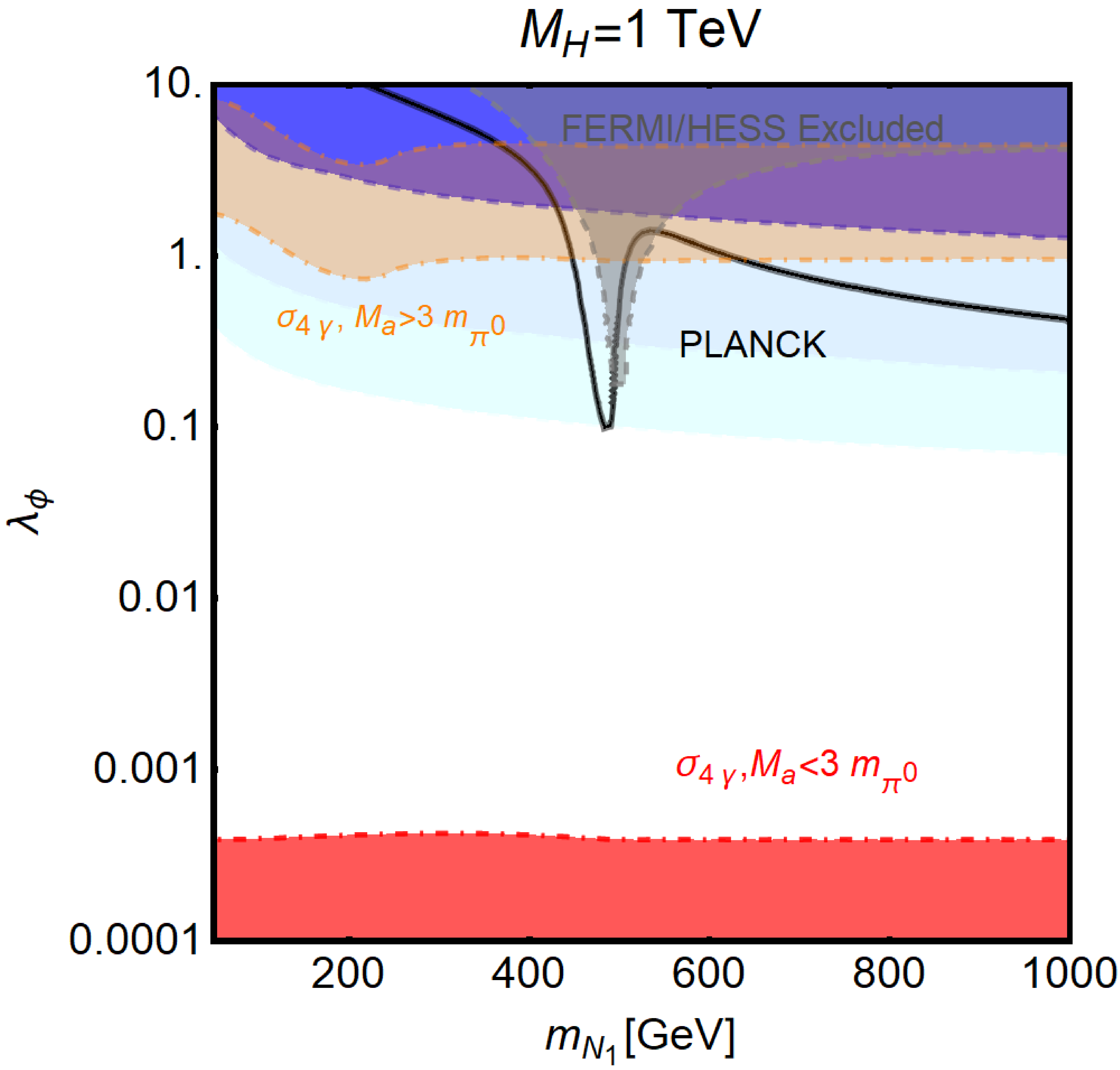}}
}
\vspace*{2mm}
\caption{Summary of constraints from DM and collider phenomenology for the model with a complex scalar singlet coupled to a sequential family of vector--like fermions. The results are shown in the $[m_{N_1},\lambda_\phi]$ plane for two assignments of $M_H$, namely 500 GeV and 1 TeV. The black contours represent the correct relic density according the WIMP paradigm. The blue (light blue/cyan) represent the current (projected) exclusion by XENON1T (XENONnT/LZ/DARWIN). The gray region is excluded by indirect searches of DM signals by FERMI/HESS. The orange (red) bands represent values of the two+four photon cross section in the $2 \sigma$ region from LHC searches of diphoton resonances by taking $M_a > 3 m_{\pi^0}$ ($M_a < 3 m_{\pi^0}$). In both plots, the masses of the charged vector--leptons are fixed to twice the DM mass while the ones of the vector--quarks are fixed to 1 TeV.}
\label{fig:FSIiAportaldyn}
\vspace*{-.2mm}
\end{figure}

In this case, it is more interesting to present the analysis of the combined
astroparticle and collider constraints in the $[m_{N_1},\lambda_\phi]$ plane for
some fixed assignments of $M_H,M_a$ and such results are shown in
Fig.~\ref{fig:FSIiAportaldyn}. The two panels of the figure differ by the
assignment of $M_H$, 500 GeV and 1 TeV in, respectively, the left and right
panel. In both cases, we have considered two values of $M_a$, one above and one
below the kinematic threshold of the $a \rightarrow 3 \pi$ decay mode (notice
that this affects only the size of the diphoton signal) while, for what concerns
the VLFs, we have set the masses of the vector--like leptons two be twice the DM
mass while the masses of the VLQs have been set to the constant value of 2 TeV. 

In both panels, the shape of the relic density contours can be explained as
follows. For $m_{N_1}\lesssim \frac12 M_H$, DM annihilation proceeds mostly
through the $gg$ (mostly relevant around the $\frac12 M_H$ pole) and the
velocity suppressed $aa$ final states. Values $\lambda_\phi >1$
are thus required to match the experimental value of the DM relic
density. For DM masses above the $\frac12 M_H$ ``pole'', DM annihilations remain
instead efficient thanks to the $s$--wave annihilation channel into $Ha$ final
states. Given the presence of a pseudoscalar mediator, the scenario under
consideration is moderately sensitive to indirect searches of $\gamma$--ray
lines. 

The strongest limits from DM searches come from direct detection as a
consequence of the spin--independent  cross section analogous to the one given
in eq.~(\ref{eq:SIgg}). As it should be clear from
Fig.~\ref{fig:FSIiAportaldyn}, the case $M_H=500\,\mbox{GeV}$ is already
excluded by present constraints. The case $M_H=1\,\mbox{TeV}$, while being at
the moment still viable, will be fully probed and eventually ruled out by
ultimate detectors like DARWIN. We finally notice a very nice complementary with
searches of diphoton resonances. This type of complementarity comes from the
observation that the cross section associated to the main annihilation channels
are proportional to $\lambda_{\phi}^2$, similarly to the production cross
section of collimated photons. 

In units of 
\beq
\sigma_H^{\rm unit} = \left( \frac{c_{\rm gg}^{\phi=H} (M_H/\sqrt{s})} {1000} \right) e\times {\left( \frac{\langle \sigma v \rangle_{aa}} {3 \times 10^{-26} {\mbox{cm}}^3 {\mbox{s}}^{-1}}\right)} \times  \left(\frac{100\,\mbox{GeV}} {m_{N_1}}\right) \times \left(\frac{M_H}{1\,\mbox{TeV}}\right)^2\, , 
\eeq
the cross sections for the two different processes can be thus related, and one  can use the following simple analytic approximations  for them
\begin{align}
\sigma_{4\gamma}\approx 1.7\,\mbox{pb} \times \sigma_H^{\rm unit}
& \quad \mbox{for}\, M_a < 3 \pi^0,\,\, m_{N_1}< \frac12 M_H, \nonumber\\
\sigma_{4\gamma}\approx 0.65\,\mbox{fb} \times \sigma_H^{\rm unit}
& \quad \mbox{for}\, M_a > 3 \pi^0,\,\,m_{N_1}< \frac12 M_H,\nonumber\\
\sigma_{4\gamma}\approx 0.41\,\mbox{pb} \times \sigma_H^{\rm unit}
& \quad \mbox{for}\, M_a < 3 \pi^0,\,\,m_{N_1}> \frac12 M_H,\nonumber\\
\sigma_{4\gamma}\approx 0.15\,\mbox{fb} \times \sigma_H^{\rm unit}
& \quad \mbox{for}\, M_a > 3 \pi^0,\,\,m_{N_1}> \frac12 M_H \, . 
\end{align}
From the previous figures, one can notice that in the regime $M_a < 3
m_{\pi^0}$, the thermally favored value of the DM annihilation cross section
corresponds to values of the production cross section at the LHC $\sigma_{4 \gamma}$ well above the experimental limits. In the opposite case,  most of the parameter space favored by thermal production of the DM particles lies in correspondence or very close to the current LHC sensitivity.

%% file: sec-2HDM.tex
\section{Doublet extensions of the Higgs sector}

We turn now to the scenarios in which the Higgs sector of the theory incorporates
two doublet fields. We first consider the case in which both Higgs doublets
contribute to electroweak symmetry breaking, the so--called 2HDM
\cite{Branco:2011iw}. These are extended to incorporate the DM particles in a
way analogous to what has been done in section 3 and our focus will be on
scenarios in which  a full sequential family of vector--like fermions or a
singlet--doublet of heavy leptons are added to the spectrum. We then consider
the case in which only one of the Higgs doublets is responsible of electroweak
symmetry breaking, while the other doublet does not acquire a vev nor couple to
SM fermions, the so called inert doublet model or IDM in which the DM candidate
will be identified with one of the neutral components of the inert field.  As a
final scenario, we consider the case in which the two doublets Higgs sector is
further extended to incorporate a light pseudoscalar singlet. Such a scenario is
of phenomenological interest as it allows a gauge invariant coupling between the
SM sector and a pure gauge singlet fermionic DM and represents a useful limit of
the NMSSM, which will be treated in the final section of this review. The
section is structured in an analogous manner as the two previous ones: we first
describe the models, including the related theoretical constraints, move then to
the collider constraints and prospects and conclude with an analysis of the
astrophysical aspects of the DM particle.

\subsection{The two--Higgs doublet model}

In a 2HDM, the Higgs sector incorporates two doublets of scalar fields $\Phi_1$ and $\Phi_2$ and, assuming  CP conservation, is described by the following scalar potential
\begin{align}
\label{eq:scalar_potential}
 V(\Phi_1,\Phi_2) &= m_{11}^2 \Phi_1^\dagger \Phi_1+ m_{22}^2 \Phi_2^\dagger \Phi_2 - m_{12}^2 \left(\Phi_1^\dagger \Phi_2 + {\rm h.c.} \right)
+\frac{\lambda_1}{2} \left( \Phi_1^\dagger \Phi_1 \right)^2
+\frac{\lambda_2}{2} \left( \Phi_2^\dagger \Phi_2 \right)^2  \nonumber\\ &
+\lambda_3\left(\Phi_1^\dagger \Phi_1 \right)\left(\Phi_2^\dagger \Phi_2 \right)
+\lambda_4\left(\Phi_1^\dagger \Phi_2 \right)\left(\Phi_2^\dagger \Phi_1 \right)
+\frac{\lambda_5}{2}\left[ \left(\Phi_1^\dagger \Phi_2 \right)^2 + {\rm h.c.} \right] \, .
\end{align}
We  have assumed from the start the presence of a discrete
symmetry~\cite{Davidson:2005cw} which forbids the appearance of two additional
couplings $\lambda_{6}$ and  $\lambda_{7}$. The electroweak symmetry is broken by the vevs $v_1$ and $v_2$ acquired by the fields $\Phi_1$ and $\Phi_2$, respectively. The vevs satisfy the relation $\sqrt{v_1^2+ v_2^2}=v \simeq 246$ GeV and their ratio defines the parameter $\tan\beta \equiv t_\beta=v_2/v_1$ which will play a most important role in the model. After electroweak symmetry
breaking, the two doublet fields can be decomposed as
\begin{equation}
\Phi_i=
\begin{pmatrix} \phi_i^+ \\ (v_i+\rho_i +i \eta_i)/\sqrt{2} \end{pmatrix}~,
\qquad
i=1,2,
\end{equation}
\noindent
and lead to five physical states: two CP--even states $h$ and $H$, a CP--odd state $A$ and two charged Higgs bosons, which are defined through the transformations
\begin{equation}
\label{eq:rotation2}
\left(
\begin{array}{c} \phi_1^+ \\ \phi_2^+ \end{array} \right) = \Re_\beta 
\left( \begin{array}{c} G^+ \\ H^+ \end{array} \right), \ \  
\left( \begin{array}{c} \eta_1 \\ \eta_2 \end{array} \right)= \Re_\beta 
\left( \begin{array}{c} G^0 \\ A \end{array} \right), \ \
\left( \begin{array}{c} \rho_1 \\ \rho_2 \end{array} \right)= \Re_\alpha 
\left( \begin{array}{c} H \\ h \end{array} \right) \, , 
\end{equation} 
with $\Re_X$ the rotation matrix of angle $X$ given in eq.~(\ref{eq:rotation}) 
and $G^0,G^+$ the Goldstone bosons that become the longitudinal degrees of
freedom of the SM gauge bosons. The angle $\alpha$ describes the mixing between
the two CP--even states $h$ and $H$, the former being again conventionally identified with the 125 GeV Higgs boson observed at the LHC, while the $H$ boson will be considered to be heavier in our context (although there is still a tiny possibility that a scalar boson lighter than 125 GeV is present in the spectrum~\cite{Cacciapaglia:2016tlr}).  

The quartic couplings of the scalar potential eq.~(\ref{eq:scalar_potential}) can be expressed as functions of the masses of the physical states and, introducing $M \equiv m_{12}/(\sin {\beta} \cos {\beta})$, they read
\begin{align}
\label{eq:quartic_physical}
\lambda_1 &= \frac{1}{v^2} \left[- M^2 \tan^2 \beta +\frac{\sin^2 \alpha}{\cos^2 \beta} M_h^2 +\frac{\cos^2 \alpha}{\cos^2\beta}M_H^2 \right], \nonumber \\
\lambda_2 &= \frac{1}{v^2} \left[ -\frac{M^2 }{\tan^2 \beta}+\frac{\cos^2 \alpha}{\sin^2 \beta}M_h^2+\frac{\sin^2 \alpha}{\sin^2 \beta}M_H^2 \right], \nonumber \\
\lambda_3 &= \frac{1}{v^2} \left[-M^2 +2 M_{H^{\pm}}^2 +\frac{\sin 2\alpha}{\sin 2\beta}( M_H^2-M_h^2) \right], \nonumber \\
\lambda_4 &= \frac{1}{v^2} \left[ M^2 + M_A^2 - 2 M_{H^{\pm}}^2 \right],  \nonumber \\
\lambda_5 & = \frac{1}{v^2} \left[ M^2 - M_A^2 \right]. 
\end{align} 
They should comply with a series of constraints which, with the help of eq.~(\ref{eq:quartic_physical}), translate into bounds on the masses $M_A,M_H,M_{H^{\pm}}$ as functions of the angles $\alpha$ and $\beta$. The most
relevant bounds are, as in the singlet Higgs  case discussed before,  
as follows~\cite{Kanemura:2004mg,Becirevic:2015fmu}:

\begin{itemize}

\item the scalar potential should be bounded from below:
\begin{equation}
\label{eq:up1}
\lambda_{1,2} > 0, \; \lambda_3 > -\sqrt{\lambda_1\lambda_2} \; \; {\rm and} \; \lambda_3 + \lambda_4 - \left|\lambda_5\right| > -\sqrt{\lambda_1\lambda_2} \; ;
\end{equation}
\item $s$--wave tree--level unitarity should be satisfied:
\begin{equation}
\label{eq:up2}
\left| a_{\pm} \right|, \left| b_{\pm} \right|, \left| c_{\pm} \right|, 
\left| d_\pm \right| , \left| e_\pm \right| , 
\left| f_{\pm} \right|  < 8\pi,
\end{equation}
where:\vspace*{-1cm}
\begin{align}
a_{\pm} &= \frac{3}{2}(\lambda_1 + \lambda_2) \pm \sqrt{\frac{9}{4}(\lambda_1-\lambda_2)^2 + (2\lambda_3 + \lambda_4)^2}, \notag \\
b_{\pm} &= \frac{1}{2}(\lambda_1 + \lambda_2) \pm \sqrt{(\lambda_1-\lambda_2)^2 + 4\lambda_4^2}, \notag \\
c_{\pm} &= \frac{1}{2}(\lambda_1 + \lambda_2) \pm \sqrt{(\lambda_1-\lambda_2)^2 + 4\lambda_5^2}, \notag \\
d_{\pm} &= \lambda_3 + 2\lambda_4 \mp 3\lambda_5, \ e_\pm = \lambda_3 \mp \lambda_5, \  f_\pm = \lambda_3 \pm  \lambda_4 \; ; 
\end{align}
\item $v_1$ and $v_2$ should be global minima  for the scalar potential~\cite{Barroso:2013awa}:
\begin{equation}
\label{eq:vacuum_2HDM}
m_{12}^2 \left(m_{11}^2-m_{22}^2 \sqrt{\lambda_1/\lambda_2}\right)\left(\tan\beta-\sqrt[4]{\lambda_1/\lambda_2}\right)>0 \; ; 
\end{equation}
\noindent
\item the electroweak vacuum should remain stable:
\begin{align}
& m_{11}^2+\frac{\lambda_1 v^2 \cos^2\beta}{2}+\frac{\lambda_3 v^2 \sin^2\beta}{2}=\tan\beta \left[m_{12}^2-(\lambda_4+\lambda_5)\frac{v^2 \sin 2\beta}{4}\right] , \nonumber\\
& m_{22}^2+\frac{\lambda_2 v^2 \sin^2\beta}{2}+\frac{\lambda_3 v^2 \cos^2\beta}{2}=\frac{1}{\tan\beta} \left[m_{12}^2-(\lambda_4+\lambda_5)\frac{v^2 \sin2\beta}{4}\right] . 
\end{align}
\end{itemize}

The mass parameter $m_{12}$ enters only in the quartic couplings among the Higgs bosons, 
\begin{eqnarray}
 \lambda_{\phi_i \phi_j \phi_k}= g^\text{2HDM}_{\phi_i \phi_j \phi_k}/g^\text{SM}_{HHH}  = f(\alpha, \beta, m_{12}) . 
\end{eqnarray}

The mixing in the CP--even Higgs sector makes that the neutral $h$ and $H$ bosons share the coupling of the SM Higgs particle to the massive gauge bosons
$V=W,Z$  
\begin{eqnarray}
g_{hVV}= g^\text{2HDM}_{hVV}/g^\text{SM}_{HVV}= \sin(\beta-\alpha) \ , \quad 
g_{HVV}= g^\text{2HDM}_{HVV}/g^\text{SM}_{HVV}= \cos(\beta-\alpha) , 
\end{eqnarray}
while, by virtue of CP invariance,  there is no coupling of the CP--odd $A$ to vector bosons, $g_{AVV}=0$. There are also couplings between two Higgs and a vector boson which, up to a normalization factor, are complementary to the ones given above. For instance, one has
\begin{eqnarray}
g_{hAZ} = g_{h H^\pm W}=  \cos(\beta-\alpha) \ , \quad 
g_{HAZ} = g_{H H^\pm W} =  \sin(\beta-\alpha) . 
\label{eq:cplg-HHV1}
\end{eqnarray}
Finally, there are additional bosonic couplings of the charged Higgs boson which are 
simply 
\begin{eqnarray}
g_{A H^\pm W}= 1 \, ,~~ g_{H^+ H- \gamma} =  -e\, , \   g_{H^+ H^- Z} = -e \cos2\theta_W/(\sin\theta_W \cos\theta_W) . 
\label{eq:cplg-HHV2}
\end{eqnarray}

The couplings of the various Higgs bosons to the SM fermions are slightly more complicated and are described by the following Yukawa Lagrangian
\begin{align}
-{\cal L}_{\rm Yuk}^{\rm SM}& =\sum\limits_{f=u,d,l} \frac{m_f}{v} \left[g_{hff} \bar{f}f h +g_{Hff} \bar{f}f H-i g_{Aff} \bar{f} \gamma_5 f A \right] \notag \\
&- \frac{\sqrt{2}}{v} \left[ \bar{u} \left(m_u g_{Auu} P_L + m_d g_{Add} P_R \right)d H^+ +  m_l g_{All} \bar \nu  P_R \ell H^+  + \mathrm{h.c.} \right],
\end{align}
where $P_{L/R}=\frac12(1\mp \gamma_5)$ and  $g_{\phi ff}$ are the reduced couplings of the $\phi$ boson to up-- and down--type quarks and charged leptons normalized to the SM couplings, $g_{\phi ff}=g^\text{2HDM}_{\phi ff}/g^\text{SM}_{H ff}$. 

\begin{table}[h!]
\renewcommand{\arraystretch}{1.55}
\begin{center}
\begin{tabular}{|c|c|c|c|c|}
\hline
~~~~~~ &  ~~Type I~~ & ~~Type II~ & Lepton-specific & Flipped \\
\hline\hline 
$g_{huu}$ & $ \frac{\cos \alpha} { \sin \beta} \rightarrow 1$ & $\frac{ \cos \alpha} {\sin \beta} \rightarrow 1$ & $\frac{ \cos \alpha} {\sin\beta} \rightarrow 1$ & $ \frac{ \cos \alpha}{ \sin\beta}\rightarrow 1$ \\ \hline
$g_{hdd}$ & $\frac{\cos \alpha} {\sin \beta} \rightarrow 1$ & $-\frac{ \sin \alpha} {\cos \beta} \rightarrow 1$ & $\frac{\cos \alpha}{ \sin \beta} \rightarrow 1$ & $-\frac{ \sin \alpha}{ \cos \beta} \rightarrow 1$ \\ \hline
$g_{hll} $ & $\frac{\cos \alpha} {\sin \beta} \rightarrow 1$ & $-\frac{\sin \alpha} {\cos \beta} \rightarrow 1$ & $- \frac{ \sin \alpha} {\cos \beta} \rightarrow 1$ & $\frac{ \cos \alpha} {\sin \beta} \rightarrow 1$  \\ \hline\hline
$g_{Huu}$ & $\frac{\sin \alpha} {\sin \beta} \rightarrow -\frac{1}{\tan\beta}$ & $\frac{ \sin \alpha} {\sin \beta} \rightarrow -\frac{1}{\tan\beta}$ & $ \frac{\sin \alpha}{\sin \beta} \rightarrow -\frac{1}{\tan\beta}$ & $\frac{ \sin \alpha}{ \sin \beta} \rightarrow -\frac{1}{\tan\beta}$ \\ \hline
$g_{Hdd}$ & $ \frac{ \sin \alpha}{\sin \beta} \rightarrow -\frac{1}{\tan\beta}$ & $\frac{\cos \alpha}{\cos \beta} \rightarrow {\tan\beta}$ & $\frac{\sin \alpha} {\sin \beta} \rightarrow -\frac{1}{\tan\beta}$ & $\frac{ \cos \alpha} {\cos \beta} \rightarrow {\tan\beta}$ \\ \hline
$g_{Hll}$ & $\frac{ \sin \alpha} {\sin \beta} \rightarrow -\frac{1}{\tan\beta}$ & $\frac{\cos \alpha} {\cos \beta} \rightarrow {\tan\beta}$ & $\frac{ \cos \alpha} {\cos \beta} \rightarrow {\tan\beta}$ & $\frac{\sin \alpha} {\sin \beta} \rightarrow -\frac{1}{\tan\beta}$ \\
\hline\hline
$g_{Auu}$ & $\frac{1}{\tan\beta}$ & $\frac{1}{\tan\beta}$ & $\frac{1}{\tan\beta}$ & $\frac{1}{\tan\beta}$
\\
\hline
$g_{Add}$ & $-\frac{1}{\tan\beta}$ & ${\tan\beta}$ & $-\frac{1}{\tan\beta}$ & ${\tan\beta}$
\\
\hline
$g_{All}$ & $-\frac{1}{\tan\beta}$ & ${\tan\beta}$ & ${\tan\beta}$ & $-\frac{1}{\tan\beta}$
\\
\hline
\end{tabular}
\caption{Couplings of the 2HDM Higgs bosons to fermions, normalized to those of the SM--like Higgs boson,  as a function of the angles $\alpha$ and $\beta$ and, in the case of the CP--even Higgs states,  their values in the alignment limit $\beta \!-\! \alpha \rightarrow \frac{\pi}{2}$.}
\label{table:2hdm_type}
\end{center}
\vspace*{-6mm}
\end{table}

In a 2HDM in which the appearance of the experimentally not observed
flavour--changing neutral currents (FCNCs) is enforced, two options are in
general  discussed for the interactions of the Higgs states with fermions
\cite{Branco:2011iw,Glashow:1976nt}: in the so--called  Type II model, the field
$\Phi_1$ generates the masses of isospin down--type fermions and $\Phi_2$ the
masses of up--type quarks. In turn, in Type I models, the field $\Phi_2$ couples
to both isospin up-- and down--type fermions.  Here, we will consider besides
these two types of models, two additional options in which the charged leptons
have a different behaviour compared to down--type  quarks, namely the lepton
specific model in which the Higgs couplings to quarks are as in Type I but those
to leptons are as in Type II, and the flipped model in which the previous
situation occurs but with Type I and Type II couplings reversed. The values of
the  fermion couplings  for these four flavour--conserving types of 2HDMs are
listed in Table~\ref{table:2hdm_type}.

\begin{figure}[!ht]
\begin{center}
\subfloat{\includegraphics[width=0.43\linewidth]{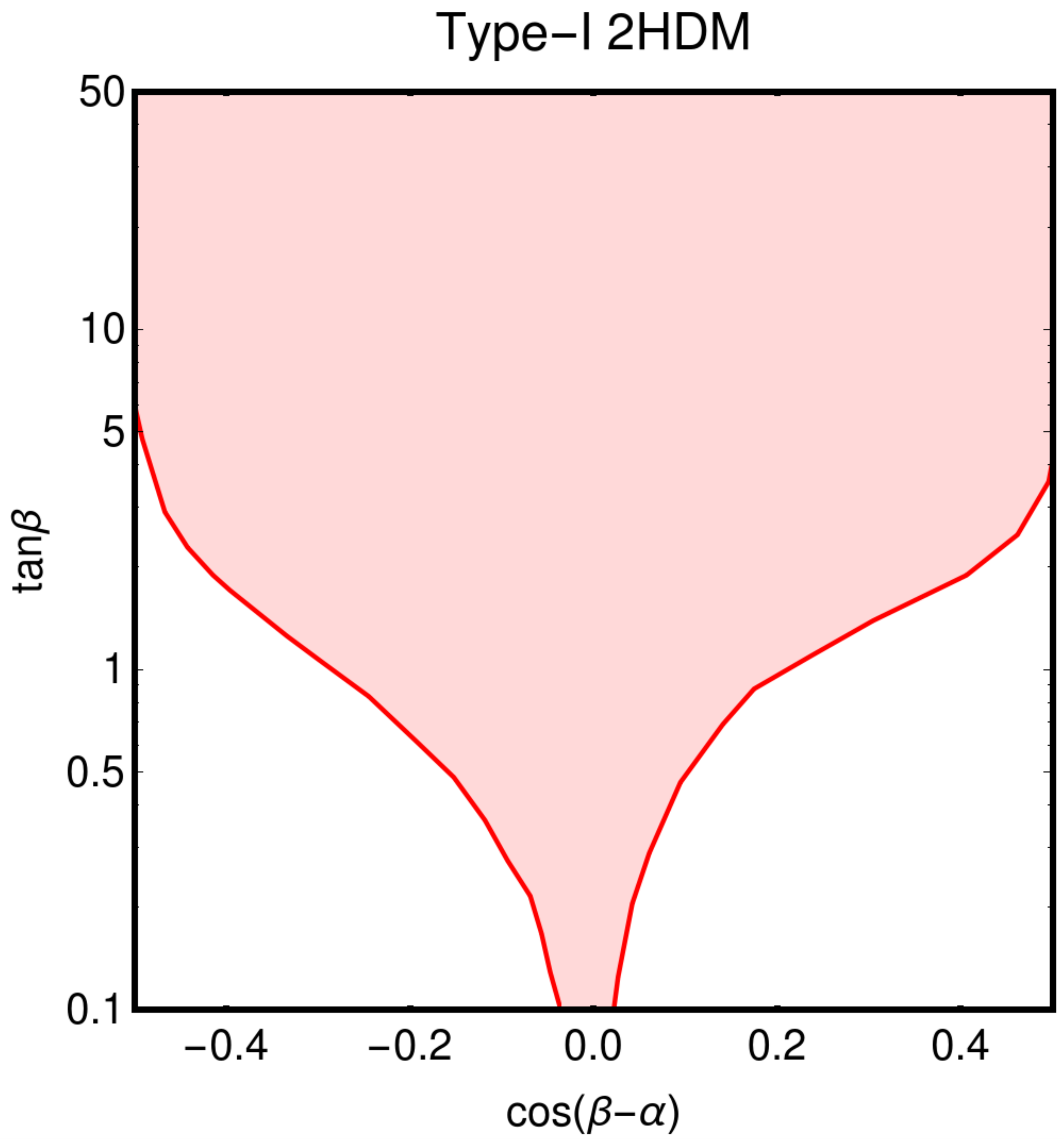}}~~~~
\subfloat{\includegraphics[width=0.43\linewidth]{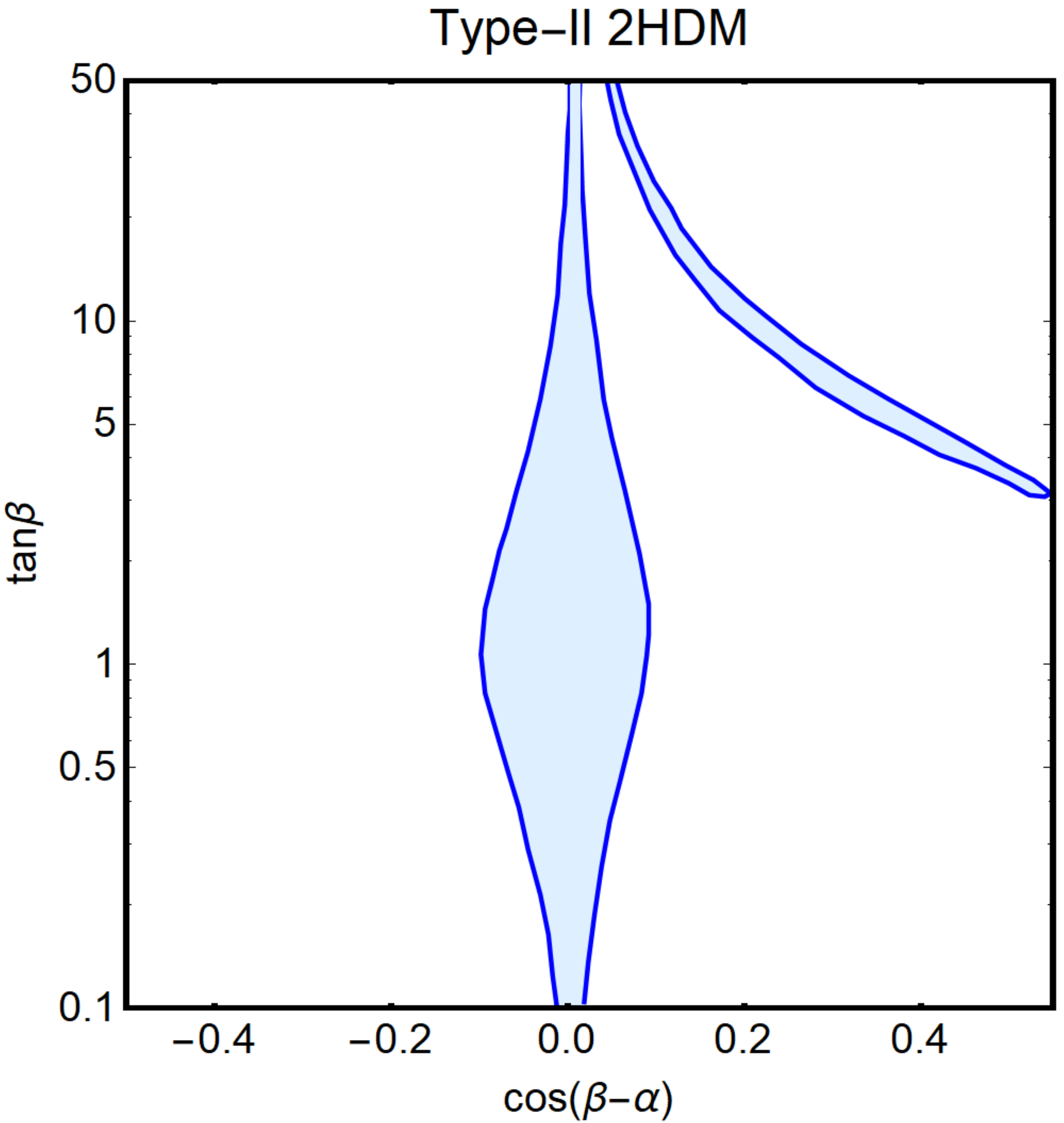}}\\
\subfloat{\includegraphics[width=0.43\linewidth]{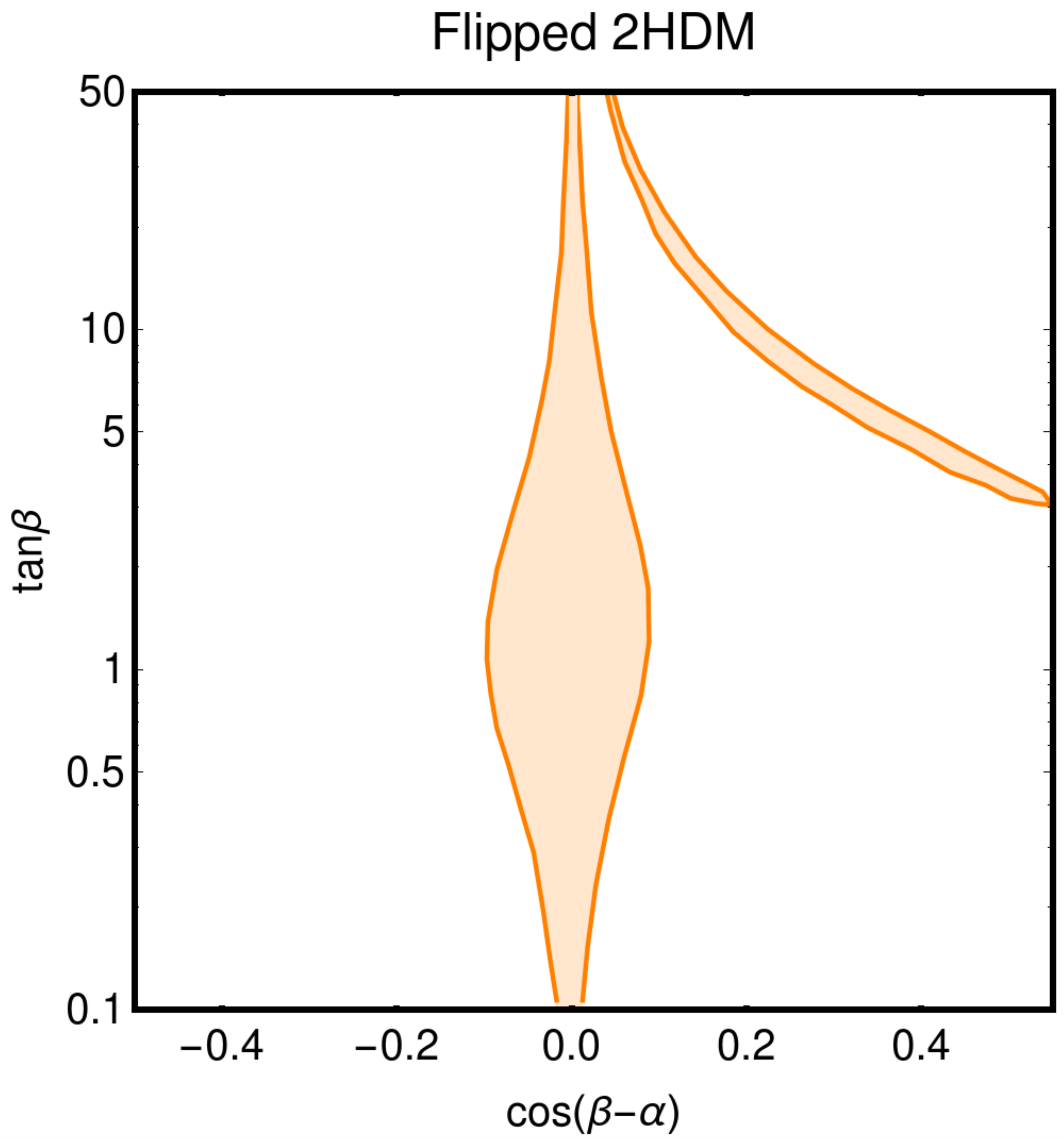}}~~~~
\subfloat{\includegraphics[width=0.43\linewidth]{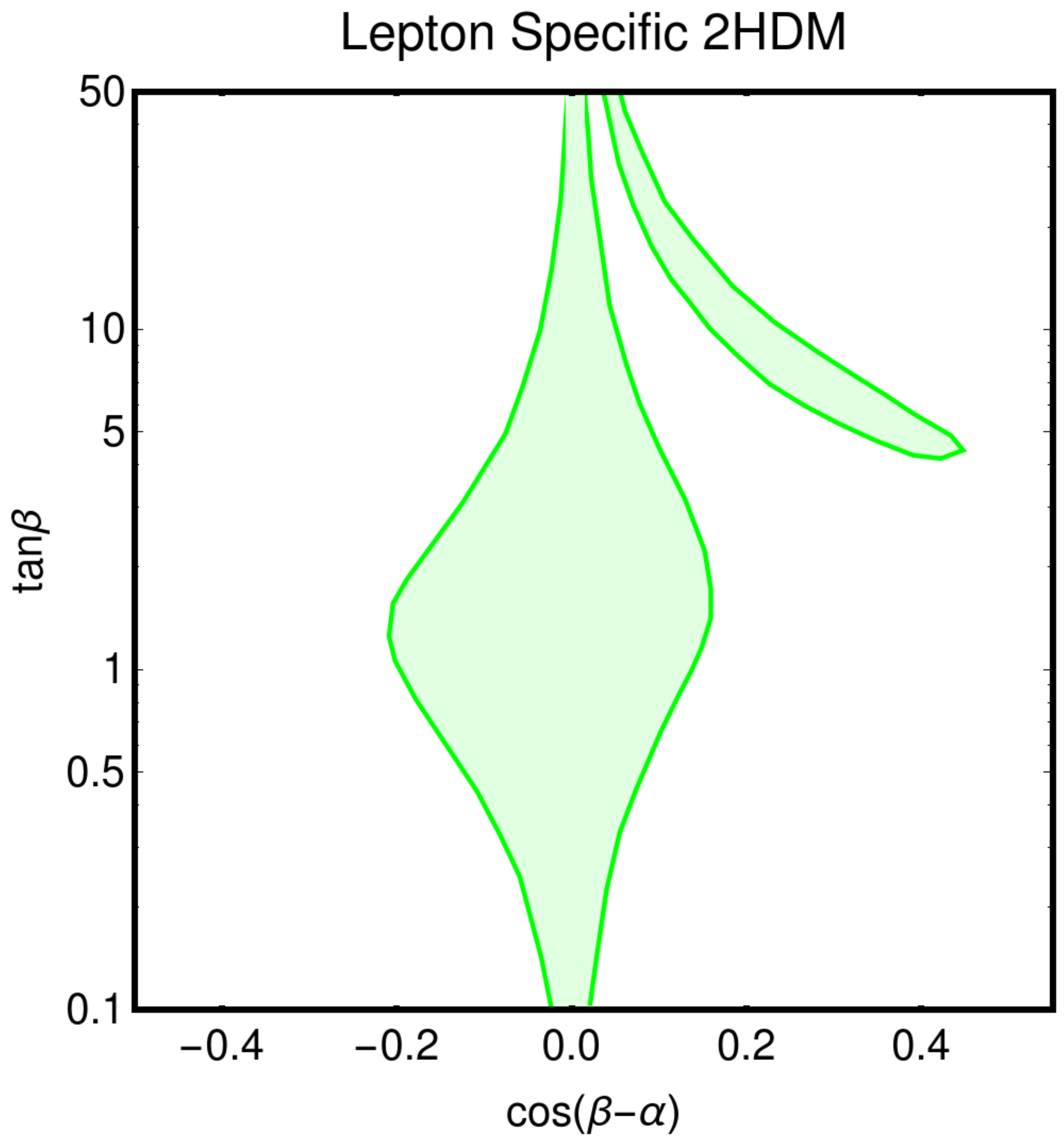}}
\end{center}
\vspace*{-5mm}
\caption{Allowed regions from the $h$ signal strengths measured at LHC 
in the plane  $[\cos({\beta-\alpha}), \tan\beta] $ for  the four types of 2HDMs that do not induce FCNCs at tree--level.}
\label{fig:higgs_constr}
\vspace*{-3mm}
\end{figure}

Let us now summarize the constraints on this model, besides the theoretical ones
on the scalar potential mentioned above.   First, as discussed in section 2,
fits of the Higgs signal strengths  favor SM--like couplings for the 125 GeV
state $h$ observed at the LHC and this implies strong constraints on the angles
$\alpha$ and $\beta$. In particular, one should have SM--like couplings of $h$
to the $W$ and $Z$ bosons so that $\kappa_V^2 \equiv \sin^2 (\beta-\alpha)$
is close to unity. We show in Fig.~\ref{fig:higgs_constr} the regions in the $[\cos(\beta-\alpha),\tan\beta]$ plane that are allowed  by the combined constraints on the Higgs signal strengths into gauge bosons, $\mu_{\gamma \gamma}, \mu_{WW}, \mu_{ZZ}$ and into bottom quark and tau lepton pairs $\mu_{bb}, \mu_{\tau \tau}$, for the four specific 2HDM realizations. 

As it should be clear from the figure, the Type I model allows for a
$\cos(\beta-\alpha)$ value  significantly  different from zero for $\tan\beta
>1$. In the other three models $\cos(\beta-\alpha)$ is, instead, forced to be
close to zero with the  exception of  narrow ``arms'' corresponding to the
so--called ``wrong-sign'' Yukawa
regime~\cite{Ferreira:2014naa,Fontes:2014tga,Ferreira:2014sld}, i.e. the case in
which the couplings of the state $h$ with the down--type quarks and/or leptons
are opposite in sign but equal in absolute values with respect to the ones of a
SM--like Higgs boson. 

All constraints from the SM--like $h$ signal strengths can be simultaneously
satisfied in the so--called  alignment limit, $\beta-\alpha = \frac{\pi} {2}$
\cite{Pich:2009sp,Craig:2013hca,Carena:2013ooa,Bernon:2014nxa}.  In this case,
the couplings of the CP--even $h$ and $H$ states to gauge bosons are such that
$g_{hVV}=1$ and $g_{HVV}=0$ and, hence,  there is no couplings of $H$ with the
$W$ and $Z$ bosons as it is automatically the case for  the $A$ state when CP
conservation in the scalar sector is assumed. The Higgs couplings to fermions in
this alignment limit are also listed in  Table~\ref{table:2hdm_type}. As can be
seen, the couplings of the  $h$ state are also SM--like, 
$g_{huu}=g_{hdd}=g_{hll} \to 1$, while the couplings of the CP--even $H$ state
reduce to those of the pseudoscalar $A$ boson. In particular, besides the fact
that there is no $H$ coupling to vector bosons, $g_{HVV} \to g_{AVV} =0$,  the
couplings to up--type fermions are $g_{Huu} = \cot \beta$ while those to
down--type fermions are, respectively,    $g_{Hdd} = \cot \beta$ and  $g_{Hdd} =
\tan \beta$ in Type I and II models, for instance. 

As for the couplings between two Higgs bosons and one gauge boson, all those
involving the $h$ state  such as $g_{hAZ}$ and $g_{hH^\pm W^\mp}$ tend to zero 
in the limit $\beta-\alpha=\frac\pi2$, while those involving the $H$
boson, such as $g_{HAZ}$ and $g_{HH^\pm W^\mp}$, tend to unity. Finally, the two most important  triple couplings among the CP--even Higgs bosons simplify to
\begin{align}
\lambda_{hhh} & = 1 \, , \  \lambda_{Hhh} = 0 \, ,  
\end{align}
meaning again that the triple $h$ coupling is SM--like, while there is no $Hhh$
coupling at the tree--level. The other triple couplings, which will depend on the additional parameter $m_{12}$, can be ignored as they do not affect our discussion here. 

The masses of the extra Higgs bosons are constrained also by the electroweak
precision observables  and we have calculated the contribution of the extended
Higgs sector to the $S,T,U$ parameters discussed in subsection 3.2. Using the
three masses $M_H,M_A,M_{H^\pm}$ as well as the two angles  $\alpha,\beta$ as
input parameters and the formalism and functions provided for example
in~Ref.~\cite{Branco:2011iw} for the various contributions to the $S,T,U$
parameters,  we have determined the excluded regions of the models via the same
$\chi^2$ fit discussed before with the data and the covariance matrix given in 
eqs.~(\ref{eq:chi2})--(\ref{eq:covariance}). As expected, the most important
corrections occur in the $T$ or $\Delta \rho$ parameters and, hence,  set strong
limits on the mass splitting between at least two of the $H,A,H^{\pm}$ states.
As already pointed out,  once the Higgs sector is coupled to the fermionic DM,
additional contributions to the $S,T,U$ parameters are generated and
consequently, one should combine in eq.~(\ref{eq:chi2}) the contributions of
both the extended scalar and fermionic sectors. We will re--discuss in more
detail the bounds from electroweak precision data  when we will introduce the
different DM models.

Finally, one has to take into account constraints from flavor physics. While the
four considered models, namely Type I, Type II, leptons specific and flipped
2HDM are free from tree--level FCNCs by construction, these are nevertheless
induced at the loop level. The strongest constraints come from processes
described at the fundamental level by $b \to s$ transitions whose rates are
mostly sensitive to the parameters $M_{H^{\pm}}$ and $\tan \beta$. The Type II
and the flipped models are the most affected ones and a lower bound associated
with the $B\rightarrow X_s \gamma$ process~\cite{Amhis:2016xyh} leads to
$M_{H^{\pm}}> 570\,\mbox{GeV}$ irrespective of $\tan\beta$
\cite{Misiak:2017bgg}. Additional constraints also come from $B$--meson decay
processes such as $B_s \rightarrow \mu^+ \mu^-$ and $B \rightarrow K \mu^+
\mu^-$~\cite{Arnan:2017lxi}. A comprehensive discussion  of flavor constraints
on 2HDMs has been presented e.g. in Ref.~\cite{Enomoto:2015wbn} and we will use 
the summary results given there in our  analysis.

Following Ref.~\cite{Arcadi:2018pfo}, we have performed a scan of the 2HDMs
over the parameter ranges,
\beq
\tan\beta \in [1,50],   \alpha \in \left[ -\frac{\pi}{2},+\frac{\pi}{2} \right], ~(M_H, M_A, M_{H^\pm})  \in [ (M_h, 20\,{\rm GeV}, 80\,{\rm GeV}), {\rm 1\,TeV}] , \hspace*{-3mm}
\eeq
where the alignment limit is not assumed  for the angle $\alpha$ at a first
stage and the Higgs masses were taken to be such that $M_H> M_h$ and
$M_{H^\pm} > M_W$ (from LEP2 searches). As already shown in the previous
section, the scenario of a light pseudoscalar mediator is very interesting for
what concerns DM phenomenology and we have consequently left the option of an
$A$ state as light as 20 GeV open (as will be shown later, the possibility of a
light pseudoscalar coupled with the SM Higgs is strongly constrained by collider
searches, hence the choice of a lower limit of 20 GeV is simply made for 
numerical convenience). In order to highlight the impact on the 2HDM parameter
space of deviations from the alignment limit, the scans have been repeated while
imposing the relation $\beta-\alpha={\pi}/{2}$.

\begin{figure}[!h]
\vspace*{-4mm}
\begin{center}
\subfloat{\includegraphics[width=0.42\linewidth]{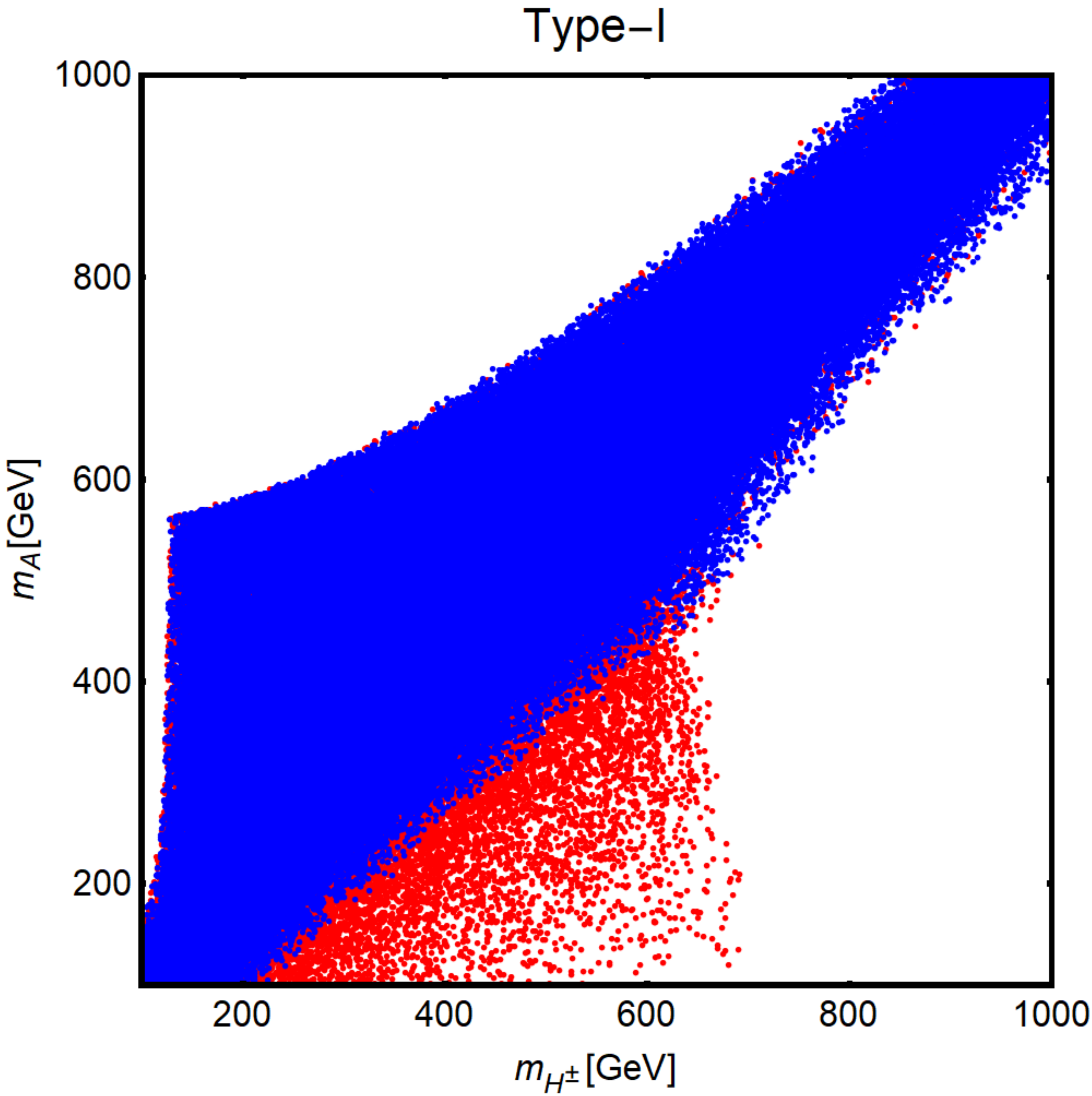}}~~~
\subfloat{\includegraphics[width=0.42\linewidth]{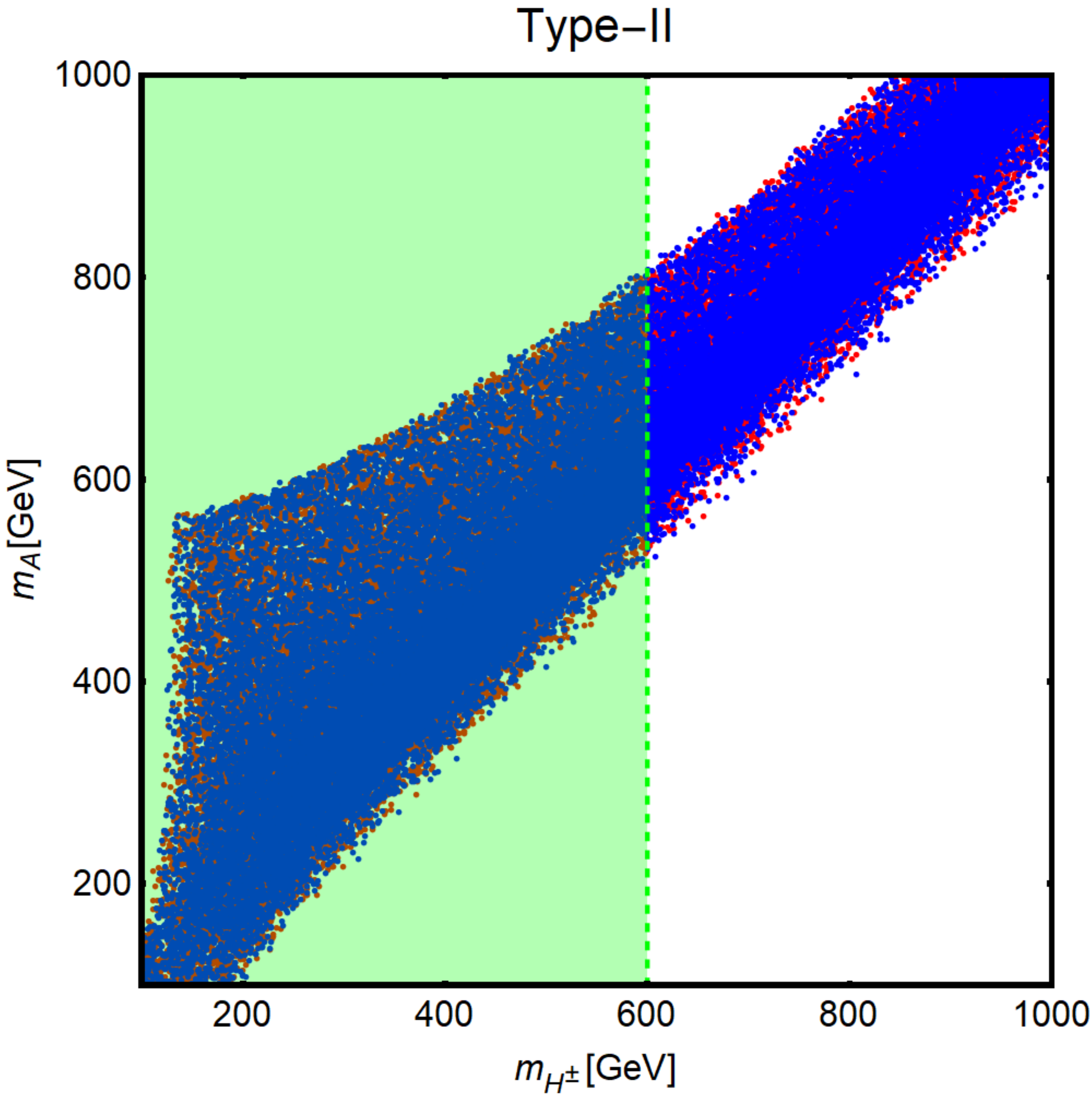}}
\end{center}
\vspace*{-5mm}
\caption{Model points in the $[M_H^{\pm},M_A]$ plane allowed by constraints on the quartic couplings, electroweak precision data and the $h$ boson signal strengths. The red points have been generated by taking $\beta$ and $\alpha$ as free parameters and the blue ones assuming $\beta-\alpha={\pi}/{2}$. The green regions are excluded by limits from flavor processes.}
\label{fig:flavor_2HDM0}
\vspace*{-3mm}
\end{figure}

\begin{figure}[!h]
\vspace*{-3mm}
\begin{center}
{\includegraphics[width=0.43\linewidth]{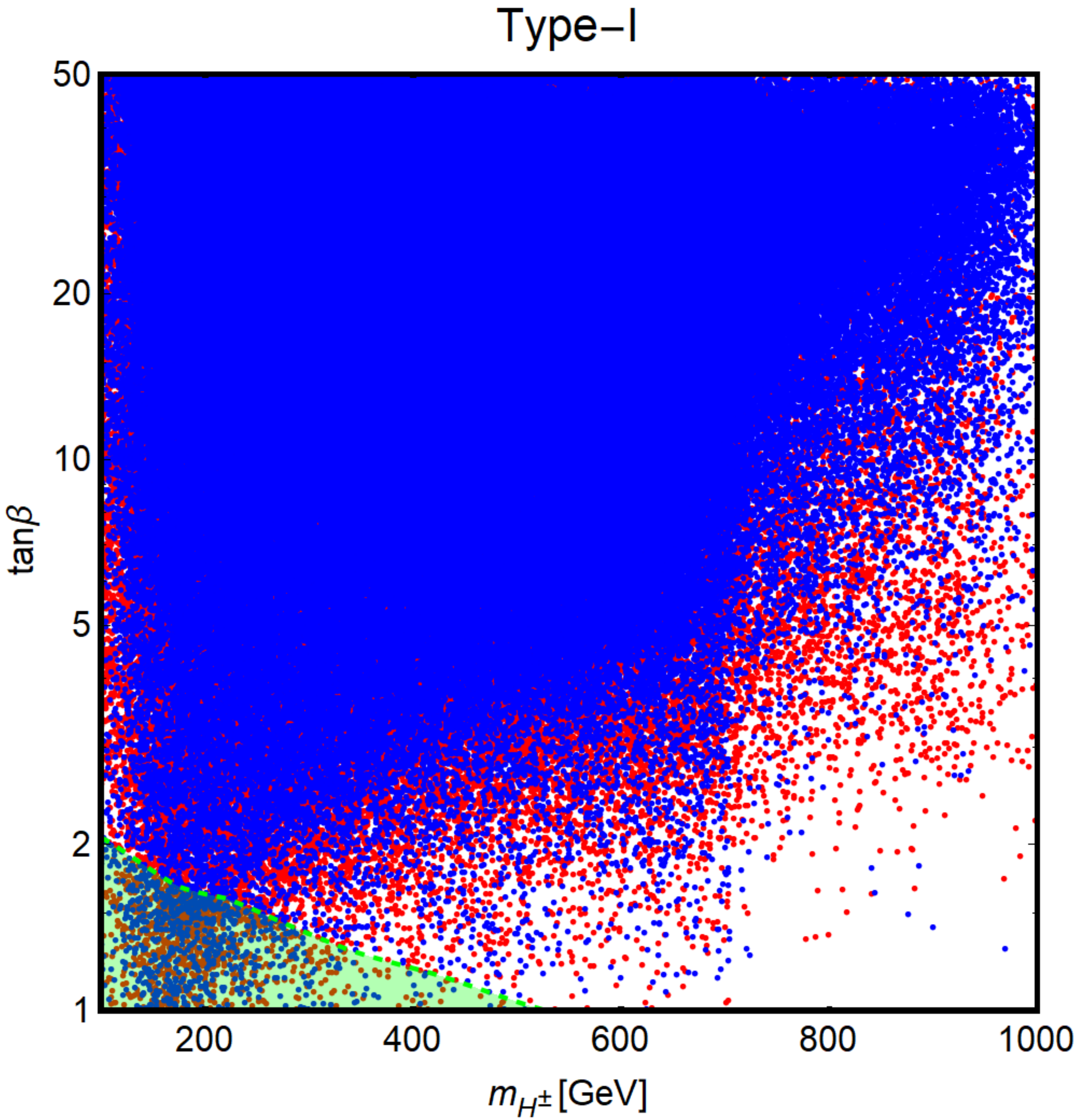}}~~~
{\includegraphics[width=0.43\linewidth]{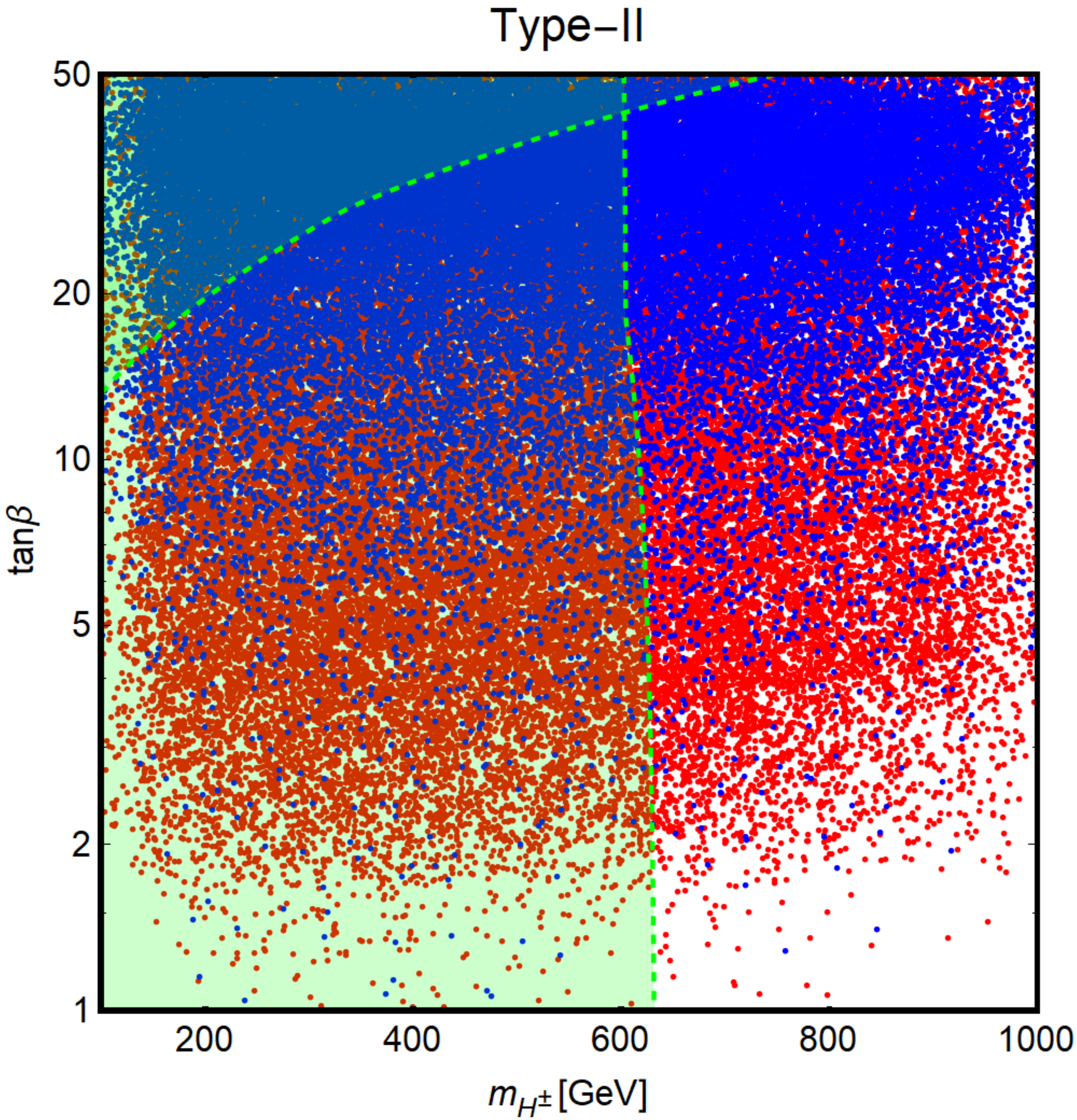}}\\[2mm]
{\includegraphics[width=0.43\linewidth]{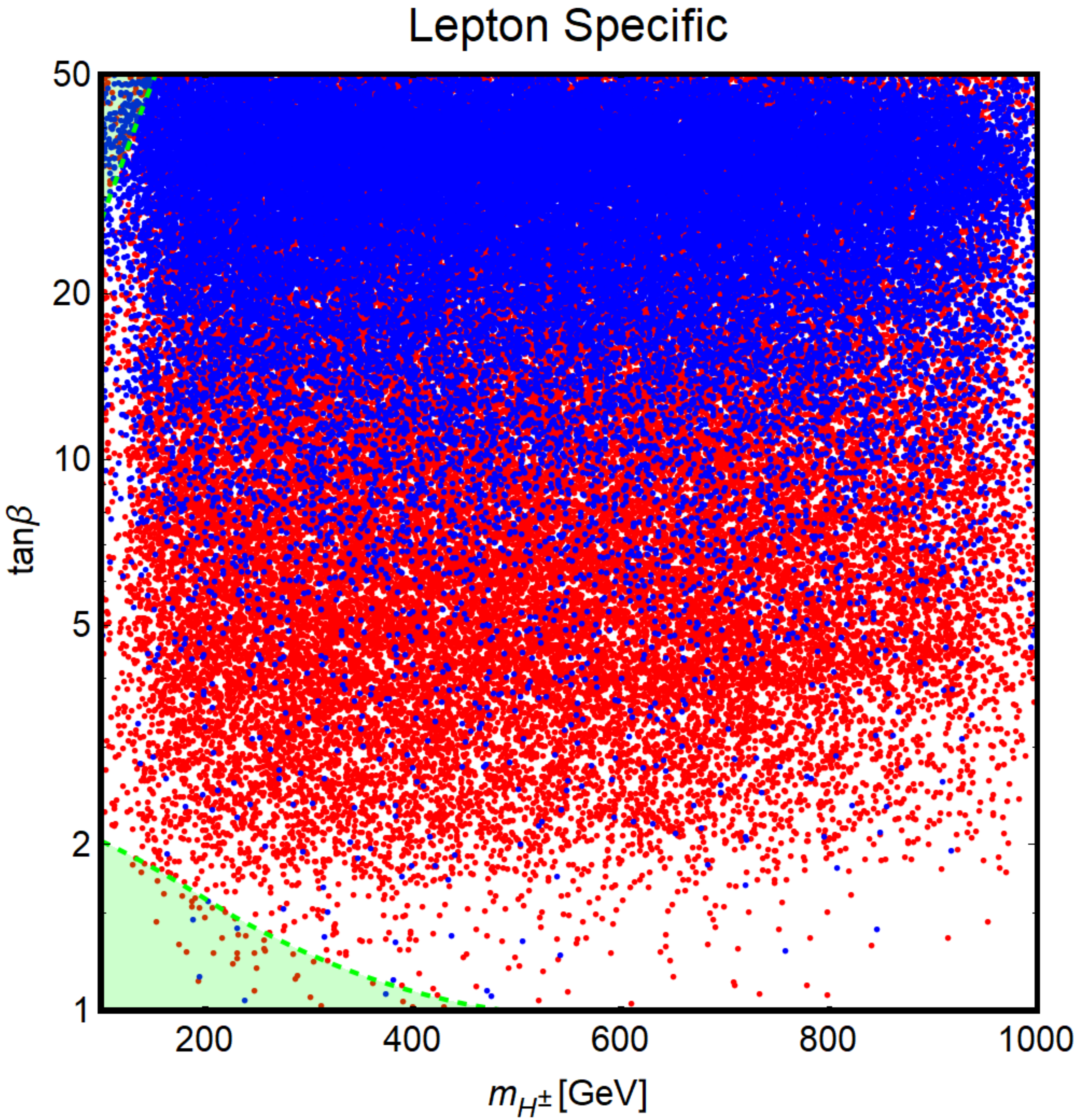}}~~~
{\includegraphics[width=0.43\linewidth]{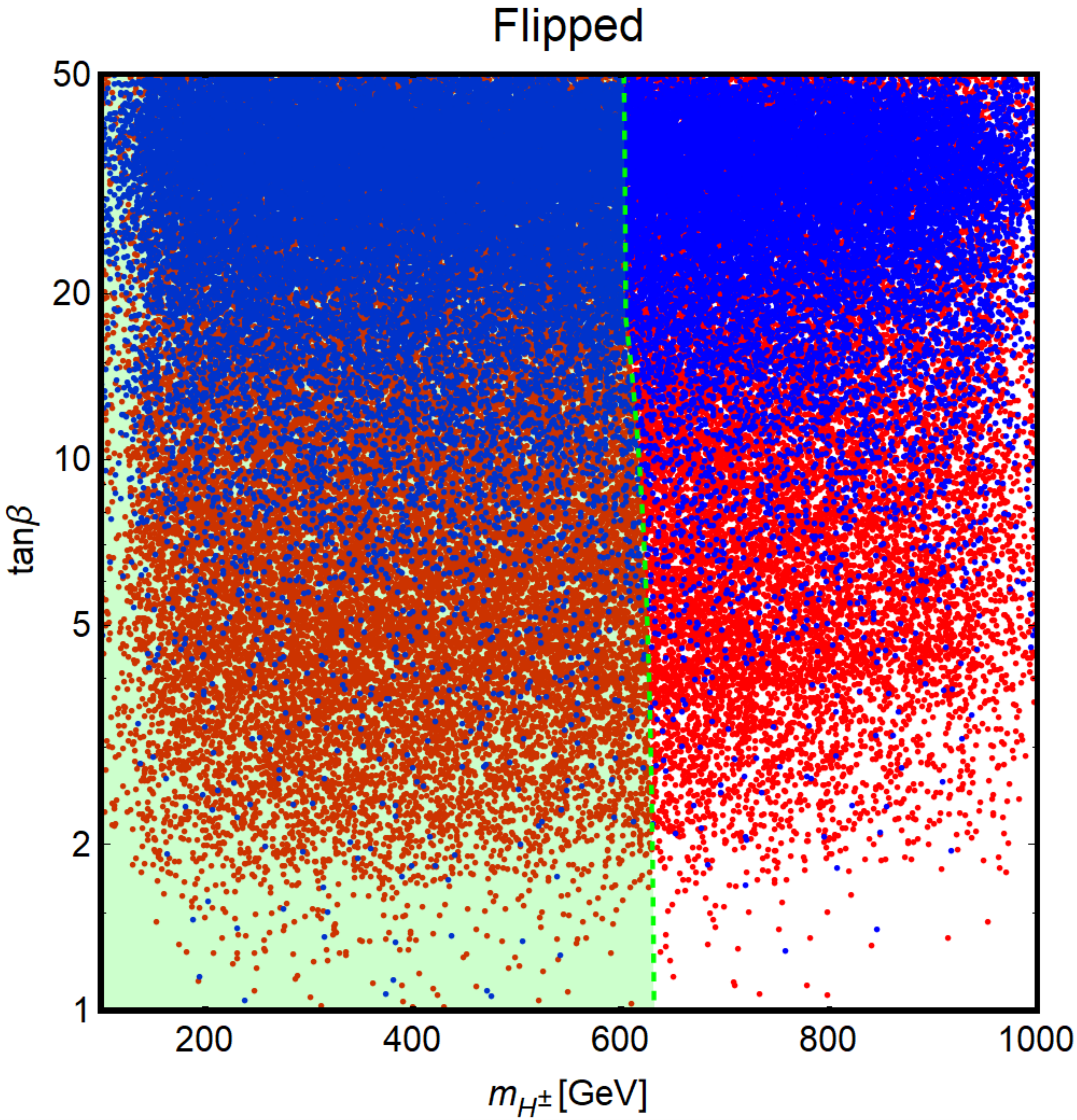}}
\end{center}
\vspace*{-5mm}
\caption{The same scan on the model points as considered in Fig.~\ref{fig:flavor_2HDM0} but reported in the $[M_{H^{\pm}},\tan\beta]$ plane; the same color code is used.}
\label{fig:flavor_2HDM}
\vspace*{-3mm}
\end{figure}

The results of our study are presented in Figs.~\ref{fig:flavor_2HDM0} and
\ref{fig:flavor_2HDM} in, respectively the $[M_{H^{\pm}}, M_A]$ and
$[M_{H^{\pm}},\tan\beta]$ planes. The figures show the model points, i.e. the
assignments of $(M_H,M_A,M_{H^{\pm}},\alpha,\beta)$, which satisfy the
theoretical constraints on the quartic couplings (i.e. a potential bounded from
below and with a proper global minimum and $s$--wave tree--level unitarity) as
well as those from the electroweak precision observables and the  observed Higgs
signal strengths.  We have distinguished using different colors, namely red and
blue, the model points for which free assignments of $\alpha,\beta$ are made
from the ones for which the alignment limit has been imposed.  The green
areas are those excluded by  the combined constraints from flavor physics as
given in Ref.~\cite{Enomoto:2015wbn}.

As can be seen from Fig.~\ref{fig:flavor_2HDM0}, the Type I model allows,
compared to the three other models, a larger mass splittings between the
$H^{\pm}$ and $A$ states (an analogous feature would have been also observed  
in the $[M_H,M_A]$ and/or $[M_H,M_{H^{\pm}}]$ planes). This is a consequence of
the less severe constraints on the $\beta-\alpha$ difference. Indeed the larger
freedom in the choice of $\alpha$ and $\beta$ translates through
eq.~(\ref{eq:quartic_physical}) into a larger freedom in the assignment of
$M_H,M_A,M_{H^{\pm}}$. On the contrary, in scenarios in which $\alpha$ and
$\beta$ lie close to the alignment limit, the mass degeneracy between the extra
Higgs states will be favored.  Fig. \ref{fig:flavor_2HDM0} shows only the
results for the Type--I and Type--II models since the outcome for the  lepton
specific and flipped 2HDM scenarios are identical to the Type--II case with the
exception that the green region would be absent for the lepton specific
model.       

Fig.~\ref{fig:flavor_2HDM} is instead intended to highlight the effects of
flavor constraints. As one can see, the scenarios in which the couplings of the
extra Higgs states are enhanced by $\tan\beta$, i.e. the Type II and the flipped
scenarios, are extremely constrained with values  $M_{H^{\pm}} \leq
570\,\mbox{GeV}$ already ruled out. In the Type II model, a further stronger
exclusion limit at high $\tan\beta$ comes from the $B_s \rightarrow \mu^+ \mu^-$
process. As shown in the right panel of Fig.~\ref{fig:flavor_2HDM0}, this
constraint impacts also the masses of the other Higgs states as they are 
related through eq.~(\ref{eq:quartic_physical}) and expected to be close to the
one of the charged Higgs.  In turn, the Type I and lepton specific models are
almost free from the flavor physics constraints except eventually in small regions of the parameter space with relatively low values of $\tan\beta$ and 
$M_{H^{\pm}}$.

\subsection{The 2HDM and the Dark Matter sector}

We now consider the Dark Matter  sector in the context of two Higgs doublet
models and discuss first two extensions that incorporate a fermionic DM
candidate which are, in fact, simply generalizations of the scenarios already
discussed in section 3:   the singlet--doublet model and a full family of
vector--like fermions. The inert doublet model and a scenario with an additional pseudoscalar field will be then analyzed. 

\subsubsection{The single--doublet fermion extension}

 The singlet--doublet model, introduced in the context of the SM Higgs sector in section 3, can be straightforwardly extended to the case of two doublet Higgs fields \cite{Berlin:2015wwa,Arcadi:2018pfo}. It can be described by the following Lagrangian 
\begin{equation}
\mathcal{L}=-\frac{1}{2}M_N N^{'\,2}-M_L L_L L_R -y_1 L_L \Phi_a N^{'}-y_2 L_R \widetilde{\Phi}_b N^{'}+\mbox{h.c.},
\end{equation}
with $a,b=1,2$. As will be made clear later, it is appropriate not to
assume arbitrary couplings of the new fermions with both the  $\Phi_1$ and $\Phi_2$  doublets. The physical mass eigenstates are obtained by diagonalizing a mass matrix analogous to eq.~(\ref{eq:SD_mass_matrix}) but with $v$ appropriately replaced by $v_{a,b}$.  In the physical basis for both the fermionic and the scalar sector, the relevant interaction Lagrangian for the  fermionic states reads
\begin{align}
 \mathcal{L} &=\overline{E^-} \gamma^\mu \left(g^V_{W^{\mp}E^{\pm}N_i}-g^A_{W^{\mp}E^{\pm}N_i}\gamma_5\right)N_i W_\mu^{-}+\mbox{h.c.}
+\frac{1}{2}\sum_{i,j=1}^3 \overline{N_i}\gamma^\mu \left(g_{Z N_i N_j}^V-g_{Z N_i N_j}^A \gamma_5\right) N_j Z_\mu \nonumber\\
& +\frac{1}{2}\sum_{i,j=1}^{3}\overline{N_i}\left(y_{h N_i N_j}h+y_{H N_i N_j}H+y_{A N_i N_j}\gamma_5 A\right)N_j +\overline{E^-} \left(g^S_{H^{\pm}EN_i}-g^P_{H^{\pm}EN_i}\gamma_5\right)N_i H^{-}+\mbox{h.c.}\nonumber\\
& -e A_\mu \overline{E^{-}}\gamma^\mu E^{-}-\frac{g}{2 c_W}(1-2 s^2_W) Z_\mu \overline{E^{-}}\gamma^\mu E^{-}+\mbox{h.c.} , 
\end{align}
\noindent
where the couplings in the case of $\phi=h,H,A$ and $H^\pm$ are given by
\begin{align}
\label{eq:SD2HDM_couplings}
& y_{ \phi N_i N_j}=\frac{\delta_\phi}{2\sqrt{2}}\left[U_{i1}\left(y_1 R_a^\phi U_{i2}+y_2 R_b^\phi U_{i3}\right)+(i \leftrightarrow j)\right] ,  \nonumber\\
& g^{S/P}_{H^{\pm}EN_i}=\frac{1}{2}U_{i1}\left(y_1 R_1^{H^{\pm}} \pm y_2 R_2^ {H^{\pm}}\right) ,
\end{align}
with $\delta_h=\delta_H=-1$ and $\delta_A=-i$ and we have considered the following decomposition of the $\Phi_{1}$ and $\Phi_{2}$ doublets in terms of the physical $h,H,A,H^{\pm}$ Higgs states:
\begin{equation}
\Phi_{1,2}=\frac{1}{\sqrt{2}}
\left(
\begin{array}{c}
\sqrt{2}R_{1,2}^+ H^+ \\
v_{1,2}+R_{1,2}^h h+ R_{1,2}^H H+i R_{1,2}^A A \, , 
\end{array}
\right)
\end{equation}
with the parameters $R_{1,2}$ being the elements of the rotation matrices $\mathcal{R}_{\alpha,\beta}$ defined in eqs.~(\ref{eq:rotation}) and (\ref{eq:rotation2}).

From a bottom--up perspective, there are four possible configurations for the
assignments of the couplings of the new fermions to the doublets $\Phi_1$ and
$\Phi_2$. We will nevertheless focus here simply on two of the cases which arise
once one extends to the DM sector the extra symmetries which define the four
flavor conserving 2HDMs (see next section for a more detailed account). The
two configurations correspond to the cases in which the new fermions couple
exclusively either with the $\Phi_1$ or with the $\Phi_2$ doublet. 

In order to have a better insight on the DM phenomenology, it is useful to write the analytical expressions for the DM--Higgs couplings $y_{\phi N_1 N_1}$ in these two scenarios $i=1,2$, as given for instance in Ref.~\cite{Berlin:2015wwa} 
\begin{align}
\label{eq:coupling_uu}
    & y_{h N_1 N_1}= y^2 v a_i^h \, (m_{N_1}+M_L \sin 2 \theta)/D_i
\, ,      \nonumber\\
    &  y_{H N_1 N_1}=y^2 v a_i^H \, (m_{N_1}+M_L \sin 2 \theta)/D_i 
\, ,     \nonumber\\
    &  y_{A N_1 N_1}=y^2 v a_i^A \, m_{N_1} \cos 2 \theta/ D_i \, , 
\end{align}
where we have used the abbreviations 
\bea
\label{eq:coupling_dd}
D_i= 2 M_L^2+ 4 M_N m_{N_1}-6 m_{N_1}^2+y^2 v^2 a_i^h \, , \hspace*{3cm}
\nonumber \\
a_1^h= \cos^2 \beta , a_1^H= a_1^A = \cos \beta \sin\beta ~~{\rm and}~~
a_2^h= \sin^2 \beta , a_2^H= a_2^A = - \cos \beta \sin\beta \, . 
\eea
In order to reduce the number of free parameters, we have assumed the alignment limit. 

As already pointed out, in this singlet--doublet model, the impact of the new
fermionic sector is rather modest and the dominant constraints apply mainly to
the scalar sector of the theory and, hence, coincide with the ones discussed in
the previous subsection.

\subsubsection{The vector--like family extension}

Turning to the case in which the 2HDM is linked with an entire family of vector--like fermions,  the most general coupling with the two Higgs doublets is described by the following Lagrangian where a sum over $i=1,2$ is implicit
\begin{align}
\label{2HDM_VL_Lag}
-{\cal L}_{\rm VLF} & =  y_i^{U_R} \overline{{\cal D}_L} 
\tilde{\Phi}_i U^\prime_R + y^{U_L}_i \overline{U^\prime_L} \tilde{\Phi}_i^\dagger {\cal D}_R
+y^{D_R}_i \overline{{\cal D}_L} \Phi_ i D^\prime_R + y^{D_L}_i \overline{D^\prime_L} \Phi_ i^\dagger {\cal D}_R 
\notag \\
&+M_{\mathcal{D}} \overline{{\cal D}_L} {\cal D}_R 
+ M_U \overline{U^\prime_L} U^\prime_R +M_D \overline{D^\prime_L} D^\prime_R + {\rm h.c.} \, . 
\end{align}
 The mass eigenstates are obtained through the same bi--diagonalization procedure illustrated in the previous section once one defines the Yukawa couplings in the Higgs mass eigenstate basis. Using the superscript $X = U_{L/R}$ or $D_{L/R}$, one has 
 \begin{align}
& \begin{pmatrix} y_h^X \\ y_H^X \end{pmatrix} = 
\begin{pmatrix} \cos {\beta} & \sin {\beta} \\ \sin {\beta} & -\cos {\beta} \end{pmatrix} \begin{pmatrix} y_1^X \\ y_2^X \end{pmatrix} . 
\label{eq:Higgs_basis}
\end{align}
As already discussed, the minimal embedding for a DM candidate consists into the addition of one family with hypercharge $Y=0$,
\begin{align}
\label{eq:lag_VLL_gen}
-{\cal L}_{\rm VLL}&= y_{N_R} \overline{L}_L \tilde{\Phi_i} N_R^\prime + y_{N_L} \overline{N}^\prime_L \tilde{\Phi_i}^\dagger L_R
+y_{E_R} \overline{L}_L \Phi_i E^\prime_R + y_{E_L} \overline{E}^\prime_L \Phi^\dagger_i L_R,\nonumber\\
& +M_L \bar L_L L_R+M_N \bar{N^{'}}_L N^{'}_R+M_E \bar{E^{'}}_L E_R + \mathrm{h.c.}\, .
\end{align}

For our analysis, we will consider the case of generic couplings of the vector--like leptons with both Higgs doublets and the one in which the new leptons are charged under the same $\mathbb{Z}_2^{\rm 2HDM}$ which defines the Type I, Type II, lepton specific and flipped 2HDMs, eq.~(\ref{eq:lag_VLL_gen}),
so that they couple selectively with the doublets $\Phi_{1}$ and $\Phi_{2}$. In this last case the interaction Lagrangian of the new leptons reduces to (for simplicity from now on, we omit mass terms)
\begin{align}
-{\cal L}_{\rm VLL}&= y_{N_R} \overline{L}_L \tilde{\Phi_2} N_R^\prime + y_{N_L} \overline{N}^\prime_L \tilde{\Phi_2}^\dagger L_R
 +y_{E_R} \overline{L}_L \Phi_i E^\prime_R + y_{E_L} \overline{E}^\prime_L \Phi^\dagger_i L_R + \mathrm{h.c.}\, . 
\end{align}
As can be seen, the vector--like doublet and the singlet $N^{\prime}_{L,R}$, interpreted as ``up--type'' vector fermions, are coupled only to the $\Phi_2$ doublet. This leads to two possibilities for the couplings of the remaining new leptons:
$i)$ $E^{\prime}_{L,R}$ is also even under $\mathbb{Z}_2^{\rm 2HDM}$, meaning
that all vector leptons  couple to $\Phi_2$ and 
$ii)$  $E^{\prime}_{L,R}$ is odd under $\mathbb{Z}_2^{\rm 2HDM}$, which implies
that vector--like electrons couple to $\Phi_1$, while their partner neutrinos couple to $\Phi_2$.

In the following, these two setups will be referred to as ``model 1'' and
``model 2''. We note that the symmetry $\mathbb{Z}_2^{\rm 2HDM}$ is in general
distinct from $\mathbb{Z}_2^{\rm VLL}$ responsible for the stability of the DM
particle. Indeed, while all the vector leptons should have the same charge under the latter symmetry, they can have different charges under  $\mathbb{Z}_2^{\rm
2HDM}$. 

In the physical basis, the interactions of the vector--like neutrinos with the neutral Higgs bosons are the same for both ``model 1'' and ``model 2'' and read 
\begin{align}
& -\sqrt{2} \mathcal{L}_{\phi NN} \!= \! \begin{pmatrix} N_L^{\dagger}\! & \! N_L^{\prime\dagger} \end{pmatrix} \begin{pmatrix} 0 \! & \! 
y_{N_R} (s_{\beta} h \! - \! c_\beta H \! - \! i c_\beta A ) \\ 
y_{N_L} (s_{\beta} h \! - \! c_\beta H \! + \! i c_\beta A ) \! & \! 0  \end{pmatrix}
\begin{pmatrix} N_R \\ N_R^{\prime} \end{pmatrix} \! + \! \mathrm{h.c.} \, .   \nonumber 
\end{align}
In turn, in the case of vector--like electrons we have for ``model 1'' and for ``model 2'' 
\begin{align}
& -\sqrt{2} \mathcal{L}_{\phi EE}^{(1)} \!=  \! \begin{pmatrix} E_L^{\dagger} \! & \! E_L^{\prime\dagger} \end{pmatrix} \begin{pmatrix} 0 \! & \! y_{E_R} 
(c_{\beta} h \! + \! s_\beta H \! - \! is_\beta A)  \\ y_{E_L} (c_{\beta} h \! + \! s_\beta H \! + \! i s_\beta A) \! & \! 0 \end{pmatrix} \begin{pmatrix} E_R \\ E_R^{\prime} \end{pmatrix} & + \mathrm{h.c.}, \nonumber
\end{align}
\begin{align}
& -\sqrt{2} \mathcal{L}_{\phi EE}^{(2)} \!= \! \begin{pmatrix} E_L^{\dagger} \! & \! E_L^{\prime\dagger} \end{pmatrix} \begin{pmatrix} 0 \! & \! y_{E_R} (s_{\beta} h \! - \! c_\beta H \! + \! i c_\beta A) \\ y_{E_L} (s_{\beta} h \!- \! c_\beta H \! - \! i c_\beta A) \! & \! 0 \end{pmatrix}
\begin{pmatrix} E_R \\ E_R^{\prime} \end{pmatrix} \! + \! \mathrm{h.c.} .  \nonumber
\end{align}

Concerning the couplings with the charged Higgs boson we have instead
\begin{align}
&\mathcal{L}_{H^{\pm} NE}^{(1)}\! = \!H^+ \begin{pmatrix} N_L^{\dagger} \!&\! N_L^{\prime\dagger} \end{pmatrix} \begin{pmatrix} 0 \!&\! y_{E_R} s_{\beta} \\ y_{N_L} c_{\beta} \!&\! 0 \end{pmatrix} \begin{pmatrix} E_R \\ E_R^{\prime} \end{pmatrix} \!+\! H^- \begin{pmatrix} E_L^{\dagger} \!&\! E_L^{\prime\dagger} \end{pmatrix} \begin{pmatrix} 0 \!&\! y_{N_R} c_{\beta} \\ y_{E_L} s_{\beta} \!&\! 0 \end{pmatrix} \begin{pmatrix} N_R \\ N_R^{\prime} \end{pmatrix} \!+\! \mathrm{h.c.}, \nonumber\\
&\mathcal{L}_{H^{\pm} NE}^{(2)} \!= \!H^+ \begin{pmatrix} N_L^{\dagger}\! & \! N_L^{\prime\dagger} \end{pmatrix} \begin{pmatrix} 0 \! & \! \!-\!y_{E_R} c_{\beta} \\ y_{N_L} c_{\beta} \!& \! 0 \end{pmatrix} \begin{pmatrix} E_R \\ E_R^{\prime} \end{pmatrix}\!+\! H^- \begin{pmatrix} E_L^{\dagger} \! & \! E_L^{\prime\dagger} \end{pmatrix} \begin{pmatrix} 0 \! & \! y_{N_R} c_{\beta} \\ \!-\!y_{E_L} c_{\beta} \!& \! 0 \end{pmatrix} \begin{pmatrix} N_R \\ N_R^{\prime} \end{pmatrix}\!+\! \mathrm{h.c.}\nonumber
\end{align}
It is important to remark that once flavour conserving configurations are adopted, the couplings of the vector--like leptons are sensitive to the value of $\tan\beta$. This would not be the case if each of them can arbitrarily couple with both Higgs doublets.

Analogously to the previous scenarios, renormalization group evolution strongly
constrain the size of the Yukawa couplings of the new fermions. As before, we
will keep the focus of the discussion on the quartic coupling of the scalar
potential, as it is the most sensitive to these effects. In the case of the
2HDM+VLF model, the system of equations to solve is particularly complicated as
it involves five quartic and multiple Yukawa couplings. Assuming for simplicity
the presence of a single family of vector fermions, the evolution equations for
the five quartic couplings $\lambda_{i=1,5}$ are given in Appendix C. 
Ref.~\cite{Angelescu:2016mhl}. These equations should be solved in combination with those of the new Yukawa couplings, as well as the one of the top quark and those of the SM gauge couplings. 

\begin{figure}
\begin{center}
\hspace*{-8mm}
\subfloat{\includegraphics[width=0.5\linewidth]{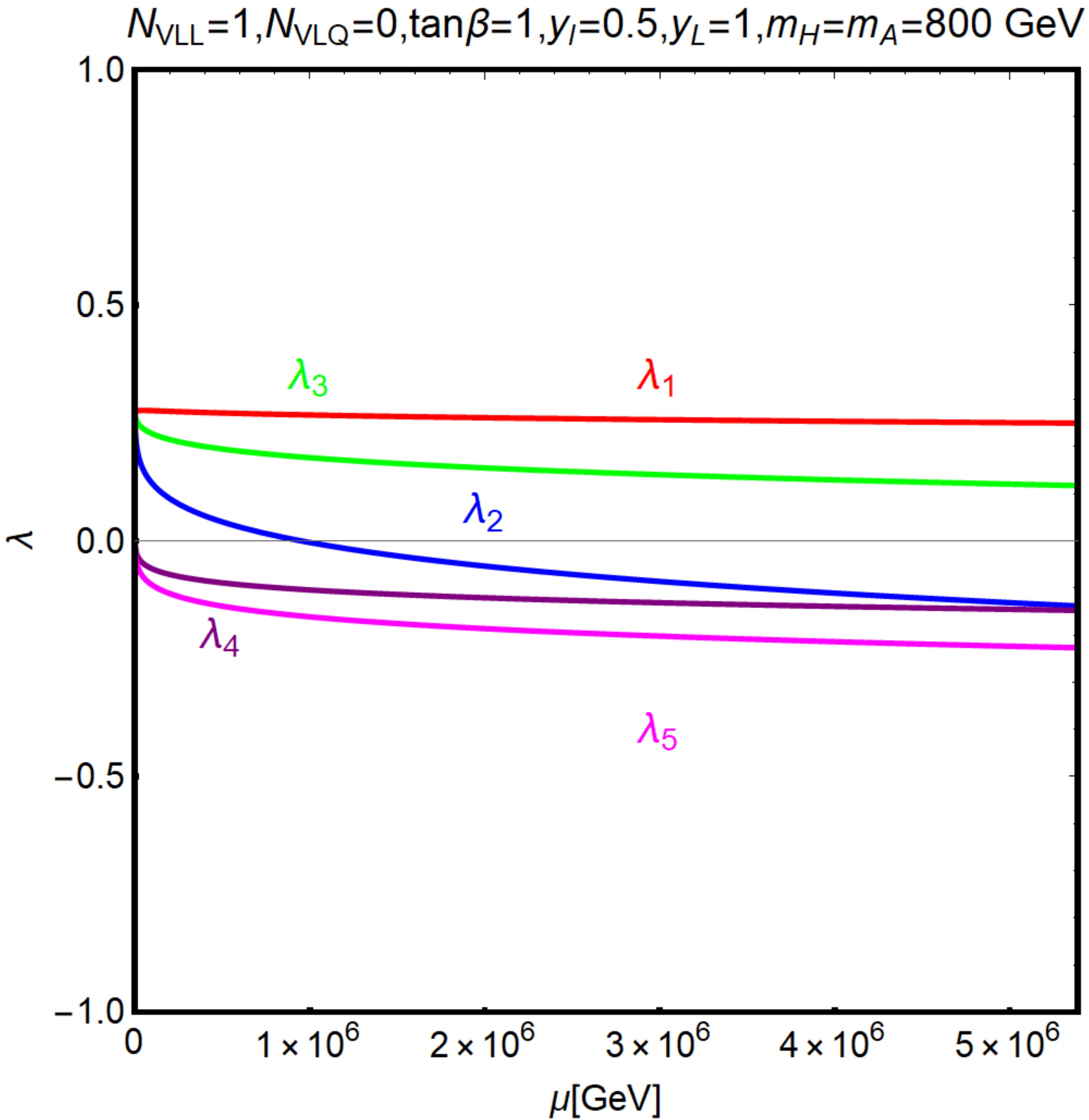}}
\subfloat{\includegraphics[width=0.52\linewidth]{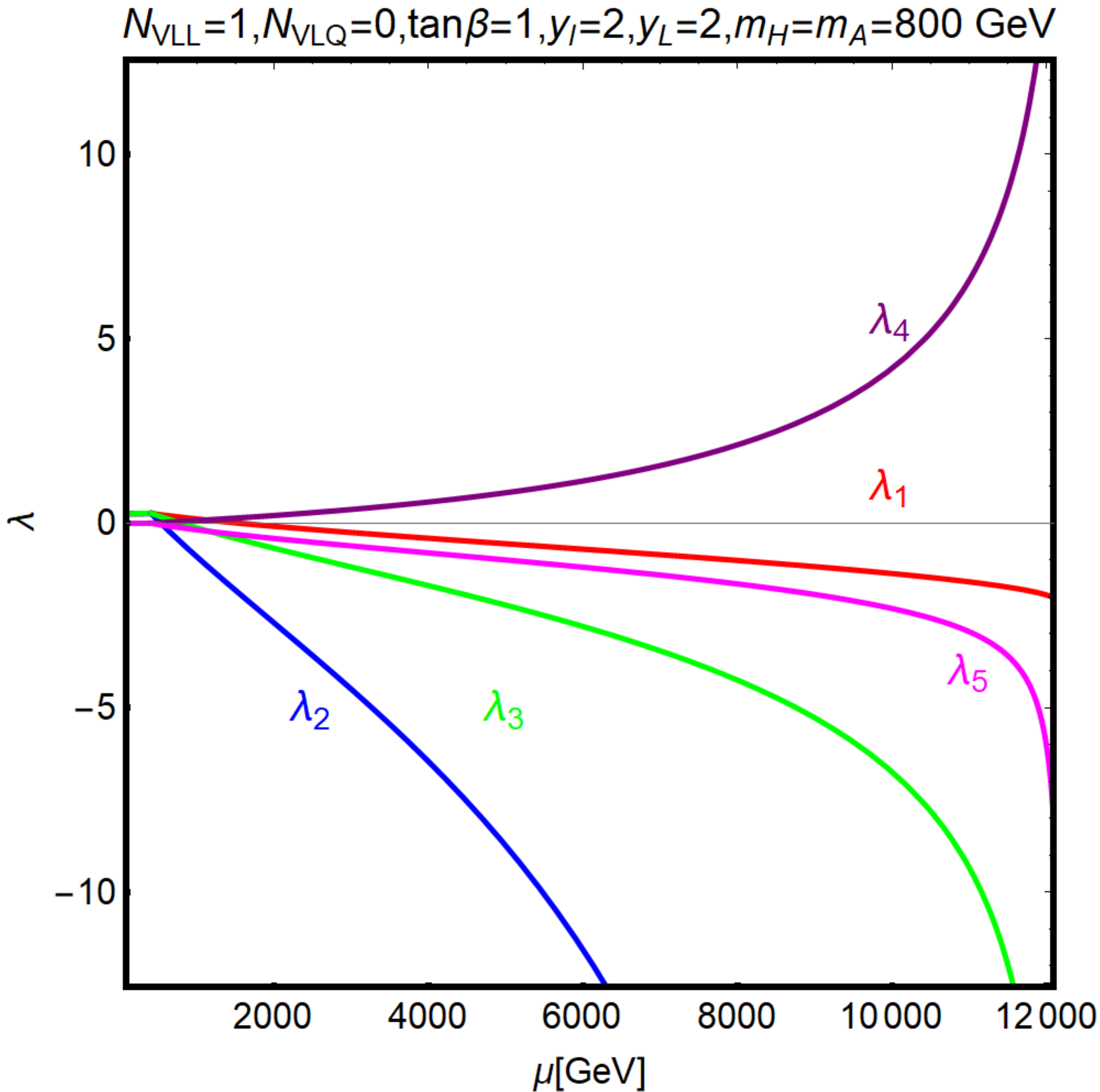}}\\[3mm]
\hspace*{-8mm}
\subfloat{\includegraphics[width=0.52\linewidth]{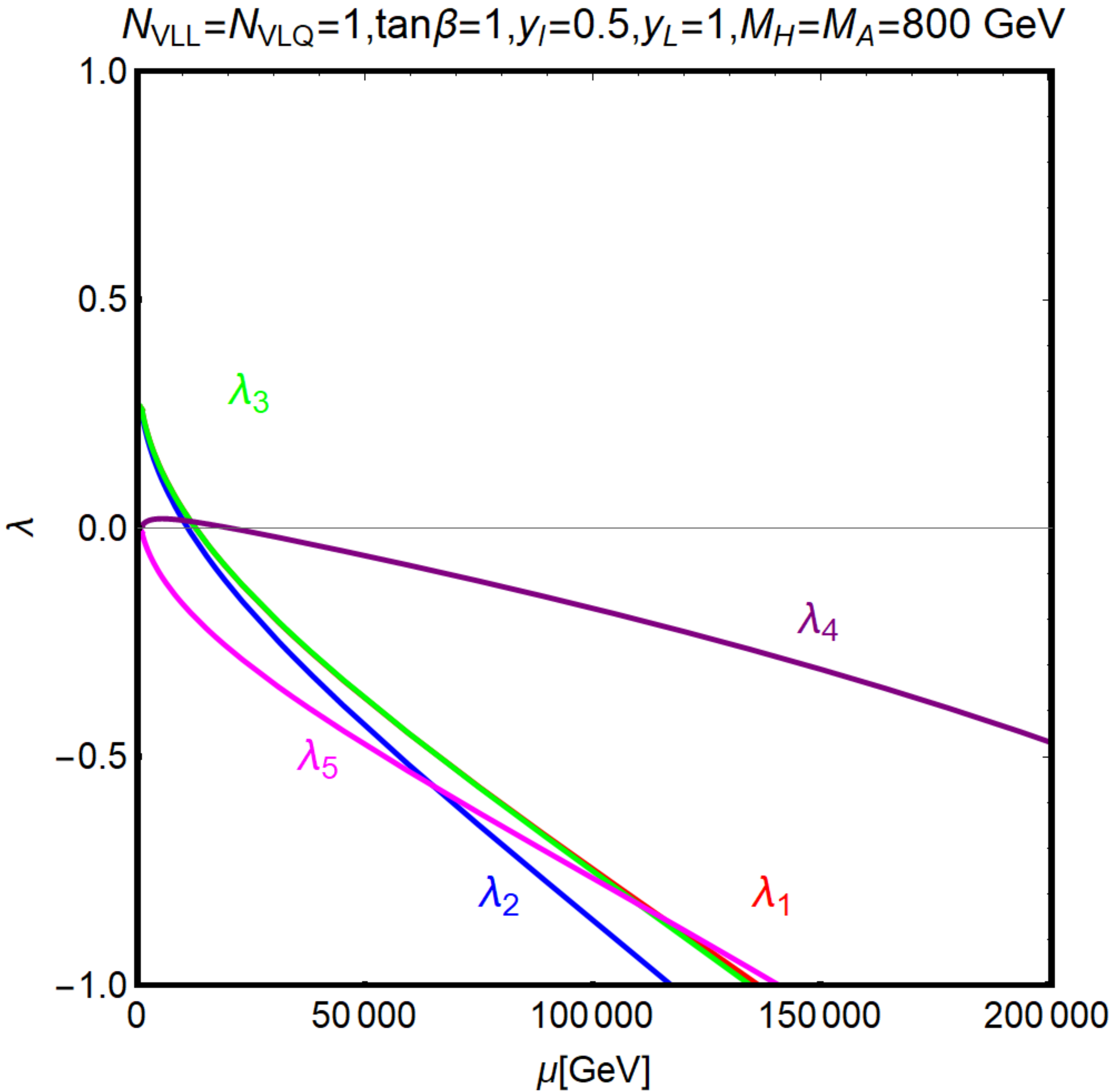}}
\subfloat{\includegraphics[width=0.5\linewidth]{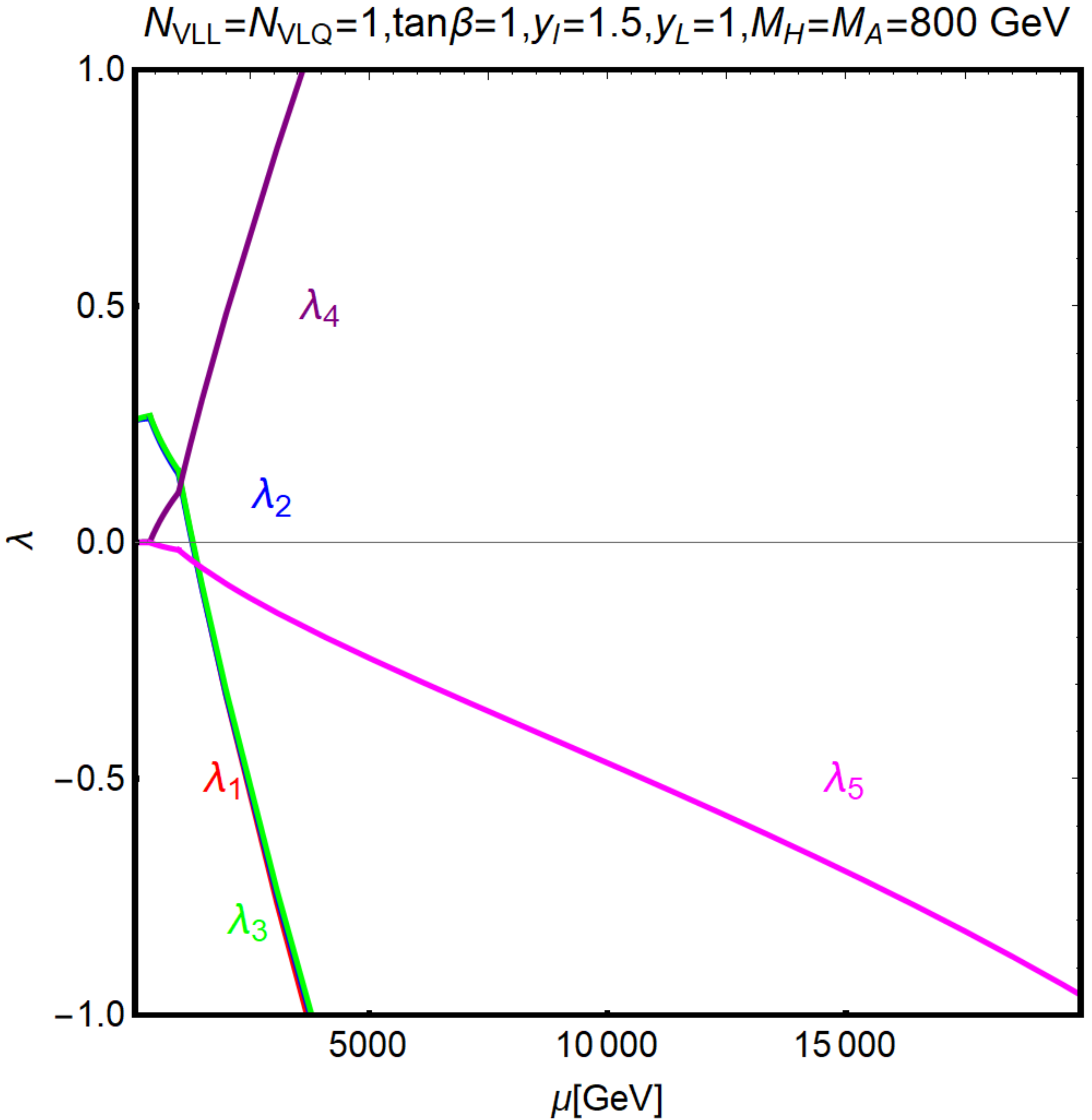}}
\end{center}
\caption{Examples of resolution of the renormalisation group equations for
the 2HDM quartic couplings $\lambda_{1\!-\!5}$ for $\tb\!=\!1$, $M_H\! =\! M_A\! = \! M_{H^\pm} \! =\! 800$ GeV. The upper panels refer to extensions of the 2HDM with only vector--like leptons with $y_l\!=\! 0,5$ and $y_L\!=\!1$ (left panel) and  $y_l\!=\!y_L\! =\! 2$ (right panel). The plots in the bottom panels refer to the case of 2HDM coupled with a full sequential family of vector--like fermions. The two benchmarks have $y_l\!=\! 0,5$ and $y_L\!=\!1$ (left panel) and $y_l\!=\! 1,5$ and $y_L\!=\!1$ (right panel). See main text for the definition of the $y_{l,L}$ couplings.}
\label{fig:RGEexamples}
\end{figure}

Examples of the evolution of the five quartic couplings with energy are shown in
Fig.~\ref{fig:RGEexamples}, distinguishing the $N_{\rm VLL}=1, N_{\rm VLQ}=0$ and
$N_{\rm VLL}=N_{\rm VLQ}=1$ cases, for the Higgs sector parameters $\tb=1$ and 
$M_H=M_A=M_{H^\pm}=800$ GeV. In the left top (bottom) panel, the initial values of
the Yukawa couplings, $y_h^{E_L}(=y_h^{B_L}=y_h^{T_L})=y_l=0.5$ and
$y_L=y_H^{E_L}=-y_H^{E_R}=-y_H^{N_L}=y_H^{N_R}=(=y_H^{B_L}=-y_H^{B_R}=-y_H^{T_L}=y_H^{T_R})=1$, 
are sufficiently  small such that the conditions
eqs.~(\ref{eq:up1})--(\ref{eq:up2}) are satisfied up to energy scales of the order
of $10^6 (3 \times 10^4)\,\mbox{GeV}$. In the right top (bottom) panel, the large
Yukawas, $y_l=2(1.5),y_L=2(1)$, cause instead the couplings $\lambda_{1,2}$ to
become negative, hence violating the first of the conditions eq.~(\ref{eq:up1}), in
proximity of the energy thresholds corresponding to the masses of th VLF and all
couplings $\lambda_{1\!-\!5}$ to become too large, possibly non perturbative, at
scales of the order of 10 TeV.

\begin{figure}
\begin{center}
\hspace*{-3mm}
\subfloat{\includegraphics[width=0.5\linewidth]{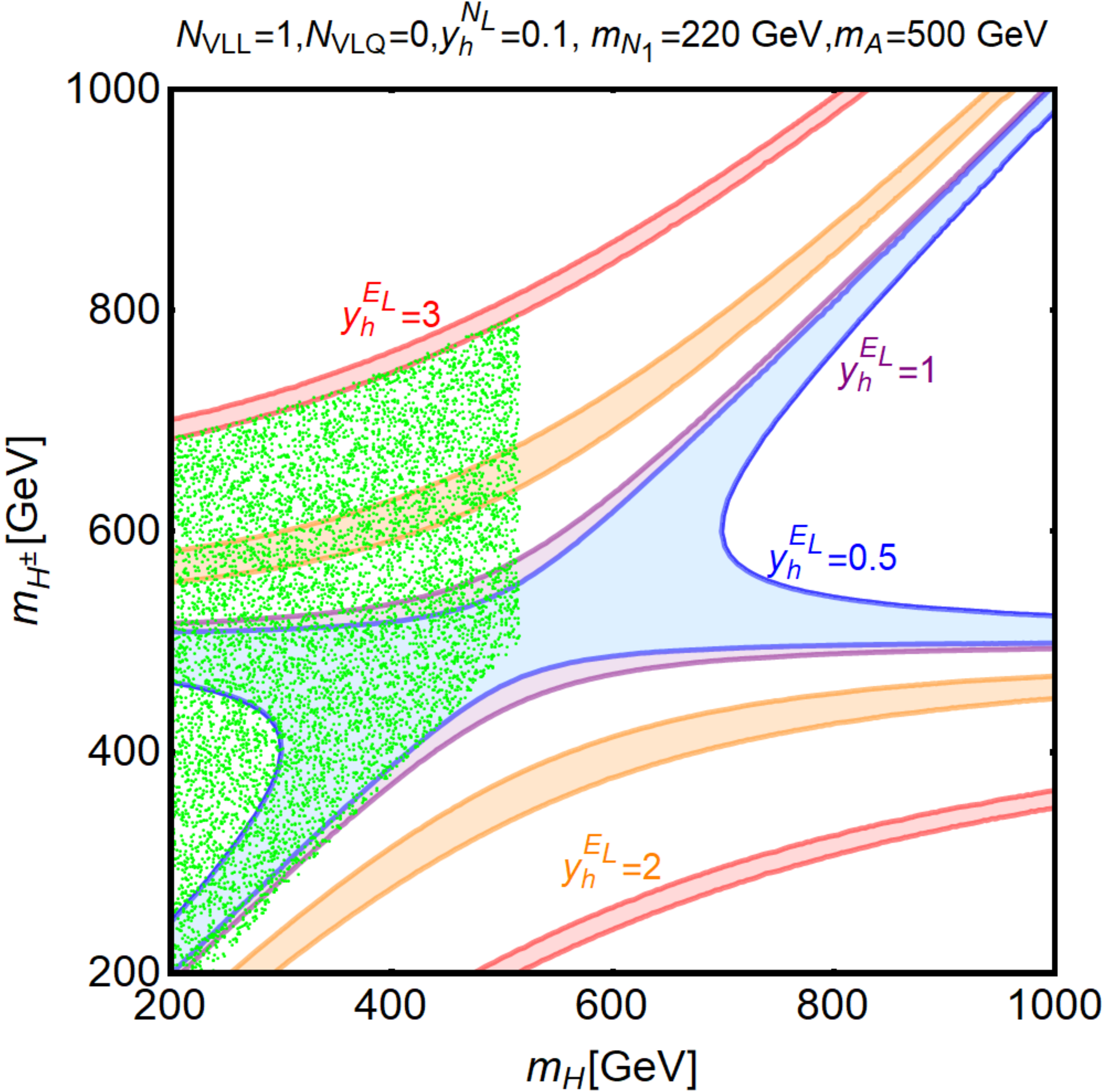}}
\subfloat{\includegraphics[width=0.5\linewidth]{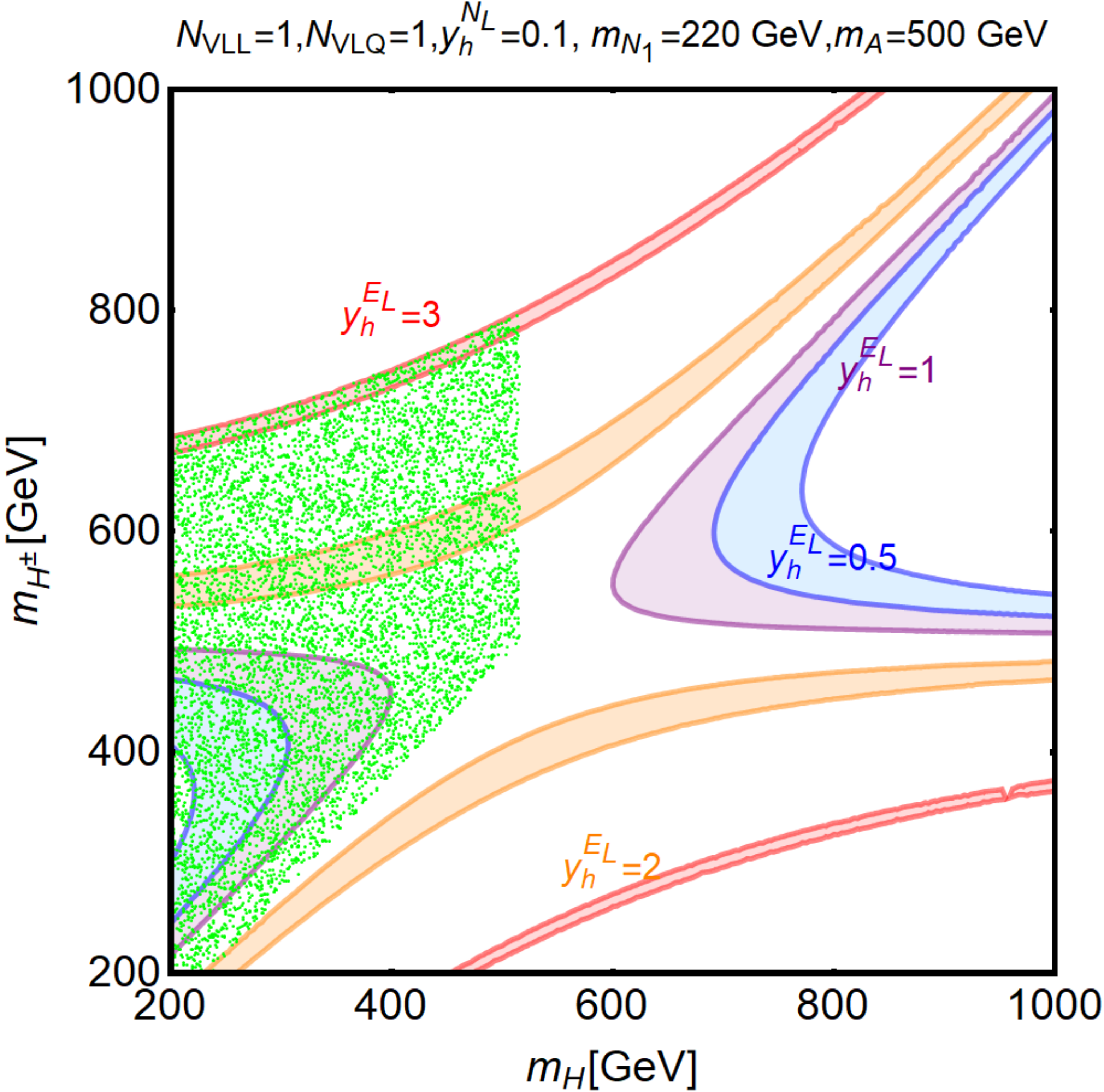}}\\[3mm]
\hspace*{-3mm}
\subfloat{\includegraphics[width=0.5\linewidth]{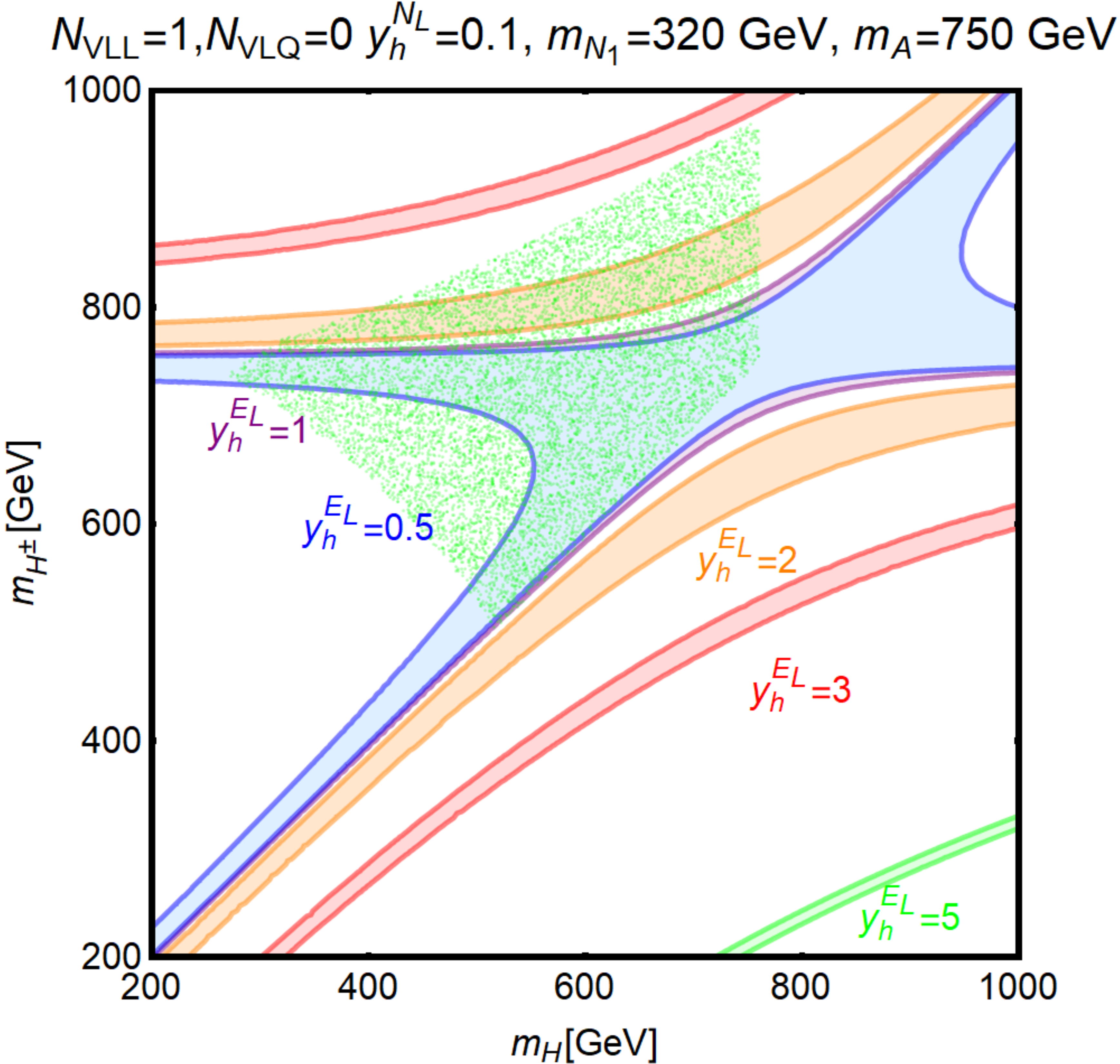}}
\subfloat{\includegraphics[width=0.5\linewidth]{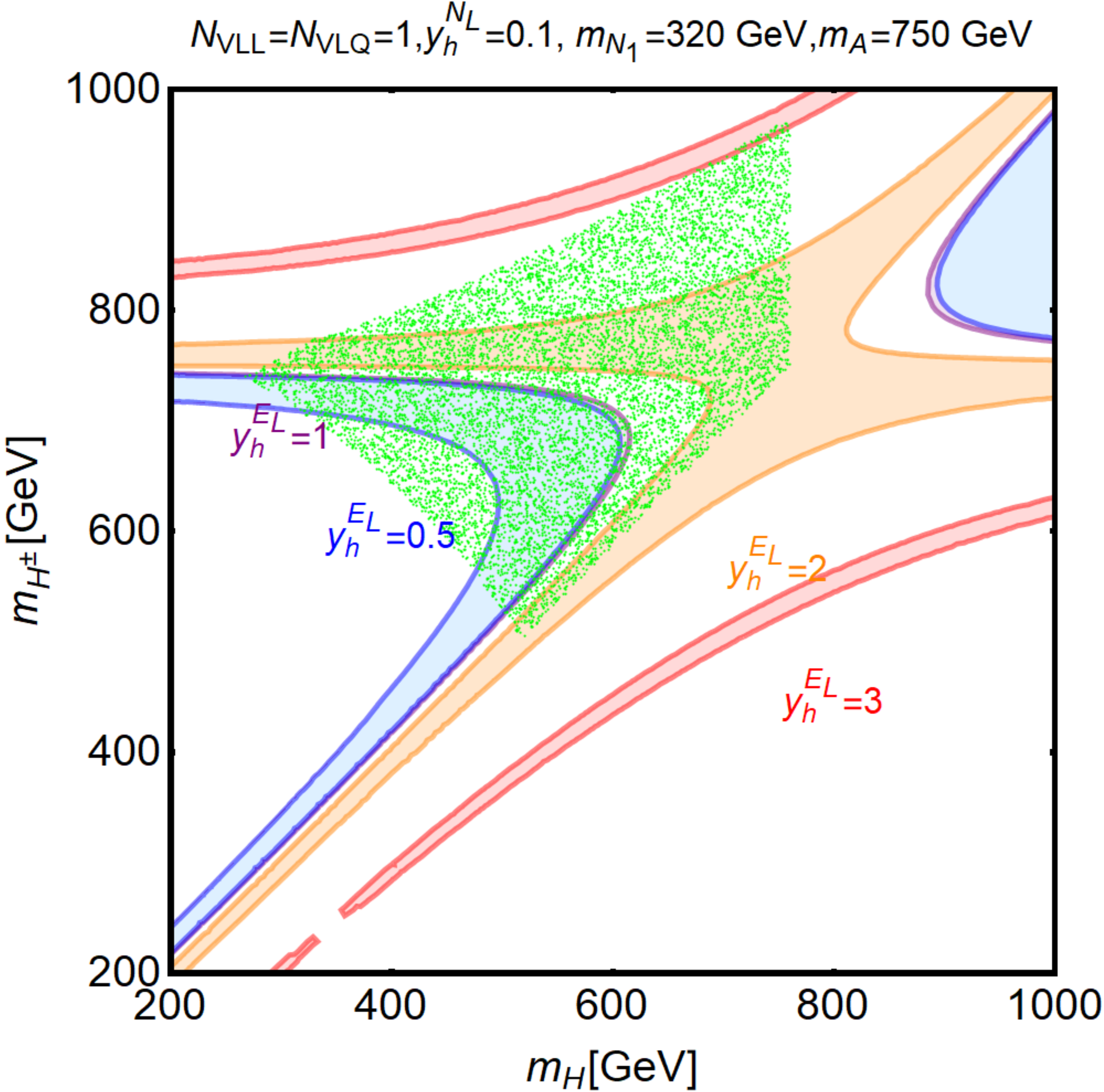}}
\end{center}
\caption{Allowed regions (the colored ones) by electroweak precision data in the plane $[M_H,M_{H^{\pm}}]$ with a vector--like fermionic content $N_{\rm VLL}\!=\! 1, N_{\rm VLQ}\! =\! 0$ (left panel) and  $N_{\rm VLL}\! =\! N_{\rm VLQ}\!=\! 1$ (right panel). For the upper (lower) panels, we have taken:  $M_A\!=\! 500\,(750)\,\mbox{GeV}$, $m_{N_1}\!=\!220\,(320)\,\mbox{GeV}$   $m_{E_1}\! =\! 250\,(375)\,\mbox{GeV}$ and $m_{Q_1}=1$ TeV. The blue, purple, orange and red regions represent the allowed parameter space for Yukawa couplings of, respectively, $y_h^{E_L} \! = \! y_h^{B_L} \! = \! y_h^{T_L}\! =\! 0.5,1,2,3$. The green points represent the configurations allowed by the theoretical constraints discussed in the text.
}
\label{fig:EWPT_2HDM_l}
\end{figure}

The size of the Yukawa couplings of the new fermions is, as already mentioned,
also constrained by the electroweak precision data. In the case of a 2HDM, an
assessment concerning the corresponding limits is complicated by the fact that
the masses of the new scalar bosons affect as well the electroweak data. We
show  in Fig.~\ref{fig:EWPT_2HDM_l} an example of the  allowed regions of the
parameter space in the case of the simultaneous presence of extra Higgs bosons
and vector fermions. 

In the figure, these regions are represented as coloured strips in the
bidimensional plane $[M_H,M_{H^{\pm}}]$ for two values of the pseudoscalar Higgs
mass $M_A \!= \!500$ GeV (top) and 750 GeV (bottom). For these two mass values,
we have considered two configurations for the vector--like fermions, namely
$N_{\rm VLL}\!=\!1,N_{\rm VLQ}\!=\!0$ (left) and $N_{\rm VLL}\!=\! N_{\rm
VLQ}\!=\!1$ (right) and, for each of these, different assignments of the Yukawa
couplings  $y_h^{E_L}\! = \! y_h^{B_L} \! =\! y_h^{T_L}$, ranging from 0.5 to 1,
while keeping fixed the other parameters. In particular, we have assumed very
suppressed values of $y_h^{N_L}$ in order to comply with constraints from direct
DM searches to be discussed later. As it can be seen from the figures, the most
favoured configurations consist of vector--fermion Yukawa couplings below unity,
implying that the dominant contribution to electroweak observables  comes from
the scalar sector. Higher values of the Yukawa couplings, up to three, can be
nevertheless allowed by invoking cancellations between the fermionic and scalar
contributions. This cancellations occur in rather narrow strips of the
bidimensional plane $[M_H,M_{H^{\pm}}]$ and, in particular, require that the
mass spectrum of the new scalars is not degenerate. 

The  regions allowed  by electroweak observable have been overlapped  with the
outcome of a scan on the parameters of the scalar sector, including the constraints
eqs.~(\ref{eq:up1})--(\ref{eq:up2}). As can be seen, one can achieve a mass
spectrum for the new Higgs bosons, compatible with
eqs.~(\ref{eq:up1})--(\ref{eq:up2}) as well as electroweak data, up to $y_h^{E_L}
\approx 3$. As shown above, values $y_h^{E_L} \gtrsim 1$ are nevertheless
disfavored by stability of the scalar potential under RG evolution.

\subsubsection{The inert Higgs doublet model}

In principle, the so--called inert Higgs doublet model 
\cite{Deshpande:1977rw,LopezHonorez:2006gr,Barbieri:2006bg,Ma:2006km,Arhrib:2013ela} should have been
discussed in section 3, since it leads to a SM--like Higgs sector, but we
analyze it here as it can be described with a formalism that is very close to
the one of the 2HDM. Indeed, the scalar potential of the model involving the two
doublets $\Phi$ and $\Phi'$ is similar to then one given in 
eq.~(\ref{eq:scalar_potential}):
\beq
V \! = \!  \mu^2 |\Phi|^2\! +\! \mu'^2 |\Phi'|^2\! + \! \lambda_1 |\Phi|^4 \! \! \! +\! \lambda_2 |\Phi'|^4 \!  +\!  \lambda_3 |\Phi|^2 |\Phi'|^2 \! + \! \lambda_4 |\Phi^{\dagger}\Phi'|^2 \!  + \!  \frac{\lambda_5}{2}\left[ (\Phi^{\dagger}\Phi')^2 \! +\!  \mbox{h.c.} \right]. ~
\label{eq:VIDM}
\eeq
However, in the case of the inert doublet, the field $\Phi'$ does not acquire a vev and, hence, does not participate to electroweak symmetry breaking. This is left to  the doublet $\Phi$ only, which then coincides with the SM Higgs doublet. After electroweak  symmetry breaking, the doublet $\Phi'$ can be then simply decomposed as
\begin{equation}
\Phi ' = \begin{pmatrix} H^+ \\ \frac{1}{\sqrt{2}} (H+ iA ) \end{pmatrix}~,
\end{equation}
where, in terms of the SM vev $v$, the SM--Higgs field has a mass given by 
$M_h^2=\mu^2+3 \lambda_1^2 v^2$ while the two electrically charged  $H^{\pm}$ and the two electrically neutral $H$ and $A$ states have masses given by 
\bea
\label{eq:IDM_masses}
 M_{H^{\pm}}^2 \hspace*{-2mm} &&=\mu'^2+\frac{\lambda_3 v^2}{2},\nonumber\\
 M_{H}^2&&=\mu'^2+\frac{1}{2}(\lambda_3+\lambda_4+\lambda_5)v^2,\nonumber\\
 M_{A}^2&&=\mu'^2+\frac{1}{2}(\lambda_3+\lambda_4-\lambda_5)v^2.
\eea

Hence, the phenomenology of the model will depend on four parameters, the three scalar masses and one quartic coupling or on four quartic couplings or their combinations,  for instance,  $\lambda_2,\lambda_3$ and
\beq
\lambda_{L/S}= \frac12 ( \lambda_3 + \lambda_4 \pm \lambda_5), 
\label{eq:cplg:labdaL}
\ee
which, respectively, correspond the couplings of the $HH$ and $AA$ pairs to the SM--like Higgs boson $h$. Similarly to the conventional 2HDM, as introduced in the previous subsection, it is possible to use the relations illustrated above to identify as free input parameters for the IDM the four physical masses $M_h,M_A,M_H,M_{H ^{\pm}}$ and the two quartic couplings $\lambda_L$ and $\lambda_2$. The coupling $\lambda_2$ does not actually explicitly appear in the relevant interactions rates for DM phenomenology. It plays nevertheless an important role since it influences the one--loop corrections to the masses of the Higgs states which are crucial to properly determine the DM relic density in the coannihilation regime~\cite{Goudelis:2013uca}.

To have a viable DM sector, one first assumes that the field $\Phi'$ is odd 
under a discrete $\mathbb{Z}_2$ symmetry,  while the SM fermions are even with
respect to it. In such a way, it is possible to forbid direct coupling between
$\Phi'$ and pairs of SM fermions. The lightest of the neutral scalar $H$ and $A$
states would be then the DM particle and, here, we will restrict to the case 
where $H$ is the DM candidate. 

Concerning the present constraints on the inert doublet model, one has first the usual ones on the quartic couplings from  the requirement of the stability of the electroweak vacuum, which imposes the tree--level relations
\bea
 \lambda_{1,2} >0 \, , \quad 
 \lambda_3, \lambda_3+\lambda_4-|\lambda_5| >-2 \sqrt{\lambda_1 \lambda_2} \, . 
\eea
In addition, one needs small couplings $\lambda_i < 4 \pi$ from the requirement
of  perturbativity. These requirements should not only hold at the weak scale but also at high enough energy to have a consistent DM and collider phenomenology. The $\beta$ functions for the five $\lambda_i$ couplings coincide with the ones given in eq.~(\ref{eq:lambda1RGE}) for $y_1=y_2=0$. 

Similarly to the conventional 2HDM, the second doublet $\Phi^{'}$ impacts
electroweak precision data which constrain the mass splitting of the extra Higgs states. The model contributions to the $S$ and $T$ parameters read for $x_A=M_A^2/M^2_{H\pm} > x_H=M_H^2/M^2_{H\pm}$ \cite{Barbieri:2006bg}:
\bea
S&=& \frac{1}{72\pi}  \frac{1}{(x_A^2- x_H^2)^3} \big[x_A^6 f_a(x_A)  - x_H^2 f_a(x_H) + 9 x_A^2 x_H^2 (x_A^2 f_b(x_A) - x_H^6 f_b(x_H) )\big]\, , \nonumber \\
T&=& \frac{1}{32\pi^2v^2 \alpha} \big[ f(M^2_{H\pm},M^2_H)+f(M^2_{H\pm},M^2_A)
-f(M^2_A,M^2_H) \big]\, , 
\eea
with $f$ given in eq.~(\ref{eq:f-deltarho}), while $f_a(x)=-5+12\log(x)$
and $f_b= 3-4\log(x)$. 

Furthermore,  there are collider bounds: $M_{H}+M_A \gsim M_Z$ from the invisible $Z$ boson width, and from LEP2 searches ~\cite{Pierce:2007ut}: $M_{H^{\pm}}> 70\!-\!90\,\mbox{GeV}$ on charged Higgs and $M_A> 100\,\mbox{GeV},M_H > 80\,\mbox{GeV}$ from $e^+e^-\to HA$ provided that $M_A-M_H >
8\,\mbox{GeV}$~\cite{Lundstrom:2008ai}. 

\subsubsection{The 2HDM plus a pseudoscalar portal}
 
Another scenario which gained some interest recently is the 2HDM plus a lighter
pseudoscalar state. Indeed,  this model offers the possibility to induce in a 
gauge invariant manner a coupling of the form $a \bar f \gamma_5 f$ between a 
singlet pseudoscalar $a$ and the SM fermions, via the mixing of $a$ with the
pseudoscalar $A$ state of the 2HDM\cite{Ipek:2014gua,Goncalves:2016iyg,Bauer:2017ota,Tunney:2017yfp,Abe:2018bpo}. The most general scalar potential for such a model is given by~\cite{Abe:2018bpo}:
\begin{equation} 
V = V(\Phi_1,\Phi_2) + \frac{1}{2} m_{a_0}^2 a_0^2+\frac{\lambda_a}{4}a_0^4
+\left(i \kappa a_0 \Phi^{\dagger}_1\Phi_2+\mbox{h.c.}\right)+\left(\lambda_{1P}a_0^2 \Phi_1^{\dagger}\Phi_1+\lambda_{2P}a_0^2 \Phi_2^{\dagger}\Phi_2\right), 
\end{equation} 
where $V(\Phi_1,\Phi_2)$ denotes the usual potential of the two Higgs doublet 
fields given in eq.~(\ref{eq:scalar_potential}). $\kappa,\lambda_{1P}, \lambda_{2P}$  are the new couplings, assumed here to be real,  between the two doublets and the pseudoscalar $a_0$ state. 

In this context, we will consider that the DM particle is a fermion $\chi$, singlet under the SM gauge group, which couples with the field $a_0$ according to \begin{equation}
\mathcal{L}=i g_\chi a_0 \bar \chi i \gamma^5 \chi\;.
\end{equation}

After symmetry breaking, the scalar sector of the theory will consist of two CP--even $h,H$, two CP--odd $a_0,A_0$ and two charged $H^{\pm}$ states. 
In addition to the usual mixing angles $\alpha$ and $\beta$ of a 2HDM, there is an extra  mixing angle $\theta$ which allows to move from the $(A_0,a_0)$ current states to the basis $(A,a)$ of physical CP--odd eigenstates
\begin{equation}
\left(
\begin{array}{c} A_0 \\ a_0 \end{array} \right)= \Re_\theta \left(
\begin{array}{c} A \\ a \end{array} \right) \, \quad 
{\rm with} \quad 
\tan2\theta=\frac{2 \kappa v}{M_{A}^2-M_{a}^2}\;.
\eeq

Similarly to the previous cases, several variants of this model, depending of the configurations of the couplings of the Higgs doublets to the SM fermions, can be considered. We will simply focus here on the specific case of the Type II model and impose the alignment limit $\beta-\alpha=\frac12 \pi $, as well as the  mass degeneracy for the $H,A,H^{\pm}$ states. In this setup, the Lagrangian of the model in the mass basis can be decomposed into three main contributions (we omit here the terms involving only the $h,H,A,H^{\pm}$ states which are not relevant to  our discussion)
\begin{equation}
    \mathcal{L}=\mathcal{L}_{\rm DM}+\mathcal{L}_{\rm Yuk}+\mathcal{L}_{\rm scalar},
\end{equation}
where $\mathcal{L}_{\rm DM}$ is the DM Lagrangian
\begin{equation}
\mathcal{L}_{\rm DM}=g_\chi \left(\cos\theta a+\sin\theta A\right) \bar \chi i \gamma_5 \chi , 
\end{equation}
while $\mathcal{L}_{\rm Yuk}$ contains the Yukawa interactions with the SM fermions
\begin{equation}
\mathcal{L}_{\rm Yuk}=\sum_f \frac{m_f}{v}\bigg[ g_{hff} h \bar f f+g_{Hff}
H\bar f f- i g_{Aff}  \bar f \gamma_5 f-i g_{aff} a \bar f \gamma_5 a \bigg] \, , \end{equation}
where the couplings $g_{\phi ff}$ for the 2HDM CP--even and charged fields are given in Table \ref{table:2hdm_type}, while the Yukawa couplings of the pseudoscalar states are given by 
\begin{align}
& g_{Auu}=~{\cos\theta}/{\tan\beta},\,\,\,\, g_{Add}=g_{Aee}=~\cos\theta \tan\beta ,\nonumber\\
& g_{auu}=-{\sin\theta}/{\tan\beta},\,\,\,\,g_{add}=g_{aee}=-\sin\theta \tan\beta .
\end{align}
Finally, $\mathcal{L}_{\rm scal}$  contains the trilinear interactions between the CP--even Higgs states and two (pseudo)scalar fields:
\begin{align}
& ~~~~~~~~~~~~~~~~~~~~~~~~~   \mathcal{L}_{\rm scal}=\lambda_{haa}h aa+\lambda_{aAh}h aA+\lambda_{AAh}h AA\, ,\nonumber\\
&   \lambda_{haa}=\frac{1}{M_h v}\left[\left(M_h^2+2 M_H^2-2 M_a^2-2 \lambda_3 v^2\right)\sin^2 \theta-2 \left(\lambda_{P1} \cos^2 \beta+\lambda_{P2}\sin^2 \beta\right)v^2 \cos^2 \theta \right] \, , \nonumber\\
& \lambda_{hAa}=\frac{1}{M_H v}[M_h^2+M_H^2-M_a^2-2 \lambda_3 v^2+2 \left(\lambda_{P1}\cos^2 \beta+\lambda_{P2}\sin^2 \beta\right)v^2]\sin\theta \cos\theta \, , \nonumber\\
& \lambda_{hAA}=\frac{1}{M_H v}\left[\cot2\beta\left(2 M_h^2-2 \lambda_3 v^2\right)\sin^2 \theta+\sin 2\beta\left(\lambda_{P1}-\lambda_{P2}\right)v^2 \cos^2 \theta\right].
\end{align}
In the alignment limit, the pseudoscalars are coupled only with the SM--like Higgs state $h$.

Concerning the theoretical constraints, one should impose the usual conditions on the quartic coupling of the potential. Assuming $\lambda_{P1},\lambda_{P2}>0$, these are analogous to the ones that apply to the 2HDM and which are summarized in eqs.~(\ref{eq:up1})--(\ref{eq:up2}). It is nevertheless useful to explicitly discuss the requirements on the coupling $\lambda_3$ in order to have a scalar potential bounded from below
\begin{align}
 \lambda_3 > 2 \lambda,\,\,\,\,\,\,\lambda=\frac{M_h^2}{2 v^2} , \
 \lambda_3 > \frac{M_A^2-M_a^2}{v^2}\sin^2 \theta -2 \lambda \cot^2 2\beta \, ,
\end{align}
where the last term has been obtained under the assumption $M_A \gg M_a$. Combining these equation with the perturbativity requirement $\lambda_3 < 4 \pi$ tells that it is not possible to have, for $\sin\theta \neq 0$, an arbitrary mass splitting between the $a$ and $A$ states. The non decoupling of the heavy scalar sector is further enforced by the requirement of perturbative unitarity for the $aa, aA$ and $AA$ scattering amplitudes into gauge bosons~\cite{Goncalves:2016iyg}
\begin{align}
\label{eq:uni}
|\Lambda_{\pm}| \leq 8 \pi \nonumber, \mbox{ where } 
\Lambda_{\pm} v^2 = \Delta_H^2 - \Delta^2_a (1-\cos 4\theta)/8 \pm \sqrt{ {\Delta_H^2}{v^2}+ \Delta_a^4 (1-\cos4 \theta)/8  } , 
\end{align}
where $\Delta_a=M_A^2-M_a^2$ and $\Delta_H =M^2-M_{H^{\pm}}^2+2 M_W^2- \frac12
M_h^2$ with $M=M_A=M_{H^{\pm}}$. It can be seen that in the limit  $M\gg M_a$ and with maximal mixing  $\sin2\theta=1$, there is an upper bound on $M_A$ of about 1.4 TeV, which is weakened by lowering the values of $\sin2\theta$. We recall that in the considered setup, the severe lower bound $M>
570\,\mbox{GeV}$~\cite{Misiak:2017bgg} which comes from the constraints on the
mass of the charged Higgs boson from flavor transitions, is also present.

There are also searches for the production of the light $a$ state in association with a $Z$ and an $h$ boson that constrain parts of the parameter space \cite{Bauer:2017ota,Tunney:2017yfp}. Finally, for $M_a \leq \frac12 M_h$, large couplings between the light $a$ and the SM--like $h$ boson would lead to a decay  $h\rightarrow a a $ with a large rate given by~\cite{Ipek:2014gua}
\begin{align}
\Gamma(h \rightarrow a a)= \frac{|g_{haa}|^2 M_a}{8 \pi} \sqrt{1- 4 M_a^2/M_h^2 }, 
\end{align}
and which is constrained both by direct searches of light pseudoscalar Higgs
states at the LHC in the  $4b$, $2b 2\ell$ and $4\ell$ (with $\ell=\mu$ or
$\tau$) modes \cite{Khachatryan:2017mnf} and by the Higgs signal strengths and
invisible Higgs decays as discussed in section 2. 


\subsection{Constraints and expectations at colliders}

\subsubsection{Higgs cross sections and branching ratios}

We come now to the collider phenomenology of the 2HDM scalars and in particular,
that of the heavier states since the lightest $h$ boson behaves essentially 
like the SM Higgs boson. We will adopt for simplicity the benchmark scenario
introduced at the end of section 5.1, namely we assume the alignment  limit
$\alpha=\beta- \frac\pi2$ which makes the $h$ boson SM--like and a near mass
degeneracy for the $H,A,H^\pm$ states, $M_H \approx M_A \approx M_{H^\pm}$. In
the case of the Type II model,  the pattern in this benchmark is  similar to
that of the MSSM  which will be discussed later.  Here, we briefly summarize the
main features in this particular scenario and then point out the main
differences in the other possible scenarios. 

The phenomenology crucially depends on the parameter $\tan\beta$. At high 
values, $\gsim 10$, the couplings of the neutral $\Phi=H,A$ and charged  $H^\pm$
bosons to top quarks,  $\propto 1/\tb$, are strongly suppressed while those to
bottom quarks, $\propto \tb$, are enhanced. The  neutral states will then decay
almost exclusively into  $b\bar b$ and  $\tau^+\tau^-$ pairs,  with branching
ratios BR$(\Phi \to b \bar b) \approx 90\%$ and BR$(\Phi \to \tau \tau) \approx 
10\%$ as a result of the color factor  and the mass hierarchy $m_\tau/\bar m_b$,
since on has $m_\tau= 1.78$ GeV and $\bar m_b \simeq 3$ GeV for the
$\overline{\rm MS}$ $b$--mass at the scale of the Higgs masses. All other $H/A$
decays are  strongly suppressed, including those into $t\bar t$ pairs despite of
the large top quark mass value.  Similarly, the charged Higgs boson will decay
into $t\bar b$ and $\tau \nu$  final states with branching ratios of 90\% and
10\% respectively. 

The situation is drastically different at low values of $\tb$, say $\tb \lsim
3$.  When the $H,A,H^\pm$ states are heavy enough to be allowed by kinematics to decay into top quarks, namely $M_H\! \approx \! M_A   \! \gsim \! 2m_t$ and
$M_{H^\pm} \gsim m_t$, the modes $\Phi \to t\bar t$ and $H^+ \to tb$ become
almost exclusive and have branching ratios close to one. At intermediate values,
$3 \lsim \tb \lsim  10$,  the suppression of the $\Phi tt$ coupling starts to be
effective while the $\Phi bb$ coupling is not yet strongly enhanced, resulting
into a competition between the $b\bar b$ and $t \bar t$ decay channels. 

In principle, other Higgs decay modes can be considered. First of all we might
have $H\to WW,ZZ$ ($A$ does not possess such decays by virtue of CP--invariance)
and $A\to hZ, H^\pm \to hW$. Their rates are proportional to 
$\cos^2(\beta-\alpha)$ and thus vanish in the alignment limit to which the
Type-II 2HDM must lie close to comply with bounds from the $h$ signal strengths.
Channels like $H \to AZ, H^\pm W$,  $A\to HZ, H^\pm W$ or  $H^\pm \to AW, HW$
have phase--space suppressed rates or are kinematically forbidden since the
requirement of the alignment limit and the compatibility with electroweak data
imply a near mass degeneracy $M_H \! \approx \! M_A \!  \approx \!  M_{H^\pm}$.
The Higgs to Higgs decay $H\to hh$ features also a vanishing rate in the
alignment limit.   Note finally that compared to the SM,  the loop induced
decays of the neutral states into $gg$ (the top loop is suppressed for $\tb>1$)
and $\gamma\gamma$ (for which the $W$ loop contribution is absent or suppressed)
are much smaller. Hence, only the fermionic decays above are relevant in
general. 

As an example, we show in the left--hand side of Fig.~\ref{Fig:BR-phi} the decay
branching ratios of the  neutral $\Phi=H,A$ bosons into the various possible
final states, as a function of $\tb$ and for the common mass value 
$M_\Phi=M_H=M_A=750$ GeV; the alignment limit is assumed. In the right--hand
side, we display as a function of $\tb$ the total decay width  of the two states
which grows like $M_\Phi$ and $(m_t/\tb)^2$ or $(\bar m_b \tb)^2$. It is very
large at low and high $\tb$ values, being $\Gamma_\Phi \! \approx \! 50$ GeV for
$\tb \approx 1$ and $60$ and is minimal at the intermediate value $\tb \approx
\sqrt{m_t / \bar m_b} \approx 7$ as $m_t\simeq 173$ GeV and $\bar m_b \simeq 3$
GeV.

\begin{figure}[!h]
\vspace*{-25mm}
\centerline{\hspace*{-1.cm} \includegraphics[scale=0.76]{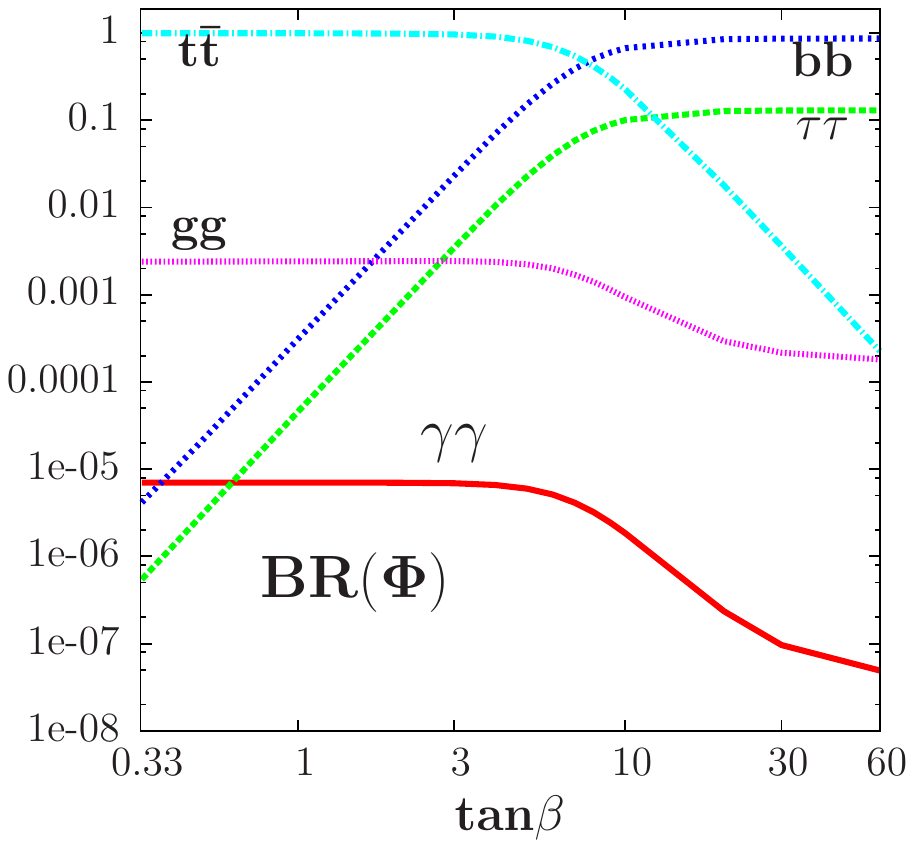}
            \hspace*{-8.2cm} \includegraphics[scale=0.76]{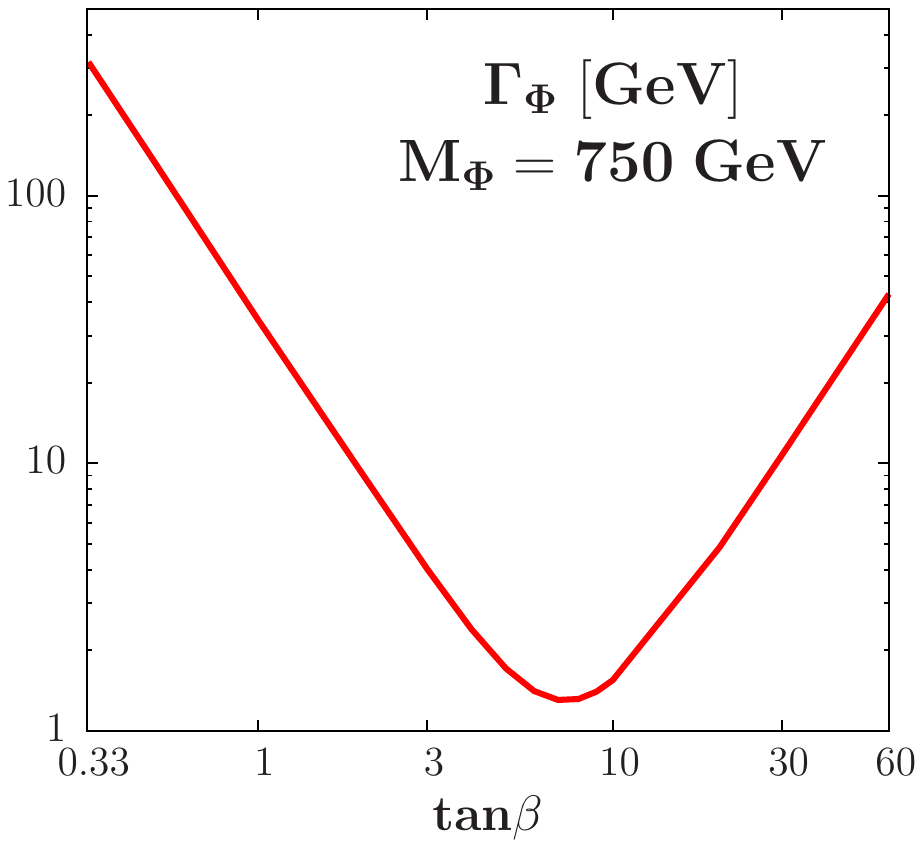}}
\vspace*{-13.5cm}
\caption{The branching ratios for $\Phi = H/A$ decays into various final states for $M_\Phi=750$ GeV as functions of $\tb$ (left) and the corresponding total decay  width in GeV (right).} 
\label{Fig:BR-phi}
\vspace*{-.2mm}
\end{figure}

In the Type I scenario, All couplings of the $\Phi=H,A$ states to fermions are
inversely  proportional to $\tan\beta$ so that the decay pattern follows more or
less the one of Type II for $\tan\beta=1$, namely: the $b\bar b$ and $\tau \tau$
decays are dominant for masses below $M_\Phi =350$ GeV, while the $t\bar t$
channel dominates above this mass value. However, slightly below this $2m_t$
threshold the three--body decay with one top quark being off--shell, $\Phi \to
\bar t t^* \to \bar tbW$, could be important and might compete with the $b\bar
b$ decays, the suppression by phase space being compensated by the large top
Yukawa coupling \cite{Djouadi:1995gv}. 

Note that the total decay width is
smaller than in the case of Type II in general and for $\tb >3$, $A,H$ are very
narrow. The pattern in  lepton specific and lepton-flipped scenarios also follows
the  same one as in Type I and Type II, respectively, as far as the decays
into $t\bar t$ and $b\bar b$ are concerned, but the  $\Phi \to \tau\tau$
channels do not follow $\Phi \to b\bar b$ anymore.  In fact, one can make these
channels either dominant with a branching ratio close to one  (in the
lepton--specific at high values of $\tan\beta$)  or completely negligible (in
the flipped scenario again at high values of $\tan\beta$).

The previous discussion does not take into account the possibility of the
invisible Higgs decays into the DM particle or the decays into its possible
companions. When these channels are kinematically open and the couplings to the
new particles are not so small, they can significantly alter the previous
pattern and can even dominate. This could particularly be the case  for Higgs
masses $M_{\Phi} <2m_t$ and small $\tan\beta$ values where only the $b\bar b$
and $\tau \tau$ decays are present and the Higgs coupling to these light
fermions is not enhanced. Hence, the possibility of invisible or almost
invisible $H,A$ states  is a serious one in these scenarios. 

Concerning the production mechanisms of the heavy neutral Higgs bosons at hadron
colliders, the dominant one stays in all cases the gluon fusion process $gg\to
\Phi$ as for the SM--like $h$ state. It is mediated mainly by triangular loops 
of top and bottom quarks and  the amplitudes, which are different in the
CP--even and  CP--odd cases, are given in Appendix A. In Type I and II 2HDMs
(we do not consider the lepton specific and flipped scenarios as they do not
matter here) for small values of  $\tan\beta$, the dominant contribution to the
amplitudes comes from top quark loops as the $\Phi t \bar t$ coupling is
strong.  For low masses, $M_\Phi \lsim 2m_t$, one could use the effective
approach in which the heavy top quark is integrated out and include not only the
NLO QCD corrections~\cite{Djouadi:1991tka,Dawson:1990zj,Spira:1995rr} but also
the  corrections up to N$^3$LO which are known in this case
\cite{Harlander:2002wh,Anastasiou:2002yz,Ravindran:2003um}; they increase the
rate by a factor $K_{\rm N^3LO}^{\rm t\!-\!loop} \approx 2$. The effective
approach was shown to be a good approximation at NLO even  above the
$M_\Phi\!=\!2 m_t$ threshold and can be used also for the higher order
corrections. 

In the Type II scenario at high $\tb$ values, the  contribution of the
$b$--quark loop to the $gg\to \Phi$ processes (which was less than 10\% in the
SM--like Higgs case) will become the dominant one. In fact, for very high
$\tb$ values, the cross section which grows as $\tan^2\beta$  and is enhanced by
large logarithms $\log(m_b^2/M_\Phi^2)$, can be extremely larger. In this case,
as $M_\Phi \gg 2m_b$, one is in the chiral limit  in which the rates are
approximately the same in the CP--even and CP--odd Higgs cases. In this limit,
one cannot use the effective approach and  integrate out the bottom quark to
implement the contribution of the higher order terms.  The QCD corrections can
be thus included only to NLO where they have been calculated  keeping the exact
quark mass dependence \cite{Spira:1995rr}. At LHC energies, the $K$--factor is
much smaller in this case, $K_{\rm NLO}^{\rm b\!-\!loop} \approx 1.2$, than in
the case of the top loop only~\cite{Dittmaier:2011ti}.  

For intermediate $\tb$ values, $\tb \approx 3$--10 for which the suppression of
the $\Phi tt$ coupling is already effective while the $bb\Phi$ coupling is not
yet strongly enhanced, the resulting production cross sections are small.  As in
the case of the total width, one obtains a minimum of the cross section at the
value $\tb \approx \sqrt{m_t / \bar m_b} \approx 7$.   Here again, because the
top and bottom loop contributions have a comparable weight, one can include only
the NLO QCD corrections which are known exactly. 

We have evaluated the production cross sections using the program {\tt SusHi}
\cite{Harlander:2012pb,Harlander:2016hcx}, in which important higher--order
effects are included, notably the large QCD corrections and some non--negligible
electroweak ones that have been discussed in Appendix A2. The production rates
for $gg\to H$ and $gg\to A$ at proton colliders are shown in
Fig.~\ref{Fig:main-pp} as a function of the c.m. energy $\sqrt s$ in the
alignment limit. The masses $M_\Phi=750$ GeV and the value $\tan\beta=1$ for
which the cross sections are the same in all types of 2HDMs are assumed. The
MSTW2008 PDF set \cite{Martin:2009iq} has been adopted.  At the LHC with $\sqrt
s=13$ TeV, the cross sections are of the order of 1 pb, and increase with energy
to reach about 100 pb at $\sqrt s=100$ TeV. Assuming an accumulated luminosity
of  a few ab$^{-1}$, as is expected to be the case at both HL--LHC and
FCC--hh/SPPC, one could then collect from $10^6$ to $10^8$ Higgs events at
these  colliders. 

\begin{figure}[!h]
\vspace*{-2.5cm}
\centerline{\hspace*{-5mm} \includegraphics[scale=0.95]{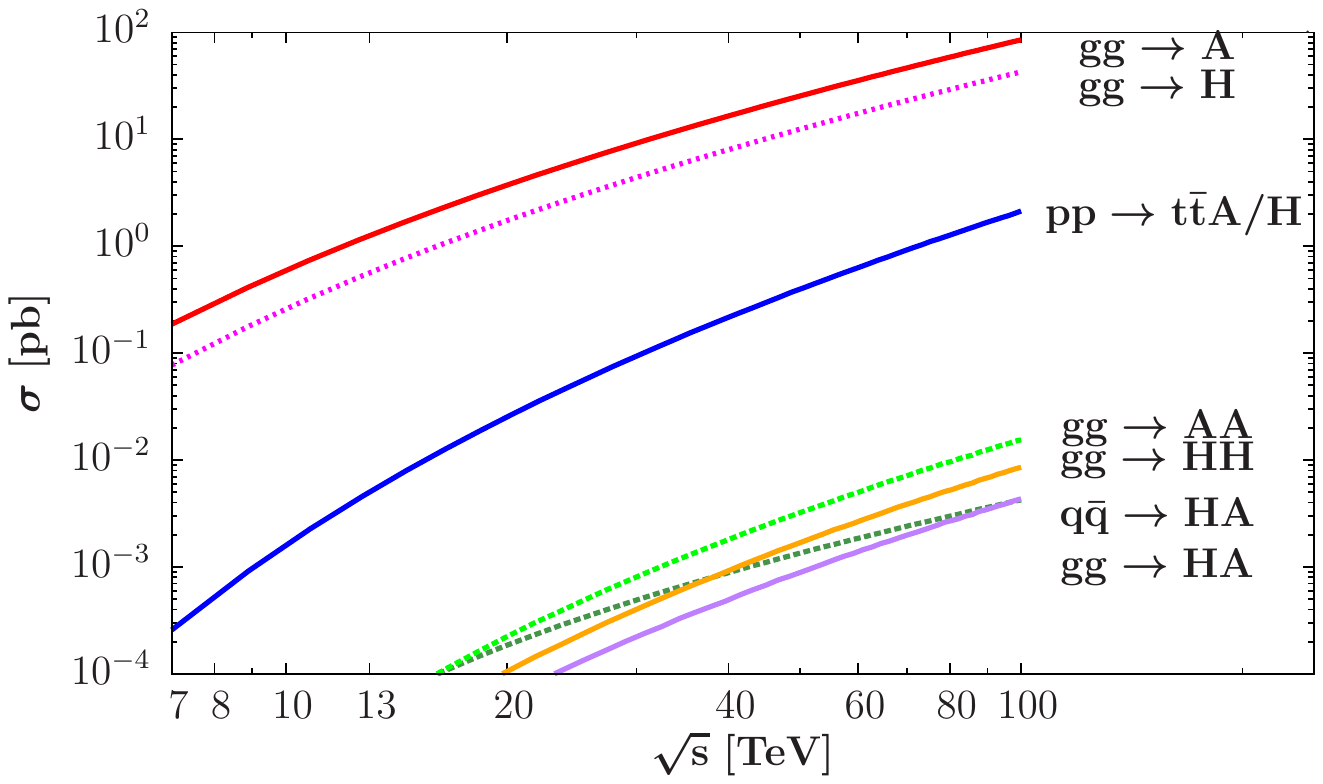}}
\vspace*{-17.9cm}
\caption{Cross sections for single, associated and pair production of the $\Phi=H,A$ bosons at $pp$ colliders as functions of the c.m. energy from $\sqrt s=7$ TeV to 100 TeV. We assume a common mass $M_{\Phi}=750$ GeV, $\tan\beta=1$ and the alignment limit. From \cite{Djouadi:2016eyy}.}
\label{Fig:main-pp}
\vspace*{-.2mm}
\end{figure}

\begin{figure}[!t]
\vspace*{-2mm}
\mbox{\hspace*{.5mm} \includegraphics[scale=0.27]{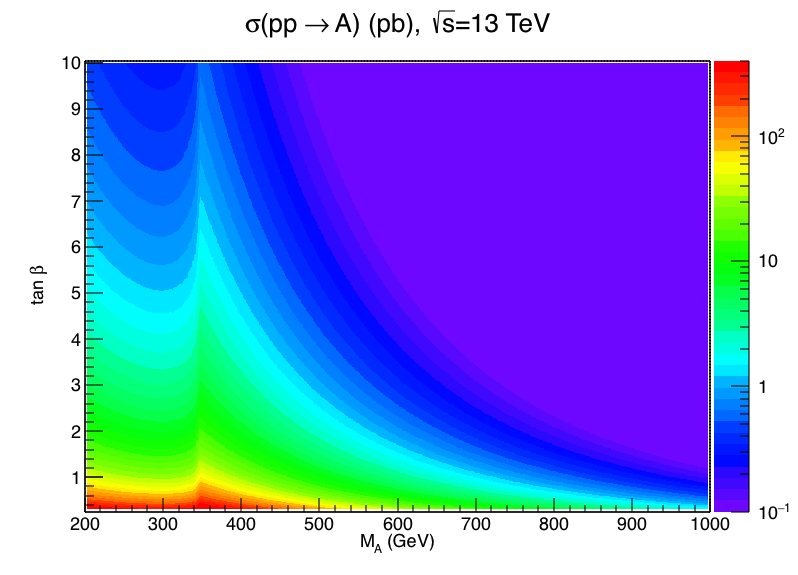}~~\includegraphics[scale=0.27]{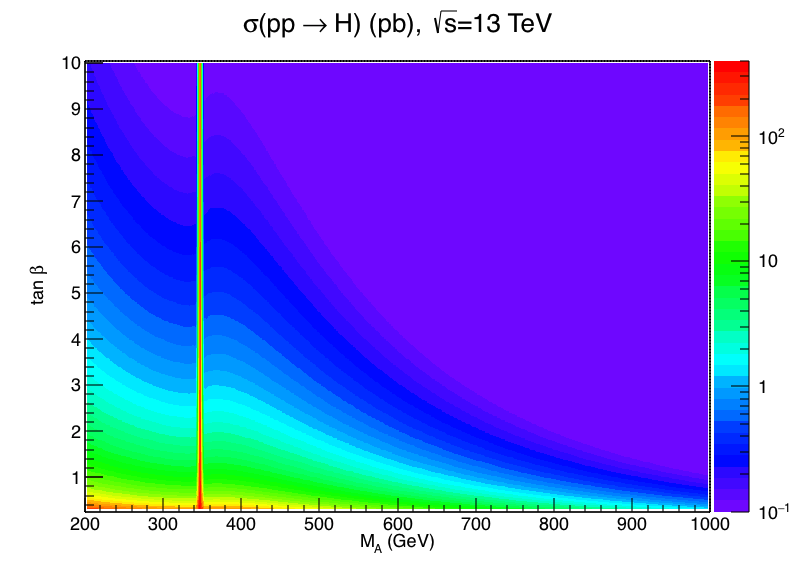} }\\[.2mm]
\mbox{\hspace*{.5mm}
\includegraphics[scale=0.27]{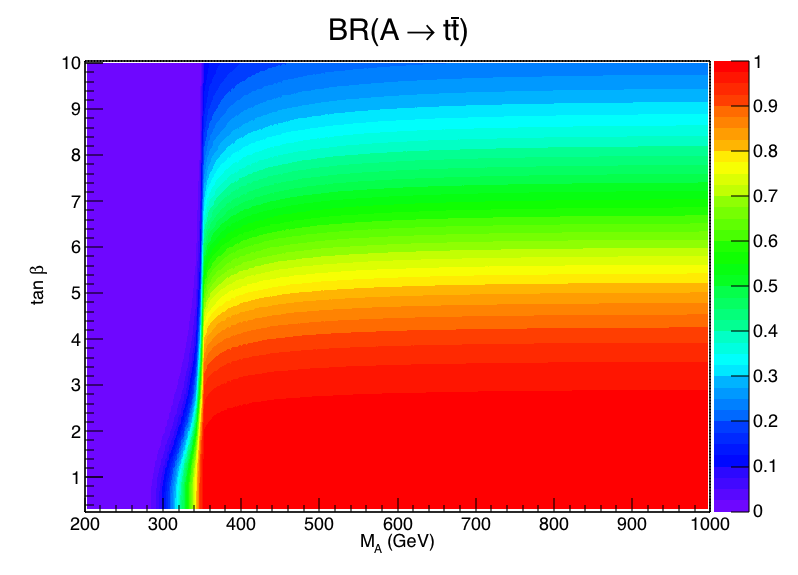}~~\includegraphics[scale=0.27]{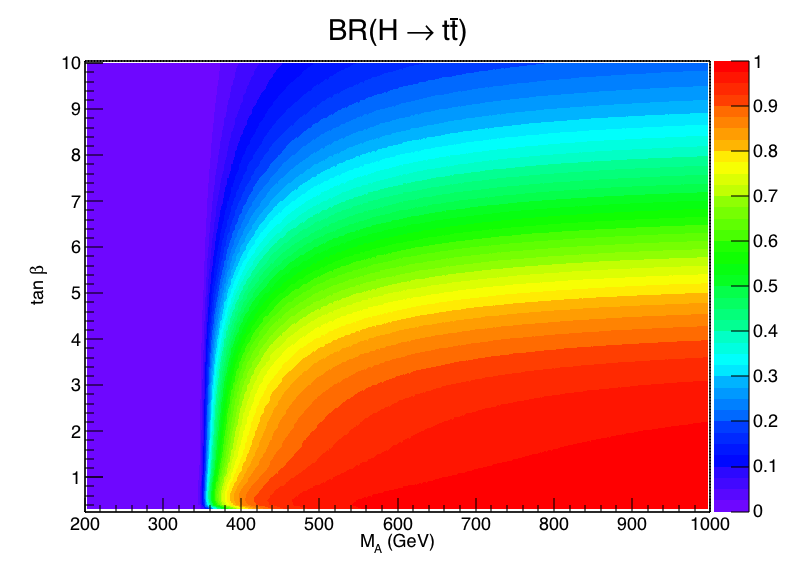} }\\[.2mm]
\mbox{\hspace*{.5mm}
\includegraphics[scale=0.27]{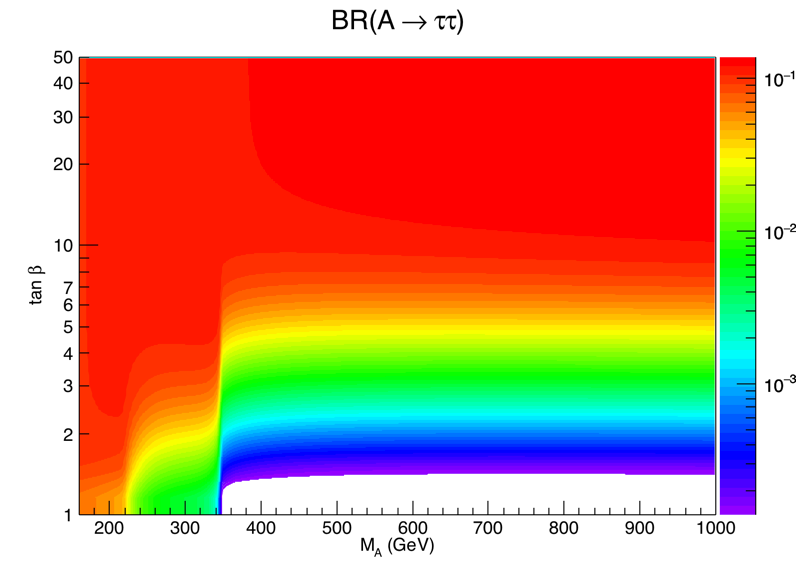}~~\includegraphics[scale=0.27]{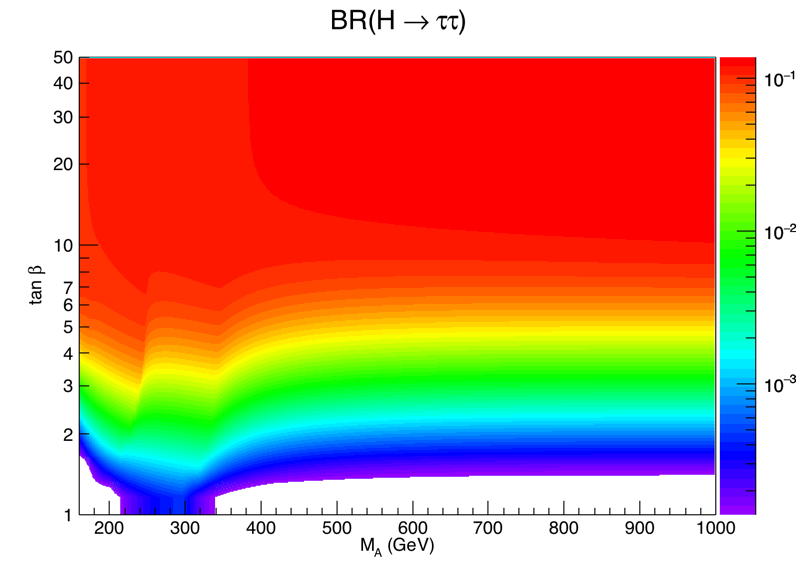} }\\[.2mm]
\mbox{\hspace*{.5mm}
\includegraphics[scale=0.27]{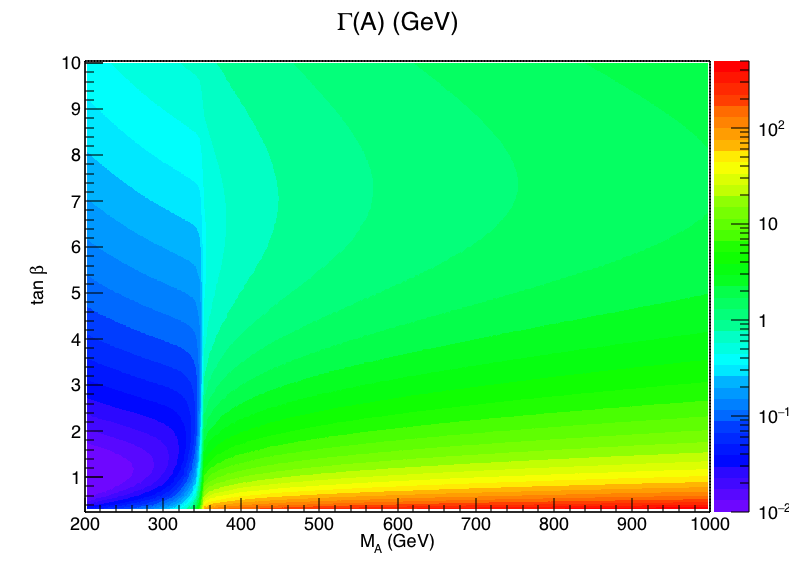}~~\includegraphics[scale=0.27]{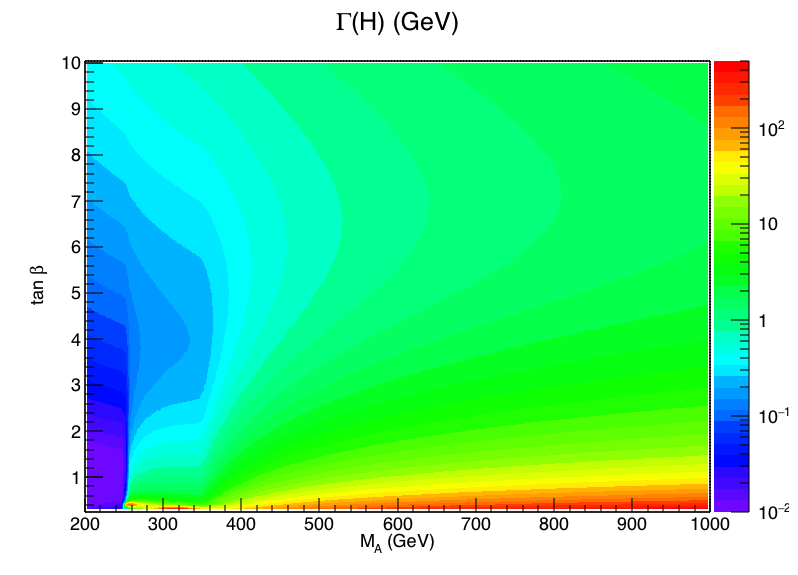} }
\vspace*{-2mm}
\caption{The $gg \rightarrow \Phi$ production cross sections at the 13 TeV
LHC,   the $\Phi \rightarrow t \bar t, \Phi \to \tau\tau$ branching  ratios and the total decay widths $\Gamma_\Phi$ of the  heavier 2HDM Higgs bosons $A$ (left) and $H$ (right) in the $[\tan\beta,M_{A}]$ plane, assuming no mass splitting $M_{H}=M_{A}$~\cite{Djouadi:2019cbm}.}
\label{rates-MSSM}
\vspace*{-3mm}
\end{figure}

Another important source for the $\Phi$ states is production in association with
heavy quark pairs, $pp \to t\bar t \Phi$ and $pp \to b\bar b \Phi$. In Type II
scenarios and at high values of $\tb$, the $gg \to b\bar b \Phi$ process is most
important and the cross sections, which are the same for $H$ and $A$ as we are
in the chiral limit $M_\Phi \gg m_b$,  are of the same order as in
gluon--fusion. In Type I and Type II models at low $\tb$ values, it is the
$pp\to t\bar t \Phi$ process that is important.  However, because of the reduced
phase space,  the production rates are at least two orders of magnitude smaller
than in the dominant $gg \to \Phi$ fusion  modes even at $\sqrt s=100$ TeV.
This  can be seen from  Fig.~\ref{Fig:main-pp} where the cross sections, that we
obtained using a modified version of the leading--order program {\tt HQQ}
\cite{Michael-web}, are shown again as a function of $\sqrt s$.

In the alignment limit, the only other possible source for the $H,A$ states
would be pair production which can occur in mainly two ways.  It first occurs in
the $q\bar q\to HA$  process with the $s$--channel exchange of a $Z$ boson that
has a maximal coupling to the $HA$ pair, $g_{ZHA}=1$. But  at high energies
where the gluon luminosity is much larger, the dominant mode becomes $gg \to
HA$, which is mediated by top quark loops at low values of $\tb$ in box or
triangular diagrams. In gluon--fusion, one can  produce in  the same way  $HH$
and $AA$ pairs.   The cross sections, evaluated at leading order using the
programs {\tt HPAIR} \cite{Michael-web} are also shown in
Fig.~\ref{Fig:main-pp}. They  are rather small, barely reaching the 10 fb level
even at $\sqrt s=100$ TeV and high luminosities will be necessary to probe
them. 

To summarize this discussion, the production cross sections $\sigma(gg\to \Phi)$
at the 13 TeV LHC,  the important branching ratios BR$(\Phi \rightarrow t \bar
t)$ and BR$(\Phi \rightarrow \tau^+  \tau^-)$ as well as the total decay widths
$\Gamma_\Phi$  are shown in Fig.~\ref{rates-MSSM} in the $[\tb, M_\Phi]$
parameter plane for $\Phi=A$ (left) and $\Phi=H$ (right).  Again, we assume a
Type II scenario in the alignment limit and a near mass degeneracy for the heavy
Higgs states.  As will be seen later, high $\tb$ values are excluded by searches
of $\tau\tau$ resonances, so we specialize sometimes in the case $\tb \lsim 5$
where one can see that for not too large values of $M_\Phi$,  the production
rate $\sigma(gg\rightarrow \Phi \rightarrow t\bar t)$ is large. This channel is
thus very important to investigate  at the LHC. 

\subsubsection{Present constraints on 2HDMs and extrapolations for the future}

Searches have been performed by the ATLAS and CMS collaborations for the neutral
Higgs bosons of the 2HDM and two of them are very important. The first one is 
the search for heavy resonances decaying into $\tau^+\tau^-$  final states 
\cite{Sirunyan:2018zut,Aaboud:2017sjh} which can be interpreted  as $\Phi=H/A$
production either singly in gluon fusion  $gg\to \Phi$ or in association with 
$b\bar b$ final states $gg\to b\bar b \Phi$ (but in fact, these quarks may not
be observable and in practise one is looking at the equivalent fusion process
$b\bar b \to \Phi$). As seen before, in Type II 2HDMs at high values of $\tb$,
the decays $\Phi \to \tau\tau$ have  a branching fraction of the order of 10\%.
In the left--hand side of Fig.~\ref{Fig:Phi-pp}, we report a search performed by
the CMS collaboration in this topology at $\sqrt s=13$ TeV with about 36
fb$^{-1}$ data. Shown are the exclusion limits at the 95\%CL from the absence of
a signal in the $[\tb,M_{A}]$ parameter space. The analysis has been done in
the  context of the hMSSM scenario to be studied in the next section, but it is
also valid in the case of a Type II 2HDM with a near mass degeneracy  of the
$H/A$ states\footnote{In the MSSM, there is a mass difference between $A$ and
$H$, but it is much smaller than the experimental resolution on the $\tau\tau$
invariant mass so that  one can also assume $M_H=M_A$ to a good approximation.}.
One can see that for $\tb\gsim 10$, the entire mass range  $M_\Phi \lsim 1$ TeV
is excluded.

\begin{figure}[!h]
\vspace*{-5mm}
\begin{tabular}{ll}
\begin{minipage}{8cm}
\hspace*{-.6cm}
\includegraphics[scale=0.35]{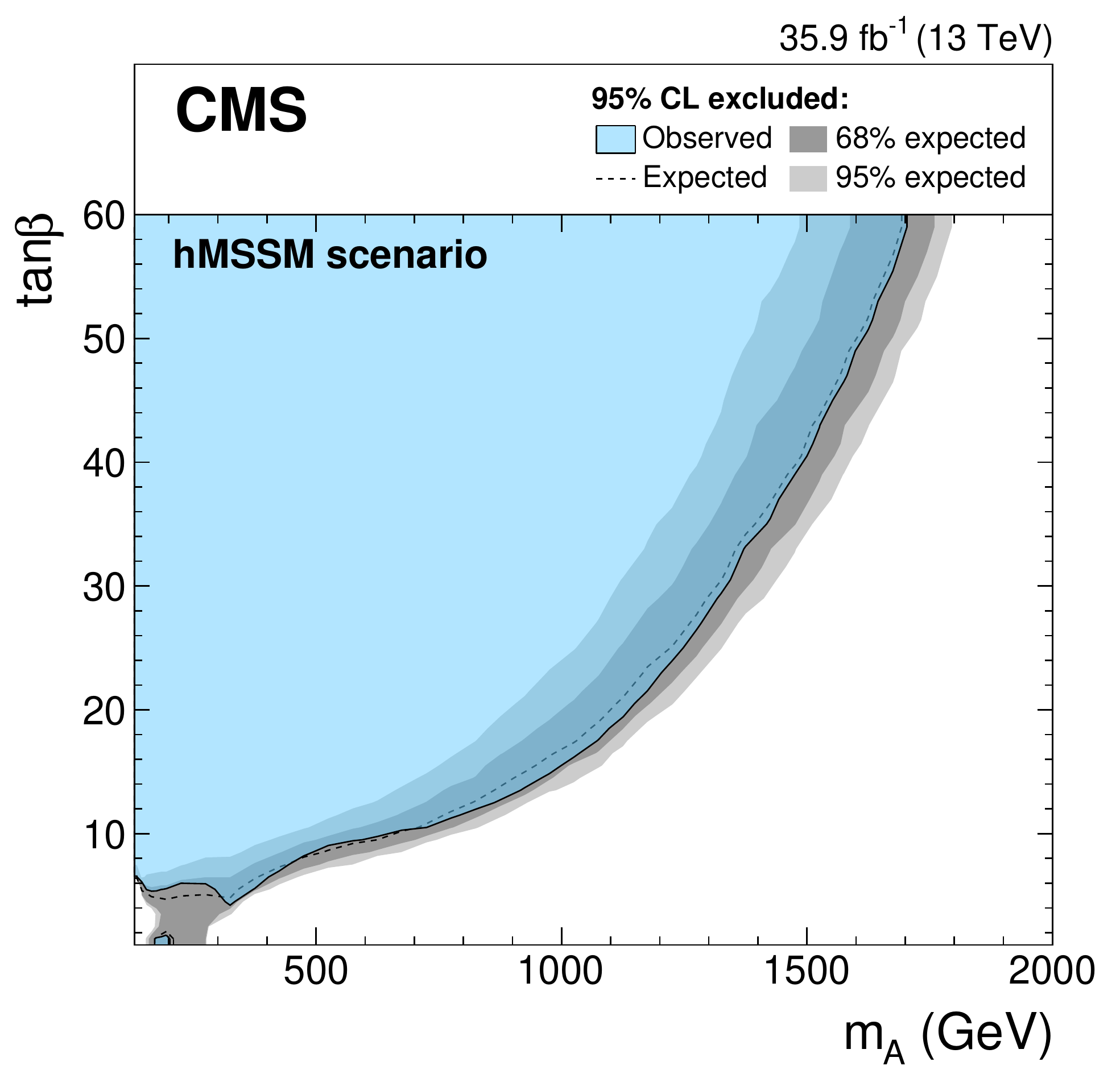}
\vspace*{-10mm}
\end{minipage}
& \hspace*{-2cm} 
\begin{minipage}{8cm}
\includegraphics[scale=0.58]{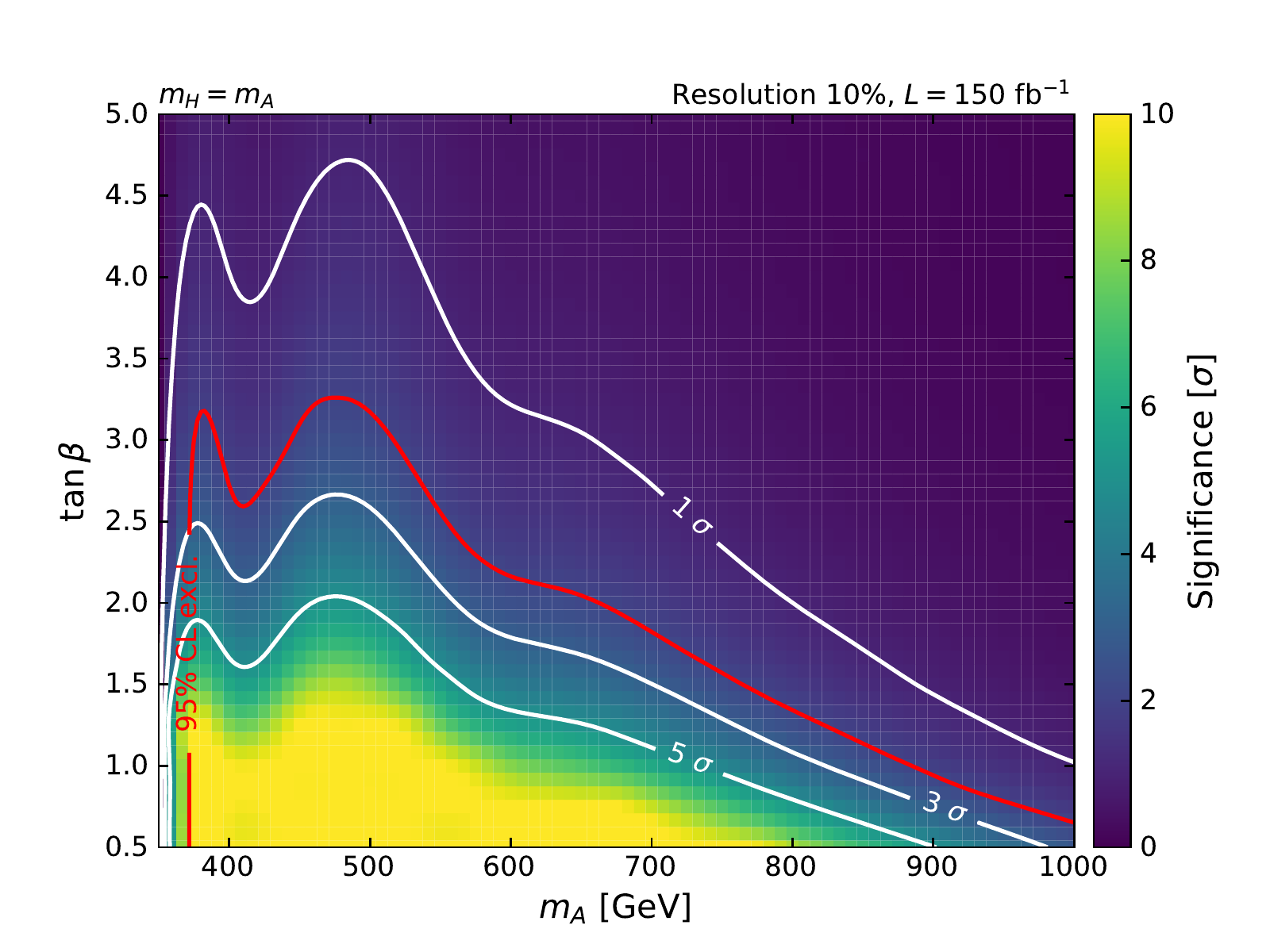}
\end{minipage}
\end{tabular}
\vspace*{.2cm}
\caption{Left: CMS exclusion limits at the 95\%CL in the $[\tb,M_{A}]$ plane from searches of high mass resonances decaying into $\tau\tau$ pairs at $\sqrt s=13$ TeV with about 36 fb$^{-1}$ data \cite{Sirunyan:2018zut}. Right: expected  significance or  exclusion  potential  for  the  aligned 2HDM with $M_H=M_A$ in the $[M_A, \tb]$ plane at the LHC with $\sqrt s=13$ TeV and 150 fb$^{-1}$ data in the channel $gg\to t\bar t$ assuming  a 10\% resolution on the $m_{t\bar t}$ invariant mass \cite{Djouadi:2019cbm}.}
\label{Fig:Phi-pp}
\vspace*{-2mm}
\end{figure}

Another important channel for $\Phi=H,A$ detection in the 2HDM is top quark pair 
production,   $pp\to gg\to \Phi \to  t\bar t$, which is relevant at low $\tan\beta$
values and for Higgs masses above the $t\bar  t$ threshold, $M_\Phi \gsim 350$ GeV.
In the case of the mass degeneracy $M_A=M_H$, one has again to take care of  both
$H$ and $A$ contributions and the interference with the large QCD continuum
background $gg\to  t\bar t$. 

This situation has also be analyzed recently at LHC
energies~\cite{Djouadi:2019cbm} taking into account the experimental environment,
along the same lines as what has been discussed in the previous section on singlet
Higgs production. Restricting to low $\tb \lsim 5$ values, which make that the
results are almost the same in Type I and II scenarios, and  assuming the alignment
limit with $M_H=M_A$, the statistical significance of observing or excluding the
Higgs bosons in this search channel is shown in the right--hand side of
Fig.~\ref{Fig:Phi-pp} in the $[M_\Phi, \tb]$ plane for $\sqrt s=13$ TeV and with a
luminosity of 150 fb$^{-1}$; we have assumed a 10\% resolution on the $t\bar t$
invariant mass. As can be seen, for low values of  $\tb$ and $M_A$, a high
significance larger than $5\sigma$ can be achieved (in the figure, values of
significance in excess  of $10\sigma$ are clipped).  Instead, a $2\sigma$
sensitivity can be obtained for $\tb \approx 1$ and  $M_A=1$ TeV or $\tb \approx 3$
and  $M_A=0.5$ TeV. A worse experimental resolution on $m_{t \bar t}$ would lead to
a degradation of the sensitivity which can, however, be compensated by an increase
in the integrated luminosity. 

Turning to the charged Higgs bosons, when light enough, i.e. $M_{H^\pm} \lsim
m_t$,  the main production channel was  the top decay mode $t \to bH^+$ with the
subsequent decay $H^- \to \tau \nu$. This channel has been searched for at the
LHC and, already at RunI with $\sqrt s=8$ TeV and 20 fb$^{-1}$ data, the absence
of a signal excluded the entire  mass range $M_{H^\pm} < 160$ GeV for any value
of $\tan\beta$  \cite{Aaboud:2018cwk}.  For larger masses, the dominant process
would be the associated $gb \to t H^\pm$ mechanism, which for $\tan\beta \approx
1$ or $\tan\beta \gg 1$,  has a large cross section at the LHC as it is shown in
the left--hand side of   Fig.~\ref{Fig:H+-pp} for  $\sqrt s=14$ TeV as a
function of $M_{H^\pm}$. In most cases, the process leads to $t\bar t  b $ final
states but at high $\tb$ values,  the $t \tau \nu$ signature is  also possible
and is easier to probe. Both topologies have been looked for at the LHC and,
for instance,  a search was performed by the ATLAS collaboration (again in the
context of the hMSSM scenario to be discussed later, but it also applies in the
Type II 2HDM case) at $\sqrt s=13$ TeV and 36 fb$^{-1}$ data. 

The outcome is displayed in the right--hand side of Fig.~\ref{Fig:H+-pp}  in the
$[\tb, M_{H^\pm}]$ plane and, as it can be seen, for Higgs masses $M_{H^\pm}
\lsim 600$ GeV, both the low  $\tb \lsim 1$ and the high  $\tb \gsim 25$ values
are excluded at the 95\%CL.  Other possible processes for charged Higgs bosons
are $q\bar q \to \gamma^*, Z^* \to H^+ H^-$ and  associated  $q\bar q \to W^*
\to HH^\pm, AH^\pm$ production but they lead to much smaller rates. The
corresponding cross sections are also shown in Fig.~\ref{Fig:H+-pp} (left) as a
function of $M_{H^\pm}$ at the c.m. energy $\sqrt s=14$ TeV and with the input
choice $\tb=1$ (but they do not depend on $\tb$ in the alignment limit).   

\begin{figure}[!h]
\vspace*{-1.6cm}
\begin{tabular}{ll}
\begin{minipage}{8cm}
\hspace*{-4.3cm}
\includegraphics[scale=0.76]{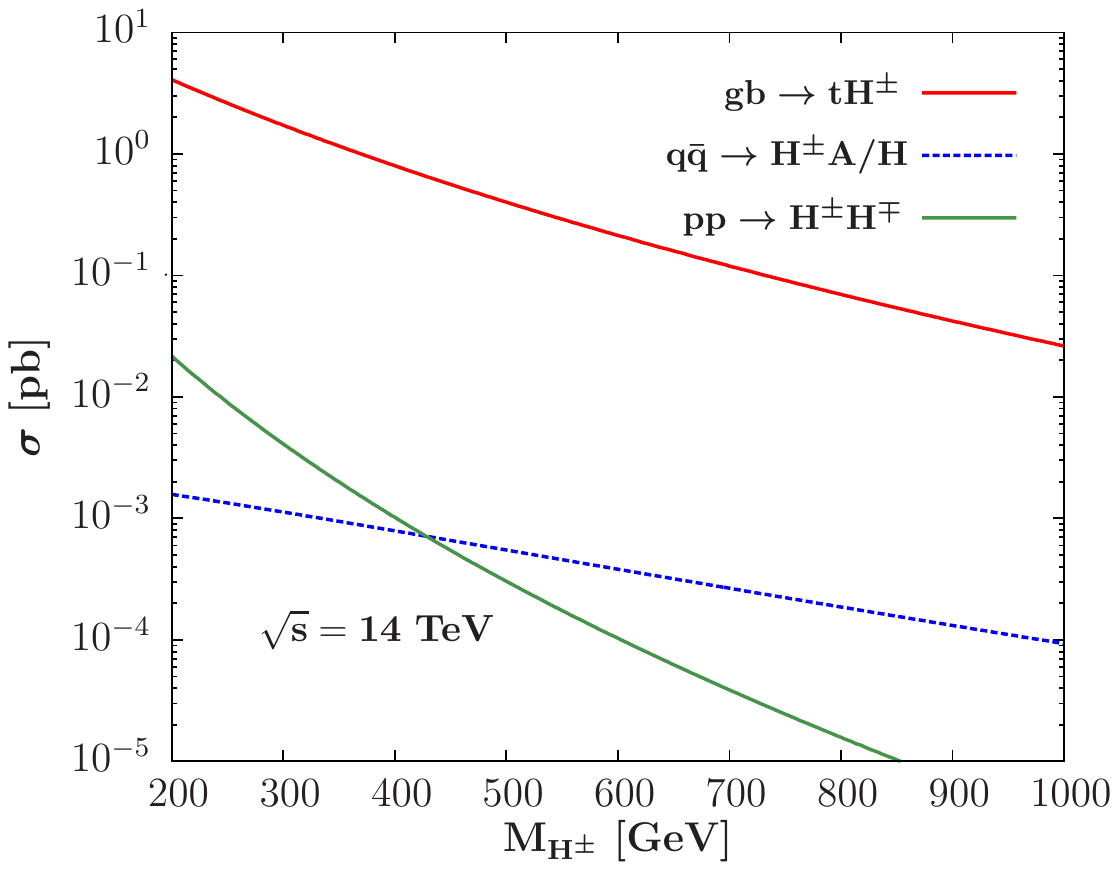}
\vspace*{-2mm}
\end{minipage}
& \hspace*{-.6cm} 
\begin{minipage}{8cm}
\vspace*{-12cm}
\includegraphics[scale=0.4]{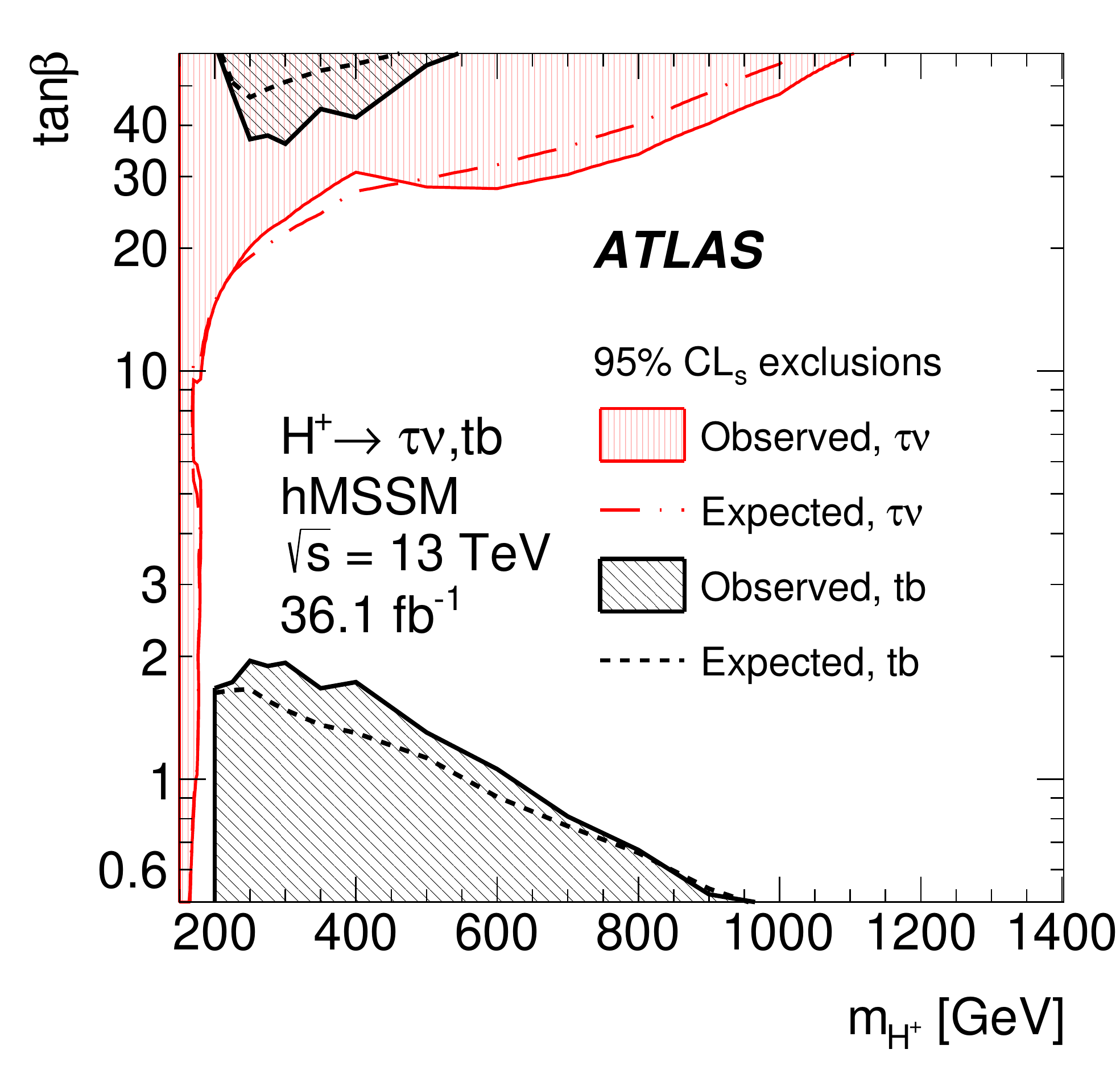}
\end{minipage}
\end{tabular}
\vspace*{-13.7cm}
\caption{Left: cross sections for the production of the charged Higgs boson at the 14 TeV LHC as a function of its mass for $\tan\beta =1$ \cite{Djouadi:2016eyy}. Right: exclusion  contours at the 95\%CL in the $[\tb,M_{H^\pm}]$ parameter space from searches of the charged Higgs boson by the ATLAS collaboration at $\sqrt s=13$ TeV with 36 fb$^{-1}$ data \cite{Aaboud:2018cwk}.}
\label{Fig:H+-pp}
\vspace*{-2mm}
\end{figure}

In the context of a Type II 2HDM, the impact of the various searches that have
been  conducted by the ATLAS and CMS collaborations can be used to constrain the
$[M_A, \tb]$ parameter space of the model if one assumes a near mass degeneracy
of the three heavy Higgs bosons, $M_{H^\pm} \approx M_H \approx M_A$.  In this
case, and if also the alignment limit is assumed, only the four fermionic
channels discussed above, namely  $H/A\to \tau\tau$ and $t\bar t$,  $H^\pm
\rightarrow \tau\nu$ and $H^+ \to tb$, need to be considered. All the
constraints from the ATLAS and CMS searches obtained at RunII  with about 36
fb$^{-1}$ data in  the $[\tb, M_{A}\! =\! M_H \! = \! M_{H^\pm}]$ plane can be
determined by combining Figs.~\ref{Fig:Phi-pp} and \ref{Fig:H+-pp} (right). The
limits are already quite impressive and a significant part of the parameter
space has been already excluded. 

The sensitivity in these channels can be vastly improved at the  HL--LHC with an
energy of $\sqrt s=14$ TeV and 3 ab$^{-1}$ data and even more at a 100 TeV
collider with the same luminosity.  Assuming that this sensitivity approximately
scales with the square root of the number of expected events, one can
extrapolate the above exclusion limits (the procedure to obtain these have been
discussed in Ref.~\cite{Djouadi:2015jea} to which we refer for the details) at
these two machines. The $2\sigma$ sensitivities are shown in
Fig.~\ref{projections_2HDM} and it can be seen that indeed, the two machines
will perform much better than presently. In the very low and very high $\tb$
regions, masses close to 1.5 TeV and 3 TeV can be probed at, respectively, the
HL--LHC and  a 100 TeV collider in the $H/A \to t\bar t$ and $H/A \to \tau^+
\tau^-$  modes. The two channels intersect at $M_A=1.5$ TeV for a 100 TeV
collider and $M_A =750$ GeV for HL--LHC,  mass values  below which the entire
Type II 2HDM parameter space is fully covered by the searches.   

\begin{figure}[!h]
\vspace*{.2cm}
\centerline{
\includegraphics[scale=0.47]{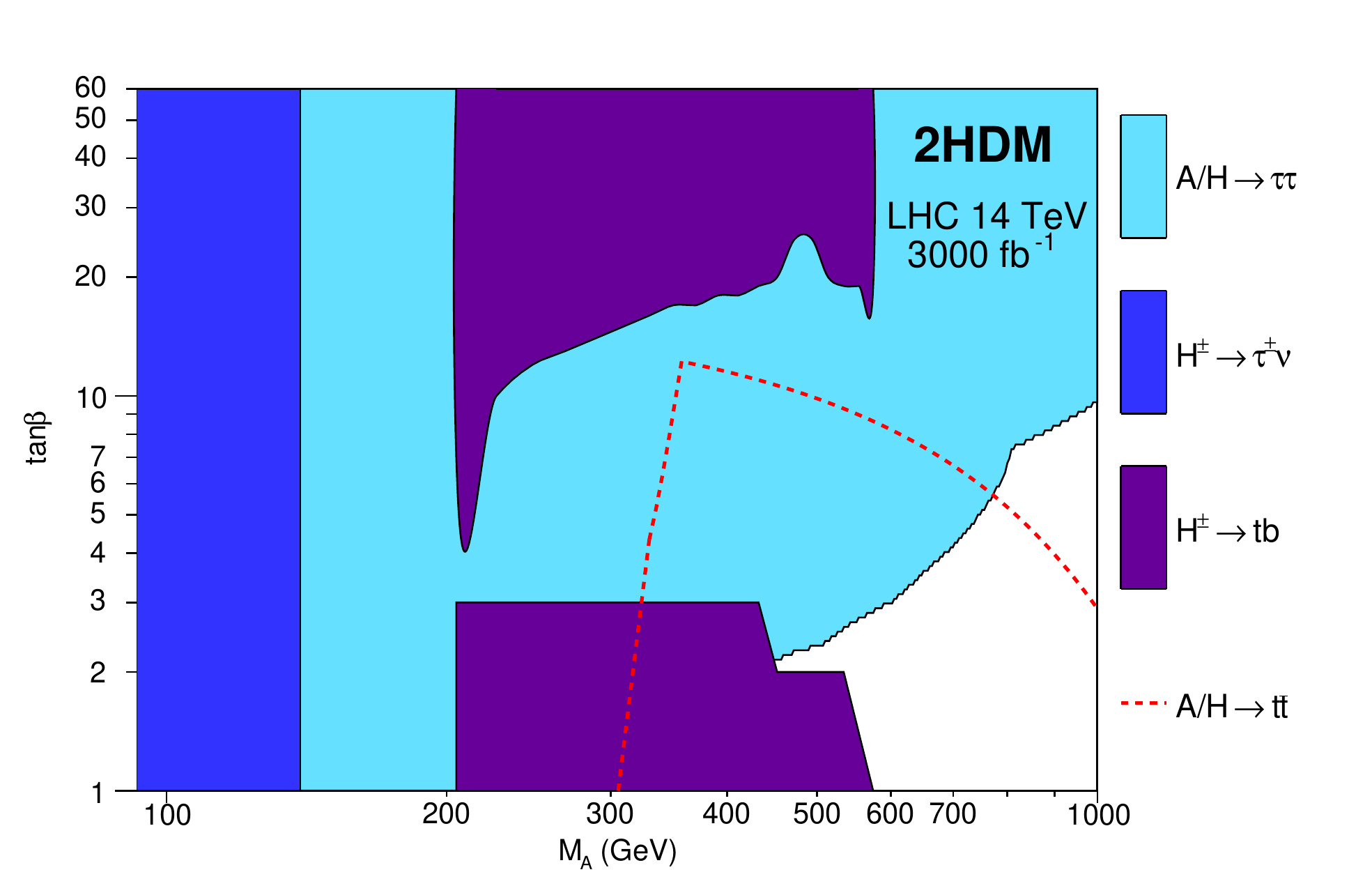} 
\hspace*{-2.1cm}
\includegraphics[scale=0.47]{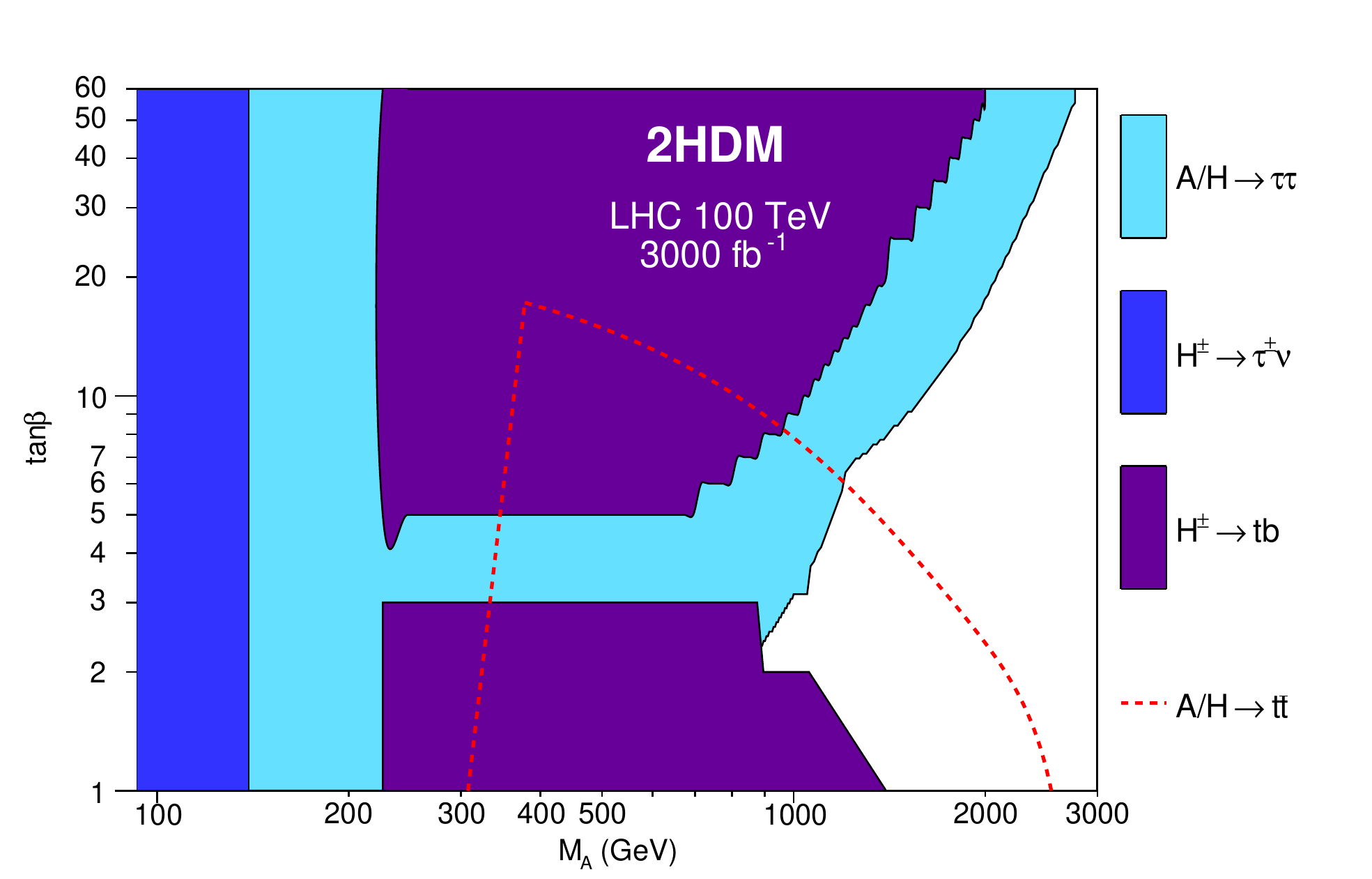} }
\vspace*{-1mm}
\caption[]{95\%CL exclusion limits or $2\sigma$ sensitivity in the   $[\tb, M_\Phi]$ plane (with $M_\Phi=M_A=M_H= M_{H^\pm})$ in the Type II 2HDM  when the combined ATLAS and CMS searches at Run1 for the $A/H/H^\pm$ states in their fermionic decays in the alignment limit are extrapolated to the HL--LHC with $\sqrt s=14$ TeV (left) and to a $\sqrt{s}= 100$~TeV machine (right) both assuming $3000$ fb$^{-1}$ of data. From Ref.~\cite{Baglio:2015wcg}.}
\label{projections_2HDM}
\vspace*{-.2cm}
\end{figure}

Before closing this discussion, let us briefly summarize the prospects for the
2HDMs  at future high--energy $e^+e^-$ colliders.  In the exact alignment limit,
the most important channel for producing the neutral Higgs bosons  is the
associated $HA$ process via $s$--channel $Z$ boson exchange, $e^+ e^- \to Z^*
\to HA$, as the coupling is maximal at the production vertex, $g_{ZHA}\to 1$.
The cross section is displayed in Fig.~\ref{Fig:ee-Phi-2HDM} (left) again for
$M_H=M_A=750$ GeV as a function of the c.m. energy $\sqrt s$. As it scales like
$1/s$, the cross section is not that large, namely ${\cal O}(1\,$fb)
significantly above the $2M_\Phi$ threshold, leading to a thousand events that
can be fully reconstructed for the anticipated luminosity of 1 ab$^{-1}$. At low
values of $\tb$ and for light $\Phi=H,A$ states, another possible channel would
be associated production with top quark pairs, $e^+ e^- \to t \bar t \Phi$
\cite{Djouadi:1991tk}, for which the combined cross sections are at the level of
0.1 fb at high enough energy as is shown in the same figure.  In all cases, the
signature for low $\tb$ values would be four top quarks in the final state,
which should have little background. At high $\tan\beta$, only the mode  $e^+
e^- \to HA\to 4b, 2b2\tau, 4\tau$ would be relevant, while at intermediate $\tb$
mixed $2t2b$ final states should also be searched for. All these final states should be easy to observe at these colliders, despite of the low rates. 

In addition to the production in the conventional $e^+e^-$ mode of future linear
colliders, the neutral  $H$ and $A$ states can be produced in the  $\gamma
\gamma$ mode as $s$--channel resonances. At low values of $\tan\beta$, the
process is mediated by a loop of top quarks with a large Yukawa coupling so that
the cross sections are significant. For $\Phi$ masses above $2m_t$, the decay
$\Phi  \to t\bar t$ is  dominant and would result into a large total decay width
$\Gamma_\Phi$.  The main background would again come from top quark pair
production, $\gamma \gamma \rightarrow t \bar t$, which is not helicity
suppressed as for light fermions.  In Fig.~\ref{Fig:ee-Phi-2HDM} (right), we 
display the cross sections  for  $\gamma \gamma \rightarrow t \bar t$,  taking
into account both the QED process and the resonance production $\gamma \gamma
\rightarrow \Phi \rightarrow t \bar t$ in the two channels with $\Phi=H+A$ and
including  the interferences. We have followed the discussion held in subsection
4.2 for the singlet scalar but  assumed $M_A=M_H=750$ GeV and $\tb=1$ which
gives Higgs total widths of $\Gamma_A=35$ GeV and  $\Gamma_H=30$ GeV.  As can be
seen, the signal is clearly standing out from the QED background. At high
$\tan\beta$, the search should be done in the $\Phi\to b\bar b$ decay mode for
which the background is suppressed for the photon helicity combination that
favors the Higgs signal.

\begin{figure}[!h]
\vspace*{-2.3cm}
\begin{tabular}{ll}
\hspace*{-.8cm}
\begin{minipage}{8cm}
\centerline{\includegraphics[scale=0.89]{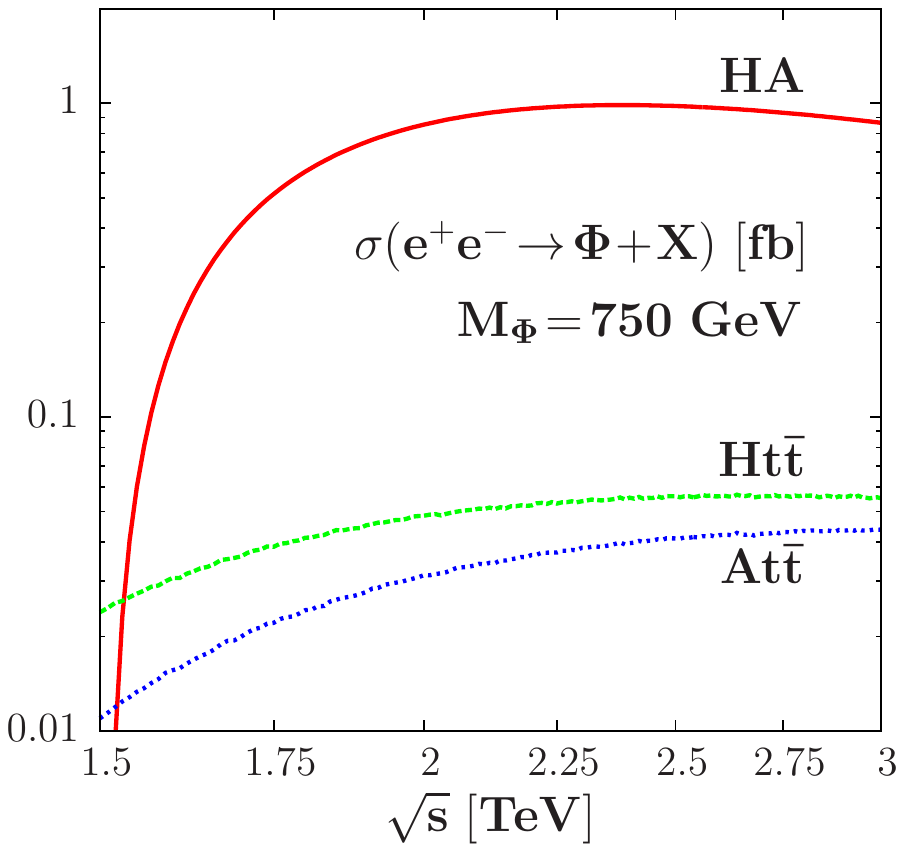} } 
\end{minipage}
& \hspace*{-.5cm} 
\begin{minipage}{8cm}
\vspace*{-1.9cm}
\centerline{\includegraphics[scale=0.80]{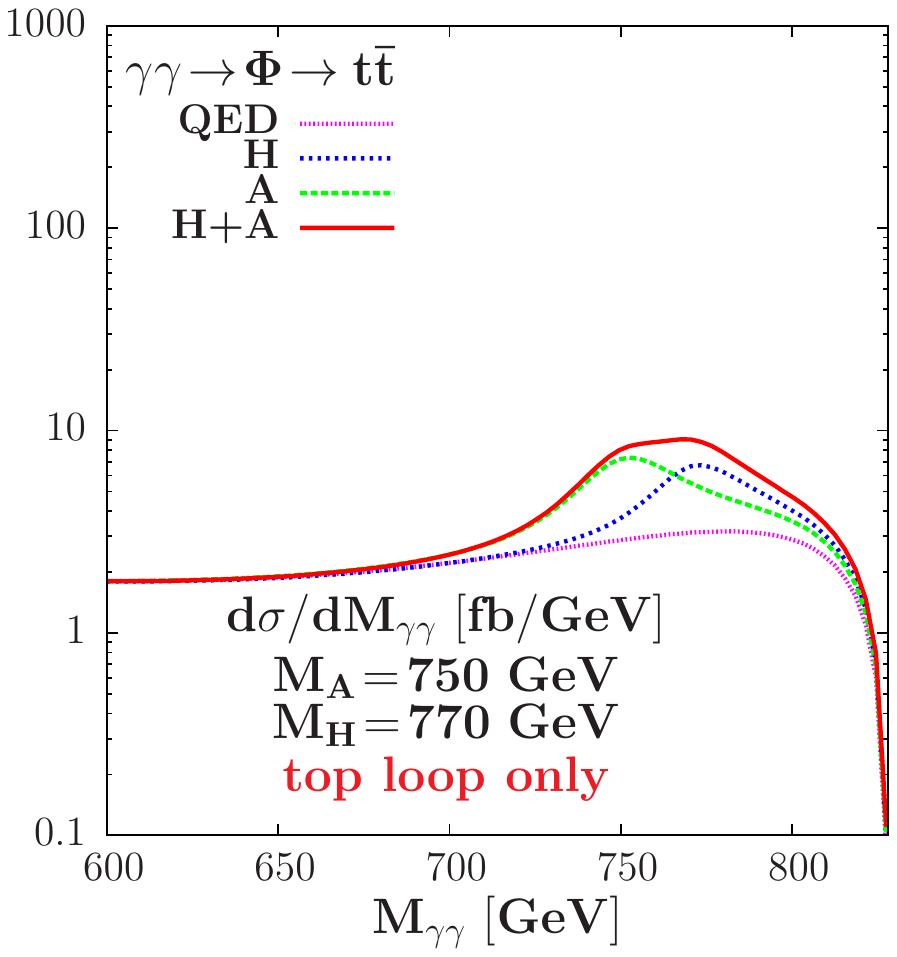} }
\end{minipage}
\end{tabular}
\vspace*{-15.6cm}
\caption{Left: the cross sections for the process $e^+e^- \to HA$ as a function of the energy $\sqrt s$ in a 2HDM in the alignment limit; the cross sections for $e^+e^- \to ttH$ and $ttA$ are also shown for $\tan\beta=1$.  Right: invariant mass distribution ${\rm d\sigma/d}M_{\gamma \gamma}$ in fb/GeV for the process $\gamma\gamma \to t\bar t$ in the photon mode of a linear $e^+e^-$ collider; shown are the pure continuum QED contribution, the additional separate contributions due to $s$--channel exchanges of the $H$ and $A$ states, and  the full set of contributions QED+$H$+$A$. In both cases, we assume $M_A=750$ GeV, $M_H=770$ GeV and $\tan\beta=1$; from Ref.~\cite{Djouadi:2016eyy}.} 
\label{Fig:ee-Phi-2HDM}
\vspace*{-2mm}
\end{figure}

\subsubsection{Constraints when including the DM sectors}

In this subsection, we will give a few  illustrations on some additional
constraints that can be imposed on the 2HDMs when the DM particles are also
involved. This issue will be again discussed, and in greater detail, in the next subsection.

First, in the 2HDM with a single--doublet DM sector, some of the constraints are
summarized in Fig.~\ref{fig:pSD_mA_tbeta} in the plane $[M_A,\tb]$ assuming our
Type II benchmark scenario with alignment and $M_H\!=\!M_{H^\pm}\!= \!M_A$. In
this plane, superimposed to the exclusion areas from LHC charged Higgs searches
in the channels $H^- \! \to \! \tau\nu$ and $H^- \! \to \! \bar tb$ as well as
from heavy neutral $H/A$ searches in the topology  $H/A \!\to \! \tau^+\tau^-$,
the isocontours of the correct DM relic density for two scenarios with $y=1$:
$M_L\! = \! 3M_N \!= \! 450$ GeV with $t_\theta\!=\!-\!6$ (black solid line) and
$M_L\! =\! 750$ GeV, $M_N\! =\! 350$ GeV with $t_\theta\!= \!-\!4$ (blue dashed
line). 

As will be clarified later, these two benchmark models have suppressed DM
scattering cross sections on nuclei as a result of the occurrence of a blind
spot. In both cases, the correct relic density is achieved until moderate values
of $\tb$, $\lesssim 20$, and close to the $m_{N_1} \sim \frac12 M_A$ poles.
Given this, the bounds from searches of the CP-odd Higgs boson $A$ are effective
in constraining the viable parameter space for DM.

\begin{figure}
    \centering
    \includegraphics[width=0.5\linewidth]{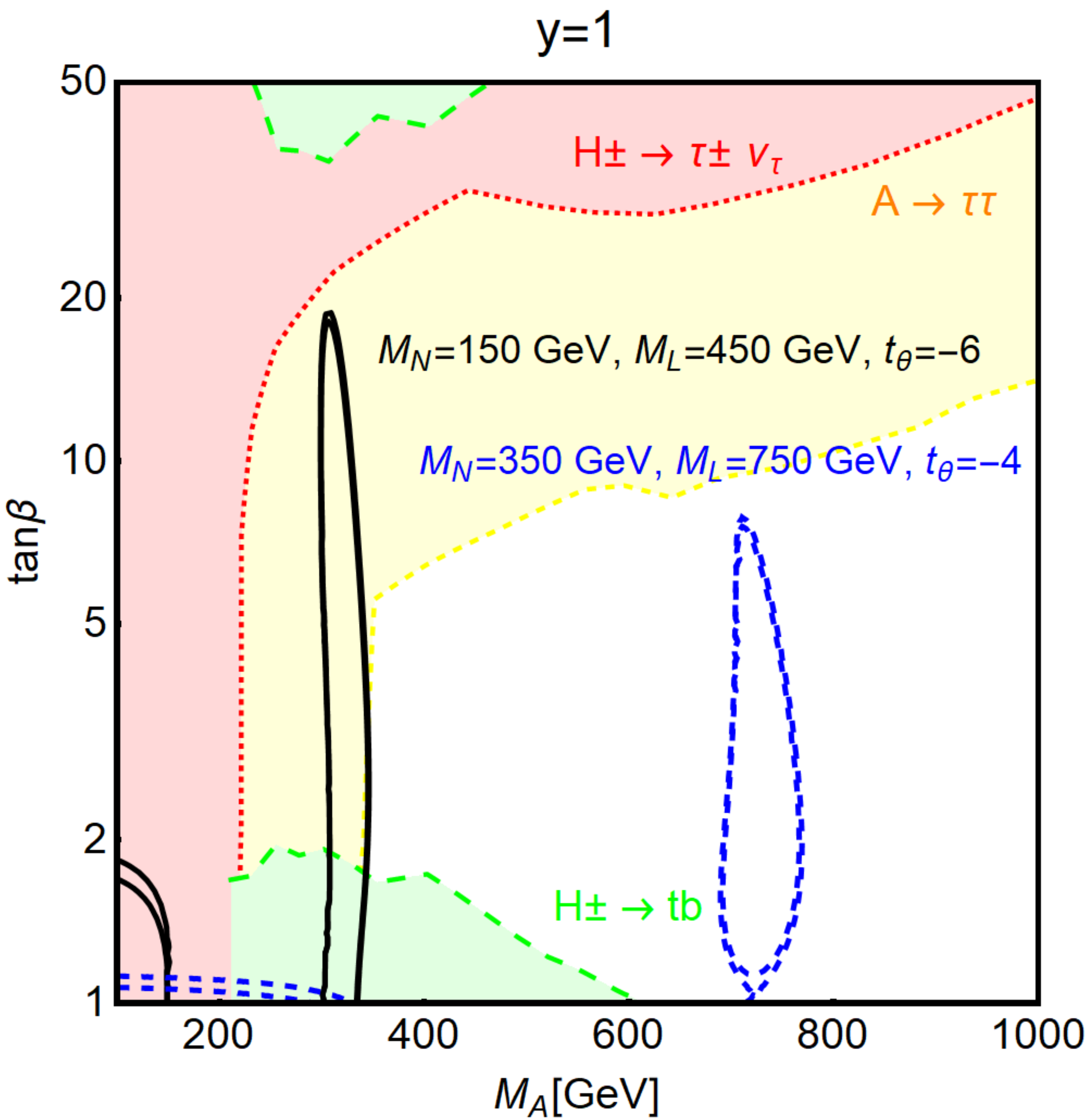}
\caption[]{Constraints in the plane $[M_A,\tb]$ of a Type II 2HDM scenario with
$M_H \!=\! M_{H^\pm} \!= \!M_A$ and the alignment limit from $H^\pm $ and $H,A$ searches at the LHC in several channels and the requirement of the correct relic density  for the DM state in two scenarios of a singlet--doublet lepton spectrum with $y\!=\!1$ as indicated in the frame.}
\label{fig:pSD_mA_tbeta}
\end{figure}

In the 2HDM with a vector--like fermion family, an interesting constraint could
be due to the search of heavy neutral Higgs resonances decaying into diphotons
\cite{Sirunyan:2018wnk,Aaboud:2017yyg}. 
The contribution from the VLF to the decay amplitude can be straightforwardly computed extending the expressions provided in section 3. Sticking for simplicity to the scenario $N_{\rm VLL}=1, N_{\rm VLQ}=0$ we have that:
\begin{align}
    & \mathcal{A}^{\rm VLL}_{H \rightarrow \gamma \gamma}=-\frac{v'}{2 m_{E_1}+v' y_h^{E_L}}\bigg\{ y_H^{E_L} \left[A_{1/2}^H (\tau_{E_1})-A_{1/2}^H (\tau_{E_2})\right]  \nonumber\\
    & +  y_H^{E_R}\left[\frac{m_{E_1}+v' y_h^{E_L}}{m_{E_1}}A_{1/2}^H(\tau_{E_1})-\frac{m_{E_1}}{m_{E_1}+v' y_h^{E_L}}A_{1/2}^H(\tau_{E_2}) \right]\bigg\}\nonumber\\
    & \mathcal{A}^{\rm VLL}_{A \rightarrow \gamma \gamma}=-\frac{v'}{2 m_{E_1}+v' y_h^{E_L}}\bigg\{ y_H^{E_L} \left[A_{1/2}^A (\tau_{E_1})-A_{1/2}^A (\tau_{E_2})\right] \nonumber\\
    & -  y_H^{E_R}\left[\frac{m_{E_1}+v' y_h^{E_L}}{m_{E_1}}A_{1/2}^A(\tau_{E_1})-\frac{m_{E_1}}{m_{E_1}+v' y_h^{E_L}}A_{1/2}^A(\tau_{E_2}) \right]\bigg\}
\end{align}
The strongest diphoton signal is obtained when the decay amplitude of the CP-odd scalar is maximized. This is achieved for $y_H^{E_L}=-y_H^{E_R}=y_L$. By redefining $y_h^{E_L}=y_l$ the expressions above simplify to:
\begin{align}
     & \mathcal{A}^{\rm VLL}_{H \rightarrow \gamma \gamma}=-\frac{v'^2 y_l y_L}{m_{E_1 }(2 m_{E_1}+v' y_h^{E_L})} \left[A_{1/2}^H (\tau_{E_1})+\frac{m_{E_1}}{2 m_{E_1}+v' y_h^{E_L}}A_{1/2}^H (\tau_{E_2})\right]\nonumber\\ 
    & \mathcal{A}^{\rm VLL}_{A \rightarrow \gamma \gamma}=-\frac{v' y_L}{m_{E_1}} \left[A_{1/2}^A (\tau_{E_1})-\frac{m_{E_1}}{2 m_{E_1}+v' y_h^{E_L}}A_{1/2}^A (\tau_{E_2})\right]
\end{align}

Similarly to above, we provide in  Fig.~\ref{fig:pgammagamma} an illustration of
the collider prospects including constraints from DM phenomenology  before a more
detailed analysis in the next subsection. In the figure, we have confronted with
the most recent limits from ATLAS and CMS,  the predicted cross section for
diphotons, $\sigma(pp \rightarrow A/H \rightarrow \gamma \gamma)$, for the model
points (see next subsections for details on their determination) that provide a
viable DM candidate with a  correct relic density and evading present constraints
from direct searches. 

We have distinguished between the different scenarios described in the previous
subsection, identified by the type of interactions with the fermions and by the
value of $\tan\beta$. As one can see, the most promising scenarios are the ones
corresponding to low $\tan\beta$ and to $\tan\beta \sim 50$ for the flipped
2HDM. These scenarios correspond, indeed, to the configurations which maximize
the production vertex of the resonance. As already emphasized, for $\tan \beta
\sim 1$, the gluon fusion process is made efficient by the strong coupling with
the top quark, while for $\tan\beta \sim 50$, the production cross section is
enhanced by $b$--quark loop contributions and the $b\bar b$ fusion process. In
the other Type I regime, the cross section quickly drops with the value of
$\tan\beta$.

\begin{figure}[!h]
\begin{center}
\includegraphics[width=0.73\linewidth]{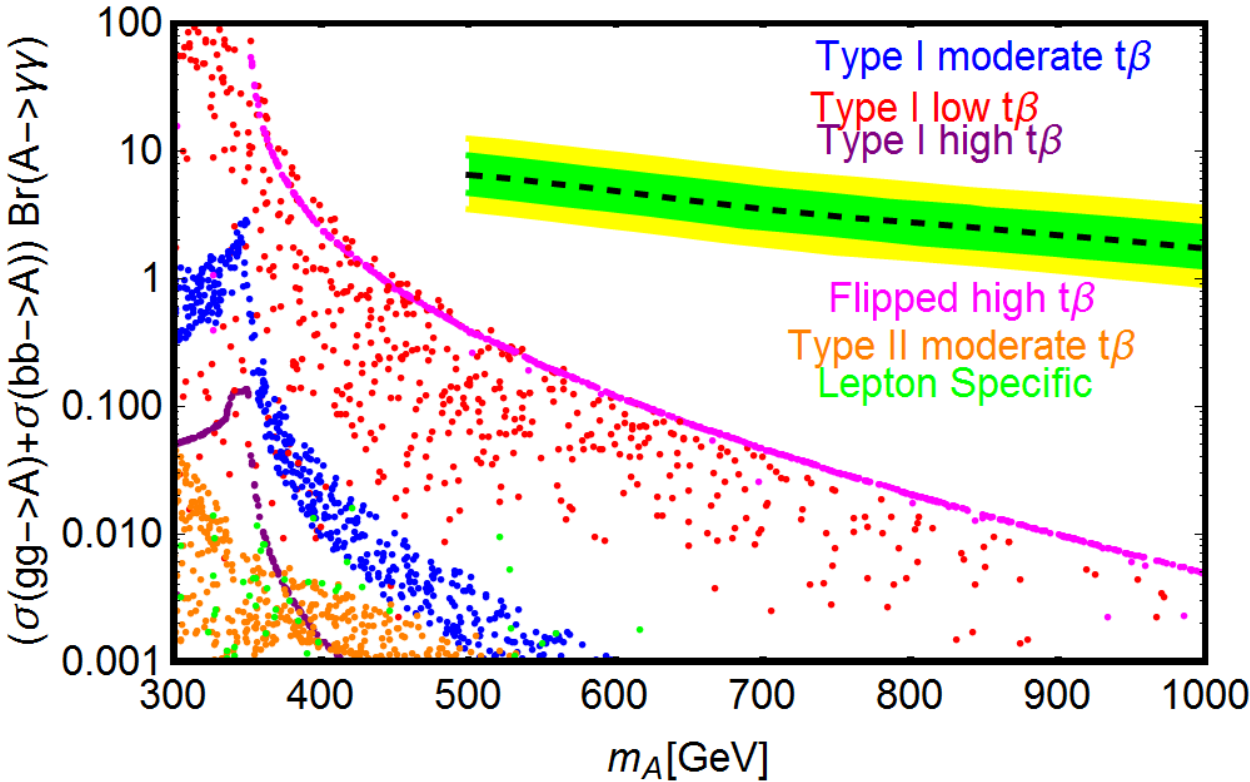}
\end{center}
\vspace*{-2mm}
\caption{Expected diphoton cross sections as function of $M_A$ for model points corresponding to the 2HDM+VLL scenario, obtained through a scan over the relevant parameters. All model points comply with theoretical constraints and correspond to a DM relic density compatible with the experimental value. The point have been divided in subsets, identified by different colors, corresponding to different combinations for the couplings of the SM fermions with the two Higgs doublets and specific ranges for $\tb$. More precisely we have considered three sets of model points with Type--I configurations but different regimes for $\tb$, i.e. low, namely $\tb=1$--5 (red points), moderate, i.e. $\tb=10$--20, (blue points) and high, $\tb> 40$ (purple points). In the high $\tb$ regime, we have also considered as set of model points with flipped couplings between the SM fermions and the Higgs doublet (magenta). The orange and green points represent, finally, model points for the Type--II configuration in the moderate $\tb$ regime and model points for the lepton specific configurations with $2 < \tb < 40$.}
\label{fig:pgammagamma}
\vspace*{-2mm}
\end{figure}

In all the considered regimes, the diphoton cross section lies below the current
experimental sensitivity and quickly drops by several orders of magnitude as the
value of $M_A$ increases. A signal in diphoton events would be hardly
observable, even in future luminosity upgrades, for values $M_A \gtrsim
700\,\mbox{GeV}$. This result is mostly due to the fact that the size of the
Yukawa couplings of the charged vector--like leptons are limited from above by
the requirement of consistency under renormalisation group  evolution and, only
for $y_h^{E_L}$, by electroweak precision data. As a consequence, no significant
enhancement of the diphoton production cross section with respect to the 2HDM
without vector--like leptons, is actually allowed\footnote{We ignore here the
possibility of a pseudoscalar resonance with a mass at exactly the
vector--lepton threshold, $M_A=2m_{\rm VLL}$, where a strong enhancement can
occur when the total resonance width is very small, as a result of Coulombic
contributions \cite{Bharucha:2016jyr}.}  We notice, in addition, that in order
to comply with limits from DM phenomenology to be discussed shortly, the
vector--like leptons should be typically heavier than the diphoton resonance.
This translates into a further suppression of the vector lepton triangle loop
contributions and, hence, a lower production rate.

Let us discuss now the case of the inert doublet model. The presence of the $\mathbb{Z}_2$ parity, which ensures the stability of the DM, forbids couplings of the DM itself and its bosonic partners with SM fermions. Consequently, the DM state and its bosonic partners can only be produced in
pairs through the exchange of a Higgs or a gauge boson. The main processes are in fact the Drell--Yan ones \cite{Drell:1970wh}:
\beq 
q\bar q \to \gamma, Z^* \to H^+ H^- , \ 
q\bar q \to Z^* \to H A , \
q'\bar q \to W^* \to H H^\pm, A H^\pm . 
\eeq
The couplings of two scalar states with $Z,W$ bosons are the same as those of the heavy Higgs bosons of the 2HDM in the alignment limit which are given in 
eqs.~(\ref{eq:cplg-HHV1})--(\ref{eq:cplg-HHV2}) with $\sin(\beta-\alpha) \to 1$.
The cross sections are thus simply those of the 2HDM processes $q\bar q \to HA$
shown in Fig.~\ref{Fig:main-pp} and $q\bar q \to HH^\pm, AH^\pm, H^\pm H^\mp$
shown in Fig.~\ref{Fig:H+-pp}, when the assumption $M_H=M_A=M_{H^\pm}$ is made. 
In all these cases, the cross sections are not that large being at the 1--10 fb
level for scalar masses in the 200 GeV range and much below for higher masses.
If the additional scalars are close in mass,  the heavier $A$ and $H^\pm$ states will decay into the lighter $H$ (our DM particle) and an off--shell gauge boson that decays into two almost massless fermions, $A\to HZ^* \to Hf\bar f$ and  $H^\pm \to HW^* \to H f\bar f'$. 

Although the IDM has not been specifically searched for at the LHC, the main
signature of the model, namely missing transverse energy together with
multi--jets and/or multi--leptons that are rather soft, is similar to the ones
searched for in other scenarios such as supersymmetric models which will be
discussed in the next section. One can thus adapt for this special case searches
made in the MSSM for higgsino--like charginos and neutralinos, namely  when the
masses of the lightest chargino $\chi_1^\pm$,  the next--to--lightest neutralino
$\chi_2^0$ and the lightest neutralino  $\chi_1^0$ (which is the stable DM state
here) are close to each other
\cite{Aaboud:2017mpt,ATL-PHYS-PUB-2017-019,Rossini:2018eai}. Indeed, the
signatures  resemble those of the IDM with a compressed $H^\pm, A$ and $H$
spectrum: besides the usual multi--lepton (or multi--jet) and missing energy
topology for a compressed spectrum, there is also the search for a disappearing
track accompanied by at least one jet with high transverse momentum from
initial--state radiation. 

\begin{figure}[!ht]
\vspace*{-.2mm}
\begin{center}
\includegraphics[width=0.66\linewidth]{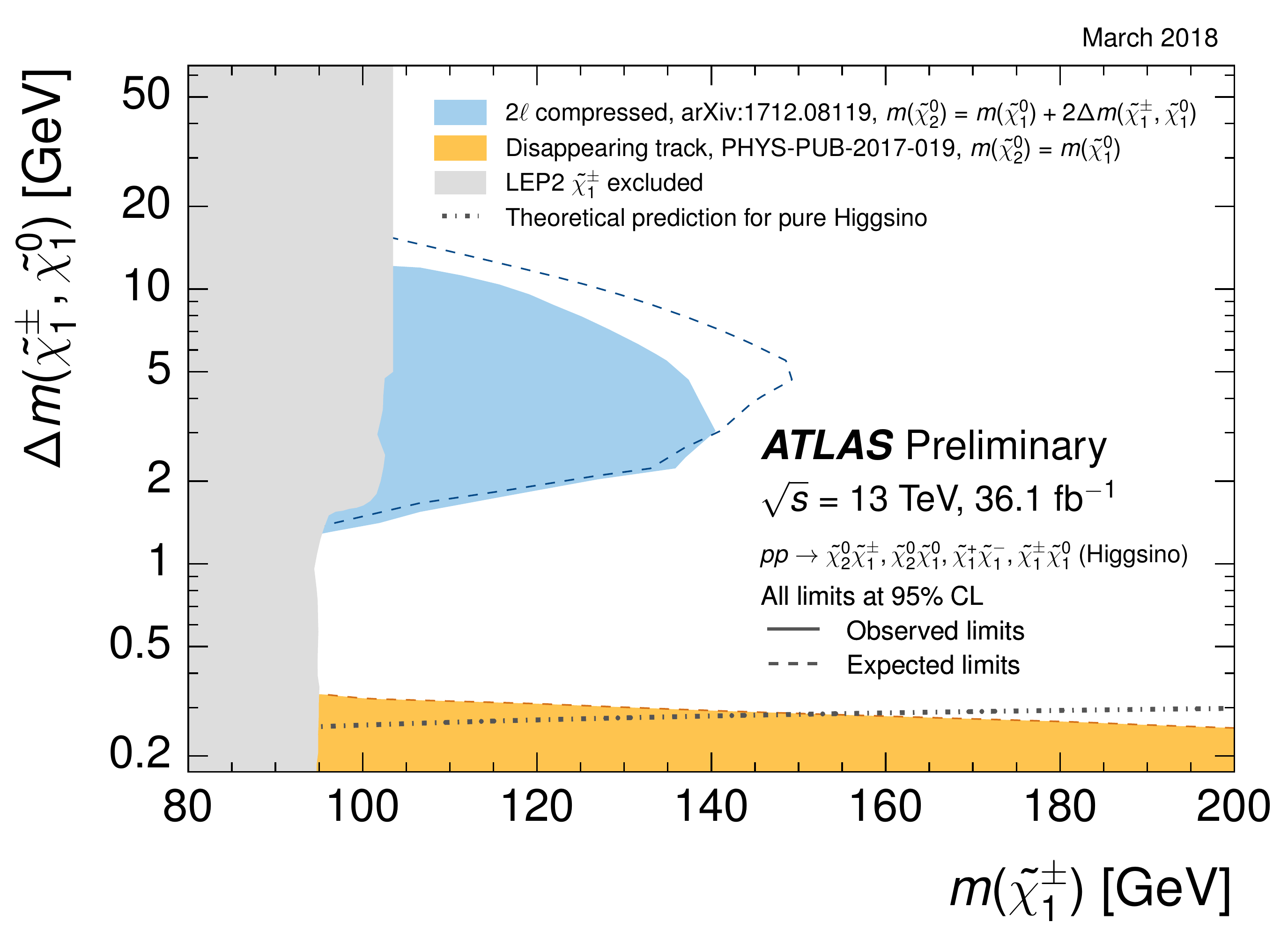}
\end{center}
\vspace*{-5mm}
\caption{95\%CL limit on the higgsinos of the MSSM from searches of a chargino that is nearly mass degenerate with the stable neutralino in an ATLAS analysis with 36 fb$^{-1}$ data collected at $\sqrt s= 13$ TeV \cite{Rossini:2018eai}; the LEP2 limit has been superimposed to the two exclusion domains from soft--leptons and a disappearing tack.}
\label{fig:ATLAS-higgsino}
\vspace*{-2mm}
\end{figure}

The constraints from such searches in the MSSM are exemplified in
Fig.~\ref{fig:ATLAS-higgsino} where an ATLAS analysis at Run II with 36
fb$^{-1}$ data is shown in the plane formed by the mass of the chargino versus
its mass difference with the stable lightest neutralino. The 95\%CL exclusion
limits are  set in the two searches above and the limits from LEP2 searches on
charginos, which give $m_{\chi_1^\pm}>95$--105 GeV depending if the mass
difference  $m_{\chi_1^\pm}-m_{\chi_1^0}$ is smaller or larger than about 1 GeV,
are superimposed to them.  

These limits have to be interpreted in the IDM\footnote{At least two main
differences have to be kept in mind when making such an interpretation. First,
the higgsinos are spin--$\frac12$ particles while we have scalars in the IDM,
and for instance the cross section  $\sigma(q\bar q \to \chi_1^+ \chi_1^-)$ is a
factor of 4 larger than $\sigma(q\bar q \to H^+H^-)$  when the couplings and the
masses of the two types of particles are the same. In addition, one has to take
into account a possible difference in their couplings to the weak bosons, in
particular for the electrically neutral states.} and this has been done for
instance in
Refs.~\cite{Dolle:2009ft,Miao:2010rg,Gustafsson:2012aj,Belanger:2015kga,Ilnicka:2015jba},
where mass values  $M_H \lesssim 40\,\mbox{GeV}$ and $M_A \lesssim
140\,\mbox{GeV}$ have been excluded. The IDM parameter space will be probed more
efficiently at the high luminosity upgrade of the LHC by eventually
complementing searches of multilepton events with searches of events with dijet
and missing energy and or with 2 jets, 2 leptons and missing
energy~\cite{Poulose:2016lvz,Datta:2016nfz,Hashemi:2016wup,Dutta:2017lny}.
Nevertheless, dedicated ATLAS and CMS analyses of these signatures in the case
of the present model would be welcome for the current LHC run.

In addition, for a light DM particle  with a mass $M_H \lsim 62$ GeV, the
invisible decay $h \to  HH$ of the 125 GeV SM--like Higgs boson produced
either through gluon or vector boson fusion, should occur and impose some
constraints. The first $h$ production processes can be probed by looking at
events with missing energy produced in association with a mono--jet, while in
the case of VBF, the DM is produced in association with two jets. The mono--jet
signature has been studied in particular in
Refs.~\cite{Belyaev:2016lok,Belyaev:2018ext}  where it has been found that the
regions of parameter space corresponding to $M_H \lesssim \frac12 M_h$ can be
probed at the HL--LHC when a luminosity $\mathcal{L}=3\,{\mbox{ab}}^{-1}$ will
be collected. Ref.~\cite{Dercks:2018wch} considered, instead, the higher
$\frac12 M_h \lesssim M_H \lesssim 100\,\mbox{GeV}$ mass range where the best
constraint is offered by searches in the VBF process which is currently
sensitive only to large values of the  coupling $\lambda_{345}=\lambda_3+
\lambda_4 +\lambda_5$, namely between 1 and 10.

In addition to these direct searches, and similarly to the case of the ordinary 2HDM, the presence of a charged state in the Higgs sector might alter in a detectable way the signal strength of the SM--like $h$ boson into diphotons.  Indeed, the $h\to \gamma \gamma$ decay rate can be written in this case as~\cite{Arhrib:2012ia,Swiezewska:2012eh}:
\begin{equation}
    \Gamma (h \rightarrow \gamma \gamma)\big|_{\rm IDM}=\frac{G_F \alpha^2 M_h^3}{128 \sqrt{2}\pi^3} \big|\mathcal{A}_{h \rightarrow \gamma \gamma}^{\rm SM}+\mathcal{A}_{h \rightarrow \gamma \gamma}^{\rm IDM} \big|^2 \, , 
\end{equation}
where the SM contribution has been discussed before, while that of the $H^\pm$ 
is given by
\begin{equation}
\mathcal{A}_{h \rightarrow \gamma \gamma}^{\rm IDM} =   \frac{\lambda_3 v^2}{2 M_{H^{\pm}}^2}A_0^h \left(\frac{4 M_{H^{\pm}}^2}{M_W^2}\right) . 
\end{equation}
By requiring that this signal strength does not conflict with the experimental measurement, it is possible to obtain bounds on the $M_{H^{\pm}}$ and $\lambda_3$ parameters of the IDM~\cite{Ilnicka:2015jba}.

Let us finally  note that the IDM scenario can be best probed at future $e^+e^-$
colliders and, in fact, it has been used as a benchmark to highlight the
capabilities of these colliders (in particular the  precise knowledge of the
beam energy) in probing missing energy and possibly  soft multi--fermion
signatures \cite{Kalinowski:2018ylg}.  A first handle could be the recoiling $Z$
boson in the $e^+e^- \to  HZ \to E_T^{\rm mis}+\ell\ell$ process but in the
worst case, one can consider the $e^+e^- \to H^+H^-$ process which has a large 
rate and, even if  the $H^\pm$ states are quasi--stable,  it allows for the
radiation of an additional photon from the initial state. 

We close this subsection by illustrating collider limits and prospects for the
2HDM+$a$ model. It has attracted interest only in rather recent times and,
consequently, its collider phenomenology has not been fully explored yet. A
series of potentially interesting signatures, mostly connected to the study of
the DM sector, has been proposed by the LHC Dark Matter working group in
Ref.~\cite{Abe:2018bpo}. These are mostly mono--X signatures, i.e. the
associated production of the pseudoscalar boson which then decays into a pair of
DM states. More precisely these are $E_T^{\rm mis}+h$, $Z$ or $W$ bosons (in
this cases the pseudoscalar might be produced in decays of resonantly produced
$H,A,H^{\pm}$ states), mono--jets and associated production of the pseudoscalar
with two heavy flavors (in this cases the $a$ state is produced through gluon
fusion). Together with these missing energy signatures, the four top signature,
i.e $pp \rightarrow a \bar t t \rightarrow \bar t t \bar t t$,  is considered as
well for this model.  A first study of these collider processes has been
conducted by the ATLAS collaboration \cite{ATLAS:2018vvx}. 

\begin{figure}[!ht]
\vspace*{-.1mm}
\centerline{\hspace*{1cm} \includegraphics[scale=0.65]{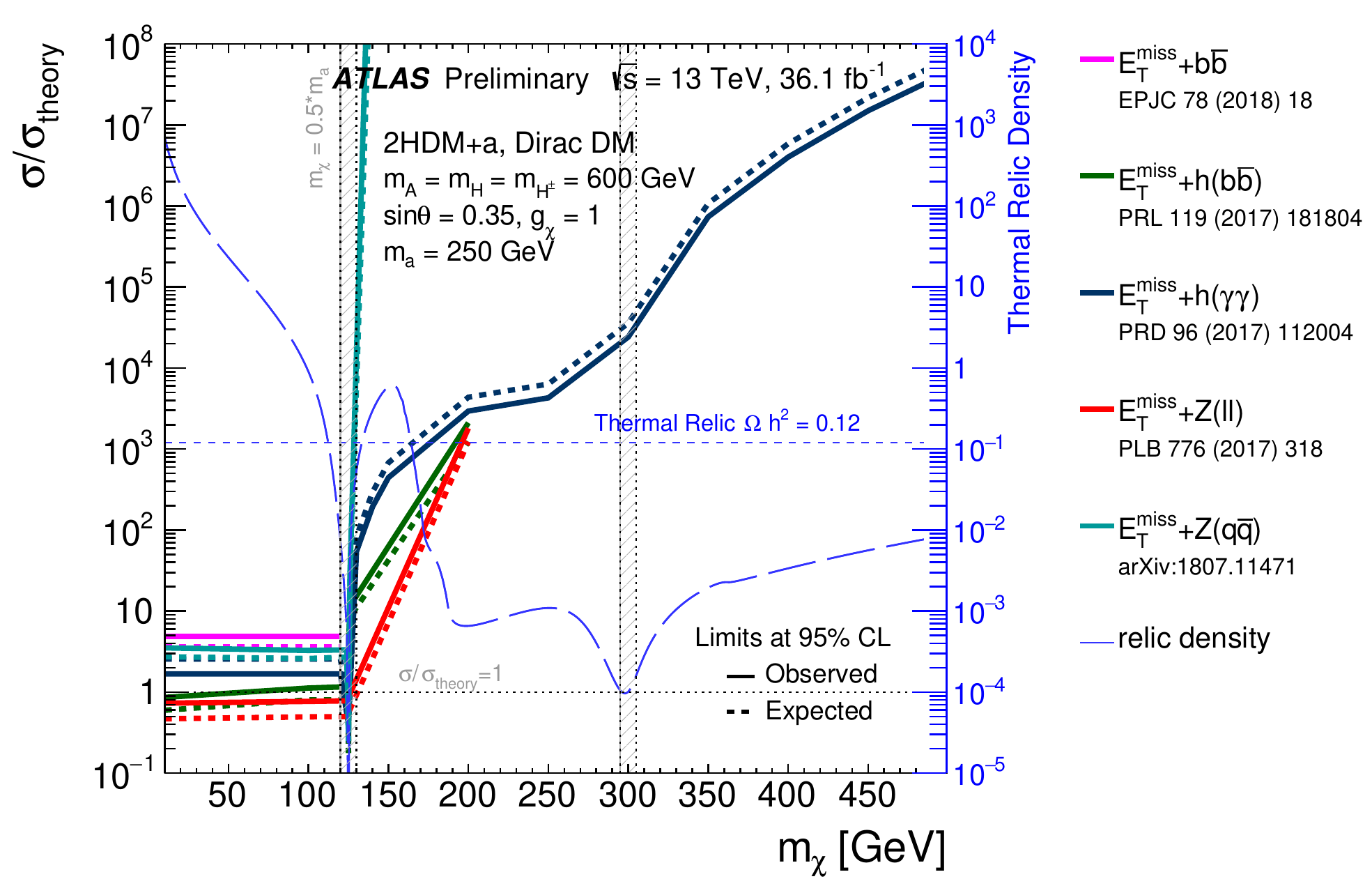}  }
\vspace*{.1cm}
\caption{95\% CL observed and expected exclusion limits for the 2HDM+$a$ model as a function of the DM mass $m_\chi$, expressed in terms of the ratio of the excluded cross section to the nominal cross section of the model. The relic density is superimposed (long-dashed blue line) and described by the right y--axis. The analysis has been done by ATLAS at $\sqrt s=13$ TeV with 36 fb$^{-1}$ data \cite{ATLAS:2018vvx}. }
\label{Fig:H2port-ATLAS}
\vspace*{-1mm}
\end{figure}

As an illustration, we show in Fig.~\ref{Fig:H2port-ATLAS}, taken from
Ref.~\cite{ATLAS:2018vvx}, the 95\,\%CL observed and exclusion limits for the
Type II 2HDM+$a$ as a function of the mass of the DM particle $m_\chi$, with the
following  choice of model parameters: $M_H\!=\!M_{H^\pm}\!=\!M_A\!=\!600$ GeV,
$\tb\!=\!1$,  $M_a\!=\!200$ GeV, $\sin\theta\!=\!0.35$ and a unit $g_\chi$
coupling. The limits are obtained in searches for transverse missing energy that
comes with a Higgs boson decaying into $b\bar b$  and $\gamma\gamma$ final
states or a $Z$ boson decaying into lepton or quark pairs; they are expressed in
terms of the ratio of the excluded cross section to the nominal one of the
model. The relic density for each DM mass is shown by the long--dashed blue line
and is described by the right y--axis. For DM masses for which the relic density
line is below $\Omega h^2 = 0.12$, the model depletes it below the thermal
value. The two valleys at $m_\chi = 125$ GeV and $m_\chi = 300$ GeV are due to
the two $a$-- and $A$--pole regions where the predicted $\Omega h^2$ is obtained
by the annihilation processes $\chi \chi \to A/a \to {\rm SM}$ particles. The
plateau in the area $m_\chi \approx 200$ GeV is due to the increase of the DM
annihilation rate close to the $ha$ and $t\bar t$ thresholds. For $m_\chi \gsim
\frac12 M_a \approx 125$ GeV, all parameters that lead to the correct density
are not constrained by the search.

A very interesting possibility for the 2HDM+$a$ model consists into a light
pseudoscalar $a$, namely $M_a < M_h$. Searches for production and decay for this kind of light state has been already performed at LEP. To mention a few, associated production of the $a$ state with  $b\bar b$ and $\tau^+\tau^-$ pairs in $Z$ decays at LEP1 should constrain the  $a b\bar b$ and $a \tau^+ \tau^-$ couplings to be extremely tiny (smaller than those of the SM Higgs boson  since $Z \to bb h$ and $Z \to h \tau^+ \tau^-$ topologies with a light SM--like $h$ boson have been searched for with no success~\cite{Djouadi:1990ms,Barate:2003sz}).  Also at LEP1, couplings of the $a$ state with gauge bosons through loops of new particles should be severely constrained by searches of the $Z \to a \gamma$ exotic decay \cite{Barate:2003sz}. Additional limits on the $Zha$ coupling come from searches in the process
$e^+ e^-  \to ha$ at LEP2. 
One should also consider limits from $B$--meson physics in the case of a non--negligible $a b\bar b$ couplings in particular for the $B_s \to \mu^+\mu^-$ decay (more details will be provided in the astroparticle section).

Concerning LHC, the most important probes are represented by the $h \rightarrow aa$ and $h \rightarrow Za$
processes~\cite{Goncalves:2016iyg,Haisch:2018kqx}. The $h \rightarrow aa$ decay
is already extensively searched for at the LHC by looking at the
$4\mu$~\cite{Khachatryan:2015wka,CMS-PAS-HIG-16-035},
$4\tau$~\cite{Khachatryan:2015nba,Khachatryan:2017mnf}, $2\mu
2\tau$~\cite{Sirunyan:2018mbx}, $2\mu 2 b$~\cite{Sirunyan:2018pzn} , $2\tau
2b$~\cite{Aad:2015sva} final states. The corresponding limits have been
interpreted in some realizations of the 2HDM+$a$ model, for example
in Ref.~\cite{Haisch:2018kqx}, and will be discussed in more detail in the
astroparticle section. No dedicated searches of the $h \rightarrow Za$ have been
yet performed by the LHC collaborations. Exclusion limits have been nevertheless
derived in Ref.~\cite{Haisch:2018kqx} by reinterpreting the result of searches
of light spin--1 bosons, i.e. $pp \rightarrow h \rightarrow Z_d Z$, into four
leptons. We postpone again a more detailed discussion to the astroparticle part
to which we turn now.    

\subsection{Astroparticle physics implications}

\subsubsection{The singlet--doublet lepton case}

Since the singlet--doublet lepton model is a direct extension of the one
presented in section 3, most of the discussions carried out there are valid also
in this case. We will therefore simply  highlight the additional
phenomenological features associated to the extension of the Higgs sector to two
doublets. 

For what concerns direct detection, the spin--independent cross section receives an additional contribution from the $t$--channel exchange of the heavy CP--even $H$ state, which will be then given by
\begin{equation}
\sigma_{\chi p}^{\rm SI}=\frac{\mu_\chi^2}{\pi}\frac{m_p^2}{v^2}\bigg| \sum_{q}f_q \left(\frac{y_{hN_1 N_1}g_{hqq}}{M_h^2}+\frac{y_{HN_1 N_1}g_{Hqq}}{M_H^2}\right) \bigg|^2 \, .
\end{equation}

As already mentioned, the pseudoscalar $A$ state can also contribute to the
spin--independent cross section, but at the one--loop level only; this 
contribution is however negligible in the setup considered here and we will
simply ignore it. 

The spin--dependent cross section is instead left unchanged with respect to the conventional singlet--doublet model, being again given by
\begin{equation}
\label{eq:SDcross}
\sigma_{\chi p}^{\rm SD}=\frac{3 \mu_{\chi p}^2}{\pi M_Z^4} \bigg| g^A_{Z N_1 N_1} \bigg|^2 \bigg[ g_u^A \Delta_u^p+g_d^A \left(\Delta_d^p+\Delta_s^p \right) \bigg]^2 \, .
\end{equation}
As can be seen from eqs.~(\ref{eq:coupling_uu})--(\ref{eq:coupling_dd}), the spin--independent cross section features again a blind spot when the condition $M_N+M_L \sin 2 \theta \simeq 0$ is met, since in this case, it corresponds to $y_{hN_1 N_1}=y_{HN_1 N_1}=0$ for both the chosen coupling configurations. We note that blind spots would be also present when the new fermions couple selectively to different Higgs doublets. However,  in such a configuration, their occurrence depends on the model parameters in a less trivial way.

For what concerns the DM relic density, the largest impact  will be due to the pseudoscalar $A$ state since, as shown in Appendix~B, it provides an additional $s$--wave contribution to the DM annihilation cross section into SM fermion final states. As can be easily argued, this additional contribution becomes important in the case of a very light $A$ boson and/or in a scenario like the Type II 2HDM, in which the $A$ coupling with some of the SM fermions is enhanced by a $\tan\beta$ factor and/or, finally, at the pole $m_{N_1} \sim \frac12 M_A$. The largest impact on the DM relic density from the Higgs sector occurs, however, when one of the extra Higgs bosons is lighter than the DM particle which implies the presence of additional annihilation channels for the latter.

As seen previously, there are many different variants of the singlet--doublet
model coupled to a 2HDM, since it is possible to chose and combine different
configurations for the couplings of the new fermions with the Higgs doublet
fields $\Phi_{1}$ and $\Phi_2$ as well as for the couplings of the latter with
the SM fermions. We will therefore focus on two phenomenologically interesting
configurations. The first configuration is simply our benchmark scenario with an
aligned 2HDM and mass degenerate heavy $H,A$ and $H^\pm$ bosons which have then
couplings to SM fermions that are either proportional or inversely proportional
to $\tb$. This benchmark model is hence defined by six parameters, namely, $M_N,
M_L,y,  \tan\theta, M_A,\tan\beta$ (this is comparable to the MSSM case to be
discussed later).

\begin{figure}[!h]
\vspace*{-2mm}
    \centering
    \subfloat{\includegraphics[width=0.47\linewidth]{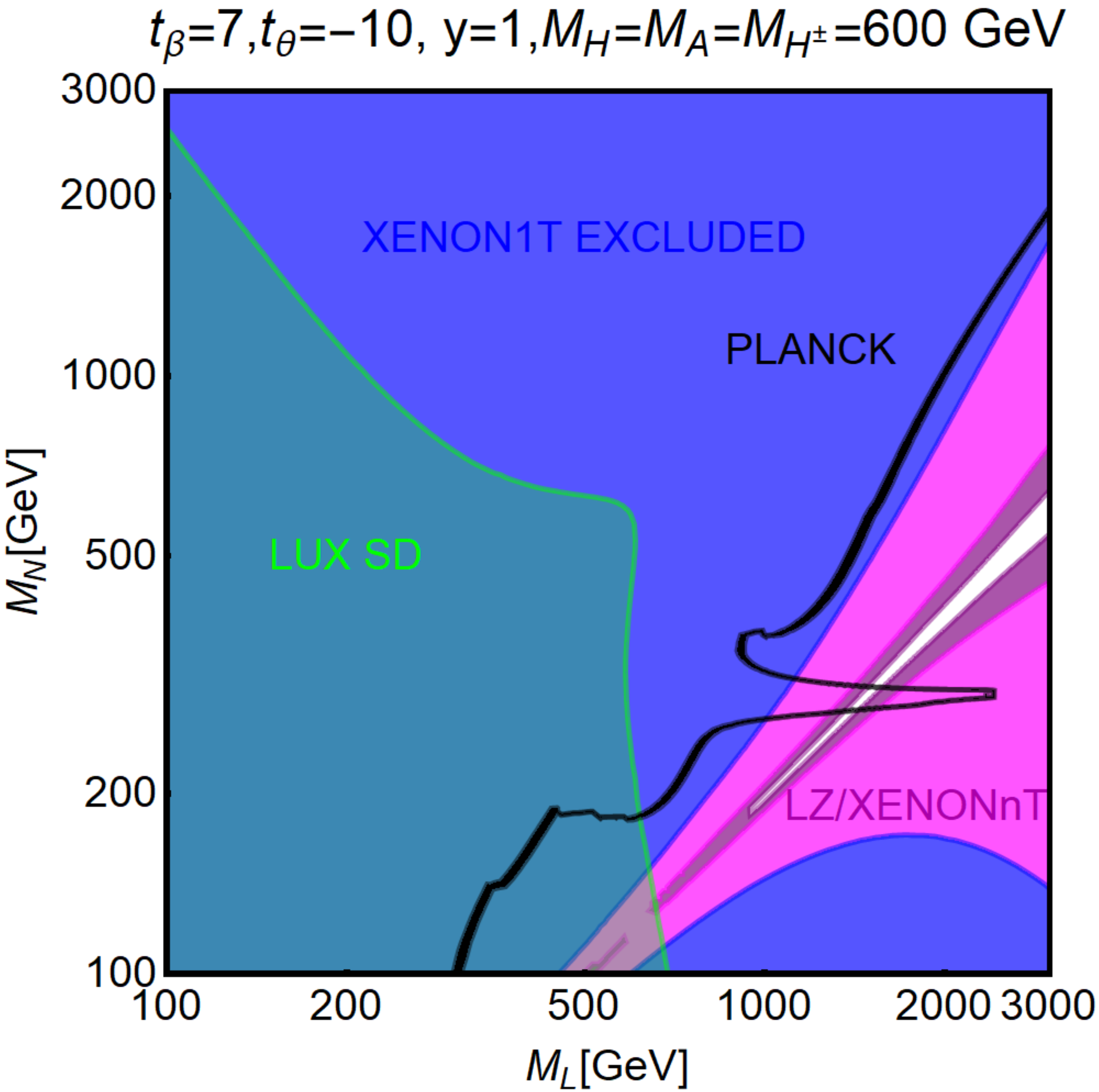}}~~
    \subfloat{\includegraphics[width=0.47\linewidth]{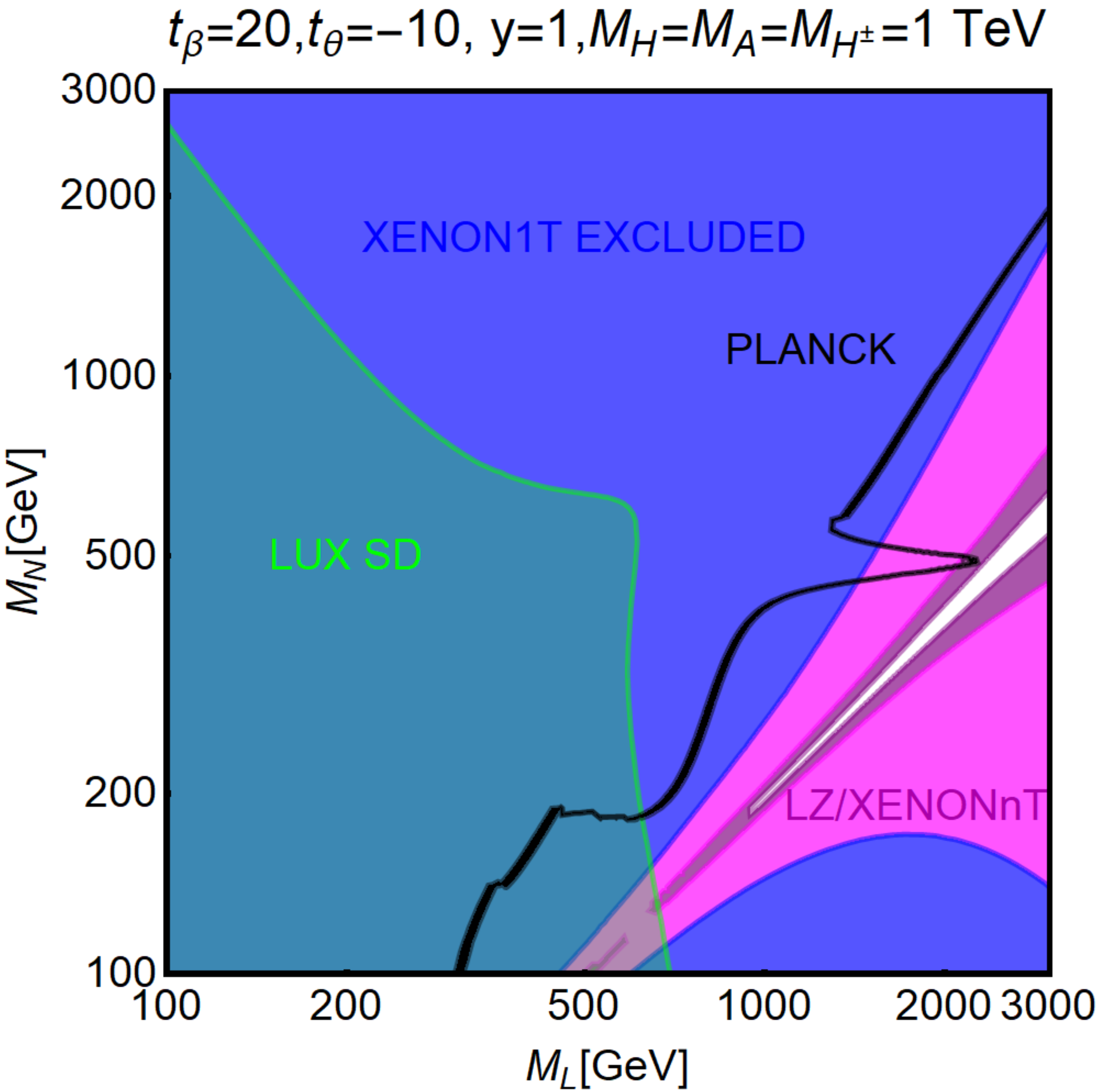}}
    \caption{Summary of DM constraints on the singlet--doublet DM model coupled to  an aligned  2HDM  with mass degenerate Higgs bosons and couplings with the SM fermions according to the Type II model. The black contours are when the correct DM relic density is achieved. The blue/magenta/purple regions represent the current/projected limits/sensitivities from direct searches of spin--independent DM interactions while the green region is excluded  by searches of spin--dependent interactions. The two plots differ by the assignment of the $(M_A,\tan\beta)$ pair as shown on top of the figures.}
    \label{fig:SDhMSSM}
\vspace*{-3mm}
\end{figure}

The combined constraints for this scenario are shown in Fig.~\ref{fig:SDhMSSM}
in the $[M_L,M_N]$ plane:  the constraint of a DM state with the correct relic
density is shown by the black isocontours, the current limits from 
spin--independent and spin--dependent DM interactions are shown in,
respectively, the blue  and green regions, while the projected sensitivities to 
spin--independent interactions from next generation experiments are shown by
the  magenta and purple regions. We have considered two high $M_A=M_H=M_{H^\pm}$
values, namely 600 GeV and 1 TeV, to comply with the constraints on
$M_{H^{\pm}}$ from $b \rightarrow s$ transitions. The values of $\tan\beta$  for
a given $M_A$ input have been chosen  in such a way that they are close to the
sensitivity of present collider searches, as can be seen in
Fig.~\ref{fig:pSD_mA_tbeta}. We have finally set $y=1$ and $\tan\theta$ to a
large negative value, $\tan\theta=-10$, to achieve blind spots for the Higgs
couplings to the DM particles. 

As can be seen from the figure, the output is not very different from what was
shown already in section 3. This is a mere consequence of the fact that the new
scalar sector is forced to be heavy by the constraints on the extra Higgs bosons
from flavor physics and LHC searches. The most relevant effect on the DM relic
density is the presence of the pole $m_{N_1} \sim \frac12 M_A$, which
corresponds to ``cusps'' in the plots, which allows to evade the constraints
from direct detection, thanks to the presence of the blind spot in the same
region of parameter space. The $s$--channel resonance regions will nevertheless
be fully probed by the next generation of direct detection experiments. 

The second benchmark scenario that we will consider features a  light
pseudoscalar, namely $M_A < M_h$. In order to be viable, this scenario requires
a sizable mass splitting between the pseudoscalar and the charged or CP--even
neutral Higgs bosons. This, in turn, requires a significant deviation from the
alignment limit which can be realized only in the Type I 2HDM. On the other
hand, the pseudoscalar Higgs boson cannot be lighter than 60 GeV so as to avoid
the decay $h \rightarrow AA$ and to comply with the fits of the SM--like Higgs
signal strengths. While it will not affect to a large extent DM direct
detection, a light pseudoscalar would alter the relic density since it  provides
additional annihilation channels to the DM particle, namely into $ZA$, $hA$
and $AA$ final states. Approximations of the corresponding cross sections are
again provided in Appendix~B. 

\begin{figure}[!ht]
    \centering
    \subfloat{\includegraphics[width=0.45\linewidth]{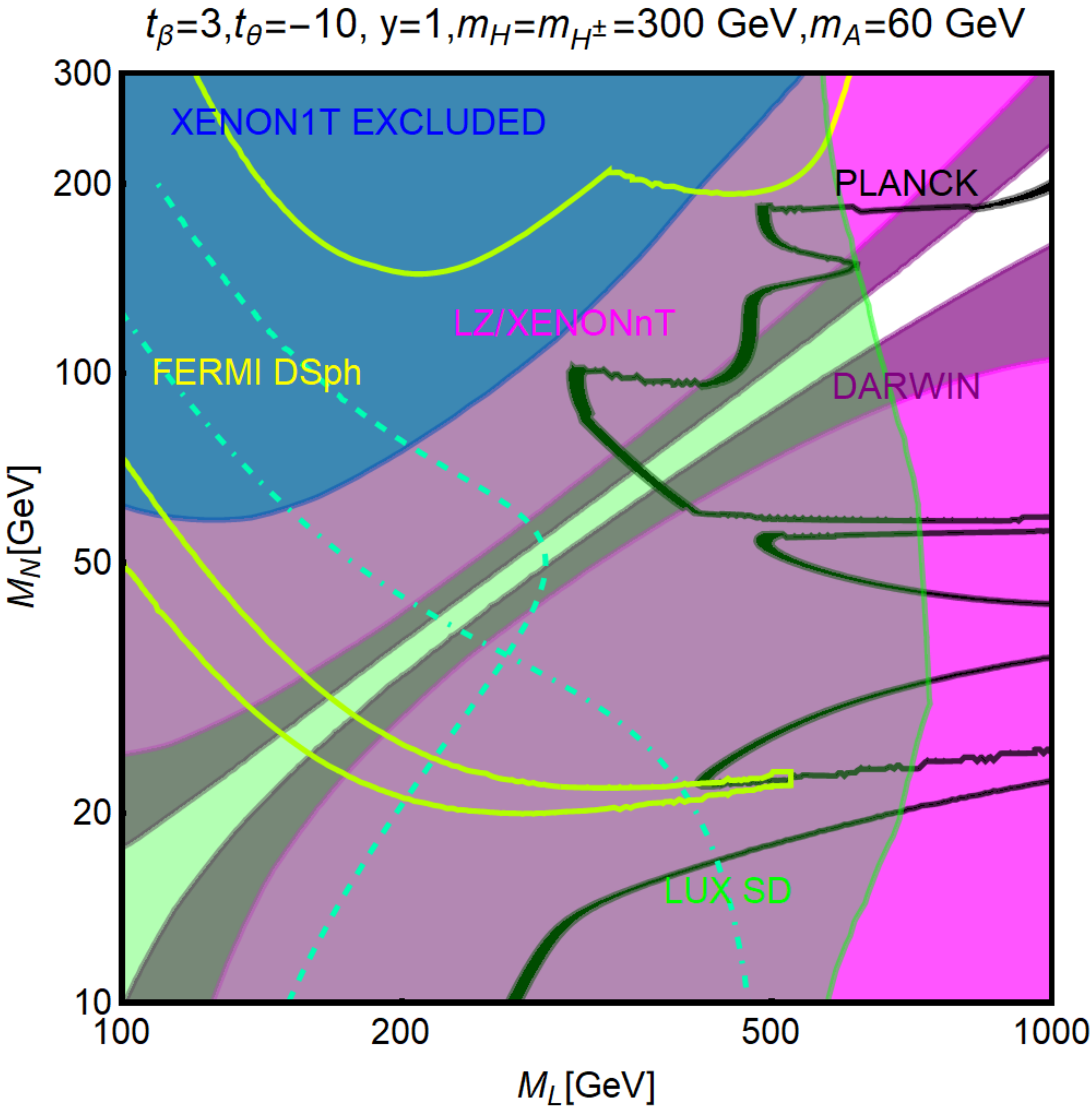}}~~~
    \subfloat{\includegraphics[width=0.45\linewidth]{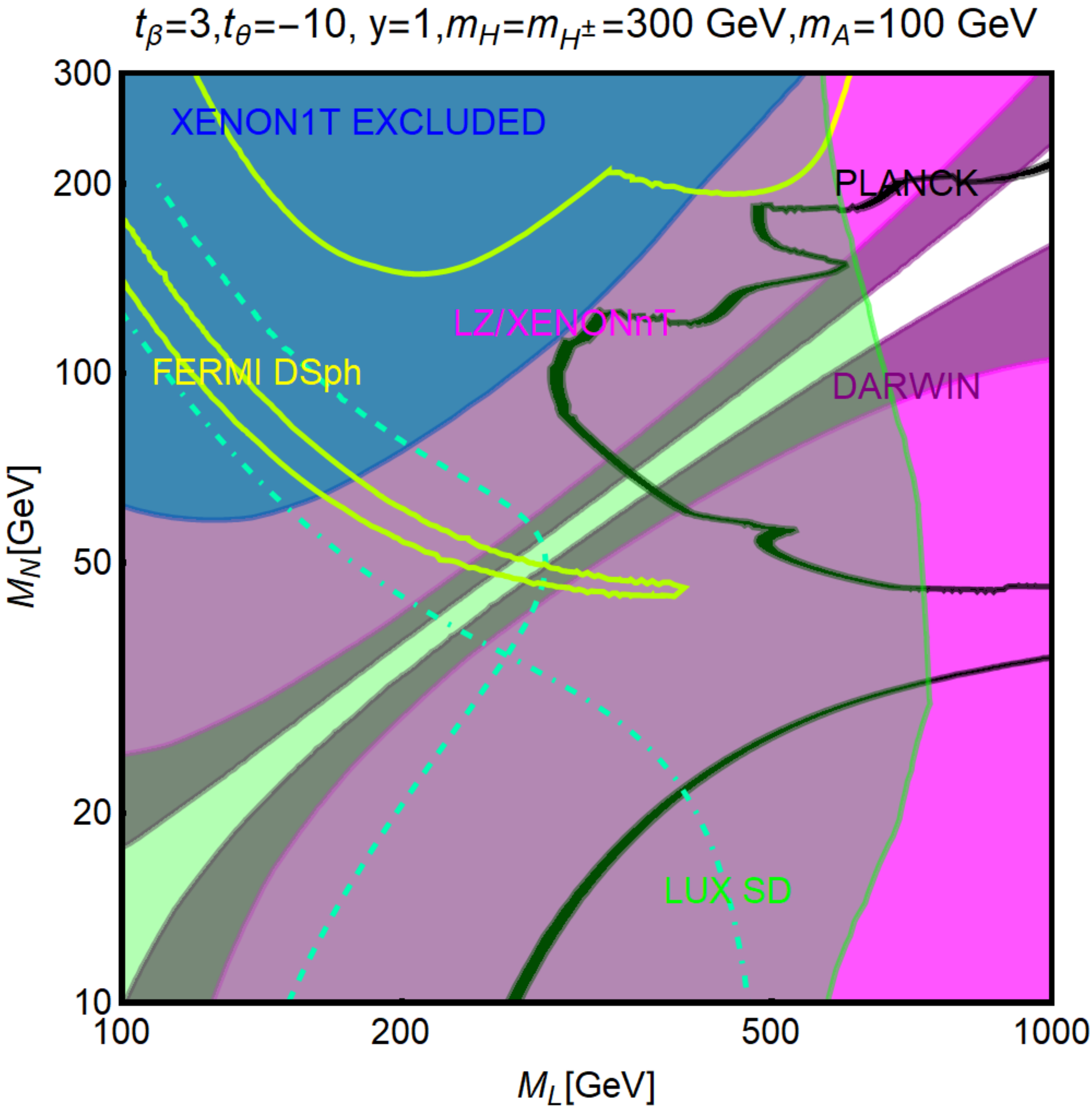}}\\
    \subfloat{\includegraphics[width=0.45\linewidth]{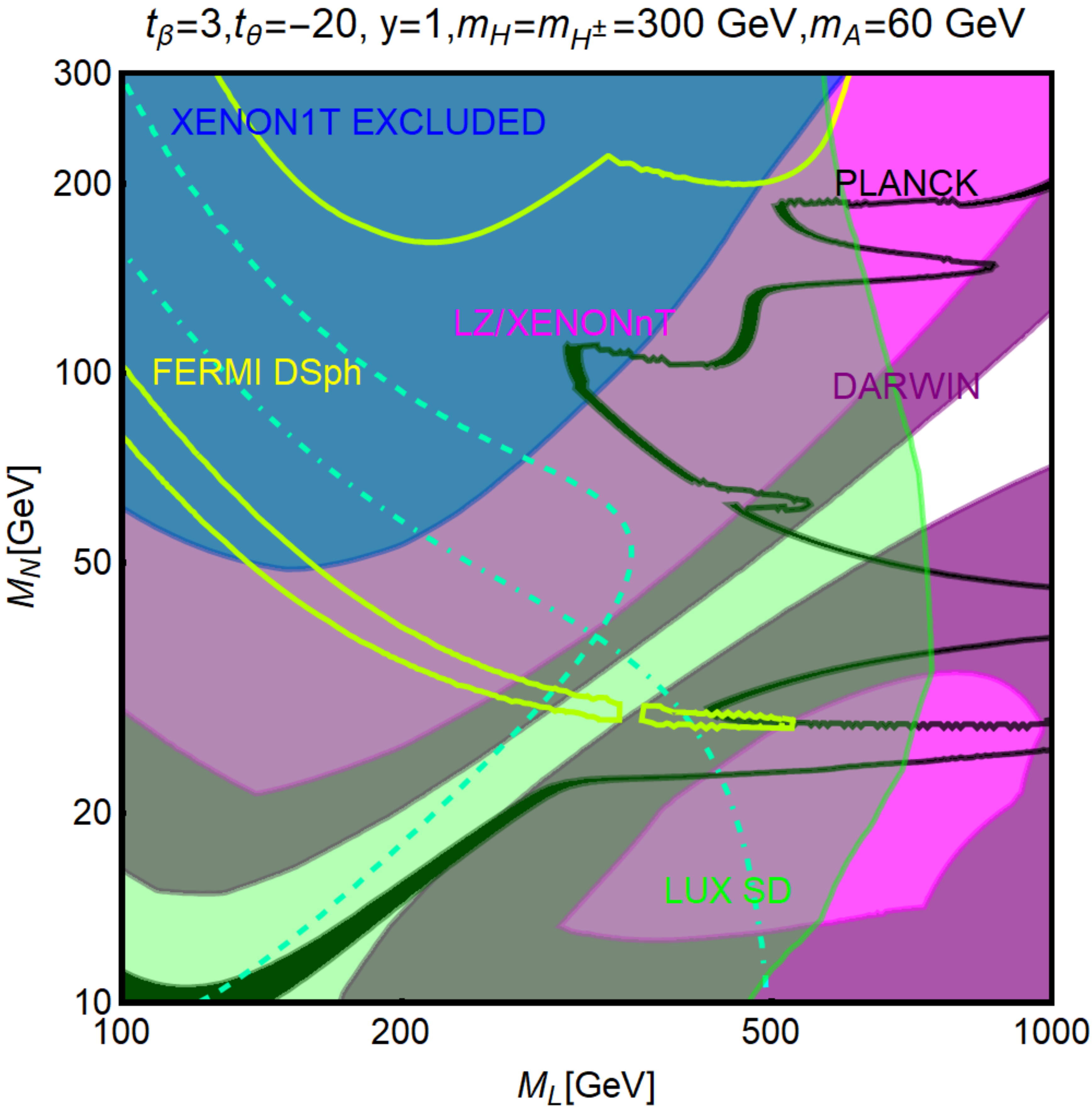}}~~
    \subfloat{\includegraphics[width=0.45\linewidth]{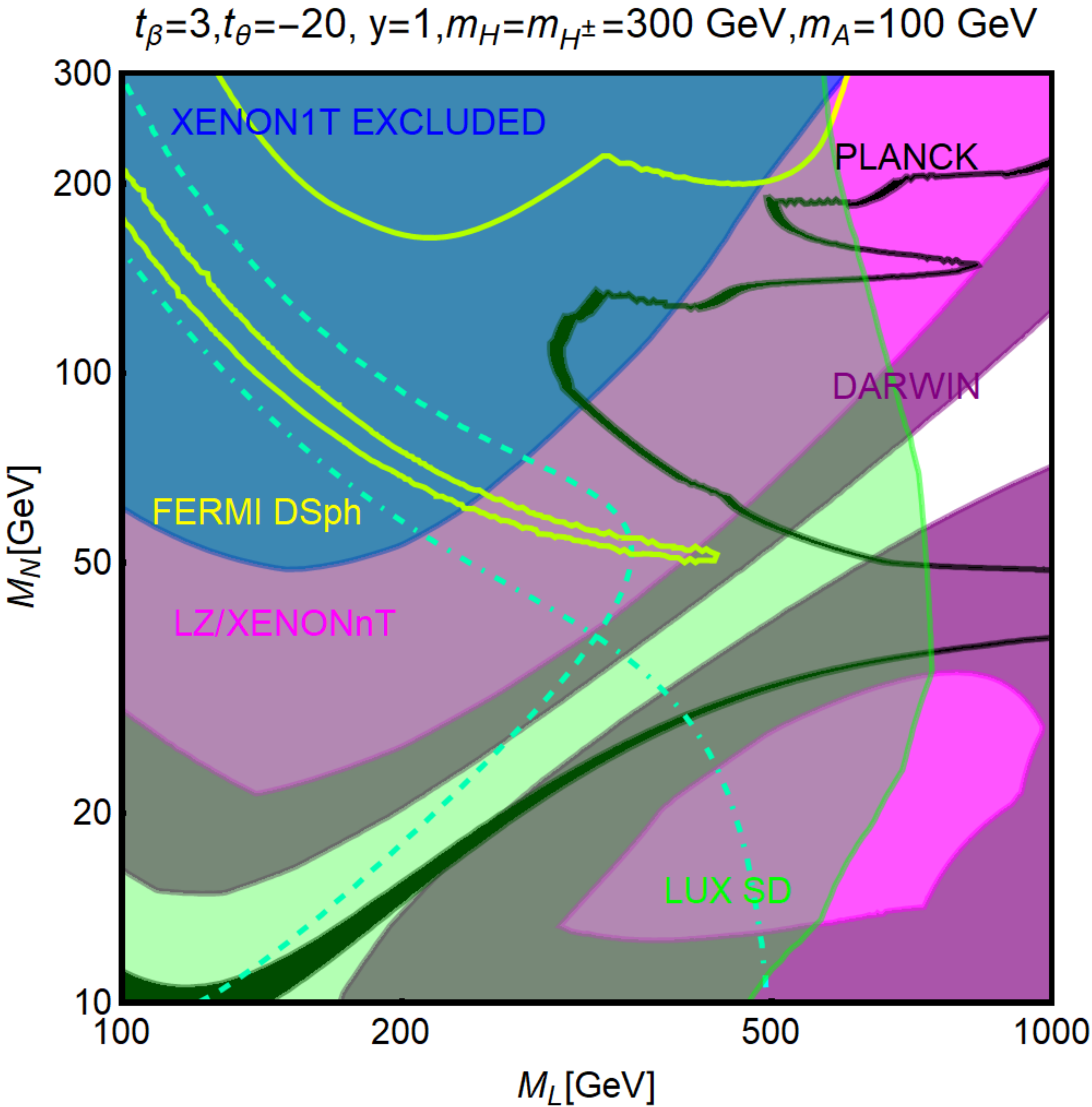}}
    \caption{Combined constraints for the singled--doublet DM model coupled to a 2HDM in the case of a light pseudoscalar; the parameter values are reported on top of the different panels. The plots differ from the assignment of $\tan\theta$, namely $-10$ (upper panels) and $-20$ (lower panels), and of $M_A$, i.e. 60 GeV (left panels) and 100 GeV (right panels). The same color code as in     Fig.~\ref{fig:SDhMSSM} is adopted.}
    \label{fig:pSDlA}
\vspace*{-2mm}
\end{figure}

The different constraints on the model with a light pseudoscalar are shown for
four benchmark scenarios in Fig.~\ref{fig:pSDlA}. In order to emphasize the
impact of the $A$ state on the relic density, as well as the complementarity
with constraints from the invisible widths of the Higgs and the $Z$ bosons, we
have selected a parameter space corresponding to a light mostly singlet--like DM
neutrino by setting the following ranges for the lepton masses $M_N,M_L$:
$10\,\mbox{GeV} < M_N < 300\,\mbox{GeV}$ and $100 \,\mbox{GeV} < M_L < 1
\,\mbox{TeV}$. We have considered two $M_A$ values, �$M_A=60$ and 100 GeV, and
assumed everywhere the low value $\tan\beta=3$ to avoid a too strong suppression
of the $A$  coupling to SM fermions.  As can be seen, the relic density curve
exhibits a rather complex shape due to the presence of multiple cusps
corresponding to the $\frac12 M_Z, \frac12M_A, \frac12 M_h, \frac12 M_H$
possible poles.

\subsubsection{The vector--like fermion family}

Turning to the scenario with a vector--like family that incorporates the DM
particle, we first describe the couplings to gauge and Higgs bosons in a more transparent way.  In the physical mass basis and after electroweak symmetry breaking, the most relevant interactions for our purpose are summarized by the following Lagrangian
\begin{align}
& \mathcal{L}=y_{h N_1 N_1}\bar N_1 N_1 h+y_{H N_1 N_1}\bar N_1 N_1 H+y_{A N_1 N_1}\bar N_1 N_1 A  
+ y_{H^+ N_1 E_1}\bar N_1 E_1 H^{+} +\mbox{h.c.} \\
&+\bar N_1 \gamma^\mu \left(y_{V,Z N_1 N_1}-y_{A,Z N_1 N_1}\gamma_5\right)N_1 Z_\mu 
+ \bar E_1 \gamma^\mu \left(y_{V,W N_1 E_1}-y_{A,W N_1 E_1}\gamma_5\right)N_1 W_\mu^{-}+\mbox{h.c.} \, , \nonumber
\end{align}
with
\begin{align}
\label{fig:DM_high_H}
& y_{H N_1 N_1}\! = \! \frac{\cos \theta_N^L \sin \theta_N^R y_H^{N_L} \!
+\! \cos \theta_N^R \sin \theta_N^L y_H^{N_R}}{\sqrt{2}}, \nonumber\\
& 
y_{A N_1 N_1} \!= \! i\frac{\cos \theta_N^L \sin \theta_N^R y_H^{N_L} \!
- \! \cos \theta_N^R \sin \theta_N^L y_H^{N_R}}{\sqrt{2}}, \\ 
& y_{H^+ N_1 E_1}=\cos \theta_N^L \sin \theta_E^R y_H^{E_L}+\sin \theta_N^L \cos \theta_E^R y_H^{E_R}
 -\cos \theta_N^R \sin \theta_E^L y_H^{N_R}-\cos \theta_N^L \sin \theta_E^R y_H^{N_L}\, . \nonumber \hspace*{-5mm}
\end{align}

In the case of arbitrary interactions of the vector--like leptons with the two Higgs doublets, the couplings $y_{h,H}^{N_{L,R},E_{L,R}}$ are in principle all free and independent. This is not the case if the symmetry $\mathbb{Z}_{\rm 2HDM}$ acts on the vector--leptons. For both model I and model II defined before, the couplings of the electrically neutral heavy leptons are as follows
\begin{equation}
y_h^{N_{L,R}} \equiv y_{N_{L,R}} \sin{\beta}, \quad y_H^{N_{L,R}} \equiv -y_{N_{L,R}} \cos{\beta} = - y_h^{N_{L,R}} \tan^{-1}{\beta},
\end{equation}
whereas for the vector--like electrons, one has
\begin{align}
& y_h^{E_{L,R}} \equiv y_{E_{L,R}} \sin{\beta}, \quad y_H^{E_{L,R}} \equiv -y_{E_{L,R}} \cos{\beta} =  -y_h^{E_{L,R}} \tan^{-1} {\beta} \,\,\,\,\, \mbox{(model I)} , \nonumber\\
& y_h^{E_{L,R}} \equiv y_{E_{L,R}} \cos{\beta}, \quad y_H^{E_{L,R}} \equiv y_{E_{L,R}} \sin{\beta} =  y_h^{E_{L,R}} \tan{\beta} \quad \quad \quad \mbox{(model II)} .
\end{align}

In order to obtain the viable parameter regions for the DM particle, we have
again to compare the constraints arising from the requirement of a correct DM
relic density with the regions excluded by negative DM searches. Concerning the
relic density, most of the considerations made in the previous subsection for
the singlet--doublet DM model are also valid in this scenario (we nevertheless
re--express the most relevant cross section in terms of the parameters of the
model in Appendix~B.).

Let us first discuss the case of arbitrary couplings of the VLLs with the Higgs doublets. Following an analogous strategy as previous sections we will first provide a simple illustration of the combined constraints in a two-parameter space, $(m_{N_1},y_h^{N_L})$ in this case and then perform a more extensive analysis through a parameter scan. 

\begin{figure}[!ht]
\begin{center}
\mbox{\subfloat{\includegraphics[width=0.45\linewidth]{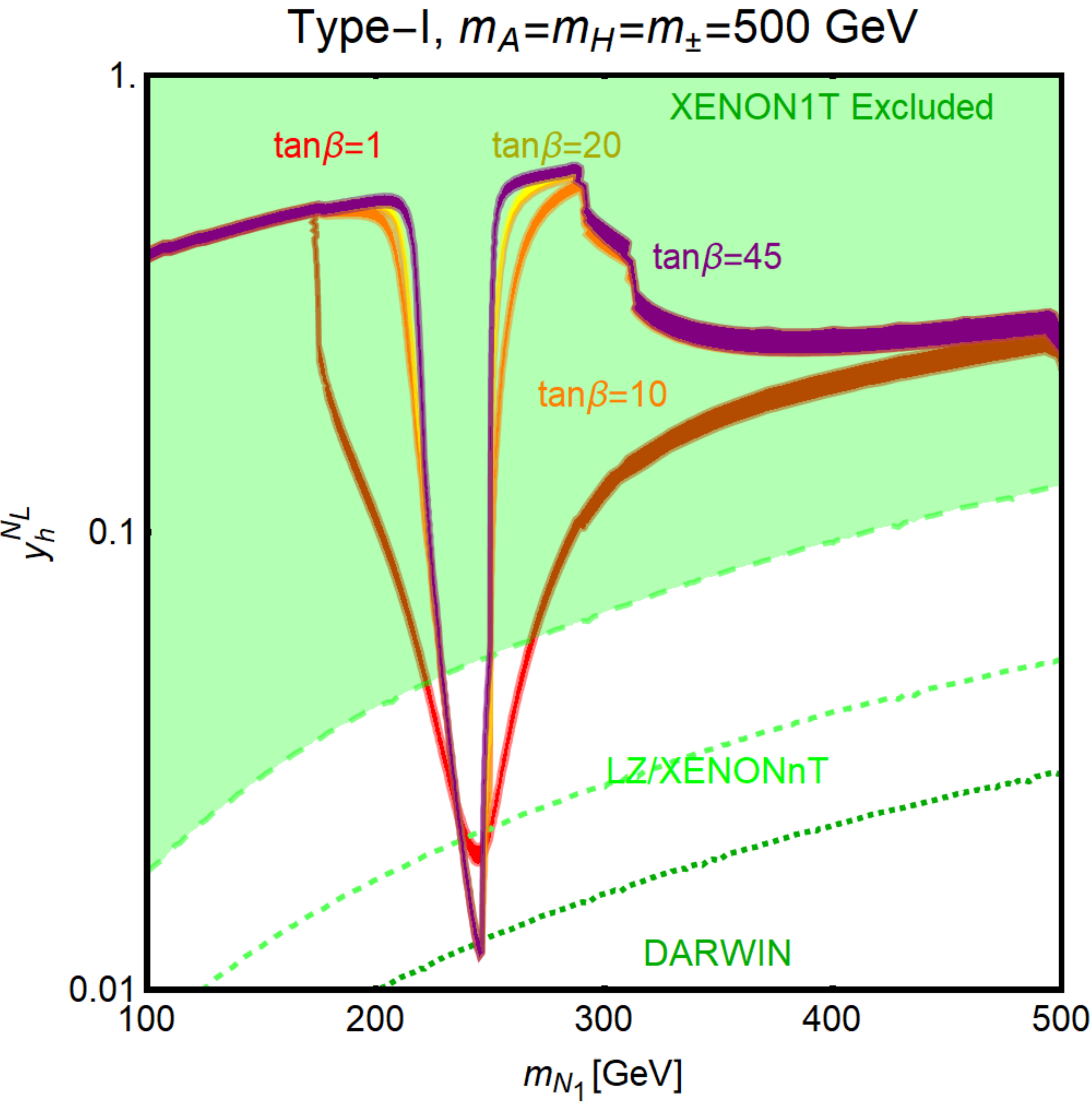}}~~
\subfloat{\includegraphics[width=0.45\linewidth]{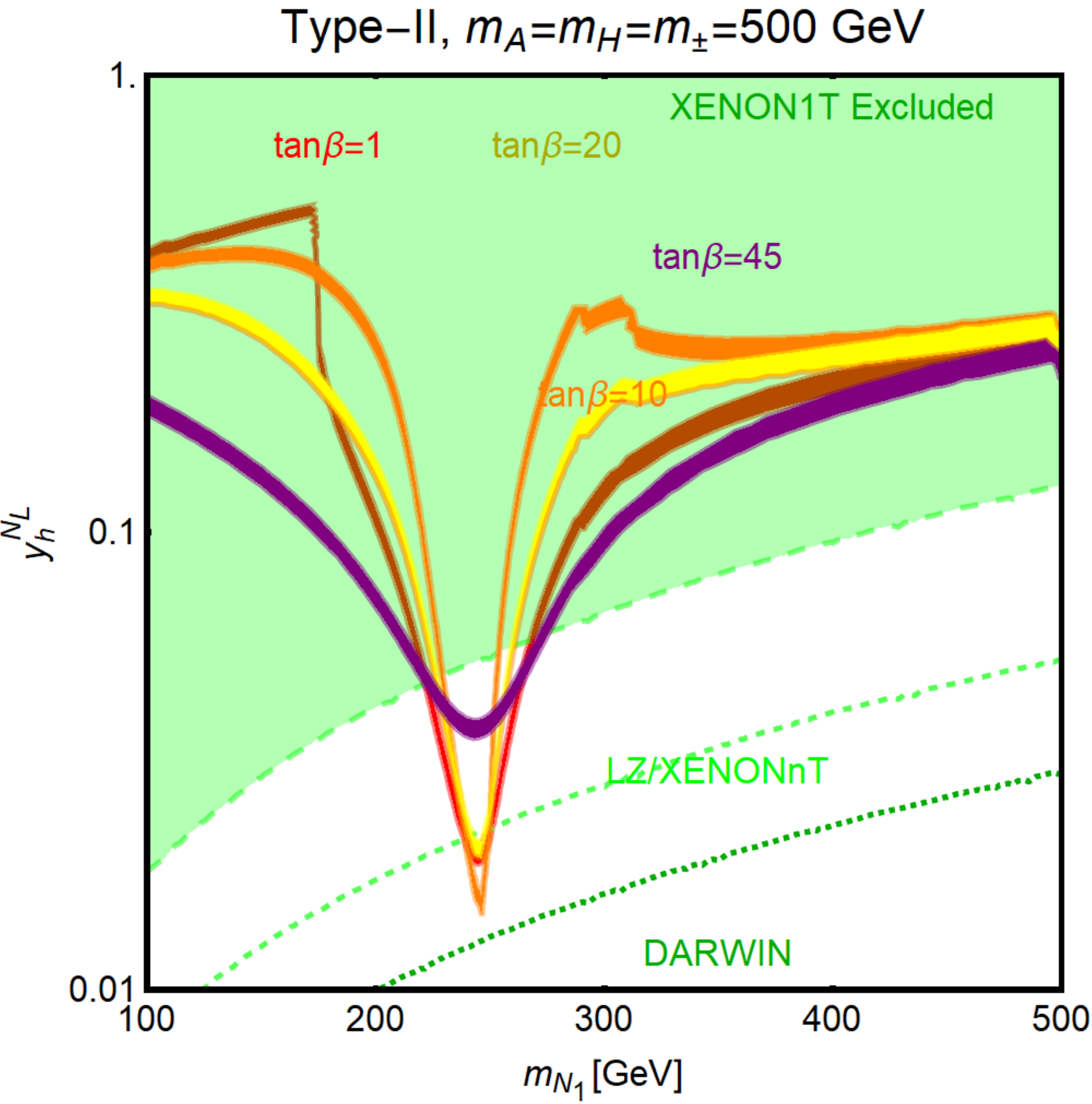}} }\\
\mbox{\subfloat{\includegraphics[width=0.45\linewidth]{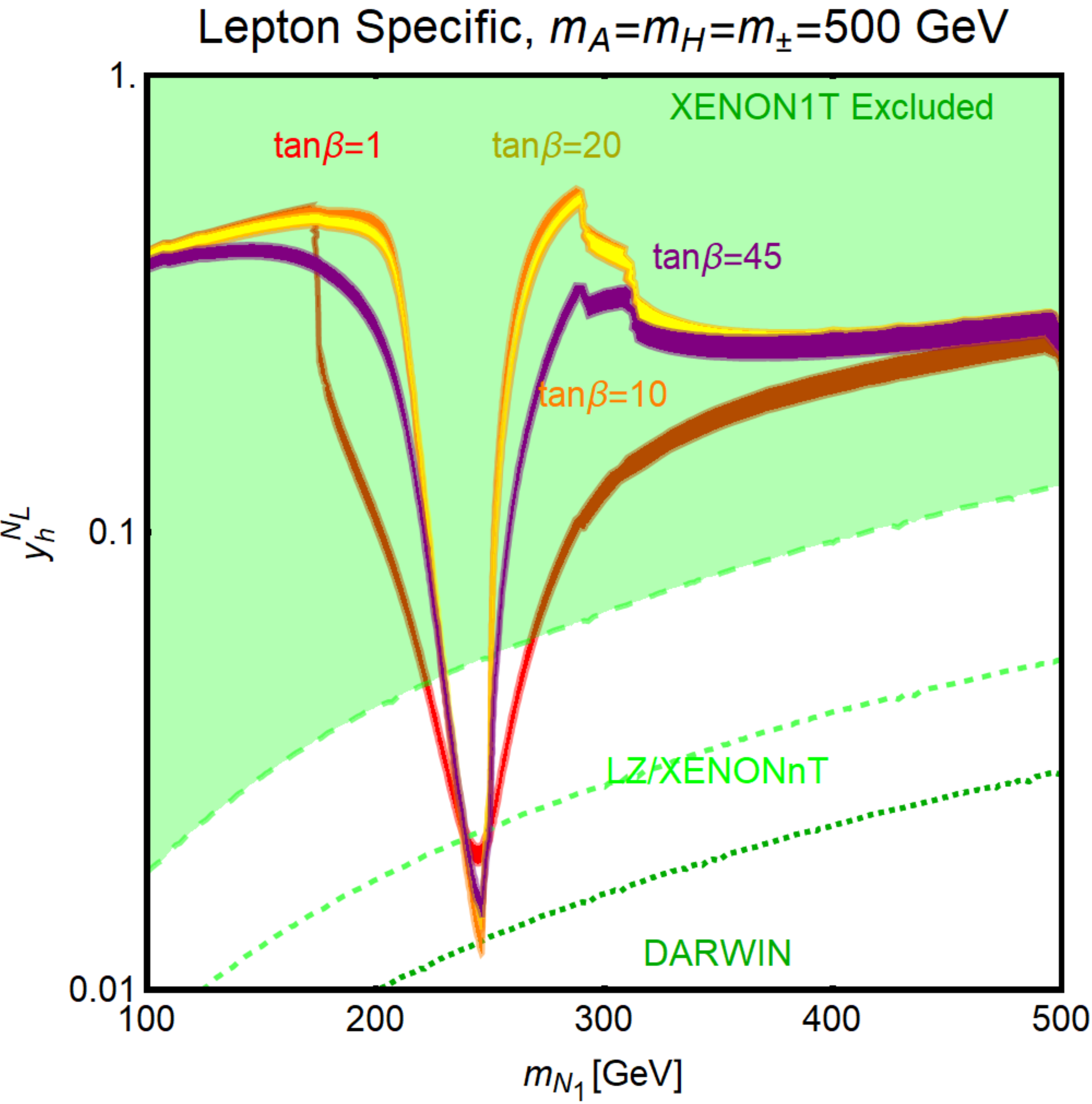}}~~
\subfloat{\includegraphics[width=0.45\linewidth]{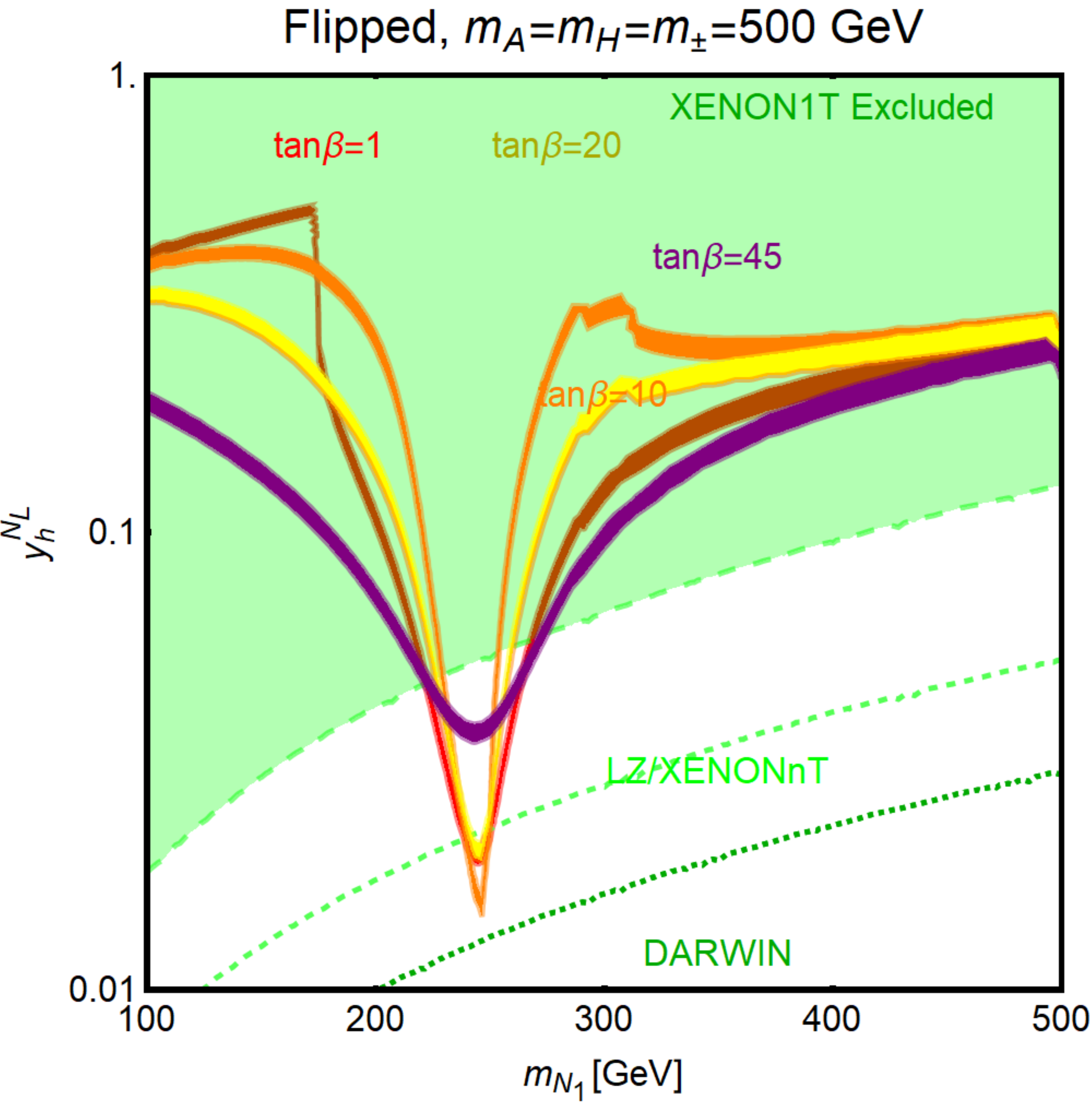}} }
\end{center}
\vspace*{-5mm}
\caption{Main astrophysics constraints on the DM particles in the four 2HDM scenarios in the alignment limit and degenerate heavy Higgs masses, 
$M_H=M_A=M_{H^\pm}=500$ GeV, when a vector--like family is present. The constraints are shown in the [$m_{N_1},y_ h^{N_L}]$ plane with the  four values $\tb=1,10,20$ and 45 for which the corresponding DM relic densities are shown. The limits from  XENON1T (in green) and the sensitivities from XENONnT/LZ (dashed lines) and DARWIN (dot--dashed lines) are also shown.}
\label{fig:pVLLheavyH}
\end{figure}

In Fig.~\ref{fig:pVLLheavyH}, we illustrate the case of heavy Higgs
bosons,  $m_{N_1}<M_{H,A,H^{\pm}}$, with arbitrary couplings to the vector--like
leptons. In the figure, represented in the bidimensional plane
$[m_{N_1},y_h^{N_L}]$, are iso-contours of the correct DM relic density for four
values $\tan\beta=1,10, 20,45$. The additional Higgs bosons are assumed to be
degenerate in mass with $M_{H}\! =\! M_A\! = \! M_{H^{\pm}} \! = \!
500\,\mbox{GeV}$.~\footnote{For a better illustration of the results we have set a same common mass for all the four types of 2HDMs. Notice however that the selected values is in tension, for the Type-II and Flipped models, with the bounds from $b \rightarrow s \gamma$ processes. As already pointed Fig.~\ref{fig:pVLLheavyH} serve just as illustration.} An assignment for the couple $(m_{N_1},y_h^{N_L})$ is
considered to be viable if it lies outside the green region, corresponding to
the present exclusion from XENON1T but will be excluded in the near future in
the case of a negative signal  by XENONnT/LZ (DARWIN) if it is above the green
(dark green) dashed curve. 

The four panels of Fig.~\ref{fig:pVLLheavyH}
correspond to the four flavour preserving configurations of the couplings of the
two Higgs doubles with SM fermions. 

As already pointed out, in the case where the vector--leptons interact with both
Higgs doublets, their couplings will not depend on $\tan\beta$; the different
behaviour of the DM iso--contours in the various panels is then due to the
enhancement or suppression, with respect to $\tan\beta$, of the Yukawa coupling
of the SM fermions with the additional Higgs states. The differences between the
four 2HDMs become particularly obvious in the vicinity of the $s$--channel
resonance regions. In the case of Type I model, because of the $1/\tan\beta$
suppression of all Yukawa couplings, the decay width of the neutral resonances
becomes increasingly small. This corresponds to strong enhancements in rather
narrow windows for the DM mass of the annihilation cross section so that the
correct relic density is met for values of $y_h^{N_L}$ down to 0.01. Far from
the resonance region, the countours of the correct relic density overlap,
indicating that its dominant contribution comes from annihilation into gauge
boson final states. In Type II models, the DM annihilation cross section is
dominated by $\bar b b$ and $\tau^+ \tau^-$ final states also far from the
resonance regions. At the same time, the enhancement of the cross section in the
``pole'' region is less pronounced with respect to  Type I model because of the
increased width of the neutral resonances.

Similarly to what has been discussed in section 3, the DM scattering cross
section on nuclei is substantially larger in magnitude than the one associated
to the singlet--doublet DM since the DM particle is of the Dirac type with
vectorial couplings to the $Z$ boson. As a consequence, the extended Higgs
sector has a negligible impact and the exclusion limits which remain practically
unchanged with respect to the ones presented in  section 3 for the case of
vector--like leptons interacting only with the SM Higgs sector.

We have then extended our results by performing a scan over the model parameters within the following ranges
\begin{align}
    & y_{h}^{N_{L,R}},y_H^{N_{L,R}} \in \left[10^{-3},1\right],\,\,\,\,\,y_{h}^{E_{L}},y_H^{E_{L,R}} \in \left[10^{-3},3\right] \, , \nonumber\\
    & M_{N,E,L} \in \left[100,1000\right]\,\mbox{GeV},\,\,\,\,\,\tan\beta \in \left[1,50\right] \, , \nonumber\\
    & M_A \in \left[100,1000\right] \,\mbox{GeV},\,\,\,\,\, M_H \in \left[M_h,1000\right] \,\mbox{GeV} \, , \nonumber\\
    & M_{H^\pm} \in \left[M_W,1000\right]\,\mbox{GeV}\,\,\,\,\,|M| \in \left[0,1\right]\,\mbox{TeV},
\end{align}
retaining only the model points complying with the bounds on the quartic
couplings, eqs.~(\ref{eq:up1})--(\ref{eq:up2}), from electroweak precision data,
low energy/flavor processes, collider searches of the extra Higgs bosons, mostly $A \rightarrow \tau^+ \tau^-$ and giving the correct DM relic density according to
the WIMP paradigm. This scan has been repeated for each of the four flavor
preserving configurations of the couplings of the two Higgs doublets with SM
fermions. As an additional hypothesis, we have enforced the mass hierarchy $M_N <M_L$, to avoid coannihilation solutions, since they would not be substantially
different from the ones discussed for the SM+VLL model. 

\begin{figure}[!h]
\vspace*{-2mm}
    \centering
    \subfloat{\includegraphics[width=0.49\linewidth]{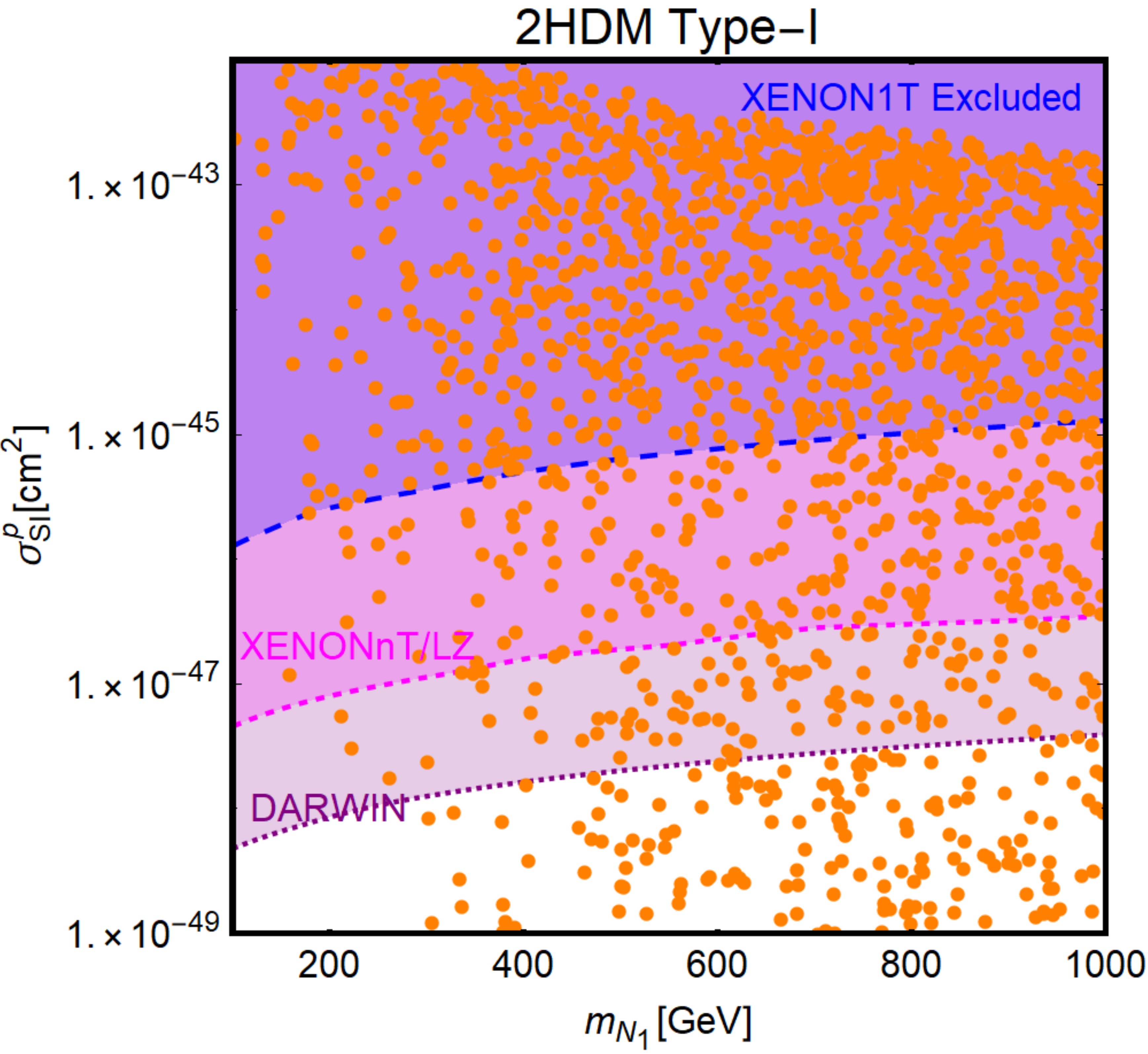}}
    \subfloat{\includegraphics[width=0.49\linewidth]{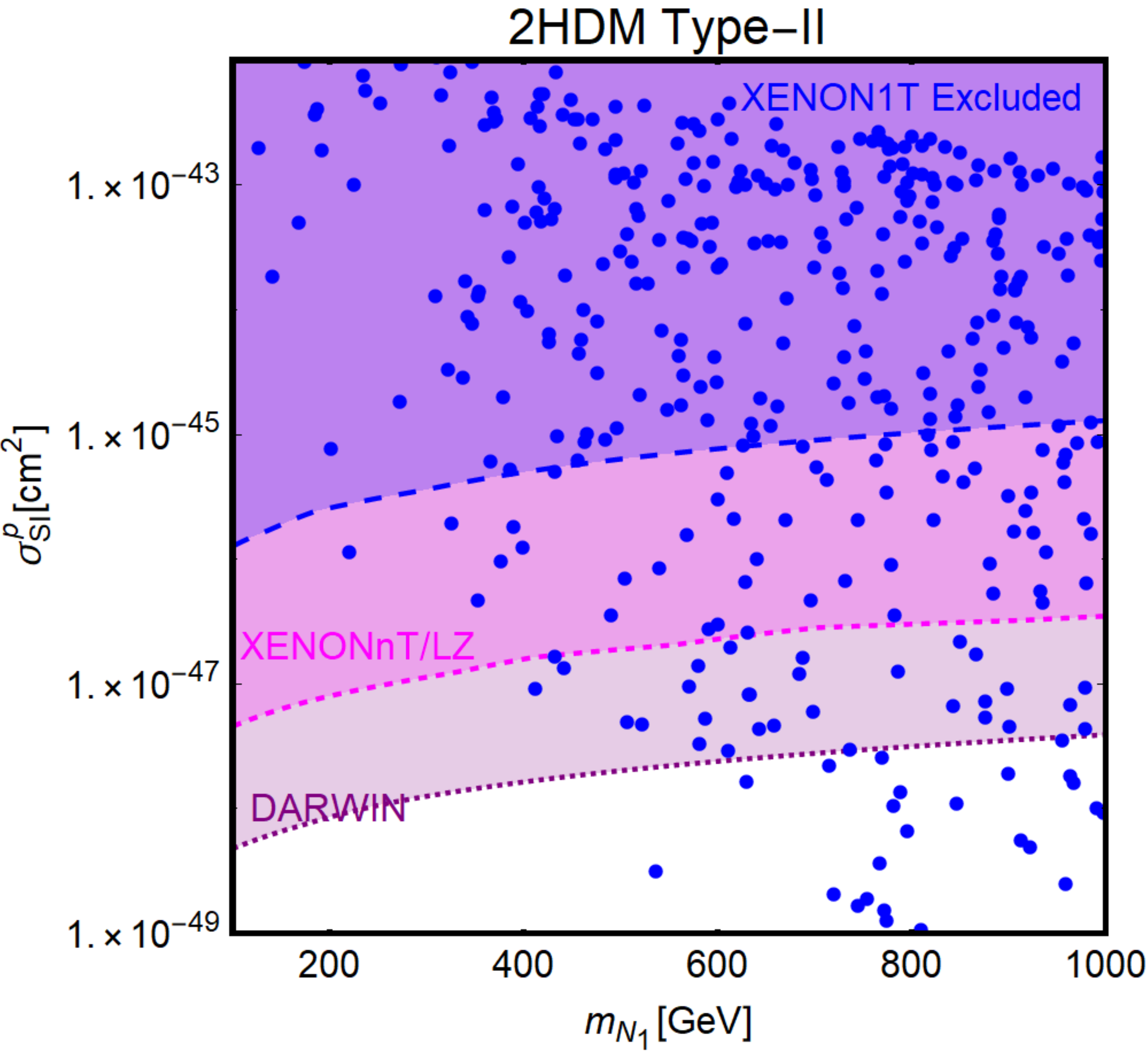}}\\
    \subfloat{\includegraphics[width=0.49\linewidth]{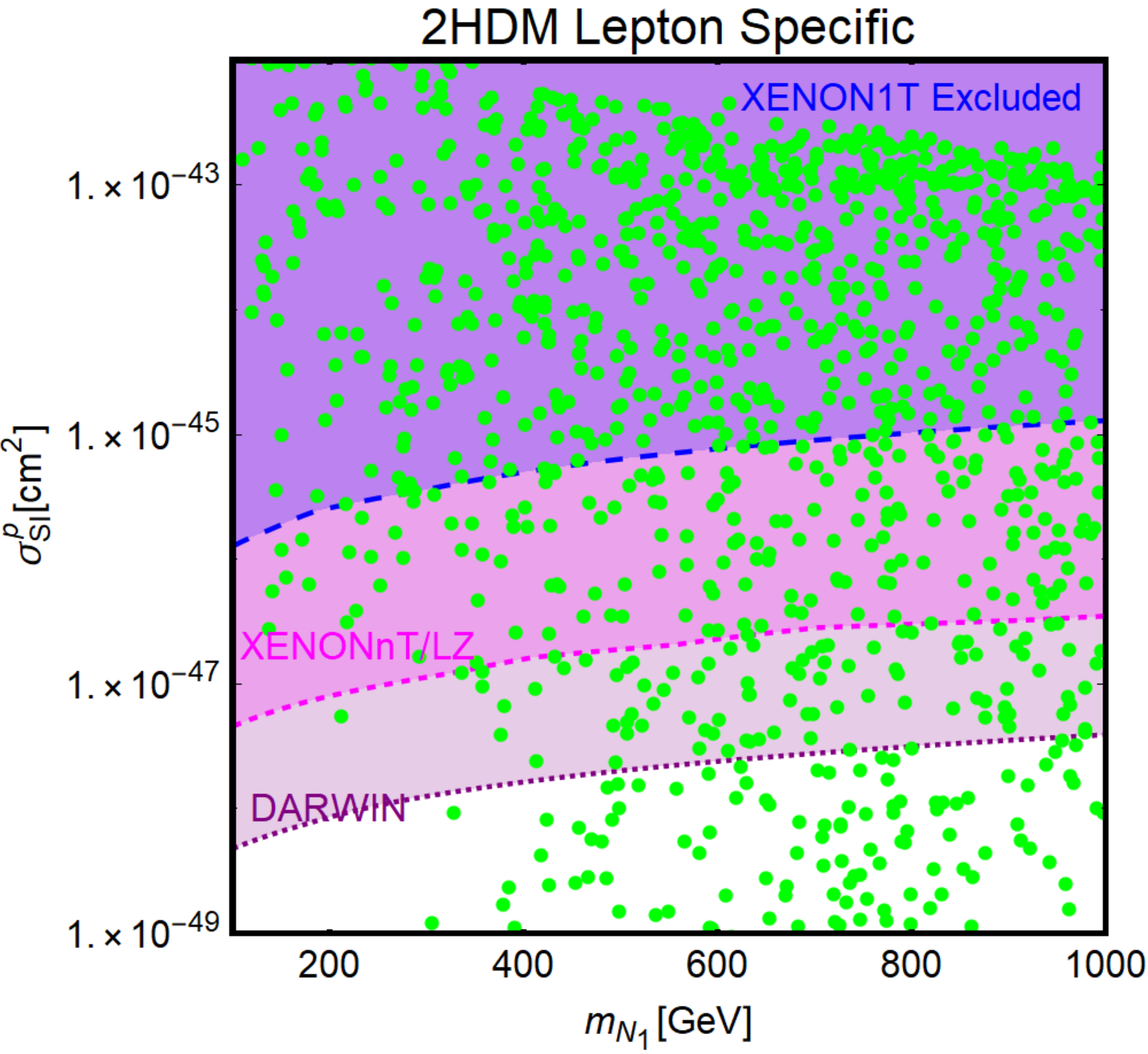}}
    \subfloat{\includegraphics[width=0.49\linewidth]{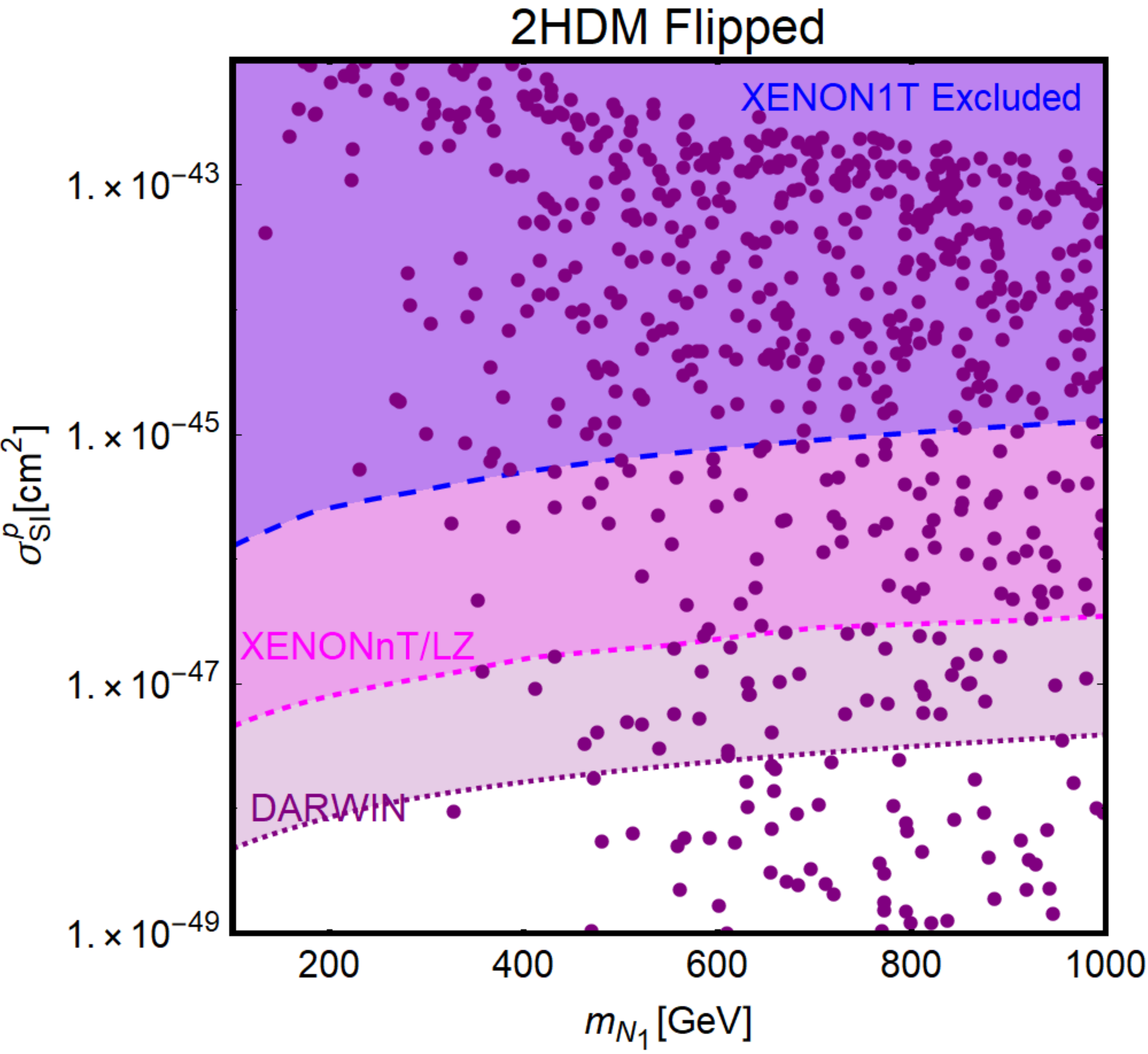}}
\vspace*{-1mm}
    \caption{Model points for the 2HDM+VLL model passing theoretical constraints and providing the correct DM relic density in the bidimensional plane $[m_{N_1},\sigma_{\rm SI}^p]$. The different panels refer to the four flavor preserving configurations for the couplings of the Higgs doublets with the SM fermions. The blue (magenta/purple) region represents present (future projected) exclusion by XENON1T (LZ/XENONnT/DARWIN).}
    \label{fig:2HDM_VLL_scans}
\vspace*{-3mm}
\end{figure}

The model points satisfying the conditions listed above have been represented in
the plane $[m_{N_1},\sigma_{\rm SI}^p]$. These points are compared with the
current limits from direct detection as given by XENON1T and the projected
sensitivities of LZ/XENON1T (magenta colored region) and DARWIN (purple colored
region). The main difference between the various 2HDM configurations is in the
allowed values of the DM masses. In the case of the Type--II and flipped models
it is possible to comply with direct detection limits only of $m_{N_1} \gtrsim
400\,\mbox{GeV}$. This is because in order to comply with direct detection
bounds, one should rely either on $s$--channel resonances or into annihilations
with Higgs bosons in the final states, in particular $H^{+}H^{-}$ and
$W^{\pm}H^{\mp}$. The reason why the latter annihilation channel are relevant is
that the corresponding rates depend on the couplings $y_H^{E_{L,R}}$, to which
direct detection is not sensitive. However, in Type--II and flipped 2HDMs there
is a very strong constraint on $M_{H^{\pm}}$, to a large extent independent on
$\tan\beta$,  from $b \rightarrow s$ transitions. In addition, in the Type--II
model, the mass of the neutral Higgs bosons, unless $\tan\beta$ is low, is
strongly constrained by the searches in the $pp \rightarrow H/A \rightarrow \tau
\tau$ channel. The presence of these lower bounds on the  masses
$M_{H^{\pm}},M_H,M_A$ hence requires, correspondingly higher masses for the DM
state. The Type--I and lepton specific 2HDMs are, instead, more loosely
constrained by searches of the Higgs bosons so that lower DM masses are still
viable. 

All the different 2HDM realizations will experience a progressive strong
reduction of the viable parameter space as bounds from direct detection will
eventually become stronger. Some model configurations would be nevertheless
capable of evading even a negative detection from the DARWIN experiment.

\begin{figure}[!ht]
\vspace*{-2mm}
\begin{center}
\subfloat{\includegraphics[width=0.47\linewidth]{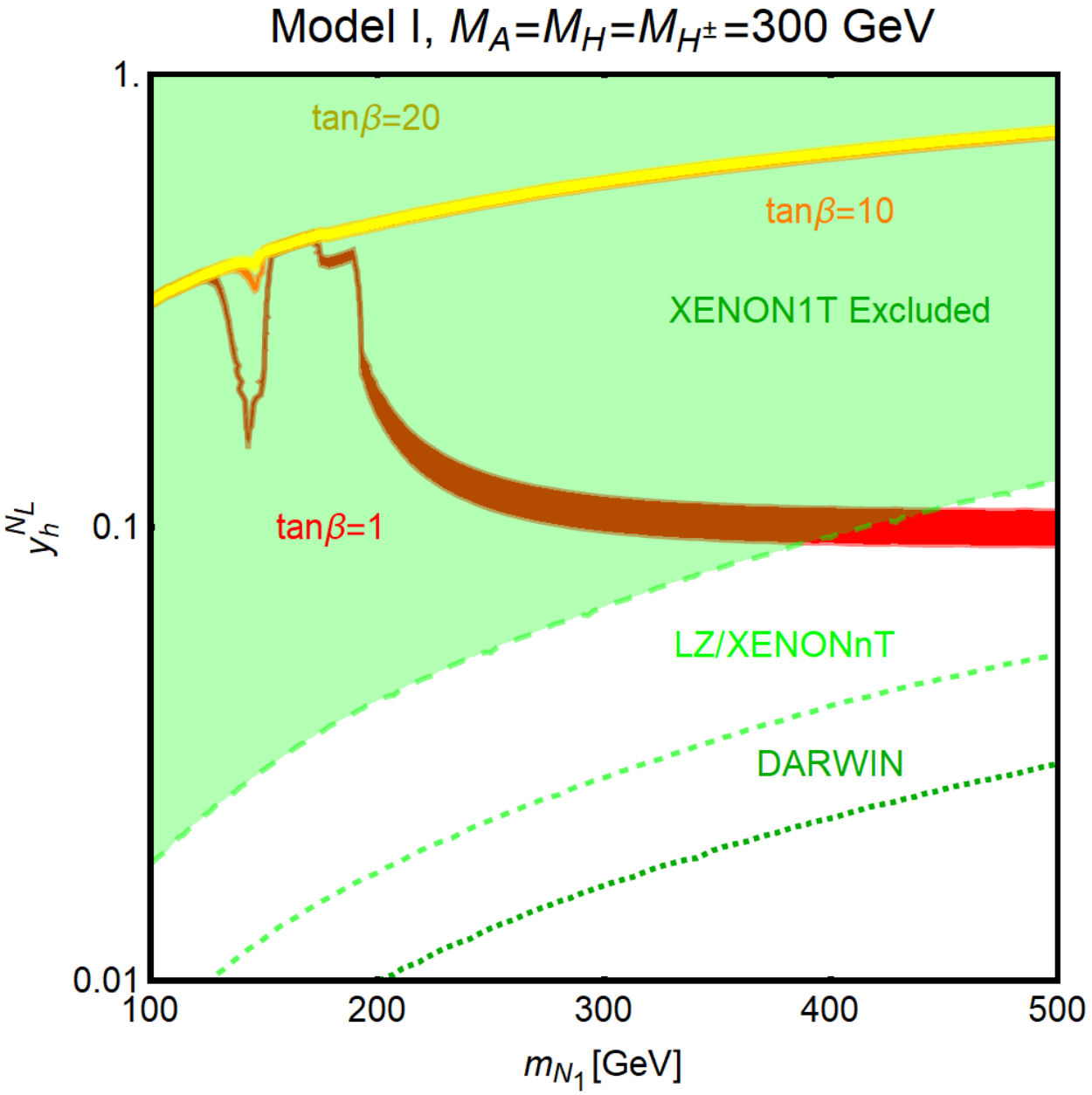}}~~
\subfloat{\includegraphics[width=0.47\linewidth]{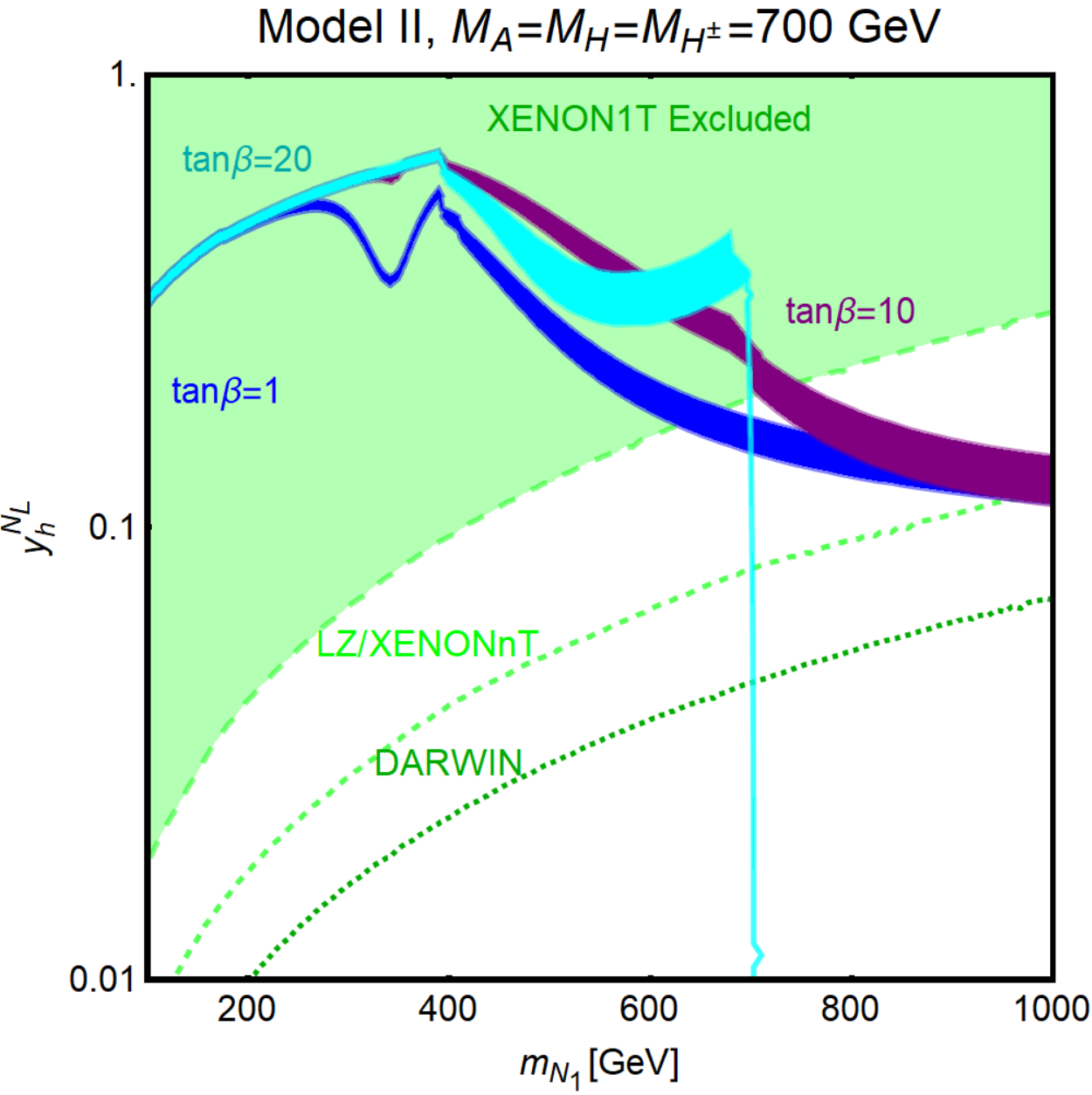}}
\end{center}
\vspace*{-5mm}
\caption{Main astrophysics constraints on the DM particles in two proposed
scenarios for a vector--like family coupled to an aligned 2HDM with  mass
degenerate heavy Higgs bosons.  In the (right) panel, we consider model I (II) with $M_A\!=\!300\,(700)$ GeV and $\tb=1,10$ and 20 for which the corresponding relic densities are shown. The limits from  XENON1T and the projected sensitivities from XENONnT/LZ and DARWIN are also shown.}
\label{fig:pFCNC}
\vspace*{-4mm}
\end{figure}

Let us now briefly consider the case in which a $\mathbb{Z}_2$ symmetry is present in the new fermionic sector, in order to enforce specific configurations for the couplings of the VLLs with the Higgs doublets. We will just limit our analysis to the study of the $[m_{N_1},y_h^{N_L}]$ plane showing two benchmark scenarios in Fig.~\ref{fig:pFCNC}. The left (right) panel of the figure is for  model I (II) with $M_H\!=\!M_A\!=\!M_{H^\pm}\! =\! 300$ GeV (700 GeV) and shows three isocontours of the correct relic density, corresponding to the values $\tan\beta\!=\! 1,10,20$.  In the case of model I, we
have chosen the low value $M_A=300\,\mbox{GeV}$ since we assume a Type I 2HDM
configuration of the couplings. For model II, associated to a Type II 2HDM 
couplings with the SM fermions, we have instead considered the higher value
$M_A=700\,\mbox{GeV}$, to comply with collider and flavor bounds. Both panels
show the excluded region (green) from XENON1T as well as the projected
sensitivities from XENONnT/LZ and DARWIN, as dashed and dot--dashed green lines,
respectively. 

The shape
of the relic density curves can be explained as follows: the couplings of the DM
state with the neutral Higgs bosons are suppressed by $\tan\beta$ and the latter
can affect the annihilation cross section into SM states only at low $\tan\beta$
values through the presence of $s$--channel ``poles''. Model I is characterized
by suppressed annihilation rates at high $\tan\beta$, also for $m_{N_1}>M_{H,A,
H^{\pm}} $ since all the Yukawa couplings of the vector leptons are suppressed
by ${1}/{\tan\beta}$. In turn, for model II, the DM annihilation cross section
into the $W^{\pm}H^{\mp}$ and $H^{\pm}H^{\mp}$ final states becomes increasingly
efficient with higher $\tan\beta$, because of the enhancement of the couplings
$y_H^{E_L}$. As can be seen from the  figure, the constraints from direct
detection are particularly severe. They can be evaded only when $m_{N_1}>
M_{H,A,H^{\pm}}$ since, in such a case, the DM annihilation rate into $H^\pm$
final states is enhanced without conflicting with direct detection constraints,
being dependent on the couplings $y_H^{E_L}$ that do no enter in the
spin--independent cross section. The viable DM region will be nevertheless ruled
out, for masses up to $m_{N_1} \approx 1\,\mbox{TeV}$, in the absence of a
signal at the next generation of direct detection experiments.

The simple illustration provided in Fig.~\ref{fig:pFCNC} has been complemented
by a scan, in order to account for the higher dimensionality of the parameter
space.  


\subsubsection{The inert doublet case}

In the inert doublet model, direct DM detection relies on the spin--independent 
interaction whose cross section is analogous to the one used for the SM--like
Higgs--portal with a spin-zero DM particle. Using the usual
conventions, the DM scattering cross section on protons can be written as
\begin{equation}
\sigma_{H p}^{\rm SI}=\frac{\mu_{H}^2}{4\pi}\frac{m_p^2}{M_{H}^2 M_h^4}\lambda_L^2 \left[f_p \frac{Z}{A}+f_n \left(1-\frac{Z}{A}\right)\right]^2,
\end{equation}
where $m_p$ is the proton mass, $\mu_{H}$ the reduced mass of the DM--proton
system and the coefficients $f_p, f_n$ have been defined  before while $A(Z)$ are the atomic mass (number) of the target nucleus. In the case where the $H$ and $A$ states are degenerate in mass, $|M_A-M_H|\lesssim 1\,\mbox{GeV}$, an additional contribution associated to the $H p \to Z^* \to A p$ process should be considered. We will not explicitly consider this scenario, though. 

More complicated is, instead, the case of DM relic density. As will be clarified
below, in large portions of the viable parameter space, the $A$ and $H^{\pm}$
bosons are very close in mass with the DM particle, so that coannihilation
processes are not negligible. Contrary to the other models, the velocity
expansion of the DM annihilation cross sections does not provide a reliable
description of the phenomenology. We will nevertheless provide some useful
expressions for them in Appendix~B.

According the studies performed for instance in
Refs.~\cite{LopezHonorez:2006gr,Hambye:2007vf,LopezHonorez:2010tb}, the correct
DM relic density can be obtained in three scenarios:\vspace*{-2mm} 

\begin{itemize}

\item For a light DM, $M_H \lesssim 50\,\mbox{GeV}$, the correct relic density
is achieved in an analogous way as in the SM Higgs--portal model, namely mostly
through  annihilation into $\bar b b$ final states via the $s$--channel exchange
of the SM Higgs boson; no coannihilation processes are expected in this regime
since the masses of the $A$ and $H^{\pm}$ states should comply with the bounds
from LEP2.\vspace*{-2mm} 

\item In the intermediate DM mass range, 50 GeV$ \lesssim M_H \lesssim
80\!-\!100\,\mbox{GeV}$, the DM annihilation cross section is enhanced by the
$HH \rightarrow WW^{*}\rightarrow Wf \bar f^{'}$ three--body final state. If
$\lambda_L >0$, the correct relic density cannot be achieved for $M_H > M_W$
because of the too efficient annihilations into $WW$ states, occurring mostly
through gauge interactions. For $\lambda_L<0$, a destructive interference among
the  different channels contributing to $H H \to WW$ annihilation occurs so that
a viable DM relic density can  be obtained for DM masses up to around 100
GeV~\cite{LopezHonorez:2010tb}. Coannihilation processes might be also relevant
in this regime.\vspace*{-2mm}

\item At high DM masses, $M_H \gtrsim 500\,\mbox{GeV}$,  the correct relic
density is achieved mostly through annihilations into  $ZZ$ and $WW$ final
states. This scenario is similar to the minimal DM models~\cite{Cirelli:2007xd}.
Coannihilations are also present in this regime since mass degeneracy among the
neutral Higgs states $H,A$ and the charged Higgs $H^{\pm}$ is needed in order to
avoid an excessive enhancement of the annihilation cross section into gauge
bosons as will be seen below.\vspace*{-2mm}   

\end{itemize}

We have now the main ingredients which will allow to summarize the most
important DM constraints in the context of this inert Higgs doublet scenario. 

\begin{figure}[!h]
\vspace*{-2mm}
\centerline{
\includegraphics[scale=0.57]{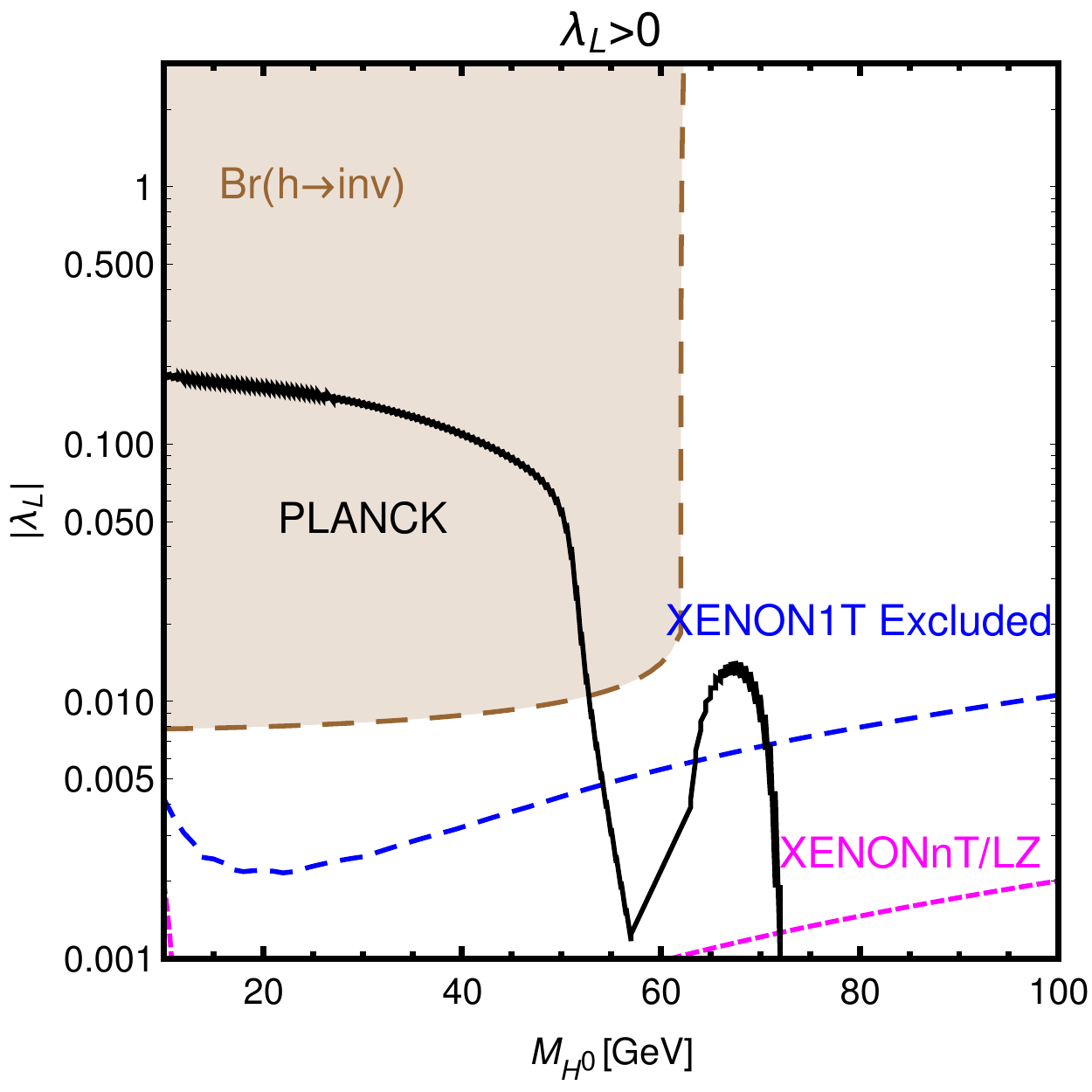}~
\includegraphics[scale=0.57]{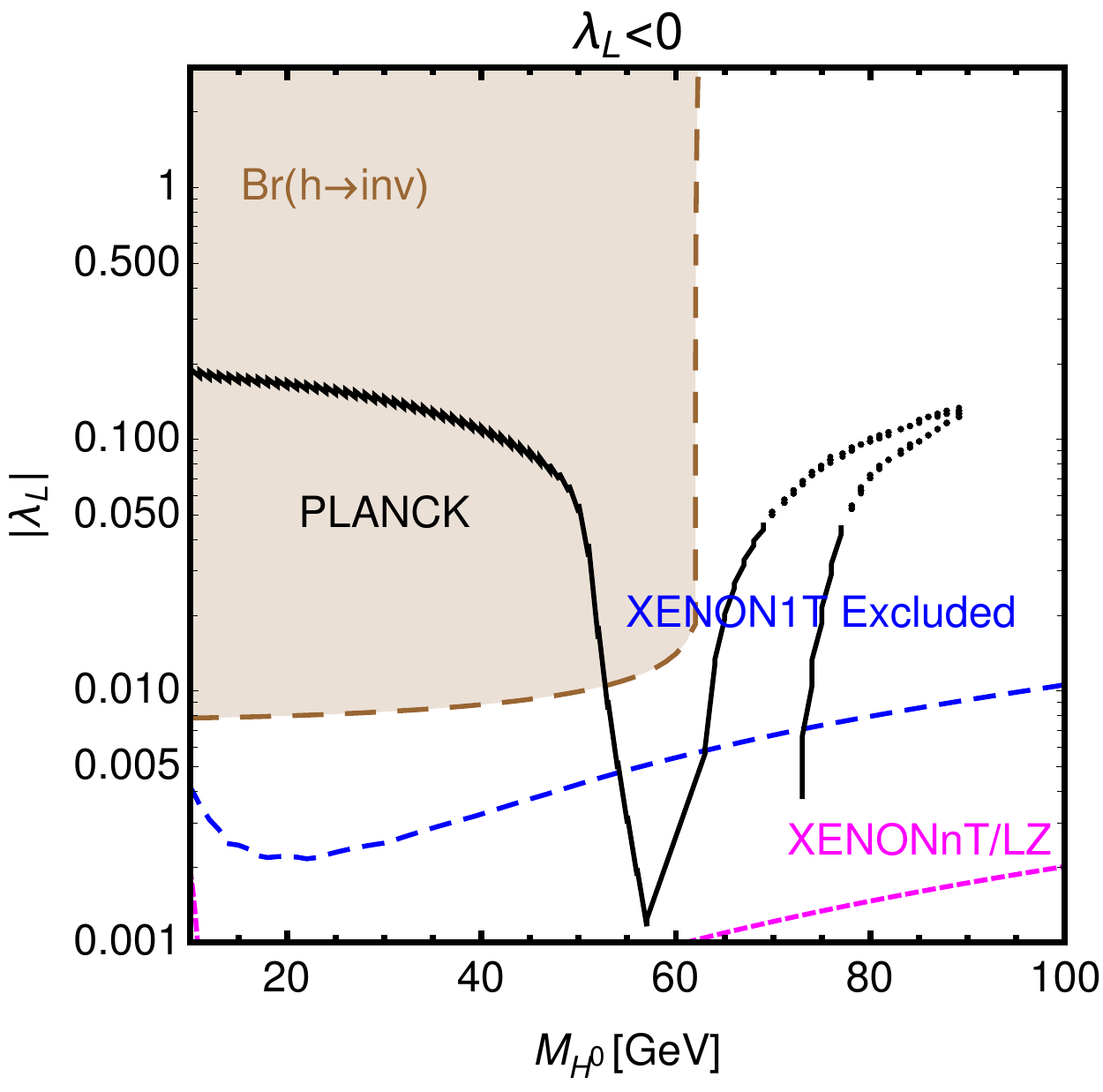}
}
\vspace*{-1mm}
\caption{Constraints on the inert doublet model in the  plane $[M_{H}, |\lambda_L| ]$  for $\lambda_L >0$ (left) and $\lambda_L<0$ (right). The black contour is when the correct DM relic density is achieved, while the regions above the blue dashed lines are those excluded by the current constraint from XENON1T. The magenta dashed lines are the expected sensitivity of the XENONnT/LZ experiment, and the brown  region is excluded by the limits on the SM--like Higgs invisible branching fraction.}
\label{fig:pIDMlight}
\vspace*{-2mm}
\end{figure}

First, to make a comparison between the IDM and the effective SM Higgs--portal
scenario,  we show the DM constraints including the ones from the invisible Higgs width in Fig.~\ref{fig:pIDMlight} in the plane $[M_{H},|\lambda_L|]$ for positive (left) and negative (right) values of the  coupling $\lambda_L$. We  have focused on the low DM mass regime $M_{H}  <100$ GeV and considered a
large enough mass splitting between the DM $H$ state and the $A$ and $H^{\pm}$ bosons in order to neglect coannihilation effects, but still not too large as to avoid tensions with electroweak precision data. As already anticipated, for $M_{H} \lesssim \frac12 M_h$, the pattern is almost  the
same as in the SM Higgs--portal scenario: a  significant region of the parameter
space is excluded by the combination of direct detection limits and the LHC
constraints on the invisible decays of the 125 GeV Higgs boson, except for the
pole region $M_{H} \sim \frac12 M_h$ which  can be probed only at the next
generation of direct detection experiments.  

For $M_{H}>\frac12 M_h$, a viable region for the relic density opens up as a
result of the $H H \rightarrow Wf \bar{f}^{'}$ annihilation channel. As already
mentioned, this process can occur only through gauge interactions, hence the
correct relic density can be achieved for very small values of $|\lambda_L|$,
implying a suppressed scattering rate of the DM on nuclei. For $\lambda_L >0$,
the correct relic density is achieved only for $M_H \lesssim M_W$, as
annihilation into two on--shell $W$ bosons is too efficient at low $M_H$. For
$\lambda_L$, the region corresponding to the correct relic density can be
extended in a small portion of the $M_H > M_W$ region. This is, however, at the
price of a higher value of $|\lambda_L|$, in tension with present constraints
from XENON1T. 

In order to properly account for the DM phenomenology for $M_H \gtrsim 100\,\mbox{GeV}$, we need to include the possibility of mass degeneracy between the DM and the other extra Higgs bosons. Similarly to what occurred for the model with vector--like DM, we have performed a scan on the input parameters of the IDM in the following ranges
\bea
M_{H} \in \left[10,1000\right]\,\mbox{GeV}, \ \
M_{A}- M_H, M_{H^{\pm}}- M_H \in \left[1,100\right]\,\mbox{GeV}, \ \ 
|\lambda_L| \in \left[10^{-6},1\right] .  
\eea

\begin{figure}[!h]
\vspace*{-2mm}
\begin{center}
\subfloat{\includegraphics[scale=0.66]{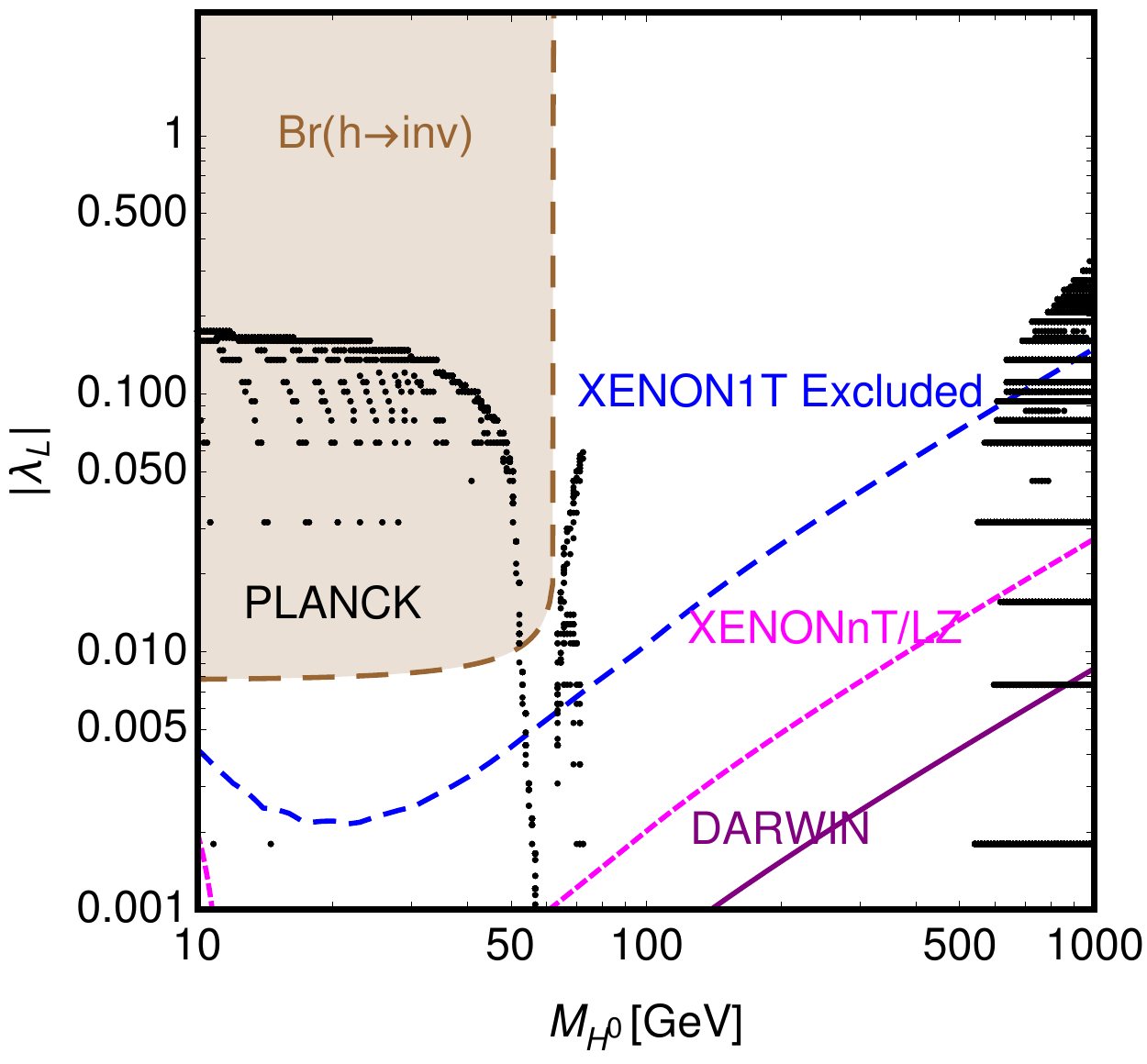}}
\end{center}
\vspace*{-5mm}
\caption{Model points points passing theoretical constraints on the inert doublet model and giving the correct relic density according to the WIMP paradigm in the bidimensional plane $M_H,\lambda_L$. The brown region is excluded by searches of invisible decays of the SM Higgs while the regions above the blue (magenta/purple) curves are (will be) excluded by XENON1T (LZ/XENONnT/DARWIN).}
\label{fig:pIDMtot}
\vspace*{-3mm}
\end{figure}

The results are shown in Fig.~\ref{fig:pIDMtot} again in the plane $[M_{H}, |\lambda_L|]$, with the same color code as previously.   It can be seen
that the coannihilation channels do not allow to evade the strong constraints in
the low DM mass regime. No model points with the correct relic density are
present for 100 GeV$\lesssim M_H \lesssim 500\,\mbox{GeV}$. A new viable region,
covering a wide range of values of $|\lambda_L|$, and compatible with the
exclusion limits from the next generation of direct detection experiments is
instead present at higher masses. This result can be explained by the fact that
the cross sections of the processes $H H \rightarrow WW$ and $H H \rightarrow ZZ$ depend on the coupling combinations $\lambda_4+\lambda_5$ and $\lambda_5$,
respectively. This corresponds to enhancement factors
\begin{align}
& \frac{\alpha^2}{M_Z^2
s_W^2}{\left(M_A^2-M_H^2\right)}^2\hspace*{4.2cm} {\rm for}~WW,\nonumber\\
& \frac{\alpha^2}{M_Z^2s_W^2} \left[{\left(M_A^2-M_H^2\right)}^2+{\left(M_{H^{\pm}}^2-M_H^2\right)}^2\right]\,\,\,\,\,\, {\rm for~}ZZ\, . 
\end{align}

In order to avoid large values for these enhancement factors, that would lead to
an underabundant DM particle, one needs small differences between the masses of
the $H,A,H^{\pm}$ states. In the exact $M_H=M_A=M_{H^{\pm}}$ limit, the DM
annihilation cross section would scale as ${\alpha^2}/{M_H^2}$ and would match
the thermally favored value for $M_H \simeq 500\,\mbox{GeV}$. A further
implication of these small mass splittings is the fact that coannihilation
processes are unavoidable. They lead to a rather interesting mechanism for the
DM relic density described in detail in Ref.~\cite{Queiroz:2015utg}. In presence
of only self--annihilations, the DM relic density after freeze--out is normally
much lower with respect to the experimentally favored value. Strongly mass
degenerate $A,H^{\pm}$ states would have annihilation cross sections of
analogous size as the DM, since they belong to the same SU(2) doublet, and
decouple slightly after the DM. Their subsequent decay into the DM state would
enhance its relic density, leading to the correct value for  $M_H$ up to order
of 2 TeV. 

This peculiar feature of the IDM has a further implication for indirect DM 
detection. Coannihilation channels are indeed absent at present times since the
involved particles decoupled from the thermal bath and then decayed  back into
DM states. We are thus left with the DM self--annihilations into gauge bosons at
very high rates, up to $10^{-25}\,{\mbox{cm}}^3\,{\mbox{s}}^{-1}$ or above,
which fall within the sensitivity of the future telescope
CTA~\cite{Garcia-Cely:2015khw,Queiroz:2015utg}.

Concerning DM indirect detection, other signals are possible within the IDM. In
the high $M_H$ regime, an additional interesting signal is associated to the
process $HH \rightarrow W^+ W^- \gamma$. While the corresponding rate is below
the present experimental sensitivity, it can fall within the reach of future
detectors like CTA~\cite{Garcia-Cely:2013zga}. 
In the low $H$ mass regime, a potentially relevant signature is represented by
$\gamma$--ray lines emerging from the loop induced annihilation processes $HH
\rightarrow \gamma \gamma$ and $HH \rightarrow
Z\gamma$~\cite{Gustafsson:2007pc}. Besides this, as the DM state is a scalar,
its annihilation cross section is $s$--wave dominated and can then be probed by
indirect detection experiments. Similarly to the case of the effective SM 
Higgs--portal, the corresponding limits are not competitive with the ones from
DM direct detection and the constraints from the Higgs invisible decay width.

\subsubsection{2HDM and a pseudoscalar portal}

As already mentioned, the 2HDM+light pseudoscalar model is a gauge invariant
embedding of a pseudoscalar portal for a SM singlet DM. Its phenomenology
presents remarkable differences with respect to the other scenarios of fermionic
DM connected to the Higgs sector. First of all, the absence of coupling between
the DM state and the CP--even Higgs bosons forbids at tree level
spin--independent interactions for the DM. The latter arise nevertheless at the
one--loop level from diagrams such as the ones shown in Fig~\ref{fig:feynloop}.
For simplicity, the figure shows only the diagrams for $a$ exchange but all
possible combinations of exchanges of the $a,A$ states should be included.

\begin{figure}[!ht]
\vspace*{-3mm}
    \centering
    \subfloat{\includegraphics[width=0.29\linewidth]{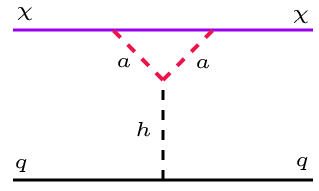}}~~
     \subfloat{\includegraphics[width=0.29\linewidth]{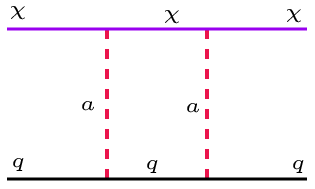}}
    \caption{Generic Feynman diagrams responsible for the loop induced scattering of the DM state on quarks in the 2HDM plus a light pseudoscalar model.}
    \label{fig:feynloop}
\vspace*{-1mm}
\end{figure}

These diagrams lead, indeed, to an effective Lagrangian of the form
\begin{equation}
\mathcal{L}=\tilde{c}_S \bar \chi \chi \bar q q,\,\,\,\,\,\tilde{c}_S=\tilde{c}_{S,\rm triangle}+\tilde{c}_{S,\rm box} \, , 
\label{eq:eff-axx}
\end{equation}
hence giving the spin--independent cross section,
\begin{equation}
    \sigma_{\chi p}^{\rm SI}=\frac{\mu_{\chi p}^2}{\pi}|\tilde{c}_S|^2 .
\end{equation}

For what concerns the DM relic density, it is essentially determined by
annihilations into SM fermions mediated by $s$--channel exchanges of the $a,A$
states, as well as annihilations into $aa$ pairs. As the DM mass increases, the
channels with $ha$ and $Za$ final states become relevant as well. For DM masses
of several hundreds of GeV, the $Aa$, $AA$, $H^{\pm}W^{\mp}$ and $HA$ final
states become kinematically accessible. Approximate analytic expressions of the
relevant cross sections can be straightforwardly derived from the ones given in
the previous subsection and will be thus reported again in Appendix B.

We finally note that the annihilation cross section into SM final states being $s$--wave dominated, the DM particle can also be probed in indirect detection.

A first illustration of the combined constraints from DM and collider
phenomenology is provided by Fig.~\ref{fig:pCOY} in  the plane $[m_\chi,M_a]$
for fixed values of $g_\chi,\theta$ and $\tb$. We have chosen a high value  for
the mass of the 2HDM  $A$ state, $M_A=600$ GeV, so that it has a marginal impact
on DM phenomenology (we recall that we cannot set its mass to an arbitrarily
high value because of the bounds on perturbative unitarity discussed
previously).

\begin{figure}[!h]
\vspace*{-1mm}
    \centering
    \subfloat{\includegraphics[width=0.45\linewidth]{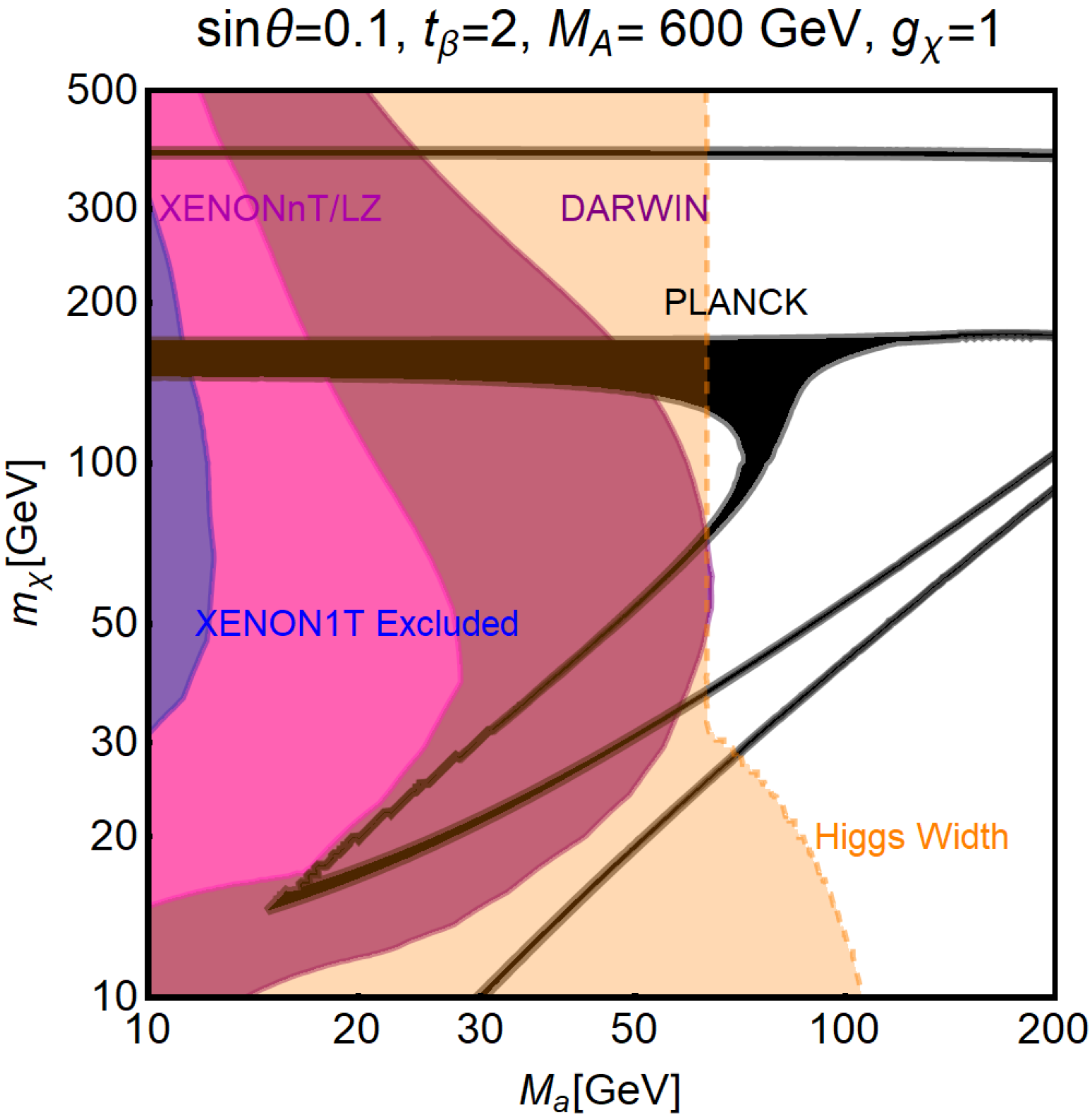}}~~
 \subfloat{\includegraphics[width=0.45\linewidth]{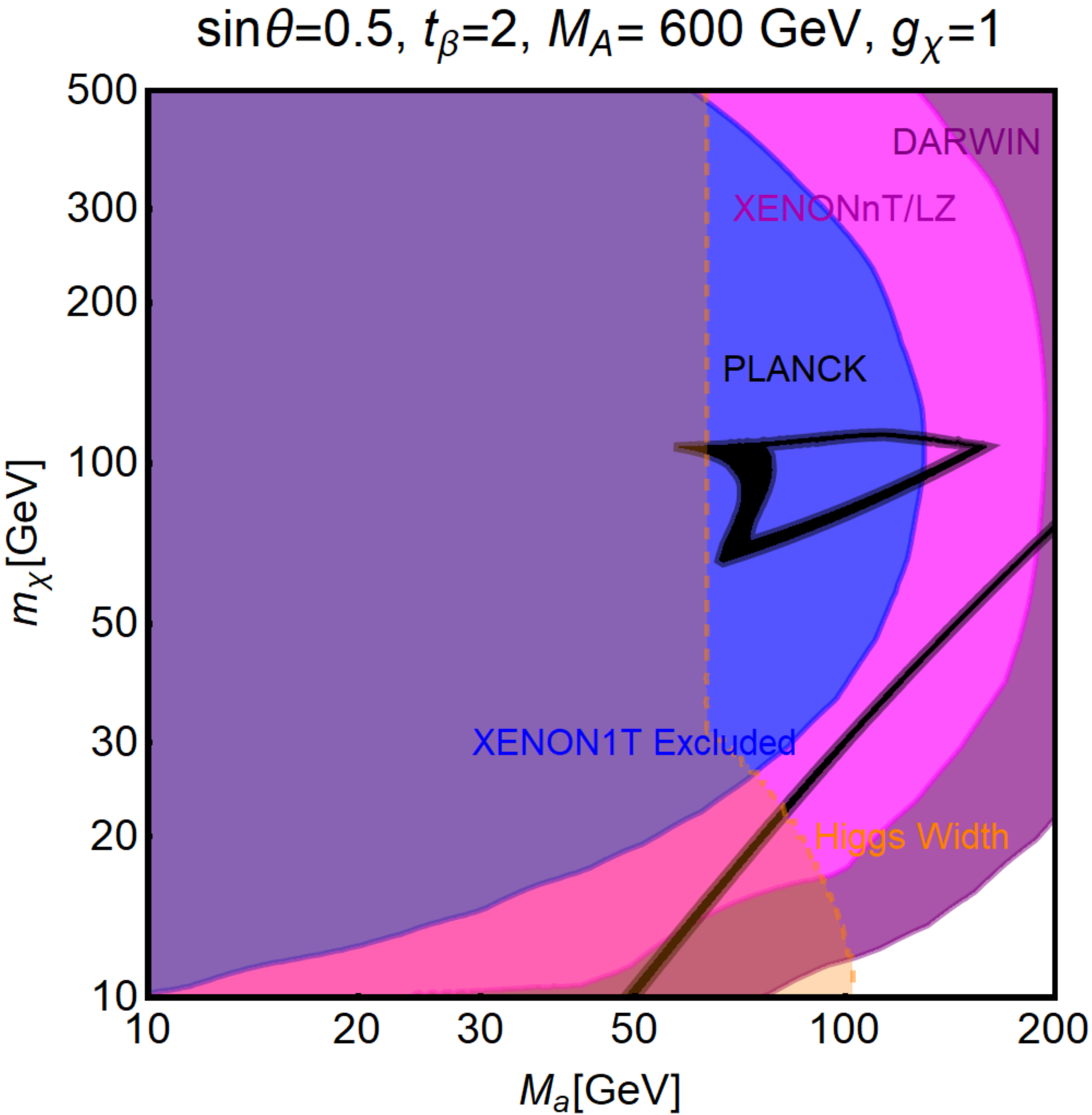}}\\[.2mm]
     \subfloat{\includegraphics[width=0.45\linewidth]{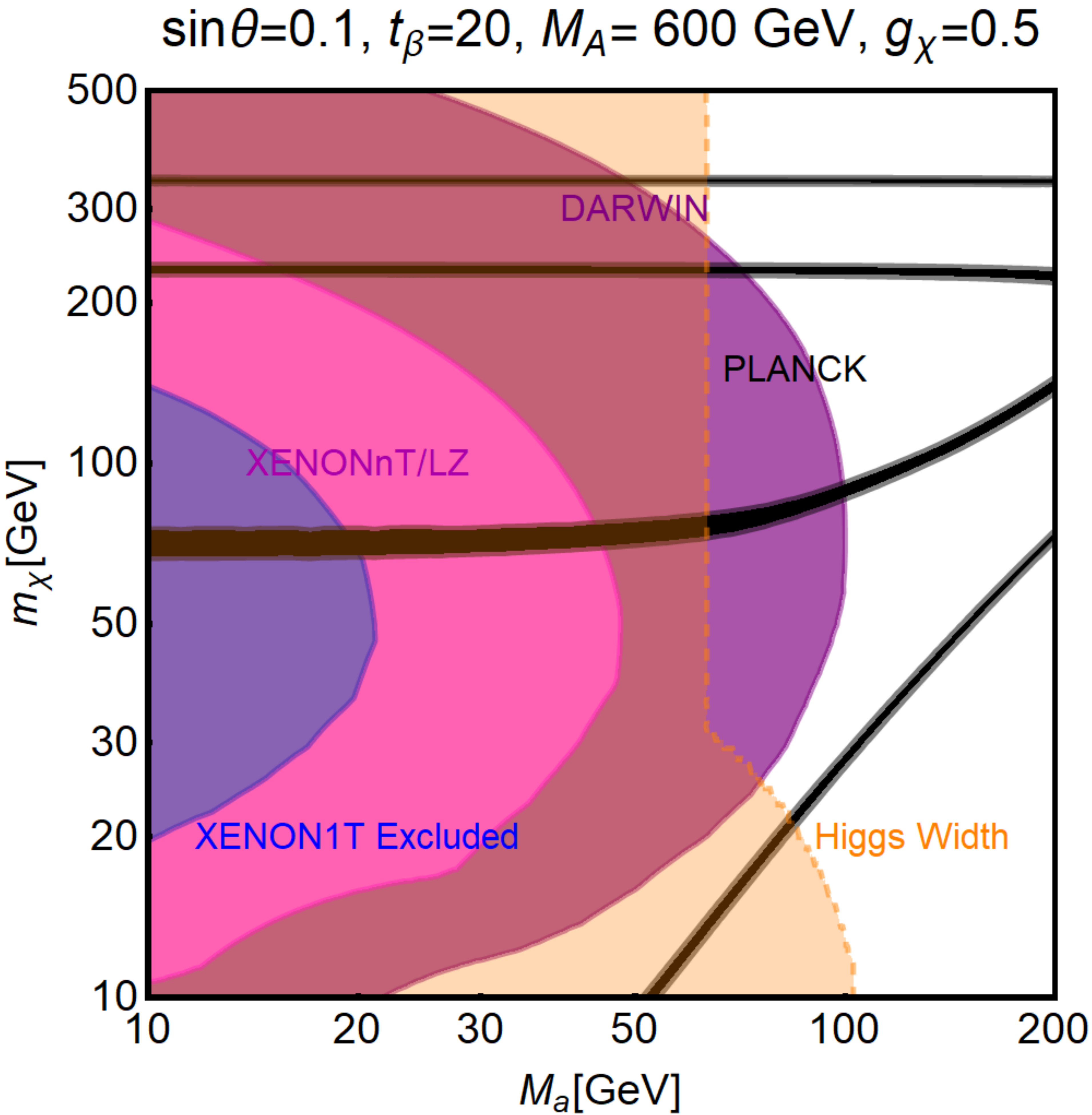}}~~
    \subfloat{\includegraphics[width=0.45\linewidth]{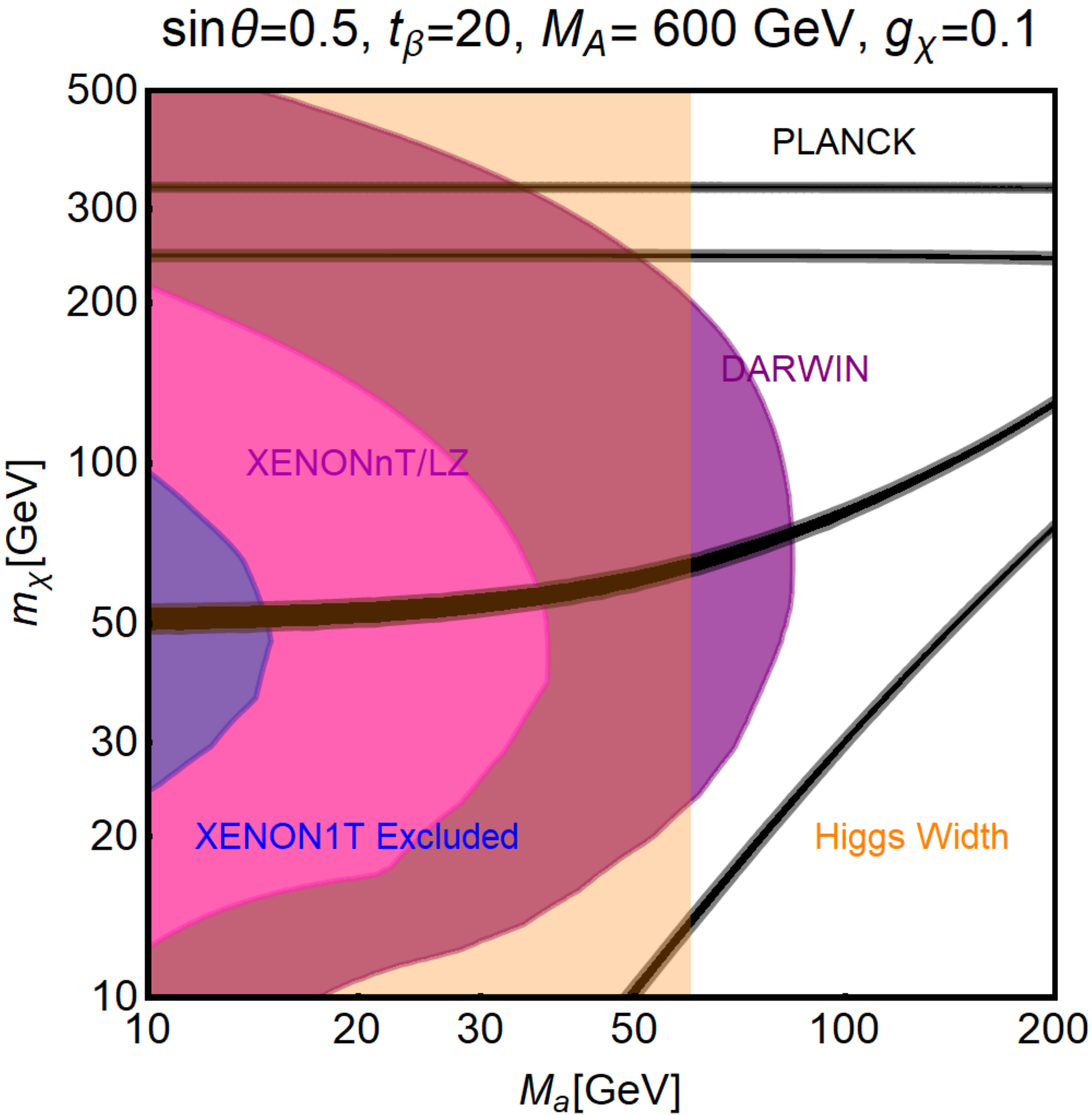}}
\vspace*{-1mm}
    \caption{Summary of constraints on the 2HDM+$a$ model in the plane $[m_\chi,M_a]$, for four fixed values of $g_\chi,\theta,\tb,M_A$. The black contours are for the correct relic density, the blue region is excluded by XENON1T while the magenta and purple regions represent the projected sensitivities of LZ/XENONnT and DARWIN, respectively. The orange region is excluded by searches of invisible Higgs decays.}
    \label{fig:pCOY}
\vspace*{-2mm}
\end{figure}

As is clear from the figure, most of the parameter space corresponding to the
correct DM relic density evades current direct detection constraints from
XENON1T unless high values of the mixing angle, like $\sin\theta=0.5$ in the
upper right panel of Fig.~\ref{fig:pCOY}, are considered. On the contrary, the
increase in sensitivity at future experiments will allow to probe more
efficiently the viable DM parameter space. For $M_a \gtrsim 100\,\mbox{GeV}$,
the scattering rate of the DM candidate lies well below the sensitivity of the
DARWIN experiment and even below the irreducible background represented by the
$Z$--mediated coherent scattering of SM neutrinos on nucleons, the so--called
neutrino floor\cite{Arcadi:2017wqi}. An efficient complementary bound comes
nevertheless from the invisible branching ratio of the 125 GeV Higgs boson which
extends, at a light DM particle, up to $M_a$. This is due to the process $h
\rightarrow aa^{*} \rightarrow 4 \chi$ with $a^*$ representing an off--shell $a$
boson. This process is kinematically allowed provided that $M_a > 2 m_\chi$ and
$M_h > M_a+2 m_\chi$ \cite{Abe:2018bpo}. 

\begin{figure}[!h]
\vspace*{-2mm} 
    \centering
    \subfloat{\includegraphics[width=0.45\linewidth]{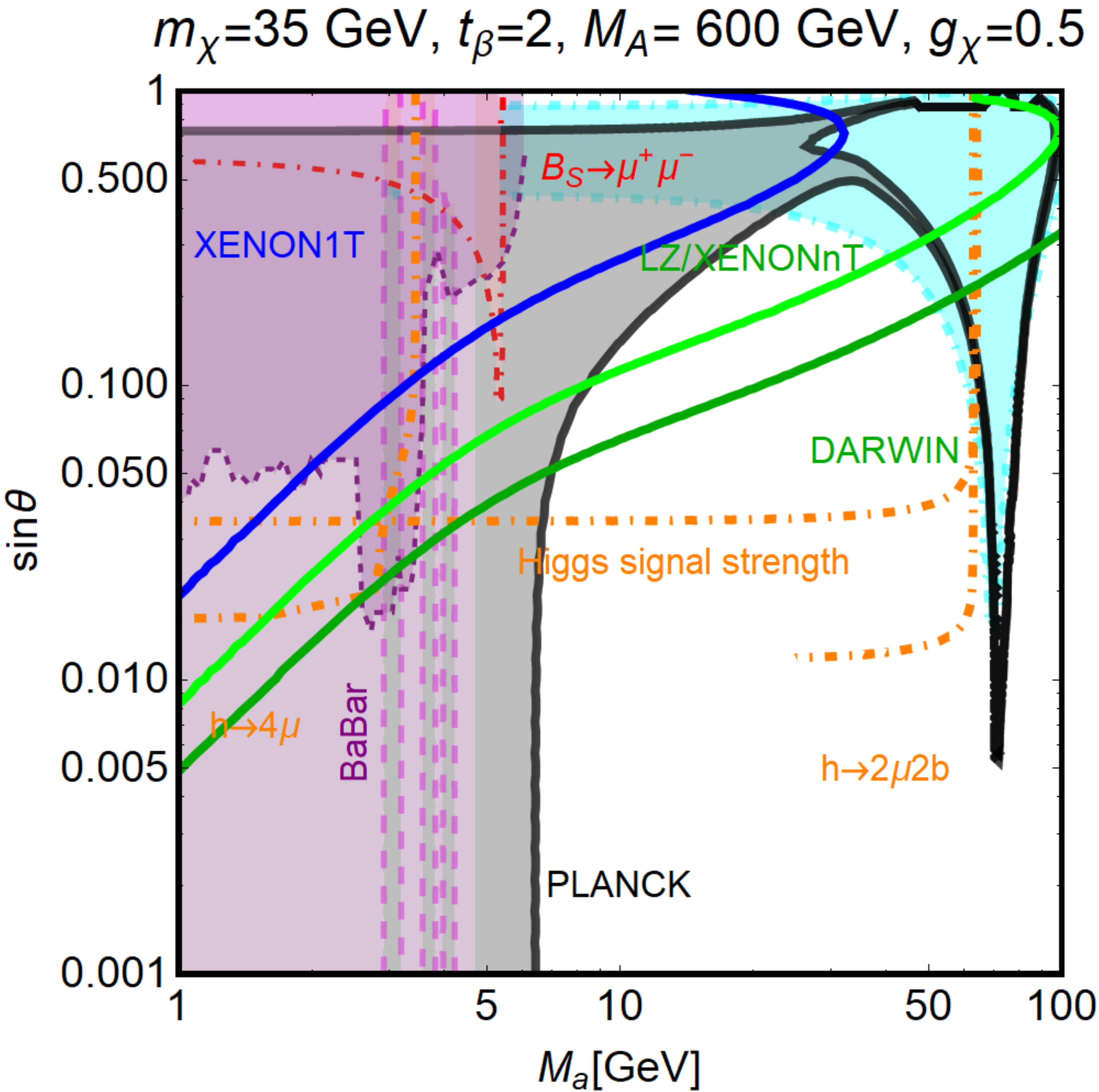}}~~
    \subfloat{\includegraphics[width=0.45\linewidth]{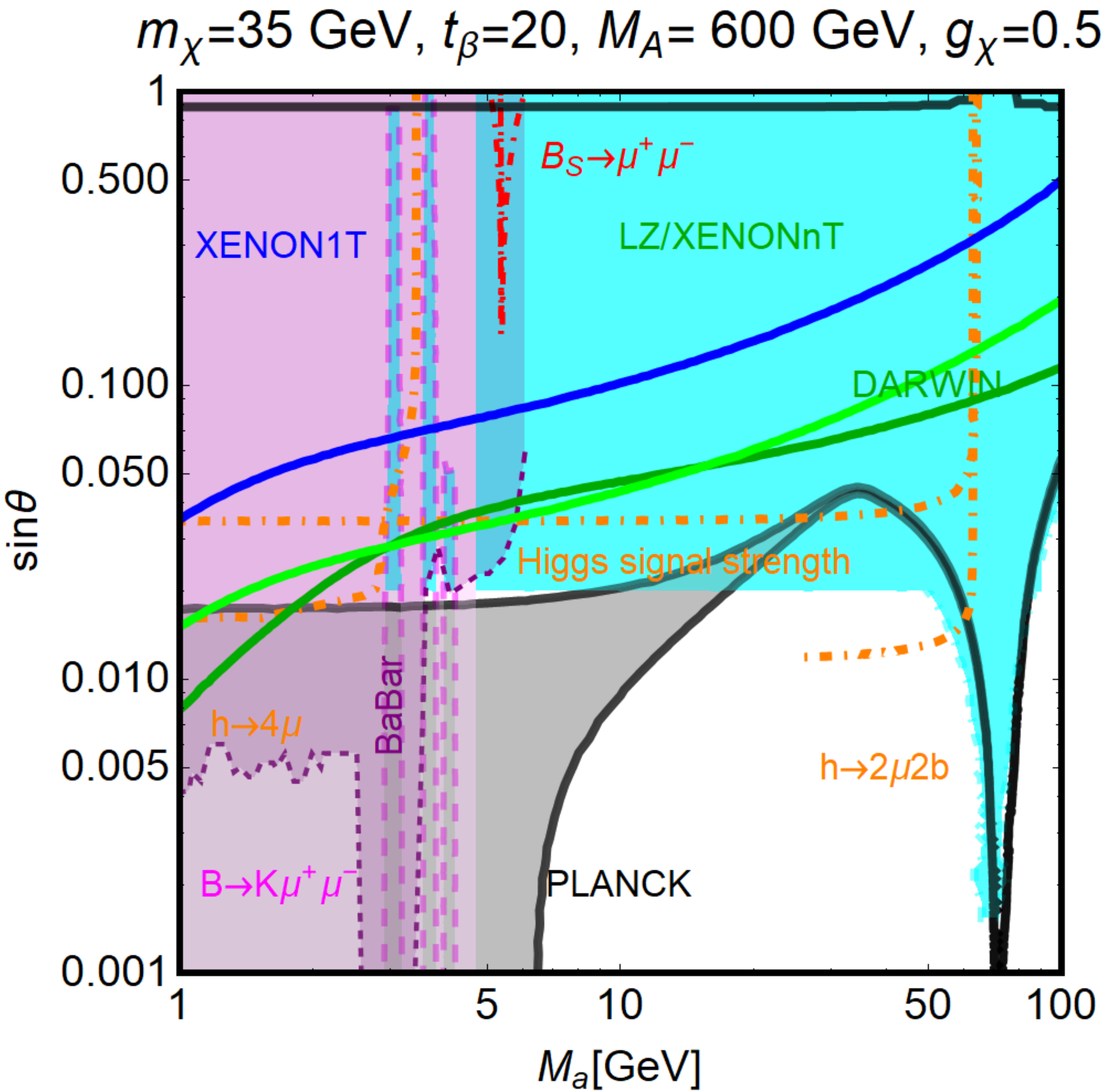}}\\[.2mm]
 \subfloat{\includegraphics[width=0.45\linewidth]{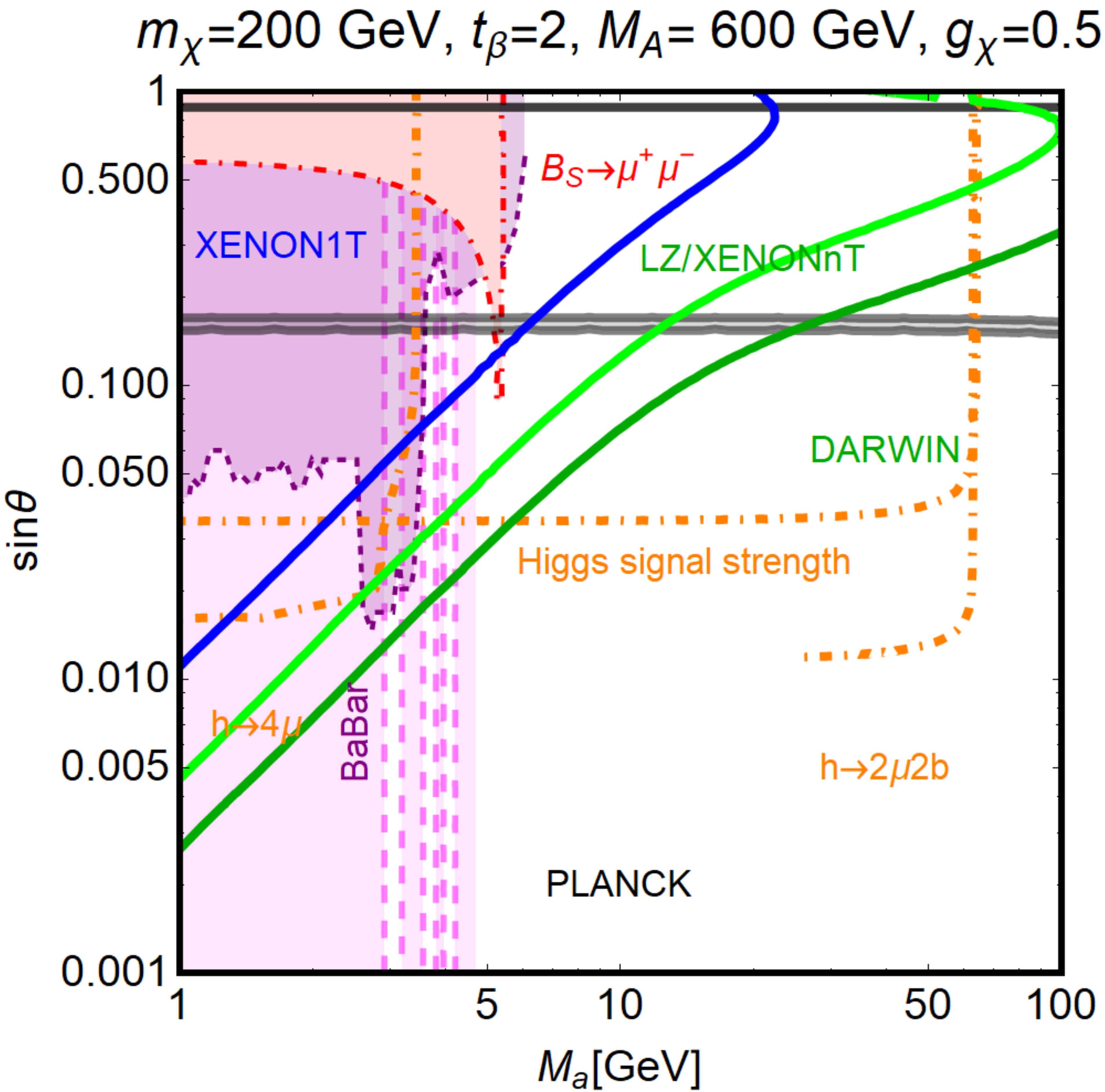}}~~
\subfloat{\includegraphics[width=0.45\linewidth]{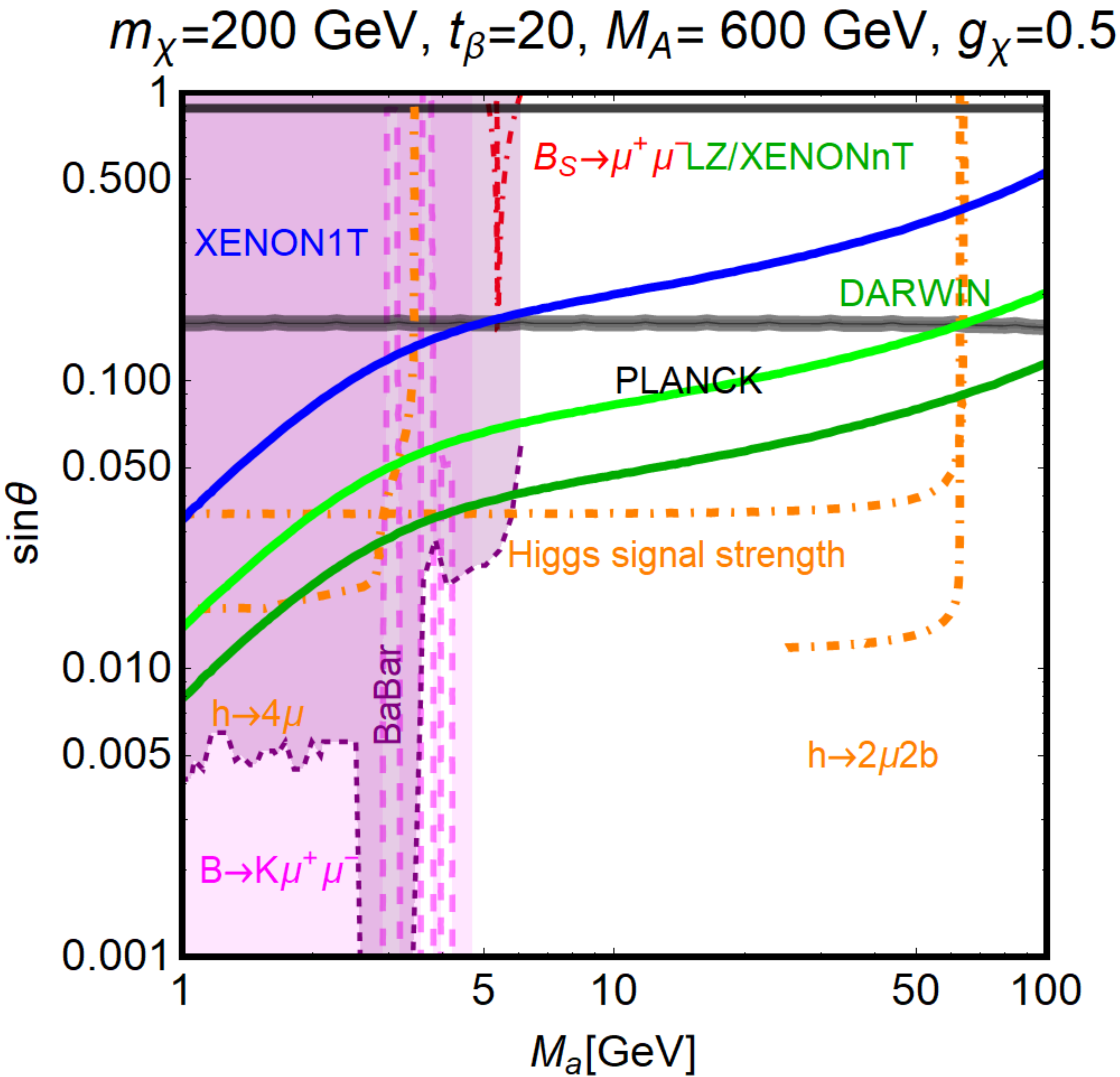}}
    \caption{Combined constraints on the 2HDM+light pseudoscalar model, for four benchmarks scenarios, from the DM relic density, direct detection, indirect detection, low energy experiments and searches for exotic decays of the 125 GeV Higgs boson. The various benchmark differ by the assignment of $m_\chi$, being 35 and 200 GeV for the upper and lower panels respectively, and of $\tan\beta$, 2 and 20 for the left and right panels respectively.}
    \label{fig:pCOY1}
\vspace*{-3mm}
\end{figure}

Fig.~\ref{fig:pCOY1} illustrates, instead, the scenario of a very light
singlet--like pseudoscalar $a$. For this reason, we present our results in the
bidimensional plane $[M_a,\sin \theta]$ for two DM masses, $m_\chi=35$ and
$m_\chi=200$ GeV and two values of $\tan\beta$, 2 and 20. In all cases, we set 
$M_A=M_H=M_{H^\pm}=600 \,\mbox{GeV}$ close to the lower limit imposed by flavor
physics and  $g_\chi= 0.5$ for the $a$ coupling to the DM $\chi$ states. As can
be seen, in the case of $m_\chi=35\,\mbox{GeV}$, the correct relic density is
achieved in a rather large region of the parameter space (gray region enclosed
within the black lines). This is due to the fact that for $M_a < m_\chi$, the DM
relic density is mostly due to the $\chi \chi \rightarrow aa$ process, whose
annihilation rate depends on $\cos\theta \sim 1$. For $m_\chi > M_a$, the main
annihilation channel is into $\bar b b$ final states, with a rate dependent on
$\sin^2 \theta$, so that the correct relic density is achieved only in narrow
contours, exhibiting the expected pole at $m_\chi \sim \frac12 M_a$. Due to the
$\tan\beta$ enhancement of the $ab\bar b$ coupling, the relevant DM region
shifts towards lower values of $\theta$ as $\tan\beta$ increases. Given the fact
that the cross section for DM annihilation  into $\bar b b$ is $s$--wave
dominated, it becomes subject to strong constraints from Fermi--LAT, as shown by
the cyan region in Fig.~\ref{fig:pCOY1}.

Concerning direct detection, the limits/projected sensitivities only show a
modest dependence on $\tan\beta$. The change of shape of the curves with $M_a$
is due to the fact that at low $M_a$, the DM scattering cross section is
dominated by the box diagram while at high $M_a$, the triangular loop
gives the largest contribution. At large values of $\tan\beta$, the regions with
a viable relic density move increasingly away from the sensitivity of the
experiments. 

A further effective constraint is due to the $h \rightarrow aa$ decay.
Considering only the limits from the Higgs signal strengths, values of
$\sin\theta$ greater than 0.05 are excluded within the full kinematical range of
this $h$ decay. Stronger bounds are obtained for more limited ranges of $M_a$,
when one considers searches of specific final states. Note that for the chosen
DM mass, the $h\rightarrow aa^*$ decay is not relevant and only the
region $M_a < \frac12 M_h$ is constrained.  Once this bound is enforced,
the correct DM relic density for $m_\chi = 35$ GeV can only be achieved in
regions of parameter space that are out of reach of direct detection
experiments. 

For $M_a \lesssim 10\,\mbox{GeV}$ the dominant bounds come, however, from low
energy processes. A light pseudoscalar can be emitted on--shell in $b
\rightarrow s$ and $s \rightarrow d$
transitions\cite{Fayet:2007ua,Andreas:2010ms}, altering the rates of meson
decays. In the mass range $1\,\mbox{GeV} \lesssim M_a \lesssim 10\,\mbox{GeV}$,
the most relevant bounds come from the $\Upsilon \rightarrow a
\gamma$\cite{Lees:2011wb,Lees:2012iw,Lees:2012te} (the excluded region has been
marked in purple and labelled as BaBar in Fig.~\ref{fig:pCOY1}), $B_s
\rightarrow \mu^+ \mu^-$ (the excluded region is marked in red in
Fig.~\ref{fig:pCOY1})\cite{CMS:2014xfa} and $B \rightarrow K \mu^+
\mu^-$\cite{Aaij:2014zsa} (the excluded region is marked in magenta in
Fig.~\ref{fig:pCOY1}). We refer to Ref.~\cite{Arcadi:2017wqi} for details on the
determination of this bounds and to Ref.~\cite{Dolan:2014ska} for a more
extensive review.

In the case of a heavier DM, $m_\chi\!=\!200\,\mbox{GeV}$, a slight decrease
of the sensitivity from direct detection experiments is observed. It has two
advantages though. First, the DM annihilation cross section is enhanced by the
$\bar t t$, $ha$ and $Zh$ channels. If $m_\chi \!\gg \! M_a$, the corresponding
annihilation rates do not depend on $M_a$. The isocontours of the relic density
are then just horizontal lines corresponding to specific values of $\sin\theta$.
More precisely, the lines in the lower panels of Fig.~\ref{fig:pCOY1} at
$\sin\theta\simeq 1$ correspond to a relic density mostly determined by
annihilations into $\bar t t$ pairs while the ones at $\sin\theta \sim
0.1$ correspond to a dominant contribution from the $ha$ final state.
Furthermore, bounds from indirect experiments are evaded as they cannot probe
thermal DM particles with such a high mass yet.

%% file: sec-MSSM.tex
\section{Supersymmetric extensions of the SM} 

\subsection{The MSSM}

\subsubsection{SUSY and the pMSSM}

Supersymmetry (SUSY)
\cite{Wess:1974tw,Golfand:1971iw,Drees:2004jm,Baer:2006rs,Martin:1997ns} was
widely considered as the most attractive extension of the SM. The main reason
was that  it solves, at least technically, the hierarchy and naturalness
problems and prevents the Higgs boson mass from acquiring very large radiative
corrections  unless an unnatural and fine adjustment of parameters is
performed.  Later on, two other  motivations for low energy SUSY  were 
recognized: the satisfactory unification of the  three gauge couplings of the SM
at the GUT scale and the fact that one can naturally  arrange so that the
lightest SUSY particle (LSP) is massive, electrically neutral, weakly
interacting and absolutely stable, in short, the ideal candidate for the DM
\cite{Goldberg:1983nd,Ellis:1983ew,Drees:1992am}. The most intensively studied
low energy SUSY extension of the SM is the  most economical one, the MSSM
\cite{Martin:1997ns,Haber:1984rc,Djouadi:1998di,Chung:2003fi} that we briefly
summarize below. 

In the MSSM,  one first assumes the SM  gauge group only and associates a 
spin--$\frac12$ gaugino to each  gauge boson, a bino $\tilde B$, three winos
$\tilde W_{i}$ and the gluinos $\tilde g$ that correspond to the U(1), SU(2) and
SU(3) groups respectively. One also assumes  the minimal particle content, i.e.
only three generations of fermions without right--handed neutrinos and their
spin--zero partners, the left-- and right handed sfermions $\tilde f_L$ and
$\tilde f_R$, which mix to give the physical states  $\tilde f_1$ and $\tilde
f_2$.  For consistency reasons to be discussed later, one has to introduce two
doublets of Higgs fields $\Phi_1$ and $\Phi_2$  and their spin--$\frac12$
partners, the higgsinos $\tilde H_i$, these will mix with the gauginos  to
produce the two chargino $\chi_i^\pm$  and the four neutralino $\chi_i^0$
physical states. Then, one introduces a discrete and multiplicative symmetry
called $R$--parity under which the SM particles are even and the SUSY ones are
odd \cite{Farrar:1978xj}. Assuming that this symmetry is conserved makes the
lightest SUSY particle absolutely stable. 

The most general globally supersymmetric superpotential, compatible with  gauge
invariance, renormalizability and $R$--parity writes  in terms of hatted
superfields, that contain both the SM particle fields and those of their superpartners, as 
\beq
{\cal W}=\sum_{\rm i,j= gen.} - Y^u_{ij} \, {\widehat {u}}_{Ri} \widehat{\Phi_2} \!  \cdot  \! {\widehat{ Q}}_j+ Y^d_{ij} \, {\widehat{ d}}_{Ri} \widehat{\Phi}_1   \! \cdot  \! {\widehat{ Q}}_j+Y^\ell_{ij} \,{\widehat{\ell}}_{Ri} \widehat{\Phi}_1  \! \cdot  \! {\widehat{ L}}_j+ \mu \widehat{\Phi}_2  \! \cdot  \! \widehat{\Phi}_1 \, . 
\label{defW}
\eeq
The product between SU(2)$_{\rm L}$ doublets for Higgs, quark and lepton fields
reads $\Phi \cdot Q \equiv \epsilon_{a b} \Phi^a Q^b$; etc... where $a, b$ are
SU(2)$_{\rm L}$ indices and $ \epsilon_{12}=1 = - \epsilon_{21}$; 
$Y^{u,d,\ell}_{ij}$ are Yukawa couplings among families. The first three terms
are a generalization of the SM Yukawa interactions and the last term is a
globally supersymmetric Higgs mass term.

In order to explicitly break SUSY, one then adds a collection of soft terms that
do not reintroduce quadratic divergences \cite{Girardello:1981wz}: mass terms
for the gauginos $\sum_{i=1,2,3} \frac12 M_i V_i^\mu V_{i\mu}$, mass terms for
the  sfermions $\sum_i m^2_{{\tilde {f}}_i} {\tilde{f}}_i^{\dagger}{\tilde{f
}}_i$, and also mass and bilinear terms for the Higgs bosons and  trilinear
couplings between sfermions and Higgs bosons:  
\begin{align}
\label{eq:LHiggs-MSSM}
-{\cal L}_{\rm Higgs} &= m^2_{\Phi_2} \Phi_2^{\dagger} \Phi_2+m^2_{\Phi_1}  \Phi_1^{\dagger}  \Phi_1 + B \mu (\Phi_2 \! \cdot  \! \Phi_1 + {\rm h.c.} ) 
\nonumber \\ 
&+ {\sum_{\rm i,j=gen} { \left[ A^u_{ij} Y^u_{ij}  {\tilde{u}}^*_{R_i} \Phi_2  \! \cdot  \! {\tilde{Q}}_j+ A^d_{ij} Y^d_{ij}  {\tilde{d}}^*_{R_i} \Phi_1  \! \cdot  \!  {\tilde{Q}}_j +A^l_{ij} Y^\ell_{ij} {\tilde{\ell}}^*_{R_i} \Phi_1  \! \cdot {\tilde{L}}_j \ + \ {\rm h.c.} \right] }}\, .~~ 
\end{align}

Although incomplete, since it does not have right--handed (s)neutrinos and has a
problem with the $\mu$ parameter, this model served as a benchmark for SUSY
phenomenology. Nevertheless, it has a too large number of free parameters, 105
in addition to the 19 parameters of the SM which, for generic values, lead to
severe problems with FCNCs, additional CP--violation,  color and charge breaking
minima, etc. To cure these, a less general scenario  was introduced, the
phenomenological MSSM (pMSSM) \cite{Djouadi:1998di,Djouadi:2002ze} in which one
assumes:  $(i)$ all soft SUSY--breaking parameters are real leading to the
absence of new sources for CP--violation; $(ii)$ the sfermion mass and trilinear
coupling matrices are all diagonal,  implying the absence of FCNCs at
tree--level;  $(iii)$ equal soft masses and trilinear couplings of the first and
second sfermion generations  to cope with constraints from heavy
flavors.  

Making these three assumptions lead to the pMSSM with only 22 input parameters,
namely:  the ratio of vevs of the two--Higgs doublet fields $\tan \beta$, the
two Higgs mass parameters squared $m_{\Phi_1}^2, m_{\Phi_2}^2$ (which can be
traded against one Higgs mass $M_A$ and the parameter $\mu$); the gauginos mass
parameters  $M_1, M_2, M_3$, the common first/second and the third generation
sfermion mass parameters $m_{\tilde{q}}, m_{\tilde{u}_R}, m_{\tilde{d}_R}, 
m_{\tilde{l}}, m_{\tilde{e}_R}$ and trilinear couplings $A_u, A_d, A_e$. Such a
model is more predictive and easier to investigate phenomenologically.  

One can further constrain this model by requiring that the soft SUSY--breaking
parameters obey a set of universal boundary conditions at the GUT scale. The
minimal supergravity model (mSUGRA)
\cite{Chamseddine:1982jx,Barbieri:1982eh,Hall:1983iz} is the most celebrated one
and manages to lead to  acceptable spectra with only three parameters besides
$\tan\beta$ and the sign of $\mu$ ($|\mu|^2$ and the parameter $B$ are fixed by
the requirement of proper EWSB):  the common soft SUSY--breaking terms of all
scalar masses $m_0$, gaugino masses $M_{1/2}$ and trilinear scalar  interactions
$A_0$ defined at the GUT scale. The various parameters at the low scale are
obtained through renormalization group evolution. This model is, however,
severely constrained by present data and requires e.g. a very heavy sfermion
spectrum to be viable.  

In our study here, since we are mostly interested in the MSSM Higgs sector 
\cite{Gunion:1989we,Djouadi:2005gj,Carena:2002es,Heinemeyer:2004gx} serving as a
portal to the DM particles, we will analyze a SUSY scenario which is half way
between mSUGRA and the pMSSM. On the one hand, we assume a common mass and
trilinear coupling for all the sfermions which makes that this sector can be
simply described by two parameters only, instead of  16 in the pMSSM: a common
scalar mass $M_S$, which will be identified with the SUSY scale taken to be the
geometric average  of the two stop squark masses $M_S = \sqrt{  m_{\tilde t_1},
m_{\tilde t_2} }$  and the trilinear coupling in the top/stop sector $A_t$ that
we will assume to be such that $A_t = \sqrt {6} M_S$ to maximize the radiative
corrections in the Higgs sector (as will be seen later). We will then have a
model with only 6 free parameters besides the sfermionic ones $M_S$ and $A_t$
above, namely: $\tan\beta$ and $M_A$ for the Higgs sector and  $M_1, M_2, M_3$
and $\mu$ for the gaugino and higgsino sectors. In addition, we will make a few
simplifying and realistic assumptions in the scenario that we will mostly
consider here:\vspace*{-1mm}

-- $M_S$ will be taken to be large  $M_S \gg M_Z$ so that it has no
phenomenological impact, i.e. the sfermion will decouple from the low energy
spectrum and are integrated out
\cite{ArkaniHamed:2004fb,Giudice:2004tc,ArkaniHamed:2004yi,Wells:2004di,Bernal:2007uv}.\vspace*{-1mm}

-- While the gluino mass parameter is kept free but large, $m_{\tilde g}\!
\approx \! M_3 \! \gsim\! 2\,$TeV, as a result of the negative LHC searches, we 
keep the GUT relation between the wino and bino mass parameters $M_1 \simeq
\frac12 M_2$, hence reducing the number of inputs in this  sector.\vspace*{-1mm}

-- From GUT restrictions, $\tb$ will be assumed in the range $1 \lsim \tb \lsim
m_t/\bar m_b$ with the lower and upper ranges favored by Yukawa coupling
unification at $M_{\rm GUT}$ \cite{Carena:1994bv,Barger:1992ac}.

We have now all elements to study the MSSM Higgs and DM sectors and their interplay. 

\subsubsection{The Higgs sector and the hMSSM}

In the MSSM, two doublets of complex scalar fields of opposite hypercharge,  
$\Phi_1$ and $\Phi_2$, are required to break spontaneously the electroweak
symmetry. This is necessary first, for the cancellation of chiral anomalies, as a unique Higgs doublet would have introduced a charged higgsino that would spoil
this cancellation. A second reason is that one cannot generate as in the SM the
masses of the isospin  $-\frac12$ fermions with the doublet $\Phi$ and those of
isospin $+\frac12$ fermions with its conjugate field (i.e. with opposite
hypercharge) $\tilde{\Phi}= i\tau_2 \Phi^*$ since conjugate fields are not
allowed in the superpotential. Hence, a second doublet $\Phi_2$ has to be
introduced to play this role and the Higgs Lagrangian can be then written as in
eq.~(\ref{eq:LHiggs-MSSM}). The resulting scalar Higgs potential $V_\Phi$  when 
all terms from various sources are added up, can be written as
\cite{Gunion:1989we,Djouadi:2005gj,Carena:2002es,Heinemeyer:2004gx}
\begin{eqnarray}
V_\Phi \!=\! \bar m_{1}^2|\Phi_1|^2 \! + \!  \bar m_{2}^2|\Phi_2|^2
\! - \!  \bar m_3^2\epsilon_{ij} (\Phi_1^i \Phi_2^j \! +\!  {\rm h.c.})
\! +\!  {g^2+g'^2 \over 8} (|\Phi_1|^2 \! -\!  |\Phi_2|^2)^2 \! +\!  {g^2 \over 2} |\Phi_1^\dagger \Phi_2|^2 \, ,~~ 
\end{eqnarray}
where $\bar m_{1}^2 \! =\! \mu^2 \!+\! m^2_{\Phi_{1}},  \bar m_{2}^2\! = \!
\mu^2\! +\! m^2_{\Phi_{2}}$ and $\bar m_3^2\! = \! B\mu$ are the Higgs soft
terms,  and $g,g'$ the electroweak gauge couplings.  Some interesting features
of the MSSM already emerge at this stage: $i)$ Its Higgs sector is a 2HDM
of Type II as the field $\Phi_1$ generates the masses  of up--type quarks while
$\Phi_2$ generates those of down--type quarks and leptons. $ii)$ The quartic
Higgs couplings are fixed in terms of the gauge couplings and hence, contrary to
a general 2HDM \cite{Branco:2011iw} which has at least 6 free parameters, one
has only 3 free parameters in  the MSSM, $\bar m_{1}^2, \bar m_{2}^2$ and
$\bar m_{3}^2$. $iii)$  While the combinations  $\bar m_{1,2}^2$ are real, 
$\bar m_{3}^2= B\mu$ can be complex but any phase there can be absorbed into
those of the fields $\Phi_1$ and $\Phi_2$ resulting into a CP conserving MSSM
scalar potential at tree--level. $iv)$ For proper electroweak symmetry breaking,
i.e. to have  stable minimum, a potential that is bounded from below and has a
saddle point at the minimum, several conditions have to fullfiled which lead to 
$m_{\Phi_1}^2 \neq m_{\Phi_2}^2$ and hence for electroweak breaking  to occur,
one needs SUSY breaking\footnote{This provides a nice connection between EWSB
and SUSY--breaking. Note that from renormalization group running, one can obtain
$m_{\Phi_2}^2<0$ or $m_{\Phi_2}^2 \ll m_{\Phi_1}^2$ which then  triggers
electroweak symmetry breaking  (the so--called radiative breaking) \cite{Ibanez:1982fr}, making it more natural in SUSY models than in the SM.}.   

To obtain the Higgs spectrum, one  requires that the minimum of the potential $V_\Phi$  breaks the  ${\rm SU(2)_L \times U(1)_Y}$ group while preserving the electromagnetic symmetry  U(1)$_{\rm Q}$. The neutral components of the two Higgs fields develop vacuum expectations values 
\bea
\Phi_1 \! = \! (H_1^0,H_1^-) \! \to \! \frac{1}{\sqrt{2}} \left( v_1 \!+ \! H_1^0 \! + \! i P_1^0 , H_1^- \right) \, ,\nonumber \\
\Phi_2\!= \! (H_2^+,H_2^0) \! \to \!\frac{1}{\sqrt{2}} \left( H_2^+ , v_2\! + \! H_2^0 \! + \! i P_2^0 ~~ 
\label{eq:decomp-field}
\right)  \, , 
\eea
where $\langle H_{1,2}^0 \rangle =  v_{1,2}/\sqrt 2$ with $(v_1^2+v_2)^2=v^2=(246~{\rm GeV})^2$ and $\tb= v_2/v_1$. Minimizing the potential at the electroweak minimum, one obtains the two conditions:
\bea
2B\mu \!&=& \! (m_{\Phi_1}^2 \! - \! m_{\Phi_2}^2) \tan 2\beta \!+ \! M_Z^2 \sin2\beta \ , \nonumber \\ 
\mu^2 \cos\beta \! &=& \! (m_{\Phi_2}^2\sin^2\beta \! - \! m_{\Phi_1}^2 \cos^2 \beta) \! - \! \frac12 M_Z^2\cos2\beta, 
\label{min-conditions}
\eea
which lowers the number of parameters needed in the MSSM  Higgs sector to only two. To obtain the Higgs physical states and their masses, after developing 
the fields eq.~(\ref{eq:decomp-field}) into real and imaginary parts which 
correspond to, respectively, the CP--even Higgs bosons and the CP--odd Higgs and Goldstone bosons, and diagonalize the mass matrices 
\begin{eqnarray}
\label{eq:matrix-tree}
{\cal M}_R^2 \!= \! \left[ \! \begin{array}{cc} \!- \! \bar{m}_3^2 \tb \! + \! M_Z^2 \cos^2 \beta  &  \bar{m}_3^2 \! - \! M_Z^2 \sin \beta \cos \beta \\
\bar{m}_3^2 \! - \! M_Z^2 \sin \beta \cos \beta & \! -\! \bar{m}_3^2 {\rm cot} \beta \! + \! M_Z^2 \sin^2 \beta \end{array} \! \right] , \,  
{\cal M}_I^2 \! = \! \left[ \! \begin{array}{cc} \! -\! \bar{m}_3^2 \tb  &  \bar{m}_3^2 \\ \bar{m}_3^2 & \! -\bar{m}_3^2 {\rm cot} \beta \! \end{array} \! \right]\! . ~ 
\end{eqnarray}  
The pseudoscalar and charged Higgs masses are simply obtained by a rotation of angle $\beta$
\beq
M_A^2= - \bar{m}_3^2 (\tb + {\rm cot} \beta) = - 2 \bar{m}_3^2/
\sin 2\beta \, , \  \ \ \ 
M_{H^\pm}^2= M_A^2 + M_W^2 , 
\label{H+Amass:tree}
\eeq
while those of CP--even Higgs bosons are obtained from a rotation of angle $\alpha$
 \beq
M_{h,H}^2= \frac{1}{2} \left[ M_A^2+M_Z^2 \mp \sqrt{ (M_A^2+M_Z^2)^2 -4M_A^2
M_Z^2 \cos^2 2\beta } \right] , 
\label{Hmasses:tree}
\eeq
where the mixing angle $\alpha$ is given in compact form by
\beq
\alpha  = \frac{1}{2} {\rm arctan} \bigg({\rm tan} 2\beta \, \frac{M_A^2 
+ M_Z^2}{ M_A^2-M_Z^2} \bigg)\ , \ \ - \frac{\pi}{2} \leq \alpha \leq 0 \, .
\label{alpha:tree}
\eeq
Thus, the supersymmetric structure of the theory has imposed strong  constraints
on the Higgs spectrum and, out of the six parameters which  describe a 2HDM,
only two parameters, taken as $\tb$ and $M_A$, are free at
tree--level. In addition, a strong hierarchy is imposed on the Higgs mass
spectrum and, besides the relations $M_H > {\rm max} (M_A,M_Z)$ and $M_{H\pm} 
>M_W$ derived from the equations above, we have the very important tree--level
constraint on the lightest $h$ boson mass 
\beq
M_h  \leq {\rm min} (M_A, M_Z) \cdot |\cos2\beta|  \leq M_Z . 
\label{treelevelMh}
\eeq

Also, if $M_A \! \gg \! M_Z$, one obtains the equalities  $M_H \! \approx \! M_A
\! \approx \! M_{H^\pm}$ and $M_h \!= \! M_Z|\cos2 \beta|$. 

Turning to the MSSM Higgs couplings, those to the massive gauge bosons  $V=W,Z$
obtained from the kinetic terms of the $\Phi_1$ and $\Phi_2$ fields,
follow a simple pattern. They are proportional to either $\sin(\beta-\alpha)$ or
$\cos(\beta- \alpha)$ and are thus complementary, the sum of  their squares
being the square of the SM Higgs coupling $g_{HVV}^{\rm SM}$.  For large $M_A$
values, one can expand the Higgs--VV  couplings in powers of $M_Z/M_A$ to obtain
\begin{eqnarray}
g_{HVV} & \propto  & \cos(\beta-\alpha)  \stackrel{M_A \gg M_Z} \longrightarrow 
\frac{M_Z^2} {2M_A^2} \sin4 \beta \quad \ \stackrel{\tb \gg 1} 
\longrightarrow  \quad - \frac{2 M_Z^2} {M_A^2 \tb} \ \to 0 \nonumber \\
g_{hVV} & \propto & \sin(\beta-\alpha)   \stackrel{M_A \gg M_Z} \longrightarrow 1- \frac{M_Z^4} {8M_A^4} \sin^2 4\beta  \stackrel{\tb \gg 1} \longrightarrow 1- \frac{2 M_Z^4} {M_A^4 \tan^2 \beta} \to 1 
\label{gHVVdecoup}
\end{eqnarray}
where we have also displayed the limits at large  $\tb$. One sees that for
$M_A\gg M_Z$, $g_{HVV}$ vanishes while $g_{hVV}$ reaches unity, i.e. the SM
value; this occurs more  quickly if $\tb$ is large. As for the couplings of a
$Z$ boson to two Higgs states, because of CP--invariance the two scalars must
have opposite parity and therefore  there are no $Zhh,ZHh, ZHH,ZAA$ couplings
and only the $ZhA$ and $ZHA$ couplings are allowed. The latter follow,
respectively, the $HZZ$ and $hZZ$ couplings, namely $g_{ZhA} =
\cos(\beta-\alpha)$ and  $g_{ZHA} = \sin(\beta-\alpha)$. 

All this is similar to the case of 2HDMs that we discussed in the previous
section. It turns out that the MSSM Higgs couplings to fermions are exactly
those of a Type II 2HDM  and which have been given in Table
\ref{table:2hdm_type}. The limiting values of these couplings, as obtained in
the alignment limit of the 2HDM $\alpha= \beta-\frac\pi2$ and which have been
also given in the Table,  are the same as the couplings that are obtained  in
the MSSM in the limit of a very heavy $A$ boson which, from
eq.~(\ref{alpha:tree}), also gives $\alpha= \beta-\frac\pi2$. In particular the
$h$ couplings become SM--like, $g_{hff}\to 1$,  and the couplings of the
$H,A,H^\pm$ states to isospin down--type fermions are proportional to $\tb$,
while those to up--type quarks are proportional to $\cot \beta$.  

In fact, for $M_A \gg M_Z$ we are in the so--called decoupling limit of the MSSM
\cite{Gunion:2002zf} in which the $h$ boson reaches its maximal mass value and
its couplings to fermions and gauge bosons  as well as its self--couplings
become SM--like. The heavier $H,A$ and $H^\pm$ states become degenerate in mass,
decouple from massive gauge bosons and couple to fermions in a similar fashion.
The decoupling  regime, which is controlled by  $\cos^2 (\beta-\alpha)$ being
close to zero or the angle $\alpha$ being close to $\beta-\frac\pi2$,  is
similar to the alignment limit which occurs in the 2HDM and the two models are
almost  identical in the 2HDM benchmark scenario that we have adopted in section
5 where we also assumed $M_H \! \approx \! M_A \! \approx \! M_{H^\pm}$.  

The above simple picture of the MSSM Higgs sector at tree--level, with only the
two inputs $M_A$ and $\tb$ needed, is nevertheless spoilt by large radiative
corrections: at higher orders, almost all parameters of the MSSM will in
principle enter the determination of the Higgs masses and couplings 
\cite{Okada:1990vk,Ellis:1990nz,Haber:1990aw,Chankowski:1991md,Heinemeyer:1998np,Degrassi:2002fi,Carena:2002es,Allanach:2004rh,Carena:2005ek,Carena:2013ytb,Bahl:2018qog}. These corrections can be described by introducing a general $2\times
2$ matrix  $\Delta {\cal M}_{ij}^2$ that corrects the CP--even Higgs mass matrix
${\cal M}_{R}^2$ of eq.~(\ref{eq:matrix-tree}) and which involves the various
MSSM contributions. Fortunately, the problem can be simplified by considering
only the by far leading  radiative corrections to the  mass matrix that are
controlled by the top Yukawa coupling,  $\lambda_t =  m_t/v \sin\beta$, and
which appears with the fourth power 
\cite{Okada:1990vk,Ellis:1990nz,Haber:1990aw,Chankowski:1991md}. In this case, only a few additional
parameters, such as the stop masses $m_{\tilde t_1}, m_{\tilde t_2}$  and the
trilinear coupling $A_{t}$ will enter the Higgs sector. If only these large
contributions are considered, one obtains a very simple analytical expression
for the correction  matrix $\Delta {\cal M}_{ij}^2$ 
\begin{eqnarray} 
\label{higgscorr}  
\Delta {\cal M}_{11}^2 \sim  \Delta {\cal M}_{12}^2 \sim 0 \, , \quad \Delta {\cal M}_{22}^2 \sim \! \frac{3 \bar{m}_t^4}{2\pi^2 v^2\sin^ 2\beta} \left[ \log \frac{M_S^2}{\bar{m}_t^2} \!+ \! \frac{X_t^2}{M_S^2} \left( 1 \! - \! \frac{X_t^2}{12M_S^2} \right) \right],  
\end{eqnarray}  
where $M_S$ is the SUSY scale $M_S\! =\! \sqrt{m_{\tilde{t}_1}m_{\tilde{t}_2}} $
and $X_{t}$ the stop mixing parameter, $X_t\!= \! A_t\! - \! \mu/\tb$; $\bar 
m_t$ is the running ${\rm \overline{MS}}$ top quark mass introduced to account
for the leading two--loop radiative corrections in a renormalisation--group
improved approach. Sub-leading contributions, such as those controlled by the
bottom Yukawa coupling $\lambda_b\! = \! m_b/v \cos\beta$ which at large values
of $\tb$ becomes relevant can be included in the component $\Delta {\cal
M}_{22}^2$ as well. Other non--leading corrections enter in all $\Delta
{\cal M}_{ij}^2$ terms of the correction matrix though, like the ones that are
proportional to $\lambda_t^2$ or $\lambda_b^2$ or those originating from the
gaugino sector, which introduce  a dependence on the parameters $M_{1,2,3}$ in
addition. However, these contributions are much smaller and can be ignored
to first approximation \cite{Degrassi:2002fi,Carena:2002es,Allanach:2004rh,Carena:2005ek,Carena:2013ytb,Bahl:2018qog}. 

In the approximation above, the maximal value $M_h^{\rm max}$ is given by  
$M_h^2 \to  M_Z^2 \cos^2 2 \beta +\Delta {\cal M}_{22}^2$ in the  decoupling
regime with a heavy $A$ state,  $M_A\! \sim \mathcal{O}$(TeV). It can be
obtained  with the following choice of parameters: $i)$  relatively high $\tb$
values, $\tb \gsim 5$, so that $\cos^2 2 \beta \approx 1$;  $ii$) heavy stops,
i.e. values $M_S\!  \gsim \!1$--3 TeV to generate large logarithmic corrections
and $iii)$ a stop trilinear coupling $X_t=\sqrt{6}M_S$, i.e. the so--called
ma\-xi\-mal  mixing scenario that maximizes the stop loops
\cite{Carena:2005ek}. If the parameters are optimized as above, the maximal
$M_h$ value can reach the one measured at the LHC, $ M_h=125$ GeV.   

It was pointed out in
Refs.~\cite{Djouadi:2013vqa,Maiani:2013hud,Djouadi:2013uqa,Djouadi:2015jea,Bagnaschi:2015hka}
that when the measured value  of $M_h$  is taken into account and only the
dominant radiative corrections are considered, the MSSM Higgs  sector can be
again described with  two free parameters such as $\tb$ and $M_A$ as at 
tree--level. Indeed, the dominant corrections that involve the SUSY parameters
will be then  fixed by $M_h$,  leading  to a rather simple parametrisation of
the MSSM Higgs sector. Hence,  if in the $\Delta {\cal M}_{ij}^2$ correction
matrix only the leading $\Delta{\cal M}^{2}_{22}$ entry  is considered, on can
trade it against the by now known $M_h$ value and obtain for the $H$ mass and
the angle $\alpha$ 
\begin{eqnarray} 
M_{H}^2 &= & \frac{(M_{A}^2+M_{Z}^2-M_{h}^2)(M_{Z}^2 \cos^2 {\beta}+M_{A}^2
\sin^{2}{\beta}) - M_{A}^2 M_{Z}^2 \cos^{2}{2\beta} } {M_{Z}^2 \cos^{2}{\beta}+M_{A}^2 \sin^{2}{\beta} - M_{h}^2}\, ,  \\
\ \ \  \alpha &=& -\arctan\left(\frac{ (M_{Z}^2+M_{A}^2) \cos{\beta} \sin{\beta}} {M_{Z}^2 \cos^{2}{\beta}+M_{A}^2 \sin^{2}{\beta} - M_{h}^2}\right)
\, , 
\label{eq:hMSSM} 
\end{eqnarray}
while the mass of the charged Higgs boson, which is not much  affected by
radiative corrections, is still given by $M_{H^\pm} \simeq \sqrt { M_A^2 +
M_W^2}$. This is called the $h$MSSM approach
~\cite{Djouadi:2013uqa,Djouadi:2015jea} which was shown to provide a very good
approximation of the MSSM Higgs sector.  We will use it in this review as it
simplifies  considerably the phenomenological analyses.

\begin{figure}[!h]
\vspace*{-1.95cm}
\centerline{
\includegraphics[scale=0.65]{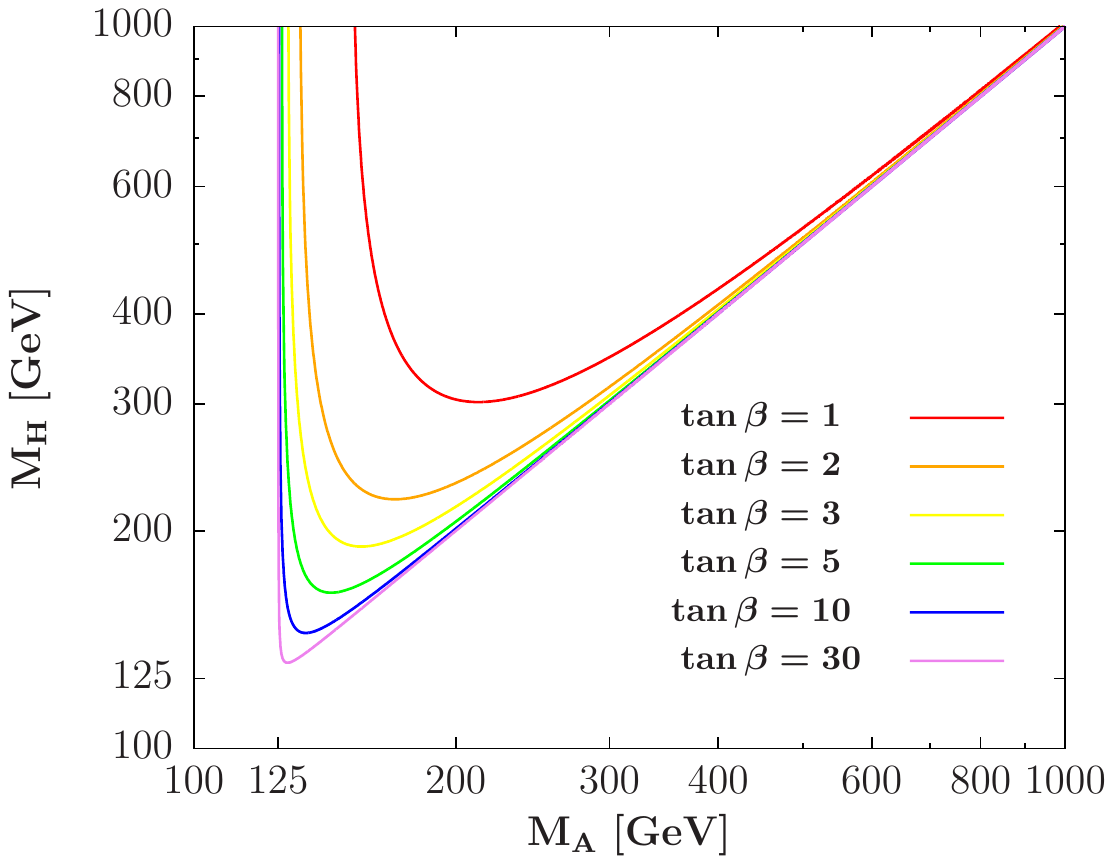}\hspace*{-6.cm}
\includegraphics[scale=0.65]{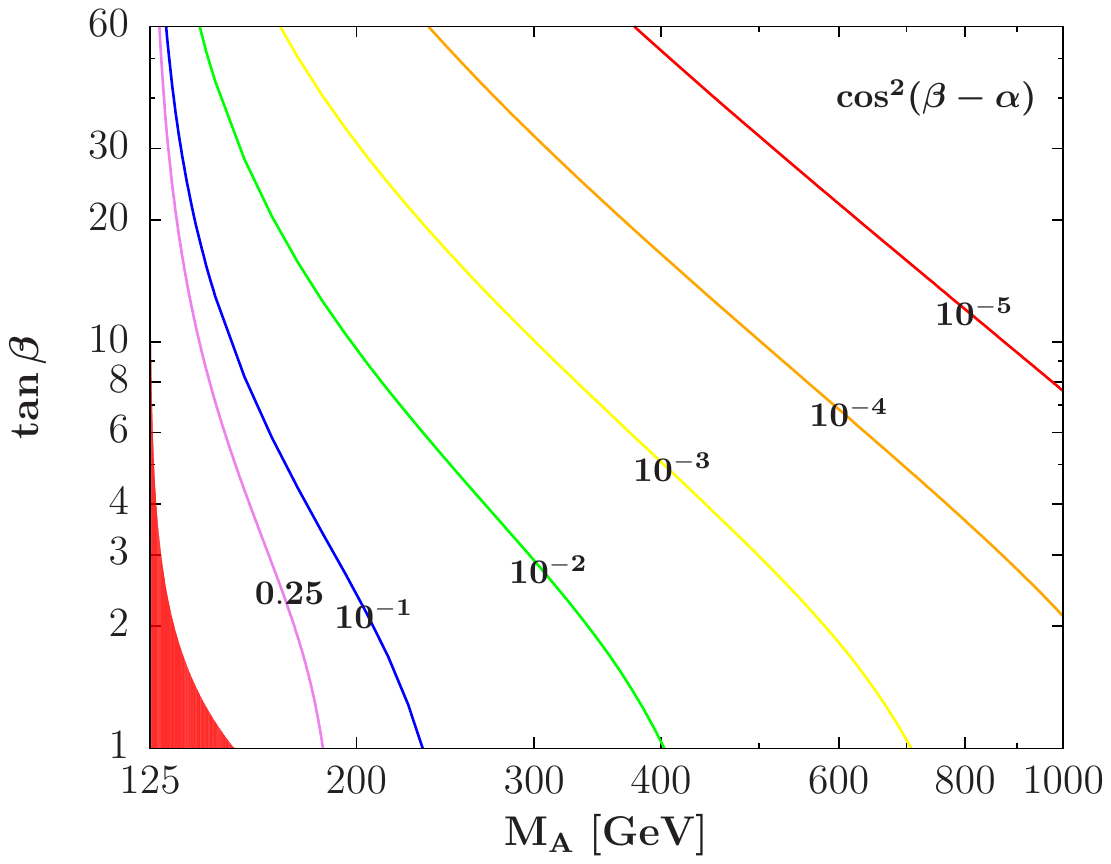} }
\vspace*{-11.9cm}
\caption{The CP--even $H$ mass as a function of $M_A$ for representative $\tan\beta$ values (left) and the coupling squared $\cos^2(\beta-\alpha)$
in the  $[M_A, \tb]$ plane (right) in the $h$MSSM \cite{Djouadi:2015jea}.
}
\vspace*{-.1cm}
\label{Fig:MH-alpha}
\end{figure}

In  Fig.~\ref{Fig:MH-alpha}, we describe the two main outputs of the $h$MSSM:
the mass $M_H$ as a function of $M_A$ for several $\tb$ values (left) and 
countours for some  $\cos^2(\beta-\alpha)$ values in the $[M_A, \tb]$ plane
(right). One sees that at high $\tb$ values, $\tb \gsim 10$,  $M_H$ becomes very
close to $M_A$ and $\cos^2(\beta-\alpha)$ close to zero, and hence we reach the
decoupling limit as soon as $M_A \gsim 200$ GeV. In turn, at low $\tb$, the mass
difference $M_H\!-\!M_A$ can be large and $\cos^2 (\beta-\alpha)$ significantly
different from zero even for $M_A \! \approx \! 400$ GeV, meaning that the
decoupling limit is reached slowly in this case. 

\begin{figure}[!h]
\vspace*{-2cm}
\centerline{ \includegraphics[scale=0.65]{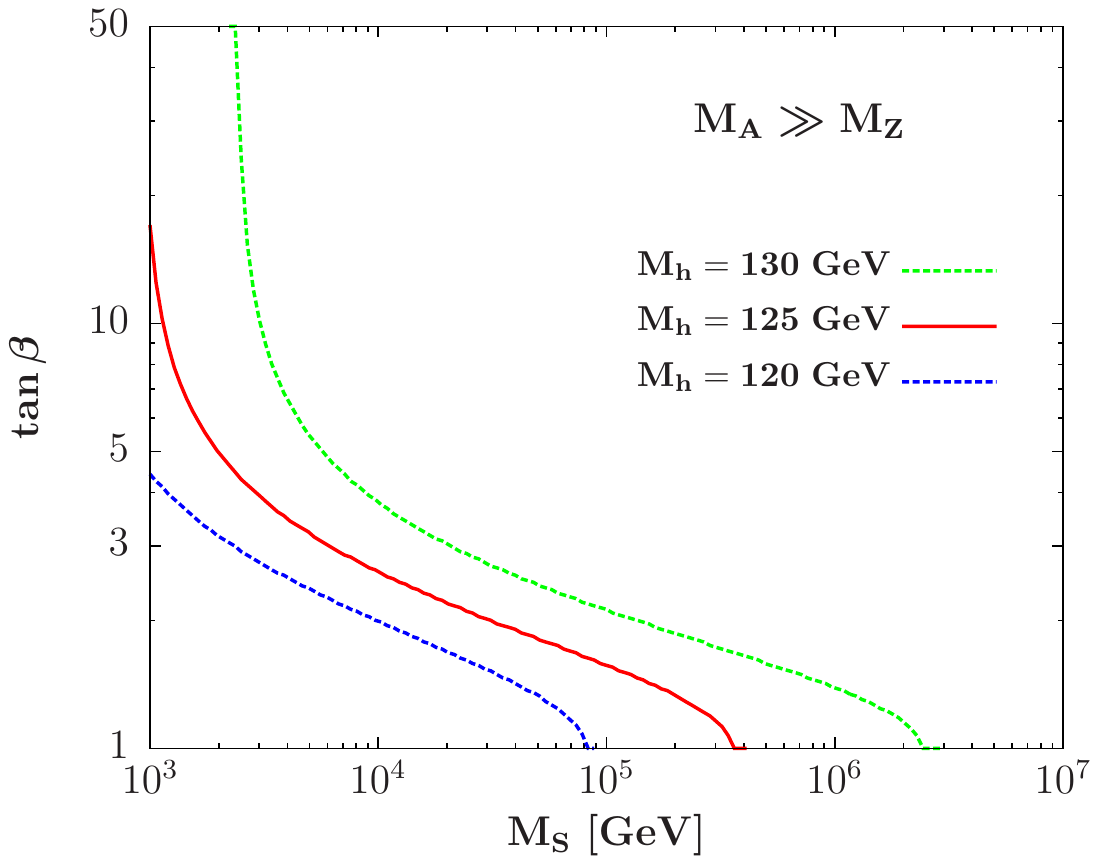} }
\vspace*{-11.8cm}
\caption{Contour plots in the $[\tan\beta, M_{S}]$ plane in which one obtains $M_h=120, 125$ and 130 GeV in the $h$MSSM; the decoupling limit and maximal stop mixing are assumed~\cite{Djouadi:2015jea}.}
\vspace*{-.3cm}
\label{Fig:tb-MS}
\end{figure}

An immediate  advantage of the $h$MSSM  is that  it allows  the possibility  to
study the low $\tb$ region of the MSSM which was  overlooked as it did not lead
to a correct $M_h$ value for reasonable SUSY spectra. This  region would be
re-opened if no assumption on the SUSY scale is made and if it can be taken as
large as possible (as in the split-SUSY scenario \cite{ArkaniHamed:2004fb,Giudice:2004tc,ArkaniHamed:2004yi,Wells:2004di,Bernal:2007uv} for instance). In
this case, values $\tb \lsim 3$ would mean extremely large $M_S$ values.  This
is shown in  Fig.~\ref{Fig:tb-MS} where we display contours
in the $[\tb, M_S]$ plane where the value $M_h\!=\!125$ GeV is obtained assuming
a (large)  uncertainty of $\pm 5$ GeV in its theoretical determination from
unaccounted subleading corrections. We have taken the limit $M_A\! \gg \!M_Z$
and assumed maximal stop mixing $X_t\!=\!\sqrt 6 M_S$ (the value of the SM
inputs are those  given in section 2.1). One sees that, indeed,  while $M_S$
values close to the TeV scale can be accommodate at high $\tb$, in the low 
$\tb$ region, extremely large and unnatural values of the scale $M_S$ are
necessary to obtain  $M_h=125\pm 5$ GeV. In this case, the sfermion spectrum 
will enter only in the radiative corrections and will not affect the
phenomenology of the Higgs sector and, in our context, it can be thus simply
ignored. 

It has been shown that the $h$MSSM approach, although very simple and
economical, provides a very good approximation in the determination of the
spectrum. There are nevertheless three cases in which it has to be used with
caution. The first one is related to the treatment of the trilinear Higgs
couplings which should be done properly. Two among these couplings are very
important, the $hhh$ and $Hhh$  couplings which, in a naive approach, would read
in the decoupling limit 
\begin{eqnarray}
\lambda_{hhh}\!=\!3M_h^2/ M_Z^2~~{\rm  and}~~ \lambda_{Hhh}= - 3 
\Delta M_{22}^2 /( 2M_Z^2)  \times \sin 2 \beta \, .
\label{eq:Hhh}
\end{eqnarray}
However, some care should be taken when including in a consistent way all
relevant corrections, in particular the direct ones in a 2HDM when the decay  $H\to hh$ is considered \cite{Chalons:2017wnz}. A proper treatment of these contributions in this context has been recently made~\cite{Liebler:2018zul}. 

Another delicate point is that in the case of $b$--quarks, additional vertex
corrections modify their tree--level couplings to the Higgs bosons: they
grow as $ \bar m_b  \mu \tan\beta$ and become very large at high  $\tb$ values.
The dominant component comes from SUSY--QCD corrections with sbottom--gluino
loops that can be approximated by \cite{Loinaz:1998ph,Babu:1998er,Carena:1999py,Ghezzi:2017enb} 
\begin{eqnarray} 
\Delta_b \simeq 2\alpha_s/(3\pi) \times \mu m_{\tilde{g}} \tb /{\rm max} (m_{\tilde{g}}^2, m_{\tilde{b}_1}^2,m_{\tilde{b}_2}^2) \, . \label{deltab} \end{eqnarray} 
These corrections are not taken care of by the $h$MSSM approach and, in
principle, they have to be added separately for the $Hb\bar b,Ab\bar b$ and $H^\pm t b$ couplings (they decouple in  the $hb\bar b$ vertex for $M_A \gg M_Z)$. But we will see later that in the most important situations, these effects are small and can be ignored in a first approximation.  

Finally, when the gaugino and higgsinos are relatively light, one should take
into account their impact on Higgs phenomenology, a feature to which we turn our
attention now. 

\subsubsection{The neutralino and chargino sectors} 

As mentioned earlier the bino, the three winos and the two higgsinos will  mix
in order to give the physical states which are the two chargino $\chi^\pm_{1,2}$
and the four neutralino  $\chi^0_{1\!-\!4}$ Majorana particles. The lightest of
the neutralino $\chi^0_{1}$ is in general the lightest  supersymmetric particle
(LSP) which, when $R$--parity is conserved, is stable and constitutes the most
favored and discussed among the DM candidates.  Let us briefly describe the
chargino-neutralino spectra and their connection with the Higgs sector. 

The chargino mass matrix, in terms of the parameters $M_2, 
\mu$ and $\tb$ reads \cite{Gunion:1989we,Haber:1984rc}
\begin{eqnarray}
{\cal M}_C = \left[ \begin{array}{cc} M_2 & \sqrt{2}M_W s_\beta
\\ \sqrt{2}M_W c_\beta & \mu \end{array} \right] \, , 
\end{eqnarray}
where we use again $s_\beta \equiv \sin \beta, c_\beta \equiv \cos \beta$. The two chargino states $\chi_1^\pm, \chi_2^\pm$ and their (positive) masses are determined via a transformation $U^* {\cal M}_C V^{-1} \!=\! {\rm diag} (m_{\chi_{1}^\pm}, m_{\chi_{2}^\pm})$, where $U,V$ are unitary matrices and  ${\rm diag} (m_{\chi_{1}^\pm}, m_{\chi_{2}^\pm})$ is the diagonal matrix. The two chargino masses can be given in analytical form  by
\begin{eqnarray}
m^2_{\chi_{1,2}^\pm} \!= \!\frac{1}{2} \left\{ M_2^2 \!+ \!\mu^2 \! + \!2M_W^2
\mp \left[ (M_2^2-\mu^2)^2+4 M_W^2( M_W^2 c^2_{2\beta} + M^2_2+\mu^2
\!+ \! 2M_2\mu s_{2\beta}) \right]^{\frac{1}{2}} \right\} . \ \ 
\end{eqnarray}
In the limit $|\mu| \gg M_2, M_W$, denoting by $\epsilon_\mu$ the sign of $\mu$,
they  reduce to
\begin{eqnarray}
m_{\chi_{1}^\pm}  \simeq   M_2 - {M_W^2}{\mu^{-2} } 
\left( M_2 +\mu s_{2\beta} \right) \ \ , \ \ 
m_{\chi_{2}^\pm}  \simeq  |\mu| + {M_W^2}{\mu^{-2}} \epsilon_\mu \left( M_2 s_{2 \beta} +\mu \right) \, .
\end{eqnarray}
For $|\mu| \to \infty$, the lightest chargino corresponds to a pure wino with a
mass $m_{\chi_{1}^\pm}  \simeq M_2$, while the heavier chargino corresponds to a
pure higgsino with a  mass $m_{\chi_{2}^\pm} = |\mu|$. In the opposite limit,
$M_2 \gg |\mu|, M_Z$,  the roles of $\chi_{1}^\pm$ and $\chi_{2}^\pm$ are
reversed.\s

In the case of neutralinos, the four--dimensional mass matrix  depends on the
same three parameters $\mu$, $M_2$ and $\tb$ as above and on $M_1$, 
when constraints such as the GUT relation $M_2 \simeq 2M_1$  are not used.  In the $(-i\tilde{B}, -i\tilde{W}_3, \tilde{H}^0_1,$ $\tilde{H}^0_2)$ basis, it  has the form  \cite{Gunion:1989we,Haber:1984rc}
\begin{eqnarray} \label{eq:chi-mass-matrix}
{\cal M}_N = \left[ \begin{array}{cccc}
 M_1 & 0  & -M_Z s_W c_\beta & M_Z  s_W s_\beta \\
 0   & M_2  & M_Z c_W c_\beta & -M_Z  c_W s_\beta \\
 -M_Z s_W c_\beta  & M_Z  c_W c_\beta  & 0 & -\mu \\
 M_Z s_W s_\beta    & -M_Z  c_W s_\beta  & -\mu & 0
\end{array} \right] \, .
\end{eqnarray}

The neutralino states $\chi_{1,2,3,4}^0$ and their masses are determined via a
transformation $Z^T {\cal M}_N Z^{-1} = {\rm diag} (m_{\chi_1^0}, m_{\chi_2^0},
m_{\chi_3^0}, m_{\chi_4^0})$ with again a unitary matrix $Z$ and the
diagonal matrix. The rather complicated expressions of the matrix elements $Z_{ij}$ with $i,j=1,..,4$ and the four masses $m_{\chi_i^0}$ simplify in two asymptotic cases.  In the limit of large $|\mu|$, $|\mu| \gg M_{1,2} \gg M_Z$, one has \cite{Djouadi:1996pj} 
\begin{eqnarray}
m_{\chi_{1}^0} & \simeq & M_1 - \frac{M_Z^2}{\mu^2} \left( M_1 +\mu s_{2\beta}
\right) s_W^2 \, ,   \nonumber \\  
m_{\chi_{2}^0} & \simeq & M_2 - \frac{M_Z^2}{\mu^2} \left( M_2 +\mu s_{2 \beta}
\right) c_W^2 \nonumber \, , \\
 m_{\chi_{3/4}^0} &\simeq & |\mu| + \frac{1}{2}\frac{M_Z^2}{\mu^2} \epsilon_\mu 
(1\mp  s_{2\beta}) \left( \mu \pm M_2 s_W^2 \mp M_1 c_W^2 \right) \, .
\hspace*{2cm}
\end{eqnarray}
Here again, two neutralinos are pure gauginos with masses $m_{\chi_{1}^0}  \simeq M_1$ and $m_{\chi_{2}^0} \simeq M_2$, while the two others are pure  higgsinos with masses $m_{\chi_{3}^0} \simeq m_{\chi_{4}^0} \simeq |\mu|$. In the opposite limit, the roles are again reversed and one has instead,  $m_{\chi_{1}^0} \simeq m_{\chi_{2}^0} \simeq |\mu|, m_{\chi_{3}^0} \simeq M_1$  and $m_{\chi_{4}^0} \simeq M_2$.

Finally, we note that the gluino mass is identified with $M_3$ at the tree
level, $m_{\tilde{g}} = M_3$, and in our discussion here, the gluinos will be
considered to be rather heavy and we will thus set  $M_3 \gg M_1, M_2$ and even
$M_3 \gg |\mu|$. 

The Higgs couplings to neutralinos and charginos come also  from several sources
such as the superpotential, in particular from  the  bilinear term, and are
affected also by the gaugino masses and couplings. They are made
more complicated by the higgsino--gaugino mixing,  the diagonalization of the
chargino/neutralino mass matrices, and the Majorana  nature of the neutralinos.
Denoting the Higgs bosons by $H_k$ with $k=1,2,3$, corresponding to $H,h, A$, respectively, and $H_4=H^\pm$ and  normalizing to the electric charge $e$, the
Higgs couplings to chargino and neutralino pairs can be written in a convenient form as \cite{Gunion:1984yn,Gunion:1986nh}
\begin{eqnarray}
g^{L,R}_{\chi^0_i \chi^+_j H_4}= g^{L,R}_{ij4} & {\rm with} &
\begin{array}{l} 
g^L_{ij4} = \frac{c_\beta}{s_W} \big[ Z_{j4} V_{i1} + \frac{1}{\sqrt{2}} 
\left( Z_{j2} + \tan \theta_W Z_{j1} \right) V_{i2} \big] \\
g^R_{ij4} = \frac{s_\beta}{s_W} \big[ Z_{j3} U_{i1} - \frac{1}{\sqrt{2}}
\left( Z_{j2} + \tan \theta_W Z_{j1} \right) U_{i2} \big]  
\label{cp:inos1}
\end{array} , \nonumber \\
g^{L,R}_{\chi^-_i \chi^+_j H_k} = g^{L,R}_{ijk} & {\rm with} & 
\begin{array}{l} 
g^L_{ijk}= \frac{1}{\sqrt{2}s_W} \left[ e_k V_{j1}U_{i2}-d_k V_{j2}U_{i1}
\right] \\
g^R_{ijk}= \frac{1}{\sqrt{2}s_W} \left[ e_k V_{i1}U_{j2}-d_k V_{i2}U_{j1}
\right] \epsilon_k 
\end{array} , 
\label{cp:inos2} \\
g^{L,R}_{\chi^0_i \chi^0_j H_k} = g^{L,R}_{ijk} & {\rm with} & 
\begin{array}{l} 
g^L_{ijk} = \frac{1}{2 s_W} \left( Z_{j2}- \tan\theta_W Z_{j1} \right) 
\left(e_k Z_{i3} + d_kZ_{i4} \right) \ + \ i \leftrightarrow j
 \\
g^R_{ijk} = \frac{1}{2 s_W}  \left( Z_{j2}- \tan\theta_W Z_{j1} 
\right) \left(e_k Z_{i3} + d_kZ_{i4} \right) \epsilon_k \ + \ i 
\leftrightarrow j 
\end{array} , \ \ \nonumber
\label{cp:inos3}
\end{eqnarray}
where $Z$ and $U/V$ are the diagonalizing matrices discussed before and
$\epsilon_{1,2}=-\epsilon_3 =1$;  the coefficients $e_k$ and $d_k$ read
(we alos give their values in the decoupling limit)
\begin{eqnarray}
e_1=+ \cos \alpha \to \sin\beta , \ e_2=- \sin \alpha \to \cos\beta ,  \ e_3=- \sin\beta , \nonumber  \\ 
d_1=  -\sin\alpha \to \cos\beta , \ d_2= -\cos\alpha \to \sin\beta ,  \ d_3= +  \cos\beta .
\label{ed-coefficients}
\end{eqnarray}
Note that the Higgs couplings to the $\chi_1^0$ DM state, for which  $Z_{11},
Z_{12}$ are the gaugino components and $Z_{13},Z_{14}$  the higgsino components,
vanish if the LSP is a pure gaugino or a pure higgsino. This statement can be
generalized to all neutralino and chargino states and the Higgs bosons couple
only to higgsino--gaugino mixtures or states\footnote{This makes that the Higgs
couplings to mixed heavy and light chargino/neutralino states are maximal in the
gaugino or higgsino  regions, while the couplings involving only heavy or light
gaugino or higgsino states are suppressed by powers of $M_2/\mu$ for $|\mu| \gg
M_2$ or powers of $|\mu|/M_2$ for $|\mu| \ll M_2$.}. The couplings of the neutral Higgs bosons to neutralinos can also accidentally vanish for certain values of $\tb$ and $M_A$ which enter in the coefficients $d_k$ and $e_k$ above.  

Finally, we will also need the couplings of the charginos and neutralinos to the
massive gauge bosons. Using the same ingredients as above, they are given by
\cite{Haber:1984rc}
\begin{eqnarray}
g^L_{\chi^0_i \chi^+_j W} =  \frac{c_W}{\sqrt{2}s_W} 
[-Z_{i4} V_{j2}+\sqrt{2}Z_{i2} V_{j1}]  & , &  
g^R_{ \chi^0_i \chi^+_j W} =  \frac{c_W}{\sqrt{2}s_W} 
[Z_{i3} U_{j2}+ \sqrt{2} Z_{i2} U_{j1}] , \nonumber \\
g^L_{\chi^0_i \chi^0_j Z} = - \frac{1}{2s_W} 
[Z_{i3} Z_{j3} - Z_{i4} Z_{j4}]   & , &
g^R_{\chi^0_i \chi^0_j Z} = + \frac{1}{2s_W} [Z_{i3} 
Z_{j3} - Z_{i4} Z_{j4} ] , \\
g^L_{\chi^-_i \chi^+_j Z} = \frac{1}{c_W} \left[\delta_{ij}s_W^2 - \frac{1}{2} 
V_{i2} V_{j2} - V_{i1} V_{j1} \right] & , &  
g^R_{\chi^-_i \chi^+_j Z} = \frac{1}{c_W} \left[\delta_{ij}s_W^2 - \frac{1}{2} 
U_{i2} U_{j2} - U_{i1} U_{j1} \right] . \nonumber
\label{cp:Z-charginos}
\end{eqnarray}

In contrast to  the Higgs bosons which couple preferentially to mixtures of
gauginos and higgsinos,  the gauge boson couplings to charginos and neutralinos
are important only for higgsino-- or gaugino--like states.  Thus, in principle,
the higgsino or gaugino--like heavier states  $\chi_2^\pm$  and $\chi_{3,4}^0$
will dominantly decay, if phase space allowed, into Higgs  bosons and the
lighter $\chi$ states as will be seen later on.

\subsection{Phenomenology at the LHC}

\subsubsection{Higgs production and decays}

The collider phenomenology of the MSSM is quite similar to that of a Type II 
2HDM if only values $\tb \gsim 1$ are considered and if the SUSY spectrum is
very heavy  
\cite{Baglio:2010ae,Christensen:2012ei,Arbey:2013jla,Craig:2015jba,Hajer:2015gka,Carena:2014nza,Djouadi:2013vqa,Djouadi:2013uqa,Djouadi:2015jea,Baglio:2015wcg}. This is particularly true at high values of $\tb$ where one is
quickly in the decoupling regime in which the lighter $h$ state is SM--like
while the heavier $H, A$ and $H^\pm$ bosons are almost degenerate in mass,
decouple from the massive gauge bosons and interact only with fermions with
coupling strengths that are enhanced by powers of $\tb$  for bottom quarks and
tau leptons and suppressed as $1/\tb$ for top quarks. This is similar to our
2HDM benchmark in which we have assumed the alignment limit and 
$M_H=M_A=M_{H^\pm}$. Ignoring again the lightest $h$ boson which has been
discussed in section 2, the neutral $\Phi=H,A$ bosons are mainly produced in
$b\bar b$ and $gg$ gluon fusion  with large rates and decay almost exclusively
into $b\bar b$ pairs with a branching ratio of 90\% and $\tau^+\tau^-$ final
states with a branching ratio of 10\%. The charged Higgs boson can be produced
in the $gb \to t H^-$ mode and would decay into $tb$ and $\tau \nu$ final states
again with branching fractions of 90\% and 10\%, respectively. 

\begin{figure}[!t]
\vspace*{-.2cm}
\centerline{
\includegraphics[scale=0.24]{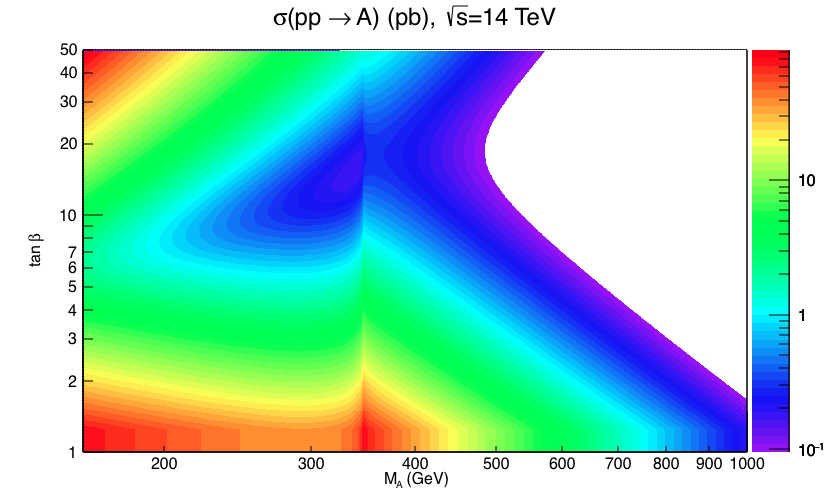}
\includegraphics[scale=0.24]{figs-MSSM/MSSM-pp_A_14TeV.png} }\vspace*{-2mm}
\centerline{
\includegraphics[scale=0.24]{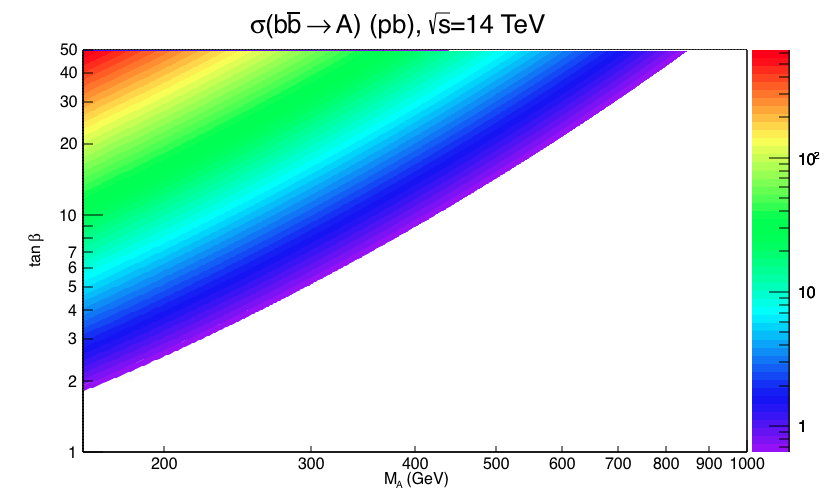}
\includegraphics[scale=0.24]{figs-MSSM/MSSM-bbA_14TeV.png} }\vspace*{-2mm}
\centerline{
\includegraphics[scale=0.24]{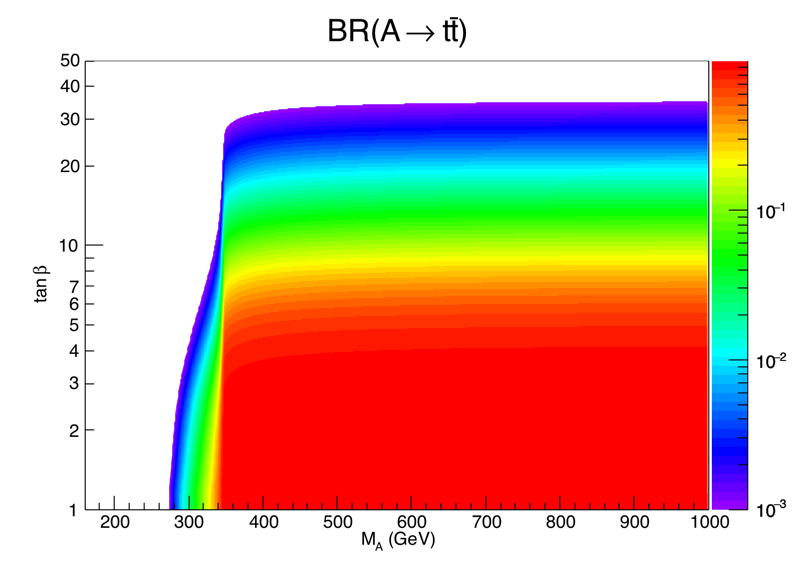}
\includegraphics[scale=0.24]{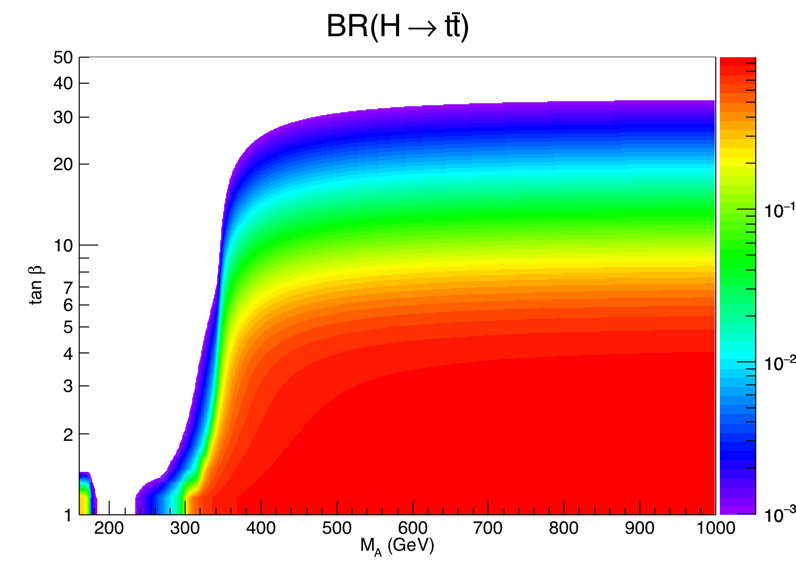} }\vspace*{-2mm}
\centerline{
\includegraphics[scale=0.24]{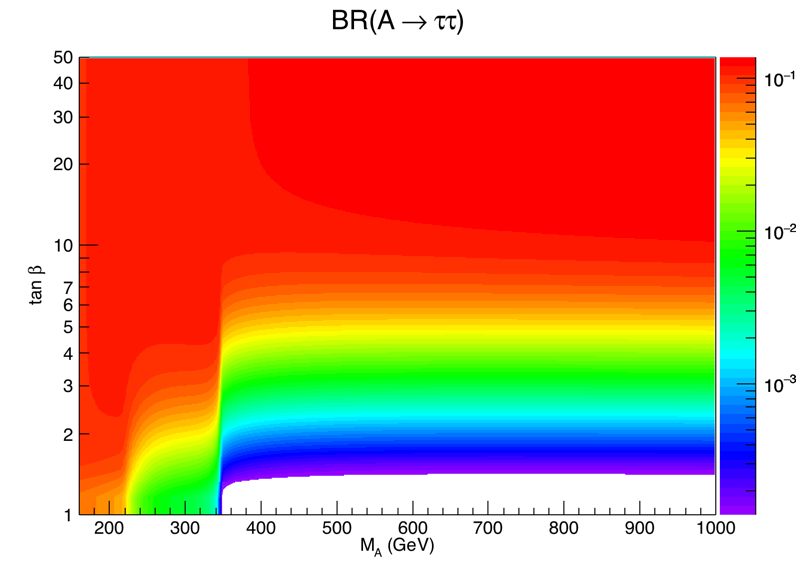}
\includegraphics[scale=0.24]{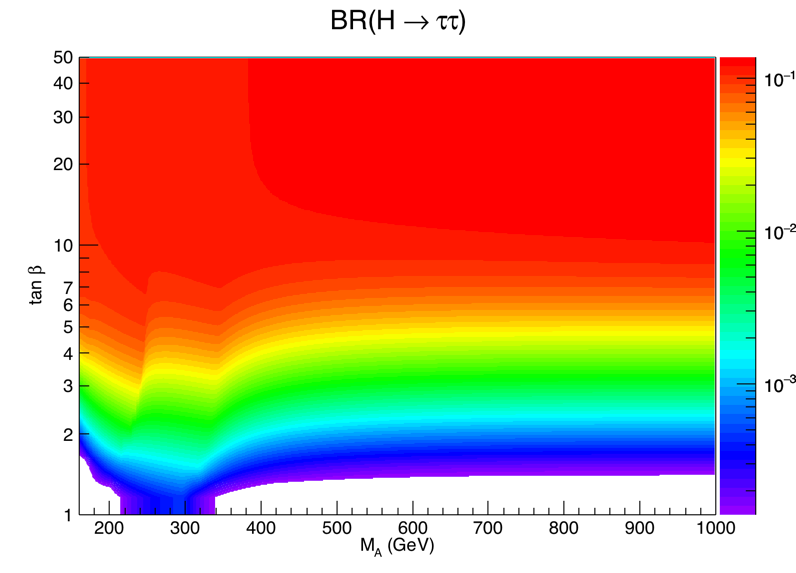} }\vspace*{-2mm}
\centerline{
\includegraphics[scale=0.24]{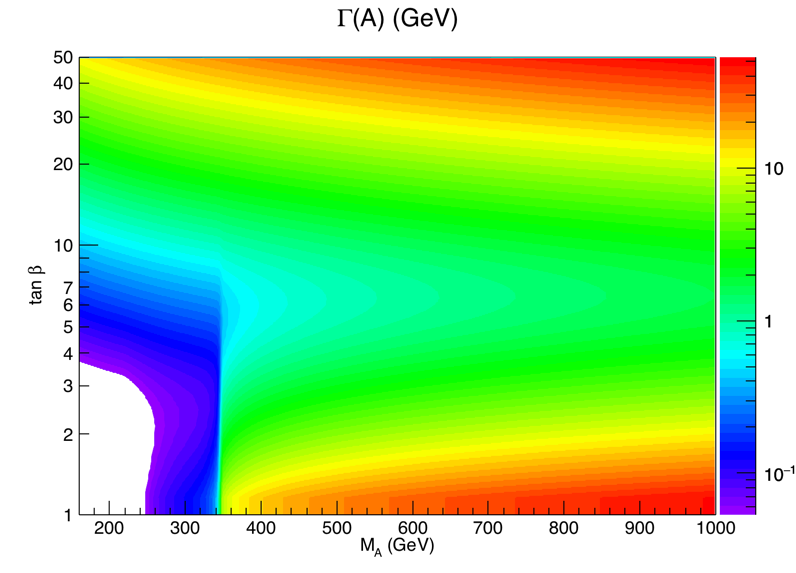} 
\includegraphics[scale=0.24]{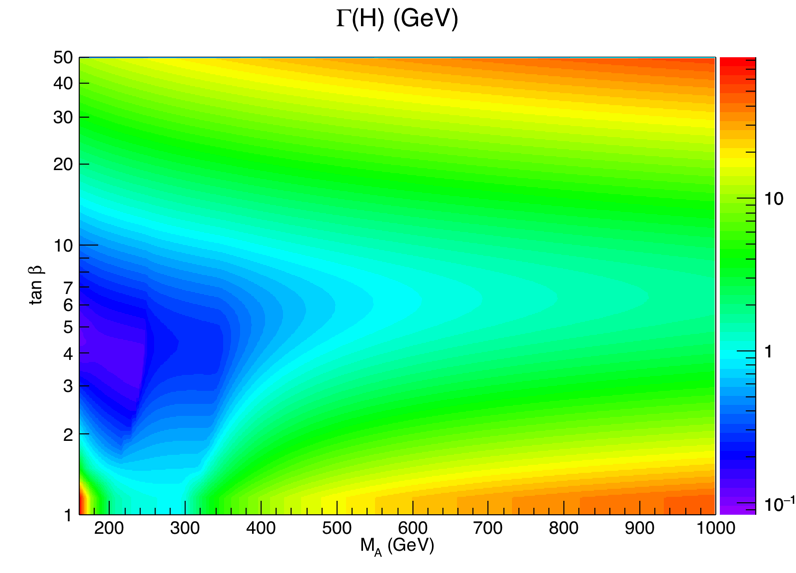}}
\vspace*{-.2cm}
\caption[]{From top to bottom: the cross sections $\sigma(gg\! \to \!\Phi), \sigma(b\bar b \! \to \!  \Phi)$ at the 14 TeV LHC, the branching ratios for $\Phi  \! \to  \! t\bar  t, \Phi   \! \to  \! \tau^+ \tau^-$ and the total widths $\Gamma( \Phi)$ for $\Phi=A$ (left panels) and $\Phi=H$ (right panels)  in the $[M_A, \tb ]$ plane; from Ref.~\cite{Djouadi:2019cbm}.}
\label{Fig:MSSM-all}
\vspace*{-.9cm}
\end{figure}

The cross section for the important production channels $gg\to \Phi$ and $b\bar
b \to \Phi$  as well as for the dominant decays in the high and low $\tb$
regimes, namely BR$(\Phi \to t\bar t)$ and BR$(\Phi \to \tau \tau)$ [the 
branching ratio for the other important decay $\Phi \to b\bar b$ is simply 
BR$(\Phi \to b\bar b) \simeq 9 \! \times!  {\rm BR} (\Phi \to \tau \tau)$] as
well as the $\Phi=H,A$ total widths are shown in Fig.~\ref{Fig:MSSM-all} in the
parameter plane $[M_A, \tb]$. The color code, indicated in the right vertical
axes, is such that the rates are more important in the red areas than  in the
blue ones.     

Nevertheless,  SUSY particles can impact these rates via the direct correction
$\Delta_b$, the leading part of which is given in eq.~(\ref{deltab}). This
correction, can be significant at large $\tb$ and $\mu$ values and modifies the
$H,A$ and $H^\pm$ couplings to $b$--quarks, $g_{\Phi bb} \approx  g_{H^\pm
tb}\approx \tb/(1+\Delta_b)$ and thus, the production and decay rates discussed
above. Nevertheless,  this correction has only a  limited impact in the main
detection of these states when the full production times decay processes are
taken into account. 

Indeed, for the $\Phi=H,A$ neutral states  the main processes to be considered 
are  $gg, b\bar b \to \Phi \to \tau \tau$ and while the cross sections are such
that $\sigma \propto (1+ \Delta_b)^{-2}$, one has for the branching ratios ${\rm
BR}(\tau\tau) =  \Gamma(\tau\tau)/[(1+ \Delta_b)^{-2}\Gamma(b\bar b)+
\Gamma(\tau\tau)]$, and the $\Delta_b$ correction  largely cancels out in the
product of the two, $\sigma \times {\rm BR} \simeq  1- \Delta_b/5$. Hence, only
when the $\Delta_b$ correction is huge  (larger than 100\% which might endanger
the perturbative series) that its impact on the $pp \to \tau\tau$ rate becomes
of the order of the theoretical (scale+PDF) uncertainty of the process which is
about 25\% \cite{Baglio:2010ae,Dittmaier:2011ti}. The same holds true for the
charged Higgs, produced in $gb$ fusion an  decaying into $\tau\nu$. Hence, the
limits set by ATLAS and CMS shown in Fig.~\ref{Fig:Phi-pp} for a Type II 2HDM, 
should not be affected by  these SUSY direct corrections. 

At low $\tb$ values, $\tb \lsim 3$, and for Higgs masses above the $2m_t$
threshold, the situation for the heavy neutral states is also rather simple. 
They will be now produced essentially in the $gg \to \Phi$ process with the top
quark loop providing the main contribution as the $g_{\Phi tt}$ is still strong
in this $\tb$ range even if suppressed compared to the SM value for $\tb >1$, 
and will almost exclusively decay into $t\bar t$ final states. The rates are 
slightly large for $A$ than  for $H$, first because the $gg\Phi$ form factor is
larger in the case of a CP--odd compared to a CP--even state and then,  the mass
$M_A$ is smaller than $M_H$ in the MSSM. In the process $gg\to \Phi \to t\bar t$
one again has to take into account both $H$ and $A$ contributions and their
interference  also with the $gg\to t\bar t$ QCD background. 

At intermediate $\tb$ values, $3 \lsim \tb \lsim 10$, the main Higgs production
mode will be still $gg \to \Phi$ (with some small additional contribution from
$b\bar b \to \Phi$); the cross section are nevertheless smaller than usual as
the coupling $g_{\Phi tt}$ is suppressed while  $g_{\Phi bb}$ is not yet
strongly enhanced. For the decays when $M_\Phi >350$ GeV,  there will be a
competition between the $\Phi \to t\bar t$ and $\Phi \to b\bar b$ modes. Any
additional Higgs decay in this regime, such as decays into charginos and
neutralinos as will be seen shortly, will impact the rates. 

A most interesting parameter region is when $\tb \lsim 3$--5 and $M_\Phi \lsim
350$ GeV.  Here, Higgs production is primarily due to the $gg\to \Phi$ process
but because we are not yet in the decoupling limit and the $HVV$ couplings is
not completely suppressed, small additional contributions from the VBF and HV
processes will be present in the case of the CP--even $H$ state, $qq \to Hqq$
and  $q\bar q \to HV$. Also, because of the not yet penalizing phase space and
the not strongly suppressed $g_{\Phi tt}$ couplings, the rates for associated
Higgs  production with $t\bar t$ final states, $pp \to t\bar t \Phi$, are not
completely negligible.  

\begin{figure}[!h]
\vspace*{-.1cm}
\mbox{
\hspace*{-3mm}
\includegraphics[width=5.5cm,height=5.5cm]{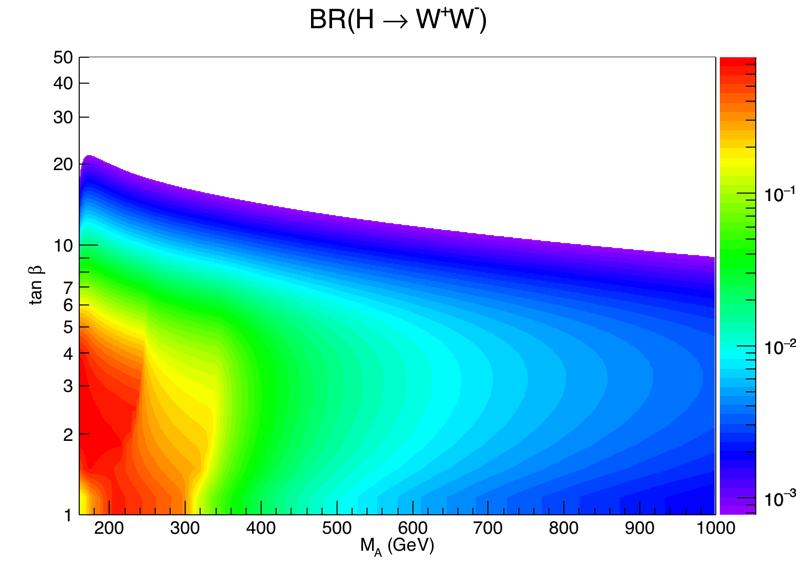}\hspace*{-2mm} 
\includegraphics[width=5.5cm,height=5.5cm]{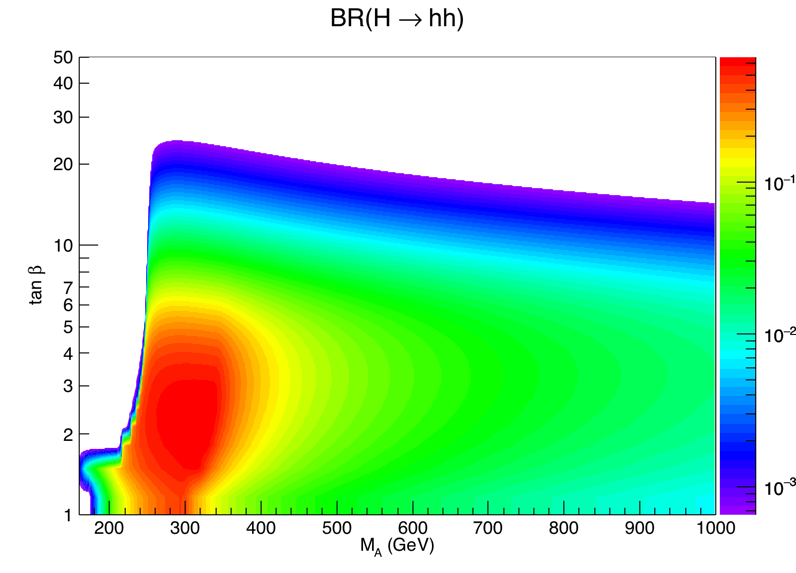}\hspace*{-2mm}  \includegraphics[width=5.5cm,height=5.5cm]{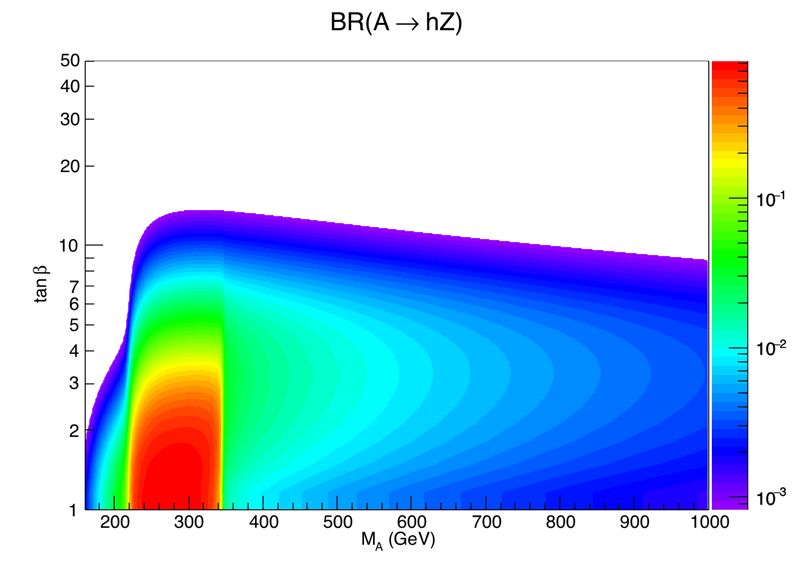}\hspace*{-2mm}  }
\vspace*{-.5cm}
\caption{The branching ratios of the neutral Higgs decays  $H\to WW,hh$
and $A\to hZ$ in the plane $[\tb, M_A]$ \cite{Djouadi:2015jea}. The rate for $H\to ZZ$ is about half that of $H\to WW$.}
\label{Fig:BR-lowtb}
\vspace*{-5mm}
\end{figure}

The situation is even more interesting on the decay side.  Because $g_{HVV}$ is
not so tiny and the longitudinal components of the vector bosons make the
partial decay widths $\Gamma(H\to VV)$ proportional to $M_H^3$ (compared to
$M_H$ only for the fermionic decays), the rates in the channels $H \to WW,ZZ$
are still important and above $M_H \gsim 200$ GeV, they can reach the 10\%
level, with  the $WW$ mode twice as large large as the $ZZ$ mode. Another
channel which is still important is the cascade decay $H\to hh$ when $2M_h \lsim
M_H \lsim 2m_t$. Outside the decoupling limit and for small $\tb$, the $Hhh$
coupling given in eq.~(\ref{eq:Hhh}) is sizeable (the correction $\Delta M_{22}$
being large) and a  rate BR$(H\to hh)$ of a few 10\%  can be reached. Finally, 
in the case of the pseudoscalar Higgs state the decay $A \! \to \! hZ$ is till
possible for  $M_h\!+\! M_Z \lsim M_A \lsim 2m_t$ and, as the coupling $g_{AZh}$
is not completely suppressed, it can also occur at the level of a few 10\%. The
branching ratios for all these decays are shown in Fig.~\ref{Fig:BR-lowtb}  in
the planes $[M_A, \tb]$  in the same configuration as for 
Fig.~\ref{Fig:MSSM-all}.

\subsubsection{Constraints from colliders and expectations}

A first constraint on the MSSM Higgs sector comes the precise determination  of
the couplings of the lightest $h$ boson at the LHC. The measurements of
the $h$ signal strengths in a given channel, such as the $h \to XX$ decay, 
gives a direct constraint on the coupling $g_{hXX}$ or its reduced form
$\kappa_X^2$ which depends on the angles $\alpha$ and $\beta$. The ATLAS and CMS
measurements given Fig.~\ref{Fig:constraints} will hence directly  constrain the
parameters $M_A$ and $\tb$. 

\begin{figure}[h!]
\vspace*{-3mm}
\begin{center}
\mbox{
\hspace*{-0.5cm}\includegraphics[width=.35\textwidth]{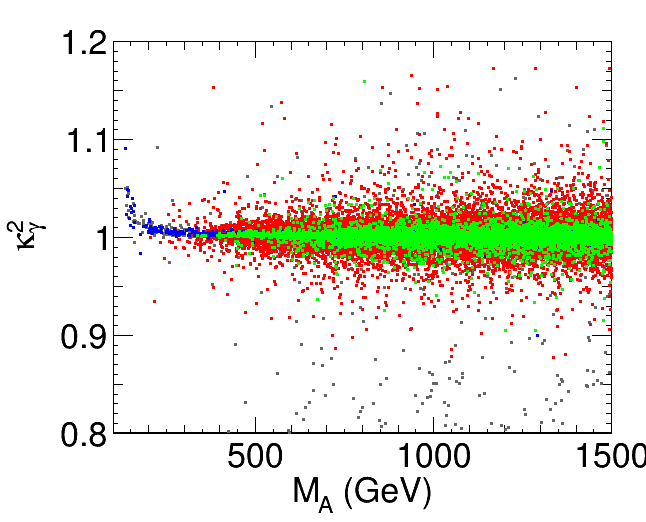}\hspace*{-0.2cm}\includegraphics[width=.35\textwidth]{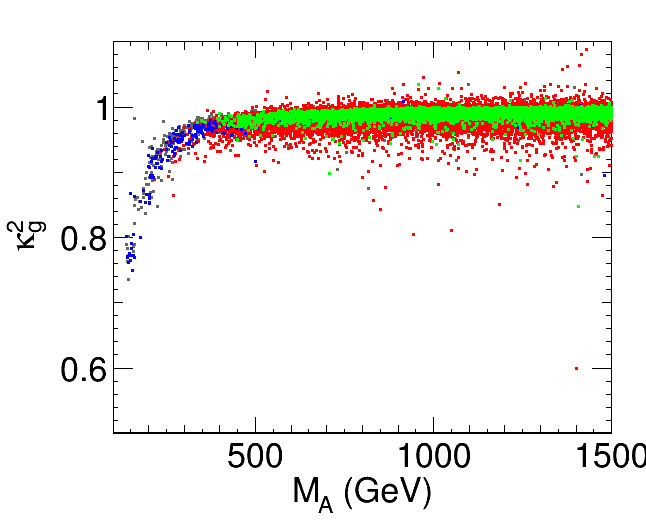}\hspace*{-0.2cm}\includegraphics[width=.35\textwidth]{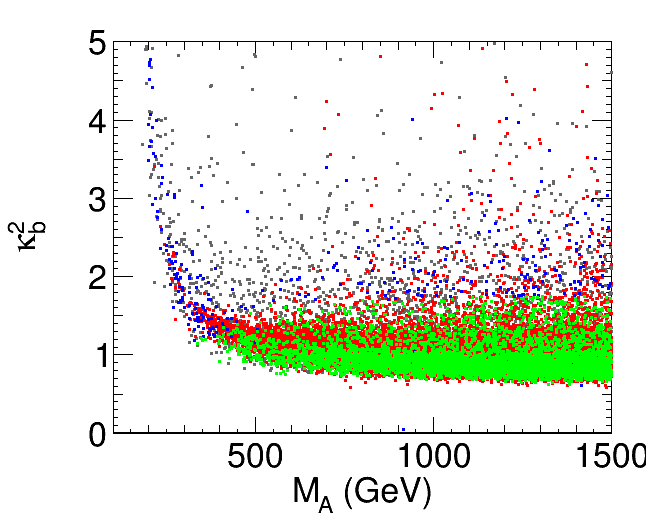}}\vspace*{-0.5cm}
\end{center}
\caption{Distributions of the squared couplings of the light $h$ boson to two photons (left), gluons (center) and bottoms (right), as a function of $M_{A}$ in the pMSSM. The color code is as explained in the text with the green points satisfying all constraints. From Ref.~\cite{Arbey:2018wjb}.}
\label{fig:kappa-MSSM}
\end{figure}

In Ref.~\cite{Arbey:2018wjb}, a scan has been performed in the pMSSM scenario
where the 22 input parameters have been varied in a wide range and all present
constraints have been imposed on the resulting spectra. The output of this scan
for the reduced couplings $\kappa_\gamma, \kappa_g$ and $\kappa_b$ are shown in
Fig.~\ref{fig:kappa-MSSM} as a function of $M_A$. The huge number of pMSSM 
points that have been generated were passed through the following filters: one
first selects those that lead to an $h$ with a mass of $M_h \sim125 \pm 3$ GeV
(grey points), one imposes to the SUSY spectra the limits from direct searches
at LEP (red points) and at the LHC (blue points); the green points are then
those that satisfy all data including  the  Higgs coupling measurements at the
LHC. As can be seen, a large  number of pMSSM points are excluded, in
particular, those  with $M_A \lsim 300$--400 GeV which lead to $\kappa^2_X$
values significantly different from one. 

\begin{figure}[!h]
\vspace*{-.2cm}
\centerline{
\includegraphics[scale=0.6]{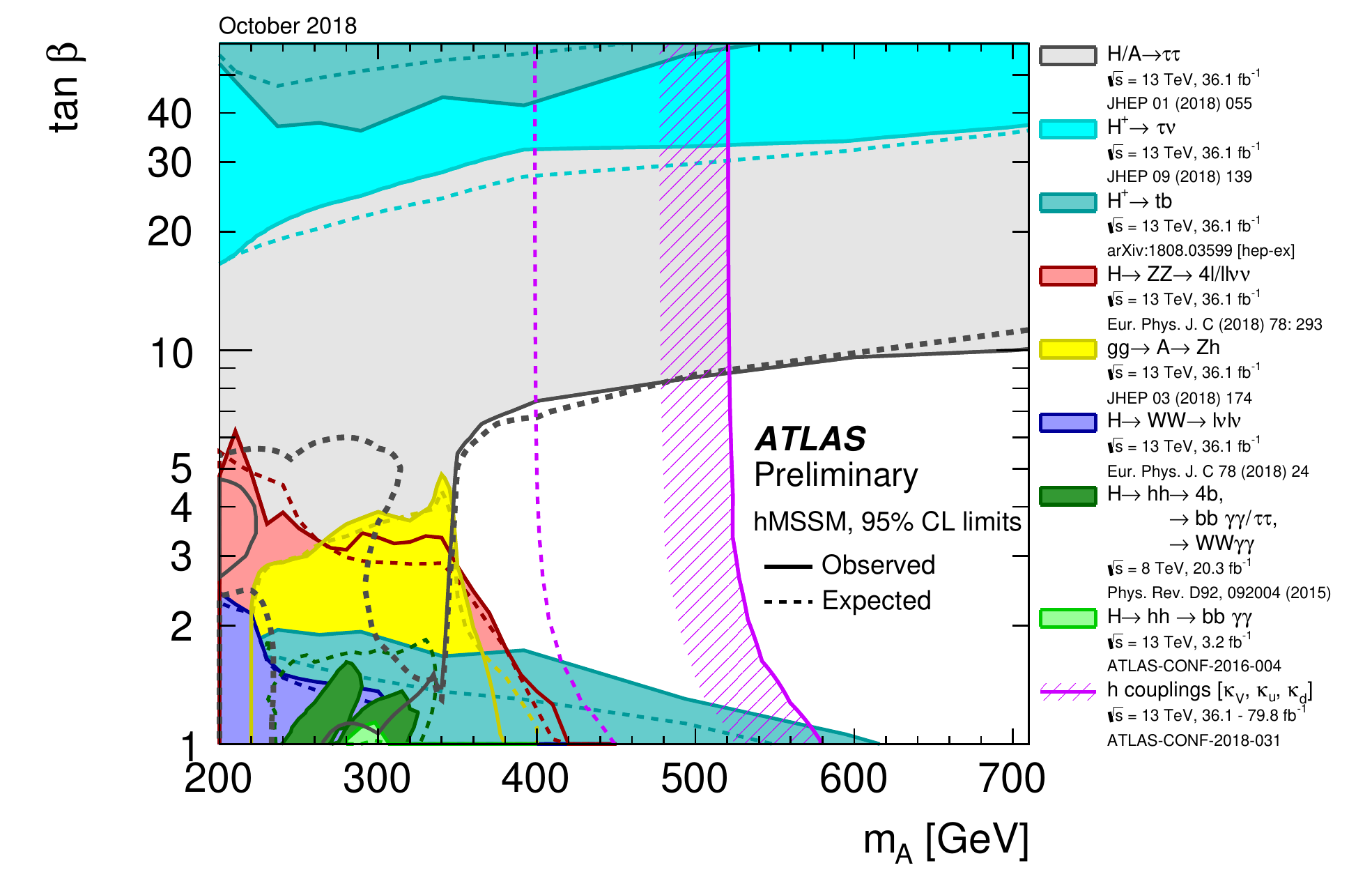} }
\vspace*{-.2cm}
\caption[]{ATLAS 95\%CL exclusion countours of the $h$MSSM using direct searches in the various detection channels described in the right--hand side of the figure (with the relevant energy, luminosity, topology and reference indicated) and a fit of the Higgs coupling measurements at $\sqrt s=13$ TeV with up to 80 fb$^{-1}$ data \cite{ATLAS:2018doi}.}
\vspace*{-.2cm}
\label{ATLAS-all-hMSSM}
\end{figure}

Furthermore, there are constraints from direct searches for the MSSM Higgs
bosons  in the various modes discussed previously,  with a very special role
played by the channels  $pp \to H/A \to \tau\tau$, $pp \to t\bar t$ with $t \to
b H^+ \to b \tau^+ \nu$ and, to a lesser extent, $H\to WW,ZZ,hh$ and $A\to hZ$.
A convenient and recent summary of these searches in the context of the $h$MSSM
has been given by ATLAS using 80 fb$^{-1}$ data at $\sqrt s=13$ TeV and the
result is displayed in  Fig.~\ref{ATLAS-all-hMSSM} in the usual $[M_A, \tb]$
plane. The 95\%CL exclusion contours from the searches above are shown and,
superposed to them, also the area of parameter space excluded by the Higgs
couplings measurements,  essentially the mass range $M_A \lsim 500$ GeV.  Small
areas at $\tb \lsim 3$--5 are excluded by the $H\to WW,ZZ$ and $A \to hZ$
searches. 

\begin{figure}[!h]
\vspace*{-3mm}
\begin{center}
\hspace*{-0.3cm}\includegraphics[width=.5\textwidth]{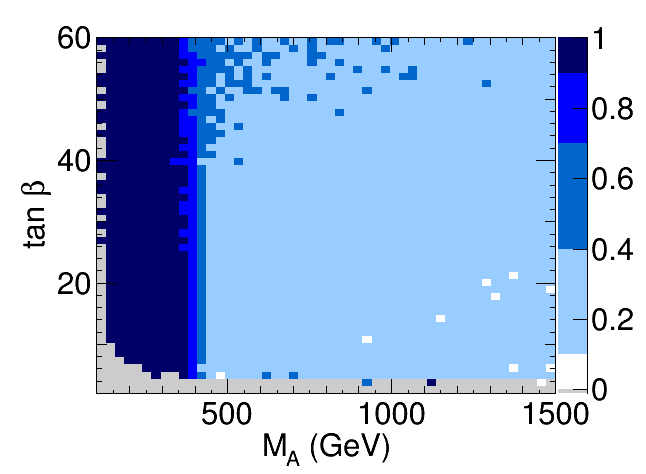}~\includegraphics[width=.5\textwidth]{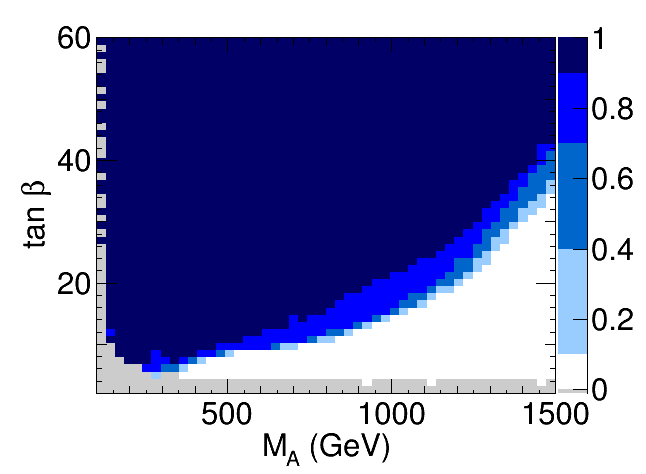}
\vspace*{-.2cm}
\caption{Fraction of pMSSM points excluded by Higgs coupling measurements (left) and heavy Higgs searches (right) at the LHC in the $[M_A,\tan\beta]$ plane as a result of a large scan of the pMSSM parameter space; from Ref.~\cite{Arbey:2018wjb}. \label{fig:tbMA-MSSM}}
\end{center}
\vspace*{-.8cm}
\end{figure}

The dedicated scan of the entire pMSSM parameter space wits 22 inputs that we
previously mentioned shows also that only a small fraction of the generated
points, less than  $\approx 2 \times 10^{-5}$, remain after imposing first
flavor constraints (the same that we discussed for the 2HDM) and  the LHC Higgs
data. The constraints are again summarized in  Fig.~\ref{fig:tbMA-MSSM} in the
$[M_A, \tb]$ plane and the most efficient ones are again the Higgs couplings
measurements (left) and the direct Higgs searches (right) in particular $pp\to
A/H \to \tau^+ \tau^-$.

\begin{figure}[!h]
\vspace*{-.2cm}
\centerline{
\includegraphics[scale=0.55]{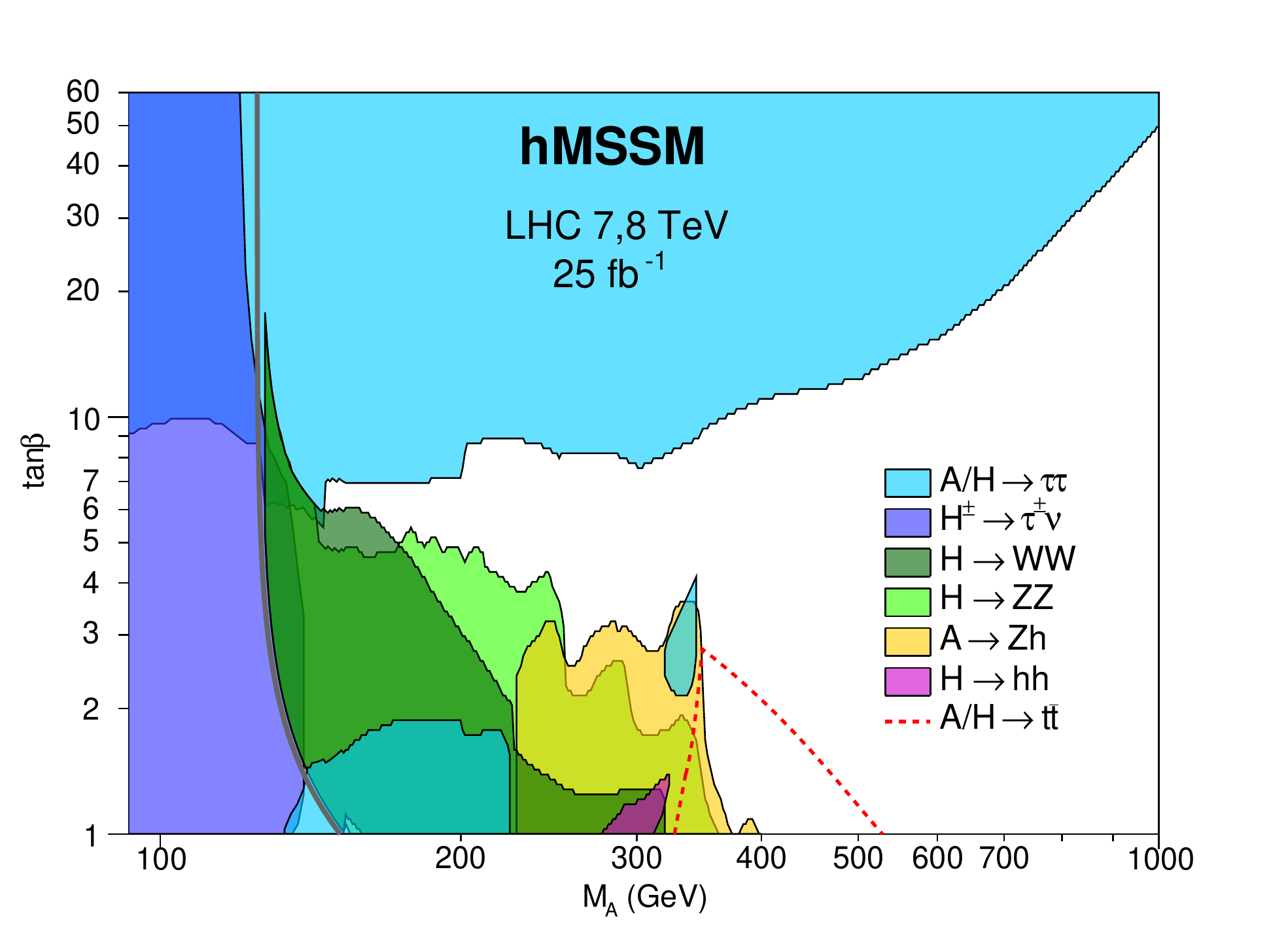} }
\centerline{
\includegraphics[scale=0.45]{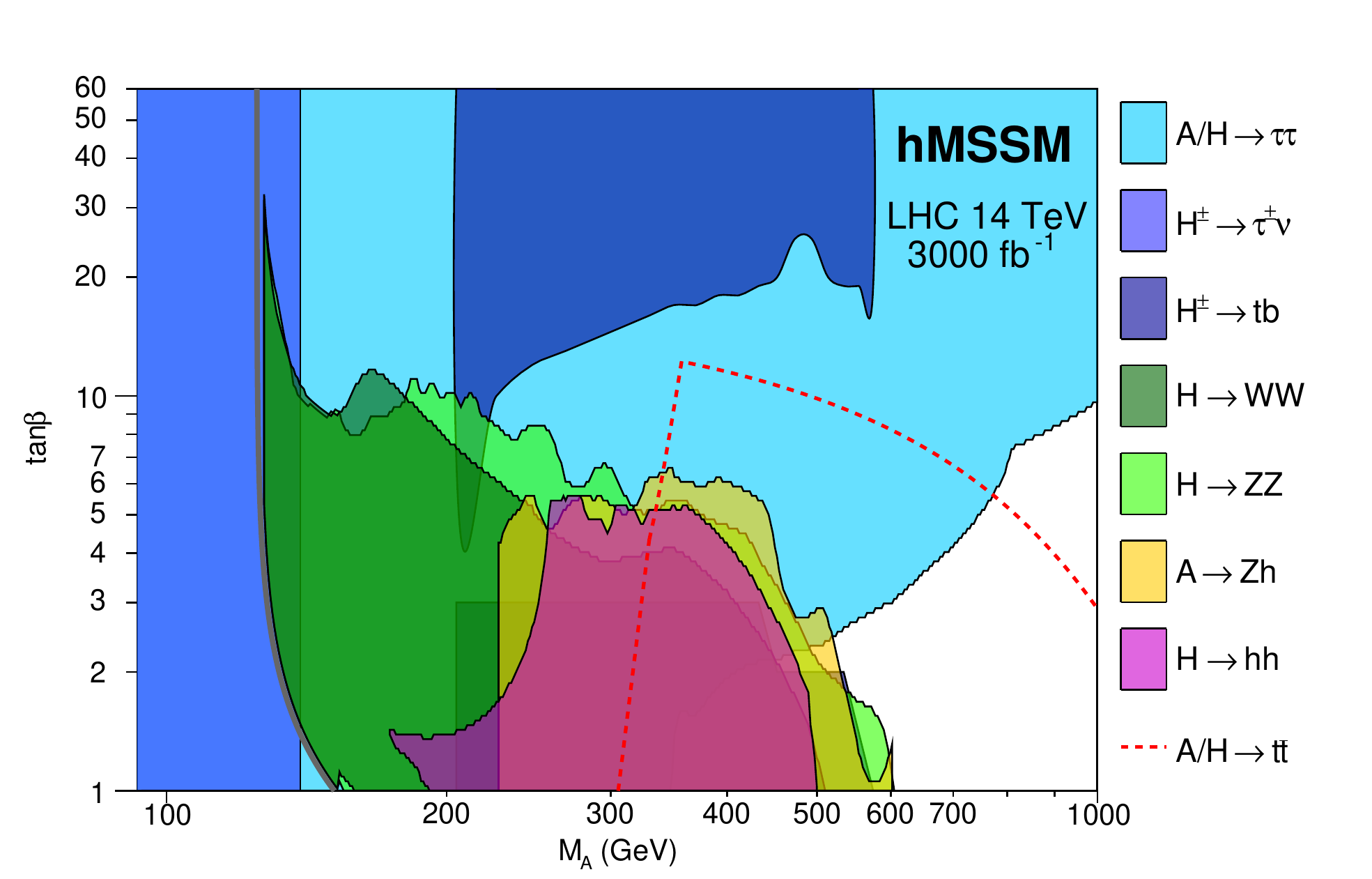}~\hspace*{-2cm} 
\includegraphics[scale=0.45]{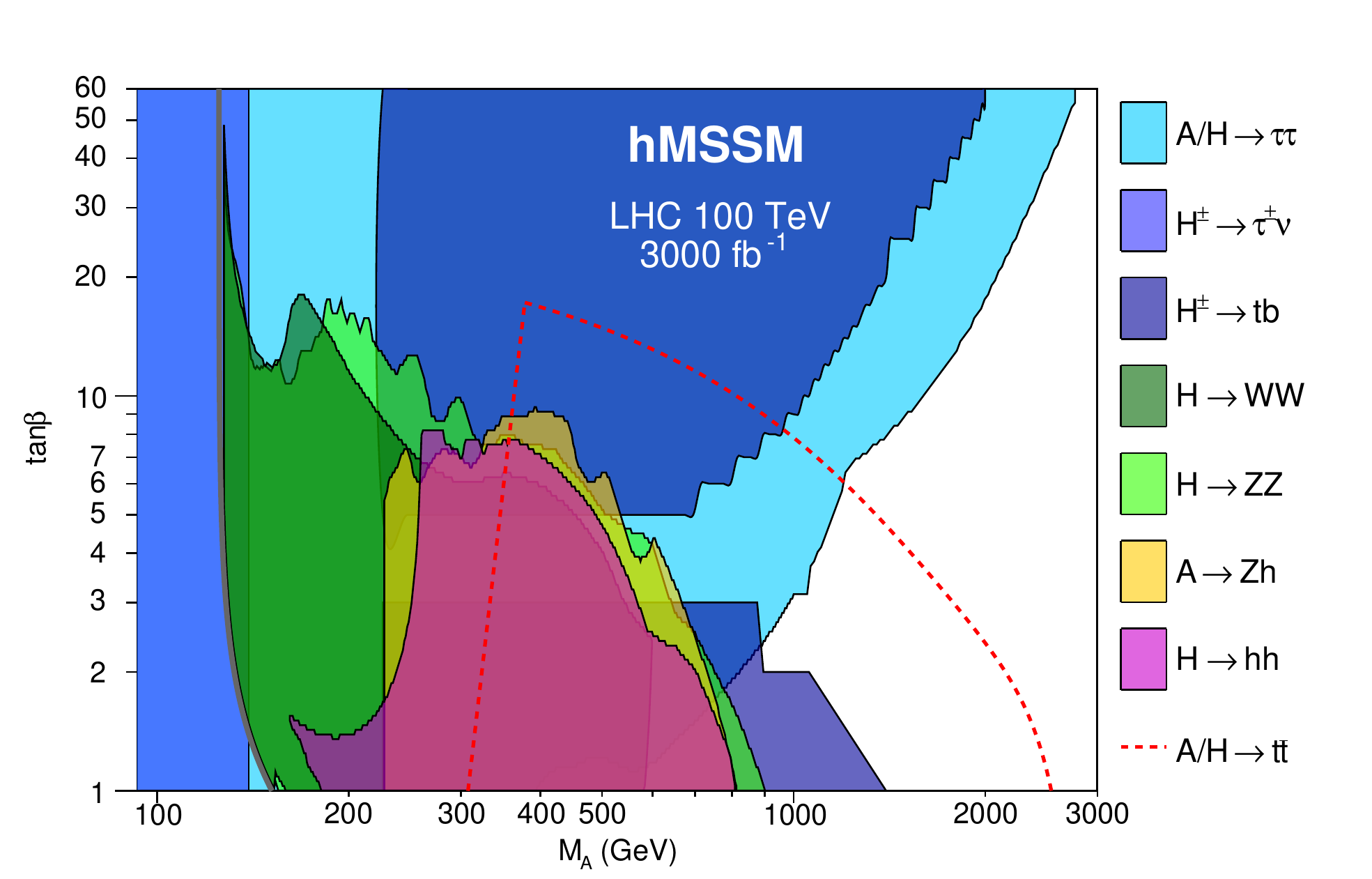} 
}
\vspace*{.1cm}
\caption[]{Top: 95\%CL countours in the $h$MSSM $[\tb, M_A]$ plane when the ATLAS and CMS searches for $A/H/H^\pm$ states in the various modes (specified in the figure with the corresponding color) at RunI are combined. Bottom: the projected  $2\sigma$ sensitivity at HL--LHC with $\sqrt s=14$ TeV (left) and at a $\sqrt{s}=100$~TeV collider with 3 ab$^{-1}$ data (right) are also shown
assuming that it scales simply with the number of events; from \cite{Djouadi:2015jea}.} 
\label{constraints_LHC8}
\vspace*{.02cm}
\end{figure}

A summary of the discussion held in this section is given in 
Fig.~\ref{constraints_LHC8} where the 95\%CL exclusions countours obtained by
the ATLAS and CMS collaborations in the Higgs searches performed in all channels
mentioned above with the full set of RunI data, are put together in  the $h$MSSM
$[M_A, \tb]$ plane. To this, added are the constraints from the $H/A \to t\bar
t$ decay channel, as analysed in Ref.~\cite{Djouadi:2015jea} in an approximate
way. The outcome is, as one can see, rather impressive.  A large portion of the
parameter space is already excluded  by the process $pp\to H/A \to \tau\tau$ at
high $\tb$ and  by the mode $pp\to H/A \to t\bar t$ at low $\tb$ as well  as by
$H\to WW,ZZ$ and $A\to hZ$ searches. Note that the area $M_A  \lsim 130$ GeV for
any  $\tb$ value is entirely excluded by the $t \to b H^\pm \to \tau \nu$
searches. 

The sensitivity will certainly improve with the 150 fb$^{-1}$ data collected so
far at  $\sqrt s= 13$ TeV but not all channels have been analysed yet. This
sensitivity will also be higher at the next HL--LHC phase with $\sqrt s=14$ TeV
and more than one order of magnitude data. Assuming naively  that the
sensitivity in the various channels simply scales with the square root of the 
number of expected events and that no additional systematical effect will
appear,  the searches in the two main channels have been extrapolated to the
HL-LHC phase with 3 ab$^{-1}$ data and to a 100 TeV pp collider with the same
luminosity. The output of these projections for the $2\sigma$ sensitivity is
presented in the $h$MSSM $[\tb, M_A]$  plane in the lower part of
Fig.~\ref{constraints_LHC8} for HL-LHC (left) and the 100 TeV machine (right). 
As can be seen, a much larger portion of the $h$MSSM  parameter spaces can be
tested and masses close to $M_A=1.5$ TeV  and 750 TeV can be probed at
respectively $\sqrt s=14$ and 100 TeV.  Below these mass values,  for which the
curves of the $H/A \to t\bar t$ and $H/A \to \tau^+ \tau^-$ channels intersect,
the entire $h$MSSM parameter space is fully covered for all $\tb$ values.   

Before turning to the superparticles, let us make a few remarks on the detection
of the MSSM heavy Higgs bosons at future $e^+e^-$ colliders
\cite{Djouadi:1994mr}. First of all,  they  can be produced in pairs,  either
$e^+ e^- \to AH$ or $e^+ e^- \to H^+ H^-$, and the cross sections are almost the
same as in the case of the 2HDM in the alignment limit and for $M_H\! = \! M_A
\! = \! M_{H^\pm}$ (a small difference occurs only at low $M_A$ and $\tb$
values).  These cross sections have been discussed  previously and were
displayed in Figs.~\ref{Fig:ee-Phi-2HDM}:  Higgs masses  close to the beam
energy $ \frac12 \sqrt s$ can be probed. Outside the decoupling regime, channels
like $e^+e^- \to HZ, H\nu \bar \nu, e^+e^-$ and  $e^+e^- \to hA$ could also  be
probed.  The neutral Higgs bosons should decay  into $tt$ or $bb,\tau\tau$ pairs
while the charged Higgs state will decay into $tb, \tau\nu$. All these final
states cannot be missed in the clean environment  of such colliders\footnote{If
the Higgs bosons happen to decay into charginos and neutralinos
\cite{Djouadi:1996pj}, the final states can be detected easily. This is  even
true in the case of invisible decays of the neutral Higgs bosons produced in the
$HA$ process, since these modes are never overwhelming and one Higgs particle
should decay visibly either into $tt,b\bar b$ or $tb$ pairs.}.  The neutral
$H/A$ states can also  be singly produced in the $\gamma\gamma$ mode of the
colliders with a mass reach that extends to 80\% of the c.m. energy of the
original $e^+e^-$ machine and the rates are again the same as the 2HDM ones
shown in Fig.~\ref{Fig:ee-Phi-2HDM}. 

\subsubsection{The superparticle sector} 

The previous discussion on the MSSM Higgs production times decay rates can be
significantly altered by the presence of superparticles. Besides contributing
virtually to the processes and altering their rates, as it was  the case for the
$\Delta_b$ correction in the $pp\to H/A \to \tau\tau$  search mode, the SUSY
particles could appear in the decays of the Higgs bosons and modify  the
branching ratios for the standard channels that are currently searched for. This
would be the case if, for instance, invisible Higgs decays  into the DM lightest
neutralinos were kinematically possible. 

\begin{figure}[!h]
\vspace*{-.1cm}
\begin{center}
\includegraphics[scale=0.38]{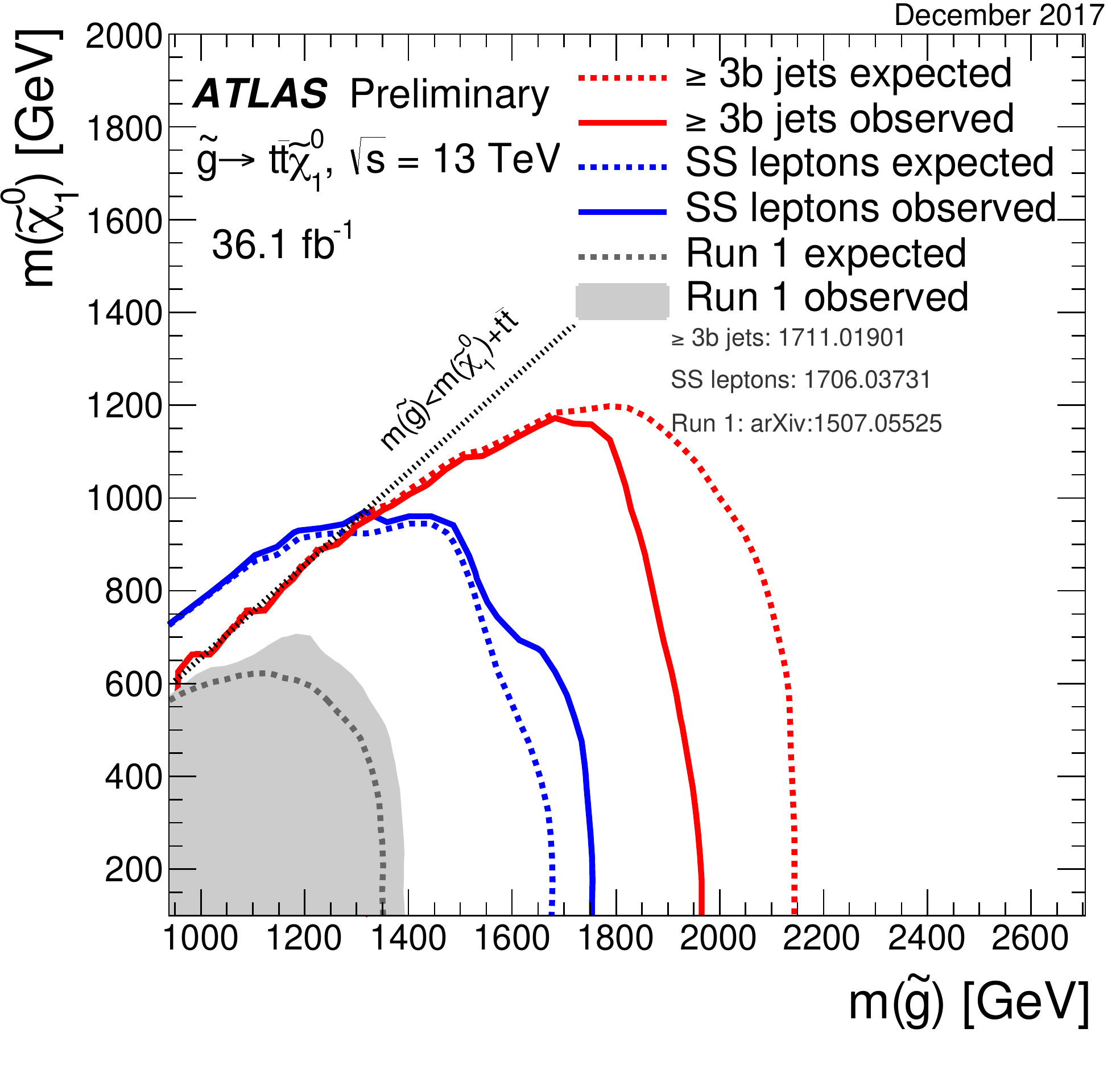}~~~
\includegraphics[scale=0.38]{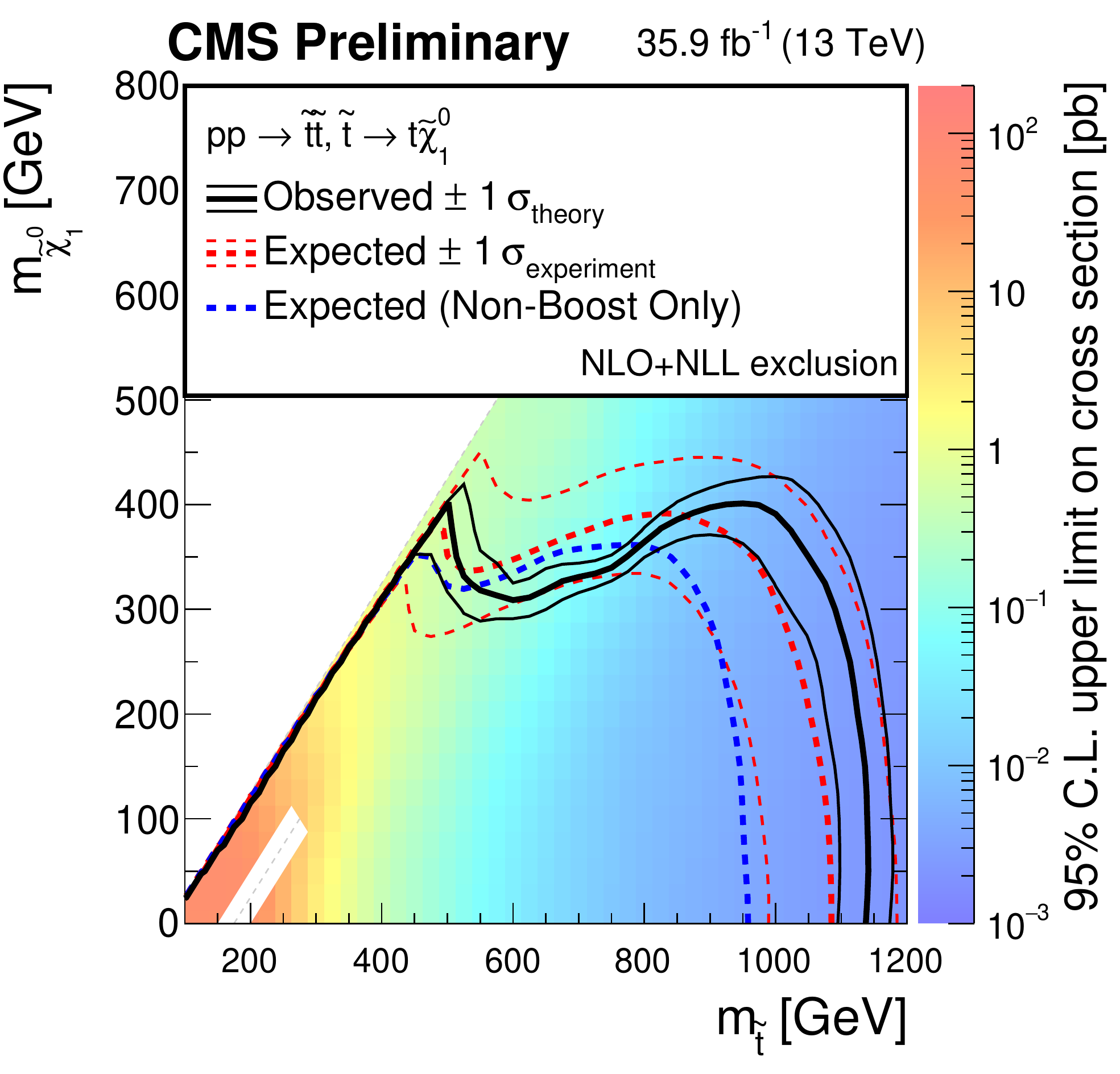}
\end{center}
\vspace*{-.6cm}
\caption{
95\%CL exclusion limits  based on 36 fb$^{-1}$ data collected at $\sqrt s \!=\!13$ TeV: for the gluino in the $[m_{\tilde g}, m_{\chi_1^0}]$  plane in a simplified ATLAS search \cite{Aad:2015iea} (left) and for the lightest stop squark in the plane $[m_{\tilde t_1}, m_{\chi_1^0}]$ in a CMS search for stop production leading to $t\bar t\!+\! E_T^{\rm mis}$ \cite{Sirunyan:2018ell} (right).
}
\label{Fig:squark-gluino}
\vspace*{-.2cm}
\end{figure}

In the current study, we have assumed from the very beginning that the sfermions
as well as the gluinos are sufficiently heavy not to impact the phenomenology of
the MSSM Higgs sector. To justify this assumption, we show in
Fig.~\ref{Fig:squark-gluino}, the  exclusion limits at the 95\%CL that were
obtained by ATLAS and CMS for the gluino (left panel) and the lightest stop
squark (right panel) using the 36 fb$^{-1}$ data collected  at $\sqrt s=13$ TeV,
plotted against the mass of the lightest neutralino. Without going into much
details, one simply notes that the resulting limits exceed 2 TeV for the gluino
and 1 TeV for the lighter stop squark\footnote{Heavy stops were anyway needed in
order to accommodate the mass of 125 GeV of the lighter $h$ boson; see e.g.
Refs.~\cite{Arbey:2011ab,Arbey:2012dq}. Note also that light stops would have
made possibly large (and negative) contributions to the $gg\to H$ production
process and significantly changed the present exclusion limits based on these
channels \cite{Djouadi:1998az}.}. The exclusion limits for the squark partners
of the SM light quarks are as severe as those on gluinos, $m_{\tilde q} \gsim
1.5$--2 TeV. The limits on the masses of sleptons from direct LHC searches are
less stringent, $m_{\tilde \ell} \gsim 500$ GeV,  but these couple in general
rather weakly to the Higgs bosons.  Hence, the only particles that could be
light with sizable enough couplings to affect Higgs phenomenology are charginos
and neutralinos. 

As discussed earlier, at very high $\tb$, the partial widths of the $\Phi \to
b\bar b, \tau^+\tau^-$ and  $H^+ \to t\bar b, \tau^+ \nu$ decays are so 
strongly enhanced, that they leave no room for SUSY channels to occur. At low
$\tb$ also, the decays $\Phi \to t\bar t$ and  $H^+ \to t\bar b$ are large when 
allowed and would be dominant. Thus, Higgs decays into charginos and neutralinos
could  play a role only for intermediate values of $\tb$ and possibly for
$M_{\Phi} \lsim 350$ GeV.  However, two conditions should be met even in this
case. First, one needs some of the $\chi$ states to be light, $M_\Phi \gsim
2m_{\chi}$,  in order to kinematically allow for some decay modes. Second, the
$\Phi \chi \chi$ couplings should be significant. These options will be
discussed in the next subsection.   

\begin{figure}[!h]
\vspace*{-.08cm}
\centerline{\hspace*{-2mm}
\includegraphics[scale=0.50]{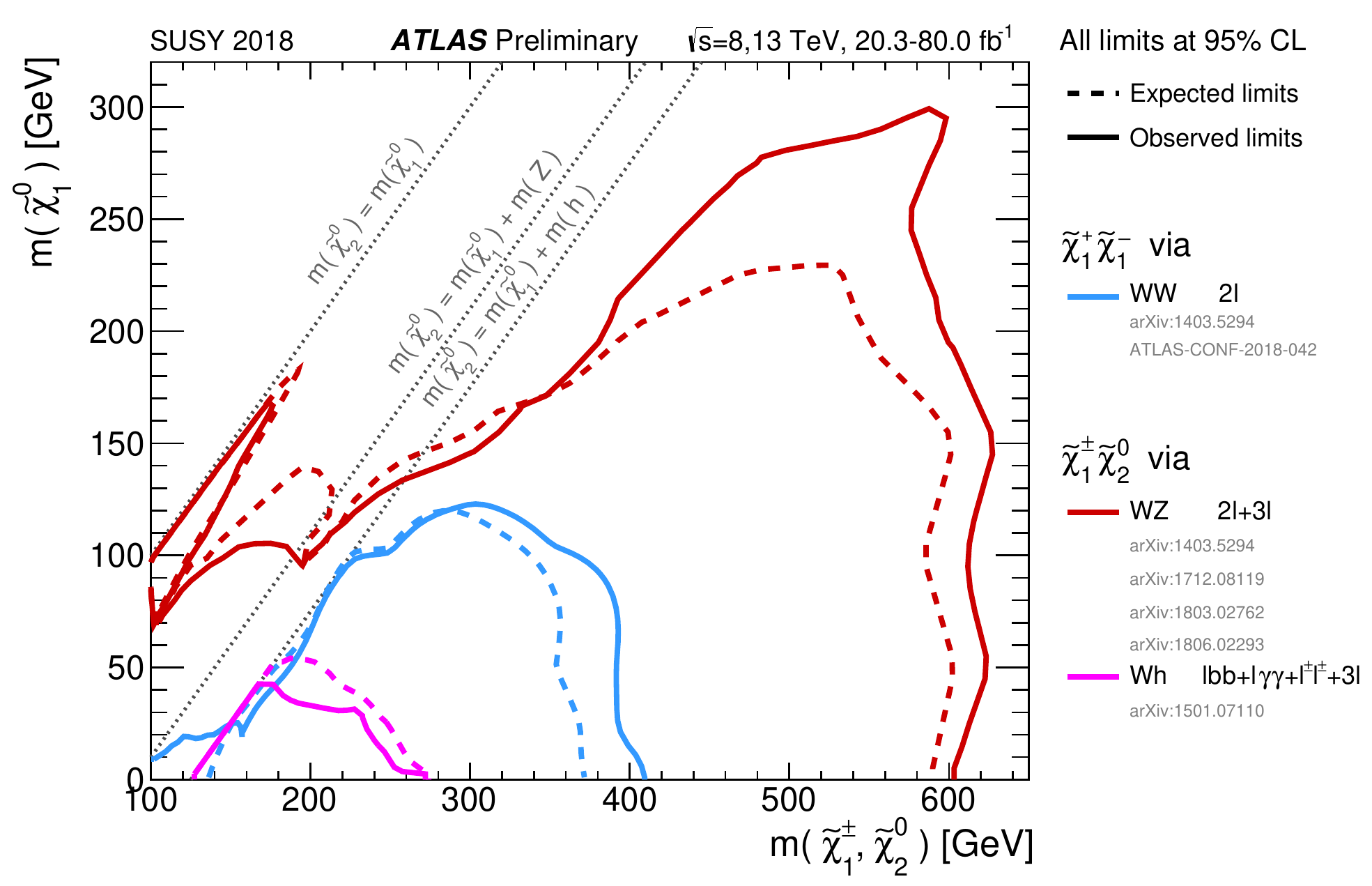}\hspace*{-2mm}
\includegraphics[scale=0.33]{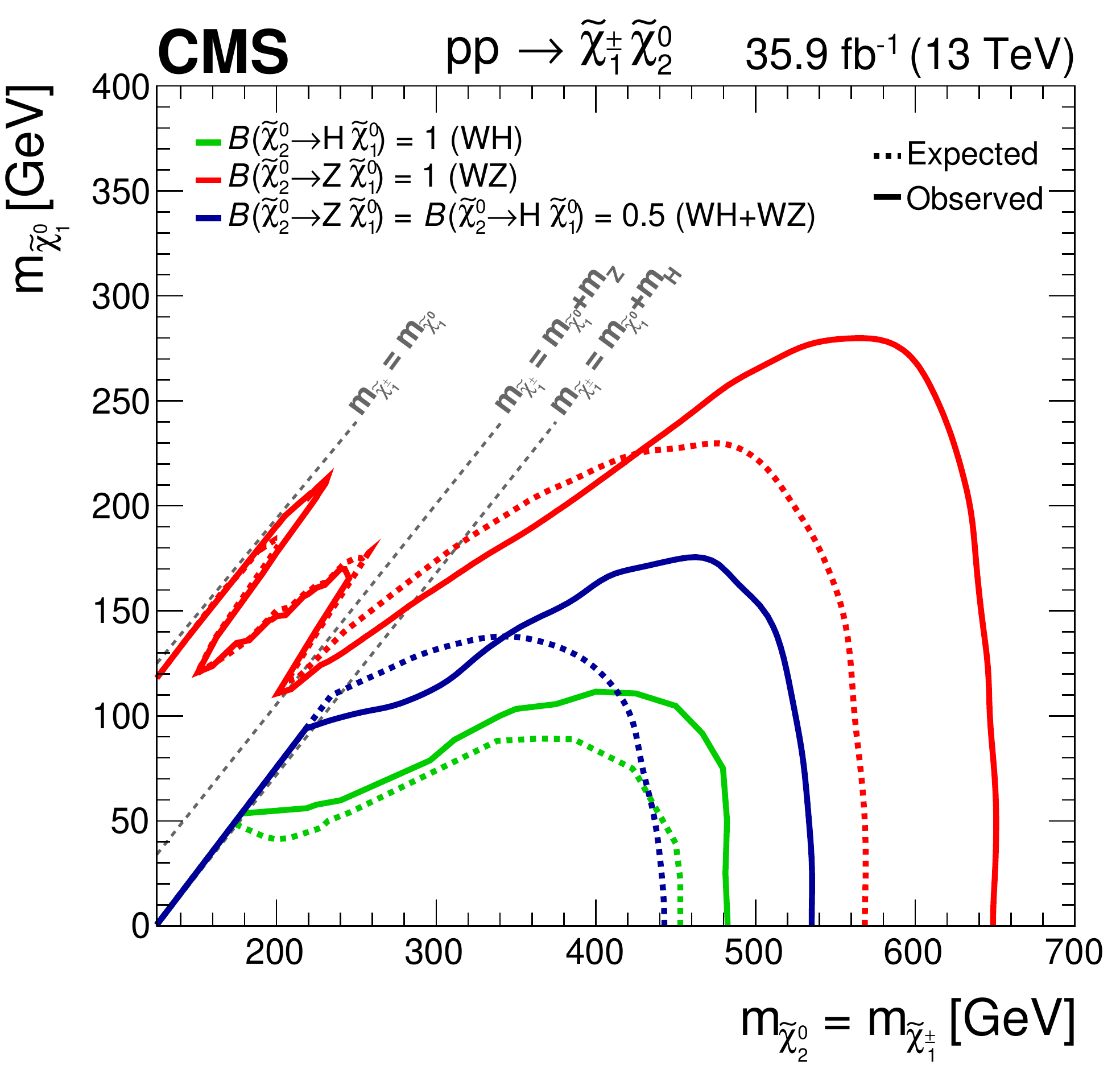}
}
\vspace*{-.1cm}
\caption{
The 95\%CL expected (solid) and observed (dashed) exclusion limits in the plane
$[m_{\chi_1^\pm}\!=\!m_{\chi_2^0}, m_{\chi_1^0}]$ in chargino--neutralino production
at the LHC using various decay modes and assumptions on their branching ratios:  an ATLAS search at $\sqrt s=13$ TeV and 36 fb$^{-1}$ data in
$pp\to {\chi_1^\pm} {\chi_2^0}$ \cite{ATLAS:2018gfq} (left) and a CMS search in the two channels $pp\to {\chi_1^\pm} {\chi_2^0}$ and $pp\to {\chi_1^\pm}  {\chi_1^\mp}$  combining 8 and 13 TeV data \cite{Sirunyan:2018ubx} (right).}
\vspace*{.1cm}
\label{Fig:LHC-ppinos}
\end{figure}

Several searches for charginos and neutralinos have been performed by ATLAS and
CMS in various channels. Of particular interest here, is direct production of
the lightest chargino and the next-to-lightest neutralino, $pp \to \chi_1^\pm
\chi_2^0$ which occurs via $W$ exchange in $q\bar q$ annihilation (we will
ignore here the additional source of charginos and  neutralinos from the cascade
decays of squarks and gluinos which, as discussed above, are assumed to be
rather heavy). The topologies that were analyzed are trileptons and missing
energy when  the decays $\chi_2^0 \to \chi_1^0 Z^{(*)} \to \chi_2^0 \ell\ell$
and  $\chi_1^\pm \to \chi_1^0 W^{(*)} \to \chi_1^0 \ell \nu$ occur. But one can
also look for the possibility $\chi_2^0 \to \chi_1^0 h$. Another interesting
channel would be $pp\! \to \! \gamma, Z \! \to \chi_1^\pm \chi_1^\mp$ leading to
two leptons and missing energy. Searches have been made by ATLAS and CMS in both
topologies and the outcome is illustrated in Fig.~\ref{Fig:LHC-ppinos}  in the
plane $[m_{\chi_1^\pm}\!= \! m_{\chi_2^0}, m_{\chi_1^0}]$. ATLAS uses the  13
TeV data while CMS combines them also with the RunI data.  

As can be seen, some  areas with masses as high as $m_{\chi_1^\pm}\!=
\!m_{\chi_2^0} \! \approx \! 600$ GeV  can be excluded but in some cases, masses
as low 200 GeV are still allowed. The limits highly depend on the mass
difference with the LSP neutralino and on the $\chi_2^0, \chi_1^\pm$ branching
fractions. 

\begin{figure}[h!]
\vspace*{-.3cm}
\begin{center}
\includegraphics[width=.49\textwidth]{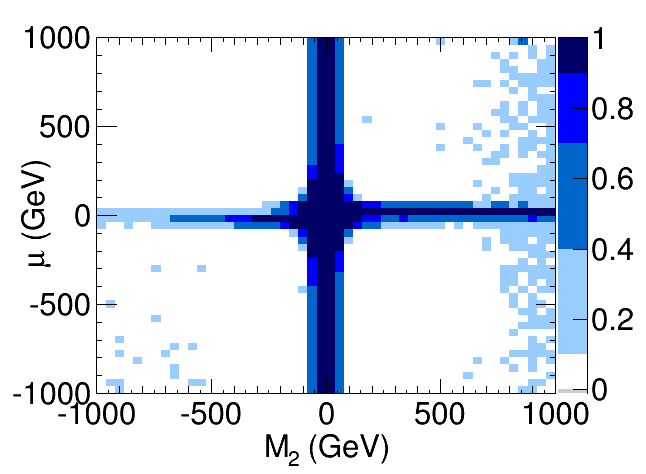}~~\includegraphics[width=.49\textwidth]{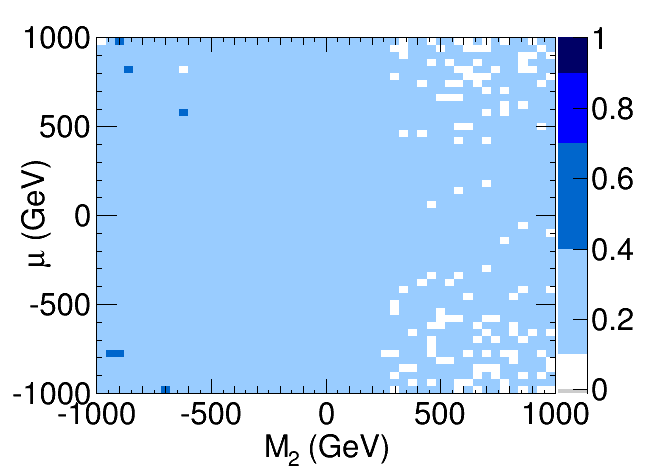}
\vspace*{-.5cm}
\caption{Fraction of excluded points by Higgs coupling measurements (left) and heavy Higgs searches (right) at the LHC, in the $[M_2,\mu]$ plane as a result of a large scan of the pMSSM parameter space \cite{Arbey:2018wjb}. \label{fig:M2-mu}
}
\end{center}

\vspace*{-1.cm}
\end{figure}

In fact, from the previously discussed wide scan of the pMSSM parameter space
when projected on  the $[M_2,\mu]$ bidimensional plane as shown in
Fig.~\ref{fig:M2-mu}, it is clear that a large portion of parameter space is
still allowed by LHC Higgs data and LEP searches. This can be seen from  the
left panel where only the narrow bands $|\mu| \approx \pm 100$ GeV and $M_2
\approx 100$ GeV are excluded, mainly from LEP2 chargino searches as these
parameters affect only little $h$ phenomenology. In turn, there is a large
impact from direct superparticle searches at the LHC and a large part of the
plane is excluded, but many points survive even for $\mu,M_2$ (and hence
$m_{\chi_1^\pm}$)  values only slightly larger than the LEP2 limit of 100 GeV.

Hence, these states can be light enough to affect Higgs phenomenology. In
particular, an interesting feature is that the charginos and  neutralinos
produced at the LHC, mainly in the pair production  modes $pp \to \chi_i^\pm
\chi_j^\mp, \chi_i^0 \chi_j^0$ and $\chi_i^\pm \chi_j^0$, can also decay into
Higgs bosons, providing an additional source for these particles besides direct
production. One example has been discussed just above: the decay channel
$\chi_2^0  \to h \chi_1^0$ which competes with the more conventional mode 
$\chi_2^0  \to Z^{(*)} \chi_1^0$, while the lightest chargino has a unique decay
channel, $\chi_1^\pm  \to W^\pm  \chi_1^0$ \cite{Aad:2015jqa,Sirunyan:2017obz}. 

To see how the decays of the $\chi_2^0$ and $\chi_1^\pm$ states behave, let us for simplicity ignore phase--space suppression and assume the decoupling limit.  The partial widths for the decay modes above, in units of $G_F M_W^2|\mu|/(8 \sqrt{2} \pi)$, are the given simply by \cite{Gunion:1987kg,Gunion:1988yc,Djouadi:1996pj,Djouadi:2001fa,Muhlleitner:2003vg}
\beq 
\Gamma( \chi_1^+ \! \to\!   \chi_1^0 W^+)  \! \approx \!   \Gamma(
\chi_2^0 \! \to \!  \chi_1^0 h ) \! \approx \! \sin^2 2\beta , \  \Gamma( \chi_2^0 \! \to \! \chi_1^0 Z ) \! \approx \!   \cos^2 2\beta  (M_2-M_1)^2/{4\mu^2}\, .~~~ 
\label{inodec-WZh} 
\eeq 
The first two are large at low $\tb$ when $\sin2\beta \approx 1$ and the last one large at high $\tb$. 

\begin{figure}[!h]
\vspace*{-.6cm}
\centerline{\hspace*{1.8cm}
\includegraphics[scale=0.56]{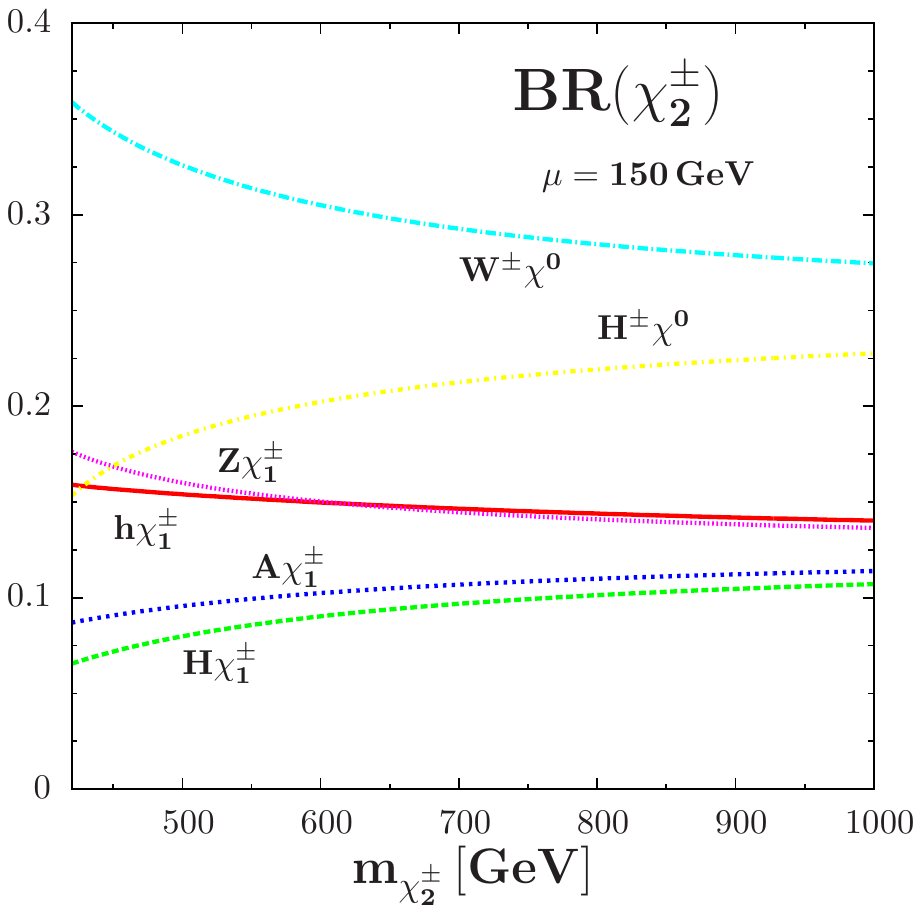}\hspace*{-6.6cm}
\includegraphics[scale=0.56]{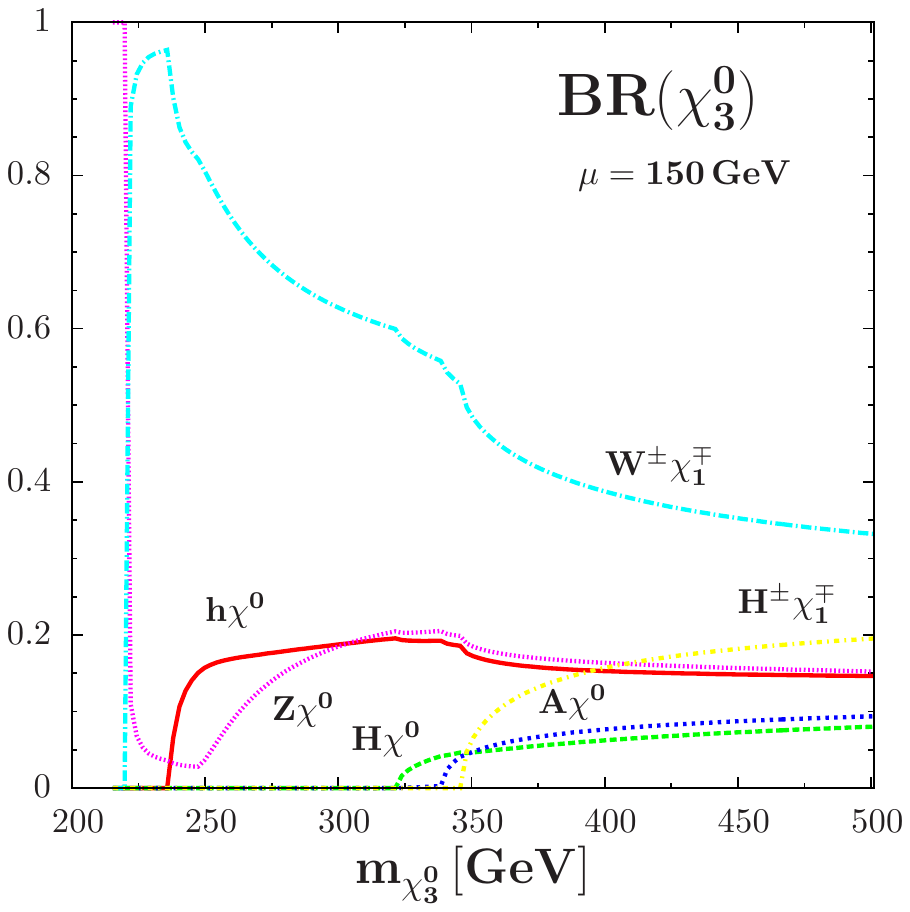}\hspace*{-6.6cm}
\includegraphics[scale=0.56]{figs-MSSM/MSSM-chi-N3.pdf}
}
\vspace*{-10.8cm}
\caption{The branching ratios for the heavier charginos and neutralinos decaying
into lighter ones and gauge or Higgs bosons for $\tb=10$, $M_A=180$ GeV and
$\mu=150$ GeV while $M_2$ is fixed by the varying mass of the decaying particle.}
\vspace*{-.2cm}
\label{Fig:inostoH}
\end{figure}

The decay pattern of the heavier charginos and neutralinos into Higgs and gauge
bosons is more involved as many other possibilities are allowed. An example of
the branching fractions that can be obtained is shown in Fig.~\ref{Fig:inostoH},
where they are given for all possible decays of  $\chi_2^\pm$, $\chi_{3}^0$ and
$\chi_{4}^0$ into the lighter charginos and neutralinos  $\chi_1^\pm,
\chi_{1,2}^0$  and gauge or Higgs bosons \cite{Datta:2001qs,Datta:2003iz}. We
have chosen a scenario in which  $\tb=10$ and $M_A= 180$ GeV; $\mu$ is  fixed at
a small value, $\mu=150$ GeV and the $M_2$ parameter is varied with the mass  of
the decaying state. This means that the lighter $\chi$ states are higgsino--like
and the heavier ones gaugino--like for the chosen $M_2 =250$--500  GeV mass
range.

As the Higgs bosons preferentially couple to gaugino--higgsino mixtures,  Higgs
couplings to mixed heavy and light chargino/neutralino states will be maximal
while those involving only heavy or light states are suppressed by powers of
$|\mu|/M_2$ for $|\mu| \ll M_2$. In turn, gauge bosons couple preferentially to
higgsino-- or gaugino--like states and should be thus suppressed. But the
corresponding partial widths receive  an extra factor of $m_{\chi_i}^2$ from the
longitudinal components of the $W,Z$ bosons which enhances them  at high masses.
This makes that, ultimately, the branching ratios for the decays  into Higgs or
gauge bosons will be of the same order but, as usual, the charged current  modes
will be more important than the neutral modes. All this can be seen from the
figure where, in addition, one can note that these decays will have in general  
individual branching fractions of the order of 10 to 30\% except when phase
space is not favorable. 

Finally, we note that charginos and neutralinos can be better detected at
$e^+e^-$ colliders.  They can be produced directly in the annihilation channels 
$e^+e^- \to \chi_i^\pm \chi_j^\pm$ and $e^+e^- \to \chi_i^0 \chi_j^0$ with a
mass reach of $m_{\chi_i^\pm} \approx \frac12 \sqrt s$ for charginos and 
$m_{\chi_i^0}+ m_{\chi_1^0} \approx \sqrt s$ for neutralinos. The LSP
neutralinos can be also produced in pairs and detected in the channel $ e^+e^-
\to \chi_1^0 \chi_1^0 \gamma$ with an initial state radiated photon, but the
mass reach is not high as the rates are small when sneutrinos (that can be
exchanged in the $t$--channel) are heavy and the $Z \chi_1^0\chi_1^0$ coupling
(which governs the $s$--channel $Z$--exchange contribution)   tiny.  Again, the
final states will be easily identifiable thanks to the clean environment and the
expected high luminosity will even allow to study their properties in great
details \cite{Choi:1998ut,Choi:2001ww}. In fact, a large number of observables
can be constructed and measured, allowing to reconstruct (even analytically) the
chargino and neutralino systems in great details
\cite{Choi:1998ei,Choi:2000ta,Choi:2001ww}.

\subsubsection{Interplay of the SUSY and Higgs sectors and the DM connection}

Let us now turn to the decays of the MSSM Higgs bosons into charginos and
neutralinos. If the Higgs states are denoted by $H_k$, with $k=1,2,3,4$ for,   
respectively, $H,h,A,H^\pm$, the partial widths of their decays into 
$\chi_i \chi_j$ pairs are  given by
\cite{Griest:1987qv,Gunion:1988yc,Djouadi:1992pu,Djouadi:1996pj}
\begin{eqnarray}
\Gamma (H_k\! \to \!\chi_i \chi_j) \!= \!\frac{G_\mu M_W^2 s_W^2}{2 \sqrt{2} \pi} \frac{ M_{H_k} \lambda_{ij}^{\frac12 } }{1\!+ \! \delta_{ij}} \left( \left[ (g_{ijk}^L)^2\! + \! (g_{jik}^R)^2 \right] (1\!-  \! 
\kappa_i^2 \! - \!  \kappa_j^2)\! -\!4 \epsilon_i \epsilon_j g_{ijk}^L g_{jik}^R \kappa_i \kappa_j  \right),~~
\end{eqnarray}
where $\kappa_i= m_{\chi_i}/M_{H_k}$ and $\delta_{ij}=0$ unless the final state
consists of two identical (Majorana) neutralinos where $\delta_{ii}=1$, 
$\epsilon_i =\pm 1$ stands for the sign of the $i$th eigenvalue of the
neutralino mass matrix  while  $\epsilon_i=1$ for charginos; $\lambda_{ij}
=1+\kappa_i^4+\kappa_j^4-2(\kappa_i^2 \kappa_j^2 +\kappa_i^2 +\kappa_j^2)$
is the phase space factor. The Higgs couplings to charginos and neutralinos were given in eqs.~(\ref{cp:inos1})--(\ref{cp:inos3}).

\begin{figure}[!h]
\vspace*{-2.5cm}
\begin{center}
\includegraphics[scale=0.8]{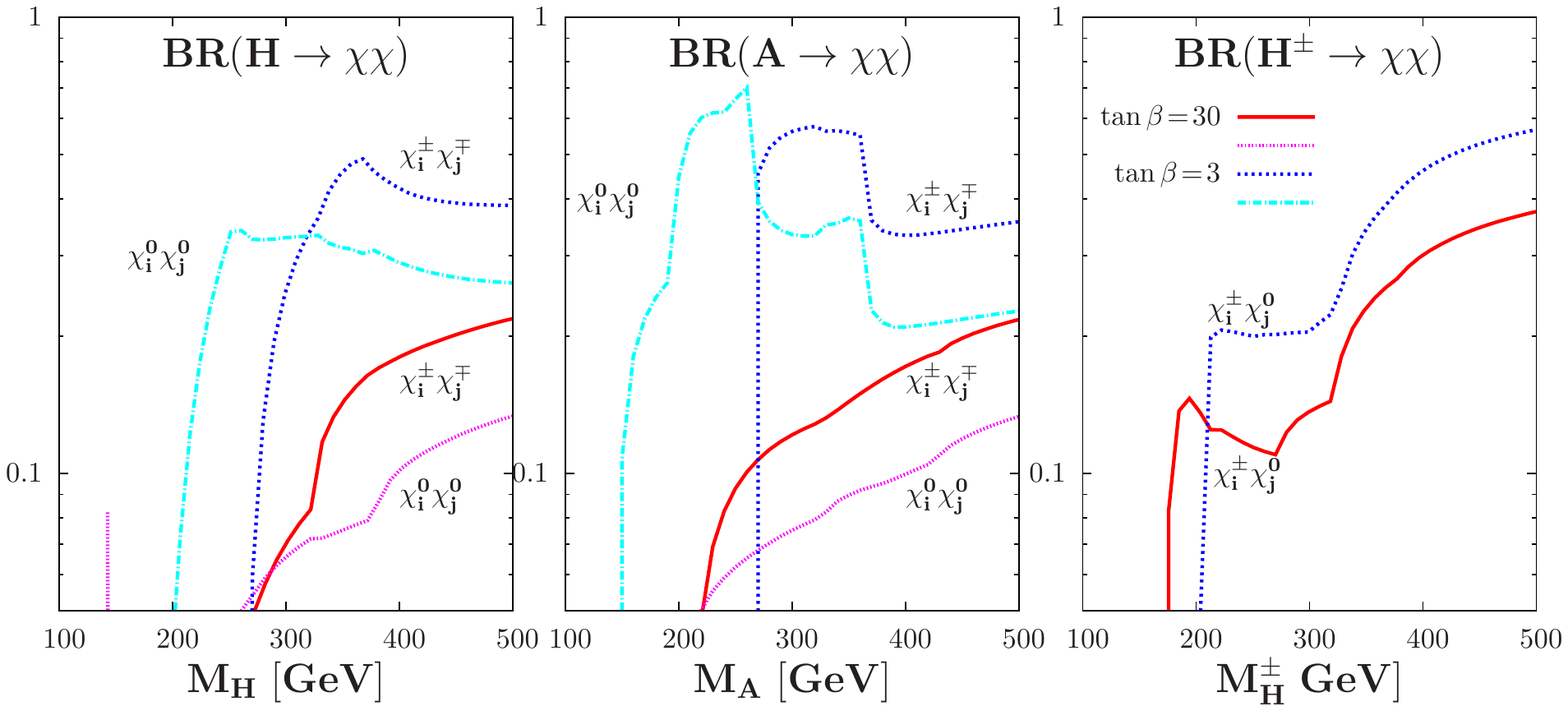}\\[-16cm]
\includegraphics[scale=0.7]{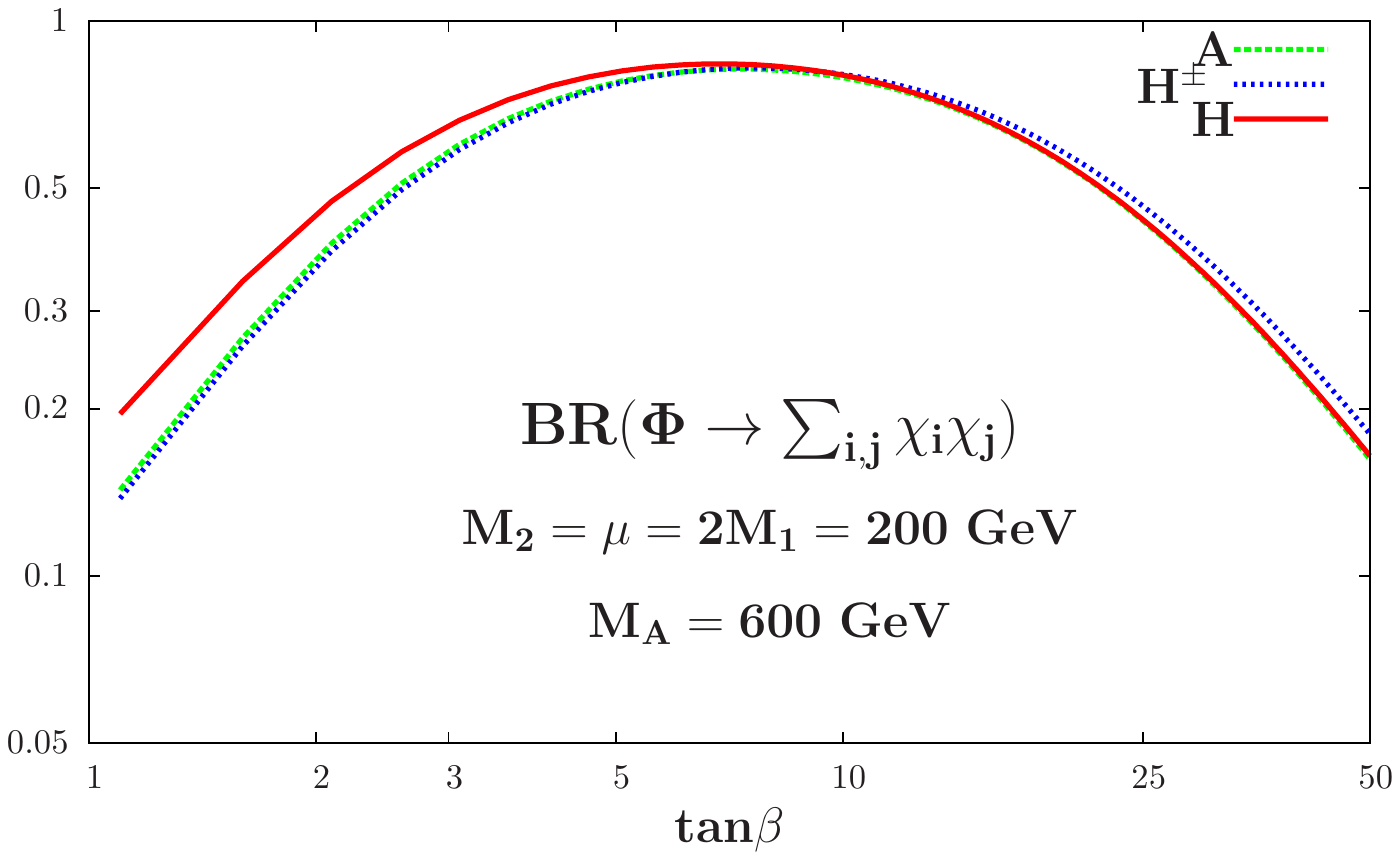}
\end{center}
\vspace*{-13.cm}
\caption{The branching ratios for Higgs decays into charginos and neutralinos 
as a function of their masses for $\tb=3,30$ and SUSY parameters $M_2=-\mu= 150$ GeV. The sum of Higgs decay branching ratios when all channels are present and summed up as a function of $\tb$ for $M_A=600$ GeV and $\mu=M_2=200$ GeV (lower plot); from Ref.~\cite{Djouadi:2006bz}.}
\label{Fig:Htoinos}
\vspace*{-.2cm}
\end{figure}

In the gaugino or higgsino limits for the lightest $\chi$ states, respectively
$|\mu| \gg M_{1,2}$ or $|\mu| \ll M_{1,2}$,  the neutral Higgs boson decays into
identical neutralinos and charginos $A/H \to \chi_i \chi_i$ as well as $H^\pm
\to \chi_{1,2}^0 \chi_1^\pm, \chi_{3,4}^0 \chi_2^\pm$ will be strongly
suppressed by the couplings but not by phase--space. Those to mixed heavy and
light states will in turn be favored by the couplings.  For instance, in the
gaugino limit and if one ignores phase--space suppression by assuming $M_{H_k}
\gg |\mu| \gg M_2$, the partial widths of the heavy Higgs decays into mixed
$\chi$ states in units of   $G_F M_W^2 M_{H_k}/(4  \sqrt{2} \pi)$ are simply
given, for $i=1,2$ and $j=3,4$, by
\bea
\Gamma(H/A \! \to \! \chi_i^0 \chi_j^0) \! =\! \delta_i [1 \pm  \sin 2
\beta ]/2 \,  , \ \ \Gamma(H/A \to \chi_1^\pm \chi_2^\mp) = 1  \, , 
\eea
with $\delta_1 = \tan^2 \theta_W, \delta_2=1$ so that one of the neutral Higgs 
decays is not suppressed when $\tb$ is either large or close to unity.  The
decays $H^\pm  \to  \chi_i^\pm \chi_j^0$ do not depend on $\tb$ in this limit
and the widths are simply either 1 or $\tan^2 \theta_W$ in the same units. The
branching ratios of the three heavy Higgs bosons decaying into the sum of
neutral and charged (or both for $H^\pm$) $\chi$ states are illustrated in
Fig.~\ref{Fig:Htoinos} as a function of the Higgs masses for two values $\tb=3$ and 30 and in the mixed gaugino--higgsino region $M_2=-\mu= 150$ GeV where all
$\chi$ states are relatively light and can appear in the decay products.

As can be seen, for large $M_{H_k}$  values, when all the channels are
kinematically open, the branching ratios are significant and sometimes even
dominant despite of the low and large values of $\tb$ which enhance  the  top
and bottom decay modes, respectively.  Once more, we note that the maximal Higgs
decay rates into these particles are obtained  at moderate $\tb$ when all
channels are kinematically accessible. In this case, as a consequence of the
unitarity of the diagonalizing $\chi$ mixing matrices, the sum of the partial
widths do not depend on any SUSY parameter when phase space effects are
neglected. One has for the total branching ratios when all decays are summed up
\cite{Gunion:1988yc,Djouadi:1996pj} 
\begin{eqnarray}  
{\rm BR}( \Phi \to \sum_{i,j} \chi_i \chi_j) = \frac{ \left( 1+\frac{1}{3} \tan^2 \theta_W \right) M_W^2 }{ \left( 1+\frac{1}{3} \tan^2\theta_W \right)  M_W^2 + \overline{m}_t^2 \cot^2 \beta + (\overline{m}_b^2 + m^2_\tau) \tan^2 \beta }   \, , 
\end{eqnarray} 
where, besides SUSY decays, only the leading $t\bar{t}$, $b\bar{b}$ and $\tau \tau$ modes  for the
neutral and the $t\bar{b}$ and $\tau \nu$ modes for the charged Higgs bosons are
included in the total widths which is indeed the case in the decoupling limit. The overall branching fraction is shown for the three MSSM Higgs bosons in the lower part of Fig.~\ref{Fig:Htoinos} as a function of $\tb$ for  $M_A=600$ GeV; the other relevant SUSY parameters are $\mu=M_2=2M_1=200$ GeV. 

One can see that the branching ratios for the three Higgses are similar and that
indeed, they do not dominate at low or  high $\tb$, being at the level of
10--20\% for both  $\tb\! =\! 1$ and $50$ ,  but exceed the level of 50\% around
the intermediate value $\tb \approx 7$ when Higgs--fermion couplings are
minimal. 

Finally, for the SUSY decays of the 125 GeV SM--like $h$ state, the experimental
bound $m_{\chi_1^\pm} \gsim 104$ GeV from LEP2 searches does not allow for any
chargino or neutralino decay mode except for the invisible decays into a pair of
the LSP neutralinos, $h \to \chi_1^0\chi_1^0$
\cite{Griest:1987qv,Gunion:1987kg,Gunion:1988yc,Djouadi:1992pu,Djouadi:1996pj}. This  is particularly true when the
universality of the gaugino masses at the GUT scale, which gives  $M_1 \sim
\frac{1}{2} M_2$ at low scales, is relaxed leading to possibly very light LSPs
while the bound on $m_{\chi_1^\pm}$ above still holds. However, as  $\chi_1^0$
should be primarily bino--like in this case, $M_1 \ll M_2, |\mu|$,  the $h
\chi_1^0 \chi_1^0$ coupling is suppressed leading to  small  invisible branching
ratios.  Nevertheless, the rate can still reach the few percent  level and,
hence, can  be revealed by future  measurements of the $h$ signal strengths or
the various direct searches for invisible Higgs decays at the HL--LHC or at
future $pp$ or $e^+e^-$ colliders.  

\begin{figure}[!h]
\vspace*{-.3cm}
\centerline{~~
\includegraphics[scale=0.35]{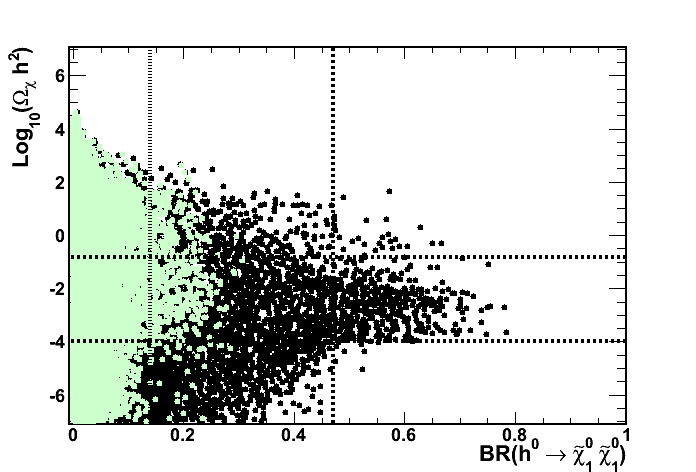}\hspace*{-.7cm}
\includegraphics[scale=0.35]{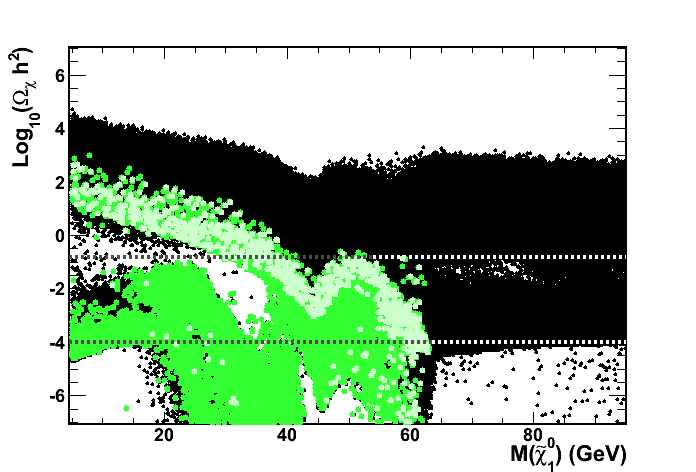}
} 
\vspace*{-.01cm}
\caption[]{The neutralino relic density $\log_{10} (\Omega_{\chi} h^2)$ as a function of BR($h \to  \chi^0_1  \chi^0_1$) (left) and $m_{\chi_1^0}$ (right) for accepted set of pMSSM points (black), those with BR$(h\to {\rm inv}) \geq 15\%$ (green) and compatible at 95\%CL with the Higgs data (light green). From Ref.~\cite{Arbey:2012bp}.}
\vspace*{-.5cm}
\label{Fig:htoLSP}
\end{figure}

In any case, such a small branching ratio allows  for the LSP to have the
required cosmological density, eq.~(1),  since it will annihilate efficiently
through the exchange of the $h$ boson. This is illustrated in
Fig.~\ref{Fig:htoLSP} where the relic density $\log_{10} (\Omega_{\chi} h^2)$,
resulting from  the previously discussed pMSSM scan \cite{Arbey:2011ab}, is
shown   as a function of the branching ratio BR($h \to  \chi^0_1  \chi^0_1$)
and  the LSP mass  $m_{\chi_1^0}$, in respectively the left and right panels.
The colored regions  indicate  the  accepted set of pMSSM points that fulfill
LEP and     flavour constraints (black dots), those with BR$(h\to \chi_1^0
\chi_1^0) \geq 15\%$ (green dots)  and those compatible at 90\%CL with the Higgs
data (light green dots). The horizontal lines show the  constraint imposed on
$\Omega_{\chi} h^2$ and the vertical lines on the panel on the right
approximatively the 90\% and 99\%CL present constraints on the invisible Higgs
branching ratio\footnote{This figure has been made in the early stage of the LHC
RunI when the luminosity was not very high and the determination of the Higgs
couplings not as precise as currently. The 68\%CL and 95\% constraints at the
time of the figure will, very roughly, correspond to the 90\% and 99\%CL present
constraints.}.  As can be seen, the area that fulfils the $\Omega_{\chi} h^2$
constraint is not very small, $ 30 \lsim m_{\chi_1^0} \lsim 60$~GeV, despite of
the strong constraints.

\subsection{Astrophysical constraints on the the MSSM}

The previous comments on the relation between the invisible Higgs branching
ratio and the DM relic density allows us to make a smooth  transition towards
the astrophysical aspects in the context of the MSSM. Let us indeed,  briefly
enumerate the traditional regions in the MSSM parameter space in which the
correct relic density for the lightest neutralino DM can be obtained (for more
detailed discussions and earlier references, we refer for instance  to
Refs.~\cite{Roszkowski:2017nbc,Bagnaschi:2017tru,Athron:2017yua,Baer:2016ucr,Arbey:2017eos,Balazs:2017ple,Barducci:2016pcb,Bagnaschi:2015eha,Roszkowski:2014iqa,Arbey:2013iza,Kadastik:2011aa}).
The DM relic density depends crucially on the composition of the lightest
neutralino.  A bino-like LSP features in general a too suppressed annihilation
cross--section as its couplings to Higgs and $Z$ bosons are strongly suppressed.
In order to achieve the correct relic density, it is then necessary to enforce
specific mass patterns in the Supersymmetric spectrum to enhance the DM
annihilation rate. On the contrary, a higgsino--like and/or wino--like LSP
features very efficient annihilation processes into gauge bosons, so that no
specific mass relations with the other SUSY particles need to be enforced. In
this case, the LSP is close in mass to some of its partners  such that
coannihilation process become important. 

Let us enumerate and comment on all the possible configurations leading to the
correct DM relic density in the MSSM.\vspace*{-2mm}

\begin{itemize}

    \item The DM relic density is mostly determined by annihilation processes
into lepton final states mediated by $t$--channel exchange of sleptons.  In
particular $\tilde \tau$'s play an important role as they are in general lighter
than the other sleptons and their fermionic $\tau$ lepton partner is the most
massive. In this regime, one has  $\Omega h^2 \propto 1/\langle \sigma v \rangle
\propto m_{\tilde \tau}^4/m_{\chi_1^0}^2$ and  the correct relic density can be
obtained for relatively light values of the DM and lightest $\tilde \tau$ 
masses, $\lesssim 150 \,\mbox{GeV}$.  This last configuration is customarily
called the ``bulk region'' \cite{Drees:1992am}. It is nevertheless nowadays
extremely constrained in the light of the negative results in searches of
superpartners at the LHC.\vspace*{-2mm}

    \item A bino--like LSP neutralino can lead to the correct relic density
through coannihilation processes, occurring when the next--to--lightest
supersymmetric particle (NLSP) is almost degenerate in mass with the DM LSP. The
most commonly considered coannihilation scenarios are with the lightest slepton
(typically the $\tau$)
\cite{Griest:1990kh,Ellis:1998kh,Ellis:1999mm,Gomez:2000sj,Baer:2002fv} or the
lightest squark (typically the lightest  stop squark)
\cite{Baer:2002fv,Boehm:1999bj,Ellis:2001nx,Arnowitt:2001yh}. A more exotic
alternative would be represented by bino--gluino coannihilations
\cite{Profumo:2004wk}. We note that in the case of coannihilation with strongly
interacting particles, like stops and gluinos, the relic density computation is
complicated by additional effects like Sommerfeld enhancement and bound state
formation \cite{Ellis:2014ipa,Ellis:2015vaa,Biondini:2018ovz}.\vspace*{-2mm}

    \item Another possibility for the correct relic density is represented by
the case in which the LSP neutralino is an appropriate admixture between bino
and higgsino components, the ``well--tempered" regime. The correct relic density
can be achieved for DM masses in the range $100\,\mbox{GeV} \lesssim
m_{\chi_1^0} \lesssim 1\,\mbox{TeV}$. In constrained and GUT inspired
realizations of the MSSM, this configuration is achieved in the so--called
hyperbolic branch or focus point region
\cite{Chan:1997bi,Feng:1999zg,Feng:2000gh}. In phenomenological realization of
the MSSM, the well tempered regime can be realized also through bino--wino
\cite{BirkedalHansen:2002am,Baer:2005jq,ArkaniHamed:2006mb} or even
bino--wino--higgsino \cite{Bae:2007pa,Feldman:2009wv} admixtures. Notice that
the well tempered bino--higgsino regime is strongly disfavored by DM direct
detection experiments, since it corresponds to enhanced spin--independent
interactions (see below).\vspace*{-2mm} 

    \item As already mentioned, the correct relic density for wino--like and
higgsino--like DM is achieved through annihilation into gauge bosons.
Coannihilations are also present because of a characteristic small mass
splitting between the DM and the lightest chargino and NLSP neutralino. The
relic density scales as $\Omega h^2 \propto m_{\chi_1^0}^2/g^4$. According to
this, the correct value would be reached for $m_{\chi_1^0} \approx 1$ and 2.5
TeV for higgsino and wino DM, respectively. The DM annihilation rate is,
however, modified by Sommerfeld factors due to the Yukawa potential originated
by the gauge bosons. While this effect is modest for the  SU(2) doublet
higgsino, it affects sensitively the SU(2) triplet wino case  such that the
correct relic density is achieved for $m_{\chi_1^0} \approx
3\,\mbox{TeV}$.\vspace*{-2mm}   

  \item  Finally, there are the Higgs pole regions and, in particular, the $A$
pole funnel in which the annihilation occurs through $s$--channel exchange of
the CP--odd Higgs boson. There, the  $A$ state can become nearly resonant, again
leading to an acceptable relic
density~\cite{Griest:1990kh,Gondolo:1990dk,Lopez:1993kr,Nath:1992ty,Drees:1995bw,Nihei:2001qs,Lahanas:2001yr,Djouadi:2001yk}.
The CP--even $H$ pole region can also be relevant for a tuned LSP texture. Also,
for very light LSPs, the Higgs pole regions can be extended to the $h$
boson~\cite{Baer:2004xx,Djouadi:2005dz} as it was discussed
above.\vspace*{-2mm}

\end{itemize}

As in this review we are mainly focusing on the connection between the DM and
the Higgs sectors, we will assume that the sfermions as well as the gluino are
very heavy (which, as seen previously is backed up by the negative searches of
these states at the LHC) and irrelevant for the DM  phenomenology. This, leaves
as unique possibilities to achieve the correct cosmological relic density, the
Higgs boson funnels, the well tempered bino--higgsino regime and the pure
higgsino-like or wino--like DM possibilities. This MSSM realization is then very
similar to the singlet--doublet model discussed at length in the previous
sections. As many aspects have been already analysed there, we will rather
briefly highlight the main differences in this section.

\begin{figure}[!h]
\vspace*{-3mm}
\centering
\subfloat{\includegraphics[width=0.47\linewidth]{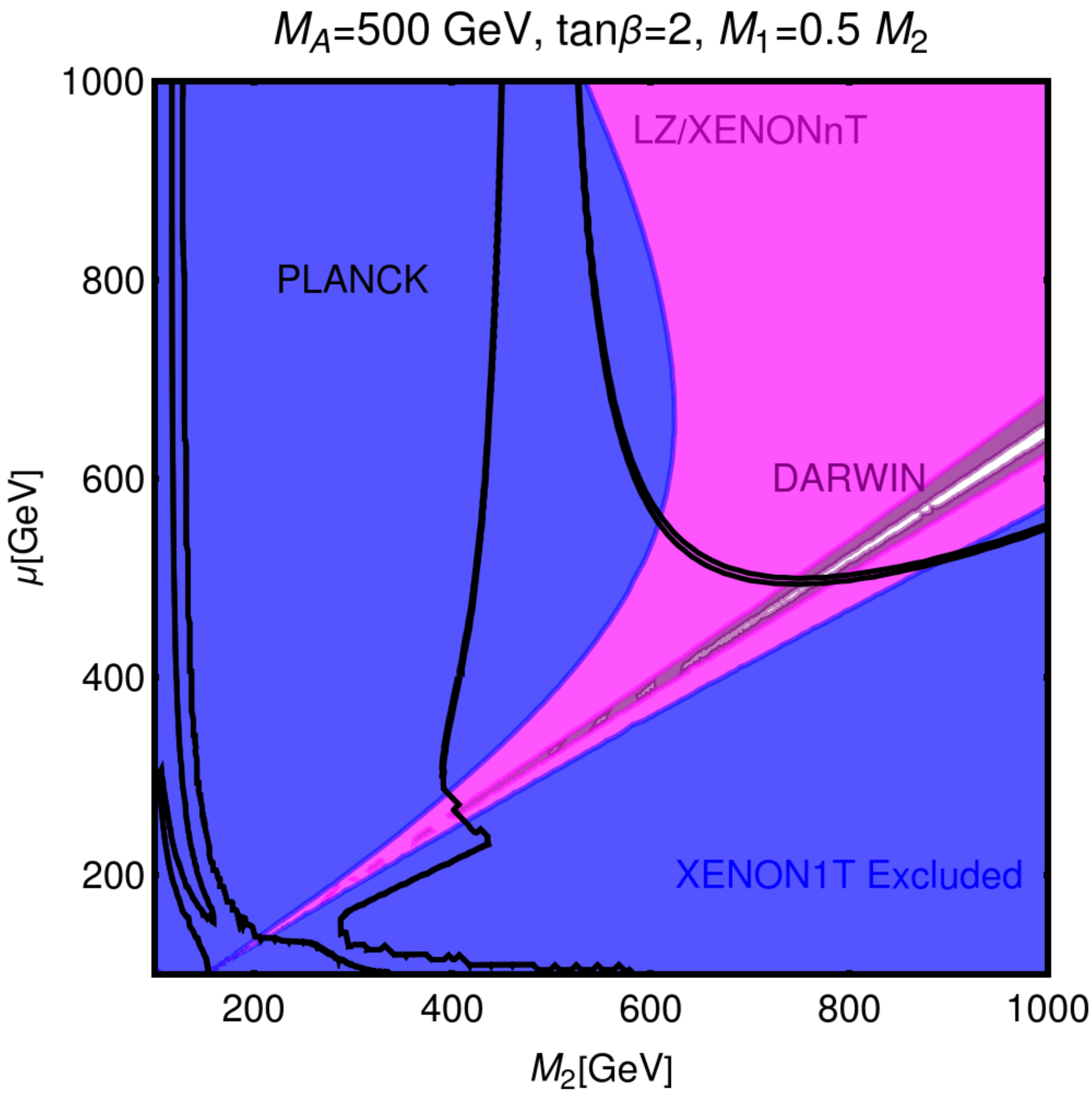}}
\subfloat{\includegraphics[width=0.47\linewidth]{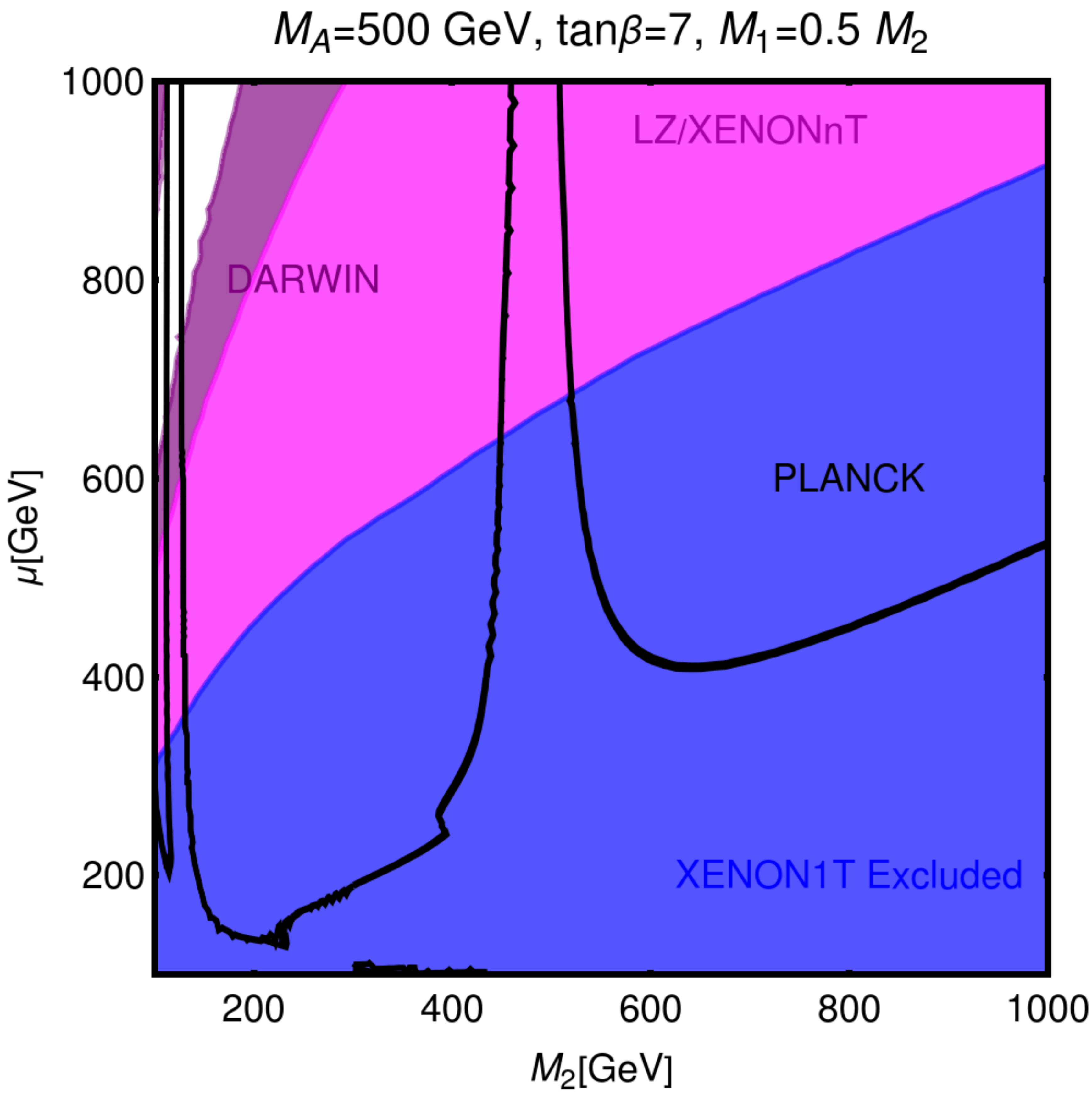}}\\
\subfloat{\includegraphics[width=0.47\linewidth]{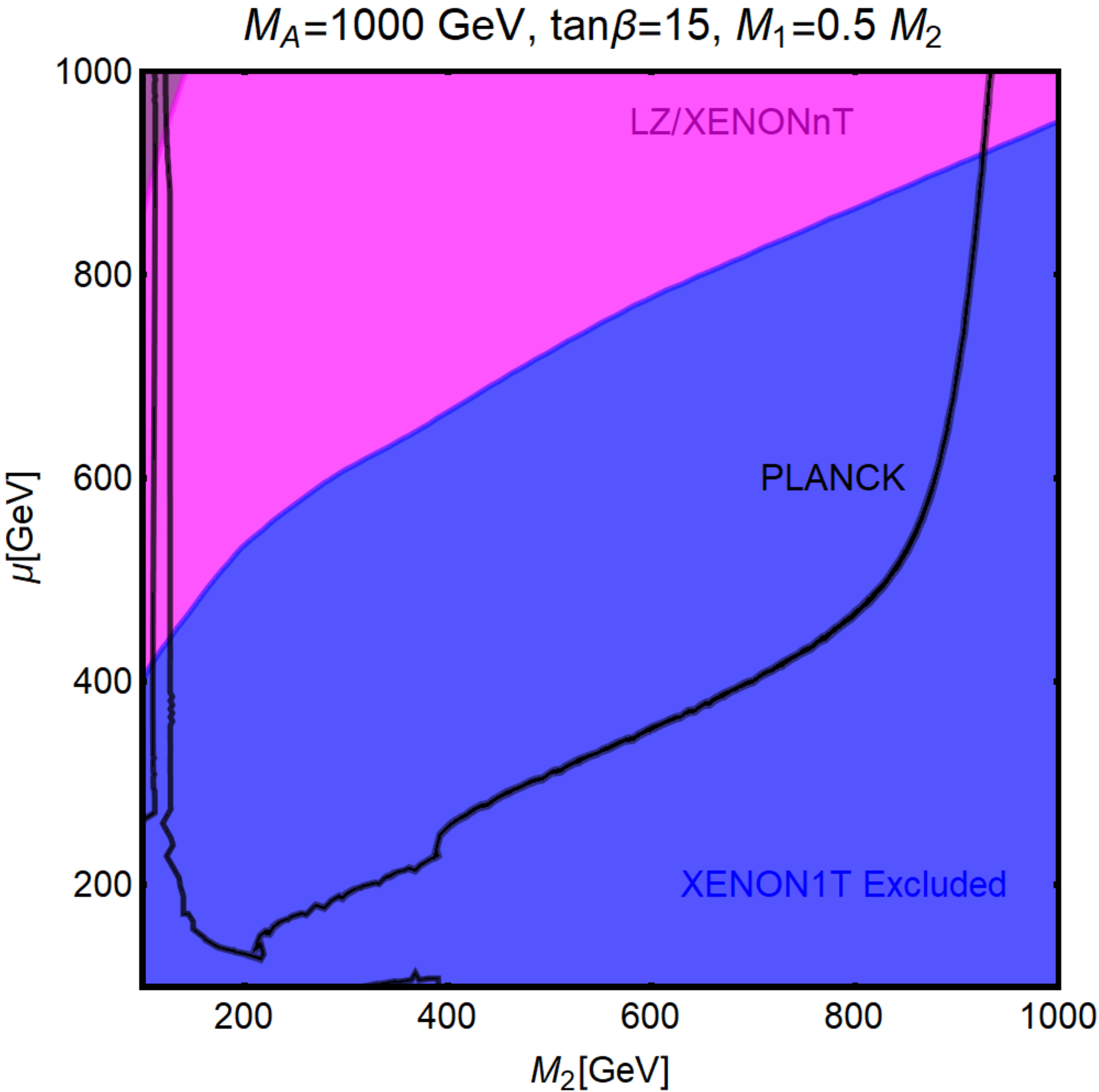}}
\subfloat{\includegraphics[width=0.47\linewidth]{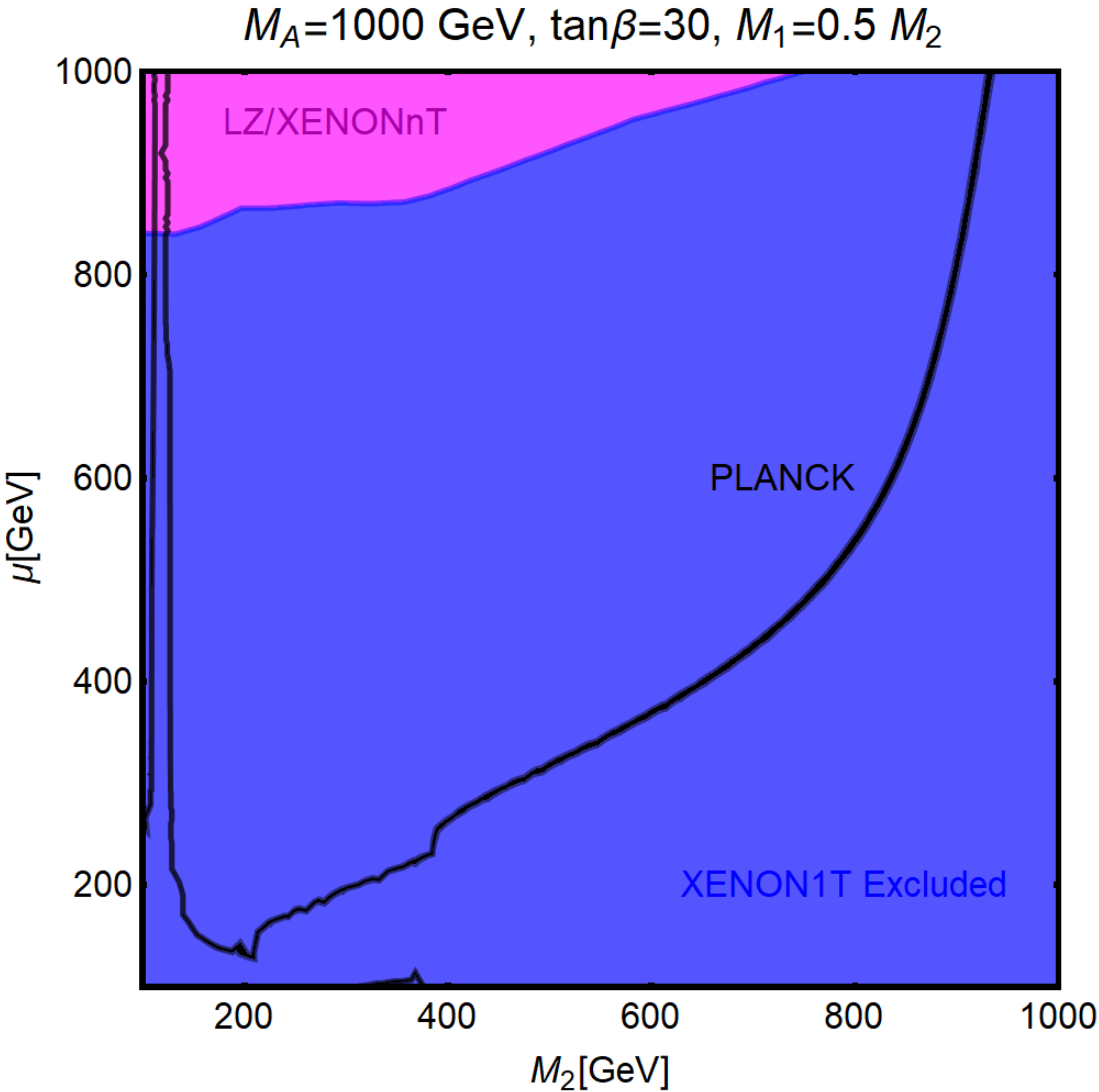}}
\vspace*{-1mm}
\caption{Combination of the main DM constraints, for the MSSM realization considered in this section, in the $[M_2,\mu]$ plane for four assignments of the $(M_A,\tb)$ parameters as reported  on top of each panel and with $M_1=\frac12 M_2$. According to the usual convention, the black contours represent the correct relic density, the blue regions are excluded by XENON1T and the magenta/purple regions will be excluded in the absence of signals at LZ/XENONnT/DARWIN.}
\label{fig:astroMSSM1}
\vspace*{-3mm}
\end{figure}

\begin{figure}[!h]
\vspace*{-3mm}
\centering
\subfloat{\includegraphics[width=0.45\linewidth]{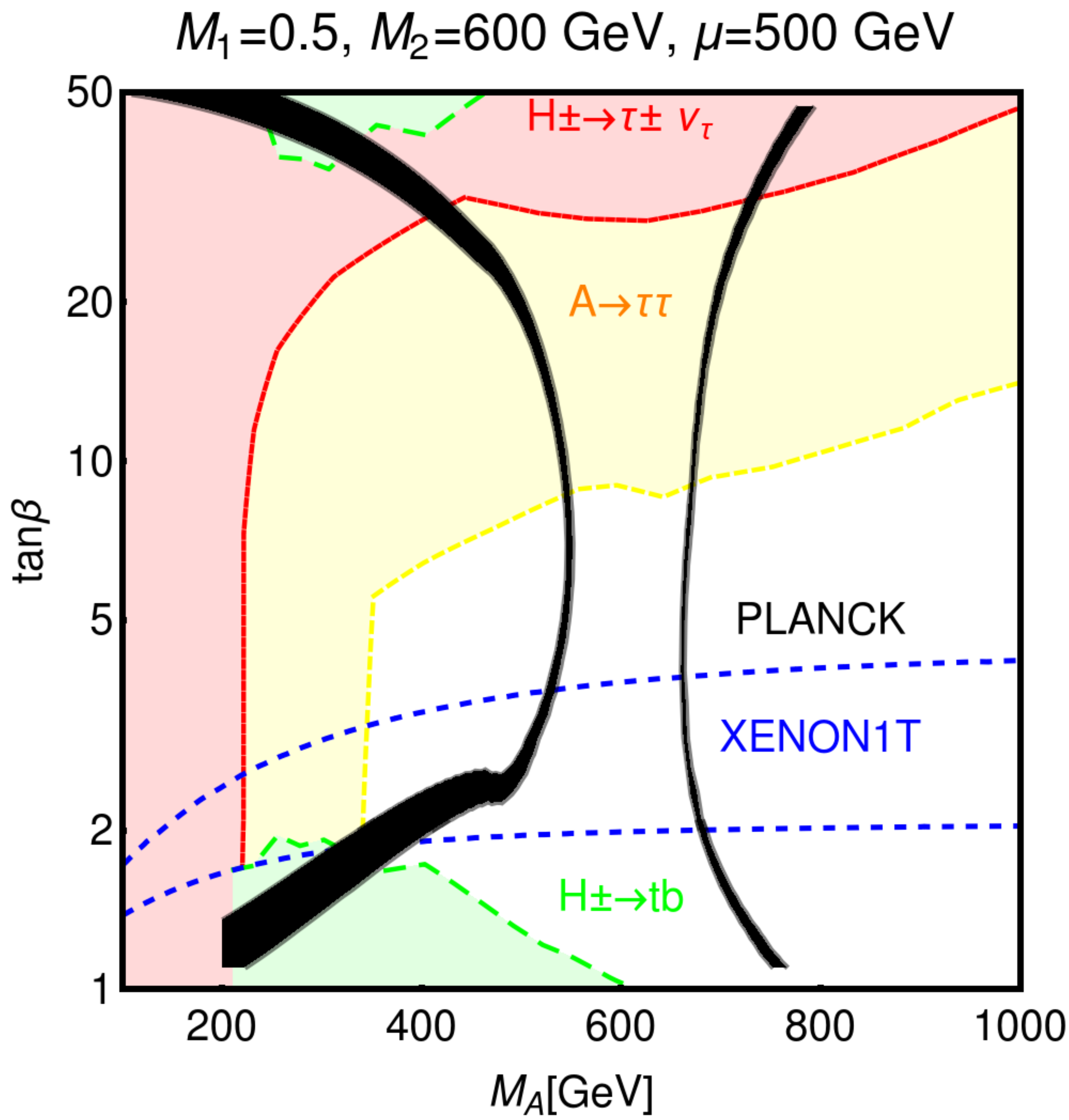}} 
\subfloat{\includegraphics[width=0.45\linewidth]{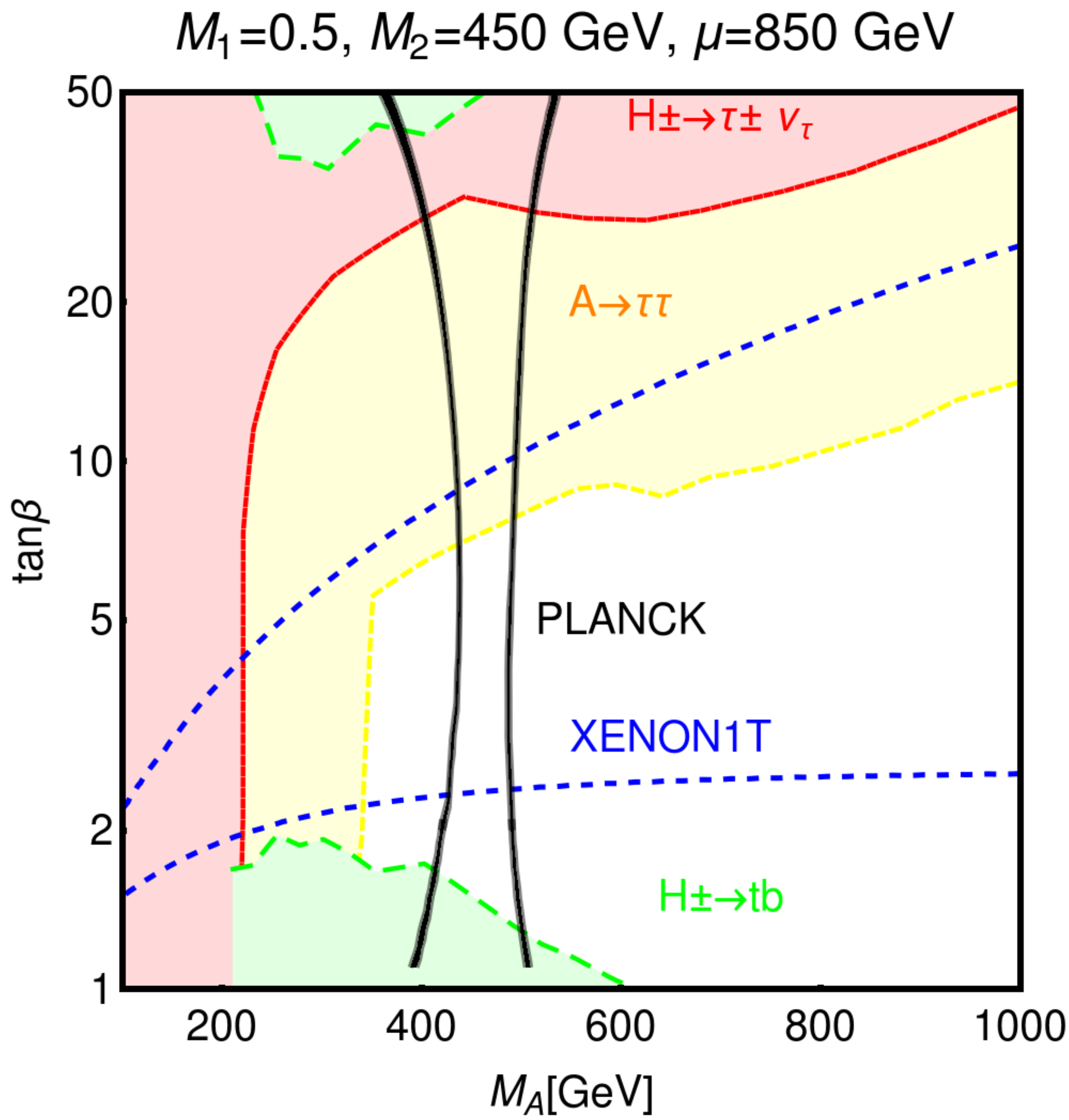}} 
\caption{Combination of collider and DM constraints in the $[M_A,\tb]$ plane for $M_2=600\,\mbox{GeV},\mu=500\,\mbox{GeV}$ (left panel) and $M_2=450\,\mbox{GeV},\mu=850\,\mbox{GeV}$ (right panel). The red/yellow/green regions are currently excluded by LHC Higgs searches. The black contours correspond to the correct DM relic density and only the regions between the dashed blue lines still evade the bounds from DM direct detection by XENON1T.}
\label{fig:astroMSSM2}
\vspace*{-1mm}
\end{figure}

The understanding of our results, obtained using the numerical package {\tt
DarkSusy}~\cite{Gondolo:2004sc,Bringmann:2018lay}, can be facilitated by 
inspecting the analytical expressions given for the singlet--doublet lepton
model. We recall that one important difference between the SDM  and the MSSM is
that in the latter, the couplings between the DM and the Higgs bosons are not
free parameters. The MSSM that we consider here features five free inputs: 
$M_1,M_2,\mu,\tan\beta$ and $M_A$ in addition to $M_S$ and the coupling $A_t$
which are taken to be very large.  However, as a further simplification, we
assume the GUT relation $M_1=\frac12 M_2$ to reduce the number of these
parameters. We thus  provide an illustration of the DM phenomenology in
Figs.~\ref{fig:astroMSSM1} and  \ref{fig:astroMSSM2} in the $[M_2,\mu]$ plane
for some fixed assignments of $[M_A,\tb]$ and vive--versa.

Similarly to the Type--II singlet--doublet model presented in section 5, one can
see from Fig.~\ref{fig:astroMSSM1} that  constraints from direct detection
experiments exclude most of the chosen  $[M_2,\mu]$ parameter space in
particular at high--$\tb$ values. The remaining viable regions will be fully
tested by future experiments, such as XENONnT and DARWIN. Hence,  ``natural" values of the LSP neutralino mass, below a few hundred GeV, are either excluded or will be soon probed.  Notice that, similarly to the case of the singlet--doublet lepton model, the DM scattering cross section can be suppressed in ``blind spot'' configurations, i.e. assignments of the model parameters corresponding to a cancellation of the coupling of the DM with the light $h$ boson or destructive interference between the contributions associated with the exchange of the two $h$ and $H$ states. An analytic expression of the blind spot condition has been provided e.g. in Ref.~\cite{Huang:2014xua} (see also Refs.~\cite{Ellis:2000ds,Ellis:2000jd,Baer:2006te}) and reads
\begin{equation}
    2 \left(m_{\chi_1^0}+\mu \sin 2 \beta \right)/{M_h^2} \simeq -\mu \tan\beta /{M_H^2} \, . 
\end{equation}
In the limit in which mass $M_H$ is large, the condition above reduces to $m_{\chi_1^0}+\mu \sin 2\beta=0$, which requires a negative $\mu$ value to be satisfied.

Moving to Fig.~\ref{fig:astroMSSM2} and to the bidimensional plane $[M_A,\tb]$,
we illustrate the combination of the DM constraints for two assignments of
$(M_2,\mu)$, namely (600,500) GeV (left panel) and (450,850) GeV (right panel).
Again, in analogy with the singlet--doublet lepton model, the strong limits
mostly coming from MSSM searches in the channel $A/H \! \rightarrow \! \tau
\tau$ which are efficient at moderate and high $\tb$ values reduce significantly
the viable DM parameter space. We have not included the constraints from the
$H/A\! \to \! t\bar t$ searches, which might be more constraining at low $\tb$
values than those from the $gb\to t H^\pm  \to t tb$ search mode that has been
included instead. While for the first benchmark bounds from DM direct detection
are by far the most competitive ones, for the second, a good complementary
between direct detection and collider bounds can be noted.


\subsection{Non minimal extensions and the NMSSM}

The Higgs sector in supersymmetric theories may be more complicated if some
basic assumptions of the CP--conserving MSSM, like the absence of new sources of
CP  violation, the presence of only two Higgs doublet fields, or the
conservation of $R$--parity, are relaxed. For instance, if CP--violation is
present in the SUSY sector, an element that is in principle required if the
model has also to explain baryogenesis at the weak scale, the new phases will
enter the MSSM Higgs sector through  the large radiative corrections and alter
the Higgs masses and couplings. In particular, the three neutral Higgs states
will not have definite CP quantum numbers and will mix with each other to
produce the physical states \cite{Lee:2003nta,Accomando:2006ga}. The Higgs
bosons that serve as portals to the DM will have both CP--even and CP--odd
components which make them rather appealing  in this respect. 

Another very interesting extension is the next--to--minimal supersymmetric SM,
the NMSSM \cite{Ellwanger:2009dp,Maniatis:2009re,Djouadi:2008uw,Baum:2017enm},
which consists of simply introducing a complex iso-scalar field which naturally
generates a weak scale value for the supersymmetric Higgs--higgsino parameter
$\mu$  \cite{Nilles:1982mp,Frere:1983ag,Kim:1983dt}. Also in this case, the
Higgs sector is extended to contain an extra CP--even and one CP--odd Higgs
particles that could be very light, generating additional portals to DM with a
quite interesting phenomenology
\cite{Ellis:1988er,Drees:1988fc,Ellwanger:2004xm}. This is the model that we
will briefly discuss here. 

\subsubsection{Basics of the NMSSM}  

The NMSSM, in which the spectrum of the MSSM is extended by one singlet
superfield, has gained a renewed interest in the last decade for three main
reasons. First, it solves in a natural and elegant way the so--called $\mu$
problem \cite{Nilles:1982mp,Frere:1983ag,Kim:1983dt} of the MSSM as in the NMSSM, it is linked to the vev
acquired by the singlet Higgs field, generating a $\mu$ value close to the
SUSY--breaking scale. Another interesting feature is that it is less
fine--tuned  as  the mass of SM--like Higgs boson receives additional
contributions at tree--level, making that a not so excessively large SUSY scale
is needed to raise it to the measured value of 125 GeV. Finally, as the Higgs
and neutralino sectors are enlarged and can be made slightly more complicated,
the present constraints from the LHC are less severe than in the MSSM. 

In the NMSSM, an additional singlet superfield $\widehat S$ is introduced in the the superpotential which then writes
\begin{eqnarray}
{\cal W} = \sum_{i,j=gen} - Y^u_{ij} \, {\widehat {u}}_{Ri} \widehat{\Phi}_2 \! 
\cdot  \! {\widehat{ Q}}_j+ Y^d_{ij} \, {\widehat{ d}}_{Ri} \widehat{\Phi}_1  
\! \cdot  \! {\widehat{ Q}}_j+Y^\ell_{ij} \,{\widehat{\ell}}_{Ri} \widehat{\Phi}_1 \! \cdot  \! {\widehat{ L}}_j + \lambda \widehat{S} \widehat{\Phi}_2 \widehat{\Phi}_1 + \frac{\kappa}{3} \, \widehat{S}^3\, , 
\end{eqnarray}
and the soft--SUSY breaking potential has additional terms besides those of the
MSSM 
\begin{eqnarray}
-{\cal L}_\mathrm{Higgs}= -{\cal L}_\mathrm{Higgs}^{\rm MSSM}+  
m_{S}^2| S |^2 + \lambda A_\lambda \Phi_2 \Phi_1 S + \frac13 \kappa  A_\kappa S^3 \, . 
\end{eqnarray}
An effective $\mu$ value is generated when the additional field $S$ 
acquires a vev,  $\mu_{\rm eff} = \lambda \langle S \rangle = \lambda v_S$
in addition to $v_1,v_2$. The input parameters of the Higgs sector are then 
\beq
\lambda, \quad \kappa, \quad \tb = v_2/v_1, \quad \mu=\lambda v_S, \quad 
A_{\kappa}, \quad M_A= 2\mu (A_{\lambda}+\kappa v_S)/ \sin 2\beta,  
\eeq
using the information for the doublets given in section 6.1.2 and the more convenient combination of Higgs fields $H_1=\cos\beta \Phi_2+\varepsilon \sin \beta \Phi_1^*$ and $H_2=\sin \beta \Phi_2+\varepsilon \cos\beta \Phi_1^*$ with $\varepsilon$ the  antisymmetric tensor in two-dimensions, one can write after symmetry breaking  \cite{Ellwanger:2009dp}:
\beq
H_1=\left( \begin{array}{c} H^+ \\ \frac{H_1^0+i P_1^0}{\sqrt{2}}\end{array}\right),~
H_2=\left(\begin{array}{c}G^+ \\ v+\frac{H_2^0 +iG^0}{\sqrt{2}}\end{array}\right),~
H_3  =  v_S +\frac{1}{\sqrt{2}} \left( H_3^0 +i P_2^0 \right). 
\eeq
In the basis $(H_1^0,H_2^0,H_3^0)$, the $3\! \times \!  3$ symmetric CP--even Higgs mass matrix $M^2_R$ reads then
\bea
M_{H_1^0 H_1^0}^2 &=&  M^2_A + (M^2_Z -\lambda^2 v^2) \sin^2 2\beta, 
\quad \quad \quad ~~~~
 M_{H_1^0 H_2^0}^2 =  -\frac{1}{2}(M^2_Z-\lambda^2 v^2) \sin4\beta, \nonumber \\
M_{H_1^0 H_3^0}^2 &=&  -\left( \frac{M^2_A \sin2\beta}{2\mu}+\kappa v_S \right) \lambda v \cos2\beta, \quad M_{H_2^0H_2^0}^2 =  M_Z^2\cos^2 2\beta +\lambda^2v^2\sin^2 2\beta, \nonumber \\
M_{H_2^0H_3^0}^2 &=&  2\lambda \mu v \left[1- \left (\frac{M_A \sin2\beta}{2\mu}
\right)^2 -\frac{\kappa}{2\lambda}\sin2\beta \right], \nonumber \\
M_{H_3^0H_3^0}^2 &=&  \frac{1}{4}\lambda^2 v^2 \left( \frac{M_A \sin2\beta}{\mu} \right)^2 +\kappa v_S A_{\kappa}+4(\kappa v_S)^2 -\frac{1}{2}\lambda\kappa v^2 \sin 2\beta . 
\eea
The CP--even Higgs boson mass eigenstates are then given by  $h_{i} = \sum_j
V_{ij} H_j^0$ with $V$ the $3\! \times \! 3$ rotation matrix that  diagonalises
the mass matrix $M^2_R$. The CP-odd eigenstates $a_1$ and $a_2$ are instead given by the diagonalization of the following matrix:
\begin{equation}
    \mathcal{M}^2_P=\left(
    \begin{array}{cc}
    M_A^2 & \lambda v \left(\frac{M_A^2}{2 \mu}\sin 2\beta-\frac{3 \kappa \mu}{\lambda}\right)  \nonumber\\
  \lambda v \left(\frac{M_A^2}{2 \mu}\sin 2\beta-\frac{3 \kappa \mu}{\lambda}\right)   & \lambda^2 v^2 \sin 2\beta \left(\frac{M_A^2}{4 \mu^2}\sin 2\beta+\frac{3 \kappa}{2 \lambda}\right)-\frac{3 \kappa A_\kappa \mu}{\lambda}
  \end{array}
  \right) \, . 
\end{equation}
We will assume $M_{h_1}\! <\! M_{h_2}\! <\! M_{h_3} $ and $M_{a_1}\! <\! 
M_{a_2} $, and the SM--like Higgs is the state dominantly made by the $H_2^0$
field. If the mixing between the $H_i^0$ fields is ignored, the mass squared of
the SM--like Higgs  receives an additional contribution $\lambda^2 v^2 \sin^2
2\beta$ compared to the case of the lighter MSSM $h$ boson. This extra
tree--level contribution  makes that this  state does not need large radiative
corrections, and hence large $M_S$ values, in order to have a mass close to 125
GeV. Furthermore, the additional sector from the singlet Higgs field is not that
constrained by experiments and masses  as low as a few GeV  or a few ten GeV are
still possible for, respectively, the lightest CP--odd $a_1$ and lightest
CP--even $h_1$ particles. 

Turning to the gaugino--higgsino sector of the NMSSM, while the charginos and
gluinos are not altered, the singlino $\tilde{S}$ will mix with the
gauginos $\tilde{B}$ and $\tilde{W}$ and the Higgsinos $\tilde{H}_1^0$ and
$\tilde{H}_2^0$ to form five neutralinos. In the basis $ \psi= (-i\tilde{B}, -i\tilde{W}^3, \tilde{H}_1^0, \tilde{H}_1^0, \tilde{S})$, the symmetric neutralino $5\times5$ mass matrix $\mathcal{M}_N$ is given by~\cite{Ellwanger:2009dp}

\beq
\mathcal{M}_N=\begin{pmatrix}
&0		&-{g' v_1}/{\sqrt{2}}	&g' v_2/\sqrt{2} &0\\
&M_2	&{gv_1}/{\sqrt{2}}	&-g v_2/\sqrt{2}	&0\\
& 	&0 &-\mu &-\lambda v_2\\
& 	&  &0 	&-\lambda v_1\\
& 	&  & 	&2\kappa v_S \end{pmatrix} \, ,
\eeq
and is diagonalised by a matrix $Z$ giving the mass eigenstates $ \chi_i^0 = \sum_{j=1}^5 Z_{ij} \psi_j$ with again $\chi_1^0$ corresponding to the lightest neutralino and, hence, the DM candidate. This neutralino can be almost singlino--like giving a distinct phenomenology for the NMSSM. 

The properties of the other superparticles, in particular their masses and
couplings, are the same as in the MSSM except when they couple to the singlet
and singlino fields. This might affect the phenomenology in a serious way and
weaken the exclusion limits on sparticle from colliders. Nevertheless, we will
assume as in the MSSM case that the sfermion spectrum is very heavy and
concentrate on the Higgs and neutralino sectors.

\subsubsection{Phenomenology of the NMSSM}

\underline{The Higgs sector.}  In a large area of the parameter space, the Higgs
sector of the NMSSM reduces to the one of the MSSM
\cite{Espinosa:1991gr,Ellwanger:1993hn,Elliott:1993bs,Ellwanger:1996gw,Miller:2003ay,Barger:2006dh,King:2012is}
but there is an interesting  possibility that is still viable: one of the
lighter neutral Higgs bosons, either the CP--even $h_1$ or the CP--odd $a_1$, is
very light, with a mass  of a few to a few ten's of GeV
\cite{Ellwanger:2005uu,Dermisek:2005ar,Djouadi:2008uw}. The SM--like CP--even
$h_2$ state could then decay into $h_1$ or $a_1$ pairs, $h_2 \to h_1 h_1$ or 
$h_2 \to a_1 a_1$ with branching ratios that are not negligible. The light Higgs
bosons would then  decay into pairs of tau leptons or $b$--quarks,  leading to
the final state topologies $h_2 \to  4b, 4\tau, 2b 2\tau$. 

In fact, the $a_1$ state can be lighter than 10 GeV so that the decay channel
into $b\bar b$ pairs is kinematically closed. In this case, decays into $\tau$
lepton pairs would be dominant but the channel $a_1 \to \mu^+ \mu^-$ has a
non--negligible branching fraction. In addition, the photonic decay will be
important as many fermions  would contribute: the top and bottom quarks and also
the tau leptons, giving a rate that is also non--negligible.  

Note that as in the MSSM, there is no coupling of the pseudoscalars to massive
gauge bosons and hence, there is no $W$ loop coupling in the $a_1 \to
\gamma\gamma$ decay for instance. Because of this feature, the $a_1$ state can
be produced only in $h_2$ decays and eventually in association with heavy
fermions if the couplings are not prohibitively small, but this is general what
occurs when the $a_1$ is mostly singlet--like. In fact, this is also the case of
the $h_1$ boson which can have suppressed couplings to the $W,Z$ bosons. This is
the reason why the experimental  limit on the $h_1$ mass from LEP2 searches is
weak  being of the order of 50 GeV or so: the cross section for the process $e^+
e^- \to h_1 Z$ is suppressed, while the states $h_2,h_3$ are heavy so that  the
processes $e^+ e^- \to h_1 h_2$ are suppressed too.  All Higgs particles can
escape detection at LEP2 and a mass as small as 50 GeV would be allowed for the
$h_1$ state.

\begin{figure}[!h]
\vspace*{-.1cm}
\centerline{ \includegraphics[scale=0.9]{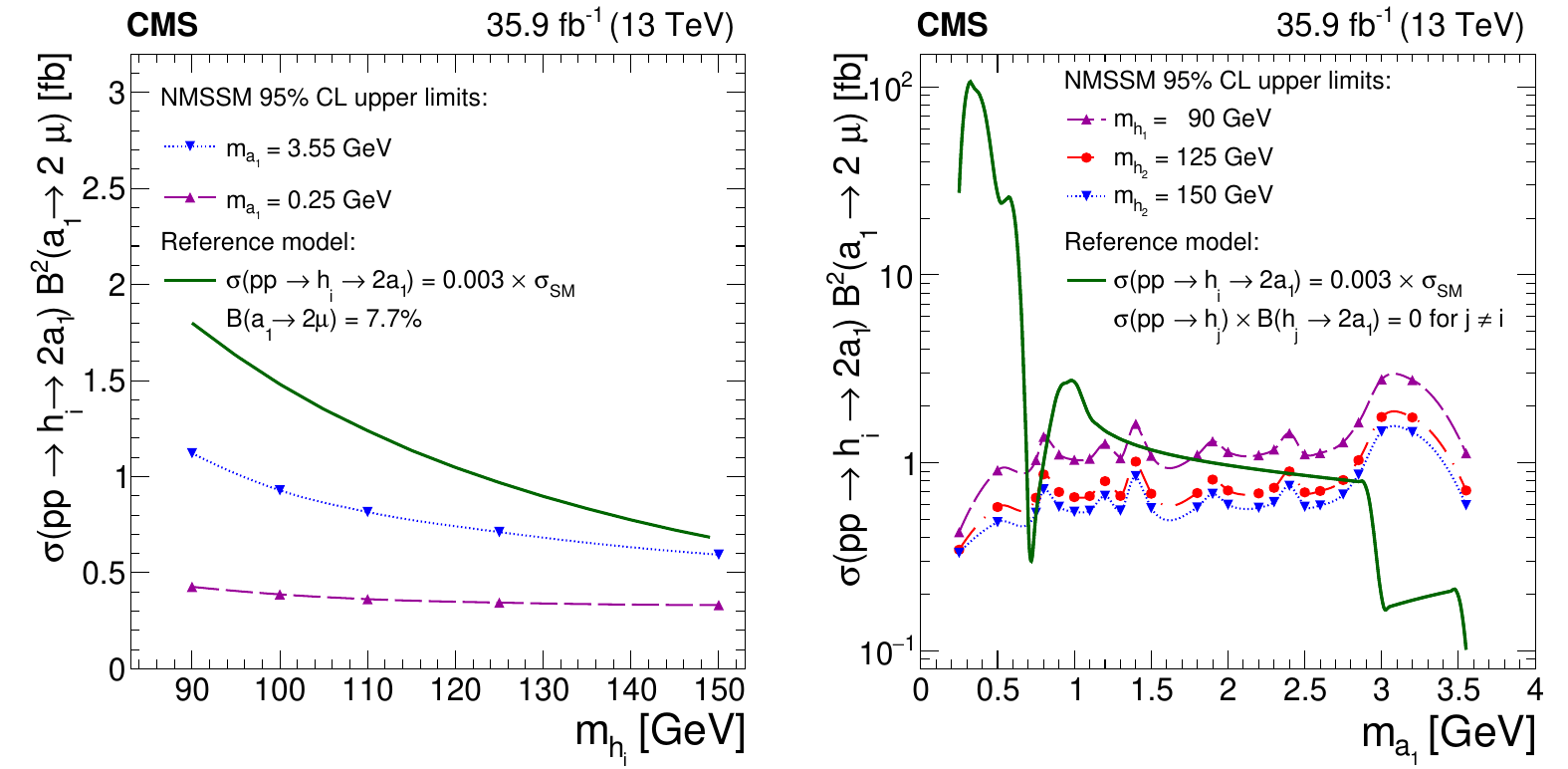} }
\vspace*{-1mm}

\caption{
Left: the 95\%CL upper limit on the rate for the process $\sigma(pp \to H_i \to 2a_1)  \times {\rm BR}^2 (a_1 \to \mu^+\mu^-)$ when compared  to a representative model (solid curve) obtained using the simplified scenario described in the text. The figure is separated into two regions: $M_{h_i}=M_{h_1}< 125$ GeV with $M_{h_2}= 125$ GeV (left) and $M_{h_1}= 125$ 
GeV with $M_{h_i}= M_{h_2}> 125$  GeV (right). The limits as compared to a representative model (solid curve) from a simplified scenario which includes gg--fusion, VBF, and VH production modes; from \cite{Sirunyan:2018mgs}. 
}
\label{CMS-NMSSM}
\vspace*{-.2cm}
\end{figure}

Searches for NMSSM Higgs bosons, in particular in the difficult topologies
described above have been performed at the LHC and an example of a  CMS analysis
at $\sqrt s=13$ TeV with 36 fb$^{-1}$ data is displayed  in
Fig.~\ref{CMS-NMSSM}.  There, the 95\%CL upper limit on the cross section for
the  process $pp \to h_1$  from all possible channels, including gluon fusion,
VBF and VH production, times the branching ratios for the decays $h_i \to a_1
a_1$ with both $a_1$ states decaying into muon pairs, $a_1 \to \mu^+\mu^-$, are
shown as a function of the $h_1$ (left) and $a_1$ (right) masses. In the left
panel, one has a very light  $a_1$ boson with $M_{a_1}= 3.55$ and $0.25$ GeV,
while  $h_2$ is the SM--like  Higgs boson and $h_1$ is lighter. The rate is
compared to a benchmark scenario in which the cross section  $\sigma(pp\to h_i)$
is the same as for SM Higgs production, while the decay branching ratios are
BR($h_i\to a_1a_1) =0.3\%$  and BR($a_1\to \mu\mu) =7.7\%$.  The same comparison
with the benchmark model is made in the right-panel, where the CP--even Higgs
masses are fixed to $M_{h_1}=90, M_{h_2}=125$ GeV and  $M_{h_3}=150$ GeV, while
$M_{a_1}$ is varied. Hence, significant  rates for such processes are already
excluded at the LHC. 

One should note that future $e^+e^-$ colliders will also be very  useful in
probing the Higgs sector of  the NMSSM with the additional CP--even and CP--odd
Higgs particles. As seen previously for the SM,  Higgs--strahlung, $e^+e^-\to Z
h_i$, allows for the detection of CP--even Higgs bosons independently of their
decay modes and thus, even if they decay  into the singlet--like light $a_1$ or
$h_1$ states or even invisibly into the LSP neutralino.  This is possible
provided that their couplings to the $Z$ boson are not prohibitively tiny but at
$\sqrt s =240$ GeV and high luminosities very small couplings could be
probed\footnote{As mentioned at some point, there are excesses of events which
point at a possible existence of a Higgs boson lighter  than the 125 GeV Higgs
boson. Indeed, there is a longstanding $2.3 \sigma$ excess observed at LEP in 
the process $e^+e^- \to ZH \to Zb\bar b$ at a mass of $98$ GeV
\cite{Barate:2003sz} and very recently, an excess of events of about 
$2.8\sigma$ (locally but only a $1.3 \sigma$ globally) has been observed in CMS 
in the 13 TeV data at a mass of  $95$ GeV in the $\gamma\gamma$ decay channel
\cite{Sirunyan:2018aui}. Its is tempting to identify this excess with the $h_1$
state of the NMSSM, the SM--like Higgs boson being $h_2$ in this case.}. 

\underline{The neutralino sector.} In the NMSSM, the singlino can well be the
LSP and if the coupling $\lambda$ is relatively small, $\lambda \lsim 10^{-2}$,
it couples very weakly  to  all  other  particles.   As a consequence,   no  
superparticle  can decay into the singlino with large rates, except for the
next-to-LSP which has no other choice if  $R$--parity is conserved. In the heavy
sfermion scenario that we are discussing here, the NLSP is generally the
next-to-lightest neutralino  $\chi_2^0$ which decays via the modes $\chi_2^0 \to
\chi_1^0+ Z, h_i, a_i$. If the mass of the singlino is very small, e.g. a few
GeV, while the NLSP is expected to  have a mass of the order of 100 GeV from
LEP2 searches, the missing energy is rather small and the usual topologies which
are looked at in the MSSM, namely the missing transverse energy signature, is
significantly reduced. The experimental limits on all SUSY particles, which will
always have the cascade decays $\tilde X \to \chi_2^0 + X \to \chi_1^0 + X'$ 
involved, will be thus weakened. 

In addition, if $\chi_1^0$ is almost a pure singlino, the $\chi_1^0$--$\chi_2^0$
couplings to the Higgs and $Z$ bosons, which could be eventually be off
mass--shell,  will be tiny and hence the partial widths will be very small.
Because of the long--lived $\chi_2^0$, into which all other sparticles  except
for $\chi_1^0$ will decay, one would have displaced vertices in the
NMSSM\footnote{Another region leading to displaced decays in the NMSSM, namely
when the lighter stau is the NLSP and is almost degenerate in mass with the
singlino LSP to achieve the correct relic density, will not be considered here
as the sleptons are assumed yo be very heavy.}.  A mostly singlino and very
light LSP will have an  impact on the DM phenomenology and make it behave
differently from what occurs in the MSSM \cite{Belanger:2005kh,Gunion:2005rw}.
Indeed, besides the regions for obtaining the correct amount of relic density 
discussed in the MSSM case, a new one will be possible. The annihilation of the
LSP neutralinos into the SM and the Higgs particles, assuming  that $h_2$ is the
SM--like Higgs boson and that $a_1$ and $h_1$ are lighter, can occur in many
new channels. First, $h_1$ and $a_1$ can now serve as additional portals and
exchanged in the $s$--channel in the annihilation process $\chi_1^0 \chi_1^0 \to
h_i, a_i  \to f\bar f$  for instance. In addition, annihilation channels such as
$\chi_1^0 \chi_1^0 \to  a_1 a_1, h_1h_1, h_1 h_2$ could open up.  Note that even
for a singlino--dominated DM, the annihilation can still occur via the usual
processes $\chi_1^0 \chi _1^0 \to W^+ W^-$ through $t$--channel exchange of the
charginos $\chi_{1}^\pm$ and $\chi_1^0 \chi _1^0 \to ZZ$ through $t$--channel
exchange of the neutralinos $\chi_{2,3}^0$. A mass splitting between $\chi_1^0$
and the $\chi_{1}^\pm$ and $\chi_{2,3}^0$ is needed in particular if the latter
are wino--like in order to avoid co--annihilation which makes the LSP
underabundant. 

Finally, let us make a remark on the probing of the NMSSM at $e^+e^-$ colliders
\cite{Weiglein:2004hn,Ellis:1988er,Djouadi:2008uj}. The scenarios with sizable
singlet--doublet mixing between the $h_1$ and $h_2$ and eventually $h_3$ states
can be probed in the Higgs--strahlung processes, $e^+e^- \to Z+ h_i$, where the
separate Higgs states can be disentangled even if they are nearly degenerate in
mass, as the resolution on the Higgs masses in this process is smaller than 100
MeV. The scenario  in which there is a light CP--even or CP--odd Higgs
particle   allowing the $h_2\to h_1h_1$ or $h_1 \to a_1 a_1$ decays to occur can
also be probed in the Higgs--strahlung processes in which the SM--like 125 GeV
Higgs boson is produced and decays into these light states, leading to a $Z$
boson and $4b, 2b2\tau$ and $4\tau$ final states.  The singlet--like CP--odd
$a_1$ boson could be also accessible in the pair production process $e^+e^- \to
h_1a_1$ with $h_1$ the 125 GeV observed state,  unless the $a_1$ mass is too
large or the coupling $Zh_1a_1$ prohibitively tiny.

\subsubsection{The NMSSM in the DM context} 

Similarly to the MSSM, we are considering an NMSSM realization in which the
sfermions are very heavy and do not affect the phenomenology. Most of the
considerations made in the MSSM concerning the DM relic density are also valid
in this case. We have to take into account though the fact that the neutralino
sector is more complicated  since the DM can feature also a singlino component.
Nevertheless, we will make here the simplified assumption that only the singlino
and eventually the bino--like neutralino are light enough to affect the DM
phenomenology. Furthermore, the DM can interact with additional scalar and
pseudoscalar Higgs states, eventually lighter than the SM--like Higgs boson.

Hence, assuming that the relevant DM phenomenology in the NMSSM is determined solely by the Higgs sector and a reduced neutralino sector, only a limited set of free parameters should be considered, namely: $\lambda,\kappa, \tan\beta, M_A, \mu, A_\kappa, M_1$. As a further simplification, we will assume the mass hierarchy $M_{a_1} \ll M_{a_2}$. To a good approximation, we can identify $M_{a_2} \simeq M_A$ and then relabel $M_{a_1} \equiv M_a$. It is at this point useful to re--express $A_\kappa$ as a function of $M_A$ and $M_a$ and adopt the latter as input parameter in place of $A_\kappa$, 
\begin{equation}
    A_\kappa\!=\!-\!\frac{\lambda}{3 \kappa \mu}\left[M_a^2 \! -\! \frac{\lambda^2 v^2 \sin 2\beta}{2 \mu}\left(\frac{M_A^2 \sin 2 \beta}{2 \mu}\! +\! 3 \frac{\kappa \mu}{\lambda} \right)\! -\! \frac{\lambda^2 v^2} {M_a^2-M_A^2}{\left(\frac{M_A^2 \sin 2 \beta}{2 \mu}\!+\! 3 \frac{\kappa \mu}{\lambda}\right)}^2\right] . 
\end{equation}
Given all the assumptions above, one can identify the heaviest scalar eigenstate $h_3$ with the heavy MSSM--like CP--even $H$ boson with $M_H \sim M_A$. Moreover, we will assume a very suppressed mixing between the remaining states $h_1,h_2$, which can be then identified, respectively, with a single--like state $h_S$ and the SM--like Higgs boson $h$.

The situation will look like the MSSM with an additional light Higgs and DM
sectors and will be also similar to to the 2HDM plus a light pseudoscalar
Higgs--portal discussed in the previous section. Many features in these two
scenarios can thus similarly occur in the simplified NMSSM that we consider and
we will thus concentrate on the new features that are specific to the present
model. In particular, for what concerns the DM relic density, we will exclude
the cases of the $A$--pole funnels and the pure higgsino DM scenarios which have
been already discussed in detail in the previous subsections. We distinguish two
interesting new scenarios: a ``well tempered'' LSP with a singlino--higgsino
admixture which can be realized for $2 {\kappa}/{\lambda} \ll 1$, and another
``well tempered" LSP featuring a bino--higgsino admixture, occurring instead for
$2 {\kappa}/{\lambda} \gg 1$.

Despite of the fact that the results presented here are based on a numerical
analysis performed with the package
NMSSMTools~\cite{Muhlleitner:2003vg,Ellwanger:2004xm,Ellwanger:2005dv,Das:2011dg}
which includes all relevant higher order effects, and given the simplifications introduced above, one can adopt to a  good
approximation the analytic expressions of the relic density provided in the 2HDM+$a$ scenario to have a rough understanding of the phenomenology of the DM state. One should nevertheless  redefine the couplings of the DM state with the scalar and
pseudoscalar mediators, as illustrated by the following expressions
\begin{align}
\label{eq:yukS}
    & y_{h_i dd}=-\frac{m_d}{\sqrt{2} v \cos\beta}S_{i,1} \, , \quad 
     y_{h_i uu}=-\frac{m_u}{\sqrt{2} v \sin\beta}S_{i,2} \ , \nonumber \\
    & y_{h_i \chi_1^0 \chi_1^0}= \big( (g'Z_{11}-g Z_{12}) Z_{13} +\sqrt{2} \lambda   Z_{14}Z_{15} \big) S_{i,1}- \big( (g'Z_{11}-g Z_{12}) Z_{14}-\sqrt{2}\lambda Z_{13}Z_{15} \big)  S_{i,2} \nonumber \\
        &  \hspace*{1.2cm} +\sqrt{2} (\lambda Z_{13}Z_{14}-\kappa Z_{15}^2 )S_{i,3} \,, 
\end{align}
where $h_i={h,H,h_S}$ and $S_{h,i}$ are the elements of the mixing matrix which diagonalizes the CP--even Higgs mass matrix, i.e. $h_i= \sum_{j=1,3} S_{i,j} H_j^0$. Under the assumption made for the scalar sector, the quantities $S_{(h,H),(1,2)}$ approximately coincide with their MSSM values:
\begin{align}
    & S_{H,1}=S_{h,2}=\sin\beta \ , \quad  S_{H,2}=-S_{h,1}=\cos\beta \, , \nonumber \\ 
    & S_{h_S,1} \sim \frac{\lambda \mu v}{M_A^2}\frac{\cos 2 \beta}{\cos \beta}  \ , \quad 
    S_{h_S,2} \sim -\frac{\lambda \mu v}{M_A^2}\frac{\cos 2 \beta}{\sin \beta} .
\end{align}
The couplings with the pseudoscalar states are instead given by 
\begin{align}
\label{eq:yukA}
    & y_{a uu}=i \frac{m_u}{\sqrt{2} v \tan\beta}P_a^{A}\ , \quad 
     y_{a dd}=i \frac{m_d \tb}{\sqrt{2} v}P_a^{A} \, , \nonumber\\
     & y_{a \chi_1^0 \chi_1^0}=i \left \{ \left[\left(Z_{14}\cos \beta -Z_{13}\sin\beta\right) [\left(g' Z_{11} -g Z_{12}\right) +\sqrt{2}\lambda Z_{15} \left(Z_{13}\cos\beta+Z_{14}\sin\beta\right)  \right]P_{a}^{A}\right. \nonumber\\
    & \hspace*{1.2cm} \left. + \sqrt{2}\left(\lambda Z_{13}Z_{14}-\kappa Z_{15}^2\right)P_a^S \right \} \, , 
\end{align}
where $P_a^{S,A}$ are defined by $M_A^2=(P_a^S)^2 M_{a_1}^2+(P_a^A)^2 M_{a_2}^2$ with $P_a^{A}=\sqrt{1-(P_a^{S})^2}$. Notice that the parameter $P_a^A$ is analogous to the quantity $\sin\theta$ in the 2HDM+$a$ model. Consequently, an even more straightforward comparison between the two models could be performed by adopting $P_a^S$ as free parameter instead of $\mu$, as proposed in Ref.~\cite{Baum:2019uzg}.

DM direct detection is mostly determined by spin--independent interactions mediated by the three CP--even Higgs bosons. The corresponding cross sections can be written as
\begin{align}
    & \sigma_{\chi^0_1 p}^{\rm SI}=\frac{4 \mu_{\chi_1^0}^2}{\pi}\left[\frac{Z}{A}f_p+\left(1-\frac{Z}{A}\right)f_n\right]
    \, , \\
    & f_{p,n}=\left(\sum_{q=u,d,s}f_q^{p,n}\frac{a_q}{m_q}+\frac{2}{27}f_{\rm TG}\sum_{q=c,b,t}\frac{a_q}{m_q}\right)m_p \, . 
\end{align}
The coefficient $a_q$ have different functional forms for up--type and down--type quarks. These are respectively given by \cite{Cheung:2014lqa}:
\begin{align}
& a_u=-\frac{g m_u}{4 M_W \sin \beta}\left[(g Z_{12}-g' Z_{11})\left \{Z_{13} \left[-\frac{S_{h_S,u}S_{h_S,d}}{M_{h_S}^ 2}-\sin\beta \cos\beta \left(\frac{1}{M_h^2}-\frac{1}{M_H^2}\right)\right]\right. \right. \nonumber\\
& \left. \left. +Z_{14}\left(\frac{\sin^2\beta}{M_h^2}+\frac{\cos^2 \beta}{M_H^2}+\frac{S^2_{h_S,u}}{M_{h_S}^2}\right)\right \} \right.\nonumber\\
& \left. +\sqrt{2}\lambda \left \{ Z_{13} Z_{14} \left(-\frac{S_{h,S}\sin\beta}{M_h^2}+\frac{S_{H,S}\cos\beta}{M_H^2}+\frac{S_{h_S,u}S_{h_S,S}}{M_{h_S}^2}\right) \right. \right. \nonumber\\
& \left. \left.+ Z_{15}\left[Z_{14} \left(\cos\beta \sin\beta \left(\frac{1}{M_h^2}-\frac{1}{M_H^2}\right)+\frac{S_{h_S,u S_{h_S,d}}}{M_{h_S}^2}\right)+Z_{13}\left(\frac{\sin^2 \beta}{M_h^2}+\frac{\cos^2 \beta}{M_H^2}+\frac{S_{h_S,u}^2}{M_{h_S}^2}\right) \right] \right \}\right.\nonumber\\
& \left. -\sqrt{2}\kappa Z_{15}^2 \left(-\frac{S_{h,S}\sin\beta}{M_h^2}+\frac{S_{H,S}\cos\beta}{M_H^2}+\frac{S_{h_S,u} S_{h_S,S}}{M_{h_S}^2}\right)\right]    \, , 
\end{align}
\begin{align}
& a_d=\frac{g m_d}{4 M_W \sin \beta}\left[(g Z_{12}-g' Z_{11})\left \{ Z_{13} \left(\frac{\cos^2 \beta}{M_h^2}+\frac{\sin^2 \beta}{M_H^2}+\frac{S_{h_S,d}^2}{M_{h_S}^2}\right) \right. \right. \nonumber\\
& \left. \left. -Z_{14} \left[\frac{S_{h_S,u} S_{h_S,d}}{M_{h_S}^2}+\cos\beta \sin\beta \left(\frac{1}{M_h^2}-\frac{1}{M_H^2}\right) \right]\right \} \right.\nonumber\\
& \left. -\sqrt{2}\lambda \left \{ Z_{13} Z_{14} \left(-\frac{S_{h,S}\cos\beta}{M_h^2}-\frac{S_{H,S}\sin\beta}{M_H^2}+\frac{S_{h_S,d}S_{h_S,S}}{M_{h_S}^2}\right) \right. \right. \nonumber\\
& \left. \left.+ Z_{15}\left[Z_{14}\left(\frac{\cos^2\beta}{M_h^2}+\frac{\sin^2 \beta}{M_H^2}+\frac{S_{h_S,d}^2}{M_{h_S}^2}\right)+Z_{13}\left(\cos\beta \sin\beta \left(\frac{1}{M_h^2}-\frac{1}{M_H^2}\right)+\frac{S_{h_S,d}S_{h_S,u}}{M_{h_S}^2}\right) \right] \right \}\right.\nonumber\\
& \left. +\sqrt{2}\kappa Z_{15}^2 \left(-\frac{S_{h,S}\cos\beta}{M_h^2}+\frac{S_{H,S}\sin\beta}{M_H^2}+\frac{S_{h_S,d} S_{h_S,S}}{M_{h_S}^2}\right)\right]    \, . 
\end{align}

\begin{figure}[!h]
    \centering
    \subfloat{\includegraphics[width=0.47\linewidth]{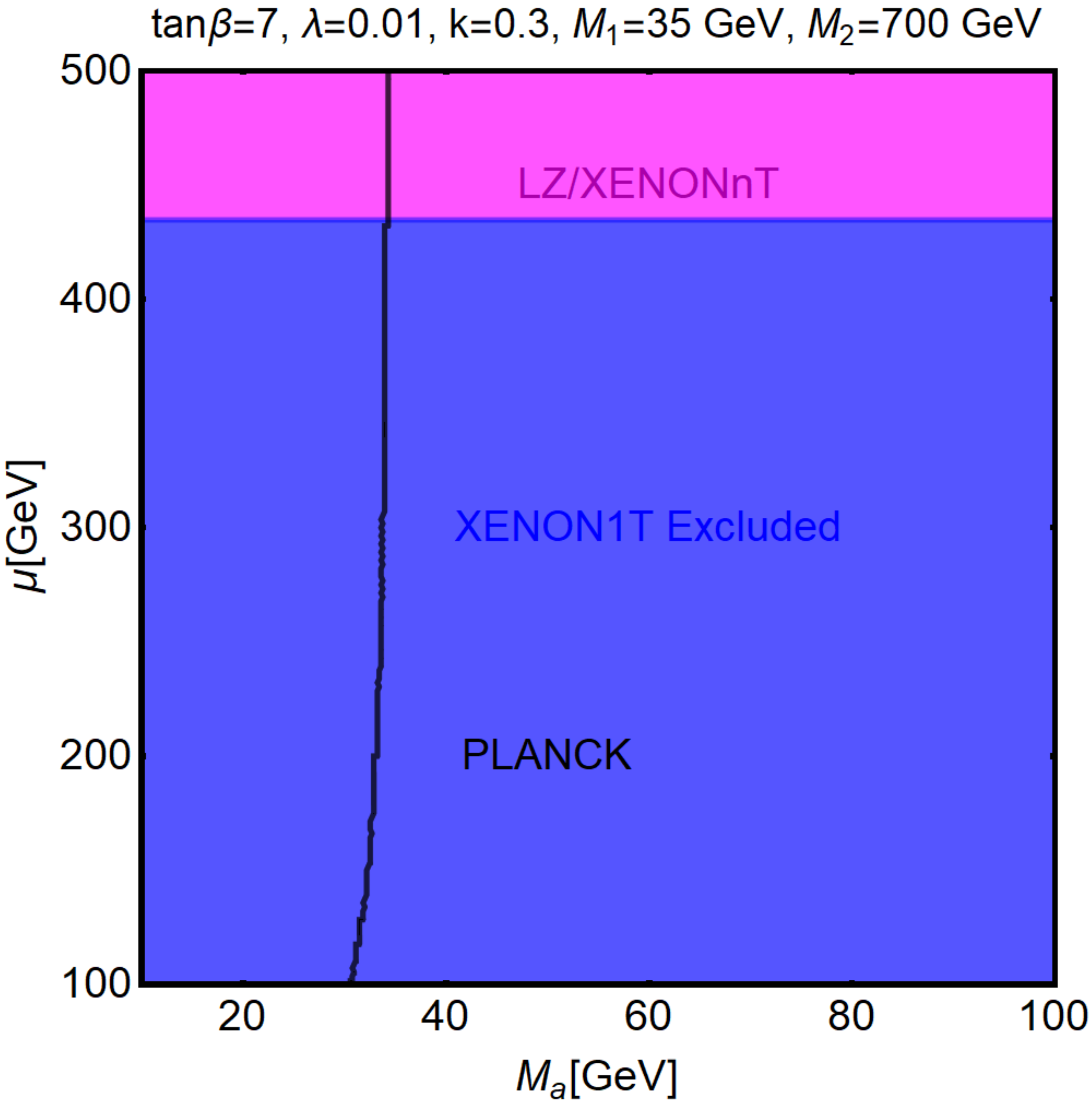}}~~~\subfloat{\includegraphics[width=0.47\linewidth]{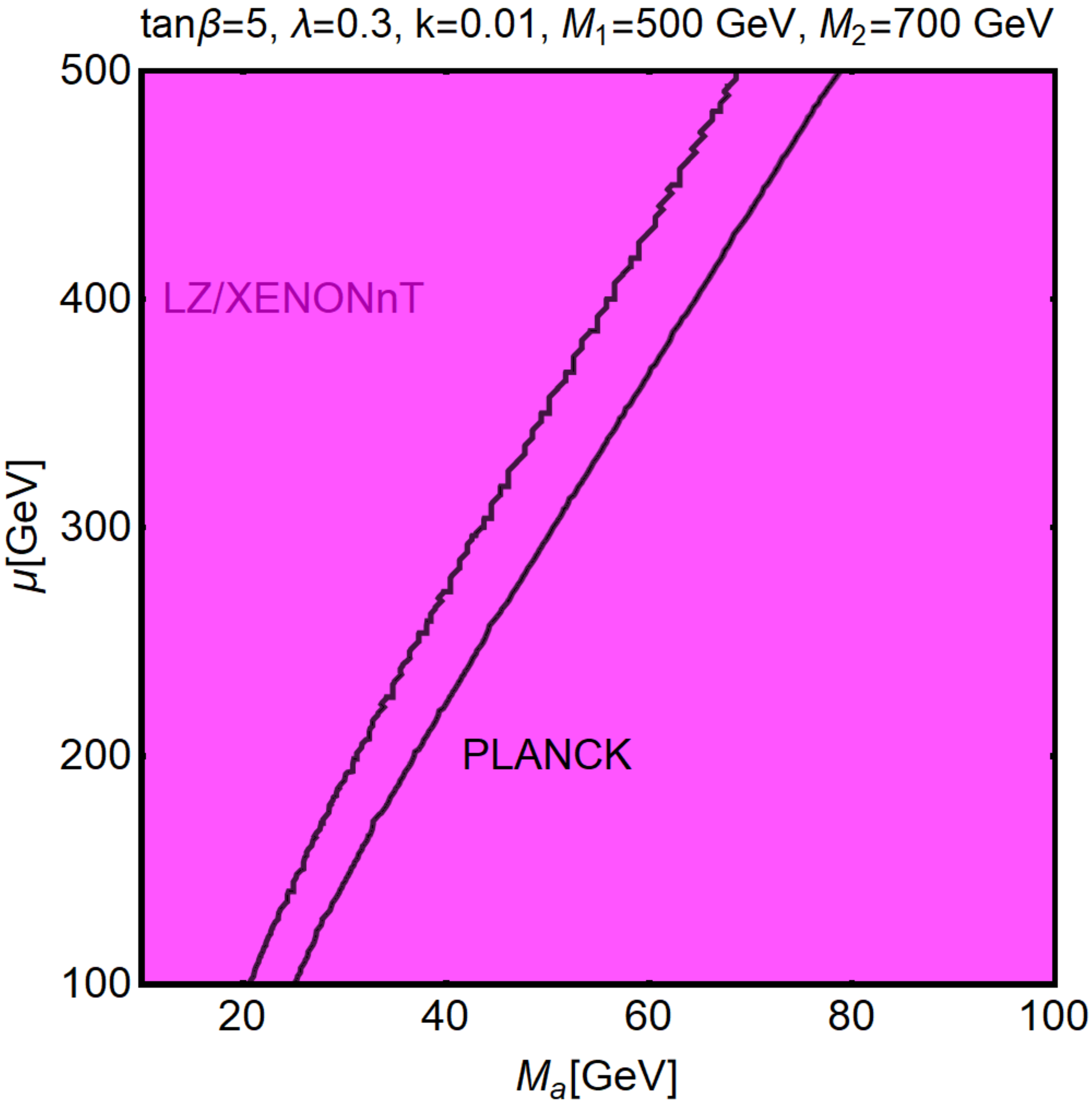}}
\vspace*{-1mm}
    \caption{DM constraints on the NMSSM in the bidimensional plane $[M_a,\mu]$
    for two benchmark scenarios with the relevant parameters reported on top of the two panels. The black contours represent the correct DM relic density, the blue region of the  parameter space is excluded by XENON1T while the magenta regions will be excluded in the absence of signals by the LZ/XENONnT experiments.}
    \label{fig:astro_NMSSM}
\vspace*{-2mm}
\end{figure}

As an illustration, we show in Fig.~\ref{fig:astro_NMSSM} the constraints in
the plane $[M_a,\mu]$ in two scenarios: in the left panel, a
bino--higgsino scenario with the choice of parameters  $\tb=7, \lambda=0.01,
\kappa=0.3$ and $M_1=35$ GeV, $M_2=700$ GeV and, in the right panel, 
a singlino--higgsino scenario with  input parameters $\tb=5,
\lambda=0.3, \kappa=0.01$ and $M_1=500$ GeV, $M_2=700$ GeV.
The first configuration is characterized by a relatively light DM, having set
$M_1=35 \,\mbox{GeV}$. In such a case the correct relic density is accounted for
mostly by DM annihilation into $\bar b b$ final states mediated by the light
pseudoscalar boson~\cite{Cheung:2014lqa}. It is also worth noticing that, in the
$\lambda/\kappa \gg 1$ limit, the singlet--like state $h_S$ becomes very heavy,
so that the CP--even Higgs sector of the theory is MSSM--like.

As already mentioned, bino--higgsino mixtures are typically characterized by
sizable direct detection cross sections but one could, nevertheless, achieve a
blind spot for negative $\mu$ values. As is made clear by the figure,  the
region of parameter space with $\mu \lesssim 450\,\mbox{GeV}$ is already
excluded by the XENON1T experiment. Negative signals from next generation
detectors, like XENONnT for instance, would rule out higher values of $\mu$. 

In the case of singlino--higgsino DM, illustrated in the right--hand side of
Fig.~\ref{fig:astro_NMSSM}, DM annihilation through the $a_1$ state
is  less efficient since the latter is mostly singlet--like in this
configuration. This suppression can be nevertheless compensated by a higher
higgsino fraction for the DM, allowing for efficient annihilation processes
through $Z$ boson exchange. This does not translate into strong bounds from
direct detection experiments. Indeed, as discussed in Ref.~\cite{Cheung:2014lqa}
for instance, for positive $\mu$ values, it is possible to suppress the DM
scattering cross section through a destructive interference between the diagrams
with the exchange of the different scalar bosons. As can be seen from the right
panel of Fig.~\ref{fig:astro_NMSSM}, this allows to evade, for the selected
benchmark, the present bounds from direct detection.  Nevertheless, this 
scenario will be  fully probed by the future LZ/XENONnT experiments.

%% file: sec-Appendix-Higgs.tex
\setcounter{section}{0}
\renewcommand{\thesection}{A}
\setcounter{equation}{0}
\renewcommand{\theequation}{A.\arabic{equation}}

\section{Appendix: Higgs decay and production at colliders}

Higgs phenomenology, and more precisely Higgs decays as well as production and
detection at the LHC and high--energy colliders in general, has been discussed
in many comprehensive reviews, including rather recent ones that take into
account the latest development in the field. For the SM Higgs boson, one can
find the relevant material in
Refs.~\cite{Gunion:1989we,Djouadi:1994mr,Spira:1997dg,Djouadi:2005gi,Baglio:2010ae,Dittmaier:2011ti,Dittmaier:2012vm,Heinemeyer:2013tqa,deFlorian:2016spz,Baglio:2015wcg,Spira:2016ztx,Dawson:2018dcd}
for instance, while Higgs bosons in extensions of the SM have been discussed in
reviews, including Refs.~\cite{Godunov:2015nea,Robens:2015gla} for singlet Higgs
models, \cite{Branco:2011iw,Bernon:2015qea,Celis:2013rcs} for two Higgs doublets
models, 
Refs.~\cite{Gunion:1989we,Spira:1997dg,Djouadi:2005gj,Baglio:2010ae,Baglio:2015wcg,Spira:2016ztx}
and
\cite{Carena:2002es,Degrassi:2002fi,Heinemeyer:2004gx,Allanach:2004rh,Bagnaschi:2018igf}
for the MSSM and \cite{Ellwanger:2009dp,Maniatis:2009re,Djouadi:2008uw} for the
NMSSM. One can also consult the various proceedings of workshops that discussed
Higgs physics at high--energy hadron 
\cite{Weiglein:2004hn,Dittmaier:2011ti,Dittmaier:2012vm,Heinemeyer:2013tqa,deFlorian:2016spz,Fcc-China,Contino:2016spe,CEPCStudyGroup:2018ghi,Cepeda:2019klc}
and lepton 
\cite{Accomando:1997wt,AguilarSaavedra:2001rg,Badelek:2001xb,Djouadi:2007ik,Battaglia:2004mw,Linssen:2012hp,Djouadi:1994mr,Baer:2013cma,Gomez-Ceballos:2013zzn,Mangano:2018mur} colliders. 

Nevertheless, to be complete and comprehensive, we will present in this Appendix
the analytical material that allows to describe the most important decay and
production channels of the neutral Higgs particles in these theories. We will
stick to the lowest order expressions in perturbation theory but summarize
briefly the impact of the sometimes very important higher order effects. In many
cases, we will take an agnostic attitude and consider both the scalar and
pseudoscalar Higgs possibilities and assume somewhat general Higgs couplings to
SM fermions and gauge bosons in order to cover most of the possibilities of
beyond the SM Higgs sectors that have been discussed in this review. 

\subsection{Higgs decays}

\underline{Decays into fermions.}

A neutral Higgs boson, which can be either a scalar or a pseudoscalar state and that we denote by $\Phi=A/H$, will decay most of the time into fermions  pairs 
with a partial decay  width given at leading order (LO)  by 
\cite{Resnick:1973vg,Ellis:1975ap}
\begin{equation}
\Gamma(\Phi \to f \bar{f} ) = N_c^f G_F m_f^2/ (4\sqrt{2} \pi ) \times 
\, g_{\Phi ff}^2 \, M_{\Phi} \, \beta^{p}_f \, , 
\label{gamma-ff}
\end{equation}
where $g_{\Phi ff}$ is the Higgs coupling normalized to the SM value,   $N_c^f$
the color factor and $\beta_f=(1-4m_f^2/M_{\Phi}^2)^{1/2}$ the fermion velocity
and  $p =3\,(1)$ for the CP--even (CP--odd) Higgs boson. For Higgs decays into
light quarks however, one has to take into account large QCD radiative
corrections, part of which can be mapped into the running of the quark masses.
In eq.~(\ref{gamma-ff}), if the $b,c$ quark masses are defined as the
$\overline{\rm MS}$ masses evaluated at the scale of the Higgs mass, giving 
$\bar m_b (M_\Phi^2) \approx 2.8$ GeV and  $\bar m_c (M_\Phi^2) \approx 0.62$
GeV for $M_\Phi=125$ GeV, one simply needs to include a multiplicative factor
$(1+ 5.67 \alpha_s(M_\Phi^2)/\pi)$ to incorporate the next-to-leading order
(NLO). Higher orders QCD corrections as well as the small electroweak
corrections (which are different for a CP--even and a CP--odd Higgs state) can
be found in Ref.~\cite{Steinhauser:2002rq,Djouadi:1995gt} for instance. 

In the case of the heavy top quarks, mass effects have to be included when
considering the NLO QCD radiative corrections to the decays $\Phi \to t\bar t$
above the $M_\Phi\! = \! 2m_t$ threshold; they can be found in
Ref.~\cite{Djouadi:1994gf}. Again, the small QCD corrections beyond NLO and the
electroweak corrections have be reviewed in
Refs.~\cite{Spira:1997dg,Djouadi:2005gi}. We should note that slightly below the
$M_\Phi\! = \! 2m_t$ threshold, the possibility of the off mass--shell $\Phi \to
tt^* \to tbW$ decays is present and can have an important impact if the $\Phi
tt$ coupling is large, typically, if it is SM--like or larger and when 300 GeV  
$\lsim M_\Phi \lsim 350$ GeV  \cite{Djouadi:1995gv}.  

Most of the time, we will not consider the case of the charged Higgs boson in
this Appendix, but we should mention at least that its main decay mode is into 
$\tau\nu$ or $tb$ pairs and the decay widths, assuming $M_{H^\pm} \gg m_\tau,
m_b, m_t$ and that the $H^\pm$ couplings are 2HDM--like and can be expressed in terms of the pseudoscalar $A$ ones, are given at LO by
\bea
\Gamma(H^- \tau \nu ) &=& G_F M_{H^\pm}/ (4\sqrt{2} \pi ) \times 
\, m_\tau^2 \, g_{H^\pm \tau\nu}^2 \,  \ , \nonumber \\
\Gamma(H^- b\bar t) &=& 3 G_F M_{H^\pm}/ (4\sqrt{2} \pi ) \times 
\, [ m_t^2 g_{A tt}^2 + m_b^2 g_{A tt}^2 ] . 
\label{H+-ff}
\eea
Again, the higher order corrections (as well as the exact expressions with mass effects included) can be found in Refs.~\cite{Djouadi:1994gf,Djouadi:2005gj}.

\underline{Decays into gauge bosons.}

The other important decays of the neutral states $\Phi$ are into massive gauge
bosons, $\Phi \to VV$ with $V=W,Z$. This is particularly true in the case of the
CP--even $H$ boson which has full strength $VV$ couplings at tree--level.  But a
pseudoscalar boson $A$ can also have induced couplings to massive gauge bosons
and, thus, also decay into these states. Here, we will assume the following
effective Lagrangians for the $\Phi VV$ interactions
\begin{eqnarray}
{\cal L}(HVV)\!=\!\left(\sqrt2 G_F \right)^{1/2}M_V^2 \, g_{HVV} H V^\mu V_\mu , \ {\cal L}(AVV)\!=\! {1 \over 4} \eta \left(\sqrt2 G_F \right)^{1/2}M_V^2 A
V^{\mu\nu} \widetilde V_{\mu\nu} ,  
\label{eq:eff-HVV-cplg}
\end{eqnarray} 
with $\widetilde V^{\mu\nu}= \epsilon^{\mu\nu\rho\sigma} V_{\rho\sigma}$ and 
$\eta$ a dimensionless factor in the case of the pseudoscalar $A$ state. For the CP--even $H$ state, the relevant reduced coupling $g_{HVV}$
is 1 in the SM, while it is suppressed by  mixing angle factors in 2HDM
extensions and, for instance,  one has $g_{hVV}=\sin(\beta-\alpha)$ for the
light SM--like state and  $g_{HVV}=\cos(\beta-\alpha)$ for the heavier
one. In the 2HDM alignment or MSSM decoupling limits, one has $g_{hVV}=1$ and $g_{HVV}=0$. 

Above the $2M_V$ thresholds, the particle decay widths for the decays of a CP--even  Higgs boson into $W$ and $Z$ bosons pairs, $H \to VV$, are given by \cite{Barger:1993wt}
\beq
\Gamma (H \to VV) = \frac{G_F M_{H}^3}{16 \sqrt{2} \pi} \, \delta_V \,
\sqrt{1-4x} \, (1-4x +12x^2) \ , \nonumber
\label{HVV-2}
\eeq
with $x= {M_V^2}/{M_{\Phi}^2}$ and  $\delta_{W}=2, \delta_Z =1$. At high Higgs
masses, the   $H$ boson mostly decays into longitudinal states whose wave
functions are linear  in the energy such that the partial widths are $\Gamma (H
\to VV) \propto M_H^3$. In the of the CP--odd $A$ state, there is no tree--level decay into $WW/ZZ$ bosons but it can be generated through the effective interaction eq.~(\ref{eq:eff-HVV-cplg}), giving a partial decay width 
$\Gamma (A \to  VV)  \propto   \eta^2 \, (1-4x)^{3/2}$.

Below the $2M_V$ threshold, the Higgs states will decay into an on--shell and an off--shell gauge boson $\Phi \to VV^*\to Vf\bar f$ with partial decay widths given by
\cite{Barger:1993wt}
\begin{align}
\Gamma (H \to VV^*) &= \frac{3 G_F^2 M_V^4}{16 \pi^3} M_H \delta_V' \bigg[
\frac{3(1-8x+20x^2)}{(4x-1)^{1/2}} \arccos \left( \frac{3x-1} 
{2x^{3/2}} \right) \nonumber \\ & 
-\frac{1-x}{2x} (2-13x+47x^2)-\frac{3}{2}(1-6x+4x^2) \log x \bigg] \, , \nonumber \\
\Gamma (A \to  VV^*) & = \frac{3 G_F^2 M_V^6}{8 \pi^3 M_A} \delta_V' \eta^2 
\bigg[ (1-7x)(4x-1)^{1/2} \arccos \left( \frac{3x-1}{2x^{3/2}}
\right)  \nonumber \\ & 
-\frac{1-x}{6}(17-64x-x^2) + \frac{1}{2}(1-9x+6x^2) \log x \bigg] \, ,
\label{HVV-3}
\end{align}
with $\delta'_W\!=\!1$, $\delta_Z' \!= \! \frac{7}{12} - \frac{10}{9}
\sin^2\theta_W+ \frac{40}{9}\sin^4\theta_W$ for massless fermions.\newpage

\underline{Decays into gluons.}

\begin{figure}[!h]
\vspace*{-2.3cm}
\centerline{\includegraphics[scale=0.7]{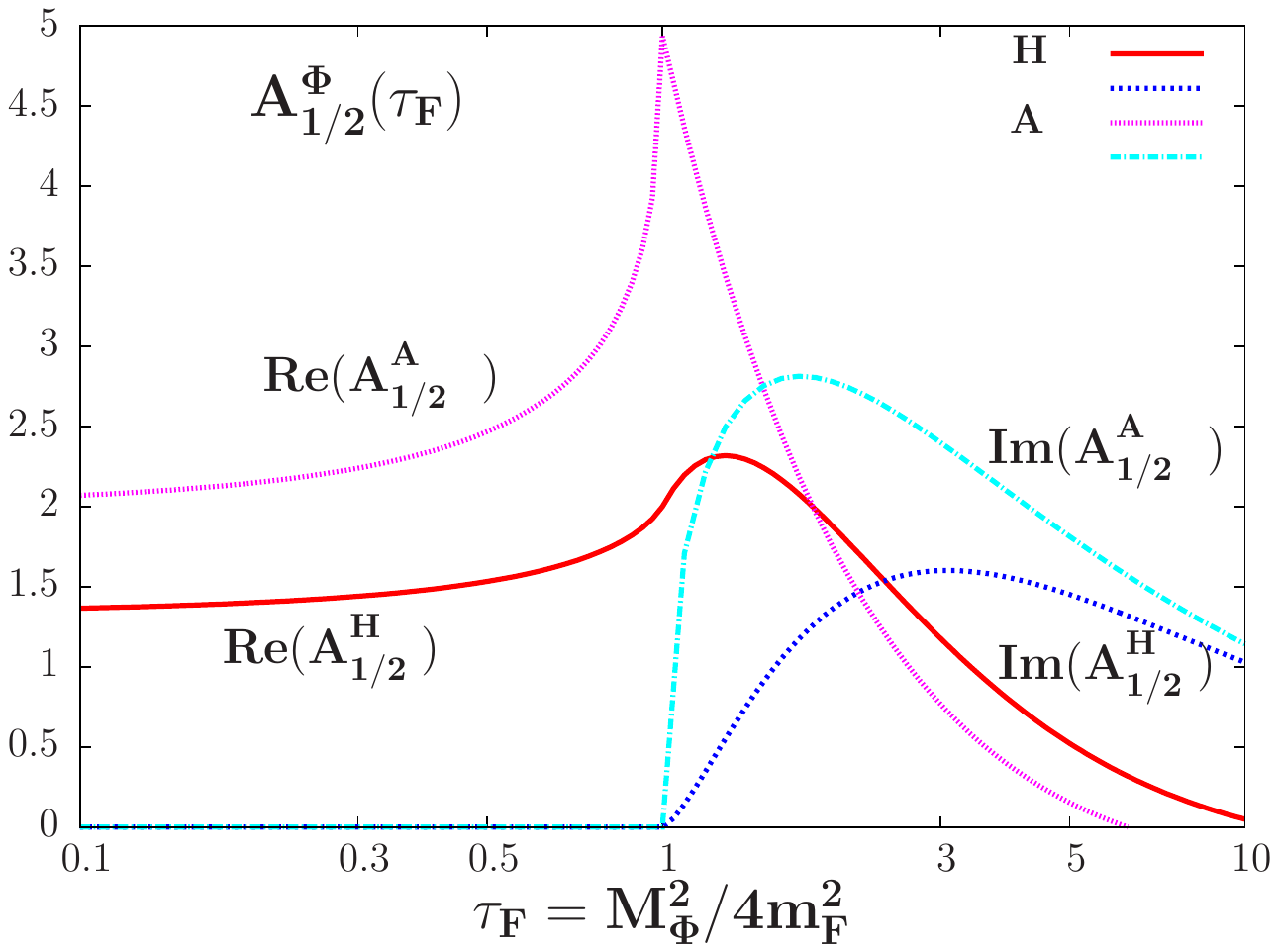} }
\vspace*{-11.8cm}
\caption{\it
The real and imaginary parts of the form factors $A^\Phi_{1/2}$ with fermion loops in the case of CP--even $H$ and CP--odd $A$ states as functions of the variable $\tau_F=M_\Phi^2/4m_F^2$.} 
\label{fig:enhancement}
\vspace*{-1mm}
\end{figure}

The $\Phi \to gg$ decay proceeds through triangular loops involving heavy strongly interacting particles that couple to the Higgs bosons. Assuming that only heavy quarks are running in the loops, the partial decay widths are give by~\cite{Ellis:1975ap} 
\begin{eqnarray}
\Gamma(\Phi  \to gg) & = &  \frac{G_F \alpha_s^2 M_\Phi^3}
{64\sqrt{2}\pi^3} \bigg| \sum_Q  g_{\Phi QQ} A_{1/2}^\Phi  (\tau_Q) 
\bigg|^2 \, , 
\label{eq:Gammagg}
\end{eqnarray}
with $g_{\Phi QQ}$ the Yukawa couplings normalised to their SM values. 
The form factors $A_{1/2}^\Phi (\tau_F)$ characterize the loop contributions of a fermion $F$ as functions of the variable $\tau_F =M_\Phi^2/4m_F^2$, which depend on the parity of the Higgs state and  are given by \cite{Gunion:1989we}
\begin{eqnarray}
A_{1/2}^{H}(\tau_F)  & = & 2 \left[  \tau_{F} +( \tau_{F} -1) f(\tau_{F})\right]  \tau_{F}^{-2} \, , \\
A_{1/2}^{A}  (\tau_F) & = & 2 \tau_{F}^{-1} f(\tau_{F}) \, ,
\label{eq:ASAP}
\end{eqnarray}
for the scalar $H$ or and pseudoscalar $A$  cases, respectively, where
\begin{eqnarray}
f(\tau_{F})= \left\{ \begin{array}{ll}  
\displaystyle \arcsin^2\frac{1}{\sqrt{\tau_F}} & {\rm for}\; \tau_{F}\geq 1\,,\\
\displaystyle -\frac{1}{4}\left[ \log\frac{1+\sqrt{1-\tau_{F}}} {1-\sqrt{1- \tau_F}} -i\pi \right]^2 \hspace{0.5cm} & {\rm for} \; \tau_{F}<1 \, .
\end{array} \right.
\label{eq:formfactors}
\end{eqnarray}
The real and imaginary parts of the  form factors for $H$ and  $A$ 
are shown in Fig.~\ref{fig:enhancement} as functions of $\tau_F$.

When the fermion in the loop is very heavy compared to $M_\Phi$, $m_F \to
\infty$, one obtains $A^{H}_{1/2}\! =\! \! \frac43$ and $A^A_{1/2}\!=\!2$ for
the form factors, and in the opposite limit of a light fermion, $m_F \to 0$, one
has instead $A_{1/2}^{\Phi} \to 0$. For $M_\Phi \le 2 m_F$ ($\tau_F \le 1$), so
that $\Phi \to {\bar F} F$ decays are forbidden, the maximal values of the form
factors are reached when $\tau_F=1$,  just at the $\Phi \to {\bar F}F$
kinematical threshold. In this case, one has the real parts
Re($A^H_{1/2})=\frac32$ and Re($A^A_{1/2})=\frac12 \pi^2 \approx 5$, and
Im($A^{\Phi}_{1/2}$) = 0 for the imaginary parts in both cases. 

In the SM, the contribution of the top quark can be approximated by setting 
$\tau_t \to 0$ giving $A^H(\tau_t) \simeq 4/3$, while the interference of the
top and bottom quarks amounts to less that 7\%. Any other heavy quark that has
SM--like couplings will double the amplitude and hence increase the decay rate
by a factor of 4. This enhancement factor will by 9 at LO in the case of a new
generation of quarks with SM--like couplings. Note that the decays are affected
by large QCD radiative corrections as discussed in
Refs.~\cite{Steinhauser:2002rq,Djouadi:2005gi}.  

In the case of many SM extensions like SUSY theories, there are also scalar
particles that contribute to the loop induced decay. These particles do not
couple to the pseudoscalar $ A$ boson because of CP invariance and the decays
occurs at higher order in this case; for scalar $H$ particles with reduced
couplings  $g_{HSS}$ to the scalar particles\footnote{In the case of the MSSM,
we are  referring to scalar squarks $\tilde Q$ that are partners of the heavy
$Q=t,b$ quarks which have couplings to the CP--even ${\cal H}=h,H$ bosons
normalized  such that $g_{ {\cal H}\tilde Q_i \tilde Q_i }= m_Q^2 g_{ {\cal
H}QQ}$ for the leading part, see Ref.~\cite{Djouadi:2005gj} for details.} $S$,
the gluonic partial decay width is given by 
\begin{eqnarray}
\Gamma({\cal H}  \to gg) & = &  \frac{G_F \alpha_s^2 M_{\cal H}^3}
{64\sqrt{2}\pi^3} \bigg| \sum_Q  g_{ {\cal H} QQ} A_{1/2}^{\cal H}  (\tau_Q) 
+ \sum_i  \frac{ g_{ {\cal H} S_iS_i} }{m_{S_i}^2}  A_{0}^{\cal H}  
(\tau_{S_i}) \bigg|^2 \, , \label{eq:GammaggS}
\end{eqnarray}
with the form--factor for scalar particles, using  the function $f$ given in 
\ref{eq:formfactors},
\begin{eqnarray}
A_{0}^{\cal H} (\tau_S) =  - \left[  \tau_{S} - f(\tau_{S})\right]  \tau_{S}^{-2} \, .  
\label{eq:ASAP-S}
\end{eqnarray}
The form factor approaches also zero for a light $S$ particle $A_0^{\cal H} \to
0$ but the asymptotic value  $A_0^{\cal H} \to \frac13$ for a heavy $S$ state;
it  reaches a maximum at near the threshold $\tau\simeq 1$ where  $\Re e
(A_0^{\cal H}) \simeq 1.5$ and $\Im m (A_0^{\cal H}) \simeq 1$. 

In principle, the higher order QCD corrections are also very large, being  of
the same order as the ones affecting the quark loops; this is at least what
occurs in the  case of squarks in the the MSSM as was discussed in
Refs.~\cite{Dawson:1996xz,Harlander:2004tp,Muhlleitner:2006wx} to which we refer for details. 

\underline{Decays into $\gamma\gamma$ and $Z\gamma$.}

Another very important Higgs decay channel is the one into two photons, $\Phi
\to \gamma\gamma$. As in the case of gluons, it is mediated  by heavy particles running in the loops, but this time all electrically charged  particles contribute. This includes the SM fermions but also the $W$ boson in the case of the CP--even Higgs particles (as already mentioned, the CP--odd $A$ has no tree level couplings to massive gauge bosons as a result of CP invariance). 
In extensions of the SM, other charged particles might contribute. This would be
the case of charged Higgs bosons in 2HDMs. In this case, the partial decay widths for CP--even $H$ and CP--odd $A$ bosons are given by 
\cite{Gunion:1989we,Shifman:1979eb}
\begin{eqnarray}
\Gamma (H\to \gamma\gamma) &=& \frac{G_F \alpha^2  M_{H}^{3}}
{128 \sqrt{2} \pi^{3}} \left| \sum_{f} N_{c}^f e_f^2 g_{Hff} A_{1/2}^H(\tau_f) 
+ A^H_1(\tau_W) + \frac{M_W^2}{2c_W^2 M_{H^\pm}^2 } g_{HH^+H^-} A^H_0(\tau_{H^\pm}) \right|^2 \, \nonumber  \\ 
\Gamma (A\to \gamma\gamma) &=& \frac{G_F \alpha^2  M_{A}^{3}}
{128 \sqrt{2} \pi^{3}} \left| \sum_{f} N_{c}^f e_f^2 g_{Aff} A_{1/2}^A (\tau_f)
\right|^2
\label{eq:Hgaga}
\end{eqnarray}
The form factors for spin--$\frac12$ and spin--0 particles have been given before and the one for spin--1 gauge bosons reads
\begin{eqnarray}
A_1^H(\tau) =  - [2\tau^2 +3\tau+3(2\tau -1)f(\tau)]\, \tau^{-2}
\label{eq:for-factor-AW}
\end{eqnarray}
The amplitudes for gauge bosons are always dominating if the $HWW$ couplings are
not suppressed:  below the $2M_W$ thresholds, the form factor is $A_1^H =-7$ for
very small Higgs masses (compared to $\frac43$ for fermions and $-\frac13$ for
scalars) to $A_1^H=-5 - 3 \pi^2/4$ at the threshold; for large Higgs masses, the
$W$ amplitude approaches $A_1^H \to -2$. In the SM, only this amplitude and the
one of the top quark are to be taken into account, the former one being
dominating and interfering destructively with the top quark amplitude. In the
presence of a fourth generation of fermions, the interference between the $W$ and fermion loops is even more
destructive and the partial decay width become much smaller than in the SM. 
Note that contrary to the two gluon decay, the amplitudes for the quark loop
contributions receive rather small QCD corrections \cite{Djouadi:1990aj}. In the case  of the SM Higgs boson, the electroweak corrections \cite{Djouadi:1997rj,Aglietti:2004nj,Degrassi:2004mx,Actis:2008ug} are  also known and are moderate. In the case of a 4th generation of fermions, the ${\cal
O}(G_Fm_{f'}^2)$ electroweak corrections are in turn very large in particular when the cancellation of the fermionic and the $W$ boson loop takes place 
\cite{Djouadi:1994ge,Djouadi:1997rj,Passarino:2011kv,Denner:2011vt}.

In SUSY extensions, two additional contributions need to be included in the case of the CP--even ${\cal H}=h,H$ states: those of the sfermions and the ones of the two charginos. In the case of the pseudoscalar $A$ state, only the latter
need to be included. The additional contributions to the two--photon decay amplitudes  are given in this case by  
\bea
{\cal A}_{\rm SUSY}^{\cal H} &= &\sum_{\chi_i^\pm} \frac{2 M_W}{ m_{\chi_i^\pm}} g_{{\cal H} \chi_i^+ \chi_i^-} A_{1/2}^{\cal H} (\tau_{\chi_i^\pm}) + 
\sum_{\tilde f_i} \frac{ g_{{\cal H} \tilde f_i \tilde f_i} }{ 
m_{\tilde{f}_i}^2} \, N_c Q_{\tilde f_i}^2 A_0^{\cal H} (\tau_{ {\tilde f}_i})
\nonumber \\
{\cal A}_{\rm SUSY}^{A} &= &
\sum_{\chi_i^\pm} \frac{2 M_W}{ m_{\chi_i^\pm}} g_{A \chi_i^+ \chi_i^-} 
A_{1/2}^{A} (\tau_{\chi_i^\pm})
\eea 
These extra contributions are suppressed by the masses of the SUSY particles and tend to zero in the limit where the latter are very heavy. 

For the other loop decay $\Phi \to \gamma Z$, the decay amplitudes are rather
involved and can be found in Refs.~\cite{Cahn:1978nz}. In the following, we will
give the expression of the partial width only in the case of the SM--like Higgs
boson which reads, 
\begin{eqnarray}
\Gamma (H\to Z\gamma ) = \frac{G^2_F M_W^2\, \alpha M_{H}^{3}} 
{64 \pi^{4}} \left( 1-\frac{M_Z^2}{M_H^2} \right)^3 \left|
\sum_{f} N_{f}^c \frac{e_f \hat{v}_f}{c_W} A_{1/2}^H(\tau_f,\lambda_f) + 
A^H_1(\tau_W,\lambda_W) \right|^2 
\label{eq:hzga}
\end{eqnarray}
with now $\tau_i= 4M_i^2/M_H^2$, $\lambda_i = 4M_i^2 /M_Z^2$. The complete
expressions of the form factors  can be again found in Refs.~\cite{Gunion:1989we,Djouadi:2005gi} but for Higgs masses below the $WW$ threshold, they can be approximated by 
\beq
A_1^H  \approx -4.6 + 0.3 M_H^2/M_W^2 , \quad 
A_{1/2}^H \approx N_t^c e_t \hat{v}_t / (3 c_W) \sim 0.3
\eeq
for the $W$--boson loop form factor and for the  one of the top quark, respectively.  The two  amplitudes interfere destructively but the top contribution is an order of magnitude smaller than the $W$ contribution for the mass value $M_H=125$ GeV.  

The same discussion given for the two--photon Higgs decay will hold also in this
case and, in fact, for $M_H \gg M_Z$ the two become identical in most cases
modulo the different photon versus $Z$ boson couplings.  In particular any
charged particle (and sometimes non--identical ones as a result of mixing) will
contribute to the $\Phi \to Z\gamma$ amplitudes. The radiative corrections are also
small in this case.  

\underline{Higgs to Higgs decays.}

Finally, in non--minimal extended Higgs models, there is the possibility of
Higgs to Higgs decays, the most common one being the decay of a heavy Higgs
boson into a lighter one and a massive gauge boson, $\Phi \to \varphi V$. In a 2HDM--like scenario such as the MSSM, the most likely possibility occurs when the final state is the lighter $h$ boson and two such decays are prominent outside the alignment or decoupling regimes of these models 
\beq
A \to hZ \quad {\rm and} \quad  H^\pm \to h W^\pm \, . 
\eeq
When the parent Higgs states are heavy enough for the decays to occur at the
two--body level, the partial widths are simply given by \cite{Gunion:1989we}
\begin{eqnarray}
\Gamma(\Phi \to \varphi  V) = \frac{G_F M_V^2}{8\sqrt{2} \pi} \, g_{\Phi 
\varphi V}^2 \, \lambda^{1/2}(M_V^2,M_\varphi^2;M_\Phi^2) \, \lambda(M_\Phi^2,
M_\varphi^2;M_V^2) \, , 
\end{eqnarray}
with $\lambda(x,y;z)=(1-x/z-y/z)^2-4xy/z^2$ being the usual two--body phase
space function. $g_{\Phi \varphi V}$ is the reduced coupling which, in the 2HDM 
or MSSM, is given by $\cos(\beta-\alpha)$ in the case of the $AhZ$ or $H^\pm hW^\mp$ couplings and, thus, vanishes in the alignment or decoupling limits. 
Other possibilities of such cascade decays involve the Heavy states only, 
\beq
H/A \to H^\pm W^\mp \, , \quad 
H^\pm \to H/A + W^\pm \, , \quad 
H \to AZ \ \ {\rm and} \ \  A \to HZ . 
\eeq
All these involve Higgs couplings $g_{\Phi \varphi V} \approx 1$ in the alignment or decoupling limits and are thus in principle favored, but they need a large enough Higgs mass  splitting, $M_\Phi - M_\varphi > M_V$, in order to occur at the two--body level. This is in general severely constrained by electroweak precision data, in particular by the $\rho$ or $T$ parameters which force the twos states to be rather close in mass. In this case, on has to resort to three--body decays $\Phi \to \varphi V^* \to \varphi f\bar f$,  which render the partial widths suppressed in most cases, in particular when the $\Phi \to t\bar t$ channel is open and/or when the $\Phi \to b\bar b$ channel is enhanced by strong couplings. The formulae for these higher--order decays can be found in Ref.~\cite{Djouadi:1995gv}.

A last possibility to be considered is heavy Higgs decays into a pair of 
lighter Higgs states. Because of CP--invariance, only  a few of such decays are
allowed in a 2HDM scenarii like the MSSM, mostly the channels $H\to hh, AA$ and $H\to
H^+H^-$. In the general case, the partial decay widths are given by
\cite{Gunion:1989we,Djouadi:2005gi}
\begin{eqnarray}
\Gamma(H \to \varphi \varphi) = \frac{G_F}{16\sqrt{2} \pi} \frac{M_Z^4}{M_H}
\left(1-4\frac{M_\varphi^2}{M_H^2} \right)^{1/2} \lambda_{H\varphi \varphi}^2
\end{eqnarray}
with $\lambda_{H\varphi \varphi}$ being a reduced  triple Higgs coupling
in units of $g_{HHH}^{\rm SM}$. However, as mentioned previously, in the benchmark scenarios that we discussed in this review (namely the MSSM and the 2HDM close to the alignment limit), the $H,A$ and $H^\pm$ masses are comparable to cope with constraints from electroweak precision observables. The decays of the $H$  bosons into pairs of $A$ or $H^\pm$ states is thus strongly disfavored by kinematics and the three--body decays $H\to \varphi \varphi^* \to b\bar b$ for instance are of higher order and thus have small branching ratios. The only possible channel is thus $H\to hh$, with $h$ being the observed 125 GeV Higgs state. While in the aligned 2HDM scenario, the $Hhh$ coupling is in general small if not zero, in the MSSM its receives large contributions from one--loop corrections as it was discussed in section 6.  The couplings is nevertheless tiny and the decay $H\to hh$ is important only in a small area of the parameter space. Note that in the case of very light pseudoscalars, the decays $h\to aa$, and eventually $H\to aa$ in the 2HDM+$a$ scenario, also occur.

\subsection{Higgs production at hadron colliders} 

At hadron colliders, there are four main channels for the single production of  the SM Higgs boson, which can be generalize to any CP--even Higgs state 
${\cal H}$
\bea
{\rm associated~production~with}~V=W/Z: & & q\bar{q} \to V + {\cal H} \, , \nonumber \\
{\rm ~vector~boson~fusion}: & & qq \to V^*V^* \to   qq+ {\cal H} \, , \nonumber  \\
{\rm gluon\!-\!gluon~fusion}: & & gg  \to {\cal H}\, ,\hspace*{2cm} \nonumber \\ 
{\rm associated~production~with~heavy~quarks}: & & gg,q\bar{q}\to Q\bar{Q}+{\cal H} \, . 
\eea
In extensions of the SM, all the four processes above also take place for CP--even ${\cal H}$ particles, but in the case of the CP--odd $A$ particle, only the gluon--gluon fusion mechanism and the associated production with heavy quarks are relevant at leading order
\bea
{\rm gluon\!-\!gluon~fusion}: & & gg  \to A \, , \hspace*{2cm} \nonumber \\
{\rm associated~production~with~heavy~quarks}:& & gg,q\bar{q}\to Q\bar{Q}+A \, .
\eea

There are in addition several mechanisms for Higgs pair production, $pp \to HH + X$, but they do not occur at LO in the electroweak or strong couplings and we will most of the time ignore them here. In 2HDM extensions, however, there are processes for the pair production of two neutral Higgs bosons (in addition to  those that are generated at higher orders such as $gg\to {\cal H}{\cal H},{\cal H}A,AA$ that we ignore here) which occur at the two--body level, 
\bea
{\rm Higgs~pair~production~}: & & q\bar{q} \to  {\cal H} A \, . 
\eea 

In this Appendix, we will give the analytical expressions of these processes at LO and briefly summarize the impact of the higher order corrections. There is one exception though: in the $gg\to \Phi$ process, we will also consider the case of Higgs production with an additional jet, $gg,q\bar q \to \Phi g$ and $gq \to \Phi q$ as  it is needed to describe DM production in the $gg$ fusion channel when an extra tagged jet is required. 

For the charged Higgs bosons, many processes are also available: top quark
decays for light $H^\pm$ states, associated production with a top quark for
heavier ones and $H^+H^-$ pair production. These have been discussed in section 
4.3.2 and, as they do not directly impact the DM issue, we refer to
Refs.~\cite{Djouadi:2005gj} for a detailed discussion. 

\underline{The Higgs--strahlung process: $q\bar{q} \to {\cal H}V$.}

The partonic process $q\bar{q} \to {\cal H}V$ proceeds through the $s$--channel exchange of a virtual $W$ or $Z$ boson and the total partonic cross section at LO is given by \cite{Glashow:1978ab}
\beq
\hat{\sigma}_{\rm LO}(q\bar{q} \to V {\cal H})\!= \! \frac{G_F^2 M_V^4}{288 \pi \hat{s}} g_{ {\cal H} VV}^2 (\hat v_q^2 + \hat a_q^2) \lambda^{1/2} (M_V^2, M_{\cal H}^2; \hat{s}) \frac{ \lambda(M_V^2, M_{\cal H}^2; \hat{s})+12 M_V^2/\hat{s}}{(1-M_V^2/\hat{s})^2}\, , 
\label{sigma-HV}
\eeq
where $\lambda$ is the two--body phase space function $\lambda(x,y;z)$
=$(1-x/z- y/z) ^2-4xy/z^2$; $g_{ {\cal H} VV}$ is the reduced ${\cal H} VV$ 
couplings that is equal to unity in the SM, and  $\hat v_f,  \hat a_f$ are the 
reduced $Vff$ couplings given in section 2.1.1. The total production cross section is obtained  by convoluting the expression above with the parton densities and summing over all contributing partons
\beq
\sigma_{\rm LO} ( pp \to V {\cal H})  = \int_{\tau_0}^1  {\rm d}\tau \,
\sum_{q,\bar{q}} \, \frac{ {\rm d} {\cal L}^{q \bar{q}} }{ {\rm d} \tau}
\, {\hat \sigma}_{\rm LO} (\hat{s}= \tau s) \, , 
\eeq
where $\tau_0= (M_V+M_{\cal H})^2/s$ with $s$ the total hadronic c.m. energy;
the parton luminosity is defined in terms of the parton densities $q_i(x_i,
\mu_F^2)$ defined at a factorization scale $\mu_F$, by  
\beq
\sum_{q,\bar{q}} \frac{ {\rm d} {\cal L}^{q \bar{q}} }{ {\rm d} \tau } =
\sum_{q_1,\bar{q}_2}   \int_{\tau}^1 \frac{{\rm d} x}{x} \, \left[ q_1 (x, 
\mu_F^2) \, \bar{q}_2 (\tau/x, \mu_F^2)  \right] .
\eeq
Note that the process can be viewed  as the Drell--Yan process \cite{Drell:1970wh} for producing a  vector boson with $q^2 \neq M_V^2$, which splits into a real vector boson and a Higgs particle. The distribution of the subprocess at LO can be then written as
\beq
\hat{\sigma} (q\bar{q} \to  {\cal H} V) =  \hat \sigma (q\bar{q} \to V^*) 
\times \frac{ {\rm d} \Gamma }{ {\rm d}q^2 }(V^* \to {\cal H} V) \, ,
\label{DY-factor}
\eeq
where, in terms of $0 \leq q^2\leq \hat{s}$ and the phase-space function $\lambda$, one has
\beq 
\frac{ {\rm d} \Gamma }{ {\rm d}q^2 } (V^* \to {\cal H} V) = \frac{ G_F M_V^4}{
2\sqrt{2} \pi^2}  \frac{\lambda^{1/2} (M_V^2, M_{\cal H}^2; q^2)}{(q^2-M_V^2)^2}
\left[1 + \frac{\lambda(M_V^2, M_{\cal H}^2;q^2)}{12M_V^2/q^2} \right] \ .
\label{HV-dGamma}
\eeq

Concerning the higher orders, the NLO QCD corrections are the pure Drell--Yan 
corrections to $q\bar q \to V^*$ \cite{Han:1991ia,Spira:1997dg,Djouadi:1999ht} 
extending up to NNLO  in the $q\bar q' \to {\cal H}W$ case\cite{Brein:2003wg,Brein:2011vx}.
In the case of ${\cal H}Z$, additional contributions at NNLO that are not
mediated by $Z$ boson exchange come from the  $gg\to {\cal H}Z$ subprocess
\cite{Brein:2003wg}. The full set of NLO+NNLO QCD corrections are moderate,
increasing the cross section by about +35\% at LHC energies, with the $gg \to
ZH$ contribution being of order of 10\% at $\sqrt s=14$ TeV. The NLO electroweak
corrections in turn reduce the cross section by an amount of about 5\% at LHC
energies \cite{Ciccolini:2003jy,Brein:2004ue}.

\underline{The vector boson fusion process: $qq \to {\cal H} qq$.}

In vector boson fusion \cite{Cahn:1983ip,Altarelli:1987ue,Kilian:1995tr}, the differential distribution of the partonic process $q_1q_2\to  q_3q_4 {\cal H}$ can be written at LO, in terms of the energy $E_{\cal H}$ and momentum 
$p_{\cal H}=\sqrt{E_{\cal H}^2-M_{\cal H}^2}$ of the Higgs boson and
the scattering angle $\theta$, as \cite{Kilian:1995tr}
\beq
\frac{ {\rm d}\hat{\sigma}_{\rm LO} } {{\rm d}E_{\cal H} {\rm d}\cos \theta } = 
\frac{G_F^3 M_V^8}{9 \sqrt2\,\pi^3 \hat s}\, g_{{\cal H}VV}^2\, \frac{p_{\cal H}}{32s_1 s_2 r}\bigg[ C_+ {\cal A}_+ + C_- {\cal A}_- \bigg] \, , 
\label{Hqq:Edistr}
\eeq
where, in terms of the reduced $Vff$ couplings given in section 2.1.1,  
\beq 
C_\pm= (\hat v_{q_1}^2 + \hat a_{q_1}^2) ( \hat v_{q_3}^2+ \hat a_{q_3}^2) 
\pm 4 \hat v_{q_1} \hat a_{q_1} \hat v_{q_3} \hat a_{q_3}  \, ,
\eeq
and 
\bea
{\cal A}_+ &=& (h_1+1)(h_2+1) \left[ \frac{2}{h_1^2-1} + \frac{2}{h_2^2-1} - \frac{6s_\chi^2}{r} + \left(\frac{3t_1t_2}{r}-c_\chi \right)
\frac{\ell}{\sqrt{r}}\right] \nonumber\\
&& \hspace*{2.5cm} - \left[ \frac{2t_1}{h_2-1} + \frac{2t_2}{h_1-1}
                + \left(t_1+t_2+s_\chi^2\right)
                \frac{\ell}{\sqrt{r}}\right] \, , \nonumber \\
{\cal A}_- &=& 	2(1 - c_\chi) \left[ \frac{2}{h_1^2-1} + \frac{2}{h_2^2-1} - \frac{6s_\chi^2}{r}  + \left(\frac{3t_1t_2}{r}-c_\chi\right)
          \frac{\ell}{\sqrt{r}}\right]		\, . 
\label{pp:VVHH}
\eea		
In these equations, we have used the following variables and abbreviations
\bea
\label{pp:VVHabrev}
  \epsilon_\nu \! = \! \sqrt{\hat s} \! - \! E_{\cal H},  \
  s_\nu \! = \! \epsilon_\nu^2 \! -\! p_{\cal H}^2, \ 
   s_{1,2} \! = \! \sqrt{\hat s}(\epsilon_\nu\pm p_{\cal H} \cos\theta), \ 
   h_{1,2} \! = \! 1 \! + \! \frac{2M_V^2}{s_{1,2}}, \
  t_{1,2} \! = \! h_{1,2} \! + \! c_\chi h_{2,1},  \nonumber \\[1mm]
   c_\chi \!= \! 1 \! - \! \frac{2 \hat s s_\nu}{s_1s_2} \!= \!1\! - \! s_\chi^2 , \ \   r \! = \! h_1^2 \! + \! h_2^2 \! + \! 2c_\chi h_1h_2 \! - \! s_\chi^2, 
   \ \  \ell \!= \! {\displaystyle \log\frac{h_1h_2 + c_\chi + \sqrt{r}}
                        {h_1h_2 + c_\chi - \sqrt{r}}}\; . \hspace*{2cm} 
\eea	
To derive the partonic total cross section, $\hat{\sigma}_{\rm LO}(qq \to qq 
{\cal H})$, the differential cross section needs to be integrated over $ -1<
\cos \theta <1$ and $M_{\cal H} < E_{\cal H} < {\sqrt{\hat s}}/{2} \times
(1+{M_{\cal H}^2}/{\hat s})$. Summing over the contributing partons, including
both the $WW$ and $ZZ$ fusion channels and folding with the parton luminosities,
one obtains the  total hadronic cross section $\sigma(pp \to V^* V^* \to qq
{\cal H})$ at leading order. The central scale in the  process is usually
chosen to be $\mu_0 = Q^*_V$, the momentum transfer of the  fusing vector bosons.

For the fully inclusive process, the NLO QCD corrections  
\cite{Han:1992hr,Figy:2003nv,Spira:1997dg,Djouadi:1999ht} increase the  total
cross section by ${\cal O}(10\%)$ and the NNLO QCD corrections (in the structure
function approach) are  below the percent level
\cite{Bolzoni:2010xr,Bolzoni:2010as} but can be large in the cross section with
cuts and in the differential distributions \cite{Cacciari:2015jma}. NLO
electroweak corrections shift the cross section by about 5\%
\cite{Ciccolini:2007ec,Figy:2010ct}. Hence, the radiative corrections (at least
to the fully inclusive cross section) are moderate and under control.

At LO, it is instructive (and will be useful later when we will discuss DM pair
production) to  display the much simpler expression for the cross section in the
longitudinal vector boson approximation, since in this case one simply needs to
calculate the cross section for the $2 \rightarrow 1$ process $V_LV_L
\rightarrow {\cal H}$ and fold it with the probabilities of emitting  a vector
boson from an energetic initial light quark, the $V_LV_L$ luminosity.  

Indeed, for large masses, the ${\cal H}$  boson is produced in the 
subprocess $VV \to {\cal H}$ mainly through the longitudinal components of the gauge bosons which give rates that grow with $M_{\cal H}$ as discussed before. The effective cross section in this case is simply given by
\beq
\sigma_{\rm eff} = \frac{16 \pi^2}{M_{\cal H}^3}  \Gamma ( {\cal H} \to V_L V_L)  \left. \frac{ {\rm d}{\cal L} } {{\rm d} \tau} \right|_{V_LV_L/qq} \, , 
\eeq
with the longitudinal vector boson luminosity defined as usual by
\beq  
 \frac{\rm d \cal L}{\rm d \tau}  \bigg \vert_{V_L V_L/pp}  =\sum_{q,q^{\prime}}
\int_{\tau}^{1} \frac{\rm d \tau^{\prime}}{\tau^{\prime}}
 \frac{\rm d {\cal L}^{qq^{\prime}}}{\rm d \tau^{\prime}}
\frac{\rm d {\cal L}}{\rm d \xi}  \bigg \vert_{V_L V_L/qq^{\prime}} \, ,
\eeq 
with $\xi=\tau/\tau^{\prime}$ and the classical quark--quark luminosity
\beq
{\rm d {\cal L}^{qq^{\prime}}}/{\rm d \tau}=\int_{\tau}^{1}
 { {\rm d} x }/ {x} \times q(x; Q^{2}) q^{\prime}(\tau/x; Q^{2}) \;.
\eeq 
If the luminosities are evaluated at the scale $Q=M_{\cal H}$, one obtains a simple expression for the longitudinal vector boson luminosity
\beq  
 \frac{{\rm d} {\cal L}}{{\rm d} \tau}  \bigg \vert_{V_L V_L/qq^{\prime}}  = \frac{\alpha^2 (\hat a_q^2 + \hat v_q^2)^2}{\pi^2} \frac{1}{\tau}
 \left[(1+\tau)\ln(1/\tau)-2(1-\tau) \right].
\label{V_lumi_spectra}
\eeq 
with $\alpha$ the fine structure constant and $\hat a_q, \hat v_q$ the reduced quark couplings  to vector bosons which have been  given before. One then finally obtains for the total partonic cross section of the VBF process
in this approximation 
\beq
\hat{\sigma}_{\rm LO} (qq \to qq {\cal H}) \simeq \frac{G_F^3 M_V^4 N_c}{128 
\sqrt{2} \pi^3} (C_+ + C_-) \, \left[ \left( 1+ \frac{M_{\cal H}^2}{\hat s} \right) \log \frac{ \hat s}{M_{\cal H}^2} -2 \left( 1- 2\frac{M_{\cal H}^2}{ \hat s} \right) \right] \, .
\label{Hqq-EWA}
\eeq
This approximation is valid only at very high energies and for not too large Higgs masses.  

\underline{The gluon fusion process: $gg \to \Phi$.}

The gluon fusion process takes place for both the CP--even and CP--odd Higgs bosons and, at lowest order, the partonic cross section are simply given by
\begin{eqnarray}
\hat\sigma_{\rm LO} (gg \to \Phi) & = & {\sigma_0^\Phi} {M_\Phi^2} \, \delta
(\hat s -M_\Phi^2)  = \frac{\pi^2}{8 M_\Phi} \Gamma_{\rm LO} (\Phi \to gg) \, \delta (\hat s -M_\Phi^2) \, , 
\end{eqnarray}
where $\hat{s}$ is the squared $gg$ invariant and the gluonic widths of the Higgs bosons have been given in eq.~(\ref{eq:GammaggS}) in the CP--even and CP--odd cases. Inserting the latter expressions in the equation above, one finds \cite{Georgi:1977gs}
\begin{equation}
\sigma_0^\Phi = \frac{G_F\alpha_{s}^{2}(\mu_R^2)}{288 \sqrt{2}\pi} \
\left| \, \frac{3}{4} \sum_{q} A_{1/2}^\Phi (\tau_{Q}) \, \right|^{2} \, ,
\end{equation}
where the form factors $A_{1/2}^\Phi (\tau_Q)$ with $\tau_Q = M_\Phi^2/4m_Q^2$  are given in eqs.~(\ref{eq:ASAP}) and are normalized such that for $m_Q \gg 
M_\Phi$, they reach the values $\frac{4}{3}$ in the CP--even $\Phi=h,H$ and 
2 in the CP--odd $\Phi=A$ cases; they both approach zero in the chiral limit 
$m_Q \to 0$. The proton--proton cross section at LO in the narrow--width 
approximation reads
\begin{equation}
\sigma_{\rm LO}(pp\to H) = \sigma_0^H \tau_H \frac{d{\cal L}^{gg}}{d\tau_H}
\ \ {\rm with} \ 
\frac{d{\cal L}^{gg}}{d\tau} = \int_\tau^1 \frac{dx}{x}~g(x,\mu_F^2)
g(\tau /x,\mu_F^2) \, 
\end{equation}
where the Drell--Yan variable is defined as usual by $\tau_H = M^2_H/s$ with 
$\sqrt s$ the collider energy. 

In the SM, the top quark loop contribution is by far dominating with the bottom 
contribution, in fact its interference with the  top quark one, not exceeding
the 10\% level. The NLO QCD corrections have been calculated not only in the
infinite top quark mass approximation $M_\Phi \! \ll \! 
2m_t$~\cite{Djouadi:1991tka,Dawson:1990zj}  but also  using the exact quark mass
dependence in the loop~\cite{Spira:1995rr}. They were found to be large  with a
$K$--factor, defined as the ratio of cross sections at the higher order to
lowest order, $K\!= \! \sigma_{\rm HO}/ \sigma_{\rm LO}$ with $\alpha_s$ and the
PDFs evaluated at the same respective orders, of around 1.7 for $M_H\!=\!125$
GeV at $\sqrt s \approx 14$ TeV. It was also shown that if the LO cross section
contains the full top quark mass dependence,  the exact and infinite mass
results approximately agree, in particular, in the Higgs mass range $M_H \lsim
2m_t$.  The NNLO QCD corrections,  computed  in the $m_t \!\to \! \infty$  limit
\cite{Harlander:2002wh,Anastasiou:2002yz,Ravindran:2003um},  lead to  an
increase of 25\% for the cross section. Recently, the N$^3$LO corrections were
evaluated \cite{Anastasiou:2015ema} and found to lead to an additional small
increase of the cross section. The electroweak corrections have been computed at
NLO in the  infinite loop mass limit $m_t,M_V\!\gg\! M_H$
\cite{Djouadi:1994ge,Djouadi:1997rj,Aglietti:2004nj} and exactly
\cite{Degrassi:2004mx,Actis:2008ug}. Approximate  mixed QCD--electroweak
corrections at NNLO are also available \cite{Anastasiou:2008tj}.  Both>
corrections amount to a few percent.

In extensions of the SM, the top--quark loop might not provide the leading
contribution and, in fact, the bottom quark loop is dominant in large areas of
the parameter space of 2HDMS, when the Higgs--$b\bar b$ couplings are enhanced
at large $\tb$ values. In this case, the cross section which grows as
$\tan^2\beta$ and is enhanced by large logarithms $\log(m_b^2/M_\Phi^2)$, can be
extremely large. In this case, as $M_\Phi \gg 2m_b$, one is in the chiral limit 
where the rates are approximately the same in the CP--even and CP--odd Higgs
cases. One cannot use anymore the infinite loop mass approximation to calculate
the higher order terms.  The QCD corrections can be thus included only to NLO
for which they are known when keeping the exact quark mass dependence
\cite{Spira:1995rr}. At LHC energies, the $K$--factors  are much smaller,
$K_{\rm NLO}^{\rm b\!-\!loop} \approx 1.2$, than in the case of the top quark
loop, $K_{\rm NNLO}^{\rm t\!-\!loop} \approx 2$.  

At leading order, the $gg$ fusion process leads to a single Higgs boson which is
invisible  when it decays into stable DM particles.  To make the process
observable experimentally, at least one extra jet, emitted from the initial
gluons or from the internal quark lines, should be produced in addition. All
these processes are in fact present when one calculates the real corrections  at
NLO for the  $gg$ fusion process \cite{Djouadi:1991tka,Dawson:1990zj} that we
briefly summarize below in the case of the SM Higgs boson. 

Adopting the effective approach  in which one only considers the dominant top
loop contribution in the limit $m_t \gg M_H$, the calculation is performed using
the dimensional regularization scheme with the coupling constant $\alpha_s$
renormalized in the $\overline{\rm MS}$ scheme with five light--quark flavors. 
When adding the virtual corrections to $gg\to H$ to the real corrections $gg \to
Hg$  the infrared singularities cancel out. The left--over  initial--state
collinear singularities in the partonic cross section are absorbed into the NLO
parton densities, also defined in the $\overline{\rm MS}$ scheme with  five
quark flavors. The remaining finite contributions  can be then cast into the form
\beq 
\sigma_{\rm LO}(pp \to H+ j) = \Delta\sigma_{gg} + \Delta\sigma_{gq} 
+ \Delta\sigma_{q\bar q} \, ,
\label{eq:gghqcd}
\eeq 
where, using $\tau_0 = M_H^2/s$, the individual contributions are given by 
\bea 
\Delta \sigma_{gg,gq,q\bar q} & = & \frac{\alpha_{s}(\mu)} {\pi} \int_{\tau_0}^1 d\tau~ \frac{d{\cal L}^{gg}}{d\tau} \int_{\tau_0/\tau}^1 \frac{dz}{z}~ \hat\sigma_{\mathrm{LO}}(Q^2 = z \tau s) {\cal F}_{gg,gq,q\bar q } \, ,\nonumber \\ 
{\cal F}_{gg} &=& - z P_{gg} (z) \log \frac{\mu_F^{2}}{\tau s}  - \frac{11}{2} (1-z)^3 + 6 [1+z^4+(1-z)^4] \left(\frac{\log (1-z)}{1-z} \right)_+ \, , \nonumber \\  {\cal F}_{gq} &=& -\frac{z}{2} P_{gq}(z) \log\frac{\mu_F^{2}}{\tau s(1-z)^2} + \frac{2}{3}z^2 - (1-z)^2  \vphantom{\frac{M^{2}}{\tau s(1-z)^2}} \, , \nonumber \\ 
{\cal F}_{q\bar q} &=& \frac{32}{27} (1-z)^3 \, .
\label{eq:individual-gg}
\eea
The functions $P_{gg}(z), P_{gq}(z)$ denote the Altarelli--Parisi splitting
functions \cite{Altarelli:1977zs}
\begin{eqnarray}
P_{gg}(z) \! = \! 6\left\{ \left(\frac{1}{1 \! - \!z}\right)_+
\! + \! \frac{1}{z} \! -\! 2 \! + \! z(1 \! - \! z) \right\} \! + \!  
\frac{23}{6}\delta(1 \! - \! z), \ \ P_{gq}(z) \! = \! \frac{4}{3} \frac{1 \! + \! (1 \! - \!z)^2}{z},
\end{eqnarray}
with five quark  flavors. The factorization scale $\mu_F$ of the parton--parton
luminosities  $d{\cal L}^{ij}/d\tau$ and the renormalization scale $\mu_R$  can
be set at the value $\mu_R=\mu_F=\frac12 M_H$.  In practice one can also include
the NLO QCD corrections to this topology, in which there are contribution with
two jets in the final state, $pp \to \chi\chi+jj$, as it can be borrowed from
the corresponding NNLO QCD corrections for Higgs production which are known
\cite{Harlander:2002wh,Anastasiou:2002yz,Ravindran:2003um}. 

\underline{Associated production with heavy quarks.}

Associated Higgs production with top quark pairs proceeds through $gg$ fusion
and $q\bar q$ annihilation, $gg,q\bar q \to t \bar t \Phi$, with the Higgs
states radiated from the top quark lines.  The processes are thus directly
proportional to $g_{\Phi tt}^2$ and provide a direct  probe of the top quark
Yukawa couplings. These are three body production process which lead to small
rates for high values of the Higgs masses and which, already at LO,  have a
rather complicated cross section
\cite{Raitio:1978pt,Ng:1983jm,Kunszt:1984ri}. The  total rates
are only slightly different for CP--even and CP--odd Higgs boson as a result of
top mass effects. The NLO QCD corrections are known to be modest provided that
the central scale value $\mu_0\!=\!\frac12 M_\Phi+m_t$ is used
\cite{Beenakker:2001rj,Beenakker:2002nc,Dawson:2002tg}.

Associated Higgs production with bottom quark pairs, $gg,q\bar q \to b \bar b
\Phi$~\cite{Raitio:1978pt,Ng:1983jm} has a rather different behavior compared 
to $t\bar t \Phi$. First, for large values of the Higgs masses one is in the
chiral limit $m_\Phi \gg m_b$ and the cross section is the same for a CP--even
and a CP--odd Higgs particle. In addition, the NLO QCD corrections  turn out to
be very large~\cite{Dittmaier:2003ej,Dawson:2003kb} as a result of the large
logarithms generated by the integration of the transverse momenta of the final
bottom quarks. These large logarithms can be re-summed by considering the bottom
quark as a massless parton and use the Altarelli--Parisi evolution \cite{Altarelli:1977zs} of
the bottom quark PDF. In practice, one then works in a five--flavor scheme  in
which the process which should be  considered at LO is simply $b\bar{b}\to
\Phi$~\cite{Dicus:1988cx}. It has a very simple expression for the cross section
at LO, given in terms of $\hat \tau= M_\Phi^2/ \hat s$,  by
\bea
b\bar b \to \Phi :  \quad \hat \sigma_{\rm LO} (\hat \tau) &=&  \frac{\pi}{12} \, \frac{ g_{\Phi b \bar b}^2}{M_\Phi^2} \, \delta(1-\hat \tau) \, . 
\eea
If ones requires a high--$p_T$ final state $b$ quark, the QCD  corrections need
to be included, with the NNLO QCD corrections leading us back to the process
$gg\to b \bar{b} H$~\cite{Harlander:2003ai}. When choosing $\mu_0 = \frac14 M_H$
for the factorization scale and if the running bottom mass at the scale of the
Higgs mass is used, the perturbative series converges rapidly. 

\underline{Higgs pair production, $q\bar q \to {\cal H}A$.}

Finally, there are processes for Higgs pair production. In the SM, there are
four such  mechanisms: gluon fusion $gg \to HH$, double Higgs--strahlung from
$Z,W$ bosons $q\bar{q} \to  V^* \to VHH$, the VBF processes $qq \to V^* V^* qq 
\to HHqq$ and associated production with heavy quarks pairs $pp \to Q\bar Q HH$
\cite{Glover:1987nx,Dicus:1987ic,Plehn:1996wb,Dawson:1998py,Djouadi:1999rca,Baglio:2012np}.
They all involve a diagram, among others,  in which an off--shell Higgs is
produced and splits into two real Higgs bosons. These processes can be
generalized to the CP--even Higgs states that appear in SM extensions. In the
case of CP--odd Higgs states, only the first and last ones are relevant as in
single Higgs production. All these processes are of high order in the
perturbative series and have low cross sections. The dominant process is $gg \to
HH$ which occurs through a triangle diagram generated by heavy quark loops and
producing an $H^*$ which splits into $HH$ final states, and a box diagram in
which both Higgs particles are emitted from the heavy quark internal lines. The
two contributions interfere destructively and lead to a cross section that is
three orders of magnitude lower than for single production. 

In some extensions of the SM such as 2HDMs, there is however a possibility to
produce two Higgs bosons at LO in perturbation theory. In the case of the neutral Higgs bosons that are of interest here, there is only one such process, $q\bar q \to Z^* \to {\cal H}A$ as CP--invariance forbids ${\cal H}{\cal H}$ and $AA$ production this way.  The partonic cross section is, up to couplings factors,  the same as associated Higgs production with a $Z$ boson 
\beq 
\hat \sigma( q\bar q \to {\cal H} A) = g_{ {\cal H} AZ}^2  \hat \sigma_{ \rm SM}( q\bar q \to {\cal H} Z) \times \frac{\lambda_{A {\cal H} }^{3} }{ \lambda_{Z {\cal H} } (\lambda_{Z {\cal H}}^2 + 12M_Z^2/ \hat s)} \, ,  \label{pp-HAxs} 
\eeq 
with an additional difference in the phase--space factor to account for the
production of two spin--zero particles. In the aligned 2HDM and the MSSM,  the
$hAZ$ coupling is small while $g_{HAZ} \approx 1$ so that only $q\bar q  \to HA$
is relevant, but the rates are small at high energies.  

Such processes occur for
the charged Higgs boson which can be either produced in pairs,  $q\bar q \to
\gamma, Z^* \to H^+H^-$ or in association with a (heavy) neutral Higgs boson, 
$q\bar q' \to W^* \to {\cal H}H^\pm, AH^\pm$. The relevant formulae can be found in Ref.~\cite{Djouadi:2005gj}.

\subsection{Higgs production at lepton colliders}

There are several mechanisms in which Higgs bosons can be produced in $e^+ e^-$ collisions. In the case of the SM Higgs boson,  these are:  
\begin{eqnarray}
{\rm Higgs\!-\!strahlung \ process} : & & e^+ e^- \rightarrow (Z^*) 
\rightarrow Z \, H , \nonumber \\
{\rm WW \ fusion \ process} : & & e^+ e^- \rightarrow \bar{\nu}\nu \, (W^*W^*) \rightarrow \bar{\nu}\ \nu \, H \nonumber \\ 
{\rm ZZ \ fusion \ process} : & & 
e^+ e^- \rightarrow e^+ e^- (Z^*Z^*) \rightarrow e^+ e^-  H \, \nonumber \\ {\rm radiation \ off \ heavy\ fermions}:& & e^+ e^- \rightarrow (\gamma^*,Z^*) \rightarrow f  \bar{f} \, H \, . 
\end{eqnarray}
There are other  higher--order processes in which Higgs particles can be
produced in $e^+ e^-$ collisions, including Higgs pair production,  with even smaller production rates and we will ignore them here. But there is one option, that we will discuss:  Higgs production as $s$--channel resonances in the 
$\gamma \gamma$ option of future $e^+ e^-$ linear colliders, 
\beq
\gamma \gamma \rightarrow H \, . 
\eeq
In 2HDMs such as the MSSM, there are also Higgs pair production processes that occur at the $2\to 2$ level and we will consider only the ones related to the neutral Higgs states, 
\begin{eqnarray}
{\rm \ Higgs\ pair \ production \ process} : & & e^+ e^- \rightarrow (Z^*) \rightarrow A \, {\cal H} .
\end{eqnarray}

\underline{The Higgs--strahlung processes.}

The production cross section for the Higgs--strahlung process \cite{Barger:1993wt} is given by
\beq 
\sigma(e^+e^-  \to ZH) = \frac{G_F^2 M_Z^4}{96 \pi s} (\hat{v}_e^2+\hat{a}_e^2) \ \lambda^{1/2} \frac{ \lambda+ 12M_Z^2/s}{(1-M_Z^2/s)^2} \, , 
\eeq
where, as usual, $\hat{a}_e=-1$ and $\hat{v}_e=-1+4s_W^2$ are the $Z$ charges 
of the electron and $\lambda^{1/2}$ the two--particle phase--space  function,
$\lambda=(1-M_H^2/s-M_Z^2/s)^2-4M_H^2M_Z^2/s^2$. The recoiling $Z$ boson in this
two--body reaction is mono--energetic, $E_Z = (s-M_H^2+M_Z^2)/ (2\sqrt{s})$, and
the Higgs mass can be derived from the energy of the $Z$ boson, $M_H^2 =s
-2\sqrt{s} E_Z +M_Z^2$, if the initial $e^+$ and $e^-$ beam energies  are
precisely known. This is very important when the Higgs decays invisibly. The
angular distribution of the process  $d\sigma/ d \cos \theta \propto \lambda^2
\sin^2\theta +8M_Z^2/s$, which at high energies  $s \gg M_Z^2$ gives the
asymptotic value  $\frac{3}{4}\sin^2\theta$, typical of the production of
spin--zero particles, since at these energies the $Z$ is longitudinally
polarized. 

The cross section  scales as the inverse of the c.m. energy, $\sigma \! \sim \! 1/s$ and for $M_H \! \approx 125 \!$ GeV, it is larger  for low energies,  the  maximal value being at $\sqrt{s} \! \sim \! M_Z \! + \!\sqrt{2}M_H \!\approx \! 240$ GeV.

\underline{The vector boson fusion process. }
 
The vector fusion channel, similar to VBF at hadron colliders,  is most important for  small values of the ratio $M_H/\sqrt{s}$, i.e. at high energies where the cross  section grows as $\sim M_V^{-2}$log$(s/M_H^2)$. The production cross section can be conveniently written as \cite{Kilian:1995tr}
\bea 
\sigma (e^+e^- \! \to \! H\ell\ell)  \!= \! \frac{G_F^3 M_V^4}{64 \sqrt{2} \pi^3} \int_{\kappa_H}^1 {\rm d}x \int_x^1 \frac{ {\rm d}y}{[1\! +\! (y\!- \!x)/ \kappa_V]^2} \left[ (\hat{v}_e^2\! +\! \hat{a}_e ^2)^2 f(x,y) \! + \! 4 \hat{v}_e^2 \hat{a}_e^2 g(x,y) \right] \, , \nonumber \label{WWxsection} 
\eea 
\vspace*{-4mm}
\bea 
f(x,y) &=& \left(\frac{2x}{y^3} -\frac{1+2x}{y^2} +\frac{2+x}{2y}
-\frac{1}{2} \right)\left[ \frac{z}{1+z} -\log (1+z) \right]
+\frac{x}{y^3}\frac{z^2(1-y)} {1+z} \, , \nonumber \\ 
g(x,y) &=& \left(-\frac{x}{y^2} +\frac{2+x}{2y} -\frac{1}{2} \right) 
\left[ \frac{z}{1+z} -\log (1+z) \right] .  
\eea
with $\kappa_H =M_H^2/s, \kappa_V=M_V^2/s , z=y(x-\kappa_H)/(\kappa_Vx)$ and 
$\hat{v}_e, \hat{a}_e$ the reduced $Vee$ couplings. 

For $M_H \approx 125$ GeV, the $WW$ fusion cross section is of about the same
magnitude as that of the bremsstrahlung  process at $\sqrt s \approx 500$
GeV; it is smaller at lower energies and larger for higher energies as it 
grows logarithmically with $s/M_H^2$ in contrast to $\sigma(HZ)$ which falls like $1/s$.  The cross section for $ZZ$ fusion is an order of magnitude smaller than the one of $WW$ fusion, a mere consequence of the fact that the neutral current couplings are smaller than the charged current couplings. In the context of DM production, the lower rate is however compensated by the more interesting signature which is observable for invisible Higgs decays and allows 
for a missing mass analysis to tag the Higgs particle.

\underline{Associated production with heavy fermions.}

In $e^+e^-$ collisions, the $Hf\bar f$ final state is generated almost exclusively through Higgs bremsstrahlung off the fermion lines, since the additional contributions when the Higgs is emitted from  the $Z$ boson line 
are very small. As both the fermion and Higgs masses should be kept non--zero, the analytical expressions of the cross section are quite involved 
\cite{Djouadi:1991tk}. However, neglecting these mass effects together with the Higgs emission off the $Z$ line give a result which approximates the total cross section at the 10\% level. In this case, the Dalitz plot density can be then written in a rather simple form \cite{Djouadi:1991tk,Djouadi:1992gp}
\bea
\frac{d \sigma}{ dx_1 dx_2}(e^+e^- \! \to \! f\bar f H)  =
\frac{\bar{\alpha }^2 g_{Hff}^2 N_c^f}{12 \pi s} \hspace*{-4mm} & & \left\{ \left[ e_e^2 e_f^2 + \frac{2 e_e e_f {v}_e {v}_f}{1-z}+\frac{( {v}_e^2
 + {a}_e^2) ({v}_f^2 + {a}_f^2)}{(1- z)^2} \right] \right. \nonumber
\\ && \hspace*{-1.8cm}\times  \frac{  x_H^2}{(1-x_1)(1-x_2)} -  \left. 2 \frac{{v}_e^2+ {a}_e^2}{(1- z)^2}  {a}_f^2 (1+x_H)  \right\} \, , 
\label{ttHxsection}
\eea
where $x_1=2 E_f/\sqrt{s}$, $x_2=2E_{\bar{f}}/ \sqrt{s}$ and
$x_H=2E_H/\sqrt{s}=2-x_1-x_2$  are the reduced energies  of the $f$, $\bar{f}$
and $H$ states and $z=M_Z^2/s$. The differential cross section has to be integrated over the allowed range of the $x_1, x_2$ variables with a boundary condition 
\begin{equation}
\left| \frac{2(1-x_1-x_2+2\mu_S-\mu_Z) + x_1x_2}
{\sqrt{x_1^2-4\mu_S} \sqrt{x_2^2-4\mu_S}} \right| \leq 1  \;.
\label{eq:dalitzbound}
\end{equation}

\underline{Higgs pair production at LO.}

In the SM, the Higgs bosons can be pair produced in the same processes that
allow for single production, with the main diagrams (that involve the important
trilinear couplings) being simply those that occur in the four channel discussed
above, but with the Higgs boson being off--shell and splitting into two reals
Higgs particles  \cite{Djouadi:1999gv}. All these are higher order processes and
lead to small cross sections which are not relevant in the DM context.  

Such processes also occur in SM extensions, but in some cases they can be generated at tree--level and lead to large rates. This is the case of the associated production of a pair of CP--even ${\cal H}$ and CP--odd $A$ states 
in 2HDMs like the MSSM. The cross sections are again simply that of SM Higgs production in Higgs--strahlung, modified to take into account the different coupling and phase--space \cite{Gunion:1989we,Djouadi:2005gi}
\beq 
\sigma( e^+e^- \to {\cal H} A) = g_{ {\cal H} AZ}^2  \sigma_{ \rm SM}( e^+ e^- \to {\cal H} Z) \times \frac{\lambda_{A {\cal H} }^{3} }{ \lambda_{Z {\cal H} } (\lambda_{Z {\cal H}}^2 + 12M_Z^2/ \hat s)} \, . \label{ee-HAxs} 
\eeq 
Here again, in the alignment or decoupling limits of 2HDM, the process is most important for the heavy $H$ state which has a coupling $g_{HAZ}\approx 1$ than 
for the lighter one with $g_{hAZ}\approx 0$ despite of the less favorable phase space. In the charged Higgs case, the pair production process $ e^+e^- \to H^+H^-$ is also important being not suppressed by mixing factors.

\underline{Higgs production in $\gamma\gamma$ collisions.}

Heavy neutral Higgs bosons $\Phi=H/A$ could be
produced as $s$--channel resonances via the $\gamma \gamma$ option of a future  parent linear $e^{+} e^{-}$ collider, see for instance Refs.~\cite{Badelek:2001xb,Ginzburg:1982yr}. Indeed, a $\gamma\gamma$ collider  can be constructed using Compton back--scattering from a laser beam via the processes~\cite{Badelek:2001xb,Ginzburg:1982yr,Djouadi:1996pb,Godbole:2002qu} 
\begin{eqnarray}
e^-(\lambda_{e^-}) \ \gamma(\lambda_{l_1}) \rightarrow e^- \ \gamma(\lambda_1) \, , \;
e^+(\lambda_{e^+}) \ \gamma(\lambda_{l_2}) \rightarrow e^+ \ \gamma(\lambda_2) \, .
\end{eqnarray}
The back--scattered laser photons will carry a large fraction of the energy of  the $e^{+}/e^{-}$ beams. Their energy spectrum and polarization depend on the laser helicities $\lambda_{l_{1}}, \lambda_{l_2}$ and of the leptons $\lambda_{e^{+}},\lambda_{e^{-}}$ and on the laser energy.  The advantage of 
such a collider is that it provides a direct access to the state in single
production and allows the opportunity to probe its CP properties. In the 
general case, one has for the production cross section
\begin{eqnarray}
\sigma(\lambda_{e^+},\lambda_{e^-},\lambda_{l_{1}},\lambda_{l_{2}},E_{b})\! 
= \! \int dx_1 dx_2 L_{\gamma\gamma}(\lambda_{e^+},\lambda_{e^-},\lambda_{l_1},\lambda_{l_2},x_1,x_2)\, \hat \sigma(\lambda_1,\lambda_2, 2 E_{b} \sqrt{x_1x_2}),~~
\end{eqnarray}
where $L_{\gamma\gamma}$  is the luminosity function for given polarizations 
of the colliding photons and  $\hat \sigma$ the cross section for the $\gamma \gamma \to \Phi \to X$  subprocess.  The invariant mass of the $\gamma\gamma$ system is given by $w\! =\! \sqrt{\hat s}\!=\!2 E_{b} \sqrt{x_{1} x_{2}}$, with $x_{1}, x_{2}$ the fractions of the beam energy $E_{b}$ carried by the  back--scattered photons.  The cross section for  $\Phi$ production via $\gamma \gamma$ fusion then reads
\begin{eqnarray}
\hat {\sigma} (w,\lambda_{1},\lambda_{2}) = 8 \pi  {\frac{\Gamma (\Phi \rightarrow \gamma \gamma) \, \Gamma (\Phi \rightarrow X)}{(w^{2} - M_{\Phi}^{2})^{2} + M_{\Phi}^{2} \Gamma_{\Phi}^{2}}} (1 + \lambda_{1} \lambda_{2}) \, ,
\label{gmgmcsec}
\end{eqnarray}
where $w$ is the $\gamma \gamma$ system c.m. energy and the factor of $(1 + \lambda_{1} \lambda_{2})$ projects out the $J_{Z} = 0$ component of the cross section, thereby maximizing the scalar resonance contribution compared to the continuum backgrounds. 

The dependence of the energies and the polarizations of the back-scattered
photons, i.e.,  $(E_{b}x_{1}, \lambda_{1})$ and  $(E_{b} x_{2}, \lambda_{2})$, 
on the electron and positron beam energy $E_{b}$ as well as on  the frequency
and the polarization of the laser \cite{Ginzburg:1982yr} are such  that the
spectrum peaks in the region of  high photon energy when $\lambda_{e}
\lambda_{l}=-1$. If, in addition, one chooses  the laser energy  $\omega_{0}$ so
that $x=4E_b \omega_0 / {m_e^2} = 4.8$,  the two-photon luminosity is peaked at
$z = 0.5 \times W/E_{b} = 0.8$. The mean helicity of the back-scattered photons
depends on their energy and for $\lambda_{e} \lambda_{l} = -1$ and $x = 4.8$, in
the region of high energy for the  back-scattered photon where the spectrum is
peaked, the  back-scattered photon also carries the polarisation of the parent
electron/positron beam. Thus, choosing  $\lambda_{e^{-}} = \lambda_{e^{+}}$
ensures that the dominant photon helicities are the  same,  which then maximizes
the Higgs signal relative to the QED background $\gamma\gamma \to f\bar f$,
leading to a luminosity $L_{\gamma \gamma} \equiv L_{\gamma \gamma}
(\lambda_{e^-},x_{1},x_{2})$. 

The total cross section for $\gamma \gamma \rightarrow \Phi$, where we write down explicitly the expression for $L_{\gamma \gamma}$ for the  previous choices of helicities,  is then
\beq
\sigma \; = \; {\frac {8 \pi^2}{M_\Phi s}}~~\Gamma (\Phi \rightarrow \gamma \gamma)~ \int_{x_1^m}^{x_1^M} {\frac{1}{x_1}} f(x_1) f(M^2_\Phi/s/y_1) \left(1 + \lambda_1(x_{1},\lambda_{e^{-}}) \lambda_2(x_{2},\lambda_{e^{+}})\right), 
\eeq
where $f(x_{i})$ denotes the probability that the back--scattered photon carries a fraction $x_{i}$ of the beam energy for the chosen laser and lepton helicities, with 
\beq
x_1^m = {M_\Phi^2}/(s x_1^M) \, \quad x_1^M = x_c/(1+x_c)~~{\rm with}~x_{c} = 4.8.
\eeq
Because of this cutoff on the fraction of the energy of the $e^{-}/e^{+}$ beam carried by the photon,  one needs a minimum energy $E_b \simeq 0.6 M_\Phi$ GeV to produce a $\Phi$ resonance with a mass $M_\Phi$. 

The results for the $\Phi$ production cross section in $\gamma \gamma$
collisions presented in  sections 4 and 5 were obtained using the above
mentioned choices of the laser energy and the helicities of $e^{-}, e^{+}$ and
those of the lasers $l_{1},l_{2}$, when the $J_{Z } =0$ contribution is made
dominant. The results include thus the folding of the expected helicities of the
backscattered photons with the cross section. The interference between the
signal and the background, which in most cases are the QED process $\gamma
\gamma \to b\bar b$ or $\gamma \gamma \to t\bar t$ should be taken into
account. 

Note again that the radiative corrections to the $\Phi \to \gamma \gamma$
partial decay and hence the $\Phi$ production rates are known and well under
control: the QCD corrections are small, being approximately $\alpha_s/\pi
\approx 4\%$  \cite{Djouadi:1990aj} while the electroweak corrections and of the
same order \cite{Djouadi:1997rj,Degrassi:2004mx,Actis:2008ug}. 


\subsection{DM pair production through Higgs exchange}

We now present the analytical expressions of the cross  sections for the
processes that lead to DM particle pair production in the continuum  at hadron
and lepton colliders~\cite{Djouadi:2012zc,Quevillon:2014owa,Baglio:2015wcg}.  From the  Lagrangians given in eq.~(\ref{Lag:DM}) for spin $0, \frac12, 1$ DM particles  that we denote collectively by $\chi$, one can write conveniently the Higgs couplings to the $\chi$ states depending on their nature,   scalar, fermionic or vectorial, as 
\begin{eqnarray} 
{g_{HSS}} = i \frac12 v \lambda_{HSS} \ , \
{g_{H ff }} = i \frac{1}{2\Lambda} v \lambda_{Hff} \ , \
{g_{H VV}} = - i \frac12 v \lambda_{HVV} \;. 
\end{eqnarray}
As the Higgs boson has spin--zero and no polarization, one can in principle 
factorize the $H^* \to \chi \chi$ subprocess for the various  DM particle spins 
and conveniently define the following three charges noted  $Q_{\chi}$ with
$\beta_\chi$  being the velocity in the center of mass frame $\beta_\chi =
\sqrt{1-4M_\chi^{2}/\hat s}$ \cite{Quevillon:2014owa}
\begin{eqnarray} 
{Q_{S}} =\left|g_{HSS} \right|^2 \; , \ \ 
{Q_{\chi}} =\left|g_{H\chi\chi} \right|^2 2s\beta_\chi^2 \; , \ \ 
{Q_{V}} =\left|g_{HVV} \right|^2 \left[2+\bigg(\frac{1+\beta_V^2}{1-\beta_V^2}\bigg)^2 
\right]. \label{def_Q}
\end{eqnarray}
This holds true in the channels where one needs to integrate over a 
phase--space of two final state particles, as  in vector boson fusion in the
longitudinal approximation for instance.  In the case where the DM particle is
produced in association with a vector boson, the final state contains three
particles, and we will use the equivalent $Q_{\chi}$ charges that will be
described explicitly. In fact, all the  discussion is similar to Higgs pair
production in which one picks only the diagrams in which an off--shell Higgs
particle is produced and splits into two real Higgs bosons, the different
structure when the DM state is not a scalar is taken care of by the charges
$Q_{\chi}$.

\underline{The Higgs--strahlung processes.} 

In the case of DM pair production in association with a $V=W,Z$ boson,  the relevant process is simply Drell--Yan production of an off--shell vector
boson $V^*$ which splits into a Higgs and a vector boson and the former
splits again into two $\chi$ particles, $pp \to q\bar q' \to V^* \to V^* H^*
\to V \chi \chi$.  The hadronic total  cross section reads
\beq 
\sigma (pp\to V^{\star} \to V \chi\chi ) = \sum_{q,\bar q'} \int_{(2M_\chi^2+M_V)^2/s}^{1} \rm d \tau  \frac{ {\rm d} {\cal L}^{q \bar q^{\prime}}} { {\rm d} \tau} \hat \sigma(q\bar q^{\prime} \to V \chi\chi\,  ; \; \hat s = \tau s)\, , 
\eeq 
where ${\rm d {\cal L}^{q \bar q^\prime} }/ {\rm d \tau}$ is the quark/antiquark
luminosities with $\tau=\hat s/s$ being the ratio of the partonic and total c.m.
energies. The partonic cross section is given by
\beq 
\hat \sigma(q \bar q^{\prime} \to V \chi\chi \,  ; \, \hat s = \tau s) =\int_{0}^{1} {\rm d} 
x_1 \int_{1-x_1}^{1} {\rm d} x_2 \frac{G_F^3 M_V^2 v^2}{192 \sqrt{2} \pi^3 s}
\frac{(\hat a_q^2 + \hat v_q^2)}{(1- \mu_V)^2}
 {\cal Z} {Q_{\chi}}\, , 
\eeq 
all elements have been defined before, expect for the reduced mass $\mu_V= M_V^2/\hat s$. Since there are three particles in the final state, we define the adequate $Q_\chi$  charges that can be written as \cite{Quevillon:2014owa}
\begin{eqnarray} 
\!&& {Q_{S}} =\left|g_{HSS} \right|^2\; , \ {Q_{\chi}} =\left|g_{H\chi\chi} \right|^2 2s\left[(1-x_3)+\mu_Z -4\mu_\chi \right], \nonumber \\ 
\! && {Q_{V}} =\left|g_{HVV} \right|^2 \frac{1}{\mu_V^2} \left[ 2\mu_V^2+\frac{1}{4}(1-x_3+\mu_Z-2\mu_V)^2 \right]\;.  
\label{def2_Q}
 \end{eqnarray}
where we have defined $\mu_X=M_X^2/\hat s$ and the reduced energies of the
$\chi$ particles, $x_1= 2E_{\chi}/\hat s, x_2=2E_{\chi^*}/\hat s$  with $x_3=
2E_V/ \hat s =2- x_1-x_2$  from energy--momentum conservation. Finally, we also
used the abbreviation ${\cal Z}$ defined as
\beq 
{\cal Z} =\frac{1}{4} \frac{ \mu_Z (x_3^2+8\mu_Z) }
{(1-x_3+\mu_Z-\mu_h)^2} \; .
\label{eq:App-Zfunction}
\eeq
The boundary for integrating the Dalitz density above is given in eq.~(\ref{eq:dalitzbound}). 
One should note that the higher order QCD corrections can be implemented in the
same  way as for simple Higgs--strahlung discussed above. In the $W\chi \chi$ mode, the corrections through NNLO are those that affect the Drell--Yan process and will lead to a $K$--factor of about 1.5. In the $Z\chi \chi$ mode, one should add the contributions from the box diagram $gg \to H^*Z \to \chi \chi Z$  at NNLO which increases the rate by $10\%$ at  the LHC and more at higher energies. 

At $e^+e^-$ colliders, the differential cross section for the pair production 
of the DM particles in association with a $Z$ boson,  $e^+e^-  \to Z\chi\chi$, after the angular dependence is integrated out, can be cast into the form: 
\beq 
\frac{{\rm d} \sigma (e^+ e^- \to Z\chi\chi)}{{\rm d} x_1 {\rm d} x_2} = 
\frac{G_F^3 M_Z^2 v^2}{96 \sqrt{2} \pi^3 s}
\frac{(\hat a_e^2 + \hat v_e^2)}{(1- \mu_Z)^2}\, {\cal Z} {Q_{\chi}}\, ,
\eeq 
where the electron--$Z$ couplings are defined as usual $\hat a_e=-1$ and $\hat v_e=-1+4 \sin^2\theta_W$ and all the variable have been defined above but with 
$\hat s$ replaced by the $e^+ e^-$ c.m. energy squared $s$. In particular, the 
${\cal Z}$ function is given in eq.~(\ref{eq:App-Zfunction}) and the $Q_{\chi}$ charges in eq.~(\ref{def2_Q}).

\underline{The vector boson fusion processes.}

At high energies, one expects that DM pair production in the vector boson fusion
channel to have a substantial cross section since the longitudinal vector bosons
have couplings to the Higgs which grow with energy. The cross section for the
full $qq^\prime \rightarrow V^*V^* \rightarrow \chi \chi  qq^\prime$ has a very
complicated structure as it involves four particles in the final states with two
of them being massive. We have therefore used  numerical tools to evaluate the
rate in this exact case.  We will nevertheless display the much simpler
expressions of the production cross section that one can obtain in the
longitudinal vector boson approximation  discussed earlier for Higgs production
and where one computes the cross section for the $2 \rightarrow 2$ process
$V_LV_L \rightarrow \chi \chi $ and fold it with the probabilities of emitting 
a vector boson from an energetic initial light quark. 

Denoting by $\beta_{V}$ and $\beta_\chi$ the velocities of the $V$ and $\chi$
particles  in the $VV$ center of mass frame, one obtains for the
$2 \rightarrow 2$ partonic cross section
\beq
\hat \sigma (V^{\star}_{L}V^{\star}_{L} \to  \chi \chi, \hat s)= 
\frac{G_F^2 M_V^4 v^2}{2\pi \hat{s}}  \frac{\beta_\chi}{\beta_V}
\left[ \frac{1+\beta_V^2}{1-\beta_V^2}  \frac{1} {(\hat{s} -M_h^2)} \right]^2 Q_\chi \, , 
\eeq
with the charges $Q_\chi$ simply given by eq.~(\ref{def_Q}) for the three spin
cases.  This last expression has to be folded with the longitudinal vector boson
luminosity  spectra in order to obtain the full $qq^\prime \rightarrow \chi \chi
qq^\prime$ partonic  cross section, which again has to be convoluted with the
parton densities to obtain the  full hadronic cross section
\beq 
\sigma (pp \to V^{\star}V^{\star}  \to  \chi\chi qq^{\prime}) \simeq \int_{4M_\chi^2/s}^{1}
{\rm d} \tau  \frac{ {\rm d} \cal{L}}{{\rm d} \tau} \bigg \vert_{V_L V_L/pp} 
\sigma (V^{\star}_{L}V^{\star}_{L} \to  \chi\chi, \hat s=\tau s) \, . 
\eeq 
The longitudinal vector boson luminosity was defined before in terms of  the
classical quark--quark luminosity, eq.~(\ref{V_lumi_spectra}).  The approximation
is valid only at very high energies and for small invariant $\chi\chi$ masses or
$H^*$ virtuality. At $\sqrt s=14$ TeV and for $M_\chi \! = \! {\cal O}(100\;$
GeV), one obtain a result that is of the order of a factor two from the exact
result. In our numerical analysis,  we will therefore use numerical tools in
order to obtain the exact cross section at leading order in QCD. To a good
approximation, one can borrow the QCD corrections from the single Higgs
production case (that are also included in our numerical analysis of the
process): they lead to a mere $\approx 10\%$ increase of the cross section at
NLO and should be  negligible at NNLO as seen before.

In $e^-e^-$ collisions, the dominant process for producing a pair of DM states is $e^-e^- \to W^* W^* \nu_e \bar \nu_e \to \chi \chi \nu_e \bar \nu_e $ which leads to a fully invisible final state. On has then to emit an additional particle like a photon in the final state to make it observable. This will significantly reduce the cross section. At high energies, one should resort 
to the $ZZ$ fusion process $e^-e^- \to Z^* Z^* e^+ e^- \to \chi \chi e^+ e^-$
which has a rate that is one order of magnitude lower. The cross section for the $2\to 2$ process $Z_LZ_L \to H^* \to \chi \chi$  in the longitudinal vector boson approximation is again given by
\beq
\sigma (Z^{\star}_{L}Z^{\star}_{L} \to  \chi \chi)= 
\frac{G_F^2 M_Z^4 v^2}{2\pi {s}}  \frac{\beta_\chi}{\beta_Z}
\left[ \frac{1+\beta_Z^2}{1-\beta_Z^2}  \frac{1} {(s -M_H^2)} 
\right]^2 Q_\chi \, , 
\eeq
and to obtain the cross section for the full process, one has to fold by the $ZZ$ luminosities 
\beq  
 \frac{{\rm d} {\cal L}}{{\rm d} \tau}  \bigg \vert_{Z_L Z_L/e^+e^-}  = \frac{\alpha^2 (\hat a_e^2 + \hat v_e^2)^2}{\pi^2} \frac{1}{\tau}
 \left[(1+\tau)\ln(1/\tau)-2(1-\tau) \right].
\label{Z_lumi_spectra}
\eeq

\underline{DM in the gluon fusion process}

At leading order, DM pair production via gluon fusion is mediated by triangle 
diagrams of heavy quarks in which an off--shell Higgs is emitted and splits into two $\chi$ particles, $gg\to H^* \to \chi\chi$. In fact, the discussion is similar to double Higgs production in which one picks up only the diagram where one has $H^* \to HH$ and replace the  final state by a $\chi$ but which is not only a spin--zero state but has a different spin. The partonic LO cross  section at a renormalization scale $\mu_R$ can be written as 
\begin{equation}
\hat \sigma_{\mathrm{LO}}(gg\to\chi\bar{\chi})
 = \int_{\hat t_-}^{\hat t_+} d\hat t \,
\frac{ \alpha_s^2(\mu_R)}{2048 (2\pi)^3}  \left| 
  \frac{A_{1/2}^H (\tau_{Q} ) }{\hat s-M_H^2+iM_H\Gamma_H}      \right|^2 Q_{\chi}.
\label{ggdmdmlo}
\end{equation}
The charges $Q_\chi$ are given by eqs.~(\ref{def_Q}) since we are dealing with a
two--body process at this stage. The Mandelstam variables for the parton process
are given by
\begin{eqnarray}
\hat s = Q^2 \ , \ \ \hat t = -\frac{1}{2} 
\left[Q^2-2M_\chi^2-\sqrt{\lambda(Q^2, M_\chi^2, M_\chi^2)}\cos\theta\right], 
\end{eqnarray}
where $\theta$ is the scattering angle in the partonic c.m.\ system with invariant mass $Q$, and $\lambda(x,y,z) = (x-y-z)^2 - 4 yz$. The integration limits of eq.~(\ref{ggdmdmlo}) read
\begin{equation}
\hat t_\pm = -\frac{1}{2} \left[ Q^2 - 2M_\chi^2 \mp
\sqrt{\lambda(Q^2, M_\chi^2, M_\chi^2)} \right] \;.
\end{equation}
In terms of the scattering angle $\theta$, they correspond to $\cos\theta=\pm
1$. The form factor $A_{1/2}^H$ is the usual one but is a function of the  scaling  variable $\tau_{Q} = 4 m_{Q}^2 /\hat{s}$. The total cross section for DM pair production  is obtained by integrating over the scattering angle and the gluon--gluon luminosity 
\beq     
  \sigma_{\mathrm{LO}} (pp \to gg \to \chi\bar{\chi}) = \int_{4 M_\chi^2 /s}^{1} d\tau \frac{d {\cal
      L}^{gg} }{d\tau} \hat{\sigma}(\hat s = \tau s) \;.
\eeq

As already mentioned, this processes would lead to an invisible final state as
the $\chi$ particles are electrically neutral and stable and to  make it
experimentally observable,  one needs an extra jet in the final state and hence,
the process $pp \to \chi\chi+j$ needs to be considered. As in the single Higgs
case, this is done by  emitting an additional gluon either from the internal
quark loop or from the initial gluon splitting into two $gg \to H^* g$; one has
in addition to add the contribution of the subleading $gq \to H^* q$ process and
to consider the $q\bar q \to  g^* \to H^*g$ subprocess. All this is similar to
the $H+j$ production channel discussed in the previous section of the Appendix
with the appropriate, and straightforward,  modifications. 

The total cross section for DM pair production for this process is thus given by 
\begin{equation}
\sigma_{\mathrm{LO}}(pp \rightarrow \chi \bar{\chi} + j) =
\Delta\sigma_{gg} + \Delta\sigma_{gq} + \Delta\sigma_{q\bar q} \, , 
\end{equation}
with the individual contributions given by eqs.~(\ref{eq:individual-gg}), but this time using $\tau_0 = 4M_\chi^2/s$.  The factorization scale $\mu_F$ of the parton--parton luminosities  $d{\cal L}^{ij}/d\tau$ as the renormalization scale $\mu_R$  should be set at the value $\mu_R=\mu_F= M_{\chi \chi}$. Again, in practice, one can also include the NLO QCD corrections to this topology, in
which there are contribution with two jets in the final state, $pp \to
\chi\chi+jj$, as it can be borrowed from the corresponding NNLO QCD corrections
for Higgs production which are also known \cite{Harlander:2002wh,Anastasiou:2002yz,Ravindran:2003um}. This is exactly what
has been done in our numerical analysis. \\

\newpage

%% file: sec-Appendix-DM.tex
\setcounter{section}{0}
\renewcommand{\thesection}{B}
\setcounter{equation}{0}
\renewcommand{\theequation}{B.\arabic{equation}}

\section{Appendix: DM interactions via the Higgs bosons}
\label{app:ann}

Despite of the fact that in our work, we have precisely determined the DM relic
density of the DM particles using numerical packages like {\tt micrOMEGAs} 
\cite{Belanger:2001fz,Belanger:2007zz,Belanger:2008sj} and {\tt DarkSusy}
\cite{Gondolo:2004sc,Bringmann:2018lay} (but other public and non--public
numerical tools also exist, see for instance 
Refs.~\cite{Baer:2002fv,Arbey:2009gu,Buchmueller:2011aa}) that include all
relevant effects,  reliable analytic estimates are often provided by the
so--called velocity expansion of the thermally averaged cross sections. We will
then provide in this Appendix, some analytic expression, useful for the
understanding of the results presented in the main text. In the most elaborated
models presented in this work, in particular the one based on two--doublet
extensions of the Higgs sector, the DM relic density is determined by a very
broad variety of annihilation channels. In such a case we report analytic
expression only for the channels which, according our numerical study,
contribute to a sizeable extent to the determination of the DM relic density.
Before that, we briefly summarize the general aspects of this expansion.

\subsection{The velocity expansion}

The velocity expansion can be formally derived by rewriting the thermally averaged cross section, $\langle \sigma v \rangle$ of eq.~(\ref{eq:general_sigmav}), as
\begin{equation}
    \langle \sigma v \rangle \simeq \frac{2 x^{3/2}}{\sqrt{\pi}}\int_0^{\infty} \left(\sigma v\right)_ {\rm lab} \epsilon^{1/2} \exp(-x \epsilon) d\epsilon\, ,
\end{equation}
where $x\equiv m_{\rm DM}/T$, $\epsilon \equiv \frac{s-4m_{\rm DM}^2}{4m_{\rm DM}^2}=(v_r/2)^2/(1-(v_r/2)^2)$ and finally, $(\sigma v)_{\rm lab}$ is defined as~\cite{Gondolo:1990dk}
\begin{equation}
    (\sigma v)_{\rm lab}=\frac{1}{64 \pi ^2 s}\frac{s}{s-2 m_{\rm DM}^2}\int d \Omega |M|^2 \, , 
\end{equation}
with $|M|^2$ being the amplitude squared of the annihilation process, averaged over the spins of the initial states and summed over those of the final ones. 
The velocity expansion is obtained by performing a Taylor expansion of $(\sigma v)_{\rm lab}$ with respect to the $\epsilon$ parameter, retaining only the two leading contributions, i.e. $(\sigma v)_{\rm lab} \simeq a+4 b \epsilon \simeq a +b v_r^2$. The thermal average is straightforwardly obtained by using the following integrals, 
\begin{align}
    & \frac{2x^{3/2}}{\sqrt{\pi}}\int_0^{\infty} \epsilon^{1/2} \exp(-x \epsilon)d \epsilon=1 \, , \nonumber\\
    & \frac{2x^{3/2}}{\sqrt{\pi}}\int_0^{\infty} \epsilon^{3/2} \exp(-x \epsilon)d \epsilon=\frac{3}{2x}\simeq \frac{1}{8}v_r^2 \, , 
\end{align}
and simply reads
\begin{equation}
    \langle \sigma v \rangle \simeq a+\frac{1}{2}b v_r^2 \simeq a +\frac{6}{x}b
\end{equation}
so that the DM relic density, using $x_{\rm fo}=m_{\rm DM}/T_{\rm fo}$,  can be finally written as
\begin{equation}
    \Omega h^2 \approx \frac{1.07 \times 10^9 (\mbox{GeV})^{-1} x_{\rm fo}}{g_{*}^{1/2}M_{\rm Pl} (a+3 b/x_{\rm fo})} \, . 
\end{equation}
In a few particle physics models, one has $a\!=\!b=\!0$ so that one has to
consider a further order in the velocity expansion, the $d$--wave term,  see
e.g.~Refs.~\cite{Giacchino:2013bta,Arcadi:2017jqd}. 

As is well known, the velocity expansion fails in three notable
scenarios~\cite{Griest:1990kh} (see also Ref.~\cite{DAgnolo:2017dbv} for a
somehow more exotic case): in the vicinity of $s$---channel resonance poles,  in the vicinity of the opening of some kinematical thresholds of new annihilation channels, and when the DM state is nearly mass degenerate with some other particles and the phenomenon of coannihilation occurs.

\subsection{The effective SM Higgs--portal} 
\label{App-DM-SM}

Starting with the case of the effective SM Higgs--portals,  the main DM
annihilation channels are into pairs of SM fermions, pairs of $Z/W$ vector
bosons and pairs of Higgs bosons (we neglect the pair annihilation into gluons
and photons that occur at the one--loop level).  Retaining only the leading
order terms in the velocity expansion, one for the  annihilation of the DM into
fermions and massive gauge boson final states  vi  $s$--channel $H$ boson
exchange, in the scalar, fermion and vector DM cases, 
\begin{align}
\label{eq:Sc_in_SM}
& \langle \sigma v \rangle_{ff}^S =  \sum_f N^c_f \frac{\lambda_{HSS}^2 m_f^2 }
{8 \pi m_{S }^3 v^2}\frac{ (m_{S }^2-m_f^2)^{3/2}}{(M_H^2-4 m_S^2)^2} \nonumber\\
&\langle \sigma v \rangle_{YY}^S =  \frac{g^2 \lambda_{HSS}^2 \delta_Y}
{16 \pi  m_{S }^3 v^2}\frac{  \sqrt{m_{S }^2-M_Y^2} }{ (M_H^2-4 m_{S }^2 )^2} \left(-4 m_{S }^2 M_Y^2+4 m_{S }^4+3 M_Y^4\right) \\ 
\label{eq:Fer_in_SM}
& \langle \sigma v \rangle_{ff}^\chi =  \sum_f N_c^f   \lambda^2_{H\chi\chi} \frac{(m_{f})^2 
\left(m_{\chi }^2-m_f^2\right){}^{3/2}}{4 \pi  m_{\chi } v^2  \left(M_H^2-4 m_{\chi }^2\right)^2}v_r^2 \nonumber\\
&\langle \sigma v \rangle_{YY}^\chi = g^2 \lambda^2_{H\chi\chi}  v_r^2 \delta_Y  \frac{\sqrt{m_{\chi }^2-M_Y^2} 
}{64 \pi   m_{\chi } v^2 
\left(M_H^2-4 m_{\chi }^2\right)^2} \left(-4 m_{\chi }^2 M_Y^2+4 m_{\chi }^4+3 M_Y^4\right) \\ 
\label{eq:Vec_in_SM}
&\langle \sigma v \rangle_{ff}^V=\sum_f N_c^f \lambda^2_{HVV} m_f^2 \frac{\sqrt{4-\frac{4 m_f^2}{m_V^2}} \left(4 m_V^2-4 m_f^2\right)}{96 \pi v^2  m_V^2  \left(4 m_V^2-M_H^2\right)^2}\nonumber\\
& \langle \sigma v \rangle_{YY}^V = 
g^2 \lambda^2_{HVV} \delta_Y \frac{ \sqrt{4-
\frac{4 M_Y^2}{m_V^2}} \left(16 m_V^4-16 m_V^2 M_Y^2+12 M_Y^4\right)}{768 \pi  m_V^2 v^2 \left(4
   m_V^2-M_H^2\right)^2} 
\end{align}
where $Y=W,Z$ with $\delta_Y$ is such that $\delta_W=2\delta_Z=1$ and 
$N_c^f$ is the color factor. For DM annihilation into Higgs bosons, one has instead, again in the three DM spin--cases, 
\begin{align}
 \langle \sigma v \rangle_{HH}^S & = \frac{1}{64 \pi m_S^2} \sqrt{1-\frac{M_H^2}{m_S^2}}  \left(2 \lambda^2_{H SS} -\frac{3  \lambda_{H SS}^2\, M_H^2}{v (M_H^2-4 m_S^2)}+\frac{81 \lambda_{HSS}^2 M_H^2}{{v^2\left(M_H^2-4  m_S^2\right)}^2} \right.\nonumber\\
& \left. 
+\frac{162\lambda_{HSS}^4 M_H^8}{{v^4 \left(M_H^2-2 m_S^2\right)}^2}
-\frac{4 \lambda_{HSS}^3\, M_H^2}{v \left(M_H^2-4 m_S^2\right) 
\left(M_H^2-2 m_S^2\right)} +\frac{4 \lambda_{H S S}^3 M_H^4 }{v^2 (M_H^2-2 m_S^2)}\right), \nonumber\\
 \langle \sigma v \rangle_{HH}^\chi & =\frac{v_r^2}{192 \pi m_\chi^2} 
\sqrt{1-\frac{M_H^2}{m_\chi^2}} \left(\frac{12\, 
M_H^2 \lambda_{H \chi \chi}^{3} m_\chi^3 (2 M_H^2-5 m_\chi^2)}{v (M_H^2-4 m_\chi^2) (M_H^2-2 m_\chi^2)^2}\right.\nonumber\\
& \left.
+ \frac{27 M_H^4 
\lambda_{H \chi \chi}^2 m_\chi^2}{v^2 (M_H^2-4 m_\chi^2)^2} +\frac{16 \lambda_{H \chi \chi}^4 \left(9 m_\chi^8
-8 m_\chi^6 M_H^2+2 M_H^8\right)}{(M_H^2-2 m_\chi^2)^4}\right)\nonumber
\end{align}
\begin{align}
 \langle \sigma v \rangle_{HH}^V & = \frac{1}{288 \pi  m_V^2} 
\sqrt{1-\frac{M_H^2}{m_V^2}}
\left[\frac{27 M_H^4 (\lambda_{HVV})^2}{4 \left(M_H^2-4 m_V^2\right)^2}+\frac{6 M_H^2 v^2 (\lambda_{HVV})^3}{(M_H^2 m_V^2-2
   m_V^4)} -\frac{9 M_H^2\, \lambda_{H VV}^2}{M_H^2-4 m_V^2} \right. \nonumber\\ &\left.  +4 v^4 \lambda^4_{HVV} \left(\frac{2}{\left(M_H^2-2 m_V^2\right)^2}+\frac{1}{m_V^4}\right)
 \right. \left. -\frac{2 v^2 \lambda^3_{HVV} \left(M_H^2-4 m_V^2\right)}{m_V^2 \left(M_H^2-2 m_V^2\right)}+\frac{3 \lambda^2_{H VV}}{4}\right]
\end{align}

As can be seen, the main difference between the various spin assignments consists into the velocity dependence of the annihilation cross sections of the fermionic DM while, on the contrary, the cross sections for spin--0 and spin--1 DM states are $s$--wave dominated.

\subsection{The SM Higgs sector plus new fermions}

\subsubsection{Singlet--doublet lepton model} 
\label{App-DM-NF-SD}

The relic density in the singlet--doublet lepton model is determined by DM annihilation processes into SM fermion pairs as well as $WW, ZZ,ZH$ and $HH$ final states, induced by $s$--channel Higgs exchange but also  by $Z$--boson exchange. Moreover, annihilation processes into bosonic final states can be mediated by $t$--channel exchange of the new fermions. Approximate expressions for the corresponding cross sections are given by
\begin{align}
 \langle \sigma v \rangle_{ff}& =\frac{1}{2\pi}\sum_f N_c^f \sqrt{1-\frac{m_f^2}{m_{N_1}^2}}\left[\frac{m_f^2}{m_Z^4}|g_{ZN_1 N_1}^A|^2 |g_{Zff}^A|^2\right. \nonumber\\
& \left. +\frac{2 v_r^2}{3 \pi}|g_{ZN_1 N_1}^A|^2 \left(|g_{Zff}^V|^2+|g_{Zff}^A|^2\right) {\left(1-\frac{m_f^2}{m_{N_1}^2}\right)}^{-1} \frac{m_{N_1}^2}{(4 m_{N_1}^2-m_Z^2)^2}\right.\nonumber\\
&\left.+\frac{v_r^2}{2\pi}|y_{hN _1 N_1}|^2 \frac{m_f^2}{v^2}\left(1-\frac{m_f^2}{m_{N_1}^2}\right)\frac{m_{N_1}^2}{(4 m_{N_1}^2-M_H^2)^2} \right],
\end{align}
where we have made a further simplification by taking the limit $m_f \ll m_{N_1},M_{Z}$ (a more complete expression can be derived from the ones reported e.g. in Refs.~\cite{Nihei:2002ij,Arcadi:2017kky}),
\begin{align}
 \langle \sigma v \rangle_{WW}& =\frac{1}{4 \pi}\sqrt{1-\frac{M_W^2}{m_{N_1}^2}}\frac{1}{M_W^4 (M_W^2-m_{N_1}^2-m_{E^{\pm}}^2)^2}  \big[ (|g_{WN_1}^V|^2+|g_{WN_1}^A|^2)^2 
\nonumber\\
& \times (2 M_W^4 (m_{N_1}^2-M_W^2))+2 |g_{WN_1}^V|^2 |g_{WN_1}^A|^2 m_{E^{\pm}}^2 (4 m_{N_1}^4+3 M_W^4-4 m_{N_1}^2 M_W^2))\big], 
\end{align}
\begin{align}
\langle \sigma v \rangle_{ZZ}& =\frac{1}{4 \pi}\sqrt{1-\frac{M_Z^2}{m_{N_1}^2}}\sum_{i=1,3}\frac{1}{(M_Z^2-m_{N_1}^2-m_{N_i}^2)^2} (|g_{ZN_1 N_i}^V|^2+|g_{ZN_1 N_i}^A|^2) \nonumber \\
& (|g_{ZN_1 N_j}^V|^2+|g_{ZN_1 N_j}^A|^2) (m_{N_1}^2-M_Z^2),
\end{align}

Contrary to the effective SM Higgs--portal, the annihilation cross section of
the fermionic state into gauge bosons is $s$--wave dominated as new
contributions arise from interactions mediated by $t$--channel exchange of the
new fermions. A final channel, is the annihilation into $ZH$ final states: 
\begin{align}
 \langle \sigma v \rangle_{ZH}& =\frac{1}{\pi}\sqrt{1-\frac{(M_H+M_Z)^2}{4 m_{N_1}^2}}\sqrt{1-\frac{(M_H-M_Z)^2}{4 m_{N_1}^2}} \frac{1}{256 m_{N_1}^2 M_Z^6}\lambda_{HZZ}^2 |g_{ZN_1 N_1}^A|^2\nonumber\\
& \times \big( M_H^4+(M_Z^2-4 m_{N_1}^2)^2-2 M_H^2 (M_Z^2-4 m_{N_1}^2)\big)
\end{align}

\subsubsection{Vector--like lepton DM}
\label{App-DM-NF-VLL}

The main annihilation channels for a vector--like DM state are the same as for the singlet--doublet lepton model.  The dominant contribution to the DM annihilation cross section into SM fermions is given by
\begin{align}
& \langle \sigma v \rangle_{ff} \approx \frac{m_{N_1}^2}{8 \pi} \frac{g^2 m_{N_1}^2}{(4 m_{N_1}^2-M_Z^2)^2+M_Z^2 \Gamma_Z^2} \sum N_c^f (|g_{Zff}^V|^2+|g_{Zff}^A|^2) |y_{Z N_1 N_1}^V|^2,
\end{align}
where we note that $y_{Z N_1 N_1}^V \propto (\sin^2 \theta_L^N+\sin^2 \theta_R^N)$, and  
\begin{equation}
g_{Zff}^V=\frac{g}{2 \cos \theta_W} (-2 q_f \sin\theta_W^2 +T^3_f),\,\,\,\,\,g_{Zff}^A=\frac{g}{2 \cos \theta_W} T^3_f \, . 
\end{equation}
Contrary to the singlet--doublet model, the annihilation is $s$--wave dominated, as a result of the vectorial interactions of the DM with the $Z$ boson. 

The other cross sections do not differ very much from the ones reported in the singlet--doublet model. We nevertheless reexpress them, more schematically, in terms of the parameters of the vector--like DM model. The $s$--wave terms in the annihilation cross sections into $WW$ and $ZZ$ final states are again due to the $t$--channel exchange of the fermionic partners of the DM particle. These can 
be written as
\begin{align}
 \langle \sigma v \rangle_{W^+ W^-} & \approx \frac{g^4 \tan\theta_W}{16 \pi M_W^2} ((\sin \theta_L^N)^2+(\sin \theta_R^N)^2)^2 \nonumber\\
& +\frac{g^4}{64}\left( \frac{1}{2 \pi} ((\sin \theta_L^N \sin \theta_L^E)^2+(\sin \theta_R^N \sin \theta_R^E)^2)^2\right.  \left.\frac{m_{N_1}^2}{(m_{N_1}^2+m_{E_1}^2)^2}\right. \notag \\
& \left.+\frac{2}{\pi}((\sin \theta_L^N \sin \theta_L^E)^2-(\sin \theta_R^N \sin \theta_R^E)^2)^2 \frac{m_{N_1}^4}{M_W^4}\right.
\left. \frac{m_{E_1}^2}{(m_{N_1}^2+m_{E_1}^2)^2}\right), \\
\langle \sigma v \rangle_{ZZ} & \approx \frac{g^4}{32 \pi \cos \theta_W^4 M_Z^2}
\left[ \frac{M_Z^2}{4 m_{N_1}^2} \left( \left|(\sin \theta_L^N)^2+(\sin \theta_R^N)^2\right|^4  +\left|(\sin \theta_L^N)^2-(\sin \theta_R^N)^2\right|^4\right) \right. \nonumber\\  
& +2 \left. \left|(\sin \theta_L^N)^2+(\sin \theta_R^N)^2\right|^2
    \left|(\sin \theta_L^N)^2-(\sin \theta_R^N)^2\right|^2\right],
\end{align}
for $WW$ and $ZZ$ final states respectively. In the expressions above we have assumed that the dominant contributions come from the exchange of the lightest fermions, i.e. the lightest charged vector--like lepton and the DM itself for respectively, $WW$ and $ZZ$. 

Finally, the annihilation cross section into $ZH$ final states takes  a very simple form
\begin{equation}
\langle \sigma v \rangle_{ZH} \approx \frac{g^2}{4 \pi v^2} |y_{V, Z N_1 N_1}|^2 \frac{M_Z^2}{m_{N_1}^2}.
\end{equation}

The cross section for the $HH$ final state can be derived from the one of the effective Higgs--portal. We have numerically checked that this final state provides a subdominant contribution to the total annihilation cross section of the DM. For this reason we do not report a detailed analytic expression.

\subsection{The Higgs sector extended with scalar singlets}

\subsubsection{SM Higgs mixed with a real scalar}

This type of scenario is a straightforward extension of the SM--like
Higgs--portal. The DM annihilates into pairs of SM fermions and massive gauge
bosons through $s$--channel exchange of both $h,H$ states, as well as into the
combinations of the $hh$, $hH$ and $HH$ final states through $t$--channel
exchange of the DM particle and $s$--channel exchange of the $h,H$ states
themselves. The corresponding cross sections can be hence derived from the one
of the effective Higgs--portal scenario in section B.2. We nevertheless
explicitly report below, the annihilation cross sections into SM fermions pairs
for the different spin assignments of the DM state
\begin{align}
    & \langle \sigma v \rangle_S=N_f^c \frac{(\lambda^S_\phi)^2 v_\phi^2 m_f^2}{8 \pi v^2}{\left(1-\frac{m_f^2}{m_S^2}\right)}^{3/2}\sin^2 \theta \cos^2 \theta \frac{(M_h^2-M_H^2)^2}{(M_h^2-4 m_S^2)^2 (M_H^2-4 m_S^2)^2},\nonumber\\
    & \langle \sigma v \rangle_\chi=N_f^c \frac{m_\chi^4 m_f^2}{4 \pi v_\phi^2 v^2}{\left(1-\frac{m_f^2}{m_\chi^2}\right)}^{3/2}\sin^2 \theta \cos^2 \theta \frac{(M_h^2-M_H^2)^2}{(M_h^2-4 m_\chi^2)^2 (M_H^2-4 m_\chi^2)^2}v_\chi^2, \nonumber\\
    & \langle \sigma v \rangle_V= N_f^c\frac{(\eta_V^H)^2 m_V^2 m_f^2}{12 \pi v}{\left(1-\frac{m_f^2}{m_V^2}\right)}^{3/2}\sin^2 \theta \cos^2 \theta \frac{(M_h^2-M_H^2)^2}{(M_h^2-4 m_V^2)^2 (M_H^2-4 m_V^2)^2} .
\label{Hmix-annihilation}
\end{align}
As already noted, spin--0 and 1 DM have $s$--wave dominated cross sections
into SM fermion pairs. We recall that in our setup, $v_\phi$ is not a free
parameter but can be expressed in terms of $M_H,\theta$ and $\lambda_{hH}$ as 
in eq.~(\ref{eq:hSrelation}). For the more complicated expression of the
annihilation rates into scalar final states,   we refer e.g.
to~Ref.\cite{Arcadi:2017kky}. 


\subsubsection{Scalar and pseudoscalar resonance coupled with gauge bosons}

In this setup the DM annihilates into the following combinations of final states: $gg$, $\gamma \gamma$, $ZZ$, $WW$ and $Z\gamma$ through $s$--channel exchange of the new singlet resonances. Rather compact expressions for the corresponding annihilation cross sections can be obtained from the generic Lagrangians of eq.~(\ref{eq:phi-couplings}). They read  \cite{Mambrini:2015wyu,DEramo:2016aee}
\begin{align}
 &   \langle \sigma v \rangle_{gg}^\phi \simeq \frac{g_{\phi N_1 N_1}^2 (c_{gg}^\phi)^2 m_{N_1}^4}{\pi (4 m_{N_1}^2-M_\phi^2)^2} \, d_\phi \ , \ \  \nonumber  \\
& \langle \sigma v \rangle_{WW}^\phi \simeq \frac{g_{\phi N_1 N_1}^2 (c_{WW}^\phi)^2 m_{N_1}^4}{2 \pi (4 m_{N_1}^2-M_\phi^2)^2} \sqrt{1-\frac{M_W^2}{m_{N_1}^2}} \left(1-\frac{M_W^2}{m_{N_1}^2}+ 3 \delta_\phi \frac{M_W^4}{m_{N_1}^4}\right) d_\phi \nonumber\\
& \langle \sigma v \rangle_{ZZ}^\phi \ \simeq \frac{g_{\phi N_1 N_1}^2 (c_{ZZ}^\phi)^2 m_{N_1}^4}{2 \pi (4 m_{N_1}^2-M_\phi^2)^2}\  \sqrt{1-\frac{M_Z^2}{m_{N_1}^2}} \left(1-\frac{M_Z^2}{m_{N_1}^2}+3\delta_\phi \frac{M_Z^4}{m_{N_1}^4}\right)  d_\phi  \nonumber \\
&  \langle \sigma v \rangle_{Z\gamma}^\phi \simeq \frac{g_{\phi N_1 N_1}^2 (c_{Z\gamma}^\phi)^2 m_{N_1}^4}{8 \pi (4 m_{N_1}^2-M_\phi^2)^2}{\left(1-\frac{M_Z^2}{ 4m_{N_1}^2}\right)}^3 d_\phi \nonumber\\
& \langle \sigma v \rangle_{\gamma \gamma}^\phi \simeq \frac{g_{\phi N_1 N_1}^2 (c_{\gamma \gamma}^\phi)^2 m_{N_1}^4}{8 \pi (4 m_{N_1}^2-M_\phi ^2)^2} d_\phi 
\end{align}
where the superscript $\phi=H,A$ refers to processes mediated by the new scalar or pseudoscalar resonance with $d_H =v_{N_1}^2, d_A =2$ and $\delta_H=1, \delta_A=0$. As can be seen, the most notable difference between the scalar and pseudoscalar scenarios is that the cross sections are $p$--wave dominated in the first case and $s$--wave dominated in the second one. From these general expressions, one can determine the relic density in the case of couplings for the mediators with a full family of vector--like fermions by just replacing the coefficients $c_{ii}^\Phi$ with eqs.~(\ref{eq:phi-couplings_2}).

In addition to the above channels, one has to consider the $t$--channel annihilation processes of the DM into the SM singlet mediators. A rather general expression for the annihilation into $HH$ final states can be written as
\begin{align}
&\langle \sigma v \rangle_{HH}=\frac{v_r^2}{192 \pi m_{N_1}^2} 
\sqrt{1-\frac{M_H^2}{m_{N_1}^2}}\left(\frac{8\, 
g_{HHH} (g_{H N_1 N_1})^{3} m_{N_1}^3 (2 M_H^2-5 m_{N_1}^2)}{(M_H^2-4 m_{N_1}^2)
(M_H^2-2 M_{N_1}^2)^2}\right.\nonumber\\
&\left. + \frac{3 (g_{HHH})^2 
(g_{H N_1 N_1})^2 m_{N_1}^2}{(M_H^2-4 m_{N_1}^2)^2}+\frac{16 (g_{H N_1 N_1})^4 \left(9 m_{N_1}^8
-8 m_{N_1}^6 M_H^2+2 M_H^8\right)}{(M_H^2-2 M_{N_1}^2)^4}\right).
\end{align}
In the expression above, $g_{HHH}$ represents a trilinear self coupling for the scalar mediator. In the scenario in which the DM couples only with a real scalar this coupling has been set, for simplicity, to zero. On the contrary, in the model in which the real scalar belongs to a complex field, one should have  $g_{HHH}=3 \sqrt{2} \sqrt{\lambda}_\Phi$. Concerning the coupling $g_{HN_1 N_1}$, it has been taken a free parameter in the real scalar mediator model while $g_{H N_1 N_1}=\sqrt{2\lambda_\Phi}\frac{m_{N_1}}{M_H}$ in the case of the complex mediator. 

The annihilation into $AA$ final states is given, assuming that the DM couples only with a pseudoscalar mediator, by
\begin{equation}
\langle \sigma v \rangle_{AA}=\frac{1}{12 \pi}(g_{A N_1 N_1})^4 
\frac{m_{N_1}^6}{{\left(M_A^2-2 m_{N_1}^2\right)}^4}{\left(1-\frac{M_A^2}{m_{N_1}^2}\right)}^{5/2}v_r^2.
\end{equation}
In the case in which the pseudoscalar $A$ is part of complex field $\Phi$, the latter cross section is substantially modified by the presence of an additional contribution, associated to the $s$--channel exchange of the scalar component of the complex field, 
\begin{equation}
\langle \sigma v \rangle_{aa}=\frac{1}{128 \pi m_{N_1}^2}\sqrt{1-\frac{M_a^2}{m_{N_1}^2}} \left(\frac{32}{3}\frac{g_{\Phi N_1 N_1}^4 m_{N_1}^4 {\left(M_a^2-m_{N_1}^2\right)}^2}{{\left(M_a^2
-2 m_{N_1}^2\right)}^4}+\frac{4 \lambda_\Phi g_{a N_1 N_1}^2 m_{N_1}^2 M_H^2}{{\left(M_H^2-
4 m_{N_1}^2\right)}^2}\right)v_r^2.
\end{equation}
We recall again that in the  considered setup, the coupling $g_{A N_1 N_1}$ is not a free parameter but is a function of $\lambda_\Phi$ and of the DM mass. 

In the model with a complex mediator, the DM features a last possible annihilation channel, namely into $Ha$ final states. Its cross section is $s$--wave dominated and reads
\begin{align}
 \langle \sigma v \rangle_{Ha}& =\frac{g_{\Phi N_1 N_1}^2 \sqrt{M_a^4-2 M_a^2 \left(4 m_{N_1}^2+M_H^2\right)
 +\left(M_H^2-4 m_{N_1}^2\right)^2}}{64 \pi  m_{N_1}^4} \nonumber\\
 & \times
 \bigg[ \frac{2   \lambda_\Phi  m_{N_1}^2 M_H^2}{\left(M_a^2-4 m_{N_1}^2\right)^2}
 +\frac{g_{\Phi N_1 N_1}^2
   \left(M_a^2+4 m_{N_1}^2-M_H^2\right)^2}{\left(M_a^2-4 m_{N_1}^2+M_H^2\right)^2} \nonumber \\
 & +\frac{2 \sqrt{2} 
   g_{\Phi N_1 N_1} \sqrt{\lambda_\Phi } m_{N_1}
   M_H \left(M_a^2+4 m_{N_1}^2-M_H^2\right)}{\left(M_a^2-4 m_{N_1}^2\right) \left(M_a^2-4 m_{N_1}^2+M_H^2\right)}
   \bigg]
   \end{align}

\subsection{The 2HDM coupled to fermionic DM}

This model is an extension of the singlet--doublet lepton and of the vector--like
DM models. We thus report below only the relevant contributions to the DM
annihilation cross section which differ from the ones presented in section B.3.

\subsubsection{Singlet--doublet lepton DM}

In the realizations of the 2HDM extension of the singlet--doublet model, the relevant annihilation channels for the DM particle are the ones into SM final states, as well as $AA$, $ZA$ and $hA$ final states, in the case where that the pseudoscalar is significantly lighter than the DM. The annihilation rate into SM final states are only slightly modified with respect to the case of coupling with the SM Higgs sector, with the exception of the one into SM fermions pairs, which receives an unsuppressed $s$--wave contribution due to the exchange of the pseudoscalar $A$ boson
\begin{align}
 \langle \sigma v \rangle_{ff}&=\frac{1}{2\pi}\sum_f N_c^f \sqrt{1-\frac{m_f^2}{m_{N_1}^2}} \big[ \frac{\big|g_{Aff} \big|^2 \big| y_{AN_1 N_1} \big|^2 m_f^2 m_{N_1}^2}{v^2 (4 m_{N_1}^2-M_A^2)^2} +\frac{m_f^2}{M_Z^4}|g_{ZN_1 N_1}^A|^2 |g_{Zff}^A|^2 \nonumber\\ 
&-2 \frac{m_f^2 m_{N_1}}{v \,M_Z^2 (4 m_{N_1}^2 -M_A^2)}
\mbox{Re}\left( g_{Aff}y_{AN_1 N_1}^{*}g_{ZN_1 N_1}^A g_{Zff}^A\right) \bigg].
\end{align}

The leading contributions to the velocity expansions of the $hA$, $AA$ and $ZA$ channels can be written as
\begin{align} 
\langle \sigma v \rangle_{ZA} & =\frac{v_{r}^2}{16 \pi M_Z^2}\sqrt{1-\frac{(M_A-M_Z)^2}{4 m_{N_1}^2}}\sqrt{1-\frac{(M_A+M_Z)^2}{4 m_{N_1}^2}}\bigg( 16 m_{N_1}^4-8 m_{N_1}^2 \nonumber\\ 
&(M_Z^2+M_A^2)+(M_Z^2-M_A^2)^2 \bigg)
\times {\left[\frac{\lambda_{hAZ}y_{hN_1 N_1}}{(4
m_{N_1}^2-m_h^2)}+\frac{\lambda_{HAZ}y_{HN_1 N_1}}{(4
m_{N_1}^2-M_H^2)}\right]}^2,
\end{align} 
\begin{align} 
& \langle \sigma v
\rangle_{hA}=\frac{1}{16\pi}\sqrt{1-\frac{(M_h+M_A)^2}{4
m_{N_1}^2}}\sqrt{1-\frac{(M_h-M_A)^2}{4 m_{N_1}^2}}
\left[\frac{\lambda_{hAA}^2 y_{AN_1 N_1}^2}{(4
m_{N_1}^2-M_A^2)^2}+\frac{1}{4}\frac{\lambda_{hAZ}^2 g_{ZN_1 N_1}^2
}{(4 m_{N_1}^2-m_Z^2)^2}\right.\nonumber\\ &
\left. \times (M_A^2-M_h^2)^2) \sum_{i,j=1,3} \frac{y_{AN_1 N_i}y_{AN_1 N_j}^{*}y_{hN_1
N_i}y_{hN_1 N_j}^{*}}{m_{N_1}^2 (M_A^2+M_h^2-2
m_{N_1}^2-m_{N_i}^2)^2 (M_A^2+M_h^2-2 m_{N_1}^2-m_{N_j}^2)^2
}\right.\nonumber\\ &\left. \times\left(M_A^4+M_h^4-8
m_{N_1}m_{N_j}M_h^2+16 m_{N_i}m_{N_j}m_{N_1}^2 -2 M_A^2 (M_h^2-4
m_{N_1}m_{N_j}) \right)\right.\nonumber\\ &\left. \times
\mbox{Re}\left[\lambda_{hAA}^{*}y_{AN_1 N_1}^{*}y_{hN_1
N_1}^{*}\lambda_{hAZ}g_{ZN_1 N_1}^A\right]\frac{(M_A^2-M_h^2)}{M_Z^2
m_{N_1}}\right.\nonumber\\
&\left.+\frac{2}{m_{N_1}^2}\mbox{Re}\left[\lambda_{hAA}^{*}y_{AN _1
N_1}^{*}y_{hN_1 N_1}^{*}y_{hN _1 N_i}y_{AN _1
N_i}\right]\frac{(M_A^2 m_{N_1}-M_h^2 m_{N_1}+4
m_{N_i}m_{N_1}^2)}{(M_A^2+M_h^2-2 m_{N_1}^2-2 m_{N_i}^2)(4
m_{N_i}^2-M_A^2)}\right.\nonumber\\ &\left.
+\frac{1}{2}\sum_{i=1,3}\mbox{Re}\left[\lambda_{hAZ}^{*}g_{ZN_1
N_1}^{*}y_{hN_1 N_i}y_{AN_1 N_i}\right] \frac{(M_A^2-M_h^2)^2+4
m_{N_1}m_{N_i} (M_A^2-M_h^2)}{m_{N_1}^2 M_Z^2(M_A^2+M_h^2-2
m_{N_1}^2-2 m_{N_i}^2)}\right] 
\end{align}
\begin{align} 
\langle \sigma v\rangle_{AA}  & =\frac{v_N^2}{128\pi}\sqrt{1-\frac{M_A^2}{m_{N}^2}}
\left[{\left( \frac{\lambda_{AAh}y_{hN_1
N_1}}{(4 m_{N_1}^2-M_h^2)}+\frac{\lambda_{AAH}y_{HN_1 N_1}}{(
m_{N_1}^2-M_H^2)}\right)}^2 +\frac{8}{3}|y_{AN_1
N_1}|^2  m_{N_1} \right.\nonumber\\ &
\left.   \bigg(2\frac{m_{N_1} (m_{N_1}^2-M_A^2)^2}{(2
m_{N_1}^2-M_A^2)^4} -\frac{(m_{N_1}^2-M_A^2)}{(2
m_{N_1}^2-M_A^2)^2} \bigg) \left(\frac{y_{hN_1 N_1}\lambda_{hAA}}{(4
m_{N_1}^2-M_h^2)}+\frac{y_{HN_1 N_1}\lambda_{HAA}}{(4
m_{N_1}^2-M_H^2)}\right)\right] ,
\end{align} 
where the trilinear couplings between the CP--even and CP--odd Higgs states are given by
\begin{align}
\lambda_{hAA} & =-\frac{1}{4v\sin 2\beta}\left \{\left[\cos(\alpha-3\beta)+3\cos(\alpha+\beta)\right]m_h^2\right.\nonumber\\
&\left. -4 \sin 2 \beta\sin(\alpha-\beta)m_A^2-4 \cos(\alpha+\beta)M^2 \right \}\nonumber\\
 \lambda_{HAA} & =-\frac{1}{4v\sin 2\beta}\left \{\left[\sin(\alpha-3\beta)+3\sin(\alpha+\beta)\right]m_H^2\right.\nonumber\\
&\left.+4 \sin 2 \beta\cos(\alpha-\beta)m_A^2-4 \sin(\alpha+\beta)M^2 \right \}
\end{align}
As can be seen, the cross sections have been expressed in terms of generic couplings. As a consequence, they can be straightforwardly adapted to the 2HDM+$a$ case as well as to the NMSSM. For this reason, we will not report analytic approximations for these two models.

\subsubsection{Vector--like DM particles}

Similarly to the previous case, we will simply illustrate the most relevant annihilation channels responsible for the DM relic density. First discussing the annihilation of the Dirac DM fermion into SM fermions pairs, the $s$--wave term of the cross section is determined by the couplings of the DM with the pseudoscalar $A$ boson as well as the vectorial coupling of the DM with the $Z$ boson. The cross section can be then written as
\begin{align}
 \langle \sigma v \rangle_{ff} & =N_c^f \sqrt{1-\frac{m_f^2}{m_{N_1}^2}}\left \{\frac{m_{N_1}^2}{8 \pi} \frac{m_f^2}{v^2}|\xi_{A}^f|^2 \frac{1}{(4 m_{N_1}^2-M_A^2)^2+M_A^2 \Gamma_A^2}|y_{A N_1 N_1}|^2 \right.\nonumber\\
 & \left. + \frac{g^2 m_{N_1}^2}{\pi [(4 m_{N_1}^2-M_Z^2)^2+M_Z^2 \Gamma_Z^2]}
 \left[\sum N_c^f (|g_{Zff}^V|^2+|g_{Zff}^A|^2) |y_{V,Z N_1 N_1}|^2\right. \right.\nonumber\\
 &\left. \left. +\frac{3 m_t^2}{2 m_{N_1}^2}(|g_{Ztt}^V|^2+|g_{Ztt}^A|^2)|y_{A,Z N_1 N_1}|^2\right] \right.\nonumber\\
 & \left. -2 \frac{m_f^2 m_{N_1}}{v \,M_Z^2 (4 m_{N_1}^2 -M_A^2)}
\mbox{Re}\left(\xi_A^f y_{AN_1 N_1}^{*}g_{ZN_1 N_1}^A g_{Zff}^A\right) \right \}
\end{align} 
The annihilation channels into $WW$ and $ZZ$ are similarly important for the DM relic density. The dominant contributions in the velocity expansion to the annihilation cross section are basically the same as the minimal singlet--doublet model and, consequently, they will not be rewritten here.

As pointed in the main text, constraints from DM direct detection can be relaxed when the DM is heavier than $(M_{H^{\pm}}+M_W)/2$ and /or $M_{H^{\pm}}$ so that the annihilation channels into, respectively, $W^{\pm} H^{\mp}$ and $H^+ H^-$ are kinematically accessible. This is due to the fact that these cross section depend on the couplings of the heavy Higgses with the charged vector leptons, which are not constrained by direct detection. For illustration we provide a simple estimate of the cross section of the $H^+ H^-$: 

\begin{align}
     \langle \sigma v \rangle_{H^+ H^-}& =\left(-\sin \theta^E_R \cos\theta_L^N y_A^{E_L}-\sin \theta_L^E \cos \theta_R^N y_A^{E_R} +\cos \theta_L^E \sin \theta_R^N y_A^{N_L} +\cos \theta_R^E \sin \theta_L^N y_A^{N_R} \right)^2 \nonumber\\ 
    & \times \left(\sin \theta^N_R \cos\theta_L^E y_A^{E_L}+\sin \theta_L^N \cos \theta_R^E y_A^{E_R} +\cos \theta_N^E \sin \theta_R^E y_A^{N_L} +\cos \theta_R^N \sin \theta_L^E y_A^{N_R} \right)^2\nonumber\\
    & \times \frac{m_{N_1}^2}{4 \pi}{\left(1-\frac{m_{H^\pm}^2}{m_{N_1}^2}\right)}^{3/2}\frac{1}{(m_{E_1}^2+m_{N_1}^2-m_{H^\pm}^2)^2}
\end{align}


\subsubsection{The inert doublet model}

Finally, we report the expressions of some relevant annihilation channels of the DM in the inert doublet model, namely into $\bar f f$, $hh$ and $WW$ final states. As pointed out in the main text, in large portions of the parameter space, and especially at high DM masses, coannihilations are unavoidable, making the velocity expansion is not entirely reliable.

Analytic approximations for the annihilation cross sections into $\bar f f$ and $hh$ final states are given by
\begin{equation}
    \langle \sigma v \rangle_{ff}=\frac{N_c^f \lambda_L^2 m_f^2}{\pi   (M_{H>}^2-M_h^2)}{\left(1-\frac{m_f^2}{M_{H}^2}\right)}^{3/2}
\end{equation}
and
\begin{equation}
    \langle \sigma v \rangle_{hh}=\frac{\lambda_L^2}{4 \pi M_{H}^2}\sqrt{1-\frac{M_h^2}{M_{H}^2}}\frac{{\left(M_h^4-4 M_{H}^4-2 M_h^2 v^2 \lambda_L+8 M_{H}^2 v^2 \lambda_L\right)}^2}{{\left(M_h^4-6 M_h^2 M_{H}^2+8 M_{H}^4\right)}^2}
\end{equation}

The expression for the $WW$ final state is, in general, rather lengthy and complicated. We will then provide it into two simplified limits, i.e. $M_{H^0} \sim M_{H^\pm}$:
\begin{align}
& \langle \sigma v \rangle_{WW}=\frac{g^4\sqrt{M_{H}^2-M_W^2}}{128 \pi  M_{H}^3
   M_W^4 \big(M_h^2-4 M_{H}^2\big)^2 \big(M_W^2-2 M_{H}^2\big)^2} \\
& \big(M_h^4 \big(4 M_{H}^8-8 M_{H}^6 M_W^2+16 M_{H}^4 M_W^4-12 M_{H}^2 M_W^6+3 M_W^8\big) -4 M_h^2   \big(8 M_{H}^{10} \nonumber \\
& +M_{H}^8 \big(8 \lambda_L v^2-16 M_W^2\big)  
+16 M_{H}^6 \big(2 M_W^4-\lambda_L M_W^2 v^2\big)+M_{H}^4 \big(22
   \lambda_L M_W^4 v^2-24 M_W^6\big) \nonumber\\
  &   +2 M_{H}^2 \big(3 M_W^8-7 \lambda_L M_W^6 v^2\big)+3 \lambda_L M_W^8 v^2\big) 
  +4 \big(16
   M_{H}^{12}-32 M_{H}^{10} \big(M_W^2-\lambda_L v^2\big) \nonumber \\
&   +16 M_{H}^8 \big(\lambda_L v^2-2 M_W^2\big)^2 
 s-8 M_{H}^6 \big(6 M_W^6-11
   \lambda_L M_W^4 v^2+4 \lambda_L^2 M_W^2 v^4\big)  \nonumber\\
&   +4 M_{H}^4 \big(3 M_W^8-14 \lambda_L M_W^6 v^2+8 \lambda_L^2 M_W^4
   v^4\big)+4 \lambda_L M_{H}^2 M_W^6 v^2 \big(3 M_W^2-4 \lambda_L v^2\big)+3 \lambda_L^2 M_W^8 v^4\big)\big) \nonumber
\end{align}
and $M_{H} \ll M_{H^\pm}$:
\begin{equation}
    \langle \sigma v \rangle \frac{g^4}{128 \pi M_{H}^2}\sqrt{1-\frac{M_W^2}{M_{H}^2}}\left(3+\frac{4 M_{H}^2 (M_{H}^2-M_W^2)}{M_W^4}\right)\frac{(4 M_{H}^0-M_h^2)+2 \lambda_L v^2)^2}{(4 M_{H}^2-M_h^2)^2}
\end{equation}
In this last case we notice that the cross section becomes suppressed if:
\begin{equation}
    \lambda_L \approx -2 \left(M_{H}^2-(M_h/2)^2\right)/v^2
\end{equation}
For $M_{H}> \frac12 M_h$, this condition is met for negative values of the coupling $\lambda_L$ and explains the viable relic density region for $M_W \lesssim M_{H} \lesssim 100\,\,\mbox{GeV}$.

We can derive an expression for the annihilation cross section into $ZZ$ final states by just replacing in the expression above, $g \rightarrow g'$, $M_W \rightarrow M_Z$ and $M_{H^\pm} \rightarrow M_A$.


%% file: sec-Appendix-RGE.tex
\section{\mbox{Appendix: Evolution of 2HDM quartic} couplings}

For completeness, we display here the renormalization group equations of the
five quartic couplings $\lambda_{i=1,5}$ of the 2HDM with and without the
contribution of a full family of vector--like leptons and quarks with the 
Lagrangian given in section 5.2.2 and with Yukawa couplings which can have a
very important impact, and that we have conveniently expressed in the
$(\Phi_1,\Phi_2)$ basis. These equations should be solved in combination with
those of the new Yukawas and the one of the top quark and the gauge couplings.

The renormalisation group equations for the five quartic couplings read
\begin{eqnarray}
\label{eq:lambda1RGE}
8 \pi^2  \beta_{\lambda_1} &\! =\!&\  \big[\lambda_1 \big(\sum_L |y_1^L|^2+ 3 \sum_Q |y_1^Q|^2\big) -\sum_L |y_1^L|^4
-3 \sum_Q |y_1^Q|^4 \big]  
 \nonumber\\ &&
+ 12 \lambda_1^2+4 \lambda_3^2+4 \lambda_3 \lambda_4 +2 \lambda_4^2+2 |\lambda_5|^2 
 \nonumber\\ &&
+\frac{3}{4}(3g^4+g^{'\,4}+2 g^2 g^{'\,2})-3 \lambda_1 (3 g^2+g^{'\,2}-4 y_t^2)-12 y_t^4 , \\
8\pi^2 \beta_{\lambda_2} & \!=\!& \big[\lambda_2 \big( \sum_L |y_2^L|^2+ 3 \sum_Q |y_1^Q|^2\big) - \sum_L |y_2^L|^4  -3 \sum_Q |y_1^Q|^4\big] 
 \nonumber\\ &&
+ 12 \lambda_2^2+4 \lambda_3^2+4 \lambda_3 \lambda_4 +2 \lambda_4^2
 \nonumber\\ && +2 |\lambda_5|^2 + \frac{3}{4}(3g^4+g^{'\,4}+2 g^2 g^{'\,2})-3 \lambda_2 (3 g^2+g^{'\,2}) , \\ 
16\pi^2 \beta_{\lambda_3} &\! =\!&  \lambda_3 \big(\sum_L (|y_1^L|^2\!+\!|y_2^L|^2)+3 \sum_Q (|y_1^Q|^2\!+\!|y_2^Q|^2)\big)
\nonumber\\ && 
(\lambda_1+\lambda_2) (6 \lambda_3+2 \lambda_4)+4 \lambda_3^2+2 \lambda_4^2+2 |\lambda_5|^2 \nonumber\\ && 
+\frac{3}{4}(3g^4+g^{'\,4}-2 g^2 g^{'\,2})
\! \!-\! 3 \lambda_3 (3 g^2\!+\!g^{'\,2}\!-\!2 y_t^2) \nonumber\\ 
&&  \!-\!2 y_1^{E_L} y_2^{E_L} y_1^{N_L}y_2^{N_L}+(|y_1^{N_L}|^2 \!+\! |y_1^{E_L}|^2)  (|y_2^{N_L}|^2+|y_2^{E_L}|^2) \nonumber \\ 
&& 
-2 y_1^{E_R} y_2^{E_R} y_1^{N_R}y_2^{N_R}+(|y_1^{N_R}|^2+|y_1^{E_R}|^2)(|y_2^{N_R}|^2+|y_2^{E_R}|^2) \nonumber\\
&& \ + 3 \big( \!-\!2 y_1^{B_L} y_2^{B_L} y_1^{T_L}y_2^{T_L}+(|y_1^{T_L}|^2 \!+\! |y_1^{B_L}|^2) (|y_2^{T_L}|^2+|y_2^{B_L}|^2) 
\nonumber \\ && 
-2 y_1^{B_R} y_2^{B_R} y_1^{T_R}y_2^{T_R}+(|y_1^{T_R}|^2+|y_1^{B_R}|^2)(|y_2^{T_R}|^2+|y_2^{B_R}|^2)\big) , \\
16 \pi^2 \beta_{\lambda_4}&\!=\!&
\lambda_4 \big( \sum_L (|y_1^L|^2+|y_2^L|^2)+3 \sum_Q (|y_1^Q|^2+|y_2^Q|^2) \big) \nonumber\\ && 
+2 (\lambda_1+\lambda_2)\lambda_4+8 \lambda_3 \lambda_4+4 \lambda_4^2+8 |\lambda_5|^2 \nonumber\\ && 
+3 g^2 g^{'\,2}-3 \lambda_4 (3 g^2+g^{'\,2}-2 y_t^2) \nonumber \\
&&  -2 y_1^{B_L} y_2^{B_L} y_1^{T_L}y_2^{T_L} +2 y_1^{B_R} y_2^{B_R} y_1^{T_R}y_2^{T_R}  \nonumber\\
&& +  (|y_1^{T_L}|^2-|y_1^{B_L}|^2)(|y_2^{T_L}|^2-|y_2^{B_L}|^2)
\nonumber\\ && 
+(|y_1^{T_R}|^2-|y_1^{B_R}|^2)(|y_2^{T_R}|^2-|y_2^{B_R}|^2) \big], \\
16\pi^2 \beta_{\lambda_5} &\!=\!&  \lambda_4 \big( \sum_L (|y_1^L|^2+|y_2^L|^2)+3 \sum_Q (|y_1^Q|^2+|y_2^Q|^2)\big) \nonumber  \\
& & -2 \sum_L |y_1^L|^2 |y_2^L|^2-6\sum_Q |y_1^Q|^2 |y_2^Q|^2\, \nonumber \\  
&&(2 \lambda_1+2 \lambda_2+8 \lambda_3+12 \lambda_4)\lambda_5  -3 \lambda_5 (3 g^2+g^{'\,2}-2 y_t^2) \, . 
\end{eqnarray}